\documentclass[a4paper]{article}

\usepackage{enumerate}
\usepackage{hyperref}
\usepackage{url}
\usepackage[nottoc, notlof, notlot]{tocbibind}
\usepackage{placeins} % pour \FloatBarrier

\usepackage{boxedminipage}
\usepackage{listings}
\usepackage{subfigure}

\usepackage{amsmath}

\usepackage{graphicx}

\usepackage{relsize}
\usepackage{xspace}

\usepackage{vmargin}
\setmarginsrb{95pt}{80pt}{95pt}{95pt}{10pt}{20pt}{100pt}{26pt}
\usepackage{boxedminipage}%%added by Akram
\usepackage{listings}%%added by Akram
\usepackage{subfigure}%%%added by Akram
\usepackage{graphicx}%%%added by Akram
\usepackage{url}%%%added by Akram
\usepackage{amsmath}%%%added by Akram
\usepackage{relsize}%%%added by Akram
\usepackage{xspace}%%%added by Akram
\usepackage{hyperref}%%%added by Akram
\usepackage{url}%%%added by Akram

\usepackage{amsfonts}%%%%added by akram

\title{Refactoring Composite to Visitor and\\ Inverse Transformation  in Java\footnote{This is the version 3 of the report. The main difference with previous versions  is the description of the computation of the minimum precondition for the base round-trip transformation,  for variations and for the use case (see appendices).}}
\author{Akram Ajouli$^1$
     \& Julien Cohen$^2$}

\date{}

\newcommand{\tab}{\phantom{--}}

\newcommand{\operation}[1]{\textbf{#1}}

\newcommand{\mparam}{\mathbb{M_P}}
\newcommand{\mwithoutparam}{\mathbb{M_W}}

\newcommand{\auxsymbol}{\mathit{aux}}

\newcommand{\intellij}{IntelliJ \textsc{Idea}\xspace}

\newcommand{\xcode}[1]{{\relsize{-1}\textsf{#1}}}

\begin{document}

\maketitle

{\small
\noindent
1: INRIA -- ASCOLA team (EMN - INRIA - LINA)\\
2: Universit\'e de Nantes -- LINA (UMR 6241, CNRS, Univ. Nantes, EMN)\\
}

We describe how to use refactoring tools to transform a Java
program conforming to the Composite design pattern into a
program conforming to the Visitor
design pattern with the same external behavior. 
We also describe the inverse transformation.
We use the refactoring tools provided by \intellij and Eclipse.
%%%%%%%%%%%%%%%%%%%%%%%%%%%%%%%%%%%%%%%%%%%%%%%%%%%%%
\setcounter{tocdepth}{2}
\tableofcontents

\section{Introduction}%%%%%%%%%%%%%%%%%%%%%%%%%%%%%%%%%%%%%%%%%%%%%%%%%%%
Composite and Visitor patterns have dual properties with respect to
modularity:
while the Composite pattern (as well as the Interpreter pattern and classic
class hierarchies) provides modularity along subtypes and leaves
operation definitions crosscut, the Visitor pattern provides
modularity along operations and leaves behavior definitions
crosscutting with respect to subtypes~\cite{Gamma:1995}.

One solution to have modularity along operations \emph{and} subtypes
would be to be able to transform automatically a program conforming
to the Composite pattern into a program with the same behavior, but
which structure would conform to the Visitor pattern, and
vice-versa~\cite{Cohen-Douence-Ajouli:2012}.

Chains of elementary refactorings can be used to make design patterns
appear~\cite{O'Cinneide:1999,Kerievsky04}, for instance to introduce
the Visitor pattern~\cite{surveyRefactoring2004, Kerievsky04},
or to replace the Visitor pattern by the Interpreter
pattern~\cite{Hills:2011}.
However, such transformations are not fully automated yet,
and our proposal of navigation between several architectures
for a same program~\cite{Cohen-Douence-Ajouli:2012} is not currently workable.

In this report we do preliminary work before automating refactoring
based Composite$\leftrightarrow$Visitor transformations:

 \begin{enumerate}

  \item We give chains of refactoring operations that provide
  Composite$\rightarrow$Visitor and Visitor$\rightarrow$Composite
  transformations for a simple Java program. Each refactoring
  operation is supported by at least one refactoring tool.

  \item We explain how to use the refactoring tools \intellij and
  Eclipse to perform the needed refactoring operations (composition of
  several operations of the tools, specific options, applying some
  operations before being able to perform another one, bugs to
  overcome, missing operations...).

  \item We study variants of the transformations for several
    variations in the implementation of the patterns.

\end{enumerate}

Our algorithms are validated on a running toy example and on the JHotDraw
program~\cite{jHotDraw}.

\section{General Approach}
\label{sec-classic-algos}
We consider the Java program of Fig.~\ref{classic-base-prog:data}. It contains a classic class hierarchy: the abstract class \emph{Graphic} has two subclasses, \emph{Square} and \emph{Ellipse}, and two methods, \emph{print} and \emph{prettyprint} implemented in the subclasses. 
We also consider that two classes \emph{Printer} and \emph{PrettyPrinter} already exist in the program: they will become visitor subclasses.

\newcommand{\visitorclass}[1]{\ensuremath{V(\mbox{#1})}}
\newcommand{\auxname}[1]{$aux($#1$)$}

\begin{center}
\includegraphics[width=6cm]{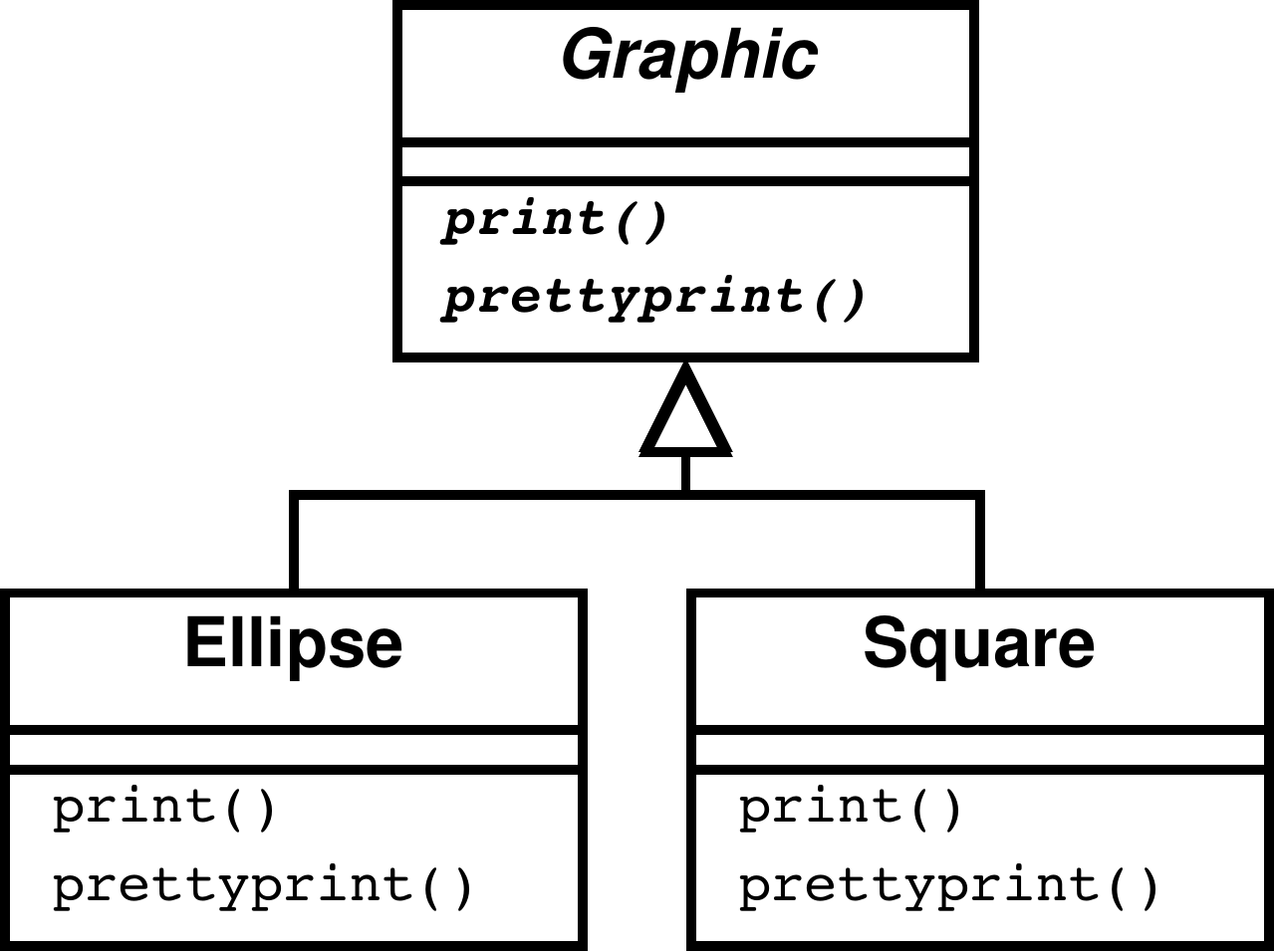}
\end{center}

\begin{figure}[!htp]
\begin{center}
  \begin{minipage}{13cm}
\begin{lstlisting}
abstract class Graphic {
	
    abstract public void print();

    abstract public void prettyprint();

}
\end{lstlisting}
\begin{lstlisting}
class Square extends Graphic {

    int l;
    
    public void print() {
    	System.out.print("Square(" + l + ")");
    }
    
    public void prettyprint(){
    	System.out.print("Square.");
    }
}
\end{lstlisting}
\begin{lstlisting}
class Ellipse extends Graphic{

    int l1, l2;

    public void print() {
        System.out.print("Ellipse: (" + l1 +"," + l2 + ")");
    }
    
    public void prettyprint(){
    	System.out.print("Ellipse.");
    }
}
\end{lstlisting}
\end{minipage}
\end{center}

\caption{Base Program (classic class hierarchy)}
\label{classic-base-prog:data}
\end{figure}

\newcommand{\LM}{\textsf{LM}\xspace}
\newcommand{\LC}{\textsf{LC}\xspace}
\newcommand{\LV}{\textsf{LV}\xspace}

In the following algorithms, we make abstraction of the class and method names and number: let \LM be the set
of traversal functions, \LC the set of \emph{concrete} classes in
the composite structure, and \textsf{S} the superclass of
the composite structure.

 Here, \LM$=\{$\textsf{print},\textsf{prettyprint}$\}$, \LC$=\{$\textsf{Ellipse},\textsf{Square}$\}$ and \textsf{S}=\textsf{Graphic}.

We also define a function $V$ that maps a name of visitor class to a name of method. We consider here \visitorclass{print} $=$ \emph{Printer} and \visitorclass{prettyprint} $=$ \emph{PrettyPrinter}.
 We also define \LV = \visitorclass{\LM} = $\{ \visitorclass{$m$}\}_{m\in \LM}$ .

\subsection{Guidelines in the Literature}%%%%%%%%%%%%%%%%%%%%%%%%%%%%%%%%%%%%%%

\begin{figure}
\begin{center}
\begin{boxedminipage}{12cm}%%%%%%%%%%%%%%%%%%%%%%%%%%%%%%%%%%%%%%%%%%%%%%%
\begin{enumerate}

\item \label{algo-mens:step-move}
{\sf ForAll m in \LM, c in \LC do\\
 \tab Let visitorname = \visitorclass{m} in\\
% \tab \tab CreateEmptyClass(visitorname)\\
\tab \tab  MoveMethodWithDelegate(c, m, visitorname)\\ % c'est la version standard de Fowler
\tab \tab  RenameMethod(visitorname, m, "visit") \\
done
}

\item \label{algo-mens:step-introduce-visitor} {\sf AddAbstractSuperClass("Visitor", \LV)}

\item \label{algo-mens:step-introduce-abstract-visit} {\sf 
  ForAll c in \LC do\\
\tab  PullUpAbstract(\LV, "visit", c, "Visitor")}

\item \label{algo-mens:step-introduce-accept} { \sf ForAll c in \LC do\\
\tab ExtractMethod(c, \LM, "accept")}

\item \label{algo-mens:step-pull-up-methods} { \sf ForAll m in \LM do\\
\tab PullUpConcrete(\LC, m, S)}

\end{enumerate}

\end{boxedminipage}
\end{center}
\caption{ Simple Class Hierarchy $\rightarrow$ Visitor transformation~\cite{surveyRefactoring2004}.}
\label{algo-mens}

\end{figure}%%%%%%%%%%%%%%%%%%%%%%%%%%%%%%%%%%%%%%%%%%%%%%%%%%%%%%%

We start by considering some guidelines given in the literature for
introducing an instance of the Visitor pattern into a typical
object-oriented class hierarchy.
We consider the guidelines of Mens and
Tourw\'e~\cite{surveyRefactoring2004},  rephrased in
Fig.~\ref{algo-mens}.

To introduce a visitor pattern, the first obvious step is to move the
business code\footnote{We call business code the code that defines the
  operations, here the bodies of \emph{print} and \emph{prettyprint}, which are
  spread over several classes (by the means of overriding).}
from the class hierarchy to visitor classes  (in this section, we consider the target classes for the moved methods already exist in the project). This is done in
step~\ref{algo-mens:step-move} (Fig.~\ref{algo-mens}).
We move the business code but,  in order not to
change the interface of the class hierarchy,
 we keep in the class hierarchy some methods with the same profiles as the original ones, which will be delegators to visitor's methods (see \emph{Move
Method} in Fowler~\cite{Fowler1999}).

The new methods in visitor classes are named \emph{visit} so
that the visitor classes will all be able to implement the abstract
class \emph{Visitor}, which is added afterward
(step~\ref{algo-mens:step-introduce-visitor}).
In visitor classes, there is one method \emph{visit} for each concrete class of
the class hierarchy \LC (with overloading). 
They are introduced as abstract methods in the \emph{Visitor} class 
(step~\ref{algo-mens:step-introduce-abstract-visit}).

To introduce the double dispatch which is typical of
the visitor pattern without changing the interface of the
class hierarchy, another delegation is introduced inside the
concrete classes of \LC
(step~\ref{algo-mens:step-introduce-accept}). The delegate
method is named \emph{accept}.

Since the initial methods are now delegators
to \emph{accept}, the overriding bodies are the same in the concrete
classes of \LC, and it can be defined once
for all in the super class
(step~\ref{algo-mens:step-pull-up-methods}).

The refactoring results in the program given in Figs.~\ref{classic-visitor-prog:data} and~\ref{classic-visitor-prog:visitor}.

\begin{figure}[!htp]
\begin{center}
  \begin{minipage}{13cm}
\begin{lstlisting}
abstract class Graphic {

    public void print() {
        accept(new PrintVisitor());
    }

    public void prettyprint() {
        accept(new PrettyPrintVisitor());
    }

    public abstract void accept(Visitor v);

}
\end{lstlisting}
\begin{lstlisting}
class Square extends Graphic {

    int l;

    public void accept(Visitor v) {
        v.visit(this);
    }

}
\end{lstlisting}
\begin{lstlisting}
class Ellipse extends Graphic{

    int l1, l2;

    public void accept(Visitor v) {
        v.visit(this);
    }

}
\end{lstlisting}
\end{minipage}
\end{center}

\caption{Program  with Visitor (classic class hierarchy)}
\label{classic-visitor-prog:data}
\end{figure}

\begin{figure}[!htp]
\begin{center}
  \begin{minipage}{13cm}
\begin{lstlisting}
public abstract class Visitor {
  public abstract void visit(Square square);

  public abstract void visit(Ellipse ellipse);
}
\end{lstlisting}
\begin{lstlisting}
public class PrintVisitor extends Visitor {
  public void visit(Square square) {
    System.out.print("Square(" + square.l + ")");
  }

  public void visit(Ellipse ellipse) {
    System.out.print("Ellipse: (" + ellipse.l1 +"," + ellipse.l2 + ")");
  }
}
\end{lstlisting}
\begin{lstlisting}
public class PrettyPrintVisitor extends Visitor {
  public void visit(Square s){
    System.out.print("Square.");
  }

  public void visit(Ellipse e){
    System.out.print("Ellipse.");
  }
}
\end{lstlisting}
\end{minipage}
\end{center}

\caption{Program with Visitor (classic class hierarchy -- visitor part)}
\label{classic-visitor-prog:visitor}
\end{figure}

\subsection{Automation}%%%%%%%%%%%%%%%%%%%%%%%%%%%%%%%%%%%

%%%%%%%%%%%%%%%%%%%%%%%%%%%%%%%%%%%%%%%%%%%%%%%%%%%%%%%%%%%%%%%%%%%%%
\begin{figure}%%%%%%%%%%%%%%%%%%%%%%%%%%%%%%%%%%%%%%%%%%%%%%%%%%%%%%%
\begin{center}
\begin{boxedminipage}{12cm}
\begin{enumerate}

\item \label{algo-mens-corrige:step-move} {\sf ForAll (m,param) in \LM, c in \LC do\\
 \tab Let visitorname = \visitorclass{m} in\\
% \tab \tab CreateEmptyClass(visitorname)\\
\tab\tab   AddParameterWithDelegate(c,m,param,visitorname)\\
\tab \tab  MoveMethod(c, m, param+visitorname, visitorname)\\ 
\tab \tab  RenameMethod(visitorname, m,param+c, "visit") \\
done
}

%idem nous
\item  \label{algo-mens-corrige:step-extract-superclass} {\sf  ExtractSuperClass(\LV, "Visitor") // with \emph{visit} abstract methods}

\item { \sf ForAll c in \LC do\\
\tab ExtractGeneralMethod(c, \LM, "accept", "Visitor") }

\item \label{algo-mens-corrige:step-pull-up-accept} {\sf PullUpAbstract(\LC, "accept", "Visitor", S)}

\item \label{algo-mens-corrige:step-pull-up-methods} { \sf ForAll m in \LM do\\
\tab PullUpConcrete(\LC, m, S)}

\end{enumerate}

\end{boxedminipage}
\end{center}
\caption{ Simple Class Hierarchy $\rightarrow$ Visitor transformation (adapted to \intellij)}
\label{algo-mens-corrige}

\end{figure}%%%%%%%%%%%%%%%%%%%%%%%%%%%%%%%%%%%%%%%%%%%%%%%%%%%%%%%
%%%%%%%%%%%%%%%%%%%%%%%%%%%%%%%%%%%%%%%%%%%%%%%%%%%%%%%%%%%%%%%%%%%

If we refer to Fowler~\cite{Fowler1999}, a
refactoring is manual with checks under the
responsibility of the programmer.
In the same way, the general guidelines given in Fig.~\ref{algo-mens} must
be \emph{interpreted} by someone which will adapt them to his
particular program.

We now consider that the programmer uses a refactoring
tool. We consider \intellij but the same is also possible
with Eclipse with small variations.

\paragraph{Prepare the move.}%%%%%%%%%%%%%%%%%%%%%%%%%%%%
A first problem occurs with the \emph{Move Method}
operation. The refactoring tool cannot move instance
methods to a class if there is no reference of
the destination class in that method (in parameters or in body).
The reason is that the receiver object cannot be inferred otherwise
(we consider \emph{instance} methods).

Before moving the methods, we have to create delegates for these methods (to keep the initial interface), then
add a parameter of the convenient visitor type to the delegates, then move them (see
Fig.~\ref{algo-mens-corrige}, step~\ref{algo-mens-corrige:step-move}).

\paragraph{Restore object type after move.}%%%%%%%%%%%%%%
In our example, the pretty-print method does not access to any
instance variables or methods (see Fig.~\ref{classic-base-prog:data}) of the receiver object. In this case, when the \emph{prettyprint}
delegate methods are moved, the tool does not make a
parameter of  type  \emph{Ellipse} or \emph{Square} appear in the resulting method.

This is problematic because we want overloaded \emph{visit} methods
(it's a design choice, here we could also use different method names) but the lack of these parameters introduces a name clash.

To solve this, it is sufficient to apply the \emph{Add
Parameter} refactoring to the methods which have been moved.
We do not make this appear into the algorithm of
Fig.~\ref{algo-mens-corrige} because we encapsulate this
behavior into the \emph{Move Method} operation. We
consider \emph{Move Method} is an abstract operation, which
can be implemented by a refactoring tool with a single
operation or with a composition/chain of several basic
operations.
We make the correspondence between abstract operation and tool operations in App.~\ref{sec-refactoring-operations} (see App.~\ref{def-move}).

\paragraph{ExtractSuperClass.}%%%%%%%%%%%%%%%%%%%%%%%
Introducing a new superclass and pulling up methods
(steps~\ref{algo-mens:step-introduce-visitor} and~\ref{algo-mens:step-introduce-abstract-visit} of Fig.~\ref{algo-mens}) is known
as \emph{Extract Superclass} in
Fowler~\cite{Fowler1999}.
That composite operations is also available in \intellij and
Eclipse. For that reason, we use it in
Fig.~\ref{algo-mens-corrige} (step~\ref{algo-mens-corrige:step-extract-superclass}).

However, in \intellij, we have had to provide an extension of that operation that applies to several classes simuleanously\footnote{\emph{Pull up method refactoring extension} plugin: \url{http://plugins.intellij.net/plugin/?idea_ce&id=6889}} (it was already possible in Eclipse).

\paragraph{Extract Method Accept.}%%%%%%%%%%%%%%%%%%%%%%%%%%%%%%%%%%%%%%

In the following code (from \emph{Square} or \emph{Ellipse}), the instruction \emph{o.visit(this)} occurs twice (with a different object \emph{o}).

\begin{verbatim}
    public void print() {
        new PrintVisitor().visit(this);
    }

    public void prettyprint() {
        new PrettyPrintVisitor().visit(this);
    }
\end{verbatim}

That instruction has to be extracted into a method \emph{accept} with \emph{o} as a parameter, and the occurrences of that expression will be replaced by \emph{accept(o)}.

The tool \intellij will accept to extract a same method for the two instances only after we introduce a same type for the receiver objects.
In practice, we first introduce a new local variable
for \emph{new PrintVisitor()} (resp. \emph{new
PrettyPrintVisitor()}), then change the type of that
variable form \emph{PrintVisitor}
(resp. \emph{PrettyPrintVisitor}) to \emph{Visitor}, and
then the extraction of the method successes (the two
instances are replaced by invocations of that method).
The operations used in \intellij are \emph{Introduce Variable} and \emph{Type Migration} (as many other refactoring operations \emph{Type Migration} checks that the change is type safe). One would may also find useful to rename the local variables or the parameter of accept to \emph{v} or \emph{visitor} (operation \emph{Rename}).

The local variables can be inlined afterward (operation \emph{Inline}).

Note that the task of making \emph{accept} act on \emph{Visitors} is left implied in the guidelines of Mens and Tourw\'e (Fig~\ref{algo-mens}). This task is not explained either by Fowler (\emph{Extract Method}~\cite{Fowler1999}). 

Again, we encapsulate these elementary changes in the \emph{ExtractGeneralMethod} refactoring operation, defined in App.~\ref{def-ExtractGeneralMethod}.

\paragraph{Pull Up.}
Note that when \emph{accept} is pulled up
(step~\ref{algo-mens-corrige:step-pull-up-accept} of
Fig.~\ref{algo-mens-corrige}), \intellij does not add
the \emph{@Override} annotation to all the subclasses, but
only in the one the operation is called on.

Also, when \emph{print} and \emph{prettyprint} are pulled up
(step~\ref{algo-mens-corrige:step-pull-up-methods} of
Fig.~\ref{algo-mens-corrige}), the tool cannot take several
classes simultaneously into account, so that the pull up
does not verifies that the code are the same in all the
concrete classes (in fact they are).
Note that for \emph{Pull Up}, Eclipse can take several classes into
account (it allows to remove overriding methods in these classes) but it
does not checks that the behavior is preserved by this change.

\paragraph{Visibility.}
In the example program, instance variables are public
(package).
If they were private or protected, we would have had to make
them public so that the moved methods can access them. This
does not depend on the way we implement the transformation,
but rather to the nature of the Visitor pattern.
Note that Eclipse \emph{Move} makes the change automatically while with \intellij you have to do it after or before the \emph{Move}.

\paragraph{Conclusion.}
We have seen that as soon as we consider a refactoring tool, \begin{enumerate}

\item the guidelines have to be adapted and

\item an algorithm can be defined (and automated).

\end{enumerate}

We have seen also that some steps are implied in the
guidelines, and that, on the opposite, some chains of
operations of the guidelines can be done with a single
tool's operation.

Finally, we have seen that we also have to adapt the chain
of operation to characteristics of the initial program. In
the following, after having studied a reverse transformation
to get the program back to its initial structure, we will
see how the algorithm is adapted to variations in the
initial program.

\section{Composite$\leftrightarrow$Visitor Transformation Scheme}%%%%%%%%%%%%%%%%%%%%%%%%%%%%%%%%%%%%%%%%%%%%%%
%%%%%%%%%%%%%%%%%%%%%%%%%%%%%%%%%%%%%%%%%%%%%%%%%%%%%%%%%%%%%%%%%%%%%%%%%%%%%
\label{transfo_sheme}

\begin{figure}[!htp]
\begin{center}
  \begin{minipage}{13cm}

\begin{lstlisting}
abstract class Graphic {
	
	abstract public void print();

	abstract public void show();

}
\end{lstlisting}
\begin{lstlisting}
class Ellipse extends Graphic{

 public void print() {
  System.out.println("Ellipse :" + this);
 }
    
 public void show(){
  System.out.println("Ellipse corresponding to the object " + this + ".");
 }
}
\end{lstlisting}
\begin{lstlisting}[firstline=3]
class CompositeGraphic extends Graphic {

    private ArrayList<Graphic> mChildGraphics = new ArrayList<Graphic>();
     
    public void print() {
    	System.out.println("Composite:");
        for (Graphic graphic : mChildGraphics) {
            graphic.print();
        }
    }
    
    public void prettyprint(){
    	System.out.println("Composite " + this + " composed of:");
    	for (Graphic graphic : mChildGraphics) {
            graphic.prettyprint();
        }	
	System.out.println("(end of composite)");
    }
 
    public void add(Graphic graphic) {
        mChildGraphics.add(graphic);
    }
 
    public void remove(Graphic graphic) {
        mChildGraphics.remove(graphic);
    }
}
\end{lstlisting}
\end{minipage}
\end{center}

\caption{Base Program (class hierarchy)}
\label{base-prog:data}
\end{figure}

We now consider an instance of the Composite pattern as the initial program (Fig.~\ref{base-prog:data}).
The difference between the classic object structure
considered before and the Composite structure is \emph{recursion}:
the data type is recursive (subclasses make references to
the superclass) and the operations are recursive (to
traverse trees of that datatype which depth in unknown).

In this section, all the business methods we handle take no parameter and do not
return any result, and the traversal process is stateless. These constraints are relaxed in Sec.~\ref{sec-variations}.

We also consider that the visitor classes are not part of the project in the Composite state (unlike in  previous section).

\subsection{Composite$\rightarrow$Visitor Transformation}%%%%%%%%%%%%%%%%%%%

Let us consider this part in the code of the CompositeGraphic class:

\begin{verbatim}
    public void print() {
        System.out.print("Composite: " + this + " with: (");
        for (Graphic graphic : childGraphics) {
            graphic.print();
        }
        System.out.println(")");
    }
\end{verbatim}

If we apply the previous transformation algorithm (Fig.~\ref{algo-mens-corrige}), after the operation \emph{AddParameterWithDelegate} (step~\ref{algo-mens-corrige:step-move}), we get the following (with \intellij):

\begin{verbatim}
public void print() {
        print(new PrintVisitor());
    }

    public void print(PrintVisitor v) {
        System.out.print("Composite: " + this + " with: (");
        for (Graphic graphic : childGraphics) {
            graphic.print();
        }
        System.out.println(")");
    }
\end{verbatim}    

We observe that the recursive invocation to
\emph{graphic.print()} in the \emph{for} loop has been left
unchanged. 
The code is still functionally correct, but it is
problematic for the following reason:
if we look at the definition of \emph{Graphic.print()}
(in the program at that moment of the
transformation, you cannot tell which instance of \emph{print()} will be invoked
because \emph{print()} is abstract in the class
\emph{Graphic}, but we know that
\emph{print()}, as a delegator, will be pulled up to the
class \emph{Graphic}), 
 we can see that each invocation of \emph{print()} will
 result in the construction of a new \emph{PrintVisitor}
 object.

Here, if possible, one would choose to use a single
\emph{PrintVisitor} object instead of creating useless new
ones.
In fact, there is a means to do this with the \intellij
refactorer, but, in order to do that, the \emph{print()}
delegator method must be pulled up,\footnote{The trick is to
  first introduce an indirection (directly in the
  superclass), then inline the delegator invocation inside
  the loop, then add the parameter to the delegate, so that
   the tool is able to insert as new parameter in invocations existing objects instead
  of using a default value.}
which impacts the rest of the algorithm (for instance, the pull-up of step~\ref{algo-mens-corrige:step-move} is already done).

This shows that, as soon as we rely on a refactoring tool,
the chain of refactoring operations depends on the
characteristics of the tool.

For this reason, here we cannot encapsulate the small change in
the transformation into a variation of one of the steps of
the algorithm, but we have to adapt the whole algorithm. Our algorithm for basic Composite$\rightarrow$Transformation is given in Fig.~\ref{fig-base-composite-visitor-transfo}.

%In Fig.~\ref{fig-base-composite-visitor-transfo}, to generate temporary names, we consider a function $aux$
%that takes a method name and returns a method name. Here,
%$aux($\textsf{print}$)=$\textsf{printAux} and
%$aux($\textsf{prettyprint}$)=$\textsf{prettyprintAux}.\footnote{Of
%  course, we should ensure that these names are not clashing
%  with other names in the project.}

We use the following notations to abstract the transfomration algorithmds :

%\begin{center}
%\noindent\begin{tabular}{|l|p{4cm}|p{3cm}|}
%\hline 
%symbol & meaning & value in the example \\
\begin{itemize}
%\hline \hline
\item $\mathbb{M}$: set of business methods, here $\mathbb{M} = $\{\xcode{print},\xcode{prettyprint}\}.
%\hlin
\item $\mathbb{C}$: set of Composite hierarchy classes except its root, here $\mathbb{C} = $\{\xcode{Ellipse}, \xcode{CompositeGraphic}\} 
%\hline

%\hline

\item $S$: root of the Composite hierarchy, here $S =$ \xcode{Graphic}.
%\hline
\item $vis$: function that generates a visitor class name from a business method name, here \visitorclass{\xcode{print}} $=$ \xcode{PrintVisitor}. 
%\hline
\item $\mathbb{V}$: set of visitor classes, here $\mathbb{V}=\{ \visitorclass{$m$}\}_{m \in \mathbb{M} } = $\{\xcode{PrintVisitor}, \xcode{PrettyPrintVisitor}\}. 
%\hline
\item  $\auxsymbol$: function used to generate names of temporary methods from business methods, here \auxname{\xcode{print}}= \xcode{printAux}. 
%\hline

\end{itemize}

%\end{tabular}
%\end{center}

%%%%%%%%%%%%%%%%%%%%%%%%%%%%%%%%%%%%%%%%%%%%%%%%%%%%%%%%%%%%%%%%%%%%%%%%%%%%%%%%%%%%%%%%%%%%%%%%%%%%%%%%%%%%%%%%%%%%%%%%%%%%%%%%%%%%%%%%%%%%%%%%%%%%%%%%%%%%%%%%%%%%%%%%%%ù

\begin{figure}
\begin{center}
\begin{boxedminipage}{\textwidth}%%%%%%%%%%%%%%%%%%%%%%%%%%%%%%%%%%%%%%%%%%%%%%%

\begin{enumerate}[1)]

\relsize{-1}
\sf

\item \label{algo-composite-visitor:create-visitors} 
ForAll m in $\mathbb{M}$ do   
  \operation{CreateEmptyClass}(\visitorclass{m})\\[-2mm]

\item   \label{algo-composite-visitor:create-indirection} 
ForAll m in $\mathbb{M}$ do 
   \operation{CreateIndirectionInSuperClass}(S,m, \auxname{m})\\[-2mm]

\item  \label{algo-composite-visitor:inline-methods} 
 ForAll m in $\mathbb{M}$, c in $\mathbb{C}$ do \operation{InlineMethodInvocations}(c, m, \auxname{m}) \\[-2mm]

\item  \label{algo-composite-visitor:add-parameter} 
 ForAll m in $\mathbb{M}$ do \operation{AddParameterWithReuse}(S, \auxname{m}, \visitorclass{m})\\[-2mm]

\item  \label{algo-composite-visitor:move} 
 ForAll m in $\mathbb{M}$, c in $\mathbb{C}$ do \operation{MoveMethodWithDelegate}(c, \auxname{m}, \visitorclass{m}, "visit")\\[-2mm]

\item  \label{algo-composite-visitor:superclass} 
 \operation{ExtractSuperClass}($\mathbb{V}$, "Visitor")\\[-2mm]

\item  \label{algo-composite-visitor:generalise-parameter} 
 ForAll m in $\mathbb{M}$ do \operation{UseSuperType}(S, \auxname{m}, \visitorclass{m}, "Visitor")\\[-2mm]

\item  \label{algo-composite-visitor:merge} 
 %Let LAUX = \{ \auxname{m} \}$_{\mbox{m} \in \mbox{$\mathbb{M}$}}$ in\\
  \operation{MergeDuplicateMethods}(S, \{\auxname{m} \}$_{\mbox{m} \in \mbox{$\mathbb{M}$}}$, "accept")

\end{enumerate}

\end{boxedminipage}
\end{center}
\caption{Base Composite$\rightarrow$Visitor transformation}
\label{fig-base-composite-visitor-transfo}
\end{figure}%%%%%%%%%%%%%%%%%%%%%%%%%%%%%%%%%%%%%%%%%%%%%%%%%%%%%%%

Note that two bugs  were encountered
with \intellij at the beginning of our work, but were solved by JetBrains, so that 
no manual intervention is needed now.

\begin{figure}
\begin{lstlisting}
abstract class Graphic {
	
	public void print() {
	    accept(new PrintVisitor());
	}

	public void prettyprint() {
		accept(new PrettyPrintVisitor());
	}

	public abstract void accept(Visitor v);

}
\end{lstlisting}
\begin{lstlisting}
class Ellipse extends Graphic{

    public void accept(Visitor v) {
		v.visit(this);
	}
}
\end{lstlisting}
\begin{lstlisting}[firstline=3]
class CompositeGraphic extends Graphic {

    ArrayList<Graphic> mChildGraphics = new ArrayList<Graphic>();
     
	public void accept(Visitor v) {
		v.visit(this);
	}
    
    public void add(Graphic graphic) {
        mChildGraphics.add(graphic);
    }
 
    public void remove(Graphic graphic) {
        mChildGraphics.remove(graphic);
    }
}
\end{lstlisting}
\caption{Program with Visitor (data classes)}
\label{base-prog:state3-data}
\end{figure}
\begin{figure}
\begin{lstlisting}
public abstract class Visitor {

	public abstract void visit(Ellipse ellipse);

	public abstract void visit(CompositeGraphic compositeGraphic);

}
\end{lstlisting}
\begin{lstlisting}

public class PrintVisitor extends Visitor {

	public void visit(CompositeGraphic compositeGraphic) {
		System.out.println("Composite:");
	    for (Graphic graphic : compositeGraphic.mChildGraphics) {
	        graphic.accept(this);
	    }
	}

	public void visit(Ellipse ellipse) {
		System.out.println("Ellipse :" + ellipse);
	}
}
\end{lstlisting}
\begin{lstlisting}

public class PrettyPrintVisitor extends Visitor {

	public void visit(CompositeGraphic compositeGraphic) {
		System.out.println("Composite " + compositeGraphic + " composed of:");
		for (Graphic graphic : compositeGraphic.mChildGraphics) {
	        graphic.accept(this);
	    }	
		System.out.println("(end of composite)");
	}

	public void visit(Ellipse ellipse) {
		System.out.println("Ellipse corresponding to the object " + ellipse + ".");
	}
}
\end{lstlisting}
\caption{Program with Visitor (visitor classes)}
\label{base-prog:state3-visitor}
\end{figure}

The result of this transformation is given in
Figs.~\ref{base-prog:state3-data}
and~\ref{base-prog:state3-visitor}.

%%%%%%%%%%%%%%%%%%%%%%%%%%%%%%%%%

\subsection{Visitor$\rightarrow$Composite Transformation}%%%%%%%%%%%%%%%%%%%

Composite$\rightarrow$Visitor transformation is based on moving business code from the data-type class hierarchy to the visitor classes.
Now we do the opposite (move business code from visitor classes to composite classes).
We proceed with three coarse steps (Fig~\ref{fig-algo-retour}):
\begin{enumerate}[i.]

\item Replace dynamic dispatch with static dispatch.

\item In-line the business code from the \textit{visitor} structure to the \textit{composite} structure.

\item Make some small changes to get the initial Composite pattern structure back.
\end{enumerate}

\begin{figure}
\begin{center}
\begin{boxedminipage}{\textwidth}
\begin{enumerate}[I )]

\relsize{-1}

\sf

% step 1

\item \label{visitor-composite-algo:specialize-accept}
   ForAll v in $\mathbb{V}$ do \operation{AddSpecializedMethodInHierarchy}(S,\\ \tab \tab"accept","Visitor",v)\\[-2mm]

\item \label{visitor-composite-algo:delete-method-in-hierarchy}  \operation{DeleteMethodInHierarchy}(S,accept,"Visitor")\\[-2mm]

\item \label{visitor-composite-algo:pushdown-visit}
       ForAll c in $\mathbb{C}$ do \operation{PushDownAll}("Visitor","visit",c)\\[-2mm]
  
\item \label{visitor-composite-algo:inline-visit}
       ForAll v in $\mathbb{V}$, c in $\mathbb{C}$ do \operation{InlineMethod}(v,"visit",c)        \\[-2mm]

\item \label{visitor-composite-algo:rename-aux}
       ForAll m in $\mathbb{M}$ do \operation{RenameMethod}(S,accept,\visitorclass{m},\auxname{m})\\[-2mm]

\item \label{visitor-composite-algo:remove-param}
       ForAll m in $\mathbb{M}$ do \operation{RemoveParameter}(S,\auxname{m},\visitorclass{m})\\[-2mm]

\item \label{visitor-composite-algo:fold}
       ForAll m in $\mathbb{M}$ do \operation{ReplaceMethodDuplication}(S,m)\\[-2mm]
      
\item \label{visitor-composite-algo:pushdown-m}
       ForAll m in $\mathbb{M}$ do \operation{PushDownImplementation}(S,m)\\[-2mm]

\item \label{visitor-composite-algo:pushdown-aux}
       ForAll m in $\mathbb{M}$ do \operation{PushDownAll}(S,\auxname{m})\\[-2mm]

\item \label{visitor-composite-algo:inline-aux}
       ForAll m in $\mathbb{M}$, c in $\mathbb{C}$ do \operation{InlineMethod}(c,\auxname{m})\\[-2mm]
 
\item \label{visitor-composite-algo:delete-visitors}
       ForAll v in $\mathbb{V}$ do \operation{DeleteClass}(v)\\[-2mm]
    
\item \label{visitor-composite-algo:delete-visitor}
      \operation{DeleteClass}("Visitor")

\end{enumerate}

\end{boxedminipage}
\end{center}
\caption{Base Visitor $\rightarrow$ Composite transformation}
\label{fig-algo-retour}
\end{figure}%%%%%%%%%%%%%%%%%%%%%%%%%%%%%%%%%%%%%%%%%%%%%%%%%%%%%%%

\paragraph{Remove Dynamic Dispatch (Fig.~\ref{fig-algo-retour}, steps~\ref{visitor-composite-algo:specialize-accept} and~\ref{visitor-composite-algo:pushdown-visit}).}%%%%%%%%%%%%%%%%%%%%%

We replace the \emph{accept(Visitor)} method by some overloaded
 methods \emph{accept}, one for each subtype
 of \emph{Visitor}. 
This removes all dynamic dispatch in \emph{visit} method invocations,
so that their invocations can be inlined afterward.
The \emph{visit} methods can also be removed from the \emph{Visitor}
class (but not from the concrete visitor classes before they are
inlined).

\begin{figure}[!htp]
\begin{lstlisting}
abstract class Graphic {
	
	public void print() {
	    accept(new PrintVisitor());
	}

	public void prettyprint() {
		accept(new PrettyPrintVisitor());
	}

    public abstract void accept(PrintVisitor v);

    public abstract void accept(PrettyPrintVisitor v);
}
\end{lstlisting}
\begin{lstlisting}
class Ellipse extends Graphic{

    public void accept(PrettyPrintVisitor v) {
        v.visit(this);
    }

    public void accept(PrintVisitor v) {
        v.visit(this);
    }
}
\end{lstlisting}
\begin{lstlisting}[firstline=3]
class CompositeGraphic extends Graphic {

    ArrayList<Graphic> mChildGraphics = new ArrayList<Graphic>();

    public void accept(PrettyPrintVisitor v) {
        v.visit(this);
    }

    public void accept(PrintVisitor v) {
        v.visit(this);
    }

    public void add(Graphic graphic) {
        mChildGraphics.add(graphic);
    }
 
    public void remove(Graphic graphic) {
        mChildGraphics.remove(graphic);
    }
}
\end{lstlisting}
\caption{Reverse-State 1 (data classes)}
\label{base-prog2:state1-data}
\end{figure}

\begin{figure}[!htp]
\begin{lstlisting}
public abstract class Visitor {

}
\end{lstlisting}
\begin{lstlisting}

public class PrintVisitor extends Visitor {

	public void visit(CompositeGraphic compositeGraphic) {
		System.out.println("Composite:");
	    for (Graphic graphic : compositeGraphic.mChildGraphics) {
	        graphic.accept(this);
	    }
	}

	public void visit(Ellipse ellipse) {
		System.out.println("Ellipse :" + ellipse);
	}
}
\end{lstlisting}
\begin{lstlisting}
public class PrettyPrintVisitor extends Visitor {

	public void visit(CompositeGraphic compositeGraphic) {
		System.out.println("Composite " + compositeGraphic + " composed of:");
		for (Graphic graphic : compositeGraphic.mChildGraphics) {
	        graphic.accept(this);
	    }	
		System.out.println("(end of composite)");
	}

	public void visit(Ellipse ellipse) {
		System.out.println("Ellipse corresponding to the object " + ellipse + ".");
	}
}
\end{lstlisting}
\caption{Reverse-State 1 (visitor classes)}
\label{base-prog2:state1-visitor}
\end{figure}

  The result of this is given in
Figs.~\ref{base-prog2:state1-data}
and~\ref{base-prog2:state1-visitor}.

\paragraph{Move Business Code (Fig.~\ref{fig-algo-retour}, step~\ref{visitor-composite-algo:inline-visit}).}%%%%%%%%%%%%%%%%%%%%%%%%%%%%%%%%%%%%

The business code in the \emph{visitor} classes is inlined:
invocations of the \emph{visit} methods in the \emph{composite}
 classes are replaced by the corresponding body (the business code) and the \emph{visit}
 methods are deleted.

\begin{figure}
\begin{lstlisting}
abstract class Graphic {
	
	public void print() {
	    accept(new PrintVisitor());
	}

	public void prettyprint() {
		accept(new PrettyPrintVisitor());
	}

    public abstract void accept(PrintVisitor v);

    public abstract void accept(PrettyPrintVisitor v);
}
\end{lstlisting}
\begin{lstlisting}
class Ellipse extends Graphic{

    public void accept(PrettyPrintVisitor v) {
        System.out.println("Ellipse corresponding to the object " + this + ".");    }

    public void accept(PrintVisitor v) {
        System.out.println("Ellipse :" + this);
    }
}
\end{lstlisting}
\begin{lstlisting}[firstline=3]
class CompositeGraphic extends Graphic {

    ArrayList<Graphic> mChildGraphics = new ArrayList<Graphic>();

    public void accept(PrettyPrintVisitor v) {
        System.out.println("Composite " + this + " composed of:");
        for (Graphic graphic : mChildGraphics) {
	    graphic.accept(v);
	}
        System.out.println("(end of composite)");
    }

    public void accept(PrintVisitor v) {
        System.out.println("Composite:");
        for (Graphic graphic : mChildGraphics) {
            graphic.accept(v);
        }
    }

    public void add(Graphic graphic) {
        mChildGraphics.add(graphic);
    }
 
    public void remove(Graphic graphic) {
        mChildGraphics.remove(graphic);
    }
}
\end{lstlisting}
\caption{Reverse-State 2 (data classes)}
\label{base-prog2:state2-data}
\end{figure}

      The result of this step is given in
Fig.~\ref{base-prog2:state2-data} (visitor classes are empty).

\paragraph{Remove Visitors and Recover Initial Structure (Fig.~\ref{fig-algo-retour}, steps~\ref{visitor-composite-algo:rename-aux} to~\ref{visitor-composite-algo:delete-visitor}). }%%%%%%%%%%%%%%%%%

Once the business code has been moved into the convenient classes, the rest of the refactoring operations are common refactoring operations allowing to recover the composite structure (the important part is done before). 
   
The result of this step is given in
Fig.~\ref{base-prog2:state3-data}.

\begin{figure}
\begin{lstlisting}
abstract class Graphic {
	
	public abstract void print();

    public abstract void prettyprint();

}
\end{lstlisting}
\begin{lstlisting}
class Ellipse extends Graphic{

    public void print() {
        System.out.println("Ellipse :" + this);
    }

    public void prettyprint() {
        System.out.println("Ellipse corresponding to the object " + this + ".");
    }
}
\end{lstlisting}
\begin{lstlisting}[firstline=3]
class CompositeGraphic extends Graphic {

    ArrayList<Graphic> mChildGraphics = new ArrayList<Graphic>();

    public void add(Graphic graphic) {
        mChildGraphics.add(graphic);
    }
 
    public void remove(Graphic graphic) {
        mChildGraphics.remove(graphic);
    }

    public void print() {
        System.out.println("Composite:");
        for (Graphic graphic : mChildGraphics) {
	    graphic.print();
	}
    }

    public void prettyprint() {
        System.out.println("Composite " + this + " composed of:");
        for (Graphic graphic : mChildGraphics) {
	    graphic.prettyprint();
	}
        System.out.println("(end of composite)");
    }
}
\end{lstlisting}
\caption{Result after Back Transformations}
\label{base-prog2:state3-data}
\end{figure}

\subsection{Result after Round Trip Transformation}%%%%%%%%%%%%%%%%%%%%%%%%%%

After this transformation, the program conforms to the Composite pattern (Fig.~\ref{base-prog2:state3-data}). 

A few more refactorings are necessary to recover exactly the original program:
 make private the fields that were made public during the
 Composite$\rightarrow$Transformation, reorder method definitions.

Note also that some comments are altered or lost during the transformation (which is not shown by our example).

\subsection{Precondition}

We use the calculus
of Kniesel and Koch~\cite{composition-of-refactorings2004} to compute the
minimum precondition  that
ensures the success of the transformation.
Our use of the calculus is described
in~\cite{Cohen-Ajouli:2013}.
The computed preconditions are given in
App.~\ref{sec-precondition-composition}.  
The formal descriptions of the refactoring operations used in
the calculus are given in
App.~\ref{sec-refactoring-operations}.

\FloatBarrier

\section{Variants of Transformations for Various Pattern Instances}%%%%%%%%%%%
\label{sec-variations}
In this section we consider some variants of either
Composite pattern or Visitor pattern and we adapt the
algorithm of transformation.

\subsection{Methods with Parameters}
    \label{methods-WithParam}

\subsubsection{Considered variation}
 We consider that  some business methods in the Composite structure have parameters, as exemplified by the following method \xcode{setColor} :

\begin{lstlisting}
// in Graphic
abstract void setColor(int c)
\end{lstlisting}
\begin{lstlisting}
// in Ellipse
int color;
void setColor(int c) { this.color = c; }
\end{lstlisting}
\begin{lstlisting}
// in CompositeGraphic
void setColor(int c) {
   for (Graphic child : children){child.setColor(c);} }
\end{lstlisting}%{verbatim}

Note that the parameter \xcode{c} of the method \xcode{setColor} is passed to each recursive call (in the class \xcode{CompositeGraphic}). 

%%%%%%%%%%%%%%%%%%%%%%%%%%%%%%%%%%%%%%%%%%%%%%%%%%%%%%%%%%%%%%%%%%%%%%%%%%%%%%%%
%%%%%%%%%%%%%%%%%%%%%%%%%%%%%%%%%%%%%%%%%%%%%%%%%%%%%%%%%%%%%%%%%%%%%%%%%%%%%%%%
\subsubsection{Target structure}

In the Visitor structure (Figs.~\ref{base-prog:state3-data} and~\ref{base-prog:state3-visitor}), the visitor object, which is
created by the interface methods of the class \xcode{Graphic}, is passed recursively as parameter
of \emph{accept} and as receiver of \emph{visit}
invocations.
So, to take the parameter \xcode{c} into account, we put it into the state of that
visitor object, so that  it is available during the traversal:

 \begin{lstlisting}
class SetColorVisitor extends Visitor{

 final int c; 

 SetColorVisitor (int c){ this.c = c ; }  
  
 void visit(Ellipse e){ e.color = c; }
 
 void visit(CompositeGraphic g){ 
  for(Graphic child : g.children){child.accept(this);} }}
\end{lstlisting}

The method \xcode{setColor} of the \xcode{Graphic} abstract class  passes the parameter \xcode{c} to the constructor of the class \xcode{SetColorVisitor}, 
then passes the resulting visitor object (with \xcode{c} in its state) to the \emph{accept} method:

\begin{lstlisting}
// in Graphic
void setColor(int c) { accept(new SetColorVisitor(c)); }
 \end{lstlisting}

The implementation of \emph{accept} in \xcode{Ellipse} and
\xcode{CompositeGraphic} is left unchanged.

\subsubsection{Composite$\rightarrow$Visitor Transformation}

The refactoring operation of step~\ref{algo-composite-visitor:add-parameter} of the basic transformation (Fig.~\ref{fig-base-composite-visitor-transfo}) add a visitor parameter to the methods that becomes \emph{accept} later.
Here, we do not want to add the visitor parameter 
to the initial method parameter (such as \xcode{c}), 
but we want to replace the initial parameter with the visitor. To do that we apply the operation  \operation{IntroduceParameterObject} (step~\ref{algo-composite-visitor:add-parameter}.A below). Note that the refactoring operation 
\operation{IntroduceParameterObject} could not be used with methods without parameters.

For that reason, we distinguish methods with parameters and methods without parameters and we introduce the following notation to introduce different treatments in the transformation algorithm: 
\begin{itemize}
  \item $\mparam$: set of methods with parameters, here $\mparam =$ \{\xcode{setColor(int c)}\}.
  \item $\mwithoutparam$: set of methods without parameters, with $\mparam \cup \mwithoutparam = \mathbb{M}$ and  $\mparam \cap \mwithoutparam = \emptyset$.
\end{itemize}

Introducing a parameter object of type \xcode{A} to a method \mbox{\xcode{m(B b)}} for example creates a class A, moves the parameter b to A as an instance variable and finally 
changes  \xcode{m(B b)} into \xcode{m(A a)}. Any old access to \xcode{b} in the body of \xcode{m} will be replaced by \xcode{a.b}. 

The initial step~\ref{algo-composite-visitor:create-visitors}  
is  omitted for methods with parameters because the operation \operation{IntroduceParameterObject} creates the new class (step~\ref{algo-composite-visitor:create-visitors}.A below replaces step~\ref{algo-composite-visitor:create-visitors}). 

Here are the deviations from the basic algorithm for this variation:\\

\noindent\begin{boxedminipage}{\columnwidth}
\begin{enumerate}[4.A)]

\relsize{-1}
\sf

\item[\ref{algo-composite-visitor:create-visitors}.A)]  ForAll m in $\mwithoutparam$ do  
 \textbf{CreateEmptyClass}(\visitorclass{m})\\ \phantom{.} \hfill (replaces step \ref{algo-composite-visitor:create-visitors}) 
 
\item[\ref{algo-composite-visitor:add-parameter}.A)]   ForAll m in $\mparam$ do\\ \tab 
       \textbf{IntroduceParameterObject}(S, \auxname{m}, \visitorclass{m})\\[-2mm]

  ForAll m in $\mwithoutparam$ do\\ \tab 
       \textbf{AddParameterWithReuse}(S, \auxname{m}, \visitorclass{m})\\ \phantom{.}\hfill  (replaces step \ref{algo-composite-visitor:add-parameter})

\end{enumerate}

\end{boxedminipage}\\

\medskip

 \subsubsection{Visitor$\rightarrow$Composite Transformation}%%%%%%%%%%
Before deleting visitor classes (step~\ref{visitor-composite-algo:delete-visitors}) we have to check that there is no references to them in the Composite hierarchy. 
For the methods without parameters, we just remove the parameters corresponding to the visitor (step~\ref{visitor-composite-algo:remove-param}.A : restriction of  step ~\ref{visitor-composite-algo:remove-param} to methods without parameters) since at this moment those methods do not  
use that parameter.
 For example, at this moment (before step~\ref{visitor-composite-algo:remove-param}), the intermediate method for \xcode{print} in \xcode{Ellipse} is as follows:
\begin{lstlisting}
// in Ellipse
void printaux(PrintVisitor v){
       System.out.println("Ellipse");}
\end{lstlisting}

For the methods with parameters, instead of deleting the visitor parameter,  we have to  inline the occurrences of visitor classes to recover the initial parameter \xcode{c}. After applying step~\ref{visitor-composite-algo:inline-aux} (before deleting visitor classes), the method \xcode{setColor} is as follows:
\begin{lstlisting}
// in Ellipse
void setColor(int c){
       this.color = new SetColorVisitor(c).c;}
\end{lstlisting}

At this point we apply the operation \operation{InlineParameterObject} which will replace 
\xcode{\mbox{new SetColorVisitor(c).c}} by \xcode{c} (step~\ref{visitor-composite-algo:delete-visitors}.A), and then we can delete visitor classes (step~\ref{visitor-composite-algo:delete-visitor}). 

Here is the extension of the back transformation:\\[-2mm]

\noindent\begin{boxedminipage}{\columnwidth}%{\textwidth}
\begin{enumerate}[AA.A]

\relsize{-1}
\sf

%\item[\ref{visitor-composite-algo:remove-param}.A] (Empty for $\mparam$) %\hfill (replaces \ref{visitor-composite-algo:remove-param})

\item[\ref{visitor-composite-algo:remove-param}.A]  
 ForAll m in $\mwithoutparam$ do \hfill (replaces step \ref{visitor-composite-algo:remove-param})\\ \tab \textbf{RemoveParameter}(S,\auxname{m},\visitorclass{m})

 \item[\ref{visitor-composite-algo:delete-visitors}.A]
 ForAll m in $\mparam$ do  \hfill (before step \ref{visitor-composite-algo:delete-visitors}) \\ \tab \textbf{InlineParameterObject}(S, \auxname{m}, \visitorclass{m})

\end{enumerate}
 
\end{boxedminipage}\\

%%%%%%%%%%%%%%%%%%%%%%%%%%%%%%%%%%%%%%%%%%%%%%%%%%%%%%%%%%%%%%%%%%%%%%%%%%%%%%%%%%%%%%%%%%%%%%%%%%%%%%%%%%%%%%%%%%%%
%%%%%%%%%%%%%%%%%%%%%%%%%%%%%%%%%%%%%%%%%%%%%%%%%%%%%%%%%%%%%%%%%%%%%%%%%%%%%%%%%%%%%%%%%%%%%%%%%%%%%%%%%%%%%%%%%%ù
\subsection{Methods with different return types}
    \label{methods-WithParam}
\subsubsection{Considered variation}
We consider now business methods with different return types. For example we consider a program 
with two methods: \xcode{Integer perimeter()} and \xcode{String  toString()}.

%Fore more significance, we use a toy program of an evaluator to deal with this variation: 

%\begin{lstlisting}
% abstract Expr{
%    Integer eval();
%    String  show();}
% class Num extends Expr{
%  int n; 
%  Num(n){ this.n = n; }
%  Integer eval(){ return n; }
%  String  show(){ return Integer.toString(n); }
% }
% class Add extends Expr {
%   Expr e1,e2;
%   Add(Expr e1, Expr e2){ this.e1 = e1; this.e2 = e2;}}

%   Integer eval() { return e1.eval() + e2.eval(); }

%   String  show() { return "(" + e1.show() + " + " + e2.show() +")";}  
%\end{lstlisting}

\subsubsection{Target Structure}

Since we have methods with different return types,  we cannot use \xcode{void} to the \emph{accept} 
method. One solution is to have an \emph{accept} method variant for each return type by the means of overloading. 
But this breaks the beauty of the Visitor pattern (one \emph{accept} 
method for each business method instead of one \emph{accept} method to implement an abstract traversal).
 To avoid that, we use generic types.
In the abstract  class \xcode{Graphic}, the \emph{accept} method becomes generic:
\begin{lstlisting}
abstract <T> T accept(Visitor <T> v)  
\end{lstlisting}

Note that the returned type is bound by the type of the
visitor class which appears as parameter. Each visitor class represents a business
method and its return type. The parameterized visitor structure is as follows:
\begin{lstlisting}
abstract class Visitor <T> {...}
\end{lstlisting}
\begin{lstlisting}
class PerimeterVisitor extends Visitor <Integer> {...} 
\end{lstlisting}
\begin{lstlisting}
class ToStringVisitor extends Visitor <String> {...}
\end{lstlisting}

%\begin{lstlisting}
%class EvalVisitor extends Visitor <Integer> {...} 
% Integer visit(Add a){
%     return a.e1.accept(this) + a.e2.accept(this);}}

%class PrettyPrintVisitor extends Visitor <String> { 
%   ...
%  String visit(Add a) { 
%   return "(" + a.e1.accept(this) + 
%               " + " + a.e2.accept(this) +")";}}
%\end{lstlisting}

\paragraph{Remark} Because of the restriction in the use of generic types in
Java, returned types which are raw types, such as \xcode{int}
or \xcode{bool}, must be converted to object types
such as \xcode{Integer} or \xcode{Boolean}.
In the case of \xcode{void}, one can use \xcode{Object} and add a \xcode{return
null} statement (we use a refactoring operation to do that).

\subsubsection{Composite$\rightarrow$Visitor Transformation}
We use the following notations in the algorithm corresponding to this variation:

\begin{itemize}
   \item $\mathbb{R}$: Set of methods and their corresponding return types, here $\mathbb{R} = $\{(\xcode{prettyprint},\xcode{String}), (\xcode{eval},\xcode{Integer})\}.
\end{itemize}
In step~\ref{algo-composite-visitor:superclass} of the basic algorithm, the operation \operation{ExtractSuperClass} creates a new abstract class and 
pulls up abstract declarations of visit methods. In the considered variation, we have to use an extension of the pull up operation that introduces generic types 
in the super class to be able to insert abstract declarations for methods with different return types.

To deal with this variation we apply the operation \operation{ExtractSuperClassWithoutPullUp} then the operation \operation{PullUpWithGenerics}\footnote{\url{http://plugins.jetbrains.com/plugin/?idea\_ce&id=6889}}
instead 
of the operation \operation{ExtractSuperClass} of the step~\ref{algo-composite-visitor:superclass} (step~\ref{algo-composite-visitor:superclass}.B).\\% (see the following algorithm).

\noindent\begin{boxedminipage}{\columnwidth}%{\textwidth}

\begin{enumerate}
\relsize{-1}
\sf

\item[\ref{algo-composite-visitor:superclass}.B]
 \textbf{ExtractSuperClassWithoutPullUp}($\mathbb{V}$, "Visitor") ;\\
          ForAll m in $\mathbb{M}$, c in $\mathbb{C}$ do \\ \tab 
            \textbf{PullUpWithGenerics}(\visitorclass{m}, "visit","Visitor")  \hfill (replaces~\ref{algo-composite-visitor:superclass}) 

%\item[...]

\end{enumerate}

\end{boxedminipage}\\

\subsubsection{Visitor$\rightarrow$Composite Transformation}

At the step~\ref{visitor-composite-algo:specialize-accept} of the base algorithm, 
we must specify the return type 
of each \emph{accept} method. The convenient return types could be identified directly from return types of visit methods existing in concrete visitors. 
This is done by the operation \operation{AddSpecialisedMethodWithGenerics} (step~\ref{visitor-composite-algo:specialize-accept}.B).\\% (see the following algorithm). 

\noindent\begin{boxedminipage}{\columnwidth}%{\textwidth}

\begin{enumerate}
\relsize{-1}
\sf

\item[\ref{visitor-composite-algo:specialize-accept}.B] ForAll v in $\mathbb{V}$ do\\ \tab
 \textbf{AddSpecializedMethodWithGenerics}(S,"accept",$\mathbb{R}$,\\"Visitor",v) 
 \hfill (replaces~\ref{visitor-composite-algo:specialize-accept}) 

%\item[...]
  
\end{enumerate}

\end{boxedminipage}\\

\subsection{Class Hierarchies with Several Levels}%%%%%%%%%%%%%%%%%%%%%%%%%%%%%%%%%%%%%%%%%%%%

\label{sec-sevLevels}
\subsubsection{Considered variation}
   We consider that the Composite hierarchy has multiple
levels, with a random repartition of business code: some
business methods are inherited, and some other are
overridden.

For example, we consider the class \xcode{Ellipse} has a subclass \xcode{ColoredEllipse} where the method \xcode{print} is  overridden whereas 
the second method \xcode{prettyprint} is inherited:
\begin{lstlisting}
class ColoredEllipse extends Ellipse{ 
 int color; 
 ColoredEllipse (int c){ this.color = c; } 

 void print{System.out.println(
       "Ellipse colored with " + color); } }
\end{lstlisting}

\subsubsection{Target Structure}

In order to have  in visitor classes one \emph{visit} method for each class
of the Composite hierarchy,
the code of the method \xcode{prettyprint()} defined in
\xcode{Ellipse} in the Composite structure and inherited
by \xcode{ColoredEllipse}, is placed in the methods
\xcode{\mbox{visit(ColoredEllipse c)}} and
\xcode{\mbox{visit(Ellipse e)}} in \xcode{PrettyPrintVisitor}:
%
%This code is similar to the code of the method
%\xcode{visit(Ellipse e)} except the object
%reference (\xcode{c} instead of \xcode{r}).
%

%
%, elsewhere, there will
%be more than one \emph{accept} method since if
%\xcode{visit(ColoredEllipse c)} is missing from
%\xcode{PrettyPrintVisitor}, we could not make an abstract
%declaration of this method on the root of Visitor structure.
%
%In this case, we must define an \emph{accept} method for
%the class \xcode{PrettyPrintVisitor} since there will not be an
%abstract traversal for this Visitor.
% And 
%in the case of a Composite that inherits a business method from its super class, we should assign the code of this method to its 
%corresponding \emph{visit} method to avoid that this one visit nothing.  

\begin{lstlisting}
class PrettyPrintVisitor extends Visitor{
 void visit(CompositeGraphic g){...}
 
 void visit(Ellipse e){
            System.out.println("Ellipse :"+ e +".");}

 void visit(ColoredEllipse c){
           System.out.println("Ellipse :"+ c +".");}}
  
\end{lstlisting}

\subsubsection{Composite$\rightarrow$Visitor Transformation}%%%%%%%%%%%%%%%%%%%%%%%%

In order to push down a duplicate of the inherited method to the right 
subclass, we apply the operation \operation{PushDownCopy}\footnote{\operation{PushDownCopy} consists in applying \operation{Extract Method}, then \operation{Push Down Method}.} (step~\ref{algo-composite-visitor:create-visitors}.C) before running the basic algorithm.

We use the following notations in the algorithm corresponding to this variation:

\newcommand{\getinherited}[1]{\ensuremath{i(\mbox{#1})}}
\newcommand{\getsuper}[1]{\ensuremath{s(\mbox{#1})}}

\begin{itemize}
\item $\getinherited{$c$}$: a function that gives the list of inherited methods of a class ; 
here $\getinherited{\xcode{ColoredEllipse}}$ = \{\xcode{prettyprint()}\}.
 \item  $\getsuper{$c$}$: a function that gives the superclass of a class.\\
%\item $\heritset$: set of inherited business methods, here $\heritset$ = \xcode{\{prettyprint\}}.
%   \item $\heritingset$ : set of daughter (sub) classes with their corresponding $\heritset$. In this set, each class is coupled with its $\heritset$, \eg 
%for a set of n classes, $\heritingset =$ \{(c1,$\heritset$1);(c2,$\heritset$2);...;(cn,$\heritset$n)\}. 
%Here $\heritingset =$ \{(\xcode{ColoredEllipse},$[ \xcode{prettyprint()}]$)\}.  

    %\item $\auxsymbol$
\end{itemize}

\noindent\begin{boxedminipage}{\columnwidth}%{\textwidth}
\begin{enumerate}
 \relsize{-1}
\sf

\item[\ref{algo-composite-visitor:create-visitors}.C] ForAll c in $\mathbb{C}$,
  ForAll m in $\getinherited{$c$}$ do  \hfill (before~\ref{algo-composite-visitor:create-visitors})\\ \tab \tab
   \textbf{PushDownCopy}(c,m,$\getsuper{$c$}$) 

%\item[...]
 
\end{enumerate}

\end{boxedminipage}\\

\subsubsection{Visitor$\rightarrow$Composite Transformation}
 First we apply the basic algorithm. Then, in order to get back the initial structure
 we delete methods (step~\ref{visitor-composite-algo:delete-visitor}.C) that were 
initially added in these classes in the step~\ref{algo-composite-visitor:create-visitors}.C of the forward transformation.\\

\noindent\begin{boxedminipage}{\columnwidth}%{\textwidth}
\begin{enumerate}[XX.X]

\relsize{-1}
\sf

\item[\ref{visitor-composite-algo:delete-visitor}.C]  ForAll (c,m) in $\mathbb{C}$,
  ForAll m in $\getinherited{$c$}$  do \\ \tab \tab 
    \textbf{DeleteMethod}(c,m)   \hfill (after~\ref{visitor-composite-algo:delete-visitor})
\end{enumerate}

\end{boxedminipage}\\

%\fbox{justifier qu'on a le droit} \fbox{que faire si le code a ete modifie' ?}

%%%%%%%%%%%%%%%%%%%%%%%%%%%%%%%%%%%%%%%%%%%%%%%%%%%%%%%%%%%%%%%%%%%%%%%%%%%%%%%%%%%%%%%%%%%%%%%%%%%%%%%%%%%%%%%%%%%%%%%%%%%%%%%%%%%%%%%%%%%%%%%%%%%%%%%%%%
%%%%%%%%%%%%%%%%%%%%%%%%%%%%%%%%%%%%%%%%%%%%%%%%%%%%%%%%%%%%%%%%%%%%%%%%%%%%%%%%%%%%%%%%%%%%%%%%%%%%%%%%%%%%%%%%%%%%%%%%%%%%%%%%%%%%%%%%%%%%%%%%%%%%%%%%%
\subsection{Interface instead of Abstract Class}
   \label{interface-variant}

\subsubsection{Considered Variation}
 We now consider that the root of the Composite hierarchy is
 not an abstract class but an interface and that
there is 
an intermediary abstract class between it and its subclasses. 
This architecture is found in real softwares: libraries are often provided by the means of an interface and compiled byte-code (\emph{Facade} pattern). 

%%%%%%%%%%%%%%%%%%%%%%%%%%%%%%%%%%%%%%%%%%%%%%%%%%%%%%%%%%%%%%%%%%%
 We suppose that there are no other subclasses 
implementing the interface. %, which is realistic in the described context. 
%
%When required, this will appear as a precondition.
%%%%%%%%%%%%%%%%%%%%%%%%%%%%%%%%%%%%%%%%%%%%%%%%%%%%%%%%%%%%%
%
%Interfaces in Java cannot provide the code of declared methods. In addition, classes that implement an interface can extend an other class.

%\fbox{Trop de discussions}\fbox{deplacer les discussions a la fin}

\begin{lstlisting}
interface Graphic{
 void print();
}
\end{lstlisting}
\begin{lstlisting}
abstract class AbstractGraphic implements Graphic {
 abstract void print();
}
\end{lstlisting}
\begin{lstlisting}
class CompositeGraphic extends AbstractGraphic{
 
ArrayList<Graphic> children = ...

 void print(){
     ...
     for(Graphic child : children){ child.print();}}
}

\end{lstlisting}
\subsubsection{Target Structure}

%In the program of Fig.\ref{composites}, we have a delegator
%in the abstract class \emph{Graphic} which ensures that the
%old interface to business methods is unchanged. With an
%interface instead of an abstract class for \emph{Graphic}, we
%cannot give the delegator body in \emph{Graphic}. 

%We propose here two candidates target structures (to be explained in the transformation description). The first target structure is same as the Visitor structure of Fig.~\ref{base-prog:state3-visitor}. The second 
%alternative is as the following:
Here is a possible target structure corresponding to the considered variation:
\begin{lstlisting}
interface Graphic{
 void print();
 void accept(Visitor v);
}
\end{lstlisting}
 \begin{lstlisting}
abstract class AbstractGraphic implements Graphic{
 void print(){accept(new PrintVisitor());}
}
\end{lstlisting}
\begin{lstlisting}
class CompositeGraphic extends AbstractGraphic{ 
 ArrayList<Graphic> children = ...
 void accept(Visitor v){v.visit(this);}
}
\end{lstlisting}
 \begin{lstlisting}
class PrintVisitor extends Visitor{
  void visit(CompositeGraphic g){
     for(Graphic child : g.children){ child.accept(this);}}
}
\end{lstlisting}

Note that the loop in \xcode{visit(CompositeGraphic)} is done on objects of type \xcode{Graphic} (not \xcode{AbstractGraphic}).

\subsubsection{Composite$\rightarrow$Visitor Transformation}

To reach the target structure, we have to create a delegator
\xcode{print()\{printaux(..)\}} in the class
\xcode{AbstractGraphic}
 and
inline the recursive call of \xcode{print} in \xcode{CompositeGraphic}
(steps~\ref{algo-composite-visitor:create-indirection} and~\ref{algo-composite-visitor:inline-methods}).
But that recursive call refers to the method
\xcode{print} declared in the \xcode{Graphic} interface whereas
the delegator is defined in the abstract class \xcode{AbstractGraphic}.
To solve that, we introduce a downcast to the class
\xcode{AbstractGraphic} in the recursive call to
\xcode{print} as follows: \xcode{((AbstractGraphic)
  child).print()}
(step~\ref{algo-composite-visitor:inline-methods}.D).
This makes the inlining by the refactoring tool possible.
%
%This step occurs just before the
%step~\ref{algo-composite-visitor:inline-methods} of base
%algorithm since inlining the recursive call \xcode{print()}
%to become later a call to \xcode{accept} requires that the
%method to be inlined must not be abstract.
%
%
This downcast is legal because we suppose that the interface
has no other implementation than the abstract class.

%To delete the downcast and recover the standard
%Visitor structure, 
%
After creating the method \xcode{accept} (step~\ref{algo-composite-visitor:merge}),
we pull up its declaration to the
interface \xcode{Graphic}, then we delete
the downcast (step~\ref{algo-composite-visitor:merge}.D).
%(step~\ref{algo-composite-visitor-delete-downcast}.D).\\

\noindent\begin{boxedminipage}{\columnwidth}%{\textwidth}
\begin{enumerate}

\relsize{-1}
\sf

\item[\ref{algo-composite-visitor:inline-methods}.D] ForAll m in $\mathbb{M}$, c in $\mathbb{C}$ do\\ \tab \tab
  \textbf{IntroduceDownCast}(c,m,S)  \hfill (before~\ref{algo-composite-visitor:inline-methods})

\item[\ref{algo-composite-visitor:merge}.D] 
 \textbf{pullupAbstractMethod}(S, "accept", I) 

%\item[\ref{algo-composite-visitor-delete-downcast}.D] \label {algo-composite-visitor-delete-downcast} 
ForAll v in $\mathbb{V}$ do\\ \tab \tab
  \textbf{DeleteDownCast}(v,"accept")  \hfill (after~\ref{algo-composite-visitor:merge})

\end{enumerate}

\end{boxedminipage}\\

\paragraph{Real practice of the transformation}
The algorithms shown above represent the ideal solution to get a Visitor structure.
In fact, there is no operation in the refactoring tools
we use to manage downcasts.
In order to automate the full transformation, we do not use
downcasts and do not inline the delegator.
As a result we get a Visitor with indirect  recursion as follows:

\begin{lstlisting}
// In Graphic
   void print();
\end{lstlisting}
\begin{lstlisting}
// In AbstractGraphic
   abstract void accept(Visitor v);
   void print(){accept(new PrintVisitor();}
\end{lstlisting}
\begin{lstlisting}
// In PrintVisitor
   void visit(CompositeGraphic g){
     for(Graphic child : g.children){ child.print();}}
\end{lstlisting}

We can see that at each recursive invocation a new instance
of a Visitor is created. The result is legal but shows a poor
use of memory. 
This problem disappears when the initial
Composite structure is recovered. 
Moreover, if needed, the downcast can be introduced manually
(or the refactoring operation can be implemented).

%To perform the practice algorithm for this solution we apply
%the base algorithms with the following deviations:
%\begin{itemize}
% \item In the Composite to Visitor pattern, we don't perform the step~\ref{algo-composite-visitor:inline-methods}. Since we don't make downcasts, we can not execute the operation \emph{InlineMethodInvocations} 
%(performed in this step) to the recursive call \xcode{child.print()} because we can not inline with abstract methods (the case of the method \xcode{print} declared in the interface \xcode{Graphic}).
%\end{itemize}

So, in practice, the variation in the algorithm is: do not apply step \ref{algo-composite-visitor:inline-methods} (nor \ref{algo-composite-visitor:inline-methods}.D); do not apply step \ref{algo-composite-visitor:merge}.D (but  step \ref{algo-composite-visitor:merge}).

\subsubsection {Visitor$\rightarrow$Composite Transformation}

After the \emph{practical} Composite$\rightarrow$Visitor
transformation, the base Visitor$\rightarrow$Composite
transformation can be applied without performing the step~\ref{visitor-composite-algo:fold}.

After the full Composite$\rightarrow$Visitor transformation
described above (with downcasts), we also have to add and
remove some downcasts to recover the Composite structure (before step~\ref{visitor-composite-algo:fold} and after~\ref{visitor-composite-algo:fold}).

\subsection{Preconditions}

The preconditions for the transformations described in this section are given in
App.~\ref{sec-precondition-composition}.

\FloatBarrier

\section{Application to JHotDraw}%%%%%%%%%%%%%%%%%%%%%%%%%%%%%%%%%%%%%%%%%%%%%
In this section we we apply our transformation to the JHotDraw framework.

% \subsection{JHotDraw  transformation}
%   \subsubsection{Overview}
   %\textit{"JHotDraw is a Java GUI framework for technical and structured Graphics. It has been developed as a "design exercise" but is already quite powerful. Its design relies heavily on some well-known design patterns. 
%JHotDraw's original authors have been Erich Gamma and Thomas Eggenschwiler"}\footnote{http://www.jhotdraw.org/}. 
%

%%%%%%%%%%%%%%%%%%%%%%%%%%%%%%%%%%%%%%%%%%%%%%%%%%%%%%%%%%%%%%%%%%%%%%%%%%%%%%%%%%%%%%%%%%%%%%%%%%%%%%%%%%%%%%%%%%%%%%%
%JHotDraw is considered as a subject for some works in the world of modularity. Ceccato et al.~\cite{Ceccato05} use JHotDraw as a software which contains many crosscutting concerns and which needs to be analyzed to locate 
%the different concerns and then preparing  it to be  migrated to aspect oriented programming. Canfora and Cerulo~\cite{Canfora05} also study the evolution of some concerns in JHotDraw in order to facilitate 
%aspect identification.
%%%%%%%%%%%%%%%%%%%%%%%%%%%%%%%%%%%%%%%%%%%%%%%%%%%%%%%%%%%%%%%%%%%%%%%%%%%%%%%%%%%%%%%%%%%%%%%%%%%%%%%%%%%%%%%%%%%%%%%%%%%%%%%%%
%\fbox{pluto dire que JHotDraw a ete utilise pour des use case dans le monde de la modularite, notamment pour les aspects, et donner les ref.}
We used a pattern detection tool to detect the Composite pattern in JHotDraw. There are  two Composite structures, a simple one (2 classes and 1 method) and a larger one with 
18 classes and 6 business methods. We consider the largest one in our study 
since it shows the four variations presented above. We aliment the transformation algorithm with the following data:

\begin{itemize}

\item  $S = $ \xcode{AbstractFigure}.
\item $\mathbb{C} = $ \{  \xcode{EllipseFigure}, \xcode{DiamondFigure}, \xcode{RectangleFigure},  \xcode{RoundRectangleFigure},  \xcode{TriangleFigure}, \xcode{TextFigure},  \xcode{BezierFigure}, 
 \xcode{TextAreaFigure}, ...
% \xcode{NodeFigure}, \xcode{SVGImage}, \xcode{SVGPath}, \xcode{DependencyFigure},  \xcode{LineConnectionFigure},  \xcode{LabeledLineConnectionFigure}, 
%\ \xcode{AbstractCompositeFigure },  \xcode{GraphicalCompositeFigure} 
\}.

\item $\mparam = $ \{ \xcode{basicTransform (AffineTransform tx)}, \xcode{contains(Point2D.Double p)},  \xcode{setAttribute(AttributeKey key,Object value)}, \xcode{findFigureInside(Point2D.Double p)},
 \xcode{addNotify(Const "Drawing d)},\xcode{removeNotify(Drawing d)}\}.

\item $\mwithoutparam = \emptyset$.

\item $\mathbb{R} = $ \{ (\xcode{basicTransform},\xcode{Void}), (\xcode{contains},\xcode{Boolean}),  (\xcode{setAttribute}, \xcode{Void}), (\xcode{findFigureInside},\xcode{Figure}),
 (\xcode{addNotify}, \xcode{Void}), (\xcode{removeNotify}, \xcode{Void})\}.

\item $\getinherited{\xcode{LineConnectionFigure}}$ =  \{\xcode{findFigureInside},  
                                    \xcode{setAttribute},\xcode{contains}\}, $\getinherited{...}$ = ...
\item $\getsuper{\xcode{LineConnectionFigure}} $=  \{\xcode{BezierFigure}\}, $\getsuper{...} $=  ...

%\item $\getinherited{\xcode{SVGPath}} = $ ...

%$\getinherited{\xcode{LabeledLineConnectionFigure}} = $ \{\xcode{basicTransform}, 
%                              \xcode{setAttribute},\xcode{findFigureInside},\xcode{contains}\}, $\getsuper{\xcode{LabeledLineConnectionFigure}} = $ \{\xcode{BezierFigure}\}
%\item $\getinherited{\xcode{DependencyFigure}} = $ \{\xcode{addNotify},
%          \xcode{basicTransform}, \xcode{setAttribute},\xcode{findFigureInside},\xcode{contains}\}, $\getsuper{\xcode{DependencyFigure}} = $ \{\xcode{LineConnectionFigure}\}

%\item $\getinherited{\xcode{NodeFigure}} = $ \{\xcode{addNotify},\xcode{basicTransform},
%                          \xcode{setAttribute},\xcode{findFigureInside},\xcode{contains}\}, $\getsuper{\xcode{NodeFigure}} = $ \{\xcode{TextFigure}\}

%\item $\getinherited{\xcode{GraphicalCompositeFigure}} = $ \{\xcode{findFigureInside}\}, $\getsuper{\xcode{NodeFigure}} = $ \{\xcode{AbstractCompositeFigure}\}

%here $\getinherited{$\textsf{ColoredRectangle}$}$ = \{\textsf{show()}\}.
% \item  $\getsuper{$c$}$:
\end{itemize}
%
%We have identified in that structure the four variations described in Sec.~\ref{sect:variation}.

%\noindent\begin{tabular}{|l|p{5cm}|p{5cm}|}
%\hline 
%Variation & Key source Code & Remarks\\

%\hline \hline
%Methods with parameters & all the six business methods have parameters &   \\
%\hline
%Interface instead of abstract class & \emph{\textbf{interface} Figure}& We have already an intermediary class between the interface and its subclasses (\emph{AbstractFigure}) \\
%\hline
%Methods with different return types &the set of business methods have different return types such as \emph{\textbf{boolean}, \textbf{Figure}, \textbf{void}}& Raw types should be converted to Object types to get Generic Visitor \\
%\hline
%Multiple levels hierarchy & \emph{DependencyFigure \textbf{extends} LineConnectionFigure} & The class \emph{LineConnectionFigure} is a subclass of a root hierarchy and the class  \emph{DependencyFigure} 
%redefines only one business method and inherits five methods. \\
%\hline

%\end{tabular}
%

 \subsection{From Composite to Visitor}
  
To switch from the Composite structure of JHotDraw to its Visitor structure we  apply the following sequence of steps: \ref{algo-composite-visitor:create-visitors}.C ; 
\ref{algo-composite-visitor:create-indirection} ; \ref{algo-composite-visitor:add-parameter}.A ; \ref{algo-composite-visitor:move} ; \ref{algo-composite-visitor:superclass}.B ; \ref{algo-composite-visitor:generalise-parameter} ; \ref{algo-composite-visitor:merge}.

 \subsection{From Visitor to Composite} To recover the initial structure, we apply the following: steps~\ref{visitor-composite-algo:specialize-accept}.B ; 
\ref{visitor-composite-algo:delete-method-in-hierarchy};
\ref{visitor-composite-algo:pushdown-visit};  \ref{visitor-composite-algo:inline-visit} ; \ref{visitor-composite-algo:rename-aux} ; \ref{visitor-composite-algo:remove-param}.A; 
\ref{visitor-composite-algo:pushdown-m} ; \ref{visitor-composite-algo:pushdown-aux} ; \ref{visitor-composite-algo:inline-aux} ; 
 \ref{visitor-composite-algo:delete-visitors}.A ; \ref{visitor-composite-algo:delete-visitors} ; \ref{visitor-composite-algo:delete-visitor} ; \ref{visitor-composite-algo:delete-visitor}.C.

\subsection{Usability of JHotDraw transformation}
Since JHotDraw Composite contains 18 classes and 6 business methods, if one add another functionality (a business method) to the program he must make 
all these classes aware about this modification. Thanks to our transformation, we could perform this evolution as a modular maintenance by applying the Composite to Visitor 
transformation and adding the new functionality as only one module in the Visitor structure. The reverse way of the 
transformation could put then the added code in the right place of the initial structure.
 
\subsection{Generated Precondition}     
The computed minimum precondition that ensures  success of the round-trip transformation is given in App.~\ref{sec-precondition-jhotdraw}.

%To validate our transformation algorithms, we apply them to 
% JHotDraw~\cite{jHotDraw}.
%

%To know on which classes to apply the transformation, we use a
%pattern detection tool.
%
%We have
%applied \linebreak \mbox{\textbf{pattern4}~\cite{Tsantalis:2006:DPD:1248727.1248777}:}
%
%it reports a Composite structure with 6 operations and it reports  the superclass and the subclass that implements the ``container''. The operations are defined by overriding methods in about 12 classes of the class hierarchy.

%The semi-automatic Composite$\rightarrow$Visitor transformation applies
%successfully with the help of variations studied in
%Sec.~\ref{sec-variations} (and with the limitations explained in these variations). 

%It takes between 8 and 9 hours to apply the whole Composite$\rightarrow$Visitor transformation interactively. Most of time is due to interaction (selection the entities to transform, selecting the refactoring operation and giving parameters). 
%The automated version takes a few minutes. We are focusing on automating the whole reversible transformation on JhotDraw.

%A second instance of the pattern is found but we have not
%transformed it since it has only one operation defined.

\section{Related work}%%%%%%%%%%%%%%%%%%%%%%%%%%%%%%%%%%%%%%%%%%%%%%%%%%%%%%%%
   \subsection{Refactoring to Patterns}%%%%%%%%%%%%%%%%%%%%%%%

Using chains of elementary refactoring
operations to introduce design patterns into programs is not new.
The idea is first proposed by Batory and Takuda~\cite{Batory:1995}.

\'O Cinn\'eide~\cite{theseOCinneide} give transformation to introduce several patterns but not the Visitor (he considers
in~\cite{theseOCinneide} that automating the introduction of a visitor
pattern is impractical).

Kerievsky~\cite{Kerievsky04} proposes two sets of guidelines
to introduce Visitor patterns.
The first one is similar to the one from Mens and
Tourw\'e~\cite{surveyRefactoring2004} described in
Sec.~\ref{sec-classic-algos}.
The second one applies to an ``external accumulation'':
instead of transforming an operation defined by overriding
methods in the class hierarchy, it applies to an operation
defined outside of the class hierarchy by a switch on the
type of an object with \emph{instanceof} and type casts.
Neither Mens and Tourw\'e~\cite{surveyRefactoring2004} nor
Kerievsky~\cite{Kerievsky04} give the inverse transformation.

Hills et al.~\cite{Hills:2011} have transformed a program
based on a Visitor pattern to introduce a Visitor pattern
instead (the Visitor pattern is similar to the Composite
pattern).
Their transformation is automated, with a  few interactions
with the user.
Since their transformation is dedicated to a specific program
and is not abstractly described, it requires some work to be
applied to other programs.

Jeaon et al.~\cite{Jeon:2002} provide automatic inference of sequence of
refactoring operations allowing to reach design pattern
based versions of programs. 
Sudan et al.~\cite{reconstruction-of-refactorings-2010}
provide an inference of a sequence of refactoring operations
allowing to pass from a given version of a program to a
second given version.
Such tools could be used to infer variations of our transformation algorithms for variations in initial programs, or to infer transformations between other patterns.

\subsection{Building Complex Refactoring Operations}%%%%%%%%%%

The transformations we aim at can be seen as
complex/composed refactoring operations.
As each refactoring operation has specific preconditions,
and as we use a large number of elementary transformations,
assistance for building such transformations would be
valuable.
Several works
provide languages to build or compose refactoring
operations.
\'O Cinn\'eide and Nixon~\cite{Cinneide00compositerefactorings} show how to compose
elementary refactoring operations with pre/post-conditions,
as well as Kniesel and
Koch~\cite{composition-of-refactorings2004}.

Verbaere et al. propose a language dedicated to building
refactoring operations~\cite{Verbaere:2006}, and Klint et al.
propose a language dedicated to program
manipulation~\cite{Klint:2009}, which they have used to build the Visitor$\rightarrow$Interpreter transformation~\cite{Hills:2011}.

\subsection{Design Patterns Discovery}%%%%%%%%%%%%%%%%%%%%%%%%%%

To provide a fully automated transformation, detection of the
occurrences of the initial design pattern must be automated. Several
work exist in that domain.
Smith and Scott provide a tool that
discovers variants of a design pattern in a given
program~\cite{Smith03spqr}.
Such tools can be used to automatically provide inputs to our transformations.

On the opposite, some tools detect pattern precursors,
anti-patterns or code smells~\cite{Rajesh:2004, Moha:2006}, but,
in our approach, we consider that the initial program has already a
good design.

\section{Conclusion}%%%%%%%%%%%%%%%%%%%%%%%%%%%%%%%%%%%%%%%%%%%%%%%%%%%%%%%%%%

In this report:\begin{itemize}

\item We have shown how to use refactoring operations to transform a
  Java program conforming to the Composite pattern (or Interpreter
  pattern) into a program (still in Java) conforming to the Visitor
  pattern and vice versa.

\item We have explained how to use a refactoring tool
  (\intellij) to perform these transformations.

\item We have discussed some variations in transformations to adapt to
  variations in the initial programs.

\item We have computed preconditions for the proposed transformations.

\end{itemize}

This work is a first step towards automation of these transformations
so that the user does not have to perform each basic refactoring with
a refactoring tool. 
On the example of the JHotDraw program, automation can reduce transformation time from 8 hours to a few minutes.
This kind of automated transformation can be used
to provide different versions of a same programs with different
properties with respect to modularity~\cite{Cohen-Douence-Ajouli:2012}.

\bibliographystyle{alpha}
\bibliography{biblio}

\begin{thebibliography}{HKVDSV11}

\bibitem[BT95]{Batory:1995}
Don Batory and Lance Tokuda.
\newblock Automated software evolution via design pattern transformations.
\newblock Technical report, University of Texas at Austin, Austin, TX, USA,
  1995.

\bibitem[CA13]{Cohen-Ajouli:2013}
Julien Cohen and Akram Ajouli.
\newblock {Practical use of static composition of refactoring operations}.
\newblock In {\em {ACM~Symposium On Applied Computing}}, Portugal, March 2013.

\bibitem[CDA12]{Cohen-Douence-Ajouli:2012}
Julien Cohen, R\'emi Douence, and Akram Ajouli.
\newblock Invertible program restructurings for continuing modular maintenance.
\newblock In {\em Soft. Maintenance and Reengineering (CSMR), 16th European
  Conf. on}, pages 347--352, 2012.

\bibitem[Fow99]{Fowler1999}
Martin Fowler.
\newblock {\em Refactoring: Improving the Design of Existing Code}.
\newblock Addison-Wesley, 1999.

\bibitem[GHJV95]{Gamma:1995}
Erich Gamma, Richard Helm, Ralph Johnson, and John Vlissides.
\newblock {\em Design patterns: elements of reusable object-oriented software}.
\newblock Addison-Wesley Longman Publishing Co., Inc., Boston, MA, USA, 1995.

\bibitem[GI]{jHotDraw}
Erich Gamma and IFA Informatik.
\newblock {JHotDraw as Open-Source Project}.
\newblock \url{http://www.jhotdraw.org/}.

\bibitem[HKVDSV11]{Hills:2011}
Mark Hills, Paul Klint, Tijs Van Der~Storm, and Jurgen Vinju.
\newblock A case of visitor versus interpreter pattern.
\newblock In {\em Proceedings of the 49th international conference on Objects,
  models, components, patterns}, TOOLS'11, pages 228--243, Berlin, Heidelberg,
  2011. Springer-Verlag.

\bibitem[JLB02]{Jeon:2002}
Sang-Uk Jeon, Joon-Sang Lee, and Doo-Hwan Bae.
\newblock An automated refactoring approach to design pattern-based program
  transformations in {J}ava programs.
\newblock In {\em Proceedings of the Ninth Asia-Pacific Software Engineering
  Conference}, APSEC '02, pages 337--, Washington, DC, USA, 2002. IEEE Computer
  Society.

\bibitem[Ker04]{Kerievsky04}
Joshua Kerievsky.
\newblock {\em Refactoring to Patterns}.
\newblock Pearson Higher Education, 2004.

\bibitem[KK04]{composition-of-refactorings2004}
G\"unter Kniesel and Helge Koch.
\newblock Static composition of refactorings.
\newblock {\em Science of Computer Programming}, 52(Issues 1-3):9--51, Aug.
  2004.
\newblock Special Issue on Program Transformation.

\bibitem[Koc02]{koch2002}
Helge Koch.
\newblock Ein refactoring-framework für {J}ava (in german).
\newblock Diploma thesis, CS Dept. III, University of Bonn, Germany, April
  2002.

\bibitem[KSV09]{Klint:2009}
Paul Klint, Tijs van~der Storm, and Jurgen Vinju.
\newblock Rascal: A domain specific language for source code analysis and
  manipulation.
\newblock In {\em Proceedings of the 2009 Ninth IEEE International Working
  Conference on Source Code Analysis and Manipulation}, SCAM '09, pages
  168--177, Washington, DC, USA, 2009. IEEE Computer Society.

\bibitem[MGL06]{Moha:2006}
Naouel Moha, Yann-Gael Gueheneuc, and Pierre Leduc.
\newblock Automatic generation of detection algorithms for design defects.
\newblock In {\em Proceedings of the 21st IEEE/ACM International Conference on
  Automated Software Engineering}, pages 297--300, Washington, DC, USA, 2006.
  IEEE Computer Society.

\bibitem[MT04]{surveyRefactoring2004}
Tom Mens and Tom Tourw\'{e}.
\newblock A survey of software refactoring.
\newblock {\em IEEE Trans. Softw. Eng.}, 30:126--139, February 2004.

\bibitem[OC00]{theseOCinneide}
Mel \'O~Cinn\'eide.
\newblock {\em Automated Application of Design Patterns: A Refactoring
  Approach}.
\newblock PhD thesis, Trinity College, Dublin, Oct. 2000.

\bibitem[OCN99]{O'Cinneide:1999}
M.~\'O~Cinn\'{e}ide and P.~Nixon.
\newblock A methodology for the automated introduction of design patterns.
\newblock In {\em Proceedings of the IEEE International Conference on Software
  Maintenance}, ICSM '99, pages 463--, Washington, DC, USA, 1999. IEEE Computer
  Society.

\bibitem[OCN00]{Cinneide00compositerefactorings}
Mel \'O~Cinn\'eide and Paddy Nixon.
\newblock Composite refactorings for {J}ava programs.
\newblock In {\em Proc. of the Workshop on Formal Techniques for Java Programs,
  ECOOP}, 2000.

\bibitem[PRSK10]{reconstruction-of-refactorings-2010}
Kyle Prete, Napol Rachatasumrit, Nikita Sudan, and Miryung Kim.
\newblock Template-based reconstruction of complex refactorings.
\newblock In {\em IEEE International Conference on Software Maintenance
  (ICSM)}, Sept. 2010.

\bibitem[RJ04]{Rajesh:2004}
J.~Rajesh and D.~Janakiram.
\newblock Jiad: a tool to infer design patterns in refactoring.
\newblock In {\em Proceedings of the 6th ACM SIGPLAN international conference
  on Principles and practice of declarative programming}, PPDP '04, pages
  227--237, New York, NY, USA, 2004. ACM.

\bibitem[SS03]{Smith03spqr}
Jason~M. Smith and David Stotts.
\newblock {SPQR}: Flexible automated design pattern extraction from source
  code.
\newblock In {\em 18th IEEE Intl. Conf. on Automated Soft. Eng.}, pages
  215--224. IEEE Computer Society Press, 2003.

\bibitem[VEdM06]{Verbaere:2006}
Mathieu Verbaere, Ran Ettinger, and Oege de~Moor.
\newblock Jungl: a scripting language for refactoring.
\newblock In {\em Proceedings of the 28th international conference on Software
  engineering}, ICSE '06, pages 172--181, New York, NY, USA, 2006. ACM.

\end{thebibliography}

\appendix

\section{Refactoring Operations}
\label{sec-refactoring-operations}
In this appendix, we define refactoring operations we use in
our transformations. For each operation, we give an example to explain its
behavior, and we tell how it is performed with \intellij or Eclipse.
We give some preconditions when an operation applies only in a specific case.
These preconditions are neither
minimal (they can be refined into weaker conditions) nor
complete (they are sufficient in our basic examples, but not
in some situations we have not considered).
Also, some preconditions dealing with name clashes are left implied.

The given \emph{backward descriptions} are used for static composition of refactorings~\cite{composition-of-refactorings2004} (see App.~\ref{sec-precondition-composition}).

\newcommand{\tools}{\paragraph{Refactoring tools.}}
\newcommand{\precond}{\paragraph{Preconditions:}}
\newcommand{\precondition}{\paragraph{Precondition.}}
\newcommand{\backdescr}{\paragraph{Backward Description.}}
\newcommand{\true}{\mathtt{TRUE}}

\subsection{CreateEmptyClass}%%%%%%%%%%%%%%%%%%%%%%%%%%%%%%%%%%%%%%%%%%%%%%%%%%%%%%%%%%%%%%%%%%
\label{def-CreateEmptyClass}

\paragraph{Overview:} \textsf{CreateEmptyClass (classname c)}: this operation is used t add a new class c.

\tools
\emph{new Class} in Eclipse and \intellij.

\precondition $\\ (\neg existsType(c))$ \\

\backdescr  
 $\\ \mathtt{ExistsClass}(c)\, \mapsto \,\top \\[2pt]
 \mathtt{ExistsType}(c)\, \mapsto \,\top \\[2pt]
 \mathtt{IsAbstractClass}(c)\, \mapsto \,\bot \\[2pt]
 \mathtt{ExistsMethodDefinition}(c,Y)\, \mapsto \,\bot \\[2pt]
 \mathtt{ExistsMethodDefinitionWithParams}(c,Y,[\,])\, \mapsto \,\bot \\[2pt]
 \mathtt{ExistsMethodDefinitionWithParams}(c,Y,[T1])\, \mapsto \,\bot \\[2pt]
 \mathtt{ExistsMethodDefinitionWithParams}(c,Y,[T1;T2])\, \mapsto \,\bot \\[2pt]
 \mathtt{ExistsMethodDefinitionWithParams}(c,Y,[T1;T2;T3])\, \mapsto \,\bot \\[2pt]
 \mathtt{ExistsMethodDefinitionWithParams}(c,Y,[T1;T2;T3;T4])\, \mapsto \,\bot \\[2pt]
 \mathtt{ExistsMethodDefinitionWithParams}(c,Y,[T1;T2;T3;T4;T5])\, \mapsto \,\bot \\[2pt]
 \mathtt{IsInheritedMethod}(c,Y)\, \mapsto \,\mathtt{IsVisible}(java.lang.Object,Y,c)\\[2pt]
 \mathtt{IsInheritedMethodWithParams}(c,Y,[\,])\, \mapsto \,\mathtt{IsVisibleMethod}(java.lang.Object,Y,[\,],c)\\[2pt]
 \mathtt{IsInheritedMethodWithParams}(c,Y,[T1])\, \mapsto \,\mathtt{IsVisibleMethod}(java.lang.Object,Y,[T1],c)\\[2pt]
 \mathtt{IsInheritedMethodWithParams}(c,Y,[T1;T2])\, \mapsto \,\mathtt{IsVisibleMethod}(java.lang.Object,Y,[T1;T2],c)\\[2pt]
 \mathtt{IsInheritedMethodWithParams}(c,Y,[T1;T2;T3])\, \mapsto \,\mathtt{IsVisibleMethod}(java.lang.Object,Y,[T1;T2;T3],c)\\[2pt]
 \mathtt{IsInheritedMethodWithParams}(c,Y,[T1;T2;T3;T4])\, \mapsto \,\mathtt{IsVisibleMethod}(java.lang.Object,Y,[T1;T2;T3;T4],c)\\[2pt]
 \mathtt{IsInheritedMethodWithParams}(c,Y,[T1;T2;T3;T4;T5])\, \mapsto \,\mathtt{IsVisibleMethod}(java.lang.Object,Y,[T1;T2;T3;T4;T5],c)\\[2pt]
 \mathtt{IsSubType}(c,X)\, \mapsto \,\bot  (condition) \\[2pt]
 \mathtt{ExtendsDirectly}(c,X)\, \mapsto \,\bot  (condition) \\[2pt]
 \mathtt{ExistsMethodDefinitionWithParams}(B,Y,[c])\, \mapsto \,\bot \\[2pt]
 \mathtt{ExistsMethodDefinitionWithParams}(B,Y,[c;T1])\, \mapsto \,\bot \\[2pt]
 \mathtt{ExistsMethodDefinitionWithParams}(B,Y,[T1;c])\, \mapsto \,\bot \\[2pt]
 \mathtt{ExistsMethodDefinitionWithParams}(B,Y,[T1;T2;c])\, \mapsto \,\bot \\[2pt]
 \mathtt{ExistsMethodDefinitionWithParams}(B,Y,[T1;c;T2])\, \mapsto \,\bot \\[2pt]
 \mathtt{ExistsMethodDefinitionWithParams}(B,Y,[c;T1;T2])\, \mapsto \,\bot \\[2pt]
 \mathtt{IsInheritedMethodWithParams}(B,Y,[c])\, \mapsto \,\bot \\[2pt]
 \mathtt{IsInheritedMethodWithParams}(B,Y,[c;T1])\, \mapsto \,\bot \\[2pt]
 \mathtt{IsInheritedMethodWithParams}(B,Y,[T1;c])\, \mapsto \,\bot \\[2pt]
 \mathtt{IsInheritedMethodWithParams}(B,Y,[T1;T2;c])\, \mapsto \,\bot \\[2pt]
 \mathtt{IsInheritedMethodWithParams}(B,Y,[T1;c;T2])\, \mapsto \,\bot \\[2pt]
 \mathtt{IsInheritedMethodWithParams}(B,Y,[c;T1;T2])\, \mapsto \,\bot \\[2pt]
 \mathtt{ExistsParameterWithName}(B,Y,[c],P)\, \mapsto \,\bot \\[2pt]
 \mathtt{ExistsParameterWithName}(B,Y,[c;T1],P)\, \mapsto \,\bot \\[2pt]
 \mathtt{ExistsParameterWithName}(B,Y,[T1;c],P)\, \mapsto \,\bot \\[2pt]
 \mathtt{ExistsParameterWithName}(B,Y,[T1;c;T2],P)\, \mapsto \,\bot \\[2pt]
 \mathtt{ExistsParameterWithName}(B,Y,[T1;T2;c],P)\, \mapsto \,\bot \\[2pt]
 \mathtt{ExistsParameterWithName}(B,Y,[c;T1;T2],P)\, \mapsto \,\bot \\[2pt]
 \mathtt{ExistsParameterWithType}(B,Y,[c],P)\, \mapsto \,\bot \\[2pt]
 \mathtt{ExistsParameterWithType}(B,Y,[c;T1],P)\, \mapsto \,\bot \\[2pt]
 \mathtt{ExistsParameterWithType}(B,Y,[T1;c],P)\, \mapsto \,\bot \\[2pt]
 \mathtt{ExistsParameterWithType}(B,Y,[T1;c;T2],P)\, \mapsto \,\bot \\[2pt]
 \mathtt{ExistsParameterWithType}(B,Y,[T1;T2;c],P)\, \mapsto \,\bot \\[2pt]
 \mathtt{ExistsParameterWithType}(B,Y,[c;T1;T2],P)\, \mapsto \,\bot \\[2pt]
 \mathtt{IsUsedMethodIn}(c,B,Y)\, \mapsto \,\bot \\[2pt]
 \mathtt{IsUsedMethod}(c,B,[T1])\, \mapsto \,\bot \\[2pt]
 \mathtt{IsUsedMethod}(c,B,[T1;T2])\, \mapsto \,\bot \\[2pt]
 \mathtt{IsUsedMethod}(c,B,[T1;T2;T3])\, \mapsto \,\bot \\[2pt]
 \mathtt{IsUsedMethod}(c,B,[T1;T2;T3;T4])\, \mapsto \,\bot \\[2pt]
 \mathtt{IsUsedConstructorAsMethodParameter}(c,B,Y)\, \mapsto \,\bot \\[2pt]
 \mathtt{IsUsedConstructorAsInitializer}(c,B,Y)\, \mapsto \,\bot \\[2pt]
 \mathtt{IsUsedConstructorAsObjectReceiver}(c,B,Y)\, \mapsto \,\bot \\[2pt]
 \mathtt{IsUsedConstructorAsMethodParameter}(B,c,Y)\, \mapsto \,\bot \\[2pt]
 \mathtt{IsUsedConstructorAsInitializer}(B,c,Y)\, \mapsto \,\bot \\[2pt]
 \mathtt{IsUsedConstructorAsObjectReceiver}(B,c,Y)\, \mapsto \,\bot \\[2pt]
 \mathtt{IsSubType}(B,c)\, \mapsto \,\bot \\[2pt]
 \mathtt{ExtendsDirectly}(B,c)\, \mapsto \,\bot \\[2pt]
 \mathtt{MethodIsUsedWithType}(B,Y,[P],[c])\, \mapsto \,\bot \\[2pt]
 \mathtt{MethodIsUsedWithType}(B,Y,[P],[c;T1])\, \mapsto \,\bot \\[2pt]
 \mathtt{MethodIsUsedWithType}(B,Y,[P],[T1;c])\, \mapsto \,\bot \\[2pt]
 \mathtt{MethodIsUsedWithType}(B,Y,[P],[T1;c;T2])\, \mapsto \,\bot \\[2pt]
 \mathtt{MethodIsUsedWithType}(B,Y,[P],[T1;T2;c])\, \mapsto \,\bot \\[2pt]
 \mathtt{AllInvokedMethodsOnObjectOInBodyOfMAreDeclaredInC}(c,T1,T2,T3)\, \mapsto \,\top \\[2pt]
 \mathtt{AllInvokedMethodsWithParameterOInBodyOfMAreNotOverloaded}(c,T1,T2)\, \mapsto \,\top \\[2pt]
 \mathtt{BoundVariableInMethodBody}(c,T1,T2)\, \mapsto \,\bot \\[2pt]
 \mathtt{ExistsField}(c,F)\, \mapsto \,\bot \\[2pt]
 \mathtt{ExistsMethodInvocation}(c,Y,T1,X)\, \mapsto \,\bot \\[2pt]
 \mathtt{ExistsAbstractMethod}(c,Y)\, \mapsto \,\bot \\[2pt]
 \mathtt{IsIndirectlyRecursive}(c,Y)\, \mapsto \,\bot \\[2pt]
 \mathtt{IsVisibleMethod}(c,Y,[T1],B)\, \mapsto \,\bot \\[2pt]
 \mathtt{IsVisibleMethod}(c,Y,[T1;T2],B)\, \mapsto \,\bot \\[2pt]
 \mathtt{IsVisibleMethod}(c,Y,[T1;T2;T3],B)\, \mapsto \,\bot \\[2pt]
 \mathtt{IsVisibleMethod}(B,Y,[B],c)\, \mapsto \,\bot \\[2pt]
 \mathtt{IsVisibleMethod}(B,Y,[B;T1],c)\, \mapsto \,\bot \\[2pt]
 \mathtt{IsVisibleMethod}(B,Y,[T1;B],c)\, \mapsto \,\bot \\[2pt]
 \mathtt{IsVisibleMethod}(B,Y,[T1;T2;B],c)\, \mapsto \,\bot \\[2pt]
 \mathtt{IsVisibleMethod}(B,Y,[T1;B;T2],c)\, \mapsto \,\bot \\[2pt]
 \mathtt{IsVisibleMethod}(B,Y,[B;T1;T2],c)\, \mapsto \,\bot \\[2pt]
 \mathtt{IsInverter}(c,Y,T1,T2)\, \mapsto \,\bot \\[2pt]
 \mathtt{IsDelegator}(c,Y,X)\, \mapsto \,\bot \\[2pt]
 \mathtt{IsAbstractClass}(c)\, \mapsto \,\bot \\[2pt]
 \mathtt{IsUsedMethodIn}(c,Y,B)\, \mapsto \,\bot \\[2pt]
 \mathtt{IsUsedMethodIn}(B,Y,c)\, \mapsto \,\bot \\[2pt]
 \mathtt{IsPrimitiveType}(c)\, \mapsto \,\bot \\[2pt]
 \mathtt{IsPublic}(c,Y)\, \mapsto \,\bot \\[2pt]
 \mathtt{IsProtected}(c,Y)\, \mapsto \,\bot \\[2pt]
 \mathtt{IsPrivate}(c,Y)\, \mapsto \,\bot \\[2pt]
 \mathtt{IsUsedAttributeInMethodBody}(c,X,Y)\, \mapsto \,\bot \\[2pt]
 \mathtt{IsGenericsSubtype}(c,[T1],B,[T2])\, \mapsto \,\bot \\[2pt]
 \mathtt{IsGenericsSubtype}(c,[T1;T2],B,[T4;T3])\, \mapsto \,\bot \\[2pt]
 \mathtt{IsGenericsSubtype}(c,[T1;T2;T3],B,[T4;T5;T6])\, \mapsto \,\bot \\[2pt]
 \mathtt{IsInheritedField}(c,F)\, \mapsto \,\bot \\[2pt]
 \mathtt{IsOverridden}(c,Y)\, \mapsto \,\bot \\[2pt]
 \mathtt{IsOverloaded}(c,Y)\, \mapsto \,\bot \\[2pt]
 \mathtt{IsOverriding}(c,Y)\, \mapsto \,\bot \\[2pt]
 \mathtt{IsRecursiveMethod}(c,Y)\, \mapsto \,\bot \\[2pt]
 \mathtt{IsRecursiveMethod}(c,Y)\, \mapsto \,\bot \\[2pt]
 \mathtt{HasReturnType}(c,Y,T1)\, \mapsto \,\bot \\[2pt]
 \mathtt{HasParameterType}(c,T1)\, \mapsto \,\bot \\[2pt]
 \mathtt{HasParameterType}(B,c)\, \mapsto \,\bot \\[2pt]
 \mathtt{MethodHasParameterType}(c,Y,T1)\, \mapsto \,\bot \\[2pt]
 \mathtt{AllSubclasses}(c,[C1;C2;C3])\, \mapsto \,\bot \\[2pt]
 \mathtt{ExtendsDirectly}(c,java.lang.Object)\, \mapsto \,\top \\[2pt]
 $

\subsection{CreateIndirectionInSuperClass}%%%%%%%%%%%%%%%%%%%%%%%%%%%%%%%%%%%%%%%%%%%%%%%%%%%%%%%%%%
\label{def-CreateIndirectionInSuperClass}

\paragraph{Overview:} \textsf{CreateIndirectionInSuperclass (classname s, subclasses [a,b], methodname m,types [t,t'], returntype q, newname n) }: this operation is used to 
create an indirection of the method  s::m to the method n in all the hierarchy.

\begin{center}
\includegraphics[scale=0.7]{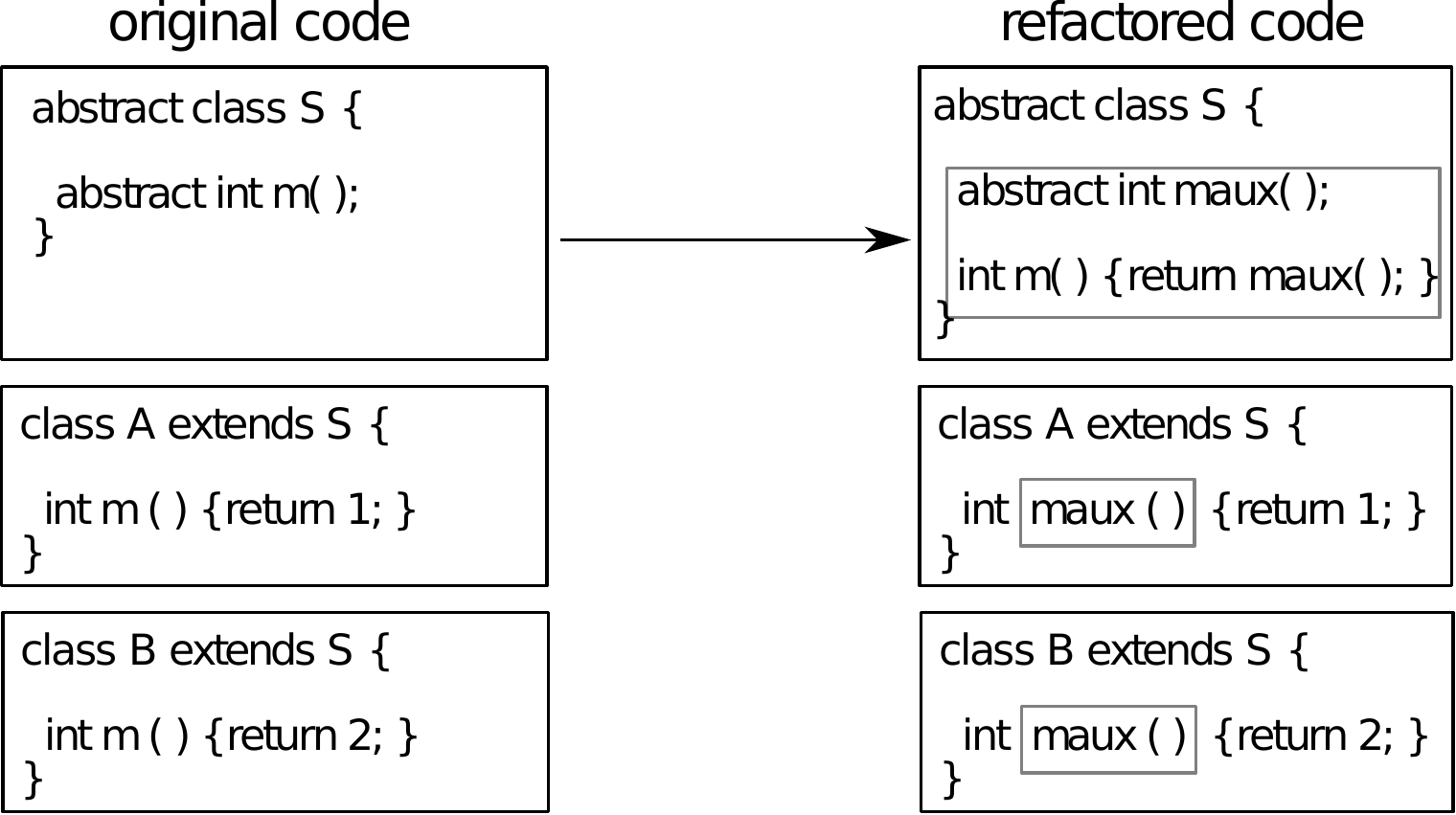}
\end{center}

\tools
With \intellij: \begin{itemize}

\item 
Use \emph{Change Signature} on the method \textsf{m} in
class \textsf{s} (select ``delegate via overloading
method'', specify the new name \textsf{n}, specify the desired visibility).
\end{itemize}

With Eclipse:\begin{itemize}

\item

Use \emph{Change Method Signature} on the method \textsf{m}
in class \textsf{s} (specify to ``keep original
method as delegate to changed method'', and specify the new
name \textsf{n}).

\end{itemize}

\precondition  
$ \\(\mathtt{ExistsClass}(s)\\ 
\wedge \mathtt{IsAbstractClass}(s)\\ 
\wedge \mathtt{ExistsMethodDefinitionWithParams}(s,m,[t;t'])\\ 
\wedge \mathtt{ExistsAbstractMethod}(s,m)\\ 
\wedge \neg \mathtt{IsInheritedMethod}(s,n)\\ 
\wedge \neg \mathtt{IsInheritedMethodWithParams}(s,n,[t;t'])\\ 
\wedge \neg \mathtt{ExistsMethodDefinitionWithParams}(s,n,[t;t'])\\ 
\wedge \mathtt{AllSubclasses}(s,[a;b])\\ 
\wedge \mathtt{HasReturnType}(s,m,r)\\ 
\wedge \neg \mathtt{IsPrivate}(s,m)\\ 
\wedge \neg \mathtt{IsPrivate}(a,m)\\ 
\wedge \neg \mathtt{IsPrivate}(b,m)\\ 
\wedge \mathtt{ExistsMethodDefinition}(s,m)\\ 
\wedge \mathtt{ExistsMethodDefinition}(a,m)\\ 
\wedge \mathtt{ExistsMethodDefinition}(b,m)\\ 
\wedge \neg \mathtt{ExistsMethodDefinition}(s,n)\\ 
\wedge \neg \mathtt{ExistsMethodDefinition}(a,n)\\ 
\wedge \neg \mathtt{ExistsMethodDefinition}(b,n))$ \\

\backdescr
$\\ \mathtt{ExistsAbstractMethod}(s,n)\, \mapsto \,\top \\[2pt]
 \mathtt{ExistsAbstractMethod}(s,m)\, \mapsto \,\bot \\[2pt]
 \mathtt{IsDelegator}(s,m,n)\, \mapsto \,\top \\[2pt]
 \mathtt{IsInheritedMethodWithParams}(s,n,[t;t'])\, \mapsto \,\bot \\[2pt]
 \mathtt{IsOverriding}(s,n)\, \mapsto \,\bot \\[2pt]
 \mathtt{ExistsType}(r)\, \mapsto \,\top \\[2pt]
 \mathtt{HasReturnType}(s,n,r)\, \mapsto \,\mathtt{HasReturnType}(s,m,r)\\[2pt]
 \mathtt{HasReturnType}(a,n,r)\, \mapsto \,\mathtt{HasReturnType}(s,m,r)\\[2pt]
 \mathtt{HasReturnType}(b,n,r)\, \mapsto \,\mathtt{HasReturnType}(s,m,r)\\[2pt]
 \mathtt{ExistsMethodDefinition}(s,n)\, \mapsto \,\top \\[2pt]
 \mathtt{ExistsMethodDefinition}(a,n)\, \mapsto \,\top \\[2pt]
 \mathtt{ExistsMethodDefinition}(b,n)\, \mapsto \,\top \\[2pt]
 \mathtt{ExistsMethodDefinitionWithParams}(s,n,[t;t'])\, \mapsto \,\top \\[2pt]
 \mathtt{ExistsMethodDefinitionWithParams}(a,n,[t;t'])\, \mapsto \,\top \\[2pt]
 \mathtt{ExistsMethodDefinitionWithParams}(b,n,[t;t'])\, \mapsto \,\top \\[2pt]
 \mathtt{ExistsParameterWithName}(s,n,[t;t'],N)\, \mapsto \,\bot \\[2pt]
 \mathtt{ExistsParameterWithName}(a,n,[t;t'],N)\, \mapsto \,\bot \\[2pt]
 \mathtt{ExistsParameterWithName}(b,n,[t;t'],N)\, \mapsto \,\bot \\[2pt]
 \mathtt{ExistsParameterWithName}(s,n,[V],N)\, \mapsto \,\mathtt{ExistsParameterWithName}(s,m,[V],N)\\[2pt]
 \mathtt{ExistsParameterWithName}(a,n,[V],N)\, \mapsto \,\mathtt{ExistsParameterWithName}(a,m,[V],N)\\[2pt]
 \mathtt{ExistsParameterWithName}(b,n,[V],N)\, \mapsto \,\mathtt{ExistsParameterWithName}(b,m,[V],N)\\[2pt]
 \mathtt{ExistsMethodDefinition}(a,m)\, \mapsto \,\bot \\[2pt]
 \mathtt{ExistsMethodDefinition}(b,m)\, \mapsto \,\bot \\[2pt]
 \mathtt{ExistsMethodDefinitionWithParams}(a,m,[t;t'])\, \mapsto \,\bot \\[2pt]
 \mathtt{ExistsMethodDefinitionWithParams}(b,m,[t;t'])\, \mapsto \,\bot \\[2pt]
 \mathtt{IsIndirectlyRecursive}(a,n)\, \mapsto \,\mathtt{IsRecursiveMethod}(a,m)\\[2pt]
 \mathtt{IsIndirectlyRecursive}(b,n)\, \mapsto \,\mathtt{IsRecursiveMethod}(b,m)\\[2pt]
 \mathtt{ExistsMethodInvocation}(a,n,s,m)\, \mapsto \,\mathtt{IsRecursiveMethod}(a,m)\\[2pt]
 \mathtt{ExistsMethodInvocation}(b,n,s,m)\, \mapsto \,\mathtt{IsRecursiveMethod}(b,m)\\[2pt]
 \mathtt{ExistsMethodInvocation}(s,m,a,n)\, \mapsto \,\top \\[2pt]
 \mathtt{ExistsMethodInvocation}(s,m,b,n)\, \mapsto \,\top \\[2pt]
 \mathtt{BoundVariableInMethodBody}(s,n,V)\, \mapsto \,\mathtt{BoundVariableInMethodBody}(s,m,V)\\[2pt]
 \mathtt{BoundVariableInMethodBody}(a,n,V)\, \mapsto \,\mathtt{BoundVariableInMethodBody}(a,m,V)\\[2pt]
 \mathtt{BoundVariableInMethodBody}(b,n,V)\, \mapsto \,\mathtt{BoundVariableInMethodBody}(b,m,V)\\[2pt]
 \mathtt{IsOverloaded}(s,n)\, \mapsto \,\bot \\[2pt]
 \mathtt{IsOverloaded}(a,n)\, \mapsto \,\bot \\[2pt]
 \mathtt{IsOverloaded}(b,n)\, \mapsto \,\bot \\[2pt]
 \mathtt{IsOverridden}(s,n)\, \mapsto \,\bot \\[2pt]
 \mathtt{IsOverridden}(a,n)\, \mapsto \,\mathtt{IsOverridden}(a,m)\\[2pt]
 \mathtt{IsOverridden}(b,n)\, \mapsto \,\mathtt{IsOverridden}(b,m)\\[2pt]
 \mathtt{IsOverriding}(a,n)\, \mapsto \,\mathtt{IsOverriding}(a,m)\\[2pt]
 \mathtt{IsOverriding}(b,n)\, \mapsto \,\mathtt{IsOverriding}(b,m)\\[2pt]
 \mathtt{IsRecursiveMethod}(s,n)\, \mapsto \,\bot \\[2pt]
 \mathtt{IsRecursiveMethod}(a,n)\, \mapsto \,\bot \\[2pt]
 \mathtt{IsRecursiveMethod}(b,n)\, \mapsto \,\bot \\[2pt]
 \mathtt{AllInvokedMethodsOnObjectOInBodyOfMAreDeclaredInC}(s,n,N,V)\, \mapsto \, \\ ~\hfill \mathtt{AllInvokedMethodsOnObjectOInBodyOfMAreDeclaredInC}(s,m,N,V)\\[2pt]
 \mathtt{AllInvokedMethodsOnObjectOInBodyOfMAreDeclaredInC}(a,n,N,V)\, \mapsto \, \\ ~\hfill \mathtt{AllInvokedMethodsOnObjectOInBodyOfMAreDeclaredInC}(a,m,N,V)\\[2pt]
 \mathtt{AllInvokedMethodsOnObjectOInBodyOfMAreDeclaredInC}(b,n,N,V)\, \mapsto \, \\ ~\hfill \mathtt{AllInvokedMethodsOnObjectOInBodyOfMAreDeclaredInC}(b,m,N,V)\\[2pt]
 \mathtt{AllInvokedMethodsWithParameterOInBodyOfMAreNotOverloaded}(s,n,N)\, \mapsto \, \\ ~\hfill \mathtt{AllInvokedMethodsWithParameterOInBodyOfMAreNotOverloaded}(s,m,N)\\[2pt]
 \mathtt{AllInvokedMethodsWithParameterOInBodyOfMAreNotOverloaded}(a,n,N)\, \mapsto \, \\ ~\hfill \mathtt{AllInvokedMethodsWithParameterOInBodyOfMAreNotOverloaded}(a,m,N)\\[2pt]
 \mathtt{AllInvokedMethodsWithParameterOInBodyOfMAreNotOverloaded}(b,n,N)\, \mapsto \, \\ ~\hfill \mathtt{AllInvokedMethodsWithParameterOInBodyOfMAreNotOverloaded}(b,m,N)\\[2pt]
 \mathtt{IsPrivate}(s,V)\, \mapsto \,\bot \\[2pt]
 \mathtt{IsPrivate}(a,V)\, \mapsto \,\bot \\[2pt]
 \mathtt{IsPrivate}(b,V)\, \mapsto \,\bot \\[2pt]
 \mathtt{IsPrivate}(s,n)\, \mapsto \,\bot \\[2pt]
 \mathtt{IsPrivate}(a,n)\, \mapsto \,\bot \\[2pt]
 \mathtt{IsPrivate}(b,n)\, \mapsto \,\bot \\[2pt]
 \mathtt{IsOverriding}(a,n)\, \mapsto \,\mathtt{IsOverriding}(a,m)\\[2pt]
 \mathtt{IsOverriding}(b,n)\, \mapsto \,\mathtt{IsOverriding}(b,m)\\[2pt]
 \mathtt{IsDelegator}(a,n,V)\, \mapsto \,\mathtt{IsDelegator}(a,m,V)\\[2pt]
 \mathtt{IsDelegator}(b,n,V)\, \mapsto \,\mathtt{IsDelegator}(b,m,V)\\[2pt]
 \mathtt{IsDelegator}(a,V,n)\, \mapsto \,\mathtt{IsDelegator}(a,V,m)\\[2pt]
 \mathtt{IsDelegator}(b,V,n)\, \mapsto \,\mathtt{IsDelegator}(b,V,m)\\[2pt]
 \mathtt{IsInheritedMethodWithParams}(a,n,[t;t'])\, \mapsto \,\mathtt{IsVisibleMethod}(s,m,[t;t'],a)\\[2pt]
 \mathtt{IsInheritedMethodWithParams}(b,n,[t;t'])\, \mapsto \,\mathtt{IsVisibleMethod}(s,m,[t;t'],b)\\[2pt]
 \mathtt{IsVisibleMethod}(s,m,[t;t'],a)\, \mapsto \,\top \\[2pt]
 \mathtt{IsVisibleMethod}(s,m,[t;t'],b)\, \mapsto \,\top \\[2pt]
 \mathtt{MethodIsUsedWithType}(a,n,[t;t'],[t;t'])\, \mapsto \,\mathtt{MethodIsUsedWithType}(a,m,[t;t'],[t;t'])\\[2pt]
 \mathtt{MethodIsUsedWithType}(b,n,[t;t'],[t;t'])\, \mapsto \,\mathtt{MethodIsUsedWithType}(b,m,[t;t'],[t;t'])\\[2pt]
 \mathtt{IsUsedMethod}(a,n,[t;t'])\, \mapsto \,\mathtt{IsUsedMethod}(a,m,[t;t'])\\[2pt]
 \mathtt{IsUsedMethod}(b,n,[t;t'])\, \mapsto \,\mathtt{IsUsedMethod}(b,m,[t;t'])\\[2pt]
 \mathtt{IsUsedMethodIn}(a,n,V)\, \mapsto \,\mathtt{IsUsedMethodIn}(a,m,V)\\[2pt]
 \mathtt{IsUsedMethodIn}(b,n,V)\, \mapsto \,\mathtt{IsUsedMethodIn}(b,m,V)\\[2pt]
 \mathtt{IsInverter}(a,n,V,V1)\, \mapsto \,\mathtt{IsInverter}(a,m,V,V1)\\[2pt]
 \mathtt{IsInverter}(b,n,V,V1)\, \mapsto \,\mathtt{IsInverter}(b,m,V,V1)\\[2pt]
 \mathtt{ExistsMethodInvocation}(a,V,V1,n)\, \mapsto \,\mathtt{ExistsMethodInvocation}(a,V,V1,m)\\[2pt]
 \mathtt{ExistsMethodInvocation}(b,V,V1,n)\, \mapsto \,\mathtt{ExistsMethodInvocation}(b,V,V1,m)\\[2pt]
 \mathtt{IsIndirectlyRecursive}(a,n)\, \mapsto \,\mathtt{IsIndirectlyRecursive}(a,m)\\[2pt]
 \mathtt{IsIndirectlyRecursive}(b,n)\, \mapsto \,\mathtt{IsIndirectlyRecursive}(b,m)\\[2pt]
 \mathtt{BoundVariableInMethodBody}(a,n,V)\, \mapsto \,\mathtt{BoundVariableInMethodBody}(a,n,V)\\[2pt]
 \mathtt{BoundVariableInMethodBody}(b,n,V)\, \mapsto \,\mathtt{BoundVariableInMethodBody}(b,n,V)\\[2pt]
 \mathtt{IsOverridden}(a,n)\, \mapsto \,\mathtt{IsOverridden}(a,m)\\[2pt]
 \mathtt{IsOverridden}(b,n)\, \mapsto \,\mathtt{IsOverridden}(b,m)\\[2pt]
 $

\subsection{AddParameter}%%%%%%%%%%%%%%%%%%%%%%%%%%%%%%%%%%%%%%%%%%%%%%%%%%%%%%%%%%%%%%%%%%%%%%%%%%%%
\label{def-AddParameter}
(\emph{Add Parameter} in Fowler~\cite{Fowler1999} et~\cite{koch2002})

\textsf{AddParameter(class c, method m, parameterType t, parameterName n, \textbf{defaultvalue} e)}:

Add a parameter of type \textsf{t} to a method \textsf{m} in class \textsf{c}. In method invocations, use the expression \textsf{e} as new parameter.

\begin{center}
\includegraphics[scale=0.7]{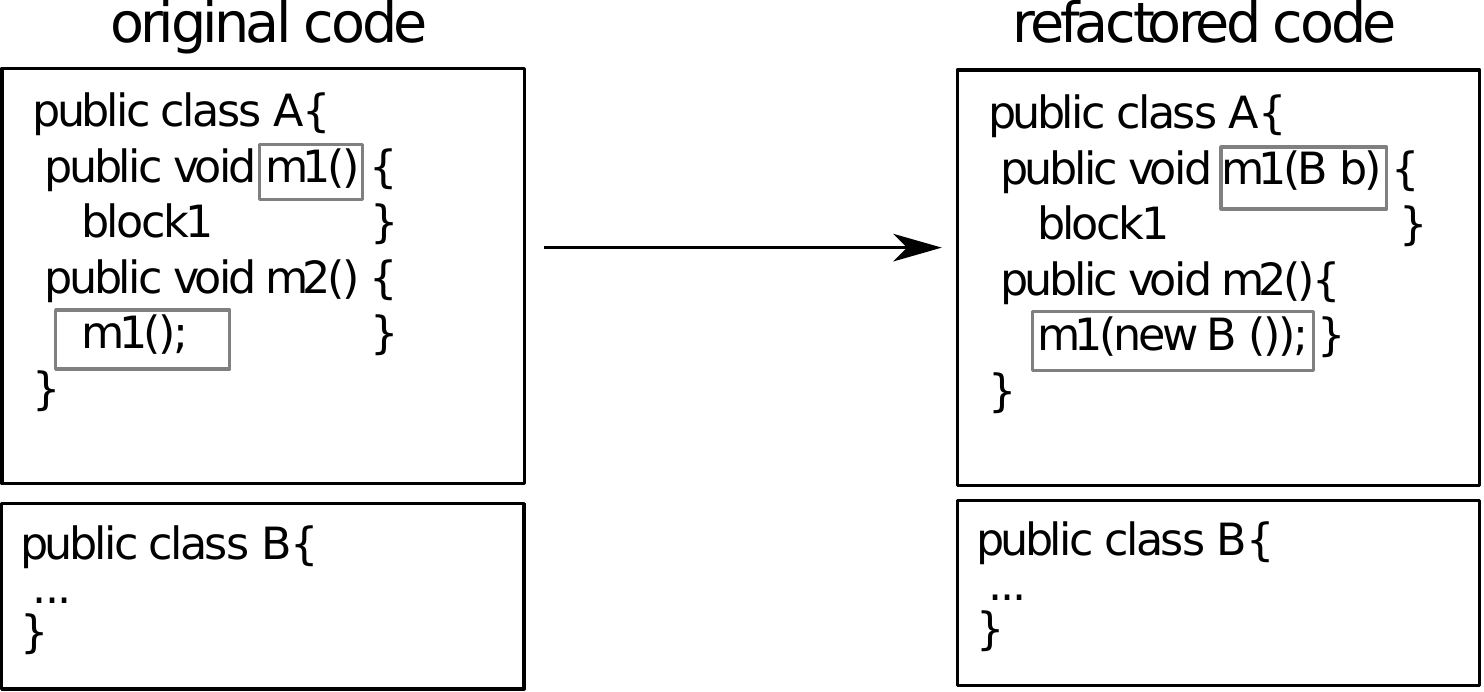}
\end{center}

\tools \emph{Change Method signature} in Eclipse tool and \emph{Change Signature} in \intellij.

\subsection{AddParameterWithReuse}%%%%%%%%%%%%%%%%%%%%%%%%%%%%%%%%%%%%%%%%%%%%%%%%%%%%%%%%%%%%%%%%%%%%%%%%%%%%

\paragraph{Overview:} \textsf{AddParameterWithReuse (classname s,subclasses [a,b], methodname m,methodparameters [] ,paramType t, paramName p,~usedvalueofparamType defaultvalue)}: this operation is used to add  the parameter p of type t to the parameters of the method s::m, a::m and b::m. 
Same as \textsf{AddParameter}, but instead of adding a default value for the additional parameter in invocations, use any value with the specified type that is visible from the invocation site. 

In \intellij, this is specified with the \emph{Any Var}
option in \emph{Change Signature}. This is not supported by
Eclipse.

Note that when several variables of the specified type are
visible, the result in unspecified. In the example of use in
this report, the type of the added parameter is a fresh
type, and in recursive methods, the only variable of this
type is the parameter being introduced so that there is not
ambiguity.

\precondition $\\ (\neg \mathtt{BoundVariableInMethodBody}(s,m,p)\\ 
\wedge \mathtt{ExistsClass}(s)\\ 
\wedge \mathtt{ExistsMethodDefinition}(s,m)\\ 
\wedge \mathtt{ExistsMethodDefinitionWithParams}(s,m,[\,])\\ 
\wedge \neg \mathtt{ExistsMethodDefinitionWithParams}(s,m,[t])\\ 
\wedge \neg \mathtt{IsInheritedMethodWithParams}(s,m,[t])\\ 
\wedge \neg \mathtt{ExistsParameterWithName}(s,m,[\,],p)\\ 
\wedge \mathtt{ExistsType}(t)\\ 
\wedge \mathtt{AllSubclasses}(s,[a;b]))$ \\ 

\backdescr
 $\mathtt{ExistsMethodDefinitionWithParams}(s,m,[\,])\, \mapsto \,\bot \\[2pt]
 \mathtt{ExistsMethodDefinitionWithParams}(s,m,[t])\, \mapsto \,\top \\[2pt]
 \mathtt{ExistsMethodDefinitionWithParams}(a,m,[\,])\, \mapsto \,\neg \mathtt{ExistsMethodDefinitionWithParams}(a,m,[\,])\\[2pt]
 \mathtt{ExistsMethodDefinitionWithParams}(b,m,[\,])\, \mapsto \,\neg \mathtt{ExistsMethodDefinitionWithParams}(b,m,[\,])\\[2pt]
 \mathtt{ExistsMethodDefinitionWithParams}(a,m,[t])\, \mapsto \,\mathtt{ExistsMethodDefinitionWithParams}(a,m,[\,])\\[2pt]
 \mathtt{ExistsMethodDefinitionWithParams}(b,m,[t])\, \mapsto \,\mathtt{ExistsMethodDefinitionWithParams}(b,m,[\,])\\[2pt]
 \mathtt{ExistsParameterWithName}(s,m,[t],p)\, \mapsto \,\top \\[2pt]
 \mathtt{ExistsParameterWithName}(a,m,[t],p)\, \mapsto \,\top \\[2pt]
 \mathtt{ExistsParameterWithName}(b,m,[t],p)\, \mapsto \,\top \\[2pt]
 \mathtt{ExistsParameterWithType}(s,m,[t],t)\, \mapsto \,\top \\[2pt]
 \mathtt{ExistsParameterWithType}(a,m,[t],t)\, \mapsto \,\top \\[2pt]
 \mathtt{ExistsParameterWithType}(b,m,[t],t)\, \mapsto \,\top \\[2pt]
 \mathtt{AllInvokedMethodsOnObjectOInBodyOfMAreDeclaredInC}(s,m,p,T)\, \mapsto \,\top \\[2pt]
 \mathtt{AllInvokedMethodsOnObjectOInBodyOfMAreDeclaredInC}(a,m,p,T)\, \mapsto \,\top \\[2pt]
 \mathtt{AllInvokedMethodsOnObjectOInBodyOfMAreDeclaredInC}(b,m,p,T)\, \mapsto \,\top \\[2pt]
 \mathtt{AllInvokedMethodsWithParameterOInBodyOfMAreNotOverloaded}(s,m,p)\, \mapsto \, \\ ~\hfill (\neg \mathtt{IsOverloaded}(s,m)\\ \wedge \neg \mathtt{IsOverloaded}(a,m)\\ \wedge \neg \mathtt{IsOverloaded}(b,m))\\[2pt]
 \mathtt{AllInvokedMethodsWithParameterOInBodyOfMAreNotOverloaded}(a,m,p)\, \mapsto \, \\ ~\hfill (\neg \mathtt{IsOverloaded}(s,m)\\ \wedge \neg \mathtt{IsOverloaded}(a,m)\\ \wedge \neg \mathtt{IsOverloaded}(b,m))\\[2pt]
 \mathtt{AllInvokedMethodsWithParameterOInBodyOfMAreNotOverloaded}(b,m,p)\, \mapsto \, \\ ~\hfill (\neg \mathtt{IsOverloaded}(s,m)\\ \wedge \neg \mathtt{IsOverloaded}(a,m)\\ \wedge \neg \mathtt{IsOverloaded}(b,m))\\[2pt]
 \mathtt{IsUsedConstructorAsMethodParameter}(t,s,m)\, \mapsto \,\top \\[2pt]
 \mathtt{IsUsedConstructorAsMethodParameter}(t,a,m)\, \mapsto \,\top \\[2pt]
 \mathtt{IsUsedConstructorAsMethodParameter}(t,b,m)\, \mapsto \,\top \\[2pt]
 $

\subsection{AddParameterWithDelegate}

\begin{center}
\includegraphics[scale=0.7]{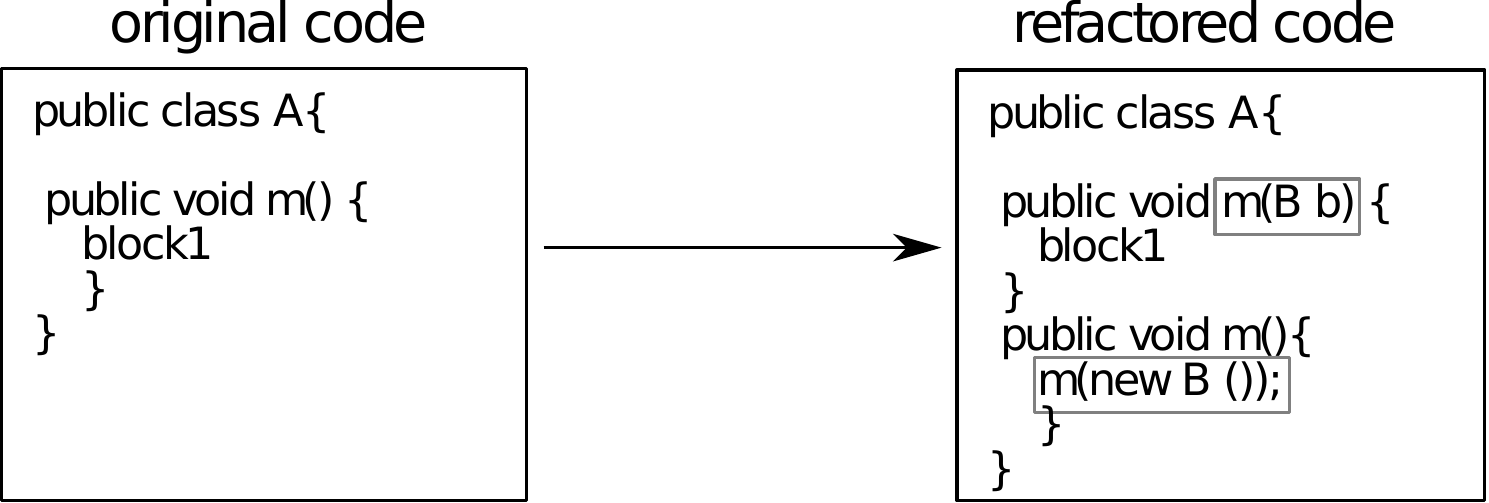}
\end{center}

\subsection{MoveMethod}%%%%%%%%%%%%%%%%%%%%%%%%%%%%%%%%%%%%%%%%%%%%%%%%%%%%%%%%%
\label{def-move}

\begin{center}
\includegraphics[scale=0.7]{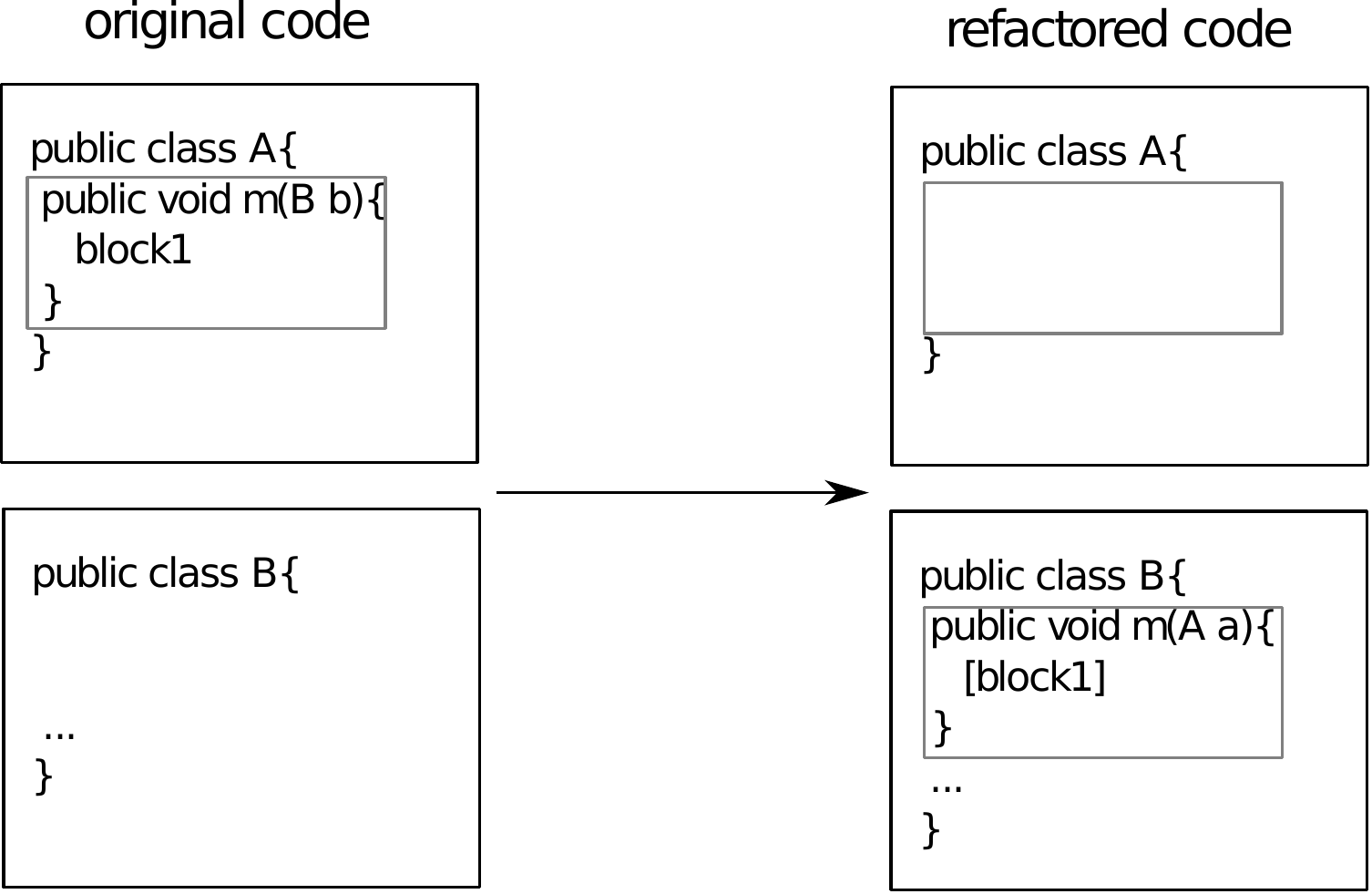}
\end{center}

\tools If the receiver object is not used in the body of the initial class, it will not be included as parameter in the destination class, so that you have to add it (see \emph{AddParameter}).

\subsection{MoveMethodWithDelegate}%%%%%%%%%%%%%%%%%%%%%%%%%%%%%%%%%%%%%%%%%%%%%
\label{def-MoveMethodWithDelegate}
 (\emph{Move Method} in Fowler~\cite{Fowler1999})

\paragraph{Overview:} \textsf{MoveMethodWithDelegate (classname s,attributes [att1,att2], targetclass a,methodtobemoved m, parameterstypes [t,a], returntype r,
movedmethod n, receivingobjectname o, newreceivingobjectname o')}: this operation is used to move the method s::m to the class a and rename it as n.

Transform a method \textsf{m} of a
class \textsf{s} into a delegator to a method \textsf{n} in an other class \textsf{a}.  The code of \textsf{m} has been moved to \texttt{n} (and adapted).

\begin{center}
\includegraphics[scale=0.7]{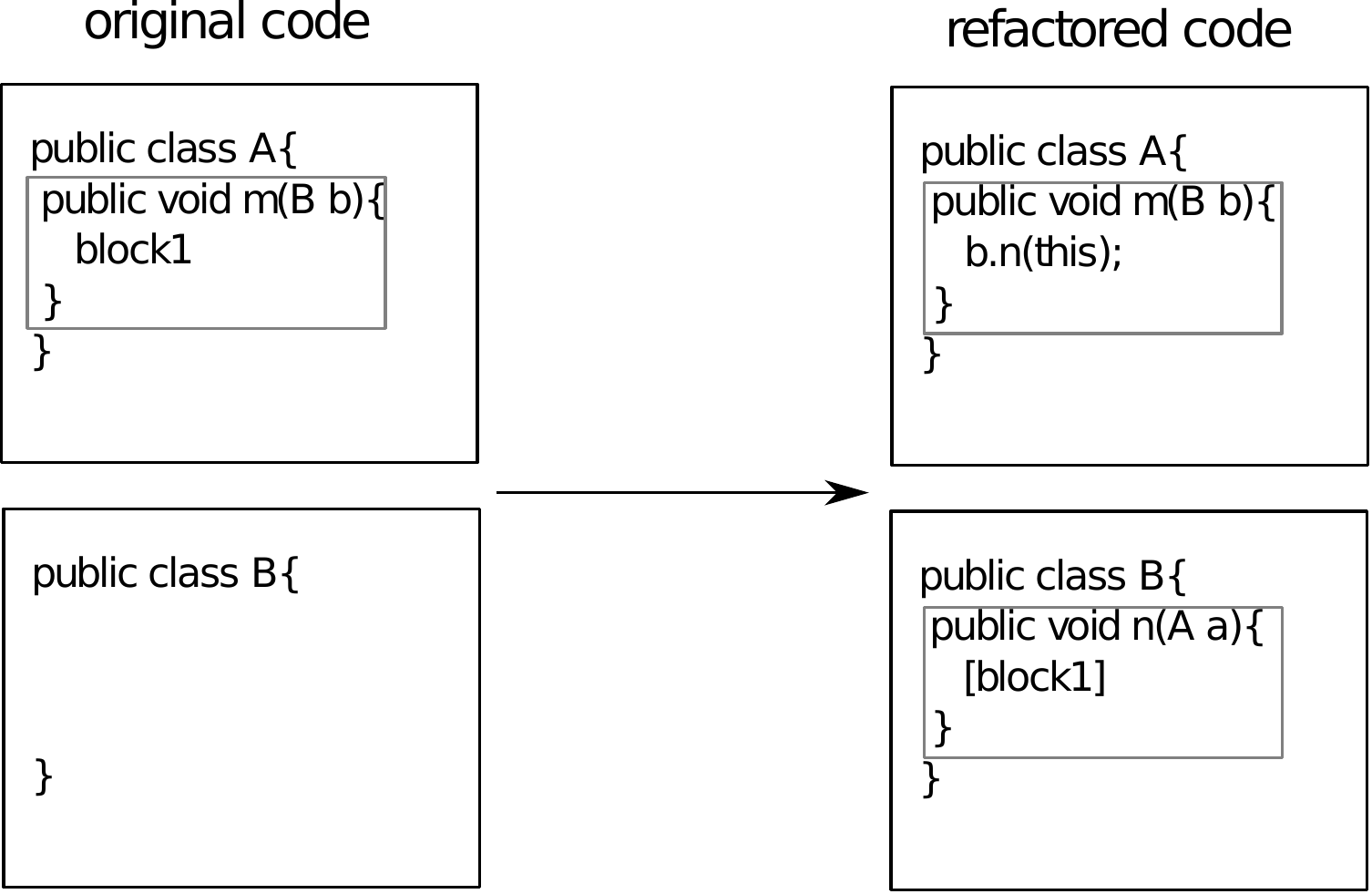}
\end{center}

\tools \emph{Move} in Eclipse tool. In \intellij, first introduce a local delegate (with \emph{Change Signature}), then \emph{Move}.

\precondition 
$\\ (\mathtt{ExistsClass}(s)\\ 
\wedge \mathtt{ExistsClass}(a)\\ 
\wedge \mathtt{ExistsMethodDefinitionWithParams}(s,m,[t;a])\\ 
\wedge \mathtt{ExistsParameterWithType}(s,m,[t;a],a)\\ 
\wedge \mathtt{ExistsParameterWithName}(s,m,[t;a],o)\\ 
\wedge \neg \mathtt{ExistsMethodDefinitionWithParams}(a,n,[t;s])\\ 
\wedge \mathtt{HasReturnType}(s,m,r)\\ 
\wedge \neg \mathtt{IsPrivate}(s,m)\\ 
\wedge \neg \mathtt{IsPrivate}(s,att1)\\ 
\wedge \neg \mathtt{IsPrivate}(s,att2))$ \\

\backdescr
$\\ \mathtt{ExistsMethodDefinitionWithParams}(s,m,[t;a])\, \mapsto \,\top \\[2pt]
 \mathtt{ExistsMethodDefinition}(a,n)\, \mapsto \,\top \\[2pt]
 \mathtt{ExistsMethodDefinitionWithParams}(a,n,[t;s])\, \mapsto \,\top \\[2pt]
 \mathtt{HasReturnType}(a,n,r)\, \mapsto \,\mathtt{HasReturnType}(s,m,r)\\[2pt]
 \mathtt{BoundVariableInMethodBody}(a,n,M)\, \mapsto \,\mathtt{BoundVariableInMethodBody}(s,m,M)\\[2pt]
 \mathtt{ExistsParameterWithName}(a,n,[t;s],N)\, \mapsto \,\mathtt{ExistsParameterWithName}(s,m,[t;a],N) (condition) \\[2pt]
 \mathtt{ExistsParameterWithName}(a,n,[t;s],o')\, \mapsto \,\top \\[2pt]
 \mathtt{ExistsParameterWithName}(a,n,[t;s],o)\, \mapsto \,\bot \\[2pt]
 \mathtt{ExistsParameterWithType}(a,n,[t;s],s)\, \mapsto \,\top \\[2pt]
 \mathtt{ExistsParameterWithType}(a,n,[t;s],a)\, \mapsto \,\bot \\[2pt]
 \mathtt{ExistsParameterWithType}(a,n,[t;s],T)\, \mapsto \,\mathtt{ExistsParameterWithType}(s,m,[t;a],T) (condition) \\[2pt]
 \mathtt{ExistsMethodInvocation}(s,m,a,n)\, \mapsto \,\top \\[2pt]
 \mathtt{IsInverter}(s,m,a,r)\, \mapsto \,\top \\[2pt]
 \mathtt{IsPrivate}(s,att1)\, \mapsto \,\bot \\[2pt]
 \mathtt{IsPrivate}(s,att2)\, \mapsto \,\bot \\[2pt]
 \mathtt{IsUsedAttributeInMethodBody}(s,att1,m)\, \mapsto \,\bot \\[2pt]
 \mathtt{IsUsedAttributeInMethodBody}(s,att2,m)\, \mapsto \,\bot \\[2pt]
 $

%\emph{(eclipse met automatiquement le champ a public, mais pas \intellij)}

\subsection{RenameMethod}%%%%%%%%%%%%%%%%%%%%%%%%%%%%%%%%%%%%%%%%%%%%%%%%%%%%%%
\label{def-Rename}
(\emph{Rename} in Fowler~\cite{Fowler1999} et~\cite{koch2002})

\textsf{RenameMethod(class c, method m, newname n)}:
Rename the method \textsf{m} of class \texttt{c} into \texttt{n}.       

\begin{center}
\includegraphics[scale=0.7]{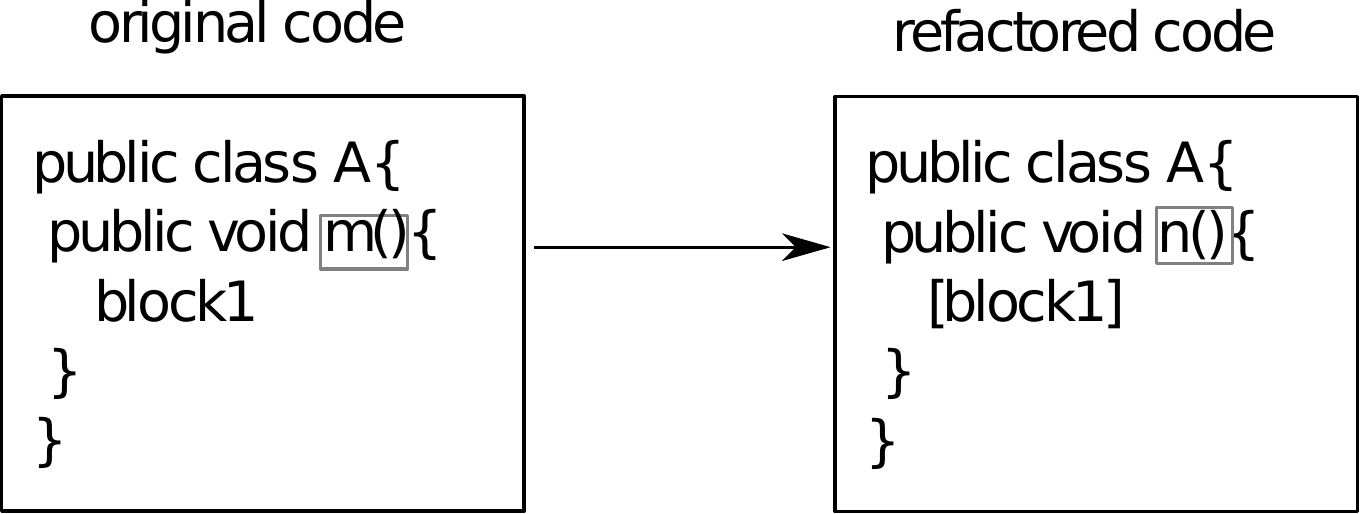}               
\end{center}       
             
\tools \emph{Rename} in Eclipse and \intellij.

We have identified two types of this operation. The first one does not accept the overloading, the second one accepts overloading. 

\subsubsection{RenameInHierarchyNoOverloading}          
  \paragraph{Overview} \textsf{RenameInHierarchyNoOverloading (class c,subclasses [a,b],method m ,types [t,t'], newname n)} : this operation is used to rename the method (c,a,b)::m into n if n does not already exist with another 
signature in the hierarchy.

\precondition 
 $\\ (\mathtt{ExistsClass}(c)\\ 
\wedge \mathtt{ExistsClass}(a)\\ 
\wedge \mathtt{ExistsClass}(b)\\ 
\wedge \mathtt{ExistsMethodDefinition}(c,m)\\ 
\wedge \mathtt{ExistsMethodDefinitionWithParams}(c,m,[t;t'])\\ 
\wedge \mathtt{AllSubclasses}(c,[a;b])\\ 
\wedge \neg \mathtt{ExistsMethodDefinition}(c,n)\\ 
\wedge \neg \mathtt{ExistsMethodDefinition}(a,n)\\ 
\wedge \neg \mathtt{ExistsMethodDefinition}(b,n)\\ 
\wedge \neg \mathtt{ExistsMethodDefinitionWithParams}(c,n,[t;t'])\\ 
\wedge \neg \mathtt{ExistsMethodDefinitionWithParams}(a,n,[t;t'])\\ 
\wedge \neg \mathtt{ExistsMethodDefinitionWithParams}(b,n,[t;t'])\\ 
\wedge \neg \mathtt{IsOverloaded}(c,m)\\ 
\wedge \neg \mathtt{IsOverloaded}(a,m)\\ 
\wedge \neg \mathtt{IsOverloaded}(b,m)\\ 
\wedge \neg \mathtt{IsInheritedMethod}(c,n))$ \\

\backdescr
$\\ \mathtt{ExistsMethodDefinition}(c,n)\, \mapsto \,\top \\[2pt]
 \mathtt{ExistsMethodDefinitionWithParams}(c,n,[t;t'])\, \mapsto \,\top \\[2pt]
 \mathtt{ExistsMethodDefinition}(a,n)\, \mapsto \,\mathtt{ExistsMethodDefinition}(a,m)\\[2pt]
 \mathtt{ExistsMethodDefinition}(b,n)\, \mapsto \,\mathtt{ExistsMethodDefinition}(b,m)\\[2pt]
 \mathtt{ExistsMethodDefinition}(c,m)\, \mapsto \,\bot \\[2pt]
 \mathtt{ExistsMethodDefinition}(a,m)\, \mapsto \,\bot \\[2pt]
 \mathtt{ExistsMethodDefinition}(b,m)\, \mapsto \,\bot \\[2pt]
 \mathtt{ExistsMethodDefinitionWithParams}(a,n,[t;t'])\, \mapsto \,\mathtt{ExistsMethodDefinitionWithParams}(a,m,[t;t'])\\[2pt]
 \mathtt{ExistsMethodDefinitionWithParams}(b,n,[t;t'])\, \mapsto \,\mathtt{ExistsMethodDefinitionWithParams}(b,m,[t;t'])\\[2pt]
 \mathtt{ExistsMethodDefinitionWithParams}(c,m,[t;t'])\, \mapsto \,\bot \\[2pt]
 \mathtt{ExistsMethodDefinitionWithParams}(a,m,[t;t'])\, \mapsto \,\bot \\[2pt]
 \mathtt{ExistsMethodDefinitionWithParams}(b,m,[t;t'])\, \mapsto \,\bot \\[2pt]
 \mathtt{IsInheritedMethod}(a,n)\, \mapsto \,\mathtt{IsInheritedMethod}(a,m)\\[2pt]
 \mathtt{IsInheritedMethod}(b,n)\, \mapsto \,\mathtt{IsInheritedMethod}(b,m)\\[2pt]
 \mathtt{IsDelegator}(c,n,V)\, \mapsto \,\mathtt{IsDelegator}(c,m,V)\\[2pt]
 \mathtt{IsDelegator}(a,n,V)\, \mapsto \,\mathtt{IsDelegator}(a,m,V)\\[2pt]
 \mathtt{IsDelegator}(b,n,V)\, \mapsto \,\mathtt{IsDelegator}(b,m,V)\\[2pt]
 \mathtt{IsDelegator}(c,V,n)\, \mapsto \,\mathtt{IsDelegator}(c,V,m)\\[2pt]
 \mathtt{IsDelegator}(a,V,n)\, \mapsto \,\mathtt{IsDelegator}(a,V,m)\\[2pt]
 \mathtt{IsDelegator}(b,V,n)\, \mapsto \,\mathtt{IsDelegator}(b,V,m)\\[2pt]
 \mathtt{IsOverloaded}(c,V)\, \mapsto \,\mathtt{IsOverloaded}(c,V) (condition) \\[2pt]
 \mathtt{IsOverloaded}(a,V)\, \mapsto \,\mathtt{IsOverloaded}(a,V) (condition) \\[2pt]
 \mathtt{IsOverloaded}(b,V)\, \mapsto \,\mathtt{IsOverloaded}(b,V) (condition) \\[2pt]
 \mathtt{IsOverriding}(a,n)\, \mapsto \,\mathtt{IsOverriding}(a,m)\\[2pt]
 \mathtt{IsOverriding}(b,n)\, \mapsto \,\mathtt{IsOverriding}(b,m)\\[2pt]
 \mathtt{IsOverridden}(a,n)\, \mapsto \,\mathtt{IsOverridden}(a,m)\\[2pt]
 \mathtt{IsOverridden}(b,n)\, \mapsto \,\mathtt{IsOverridden}(b,m)\\[2pt]
 \mathtt{ExistsParameterWithName}(c,n,[t;t'],V1)\, \mapsto \,\mathtt{ExistsParameterWithName}(c,m,[t;t'],V1)\\[2pt]
 \mathtt{ExistsParameterWithName}(a,n,[t;t'],V1)\, \mapsto \,\mathtt{ExistsParameterWithName}(a,m,[t;t'],V1)\\[2pt]
 \mathtt{ExistsParameterWithName}(b,n,[t;t'],V1)\, \mapsto \,\mathtt{ExistsParameterWithName}(b,m,[t;t'],V1)\\[2pt]
 \mathtt{ExistsParameterWithType}(c,n,[t;t'],V1)\, \mapsto \,\mathtt{ExistsParameterWithType}(c,m,[t;t'],V1)\\[2pt]
 \mathtt{ExistsParameterWithType}(a,n,[t;t'],V1)\, \mapsto \,\mathtt{ExistsParameterWithType}(a,m,[t;t'],V1)\\[2pt]
 \mathtt{ExistsParameterWithType}(b,n,[t;t'],V1)\, \mapsto \,\mathtt{ExistsParameterWithType}(b,m,[t;t'],V1)\\[2pt]
 \mathtt{IsRecursiveMethod}(c,n)\, \mapsto \,\mathtt{IsRecursiveMethod}(c,m)\\[2pt]
 \mathtt{IsRecursiveMethod}(a,n)\, \mapsto \,\mathtt{IsRecursiveMethod}(a,m)\\[2pt]
 \mathtt{IsRecursiveMethod}(b,n)\, \mapsto \,\mathtt{IsRecursiveMethod}(b,m)\\[2pt]
 \mathtt{ExistsAbstractMethod}(c,n)\, \mapsto \,\mathtt{ExistsAbstractMethod}(c,m)\\[2pt]
 \mathtt{ExistsAbstractMethod}(a,n)\, \mapsto \,\mathtt{ExistsAbstractMethod}(a,m)\\[2pt]
 \mathtt{ExistsAbstractMethod}(b,n)\, \mapsto \,\mathtt{ExistsAbstractMethod}(b,m)\\[2pt]
 \mathtt{IsInheritedMethodWithParams}(a,n,[t;t'])\, \mapsto \,\mathtt{IsVisibleMethod}(c,m,[t;t'],a)\\[2pt]
 \mathtt{IsInheritedMethodWithParams}(b,n,[t;t'])\, \mapsto \,\mathtt{IsVisibleMethod}(c,m,[t;t'],b)\\[2pt]
 \mathtt{MethodIsUsedWithType}(c,n,[t;t'],[t;t'])\, \mapsto \,\mathtt{MethodIsUsedWithType}(c,m,[t;t'],[t;t'])\\[2pt]
 \mathtt{MethodIsUsedWithType}(a,n,[t;t'],[t;t'])\, \mapsto \,\mathtt{MethodIsUsedWithType}(a,m,[t;t'],[t;t'])\\[2pt]
 \mathtt{MethodIsUsedWithType}(b,n,[t;t'],[t;t'])\, \mapsto \,\mathtt{MethodIsUsedWithType}(b,m,[t;t'],[t;t'])\\[2pt]
 \mathtt{IsUsedMethod}(c,n,[t;t'])\, \mapsto \,\mathtt{IsUsedMethod}(c,m,[t;t'])\\[2pt]
 \mathtt{IsUsedMethod}(a,n,[t;t'])\, \mapsto \,\mathtt{IsUsedMethod}(a,m,[t;t'])\\[2pt]
 \mathtt{IsUsedMethod}(b,n,[t;t'])\, \mapsto \,\mathtt{IsUsedMethod}(b,m,[t;t'])\\[2pt]
 \mathtt{IsUsedMethodIn}(c,n,V)\, \mapsto \,\mathtt{IsUsedMethodIn}(c,m,V)\\[2pt]
 \mathtt{IsUsedMethodIn}(a,n,V)\, \mapsto \,\mathtt{IsUsedMethodIn}(a,m,V)\\[2pt]
 \mathtt{IsUsedMethodIn}(b,n,V)\, \mapsto \,\mathtt{IsUsedMethodIn}(b,m,V)\\[2pt]
 \mathtt{HasReturnType}(c,n,V1)\, \mapsto \,\mathtt{HasReturnType}(c,m,V1)\\[2pt]
 \mathtt{HasReturnType}(a,n,V1)\, \mapsto \,\mathtt{HasReturnType}(a,m,V1)\\[2pt]
 \mathtt{HasReturnType}(b,n,V1)\, \mapsto \,\mathtt{HasReturnType}(b,m,V1)\\[2pt]
 \mathtt{IsInverter}(c,n,V,V1)\, \mapsto \,\mathtt{IsInverter}(c,m,V,V1)\\[2pt]
 \mathtt{IsInverter}(a,n,V,V1)\, \mapsto \,\mathtt{IsInverter}(a,m,V,V1)\\[2pt]
 \mathtt{IsInverter}(b,n,V,V1)\, \mapsto \,\mathtt{IsInverter}(b,m,V,V1)\\[2pt]
 \mathtt{ExistsMethodInvocation}(c,V,V1,n)\, \mapsto \,\mathtt{ExistsMethodInvocation}(c,V,V1,m)\\[2pt]
 \mathtt{ExistsMethodInvocation}(a,V,V1,n)\, \mapsto \,\mathtt{ExistsMethodInvocation}(a,V,V1,m)\\[2pt]
 \mathtt{ExistsMethodInvocation}(b,V,V1,n)\, \mapsto \,\mathtt{ExistsMethodInvocation}(b,V,V1,m)\\[2pt]
 \mathtt{ExistsMethodInvocation}(c,m,V,V1)\, \mapsto \,\bot \\[2pt]
 \mathtt{ExistsMethodInvocation}(a,m,V,V1)\, \mapsto \,\bot \\[2pt]
 \mathtt{ExistsMethodInvocation}(b,m,V,V1)\, \mapsto \,\bot \\[2pt]
 \mathtt{IsIndirectlyRecursive}(c,n)\, \mapsto \,\mathtt{IsIndirectlyRecursive}(c,m)\\[2pt]
 \mathtt{IsIndirectlyRecursive}(a,n)\, \mapsto \,\mathtt{IsIndirectlyRecursive}(a,m)\\[2pt]
 \mathtt{IsIndirectlyRecursive}(b,n)\, \mapsto \,\mathtt{IsIndirectlyRecursive}(b,m)\\[2pt]
 \mathtt{BoundVariableInMethodBody}(c,n,V)\, \mapsto \,\mathtt{BoundVariableInMethodBody}(c,n,V)\\[2pt]
 \mathtt{BoundVariableInMethodBody}(a,n,V)\, \mapsto \,\mathtt{BoundVariableInMethodBody}(a,n,V)\\[2pt]
 \mathtt{BoundVariableInMethodBody}(b,n,V)\, \mapsto \,\mathtt{BoundVariableInMethodBody}(b,n,V)\\[2pt]
 \mathtt{IsOverridden}(a,n)\, \mapsto \,\mathtt{IsOverridden}(a,m)\\[2pt]
 \mathtt{IsOverridden}(b,n)\, \mapsto \,\mathtt{IsOverridden}(b,m)\\[2pt]
 \mathtt{IsUsedConstructorAsMethodParameter}(V,c,m)\, \mapsto \,\bot \\[2pt]
 \mathtt{IsUsedConstructorAsMethodParameter}(V,a,m)\, \mapsto \,\bot \\[2pt]
 \mathtt{IsUsedConstructorAsMethodParameter}(V,b,m)\, \mapsto \,\bot \\[2pt]
 $

\subsubsection{RenameOverloadedMethodInHierarchy}          
  \paragraph{Overview} \textsf{RenameOverloadedMethodInHierarchy (class c,subclasses [a,b],method m ,usedconstrcutorsInM [c1,c2], newname n, types [t])}~: this operation is used to rename the method (c,a,b)::m 
into n nevertheless n will be overloaded or not.

\precondition 
$\\ (\mathtt{ExistsClass}(c)\\ 
\wedge \mathtt{ExistsMethodDefinitionWithParams}(c,m,[t])\\ 
\wedge \neg \mathtt{IsInheritedMethodWithParams}(c,n,[t])\\ 
\wedge \neg \mathtt{ExistsMethodDefinitionWithParams}(c,n,[t])\\ 
\wedge \neg \mathtt{ExistsMethodDefinitionWithParams}(a,n,[t])\\ 
\wedge \neg \mathtt{ExistsMethodDefinitionWithParams}(b,n,[t])\\ 
\wedge \neg \mathtt{IsInheritedMethodWithParams}(c,m,[t])\\ 
\wedge \mathtt{AllSubclasses}(c,[a;b]))$ \\ 

\backdescr
$\\ \mathtt{ExistsMethodDefinition}(c,m)\, \mapsto \,\bot \\[2pt]
 \mathtt{ExistsMethodDefinition}(c,n)\, \mapsto \,\top \\[2pt]
 \mathtt{IsOverriding}(c,n)\, \mapsto \,\bot \\[2pt]
 \mathtt{IsOverridden}(c,n)\, \mapsto \,\bot \\[2pt]
 \mathtt{IsPublic}(c,n)\, \mapsto \,\mathtt{IsPublic}(c,m)\\[2pt]
 \mathtt{ExistsMethodDefinitionWithParams}(c,n,[t])\, \mapsto \,\mathtt{ExistsMethodDefinitionWithParams}(c,m,[t])\\[2pt]
 \mathtt{ExistsMethodDefinitionWithParams}(a,n,[t])\, \mapsto \,\mathtt{ExistsMethodDefinitionWithParams}(a,m,[t])\\[2pt]
 \mathtt{ExistsMethodDefinitionWithParams}(b,n,[t])\, \mapsto \,\mathtt{ExistsMethodDefinitionWithParams}(b,m,[t])\\[2pt]
 \mathtt{ExistsMethodDefinition}(a,n)\, \mapsto \,\top \\[2pt]
 \mathtt{ExistsMethodDefinition}(b,n)\, \mapsto \,\top \\[2pt]
 \mathtt{ExistsMethodDefinitionWithParams}(c,m,[t])\, \mapsto \,\bot \\[2pt]
 \mathtt{ExistsMethodDefinitionWithParams}(a,m,[t])\, \mapsto \,\bot \\[2pt]
 \mathtt{ExistsMethodDefinitionWithParams}(b,m,[t])\, \mapsto \,\bot \\[2pt]
 \mathtt{ExistsMethodDefinition}(a,m)\, \mapsto \,\bot \\[2pt]
 \mathtt{ExistsMethodDefinition}(b,m)\, \mapsto \,\bot \\[2pt]
 \mathtt{isOverridingMethod}(a,n,[t])\, \mapsto \,\mathtt{isOverridingMethod}(a,m,[t])\\[2pt]
 \mathtt{isOverridingMethod}(b,n,[t])\, \mapsto \,\mathtt{isOverridingMethod}(b,m,[t])\\[2pt]
 \mathtt{ExistsParameterWithName}(c,n,[t],V)\, \mapsto \,\mathtt{ExistsParameterWithName}(c,m,[t],V)\\[2pt]
 \mathtt{ExistsParameterWithName}(a,n,[t],V)\, \mapsto \,\mathtt{ExistsParameterWithName}(a,m,[t],V)\\[2pt]
 \mathtt{ExistsParameterWithName}(b,n,[t],V)\, \mapsto \,\mathtt{ExistsParameterWithName}(b,m,[t],V)\\[2pt]
 \mathtt{ExistsParameterWithType}(c,n,[t],V)\, \mapsto \,\mathtt{ExistsParameterWithType}(c,m,[t],V)\\[2pt]
 \mathtt{ExistsParameterWithType}(a,n,[t],V)\, \mapsto \,\mathtt{ExistsParameterWithType}(a,m,[t],V)\\[2pt]
 \mathtt{ExistsParameterWithType}(b,n,[t],V)\, \mapsto \,\mathtt{ExistsParameterWithType}(b,m,[t],V)\\[2pt]
 \mathtt{IsDelegator}(c,n,V)\, \mapsto \,\mathtt{IsDelegator}(c,m,V)\\[2pt]
 \mathtt{IsDelegator}(a,n,V)\, \mapsto \,\mathtt{IsDelegator}(a,m,V)\\[2pt]
 \mathtt{IsDelegator}(b,n,V)\, \mapsto \,\mathtt{IsDelegator}(b,m,V)\\[2pt]
 \mathtt{IsDelegator}(c,V,n)\, \mapsto \,\mathtt{IsDelegator}(c,V,m)\\[2pt]
 \mathtt{IsDelegator}(a,V,n)\, \mapsto \,\mathtt{IsDelegator}(a,V,m)\\[2pt]
 \mathtt{IsDelegator}(b,V,n)\, \mapsto \,\mathtt{IsDelegator}(b,V,m)\\[2pt]
 \mathtt{IsRecursiveMethod}(c,n)\, \mapsto \,\mathtt{IsRecursiveMethod}(c,m)\\[2pt]
 \mathtt{IsRecursiveMethod}(a,n)\, \mapsto \,\mathtt{IsRecursiveMethod}(a,m)\\[2pt]
 \mathtt{IsRecursiveMethod}(b,n)\, \mapsto \,\mathtt{IsRecursiveMethod}(b,m)\\[2pt]
 \mathtt{ExistsAbstractMethod}(c,n)\, \mapsto \,\mathtt{ExistsAbstractMethod}(c,m)\\[2pt]
 \mathtt{ExistsAbstractMethod}(a,n)\, \mapsto \,\mathtt{ExistsAbstractMethod}(a,m)\\[2pt]
 \mathtt{ExistsAbstractMethod}(b,n)\, \mapsto \,\mathtt{ExistsAbstractMethod}(b,m)\\[2pt]
 \mathtt{IsInheritedMethodWithParams}(a,n,[t])\, \mapsto \,\mathtt{IsVisibleMethod}(c,m,[t],a)\\[2pt]
 \mathtt{IsInheritedMethodWithParams}(b,n,[t])\, \mapsto \,\mathtt{IsVisibleMethod}(c,m,[t],b)\\[2pt]
 \mathtt{MethodIsUsedWithType}(c,n,[t],[t])\, \mapsto \,\mathtt{MethodIsUsedWithType}(c,m,[t],[t])\\[2pt]
 \mathtt{MethodIsUsedWithType}(a,n,[t],[t])\, \mapsto \,\mathtt{MethodIsUsedWithType}(a,m,[t],[t])\\[2pt]
 \mathtt{MethodIsUsedWithType}(b,n,[t],[t])\, \mapsto \,\mathtt{MethodIsUsedWithType}(b,m,[t],[t])\\[2pt]
 \mathtt{IsUsedMethod}(c,n,[t])\, \mapsto \,\mathtt{IsUsedMethod}(c,m,[t])\\[2pt]
 \mathtt{IsUsedMethod}(a,n,[t])\, \mapsto \,\mathtt{IsUsedMethod}(a,m,[t])\\[2pt]
 \mathtt{IsUsedMethod}(b,n,[t])\, \mapsto \,\mathtt{IsUsedMethod}(b,m,[t])\\[2pt]
 \mathtt{IsUsedMethodIn}(c,n,V)\, \mapsto \,\mathtt{IsUsedMethodIn}(c,m,V)\\[2pt]
 \mathtt{IsUsedMethodIn}(a,n,V)\, \mapsto \,\mathtt{IsUsedMethodIn}(a,m,V)\\[2pt]
 \mathtt{IsUsedMethodIn}(b,n,V)\, \mapsto \,\mathtt{IsUsedMethodIn}(b,m,V)\\[2pt]
 \mathtt{HasReturnType}(c,n,V)\, \mapsto \,\mathtt{HasReturnType}(c,m,V)\\[2pt]
 \mathtt{HasReturnType}(a,n,V)\, \mapsto \,\mathtt{HasReturnType}(a,m,V)\\[2pt]
 \mathtt{HasReturnType}(b,n,V)\, \mapsto \,\mathtt{HasReturnType}(b,m,V)\\[2pt]
 \mathtt{IsInverter}(c,n,V,V1)\, \mapsto \,\mathtt{IsInverter}(c,m,V,V1)\\[2pt]
 \mathtt{IsInverter}(a,n,V,V1)\, \mapsto \,\mathtt{IsInverter}(a,m,V,V1)\\[2pt]
 \mathtt{IsInverter}(b,n,V,V1)\, \mapsto \,\mathtt{IsInverter}(b,m,V,V1)\\[2pt]
 \mathtt{ExistsMethodInvocation}(c,V,V1,n)\, \mapsto \,\mathtt{ExistsMethodInvocation}(c,V,V1,m)\\[2pt]
 \mathtt{ExistsMethodInvocation}(a,V,V1,n)\, \mapsto \,\mathtt{ExistsMethodInvocation}(a,V,V1,m)\\[2pt]
 \mathtt{ExistsMethodInvocation}(b,V,V1,n)\, \mapsto \,\mathtt{ExistsMethodInvocation}(b,V,V1,m)\\[2pt]
 \mathtt{IsIndirectlyRecursive}(c,n)\, \mapsto \,\mathtt{IsIndirectlyRecursive}(c,m)\\[2pt]
 \mathtt{IsIndirectlyRecursive}(a,n)\, \mapsto \,\mathtt{IsIndirectlyRecursive}(a,m)\\[2pt]
 \mathtt{IsIndirectlyRecursive}(b,n)\, \mapsto \,\mathtt{IsIndirectlyRecursive}(b,m)\\[2pt]
 \mathtt{BoundVariableInMethodBody}(c,n,V)\, \mapsto \,\mathtt{BoundVariableInMethodBody}(c,m,V)\\[2pt]
 \mathtt{BoundVariableInMethodBody}(a,n,V)\, \mapsto \,\mathtt{BoundVariableInMethodBody}(a,m,V)\\[2pt]
 \mathtt{BoundVariableInMethodBody}(b,n,V)\, \mapsto \,\mathtt{BoundVariableInMethodBody}(b,m,V)\\[2pt]
 \mathtt{IsOverridden}(a,n)\, \mapsto \,\mathtt{IsOverridden}(a,m)\\[2pt]
 \mathtt{IsOverridden}(b,n)\, \mapsto \,\mathtt{IsOverridden}(b,m)\\[2pt]
 \mathtt{IsUsedConstructorAsMethodParameter}(V,c,m)\, \mapsto \,\bot \\[2pt]
 \mathtt{IsUsedConstructorAsMethodParameter}(V,a,m)\, \mapsto \,\bot \\[2pt]
 \mathtt{IsUsedConstructorAsMethodParameter}(V,b,m)\, \mapsto \,\bot \\[2pt]
 \mathtt{IsUsedConstructorAsObjectReceiver}(V,c,m)\, \mapsto \,\bot \\[2pt]
 \mathtt{IsUsedConstructorAsObjectReceiver}(V,a,m)\, \mapsto \,\bot \\[2pt]
 \mathtt{IsUsedConstructorAsObjectReceiver}(V,b,m)\, \mapsto \,\bot \\[2pt]
 \mathtt{IsUsedConstructorAsMethodParameter}(V,c,m)\, \mapsto \,\bot \\[2pt]
 \mathtt{IsUsedConstructorAsMethodParameter}(V,a,m)\, \mapsto \,\bot \\[2pt]
 \mathtt{IsUsedConstructorAsMethodParameter}(V,b,m)\, \mapsto \,\bot \\[2pt]
 \mathtt{IsUsedConstructorAsMethodParameter}(V,c,n)\, \mapsto \,\mathtt{IsUsedConstructorAsMethodParameter}(V,c,m)\\[2pt]
 \mathtt{IsUsedConstructorAsMethodParameter}(V,a,n)\, \mapsto \,\mathtt{IsUsedConstructorAsMethodParameter}(V,a,m)\\[2pt]
 \mathtt{IsUsedConstructorAsMethodParameter}(V,b,n)\, \mapsto \,\mathtt{IsUsedConstructorAsMethodParameter}(V,b,m)\\[2pt]
 \mathtt{IsUsedConstructorAsObjectReceiver}(c1,c,n)\, \mapsto \,\top \\[2pt]
 \mathtt{IsUsedConstructorAsObjectReceiver}(c2,c,n)\, \mapsto \,\top \\[2pt]
 \mathtt{IsUsedConstructorAsObjectReceiver}(c1,a,n)\, \mapsto \,\top \\[2pt]
 \mathtt{IsUsedConstructorAsObjectReceiver}(c2,a,n)\, \mapsto \,\top \\[2pt]
 \mathtt{IsUsedConstructorAsObjectReceiver}(c1,b,n)\, \mapsto \,\top \\[2pt]
 \mathtt{IsUsedConstructorAsObjectReceiver}(c2,b,n)\, \mapsto \,\top \\[2pt]
 \mathtt{IsUsedConstructorAsMethodParameter}(V,c,n)\, \mapsto \,\mathtt{IsUsedConstructorAsMethodParameter}(V,c,m)\\[2pt]
 \mathtt{IsUsedConstructorAsMethodParameter}(V,a,n)\, \mapsto \,\mathtt{IsUsedConstructorAsMethodParameter}(V,a,m)\\[2pt]
 \mathtt{IsUsedConstructorAsMethodParameter}(V,b,n)\, \mapsto \,\mathtt{IsUsedConstructorAsMethodParameter}(V,b,m)\\[2pt]
 $

\subsubsection{RenameDelegatorWithOverloading }          
  \paragraph{Overview} \textsf{RenameDelegatorWithOverloading (classname s, subclasses [a,b], method m,paramtype t,paramName pn,super~typeOfparamtype t',
newname n)}~: this operation is used to rename the method (c,a,b)::m into n and accepts overloaded methods. This operation is an ad-hoc use of the operation 
\textsf{RenameOverloadedMethodInHierarchy} (we need in this use  more details about the signature of the method to be renamed). 

\precondition 
 $\\ (\mathtt{ExistsClass}(s)\\ 
\wedge \mathtt{ExistsClass}(a)\\ 
\wedge \mathtt{ExistsClass}(b)\\ 
\wedge \mathtt{ExistsMethodDefinition}(s,m)\\ 
\wedge \mathtt{ExistsMethodDefinitionWithParams}(s,m,[t])\\ 
\wedge \mathtt{AllSubclasses}(s,[a;b])\\ 
\wedge \neg \mathtt{ExistsMethodDefinitionWithParams}(s,n,[t])\\ 
\wedge \neg \mathtt{ExistsMethodDefinitionWithParams}(a,n,[t])\\ 
\wedge \neg \mathtt{ExistsMethodDefinitionWithParams}(b,n,[t])\\ 
\wedge \neg \mathtt{IsInheritedMethod}(s,n)\\ 
\wedge \neg \mathtt{ExistsMethodDefinitionWithParams}(s,n,[t]))$ \\

\backdescr
$\\ \mathtt{ExistsMethodDefinition}(s,n)\, \mapsto \,\top \\[2pt]
 \mathtt{ExistsMethodDefinitionWithParams}(s,n,[t])\, \mapsto \,\top \\[2pt]
 \mathtt{IsPublic}(s,n)\, \mapsto \,\mathtt{IsPublic}(s,m)\\[2pt]
 \mathtt{ExistsMethodDefinition}(s,m)\, \mapsto \,\bot \\[2pt]
 \mathtt{ExistsMethodDefinition}(a,m)\, \mapsto \,\bot \\[2pt]
 \mathtt{ExistsMethodDefinition}(b,m)\, \mapsto \,\bot \\[2pt]
 \mathtt{ExistsMethodDefinitionWithParams}(s,m,[t])\, \mapsto \,\bot \\[2pt]
 \mathtt{ExistsMethodDefinitionWithParams}(a,m,[t])\, \mapsto \,\bot \\[2pt]
 \mathtt{ExistsMethodDefinitionWithParams}(b,m,[t])\, \mapsto \,\bot \\[2pt]
 \mathtt{ExistsMethodDefinition}(a,n)\, \mapsto \,\mathtt{ExistsMethodDefinition}(a,m)\\[2pt]
 \mathtt{ExistsMethodDefinition}(b,n)\, \mapsto \,\mathtt{ExistsMethodDefinition}(b,m)\\[2pt]
 \mathtt{ExistsMethodDefinitionWithParams}(a,n,[t])\, \mapsto \,\mathtt{ExistsMethodDefinitionWithParams}(a,m,[t])\\[2pt]
 \mathtt{ExistsMethodDefinitionWithParams}(b,n,[t])\, \mapsto \,\mathtt{ExistsMethodDefinitionWithParams}(b,m,[t])\\[2pt]
 \mathtt{IsInheritedMethod}(a,n)\, \mapsto \,\mathtt{IsInheritedMethod}(a,m)\\[2pt]
 \mathtt{IsInheritedMethod}(b,n)\, \mapsto \,\mathtt{IsInheritedMethod}(b,m)\\[2pt]
 \mathtt{MethodIsUsedWithType}(s,n,[t],[t])\, \mapsto \,\mathtt{MethodIsUsedWithType}(s,m,[t],[t])\\[2pt]
 \mathtt{MethodIsUsedWithType}(a,n,[t],[t])\, \mapsto \,\mathtt{MethodIsUsedWithType}(a,m,[t],[t])\\[2pt]
 \mathtt{MethodIsUsedWithType}(b,n,[t],[t])\, \mapsto \,\mathtt{MethodIsUsedWithType}(b,m,[t],[t])\\[2pt]
 \mathtt{MethodIsUsedWithType}(s,m,[t],[t])\, \mapsto \,\bot \\[2pt]
 \mathtt{MethodIsUsedWithType}(a,m,[t],[t])\, \mapsto \,\bot \\[2pt]
 \mathtt{MethodIsUsedWithType}(b,m,[t],[t])\, \mapsto \,\bot \\[2pt]
 \mathtt{ExistsParameterWithName}(s,n,[t],V)\, \mapsto \,\mathtt{ExistsParameterWithName}(s,m,[t],V)\\[2pt]
 \mathtt{ExistsParameterWithName}(a,n,[t],V)\, \mapsto \,\mathtt{ExistsParameterWithName}(a,m,[t],V)\\[2pt]
 \mathtt{ExistsParameterWithName}(b,n,[t],V)\, \mapsto \,\mathtt{ExistsParameterWithName}(b,m,[t],V)\\[2pt]
 \mathtt{ExistsParameterWithType}(s,n,[t],V)\, \mapsto \,\mathtt{ExistsParameterWithType}(s,m,[t],V)\\[2pt]
 \mathtt{ExistsParameterWithType}(a,n,[t],V)\, \mapsto \,\mathtt{ExistsParameterWithType}(a,m,[t],V)\\[2pt]
 \mathtt{ExistsParameterWithType}(b,n,[t],V)\, \mapsto \,\mathtt{ExistsParameterWithType}(b,m,[t],V)\\[2pt]
 \mathtt{ExistsMethodInvocation}(s,V1,V,n)\, \mapsto \,\mathtt{ExistsMethodInvocation}(s,V1,V,m)\\[2pt]
 \mathtt{ExistsMethodInvocation}(a,V1,V,n)\, \mapsto \,\mathtt{ExistsMethodInvocation}(a,V1,V,m)\\[2pt]
 \mathtt{ExistsMethodInvocation}(b,V1,V,n)\, \mapsto \,\mathtt{ExistsMethodInvocation}(b,V1,V,m)\\[2pt]
 \mathtt{IsDelegator}(s,n,V)\, \mapsto \,\mathtt{IsDelegator}(s,m,V)\\[2pt]
 \mathtt{IsDelegator}(a,n,V)\, \mapsto \,\mathtt{IsDelegator}(a,m,V)\\[2pt]
 \mathtt{IsDelegator}(b,n,V)\, \mapsto \,\mathtt{IsDelegator}(b,m,V)\\[2pt]
 \mathtt{IsDelegator}(s,V,n)\, \mapsto \,\mathtt{IsDelegator}(s,V,m)\\[2pt]
 \mathtt{IsDelegator}(a,V,n)\, \mapsto \,\mathtt{IsDelegator}(a,V,m)\\[2pt]
 \mathtt{IsDelegator}(b,V,n)\, \mapsto \,\mathtt{IsDelegator}(b,V,m)\\[2pt]
 \mathtt{IsUsedMethod}(s,n,[t])\, \mapsto \,\mathtt{IsUsedMethod}(s,m,[t])\\[2pt]
 \mathtt{IsUsedMethod}(a,n,[t])\, \mapsto \,\mathtt{IsUsedMethod}(a,m,[t])\\[2pt]
 \mathtt{IsUsedMethod}(b,n,[t])\, \mapsto \,\mathtt{IsUsedMethod}(b,m,[t])\\[2pt]
 \mathtt{IsUsedMethodIn}(s,n,V)\, \mapsto \,\mathtt{IsUsedMethodIn}(s,m,V)\\[2pt]
 \mathtt{IsUsedMethodIn}(a,n,V)\, \mapsto \,\mathtt{IsUsedMethodIn}(a,m,V)\\[2pt]
 \mathtt{IsUsedMethodIn}(b,n,V)\, \mapsto \,\mathtt{IsUsedMethodIn}(b,m,V)\\[2pt]
 \mathtt{HasReturnType}(s,n,V)\, \mapsto \,\mathtt{HasReturnType}(s,m,V)\\[2pt]
 \mathtt{HasReturnType}(a,n,V)\, \mapsto \,\mathtt{HasReturnType}(a,m,V)\\[2pt]
 \mathtt{HasReturnType}(b,n,V)\, \mapsto \,\mathtt{HasReturnType}(b,m,V)\\[2pt]
 \mathtt{IsInverter}(s,n,V,V1)\, \mapsto \,\mathtt{IsInverter}(s,m,V,V1)\\[2pt]
 \mathtt{IsInverter}(a,n,V,V1)\, \mapsto \,\mathtt{IsInverter}(a,m,V,V1)\\[2pt]
 \mathtt{IsInverter}(b,n,V,V1)\, \mapsto \,\mathtt{IsInverter}(b,m,V,V1)\\[2pt]
 \mathtt{IsIndirectlyRecursive}(s,n)\, \mapsto \,\mathtt{IsIndirectlyRecursive}(s,m)\\[2pt]
 \mathtt{IsIndirectlyRecursive}(a,n)\, \mapsto \,\mathtt{IsIndirectlyRecursive}(a,m)\\[2pt]
 \mathtt{IsIndirectlyRecursive}(b,n)\, \mapsto \,\mathtt{IsIndirectlyRecursive}(b,m)\\[2pt]
 \mathtt{BoundVariableInMethodBody}(s,n,V)\, \mapsto \,\mathtt{BoundVariableInMethodBody}(s,n,V)\\[2pt]
 \mathtt{BoundVariableInMethodBody}(a,n,V)\, \mapsto \,\mathtt{BoundVariableInMethodBody}(a,n,V)\\[2pt]
 \mathtt{BoundVariableInMethodBody}(b,n,V)\, \mapsto \,\mathtt{BoundVariableInMethodBody}(b,n,V)\\[2pt]
 \mathtt{IsOverridden}(a,n)\, \mapsto \,\mathtt{IsOverridden}(a,m)\\[2pt]
 \mathtt{IsOverridden}(b,n)\, \mapsto \,\mathtt{IsOverridden}(b,m)\\[2pt]
 \mathtt{IsUsedConstructorAsMethodParameter}(V,s,m)\, \mapsto \,\bot \\[2pt]
 \mathtt{IsUsedConstructorAsMethodParameter}(V,a,m)\, \mapsto \,\bot \\[2pt]
 \mathtt{IsUsedConstructorAsMethodParameter}(V,b,m)\, \mapsto \,\bot \\[2pt]
 \mathtt{IsUsedConstructorAsObjectReceiver}(t,s,n)\, \mapsto \,\mathtt{IsUsedConstructorAsObjectReceiver}(t,s,m)\\[2pt]
 \mathtt{IsUsedConstructorAsObjectReceiver}(t,a,n)\, \mapsto \,\mathtt{IsUsedConstructorAsObjectReceiver}(t,a,m)\\[2pt]
 \mathtt{IsUsedConstructorAsObjectReceiver}(t,b,n)\, \mapsto \,\mathtt{IsUsedConstructorAsObjectReceiver}(t,b,m)\\[2pt]
 $

\subsection{ExtractSuperClass}%%%%%%%%%%%%%%%%%%%%%%%%%%%%%%%%%%%%%%%%%%%%%%%%%
\label{def-ExtractAbstractSuperClass}

(\emph{Extract Super Class} in Fowler~\cite{Fowler1999} and~\cite{koch2002})

\paragraph{Overview:} \textsf{ExtractSuperClass (subclasses[a,b], superclass s,methodsOfsubclasses [m,n],returntype t)}: 
this operation is used to extract a super-class s from the classes a and b and make an abstract declaration of methods a::m, a::n, b::m and b::n  in this new super-class.

\begin{center}
\includegraphics[scale=0.6]{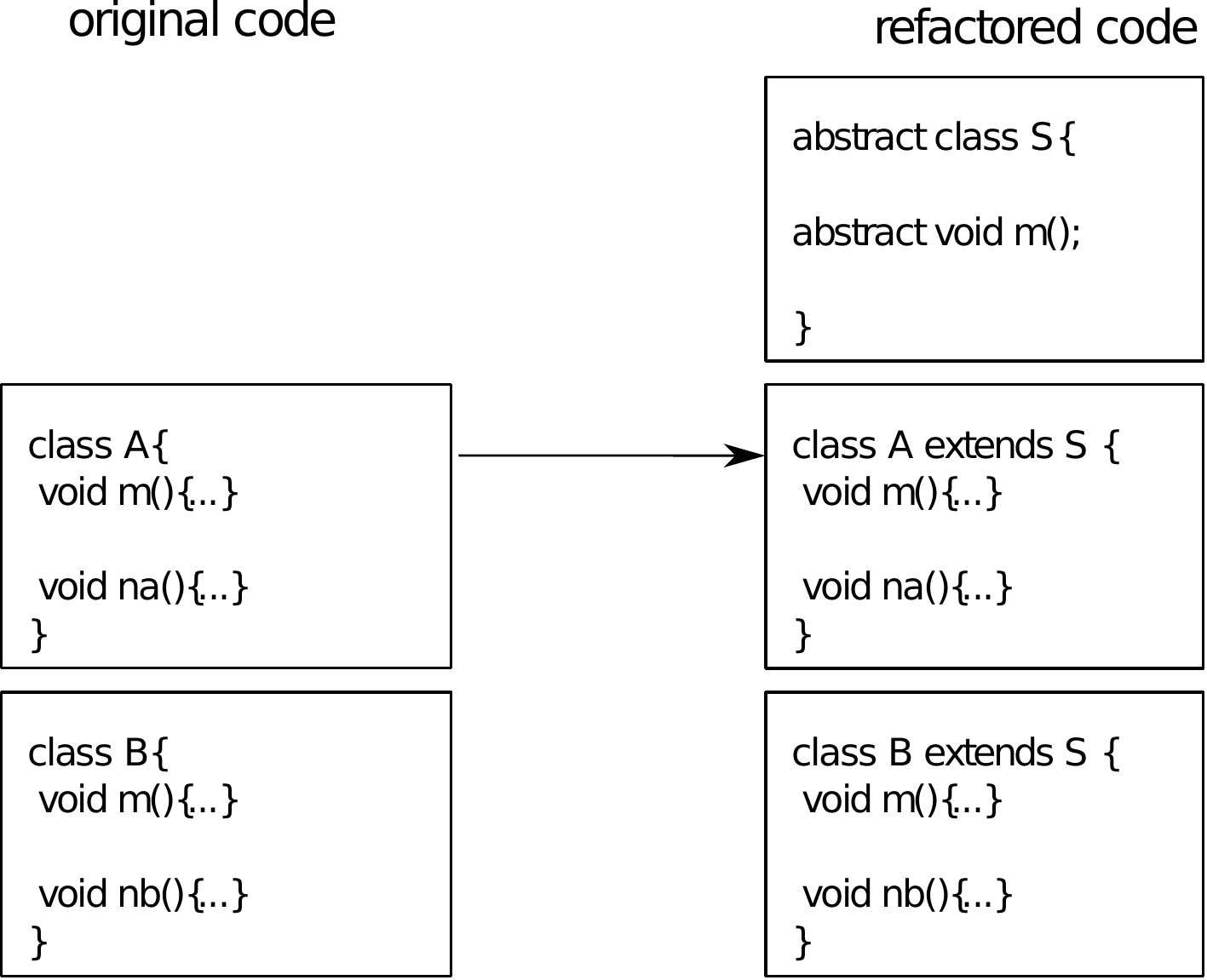}
\end{center}

\tools \emph{Extract Superclass} in Eclipse tool and
\intellij. In \intellij, the \emph{Extract Superclass}
operation cannot be applied to several classes
simultaneously, so we have maintain the code of this operation in order to run it on several classes.

\precondition 
$\\(\neg \mathtt{ExistsType}(s)\\ 
\wedge \mathtt{ExistsClass}(a)\\ 
\wedge \mathtt{ExistsClass}(b)\\ 
\wedge \mathtt{ExtendsDirectly}(a,java.lang.Object)\\ 
\wedge \mathtt{ExtendsDirectly}(b,java.lang.Object)\\ 
\wedge \mathtt{HasReturnType}(a,m,t)\\ 
\wedge \mathtt{HasReturnType}(a,n,t)\\ 
\wedge \mathtt{HasReturnType}(b,m,t)\\ 
\wedge \mathtt{HasReturnType}(b,n,t))$ \\

\backdescr
 $\\ \mathtt{IsAbstractClass}(s)\, \mapsto \,\top \\[2pt]
 \mathtt{ExistsClass}(s)\, \mapsto \,\top \\[2pt]
 \mathtt{ExistsType}(s)\, \mapsto \,\top \\[2pt]
 \mathtt{ExistsMethodDefinition}(s,X)\, \mapsto \,(\mathtt{ExistsMethodDefinition}(a,X)\\ \wedge \mathtt{ExistsMethodDefinition}(b,X))\\[2pt]
 \mathtt{ExistsMethodDefinitionWithParams}(s,X,[\,])\, \mapsto \,(\mathtt{ExistsMethodDefinitionWithParams}(a,X,[\,])\\ \wedge \mathtt{ExistsMethodDefinitionWithParams}(b,X,[\,]))\\[2pt]
 \mathtt{ExistsMethodDefinitionWithParams}(s,X,[Y])\, \mapsto \,(\mathtt{ExistsMethodDefinitionWithParams}(a,X,[Y])\\ \wedge \mathtt{ExistsMethodDefinitionWithParams}(b,X,[Y]))\\[2pt]
 \mathtt{IsUsedMethodIn}(s,X,Y)\, \mapsto \,\bot \\[2pt]
 \mathtt{ExistsMethodDefinitionWithParams}(X,Y,[s])\, \mapsto \,\bot \\[2pt]
 \mathtt{IsUsedMethod}(s,X,[Y])\, \mapsto \,\bot \\[2pt]
 \mathtt{AllSubclasses}(s,[a;b])\, \mapsto \,\top \\[2pt]
 \mathtt{MethodIsUsedWithType}(X,Y,[Z],[s])\, \mapsto \,\bot \\[2pt]
 \mathtt{IsInheritedMethodWithParams}(X,Y,[s])\, \mapsto \,\bot \\[2pt]
 \mathtt{IsUsedConstructorAsMethodParameter}(s,X,Y)\, \mapsto \,\bot \\[2pt]
 \mathtt{IsUsedConstructorAsInitializer}(s,X,Y)\, \mapsto \,\bot \\[2pt]
 \mathtt{IsUsedConstructorAsObjectReceiver}(s,X,Y)\, \mapsto \,\bot \\[2pt]
 \mathtt{IsUsedConstructorAsMethodParameter}(X,s,Y)\, \mapsto \,\bot \\[2pt]
 \mathtt{IsUsedConstructorAsInitializer}(X,s,Y)\, \mapsto \,\bot \\[2pt]
 \mathtt{IsUsedConstructorAsObjectReceiver}(X,s,Y)\, \mapsto \,\bot \\[2pt]
 \mathtt{IsPrimitiveType}(s)\, \mapsto \,\bot \\[2pt]
 \mathtt{IsSubType}(a,s)\, \mapsto \,\top \\[2pt]
 \mathtt{IsSubType}(b,s)\, \mapsto \,\top \\[2pt]
 \mathtt{IsSubType}(X,s)\, \mapsto \,\mathtt{IsSubType}(X,a)\\[2pt]
 \mathtt{IsSubType}(X,s)\, \mapsto \,\mathtt{IsSubType}(X,b)\\[2pt]
 \mathtt{IsPublic}(s,m)\, \mapsto \,\top \\[2pt]
 \mathtt{IsPublic}(s,n)\, \mapsto \,\top \\[2pt]
 \mathtt{ExistsAbstractMethod}(s,m)\, \mapsto \,\top \\[2pt]
 \mathtt{ExistsAbstractMethod}(s,n)\, \mapsto \,\top \\[2pt]
 \mathtt{IsOverriding}(s,m)\, \mapsto \,\bot \\[2pt]
 \mathtt{IsOverriding}(s,n)\, \mapsto \,\bot \\[2pt]
 \mathtt{IsOverridden}(s,m)\, \mapsto \,\top \\[2pt]
 \mathtt{IsOverridden}(s,n)\, \mapsto \,\top \\[2pt]
 \mathtt{IsPrivate}(s,m)\, \mapsto \,\bot \\[2pt]
 \mathtt{IsPrivate}(s,n)\, \mapsto \,\bot \\[2pt]
 $

\subsubsection{ExtractSuperClassWithoutPullUp}
 \label{extract-super-class-withoutPullUp}

\paragraph{Overview:} \textsf{ExtractSuperClassWithouPullUp (subclasses[a,b], superclass s)}: 
this operation is a specific variant of the operation extract super-class. It is simply used to extract a super-class without pull up the methods 
of sub-classes.

\precondition 
 $\\ (\neg \mathtt{ExistsType}(s)\\ 
\wedge \mathtt{ExistsClass}(a)\\ 
\wedge \mathtt{ExistsClass}(b)\\ 
\wedge \mathtt{ExtendsDirectly}(a,java.lang.Object)\\ 
\wedge \mathtt{ExtendsDirectly}(b,java.lang.Object))$ \\

\backdescr
 $\\ \mathtt{IsAbstractClass}(s)\, \mapsto \,\top \\[2pt]
 \mathtt{ExistsClass}(s)\, \mapsto \,\top \\[2pt]
 \mathtt{ExistsMethodDefinitionWithParams}(X,Y,[s])\, \mapsto \,\bot \\[2pt]
 \mathtt{ExistsMethodDefinitionWithParams}(s,X,[Y])\, \mapsto \,\bot \\[2pt]
 \mathtt{ExistsType}(s)\, \mapsto \,\top \\[2pt]
 \mathtt{AllSubclasses}(s,[a;b])\, \mapsto \,\top \\[2pt]
 \mathtt{IsUsedConstructorAsMethodParameter}(s,X,Y)\, \mapsto \,\bot \\[2pt]
 \mathtt{IsUsedConstructorAsInitializer}(s,X,Y)\, \mapsto \,\bot \\[2pt]
 \mathtt{IsUsedConstructorAsObjectReceiver}(s,X,Y)\, \mapsto \,\bot \\[2pt]
 \mathtt{IsUsedConstructorAsMethodParameter}(X,s,Y)\, \mapsto \,\bot \\[2pt]
 \mathtt{IsUsedConstructorAsInitializer}(X,s,Y)\, \mapsto \,\bot \\[2pt]
 \mathtt{IsUsedConstructorAsObjectReceiver}(X,s,Y)\, \mapsto \,\bot \\[2pt]
 \mathtt{IsPrimitiveType}(s)\, \mapsto \,\bot \\[2pt]
 \mathtt{IsUsedMethod}(s,X,[Y])\, \mapsto \,\bot \\[2pt]
 \mathtt{IsInheritedMethodWithParams}(X,Y,[s])\, \mapsto \,\bot \\[2pt]
 \mathtt{IsPrivate}(s,X)\, \mapsto \,\bot \\[2pt]
 \mathtt{MethodIsUsedWithType}(X,Y,[Z],[s])\, \mapsto \,\bot \\[2pt]
 \mathtt{IsSubType}(a,s)\, \mapsto \,\top \\[2pt]
 \mathtt{IsSubType}(b,s)\, \mapsto \,\top \\[2pt]
 \mathtt{IsSubType}(X,s)\, \mapsto \,\mathtt{IsSubType}(X,a)\\[2pt]
 \mathtt{IsSubType}(X,s)\, \mapsto \,\mathtt{IsSubType}(X,b)\\[2pt]
 $

\subsection{GeneraliseParameter}%%%%%%%%%%%%%%%%%%%%%%%%%%%%%%%%%%%%%%%%%%%%%%%%%%%
\label{def-GeneraliseParameter}

\paragraph{Overview:} \textsf{GeneraliseParameter (classname s, subclasses [a,b], methodname m,paramName p,type t ,supertype st)}: 
this operation is used to change the type t of the parameter p of the methods s::m, a::m and b::m into a super-type st. 
All the uses of the parameter which type is changed must be legal with the new type \textsf{st} (method invocations, object passed as parameter of other methods). The uses of the parameter which type is changed as parameter of methods must not result in a charge of the invoked code (static resolving of overloading).

\begin{center}

\includegraphics[scale=0.7]{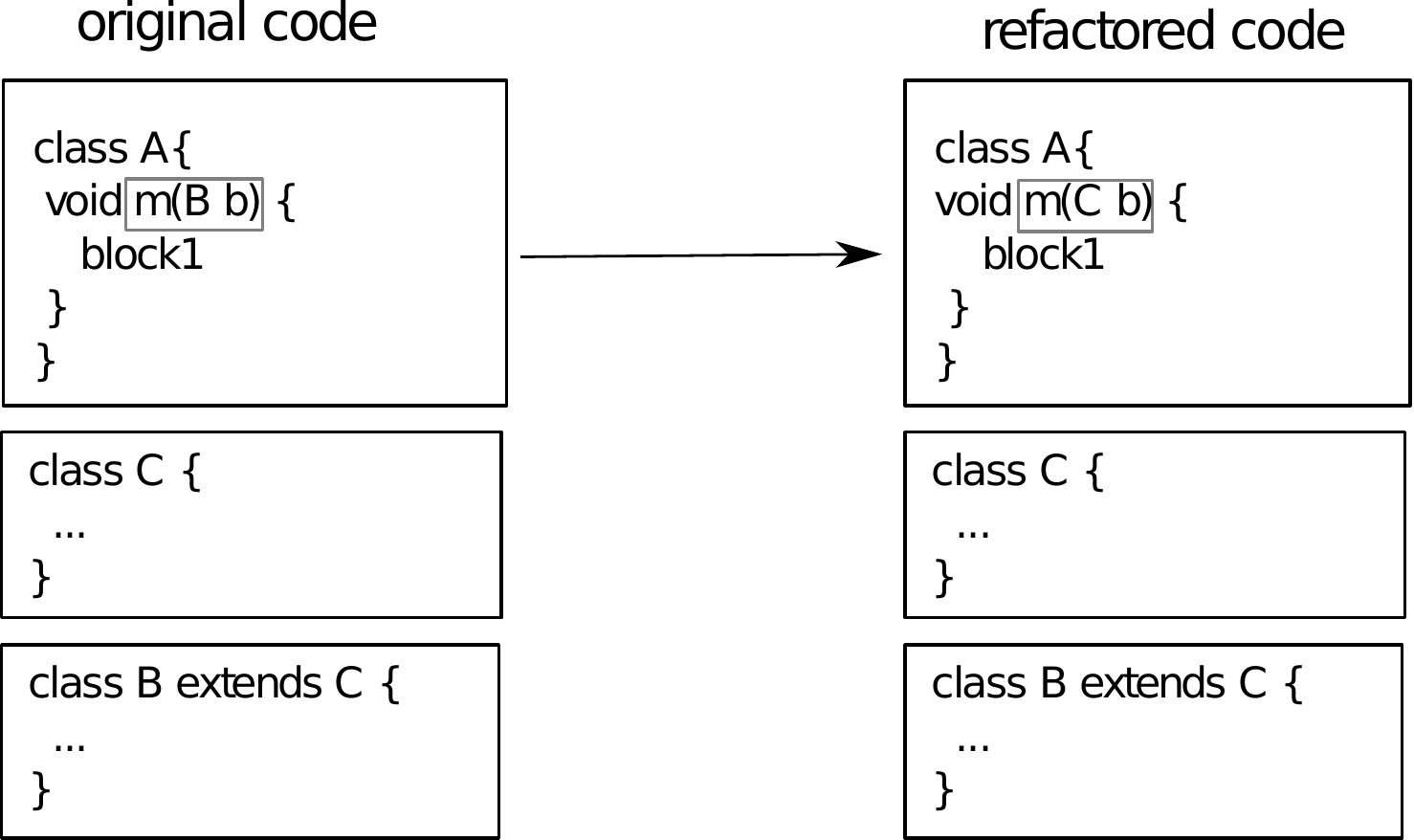}
\end{center}

\tools%%%%%%%%%
\emph{Change Method Signature} in Eclipse tool and \emph{Type Migration} in \intellij (or \emph{Change Signature}).

\precondition 
 $\\ (\mathtt{ExistsClass}(s)\\ 
\wedge \mathtt{ExistsClass}(a)\\ 
\wedge \mathtt{ExistsClass}(b)\\ 
\wedge \mathtt{ExistsMethodDefinition}(s,m)\\ 
\wedge \mathtt{ExistsMethodDefinition}(a,m)\\ 
\wedge \mathtt{ExistsMethodDefinition}(b,m)\\ 
\wedge \mathtt{IsSubType}(st,t)\\ 
\wedge \mathtt{AllSubclasses}(s,[a;b])\\ 
\wedge \mathtt{AllInvokedMethodsOnObjectOInBodyOfMAreDeclaredInC}(s,m,p,t)\\ 
\wedge \mathtt{AllInvokedMethodsOnObjectOInBodyOfMAreDeclaredInC}(a,m,p,t)\\ 
\wedge \mathtt{AllInvokedMethodsOnObjectOInBodyOfMAreDeclaredInC}(b,m,p,t)\\ 
\wedge \mathtt{AllInvokedMethodsWithParameterOInBodyOfMAreNotOverloaded}(s,m,p)\\ 
\wedge \mathtt{AllInvokedMethodsWithParameterOInBodyOfMAreNotOverloaded}(a,m,p)\\ 
\wedge \mathtt{AllInvokedMethodsWithParameterOInBodyOfMAreNotOverloaded}(b,m,p))$ \\

\backdescr
 $\\ \mathtt{IsInverter}(s,m,t,V)\, \mapsto \,\mathtt{IsInverter}(s,m,st,V)\\[2pt]
 \mathtt{IsInverter}(a,m,t,V)\, \mapsto \,\mathtt{IsInverter}(a,m,st,V)\\[2pt]
 \mathtt{IsInverter}(b,m,t,V)\, \mapsto \,\mathtt{IsInverter}(b,m,st,V)\\[2pt]
 \mathtt{ExistsMethodDefinitionWithParams}(s,m,[t])\, \mapsto \,\mathtt{ExistsMethodDefinitionWithParams}(s,m,[st])\\[2pt]
 \mathtt{ExistsMethodDefinitionWithParams}(a,m,[t])\, \mapsto \,\mathtt{ExistsMethodDefinitionWithParams}(a,m,[st])\\[2pt]
 \mathtt{ExistsMethodDefinitionWithParams}(b,m,[t])\, \mapsto \,\mathtt{ExistsMethodDefinitionWithParams}(b,m,[st])\\[2pt]
 \mathtt{ExistsMethodDefinitionWithParams}(s,m,[st])\, \mapsto \,\bot \\[2pt]
 \mathtt{ExistsMethodDefinitionWithParams}(a,m,[st])\, \mapsto \,\bot \\[2pt]
 \mathtt{ExistsMethodDefinitionWithParams}(b,m,[st])\, \mapsto \,\bot \\[2pt]
 \mathtt{IsInheritedMethodWithParams}(a,m,[t])\, \mapsto \,\top \\[2pt]
 \mathtt{IsInheritedMethodWithParams}(b,m,[t])\, \mapsto \,\top \\[2pt]
 \mathtt{IsUsedConstructorAsMethodParameter}(t,s,m)\, \mapsto \,\top \\[2pt]
 \mathtt{IsUsedConstructorAsMethodParameter}(t,a,m)\, \mapsto \,\top \\[2pt]
 \mathtt{IsUsedConstructorAsMethodParameter}(t,b,m)\, \mapsto \,\top \\[2pt]
 \mathtt{IsUsedConstructorAsMethodParameter}(st,s,m)\, \mapsto \,\bot \\[2pt]
 \mathtt{IsUsedConstructorAsMethodParameter}(st,a,m)\, \mapsto \,\bot \\[2pt]
 \mathtt{IsUsedConstructorAsMethodParameter}(st,b,m)\, \mapsto \,\bot \\[2pt]
 \mathtt{IsOverridden}(a,m)\, \mapsto \,\mathtt{ExistsMethodDefinition}(a,m)\\[2pt]
 \mathtt{IsOverridden}(b,m)\, \mapsto \,\mathtt{ExistsMethodDefinition}(b,m)\\[2pt]
 \mathtt{IsOverriding}(a,m)\, \mapsto \,\mathtt{ExistsMethodDefinition}(a,m)\\[2pt]
 \mathtt{IsOverriding}(b,m)\, \mapsto \,\mathtt{ExistsMethodDefinition}(b,m)\\[2pt]
 \mathtt{ExistsParameterWithName}(s,m,[t],p)\, \mapsto \,\top \\[2pt]
 \mathtt{ExistsParameterWithName}(a,m,[t],p)\, \mapsto \,\top \\[2pt]
 \mathtt{ExistsParameterWithName}(b,m,[t],p)\, \mapsto \,\top \\[2pt]
 \mathtt{ExistsParameterWithType}(s,m,[t],t)\, \mapsto \,\top \\[2pt]
 \mathtt{ExistsParameterWithType}(a,m,[t],t)\, \mapsto \,\top \\[2pt]
 \mathtt{ExistsParameterWithType}(b,m,[t],t)\, \mapsto \,\top \\[2pt]
 $

\subsection{MergeDuplicateMethods}%%%%%%%%%%%%%%%%%%%%%%%%%%%%%%%%%%%%%%%%%%%%%%%%%%%%
\label{def-MergeDuplicateMethods}

 \paragraph{Overview} \textsf{MergeDuplicateMethods (classname c, subclasses [a,b],mergedmethods [m,n],newmethod m2, invertedtype t, returntype q)}: 
this operation is used to merge methods m and n existing in the hierarchy to a single method m2. The formal description 
of this operation is built on the formal description of five refactoring operations since it is composed of these operations.

\paragraph{Algorithm of the operation} The operation \textsf{MergeDuplicateMethods} is based on four steps : \\ \\
%\begin{figure}
%\begin{center}
\begin{boxedminipage}{\columnwidth}%%%%%%%%%%%%%%%%%%%%%%%%%%%%%%%%%%%%%%%%%%%%%%%

\sf

 MergeDuplicateMethods (c,[a,b],[m,n],m2,t,q) =

\begin{enumerate}
\item  ReplaceMethodcodeDuplicatesInverter (c, m, [n], t,q))
\item  PullupConcreteDelegator(c, [a,b], n ,m))
\item  InlineAndDelete(c,n)) 
\item  RenameInHierarchyNoOverloading (c,  [a,b], m,[t], m2)

\end{enumerate}

\end{boxedminipage}

\begin{center}
\includegraphics[scale=0.7]{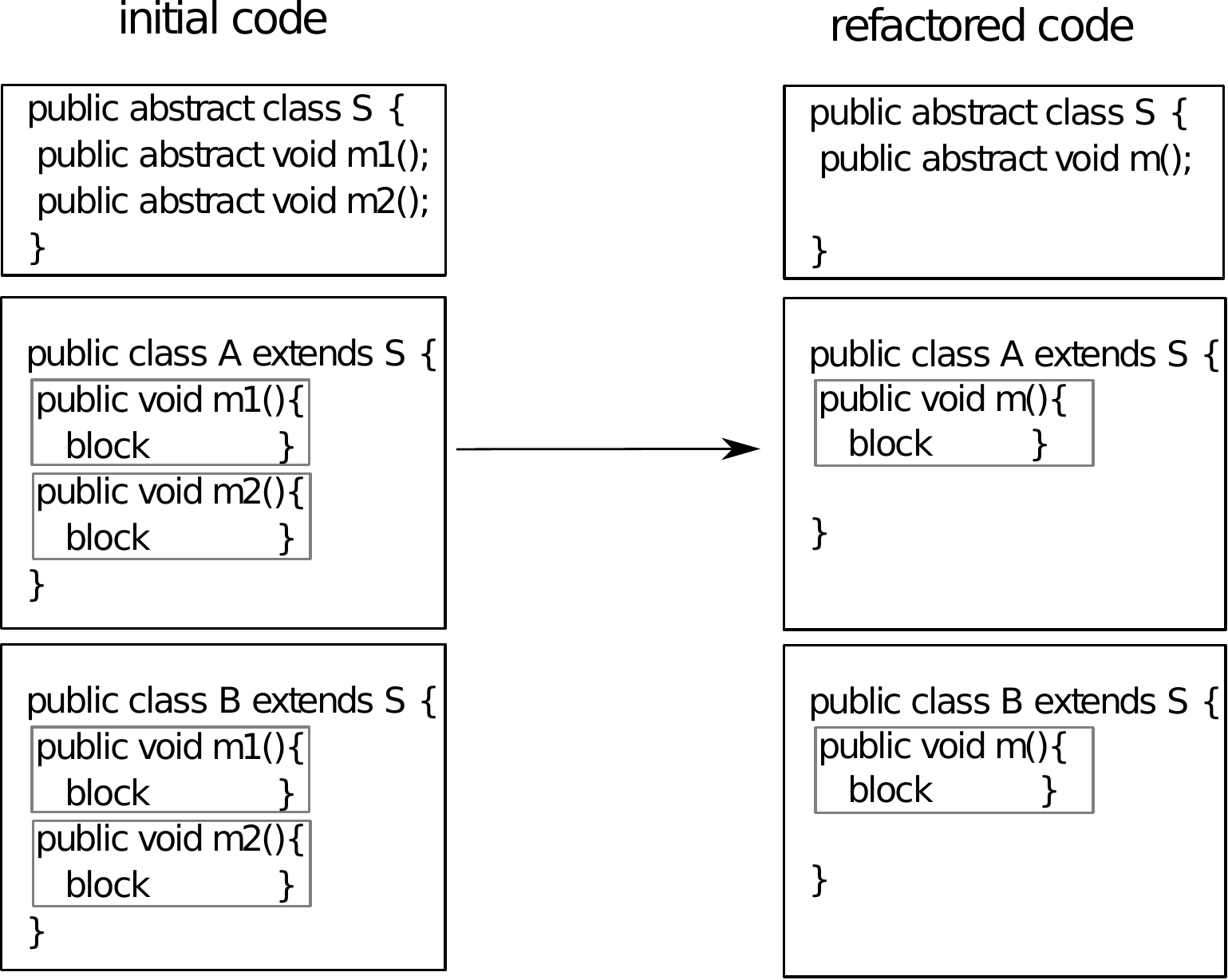}
\end{center}

\tools%%%%%%%
 \emph{Rename, Replace Method duplication, Extract Method,
   In-line} in Eclipse,
 \emph{Rename, Replace Method Code
   Duplicates, Pull Up, Inline} in \intellij:

\paragraph{Notes.}\begin{itemize}

\item 
\textsf{ReplaceCodeDuplicates} introduces a delegation.
 After that, the code for \textsf{m2} is the same in all the subclasses of \textsf{c} (a delegation).
Then the pull-up can be done without changing the semantics, which allows to inline (and remove) \textsf{m2} afterwards.

\item After the first pull-up, the \intellij pull-up warns that some code already exists. The first time, when some code replaces the abstract declaration, the refactorer manages to remove the abstract declaration. 
The next times, when a second code comes in addition of the
first one, the refactorer prefers to leave the two versions
(which are identical in this case), so that we have to
use \emph{safe delete} to remove one of them.

We could provide an extension of the \emph{pull-up} operation with the customized behavior.

\end{itemize}

\precond%%%
\begin{itemize}

\item
The two concerned methods bodies must be syntactically equals.

\item The new name must not introduce an overloading.

\item The two methods must not be overloaded.

\end{itemize}

\subsection{ReplaceMethodcodeDuplicatesInverter}

 \paragraph{Overview} \textsf{ReplaceMethodcodeDuplicatesInverter (classname c, method m, copies [n,m1],invertedtype t,returntype r)} : this operation is used 
to replace c::[n,m1] by c::m. 
\precondition 
 $\\ (\mathtt{ExistsClass}(c)\\ 
\wedge (\mathtt{ExistsMethodDefinition}(c,m)\\ 
\wedge \mathtt{ExistsMethodDefinition}(c,n)\\ 
\wedge \mathtt{ExistsMethodDefinition}(c,m1))\\ 
\wedge (\mathtt{IsInverter}(c,m,t,r)\\ 
\wedge \mathtt{IsInverter}(c,n,t,r)\\ 
\wedge \mathtt{IsInverter}(c,m1,t,r)))$ \\

\backdescr
 $\\ \mathtt{IsDelegator}(c,n,m)\, \mapsto \,\top \\[2pt]
 \mathtt{IsDelegator}(c,m1,m)\, \mapsto \,\top \\[2pt]
 \mathtt{ExistsMethodInvocation}(c,n,c,m)\, \mapsto \,\bot \\[2pt]
 \mathtt{ExistsMethodInvocation}(c,m1,c,m)\, \mapsto \,\bot \\[2pt]
 \mathtt{IsRecursiveMethod}(c,n)\, \mapsto \,\bot \\[2pt]
 \mathtt{IsRecursiveMethod}(c,m1)\, \mapsto \,\bot \\[2pt]
 $

\subsection{SafeDeleteDelegatorOverriding} 
 \paragraph{Overview} \textsf{SafeDeleteDelegatorOverriding (classname c, method m, superclass s, deleguee n)} : this operations is used to remove useless overridings.

\precondition 
$\\ (\mathtt{ExistsClass}(c)\\ 
\wedge \mathtt{ExistsClass}(s)\\ 
\wedge \mathtt{ExistsMethodDefinition}(c,m)\\ 
\wedge \mathtt{ExistsMethodDefinition}(s,m)\\ 
\wedge \mathtt{IsDelegator}(c,m,n)\\ 
\wedge \mathtt{IsDelegator}(s,m,n)\\ 
\wedge \mathtt{AllInvokedMethodsWithParameterOInBodyOfMAreNotOverloaded}(c,m,this))$ \\

\backdescr
$\\ \mathtt{ExistsMethodDefinition}(c,m)\, \mapsto \,\bot \\[2pt]
 \mathtt{IsInheritedMethod}(c,m)\, \mapsto \,\top \\[2pt]
 \mathtt{AllInvokedMethodsOnObjectOInBodyOfMAreDeclaredInC}(c,m,X,Y)\, \mapsto \,\bot \\[2pt]
 \mathtt{AllInvokedMethodsWithParameterOInBodyOfMAreNotOverloaded}(c,m,X)\, \mapsto \,\bot \\[2pt]
 \mathtt{BoundVariableInMethodBody}(c,m,X)\, \mapsto \,\bot \\[2pt]
 \mathtt{ExistsParameterWithName}(c,m,[X],Y)\, \mapsto \,\bot \\[2pt]
 \mathtt{ExistsParameterWithType}(c,m,[X],Y)\, \mapsto \,\bot \\[2pt]
 \mathtt{ExistsMethodInvocation}(c,m,X,Y)\, \mapsto \,\bot \\[2pt]
 \mathtt{ExistsMethodDefinitionWithParams}(c,m,[X])\, \mapsto \,\bot \\[2pt]
 \mathtt{IsInheritedMethodWithParams}(X,m,[Y])\, \mapsto \,\bot \\[2pt]
 \mathtt{IsIndirectlyRecursive}(c,m)\, \mapsto \,\bot \\[2pt]
 \mathtt{IsVisibleMethod}(c,m,[X],Y)\, \mapsto \,\bot \\[2pt]
 \mathtt{IsInverter}(c,m,X,Y)\, \mapsto \,\bot \\[2pt]
 \mathtt{IsDelegator}(c,m,X)\, \mapsto \,\bot \\[2pt]
 \mathtt{IsUsedMethod}(c,m,[X])\, \mapsto \,\bot \\[2pt]
 \mathtt{IsUsedMethodIn}(c,m,X)\, \mapsto \,\bot \\[2pt]
 \mathtt{IsUsedConstructorAsMethodParameter}(X,c,m)\, \mapsto \,\bot \\[2pt]
 \mathtt{IsUsedConstructorAsInitializer}(X,c,m)\, \mapsto \,\bot \\[2pt]
 \mathtt{IsUsedConstructorAsObjectReceiver}(X,c,m)\, \mapsto \,\bot \\[2pt]
 \mathtt{IsPublic}(c,m)\, \mapsto \,\bot \\[2pt]
 \mathtt{IsProtected}(c,m)\, \mapsto \,\bot \\[2pt]
 \mathtt{IsPrivate}(c,m)\, \mapsto \,\bot \\[2pt]
 \mathtt{IsUsedAttributeInMethodBody}(c,X,m)\, \mapsto \,\bot \\[2pt]
 \mathtt{IsOverridden}(c,m)\, \mapsto \,\bot \\[2pt]
 \mathtt{IsOverloaded}(c,m)\, \mapsto \,\bot \\[2pt]
 \mathtt{IsOverriding}(c,m)\, \mapsto \,\bot \\[2pt]
 \mathtt{IsRecursiveMethod}(c,m)\, \mapsto \,\bot \\[2pt]
 \mathtt{HasReturnType}(c,m,X)\, \mapsto \,\bot \\[2pt]
 \mathtt{MethodHasParameterType}(c,m,X)\, \mapsto \,\bot \\[2pt]
 \mathtt{MethodIsUsedWithType}(c,m,[X],[X])\, \mapsto \,\bot \\[2pt]
 $

\subsection{PullUpAbstract}%%%%%%%%%%%%%%%%%%%%%%%%%%%%%%%%%%%%%%%%%%%%%%%%%%%%%%%%%%%%%
\label{def-PullUpAbstract} 

\textsf{PullUpAbstract(set of classes C, method m, interface s)}

Pull up a method implemented in a set of classes \textsf{C}
to their superclass \textsf{s}: do not move the definitions,
just declare the method abstract in \textsf{s}.

\begin{center}

\includegraphics[scale=0.7]{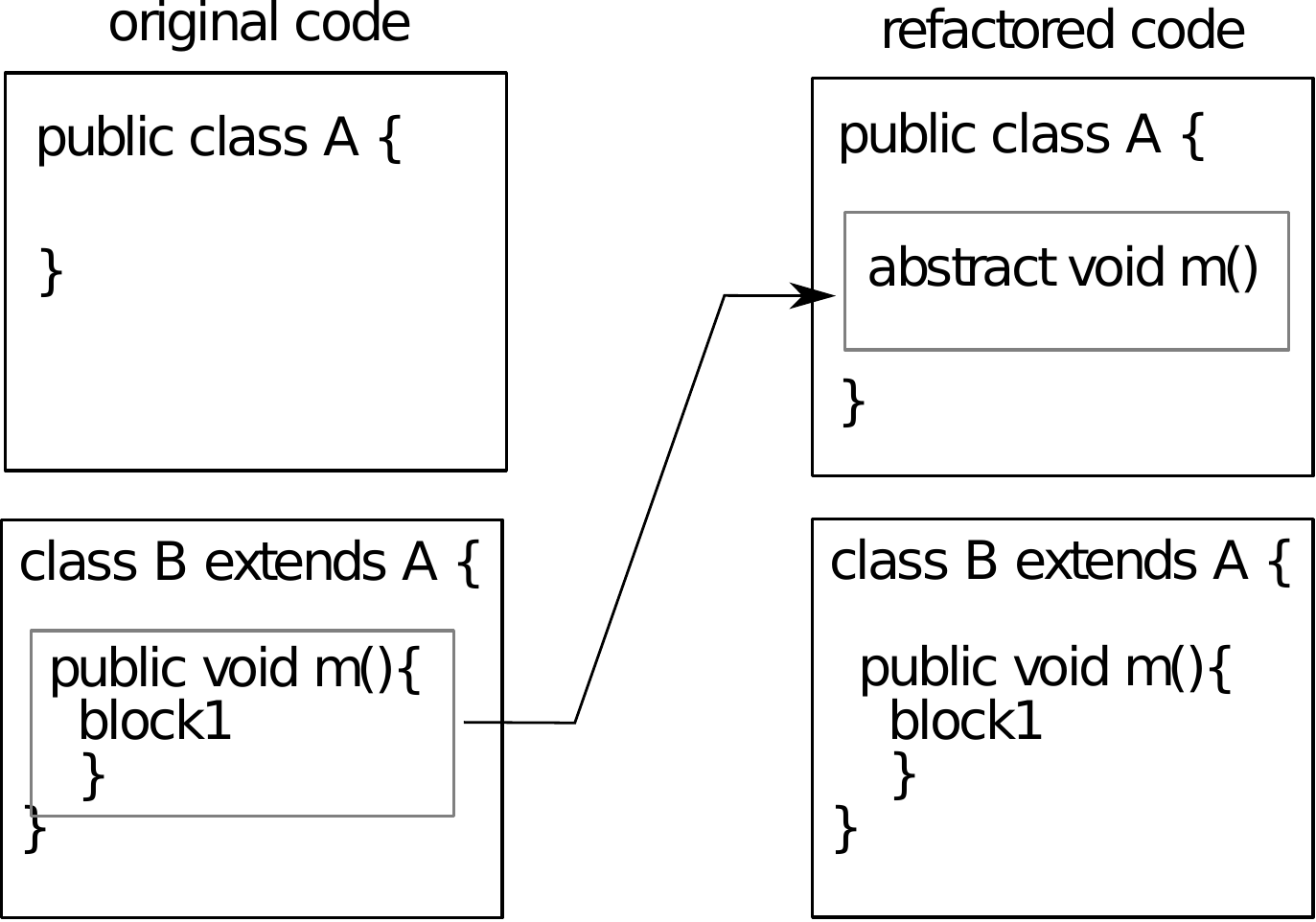}
\end{center}

\tools  \emph{Pull Up} in Eclipse tool and \intellij.

\precond%%%
\begin{itemize}           

       \item \textsf{s} is a superclass of each class in \textsf{C}.

       \item \texttt{m} is defined in all the classes of \texttt{C}
\end{itemize}

%\emph{(attention bug dans l'introduction des \@ Override)}

\subsection{PullUpImplementation}%%%%%%%%%%%%%%%%%%%%%%%%%%%%%%%%%%%%%%%%%%%%%%%%%%
\label{def-PullUpConcrete} 

\paragraph{Overview:} \textsf{PullupImplementation(a,[att1,att2],m,s)}: 
this operation is used to pull up the definition of the method a::m to s and delete it from a.

\begin{center}

\includegraphics[scale=0.7]{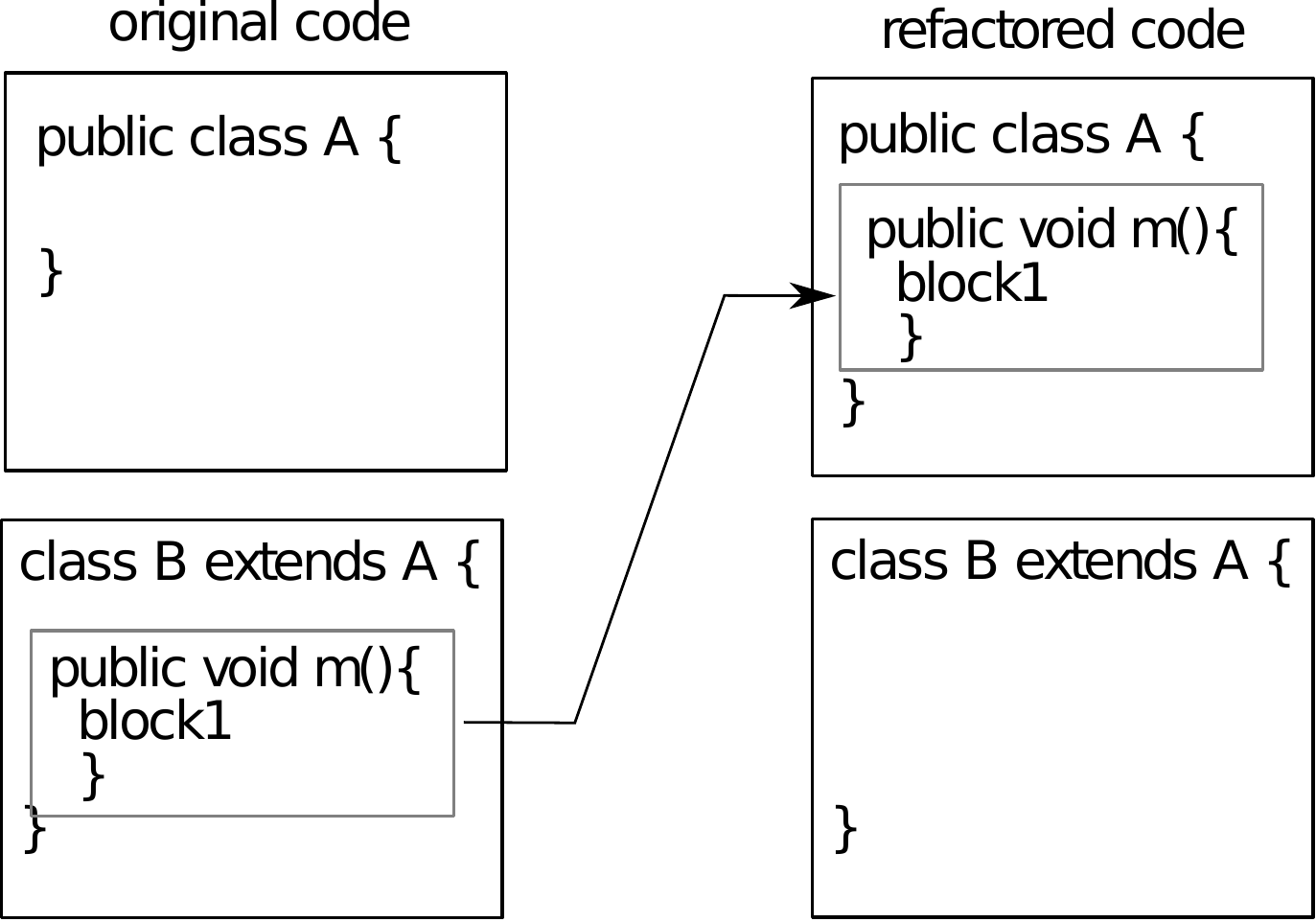}
\end{center}

\tools  \emph{Pull Up} in Eclipse tool and \intellij.

\precondition 
$(\mathtt{ExistsClass}(c)\\ 
\wedge \mathtt{ExistsClass}(s)\\ 
\wedge \mathtt{IsAbstractClass}(s)\\ 
\wedge \mathtt{ExistsMethodDefinition}(c,m)\\ 
\wedge \mathtt{ExistsAbstractMethod}(s,m)\\ 
\wedge \mathtt{AllInvokedMethodsOnObjectOInBodyOfMAreDeclaredInC}(c,m,this,s)\\ 
\wedge \mathtt{AllInvokedMethodsWithParameterOInBodyOfMAreNotOverloaded}(c,m,this)\\ 
\wedge \neg \mathtt{IsPrivate}(c,m)\\ 
\wedge \neg \mathtt{IsUsedAttributeInMethodBody}(c,att1,m)\\ 
\wedge \neg \mathtt{IsUsedAttributeInMethodBody}(c,att2,m))$ \\

\backdescr
$\\ \mathtt{ExistsMethodDefinition}(c,m)\, \mapsto \,\bot \\[2pt]
 \mathtt{ExistsMethodDefinition}(s,m)\, \mapsto \,\top \\[2pt]
 \mathtt{ExistsAbstractMethod}(s,m)\, \mapsto \,\bot \\[2pt]
 \mathtt{IsDelegator}(s,m,X)\, \mapsto \,\mathtt{IsDelegator}(c,m,X) (condition) \\[2pt]
 \mathtt{ExistsMethodDefinitionWithParams}(c,m,[X])\, \mapsto \,\bot \\[2pt]
 \mathtt{IsInheritedMethodWithParams}(c,m,[X])\, \mapsto \,\top \\[2pt]
 \mathtt{IsInheritedMethod}(c,m)\, \mapsto \,\top \\[2pt]
 \mathtt{IsVisibleMethod}(s,m,[X],c)\, \mapsto \,\top \\[2pt]
 \mathtt{IsPrivate}(c,m)\, \mapsto \,\bot \\[2pt]
 \mathtt{IsOverridden}(c,m)\, \mapsto \,\bot \\[2pt]
 \mathtt{IsOverriding}(c,m)\, \mapsto \,\bot \\[2pt]
 \mathtt{IsVisible}(s,m,c)\, \mapsto \,\top \\[2pt]
 \mathtt{IsOverloaded}(s,m)\, \mapsto \,\mathtt{ExistsMethodDefinition}(s,m)\\[2pt]
 \mathtt{IsUsedAttributeInMethodBody}(c,X,m)\, \mapsto \,\bot \\[2pt]
 \mathtt{IsOverridden}(c,m)\, \mapsto \,\bot \\[2pt]
 \mathtt{IsOverloaded}(c,m)\, \mapsto \,\bot \\[2pt]
 \mathtt{IsRecursiveMethod}(c,m)\, \mapsto \,\bot \\[2pt]
 \mathtt{HasReturnType}(c,m,X)\, \mapsto \,\bot \\[2pt]
 \mathtt{MethodHasParameterType}(c,m,X)\, \mapsto \,\bot \\[2pt]
 \mathtt{MethodIsUsedWithType}(c,m,[X],[X])\, \mapsto \,\bot \\[2pt]
 \mathtt{IsPrivate}(c,att1)\, \mapsto \,\bot \\[2pt]
 \mathtt{IsPrivate}(c,att2)\, \mapsto \,\bot \\[2pt]
 $

\subsection{PullUpWithGenerics}%%%%%%%%%%%%%%%%%%%%%%%%%%%%%%%%%%%%%%%%%%%%%%%%%%
\label{def-PullUpWithGenerics} 

\paragraph{Overview:} \textsf{PullupWithGenerics (classname s, subclassname a, [att1,att2],methodname m,returntype r ,parameterType T)}: 
this operation is used to pull up the method a::m to s and then creates the parameter type T to the class s (as shown in  the following figure). After performing 
this operation a polymorphism is created in the hierarchy (Java \emph{Generic} types).

\begin{center}

\includegraphics[scale=0.7]{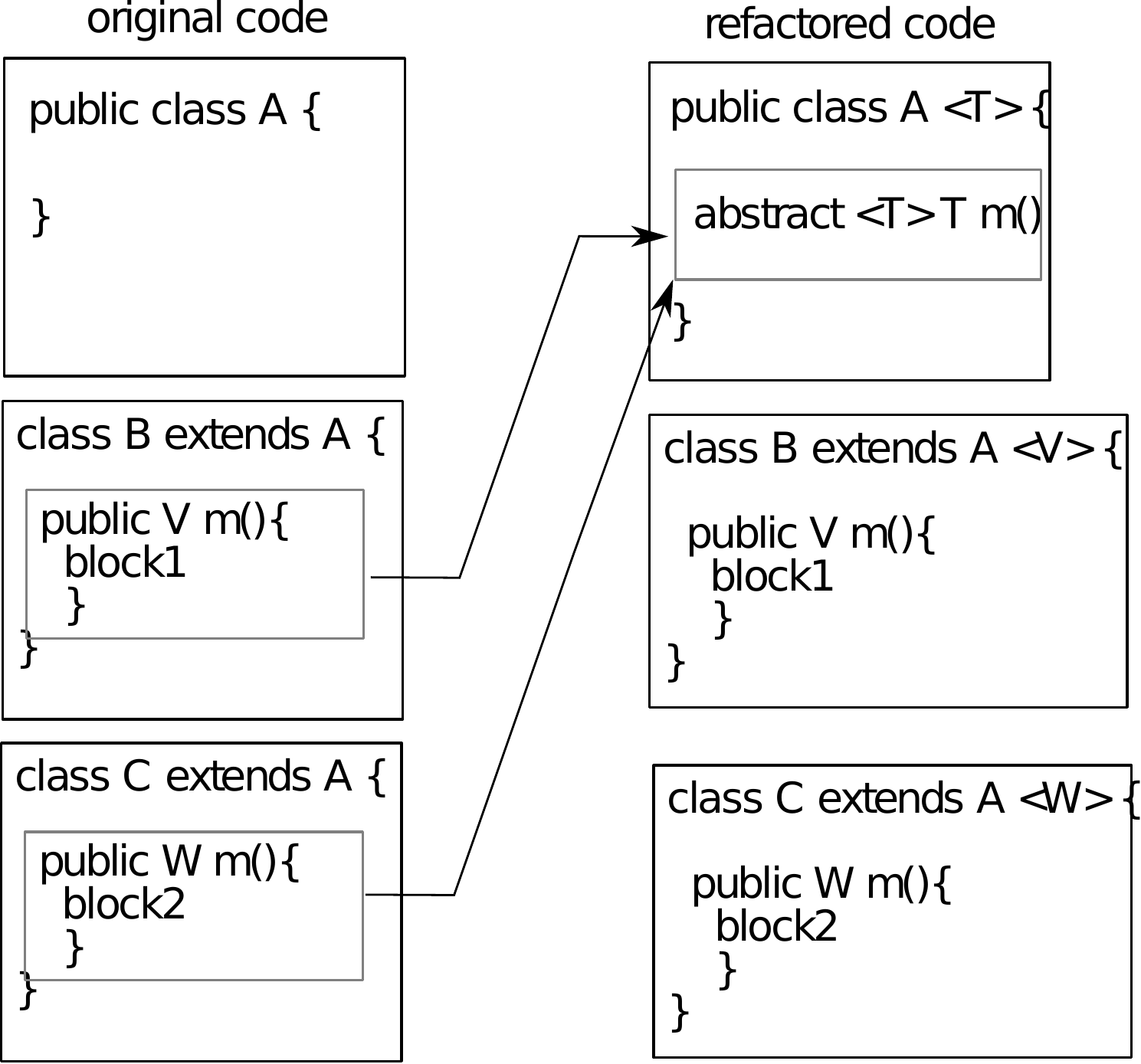}
\end{center}

\tools We provide this operation as a plugin for \intellij (\emph{Pull up method refactoring extension}: \url{http://plugins.intellij.net/plugin/?idea_ce&id=6889} ).

\precondition 
 $\\ (\mathtt{ExistsClass}(s)\\ 
\wedge \mathtt{IsAbstractClass}(s)\\ 
\wedge \mathtt{ExistsClass}(a)\\ 
\wedge \mathtt{IsSubType}(a,s)\\ 
\wedge \mathtt{HasReturnType}(a,m,r)\\ 
\wedge \neg \mathtt{ExistsAbstractMethod}(s,m)\\ 
\wedge \neg \mathtt{IsPrimitiveType}(r)\\ 
\wedge \neg \mathtt{IsPrivate}(a,m)\\ 
\wedge \neg \mathtt{HasParameterType}(a,r)\\ 
\wedge \neg \mathtt{IsPrivate}(a,att1)\\ 
\wedge \neg \mathtt{IsPrivate}(a,att2))$ \\

\backdescr
 
$\mathtt{HasReturnType}(s,m,T)\, \mapsto \,\top \\[2pt]
 \mathtt{ExistsMethodDefinitionWithParams}(s,m,[X])\, \mapsto \,\top \\[2pt]
 \mathtt{MethodHasParameterType}(s,m,T)\, \mapsto \,\top \\[2pt]
 \mathtt{HasParameterType}(s,T)\, \mapsto \,\top \\[2pt]
 \mathtt{extendsFromPrametricClass}(a,s,r)\, \mapsto \,\top \\[2pt]
 \mathtt{IsGenericsSubtype}(a,[r],s,[T])\, \mapsto \,\top \\[2pt]
 \mathtt{IsPrivate}(a,m)\, \mapsto \,\bot \\[2pt]
 $

\subsection{InlineAndDelete}%%%%%%%%%%%%%%%%%%%%%%%%%%%%%%%%%%%%%%%%%%%%%%%%%%%%
\label{def-InlineAndDelete}
(\emph{Inline Method} in ~\cite{Fowler1999})

\paragraph{Overview:} \textsf{InlineAndDelete (classname s,methodname m,types [t,t'],invocatormethod n, othermethods [m1,m2],otherclasses [a,b,c] )}:
this operation is used to replace one or all invocations of a given method by its body and delete it.

\begin{center}

\includegraphics[scale=0.7]{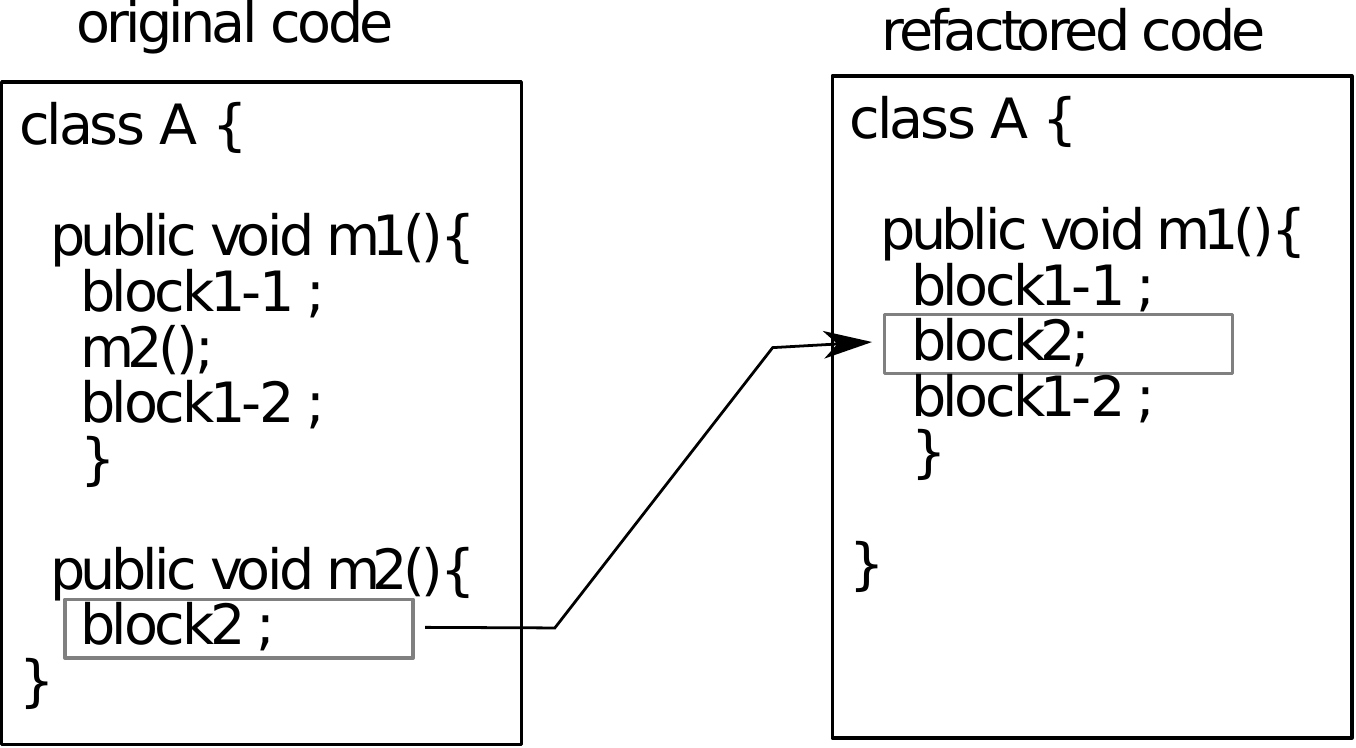}
\end{center}

\tools \emph{In-line} in Eclipse tool and \intellij.

\precondition 
$\\ (\mathtt{ExistsClass}(s)\\ 
\wedge \mathtt{ExistsMethodDefinition}(s,m)\\ 
\wedge \neg \mathtt{IsOverriding}(s,m)\\ 
\wedge \neg \mathtt{IsOverridden}(s,m)\\ 
\wedge \neg \mathtt{IsRecursiveMethod}(s,m)\\ 
\wedge \neg \mathtt{ExistsMethodInvocation}(s,m,s,m1)\\ 
\wedge \neg \mathtt{ExistsMethodInvocation}(s,m,s,m2)\\ 
\wedge \neg \mathtt{IsUsedMethodIn}(s,m,a)\\ 
\wedge \neg \mathtt{IsUsedMethodIn}(s,m,b)\\ 
\wedge \neg \mathtt{IsUsedMethodIn}(s,m,c))$ \\

\backdescr
$\mathtt{ExistsMethodDefinition}(s,m)\, \mapsto \,\bot \\[2pt]
 \mathtt{ExistsMethodDefinitionWithParams}(s,m,[t;t'])\, \mapsto \,\bot \\[2pt]
 \mathtt{AllInvokedMethodsOnObjectOInBodyOfMAreDeclaredInC}(s,m,X,Y)\, \mapsto \,\bot \\[2pt]
 \mathtt{AllInvokedMethodsWithParameterOInBodyOfMAreNotOverloaded}(s,m,X)\, \mapsto \,\bot \\[2pt]
 \mathtt{BoundVariableInMethodBody}(s,m,X)\, \mapsto \,\bot \\[2pt]
 \mathtt{ExistsParameterWithName}(s,m,[X],Y)\, \mapsto \,\bot \\[2pt]
 \mathtt{ExistsParameterWithType}(s,m,[X],Y)\, \mapsto \,\bot \\[2pt]
 \mathtt{ExistsMethodInvocation}(s,m,X,Y)\, \mapsto \,\bot \\[2pt]
 \mathtt{ExistsMethodInvocation}(s,n,Y,Z)\, \mapsto \, \\ ~\hfill (\mathtt{ExistsMethodInvocation}(s,m,Y,Z)\tab \vee \mathtt{ExistsMethodInvocation}(s,n,Y,Z))\\[2pt]
 \mathtt{IsUsedConstructorAsObjectReceiver}(X,s,n)\, \mapsto \, \\ ~\hfill (\mathtt{IsUsedConstructorAsObjectReceiver}(X,s,m)\tab \vee \mathtt{IsUsedConstructorAsObjectReceiver}(X,s,n))\\[2pt]
 \mathtt{IsIndirectlyRecursive}(s,m)\, \mapsto \,\bot \\[2pt]
 \mathtt{IsVisibleMethod}(s,m,[X],Y)\, \mapsto \,\bot \\[2pt]
 \mathtt{IsInverter}(s,m,X,Y)\, \mapsto \,\bot \\[2pt]
 \mathtt{IsDelegator}(s,m,X)\, \mapsto \,\bot \\[2pt]
 \mathtt{IsUsedMethod}(s,m,[X])\, \mapsto \,\bot \\[2pt]
 \mathtt{IsUsedMethodIn}(s,m,X)\, \mapsto \,\bot \\[2pt]
 \mathtt{IsUsedConstructorAsMethodParameter}(X,s,m)\, \mapsto \,\bot \\[2pt]
 \mathtt{IsUsedConstructorAsInitializer}(X,s,m)\, \mapsto \,\bot \\[2pt]
 \mathtt{IsUsedConstructorAsObjectReceiver}(X,s,m)\, \mapsto \,\bot \\[2pt]
 \mathtt{IsPublic}(s,m)\, \mapsto \,\bot \\[2pt]
 \mathtt{IsProtected}(s,m)\, \mapsto \,\bot \\[2pt]
 \mathtt{IsPrivate}(s,m)\, \mapsto \,\bot \\[2pt]
 \mathtt{IsUsedAttributeInMethodBody}(s,X,m)\, \mapsto \,\bot \\[2pt]
 \mathtt{IsOverridden}(s,m)\, \mapsto \,\bot \\[2pt]
 \mathtt{IsOverloaded}(s,m)\, \mapsto \,\bot \\[2pt]
 \mathtt{IsOverriding}(s,m)\, \mapsto \,\bot \\[2pt]
 \mathtt{IsRecursiveMethod}(s,m)\, \mapsto \,\bot \\[2pt]
 \mathtt{HasReturnType}(s,m,X)\, \mapsto \,\bot \\[2pt]
 \mathtt{HasParameterType}(s,m)\, \mapsto \,\bot \\[2pt]
 \mathtt{MethodHasParameterType}(s,m,X)\, \mapsto \,\bot \\[2pt]
 \mathtt{MethodIsUsedWithType}(s,m,[X],[X])\, \mapsto \,\bot \\[2pt]
 $

\subsection{InlineMethodInvocations}%%%%%%%%%%%%%%%%%%%%%%%%%%%%%%%%%%%%%%

\paragraph{Overview:} \textsf{InlineMethodInvocations(classname c,inlinedmethod m, classofinlinedmethod a, modifiedmethod n)}: this operation is used to in-line
a method invocation of the method c::m inside the method a::n.

\begin{center}

\includegraphics[scale=0.7]{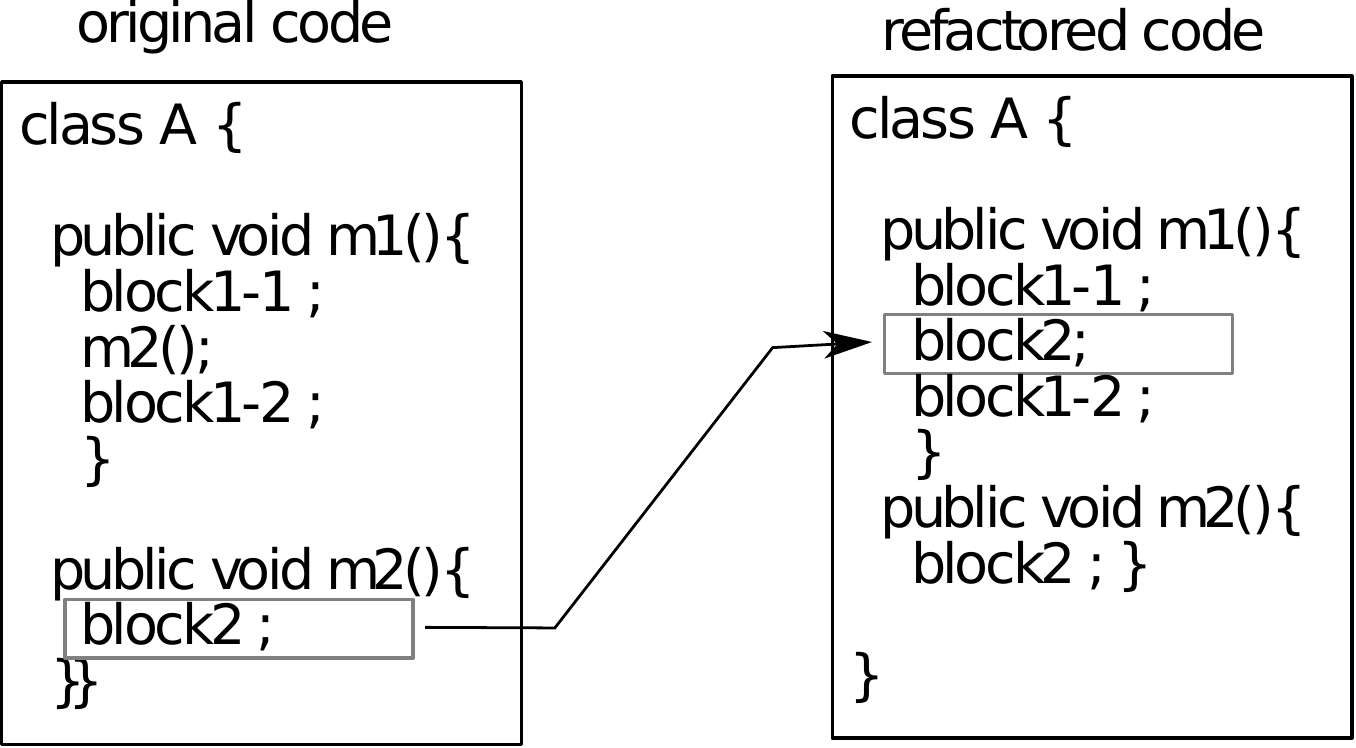}
\end{center}

\tools {Inline} in Eclipse and \intellij: select
an invocation to inline and specify you want to inline only
that one.

\precondition 
$\\ (\mathtt{ExistsClass}(c)\\ 
\wedge \mathtt{IsIndirectlyRecursive}(c,m)\\ 
\wedge \mathtt{IsRecursiveMethod}(c,n)\\ 
\wedge \neg \mathtt{IsRecursiveMethod}(c,m)\\ 
\wedge \mathtt{ExistsMethodInvocation}(c,m,a,n)\\ 
\wedge \mathtt{ExistsMethodDefinition}(c,m)\\ 
\wedge \mathtt{ExistsMethodDefinitionWithParams}(c,m,[t;t'])\\ 
\wedge \mathtt{ExistsMethodDefinition}(a,n)\\ 
\wedge \mathtt{ExistsMethodDefinitionWithParams}(a,n,[t1;t1']))$ \\ 

\backdescr
 $\mathtt{ExistsMethodDefinitionWithParams}(c,m,[t;t'])\, \mapsto \,\top \\[2pt]
 \mathtt{ExistsMethodDefinition}(c,m)\, \mapsto \,\top \\[2pt]
 \mathtt{ExistsMethodInvocation}(c,m,a,n)\, \mapsto \,\bot \\[2pt]
 \mathtt{IsRecursiveMethod}(c,m)\, \mapsto \,\mathtt{ExistsMethodInvocation}(a,n,c,m)\\[2pt]
 \mathtt{IsIndirectlyRecursive}(c,m)\, \mapsto \, \\ ~\hfill(\mathtt{ExistsMethodInvocation}(a,n,C,X)\\ \wedge \mathtt{ExistsMethodInvocation}(C,X,c,m)) (condition) \\[2pt]
 \mathtt{IsUsedMethodIn}(a,n,m)\, \mapsto \,\bot \\[2pt]
 $

\subsection{AddSpecializedMethodInHierarchy  (Composed)}%%%%%%%%%%%%%%%%%%%%%%%%%%%%%%%%%%%%%%%%%%%%%
  \label{def-AddSpecializedMethodInHierarchy}

  \paragraph{Overview:} \textsf{AddSpecializedMethodInHierarchy(class s, subclasses [a,b], methodname m, callermethods [n,o], inkovekmethods [p,q], paramtype t, paramname pn , 
subtypesOfparamtype [t1,t2],newtype t')}: this operation is used to get the method s::m(t' pn) instead of s::m(t pn). 
This new duplication takes place in \textsf{s} and in all its subclasses that override \textsf{m}.

\begin{center}

 \includegraphics[scale=0.7]{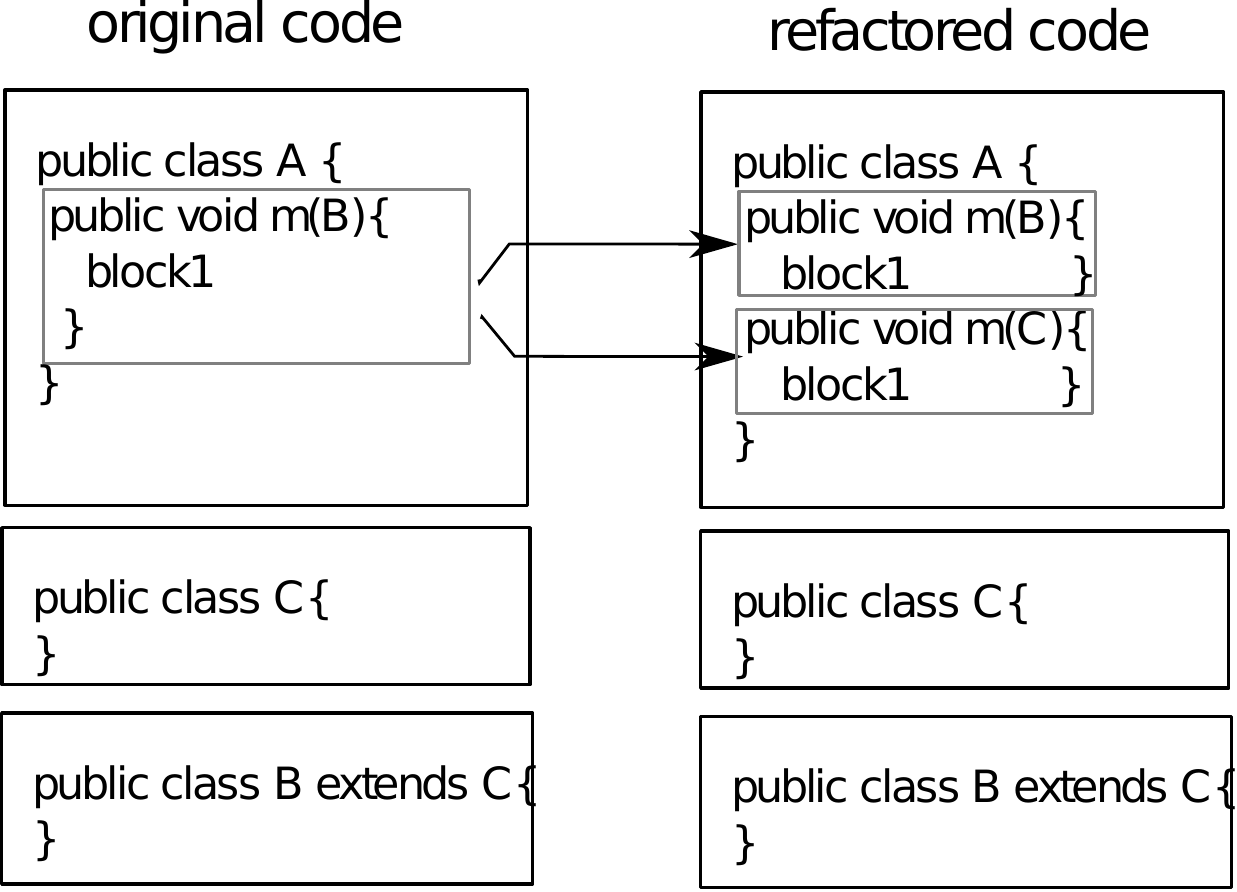}
\end{center}

\paragraph{Algorithm of the operation} The operation \textsf{AddSpecializedMethodInHierarchy} is based on three steps : \\ \\
%\begin{figure}
%\begin{center}
\begin{boxedminipage}{\columnwidth}%%%%%%%%%%%%%%%%%%%%%%%%%%%%%%%%%%%%%%%%%%%%%%%

\sf
 AddSpecializedMethodInHierarchy(class s, subclasses [a,b], methodname m, callermethods [n,o], inkovekmethods [p,q], paramtype t, paramname pn , 
subtypesOfparamtype [t1,t2],newtype t') =

\begin{enumerate}
\item  DuplicateMethodInHierarchy  s [a,b] m [p,q]  [n,o] temporaryName  [t]
\item  SpecialiseParameter s [a,b] temporaryName t pn [t1,t2] t'; 
\item  RenameDelegatorWithOverloading (s, [a,b], temporaryName,t', pn,t,m)

\end{enumerate}

\end{boxedminipage}

\tools With \intellij:

\begin{enumerate}

\item Apply \textsf{DuplicateMethodInHierarchy(c, m, temp-name)} (see~\ref{def-duplicate-method-in-hierarchy} below).

\item Apply \emph{Change Signature} on the method
  \textsf{temp-name} in the class \textsf{s}, to change the
  parameter type \textsf{t} into \textsf{t'}
  (this change is propagated into subclasses).

 Note that the behavior preservation is not guaranteed by
 this operation in general, but here we introduce a new
 method so the behavior is not changed.
  Note also, that here we cannot use the operation
  \emph{Type Migration} of \intellij : replacing a parameter
  type by one of its subtypes is not safe in general.

\item Rename \textsf{temp-name} into \textsf{m}
  in \textsf{s} with \emph{Rename}.
 That renaming introduces an overloading.  In general, this
  could change the semantics of the program, but in the case
  of this particular chain, and provided the preconditions
  given below are satisfied, the behavior is preserved (the
  two methods have the same body; some invocations may be
  dispatched on the new method, but the external behavior is
  the same).

\end{enumerate}

\subsection{DuplicateMethodInHierarchy}%%%%%%%%%%%%%%%%%%%%%%%%%%%%%%%%%%%%%%%%%%%%%%%%%%%%%%%%%%%%%
\label{def-duplicate-method-in-hierarchy}
 \paragraph{Overview:} \textsf{DuplicateMethodInHierarchy(class s, subclasses [a,b], methodname m, callermethods [m1,m2], inkovekmethods [m3,m4],newname n ,paramType [t])} : 
this operation is used to create a duplicate of the method s::m with the name n. All overriding
methods in subclasses are also duplicated in these classes.

\tools%%%%%%%%%%%%%%
With \intellij:\begin{enumerate}

\item For each implementation of the method \textsf{m} in
  the subclasses of the class \textsf{s}, duplicate \textsf{m} by
  applying \emph{Extract Method} on its body (give the new
  name, specify the desired visibility), then inline the 
   invocation of method \textsf{n} that has replaced the method's body.

\item Use \emph{Pull Members Up} to make the new method
  appear in classes where the initial method is declared
  abstract (specify that it must appear as abstract) (see \emph{PullUpAbstract}).

\end{enumerate}

\precondition 

$\\ (\mathtt{ExistsClass}(s)\\ 
\wedge \mathtt{ExistsMethodDefinitionWithParams}(s,m,[t;t'])\\ 
\wedge \mathtt{ExistsMethodDefinition}(s,m)\\ 
\wedge \neg \mathtt{ExistsMethodDefinitionWithParams}(s,n,[t;t'])\\ 
\wedge \neg \mathtt{ExistsMethodDefinitionWithParams}(a,n,[t;t'])\\ 
\wedge \neg \mathtt{ExistsMethodDefinitionWithParams}(b,n,[t;t'])\\ 
\wedge \neg \mathtt{IsInheritedMethodWithParams}(s,n,[t;t'])\\ 
\wedge \mathtt{AllSubclasses}(s,[a;b]))$ \\ 

\backdescr
 $\\ \mathtt{ExistsMethodDefinition}(s,n)\, \mapsto \,\top \\[2pt]
 \mathtt{ExistsMethodDefinitionWithParams}(s,n,[t;t'])\, \mapsto \,\top \\[2pt]
 \mathtt{AllInvokedMethodsWithParameterOInBodyOfMAreNotOverloaded}(s,n,V)\, \mapsto \,\top  (condition) \\[2pt]
 \mathtt{AllInvokedMethodsOnObjectOInBodyOfMAreDeclaredInC}(s,n,V,V1)\, \mapsto \,\mathtt{AllInvokedMethodsOnObjectOInBodyOfMAreDeclaredInC}(s,m,V,V1)\\[2pt]
 \mathtt{BoundVariableInMethodBody}(s,n,V)\, \mapsto \,\mathtt{BoundVariableInMethodBody}(s,m,V)\\[2pt]
 \mathtt{IsPublic}(s,n)\, \mapsto \,\mathtt{IsPublic}(s,m)\\[2pt]
 \mathtt{ExistsParameterWithName}(s,n,[t;t'],V)\, \mapsto \,\mathtt{ExistsParameterWithName}(s,m,[t;t'],V)\\[2pt]
 \mathtt{ExistsParameterWithType}(s,n,[t;t'],T)\, \mapsto \,\mathtt{ExistsParameterWithType}(s,m,[t;t'],T)\\[2pt]
 \mathtt{IsIndirectlyRecursive}(s,n)\, \mapsto \,\mathtt{IsIndirectlyRecursive}(s,m)\\[2pt]
 \mathtt{IsRecursiveMethod}(s,n)\, \mapsto \,\mathtt{IsRecursiveMethod}(s,m)\\[2pt]
 \mathtt{IsInverter}(s,n,T,V)\, \mapsto \,\mathtt{IsInverter}(s,m,T,V)\\[2pt]
 \mathtt{IsUsedAttributeInMethodBody}(s,V,n)\, \mapsto \,\mathtt{IsUsedAttributeInMethodBody}(s,V,m)\\[2pt]
 \mathtt{MethodHasParameterType}(s,n,V)\, \mapsto \,\mathtt{MethodHasParameterType}(s,m,V)\\[2pt]
 \mathtt{ExistsMethodDefinitionWithParams}(a,n,[t;t'])\, \mapsto \,\mathtt{ExistsMethodDefinitionWithParams}(a,m,[t;t'])\\[2pt]
 \mathtt{ExistsMethodDefinitionWithParams}(b,n,[t;t'])\, \mapsto \,\mathtt{ExistsMethodDefinitionWithParams}(b,m,[t;t'])\\[2pt]
 \mathtt{IsDelegator}(s,n,m3)\, \mapsto \,\top \\[2pt]
 \mathtt{IsDelegator}(s,n,m4)\, \mapsto \,\top \\[2pt]
 \mathtt{IsDelegator}(a,n,m3)\, \mapsto \,\top \\[2pt]
 \mathtt{IsDelegator}(a,n,m4)\, \mapsto \,\top \\[2pt]
 \mathtt{IsDelegator}(b,n,m3)\, \mapsto \,\top \\[2pt]
 \mathtt{IsDelegator}(b,n,m4)\, \mapsto \,\top \\[2pt]
 \mathtt{ExistsMethodDefinition}(s,n)\, \mapsto \,\top \\[2pt]
 \mathtt{ExistsMethodDefinition}(a,n)\, \mapsto \,\top \\[2pt]
 \mathtt{ExistsMethodDefinition}(b,n)\, \mapsto \,\top \\[2pt]
 \mathtt{MethodIsUsedWithType}(s,n,[t;t'],[t;t'])\, \mapsto \,\bot \\[2pt]
 \mathtt{MethodIsUsedWithType}(a,n,[t;t'],[t;t'])\, \mapsto \,\bot \\[2pt]
 \mathtt{MethodIsUsedWithType}(b,n,[t;t'],[t;t'])\, \mapsto \,\bot \\[2pt]
 \mathtt{MethodIsUsedWithType}(s,n,[t;t'],[T])\, \mapsto \,\bot \\[2pt]
 \mathtt{MethodIsUsedWithType}(a,n,[t;t'],[T])\, \mapsto \,\bot \\[2pt]
 \mathtt{MethodIsUsedWithType}(b,n,[t;t'],[T])\, \mapsto \,\bot \\[2pt]
 \mathtt{ExistsMethodInvocation}(s,m1,V,n)\, \mapsto \,\top \\[2pt]
 \mathtt{ExistsMethodInvocation}(s,m2,V,n)\, \mapsto \,\top \\[2pt]
 \mathtt{ExistsMethodInvocation}(a,m1,V,n)\, \mapsto \,\top \\[2pt]
 \mathtt{ExistsMethodInvocation}(a,m2,V,n)\, \mapsto \,\top \\[2pt]
 \mathtt{ExistsMethodInvocation}(b,m1,V,n)\, \mapsto \,\top \\[2pt]
 \mathtt{ExistsMethodInvocation}(b,m2,V,n)\, \mapsto \,\top \\[2pt]
 \mathtt{IsInheritedMethodWithParams}(a,n,[t;t'])\, \mapsto \,\neg \mathtt{ExistsMethodDefinitionWithParams}(a,m,[t;t'])\\[2pt]
 \mathtt{IsInheritedMethodWithParams}(b,n,[t;t'])\, \mapsto \,\neg \mathtt{ExistsMethodDefinitionWithParams}(b,m,[t;t'])\\[2pt]
 \mathtt{IsInheritedMethod}(a,n)\, \mapsto \,\neg \mathtt{ExistsMethodDefinition}(a,m)\\[2pt]
 \mathtt{IsInheritedMethod}(b,n)\, \mapsto \,\neg \mathtt{ExistsMethodDefinition}(b,m)\\[2pt]
 \mathtt{IsOverriding}(a,n)\, \mapsto \,\neg \mathtt{ExistsMethodDefinition}(a,m)\\[2pt]
 \mathtt{IsOverriding}(b,n)\, \mapsto \,\neg \mathtt{ExistsMethodDefinition}(b,m)\\[2pt]
 \mathtt{IsOverridden}(a,n)\, \mapsto \,\neg \mathtt{ExistsMethodDefinition}(a,m)\\[2pt]
 \mathtt{IsOverridden}(b,n)\, \mapsto \,\neg \mathtt{ExistsMethodDefinition}(b,m)\\[2pt]
 $

\subsection{DeleteMethodInHierarchy}%%%%%%%%%%%%%%%%%%%%%%%%%%%%%%%%%%%%%%%%%%%%%%%%%%%%%%
\label{def-DeleteMethodInHierArchy}
 
(\emph{Delete Method} in Fowler~\cite{Fowler1999} and ~\cite{koch2002})

\paragraph{Overview:} \textsf{DeleteMethodInHierarchy (classname s, subclasses [a,b], method m, invokedmethodsInm [m1,m2], paramType t)} : this operation is 
used to delete the method m from the hierarchy of classes s, a and b.

\begin{center}
\includegraphics[scale=0.7]{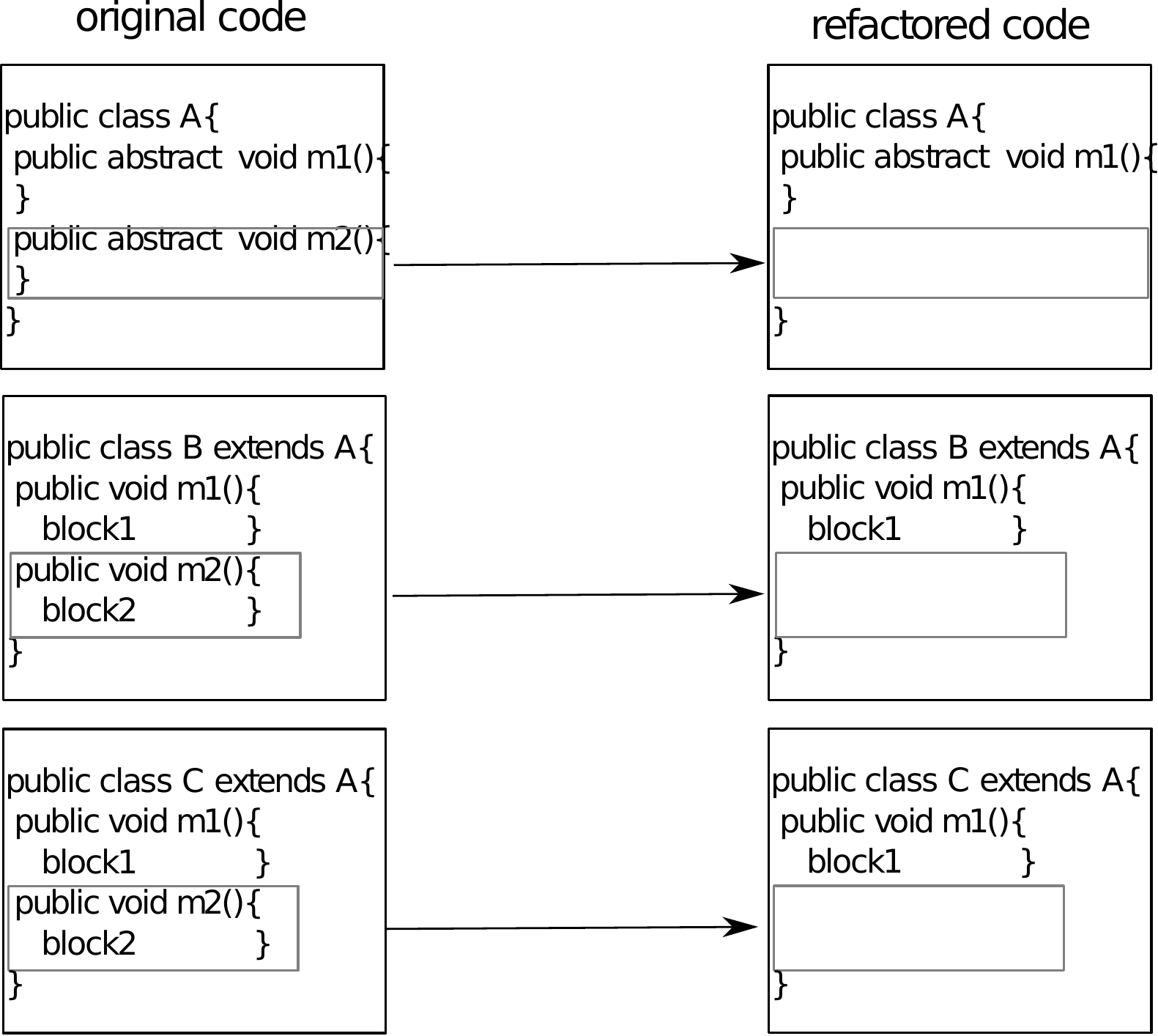}
\end{center}

\tools \emph{Safe Delete} in \intellij and \emph{Delete} in Eclipse.

\precondition 
$\\(\mathtt{ExistsClass}(s)\\ 
\wedge \mathtt{ExistsMethodDefinitionWithParams}(s,m,[t])\\ 
\wedge \neg \mathtt{MethodIsUsedWithType}(s,m,[t],[t])\\ 
\wedge \neg \mathtt{MethodIsUsedWithType}(a,m,[t],[t])\\ 
\wedge \neg \mathtt{MethodIsUsedWithType}(b,m,[t],[t])\\ 
\wedge \mathtt{AllSubclasses}(s,[a;b]))$ \\ 

\backdescr
$\\ \mathtt{ExistsParameterWithType}(s,m,[t],t)\, \mapsto \,\bot \\[2pt]
 \mathtt{ExistsParameterWithType}(a,m,[t],t)\, \mapsto \,\bot \\[2pt]
 \mathtt{ExistsParameterWithType}(b,m,[t],t)\, \mapsto \,\bot \\[2pt]
 \mathtt{ExistsMethodDefinitionWithParams}(s,m,[t])\, \mapsto \,\bot \\[2pt]
 \mathtt{ExistsMethodDefinitionWithParams}(a,m,[t])\, \mapsto \,\bot \\[2pt]
 \mathtt{ExistsMethodDefinitionWithParams}(b,m,[t])\, \mapsto \,\bot \\[2pt]
 \mathtt{ExistsMethodDefinition}(s,m)\, \mapsto \,\bot \\[2pt]
 \mathtt{ExistsMethodDefinition}(a,m)\, \mapsto \,\bot \\[2pt]
 \mathtt{ExistsMethodDefinition}(b,m)\, \mapsto \,\bot \\[2pt]
 \mathtt{IsUsedMethod}(s,m1,[V1])\, \mapsto \,\bot \\[2pt]
 \mathtt{IsUsedMethod}(s,m2,[V1])\, \mapsto \,\bot \\[2pt]
 \mathtt{IsUsedMethod}(a,m1,[V1])\, \mapsto \,\bot \\[2pt]
 \mathtt{IsUsedMethod}(a,m2,[V1])\, \mapsto \,\bot \\[2pt]
 \mathtt{IsUsedMethod}(b,m1,[V1])\, \mapsto \,\bot \\[2pt]
 \mathtt{IsUsedMethod}(b,m2,[V1])\, \mapsto \,\bot \\[2pt]
 \mathtt{IsUsedConstructorAsMethodParameter}(V1,s,m)\, \mapsto \,\bot \\[2pt]
 \mathtt{IsUsedConstructorAsMethodParameter}(V1,a,m)\, \mapsto \,\bot \\[2pt]
 \mathtt{IsUsedConstructorAsMethodParameter}(V1,b,m)\, \mapsto \,\bot \\[2pt]
 \mathtt{IsUsedConstructorAsMethodParameter}(t,s,m)\, \mapsto \,\bot \\[2pt]
 \mathtt{IsUsedConstructorAsMethodParameter}(t,a,m)\, \mapsto \,\bot \\[2pt]
 \mathtt{IsUsedConstructorAsMethodParameter}(t,b,m)\, \mapsto \,\bot \\[2pt]
 \mathtt{IsUsedConstructorAsObjectReceiver}(t,s,m)\, \mapsto \,\bot \\[2pt]
 \mathtt{IsUsedConstructorAsObjectReceiver}(t,a,m)\, \mapsto \,\bot \\[2pt]
 \mathtt{IsUsedConstructorAsObjectReceiver}(t,b,m)\, \mapsto \,\bot \\[2pt]
 \mathtt{IsInheritedMethod}(a,m)\, \mapsto \,\bot \\[2pt]
 \mathtt{IsInheritedMethod}(b,m)\, \mapsto \,\bot \\[2pt]
 \mathtt{AllInvokedMethodsOnObjectOInBodyOfMAreDeclaredInC}(s,m,V1,V2)\, \mapsto \,\bot \\[2pt]
 \mathtt{AllInvokedMethodsOnObjectOInBodyOfMAreDeclaredInC}(a,m,V1,V2)\, \mapsto \,\bot \\[2pt]
 \mathtt{AllInvokedMethodsOnObjectOInBodyOfMAreDeclaredInC}(b,m,V1,V2)\, \mapsto \,\bot \\[2pt]
 \mathtt{ExistsAbstractMethod}(s,m)\, \mapsto \,\bot \\[2pt]
 \mathtt{ExistsAbstractMethod}(a,m)\, \mapsto \,\bot \\[2pt]
 \mathtt{ExistsAbstractMethod}(b,m)\, \mapsto \,\bot \\[2pt]
 \mathtt{AllInvokedMethodsWithParameterOInBodyOfMAreNotOverloaded}(s,m,V1)\, \mapsto \,\bot \\[2pt]
 \mathtt{AllInvokedMethodsWithParameterOInBodyOfMAreNotOverloaded}(a,m,V1)\, \mapsto \,\bot \\[2pt]
 \mathtt{AllInvokedMethodsWithParameterOInBodyOfMAreNotOverloaded}(b,m,V1)\, \mapsto \,\bot \\[2pt]
 \mathtt{BoundVariableInMethodBody}(s,m,V1)\, \mapsto \,\bot \\[2pt]
 \mathtt{BoundVariableInMethodBody}(a,m,V1)\, \mapsto \,\bot \\[2pt]
 \mathtt{BoundVariableInMethodBody}(b,m,V1)\, \mapsto \,\bot \\[2pt]
 \mathtt{ExistsParameterWithName}(s,m,[t],V1)\, \mapsto \,\bot \\[2pt]
 \mathtt{ExistsParameterWithName}(a,m,[t],V1)\, \mapsto \,\bot \\[2pt]
 \mathtt{ExistsParameterWithName}(b,m,[t],V1)\, \mapsto \,\bot \\[2pt]
 \mathtt{ExistsMethodInvocation}(s,m,V1,V2)\, \mapsto \,\bot \\[2pt]
 \mathtt{ExistsMethodInvocation}(a,m,V1,V2)\, \mapsto \,\bot \\[2pt]
 \mathtt{ExistsMethodInvocation}(b,m,V1,V2)\, \mapsto \,\bot \\[2pt]
 \mathtt{IsInheritedMethodWithParams}(a,m,[t])\, \mapsto \,\bot \\[2pt]
 \mathtt{IsInheritedMethodWithParams}(b,m,[t])\, \mapsto \,\bot \\[2pt]
 \mathtt{IsIndirectlyRecursive}(s,m)\, \mapsto \,\bot \\[2pt]
 \mathtt{IsIndirectlyRecursive}(a,m)\, \mapsto \,\bot \\[2pt]
 \mathtt{IsIndirectlyRecursive}(b,m)\, \mapsto \,\bot \\[2pt]
 \mathtt{IsVisibleMethod}(s,m,[t],V1)\, \mapsto \,\bot \\[2pt]
 \mathtt{IsVisibleMethod}(a,m,[t],V1)\, \mapsto \,\bot \\[2pt]
 \mathtt{IsVisibleMethod}(b,m,[t],V1)\, \mapsto \,\bot \\[2pt]
 \mathtt{IsInverter}(s,m,V1,V2)\, \mapsto \,\bot \\[2pt]
 \mathtt{IsInverter}(a,m,V1,V2)\, \mapsto \,\bot \\[2pt]
 \mathtt{IsInverter}(b,m,V1,V2)\, \mapsto \,\bot \\[2pt]
 \mathtt{IsDelegator}(s,V1,m)\, \mapsto \,\bot \\[2pt]
 \mathtt{IsDelegator}(a,V1,m)\, \mapsto \,\bot \\[2pt]
 \mathtt{IsDelegator}(b,V1,m)\, \mapsto \,\bot \\[2pt]
 \mathtt{IsUsedMethodIn}(s,m,V1)\, \mapsto \,\bot \\[2pt]
 \mathtt{IsUsedMethodIn}(a,m,V1)\, \mapsto \,\bot \\[2pt]
 \mathtt{IsUsedMethodIn}(b,m,V1)\, \mapsto \,\bot \\[2pt]
 \mathtt{IsUsedConstructorAsInitializer}(V1,s,m)\, \mapsto \,\bot \\[2pt]
 \mathtt{IsUsedConstructorAsInitializer}(V1,a,m)\, \mapsto \,\bot \\[2pt]
 \mathtt{IsUsedConstructorAsInitializer}(V1,b,m)\, \mapsto \,\bot \\[2pt]
 \mathtt{IsUsedAttributeInMethodBody}(s,V1,m)\, \mapsto \,\bot \\[2pt]
 \mathtt{IsUsedAttributeInMethodBody}(a,V1,m)\, \mapsto \,\bot \\[2pt]
 \mathtt{IsUsedAttributeInMethodBody}(b,V1,m)\, \mapsto \,\bot \\[2pt]
 \mathtt{IsOverridden}(a,m)\, \mapsto \,\bot \\[2pt]
 \mathtt{IsOverridden}(b,m)\, \mapsto \,\bot \\[2pt]
 \mathtt{IsOverloaded}(s,m)\, \mapsto \,\bot \\[2pt]
 \mathtt{IsOverloaded}(a,m)\, \mapsto \,\bot \\[2pt]
 \mathtt{IsOverloaded}(b,m)\, \mapsto \,\bot \\[2pt]
 \mathtt{IsOverriding}(a,m)\, \mapsto \,\bot \\[2pt]
 \mathtt{IsOverriding}(b,m)\, \mapsto \,\bot \\[2pt]
 \mathtt{IsRecursiveMethod}(s,m)\, \mapsto \,\bot \\[2pt]
 \mathtt{IsRecursiveMethod}(a,m)\, \mapsto \,\bot \\[2pt]
 \mathtt{IsRecursiveMethod}(b,m)\, \mapsto \,\bot \\[2pt]
 \mathtt{HasReturnType}(s,m,V1)\, \mapsto \,\bot \\[2pt]
 \mathtt{HasReturnType}(a,m,V1)\, \mapsto \,\bot \\[2pt]
 \mathtt{HasReturnType}(b,m,V1)\, \mapsto \,\bot \\[2pt]
 \mathtt{MethodHasParameterType}(s,m,V1)\, \mapsto \,\bot \\[2pt]
 \mathtt{MethodHasParameterType}(a,m,V1)\, \mapsto \,\bot \\[2pt]
 \mathtt{MethodHasParameterType}(b,m,V1)\, \mapsto \,\bot \\[2pt]
 \mathtt{MethodIsUsedWithType}(s,m,[t],[t])\, \mapsto \,\bot \\[2pt]
 \mathtt{MethodIsUsedWithType}(a,m,[t],[t])\, \mapsto \,\bot \\[2pt]
 \mathtt{MethodIsUsedWithType}(b,m,[t],[t])\, \mapsto \,\bot \\[2pt]
 $

\subsection{PushDownAll}%%%%%%%%%%%%%%%%%%%%%%%%%%%%%%%%%%%%%%%%%%%%%%%%%%%%%%%%%%%%%%
\label{def-PushDownAll}

\paragraph{Overview:} \textsf{PushDownAll (classname s, attributes [att1,att2], subclasses [a,b], method m,paramType [t])} : this operation is used to 
push down the method s::m to its subclasses and delete that
method from s (in \emph{Push Down Method} by
Fowler~\cite{Fowler1999}, methods are not necessarily pushed
down to all the subclasses).

\begin{center}
\includegraphics[scale=0.7]{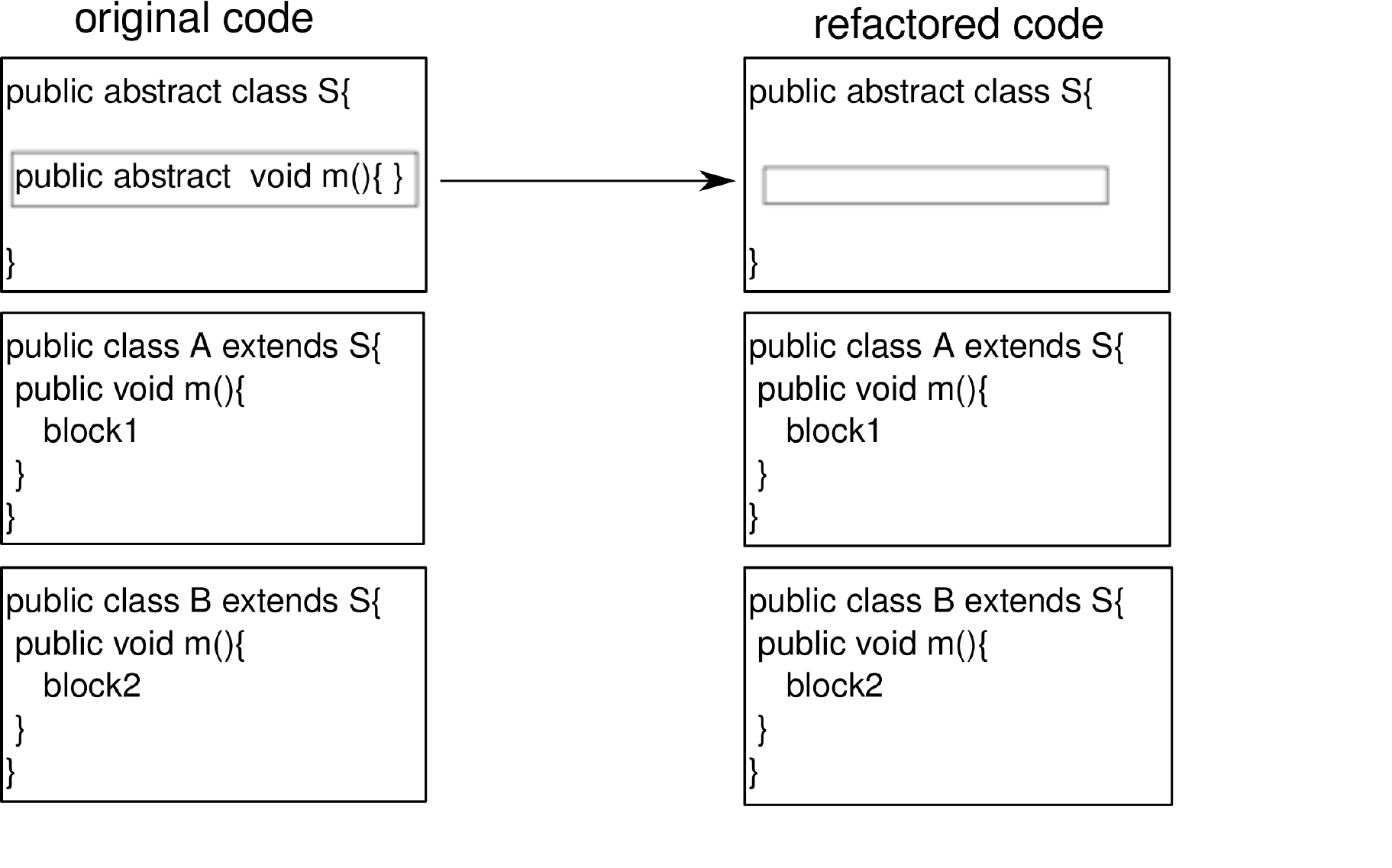}
\end{center}

Variation for non-abstract methods:

\begin{center}
\includegraphics[scale=0.7]{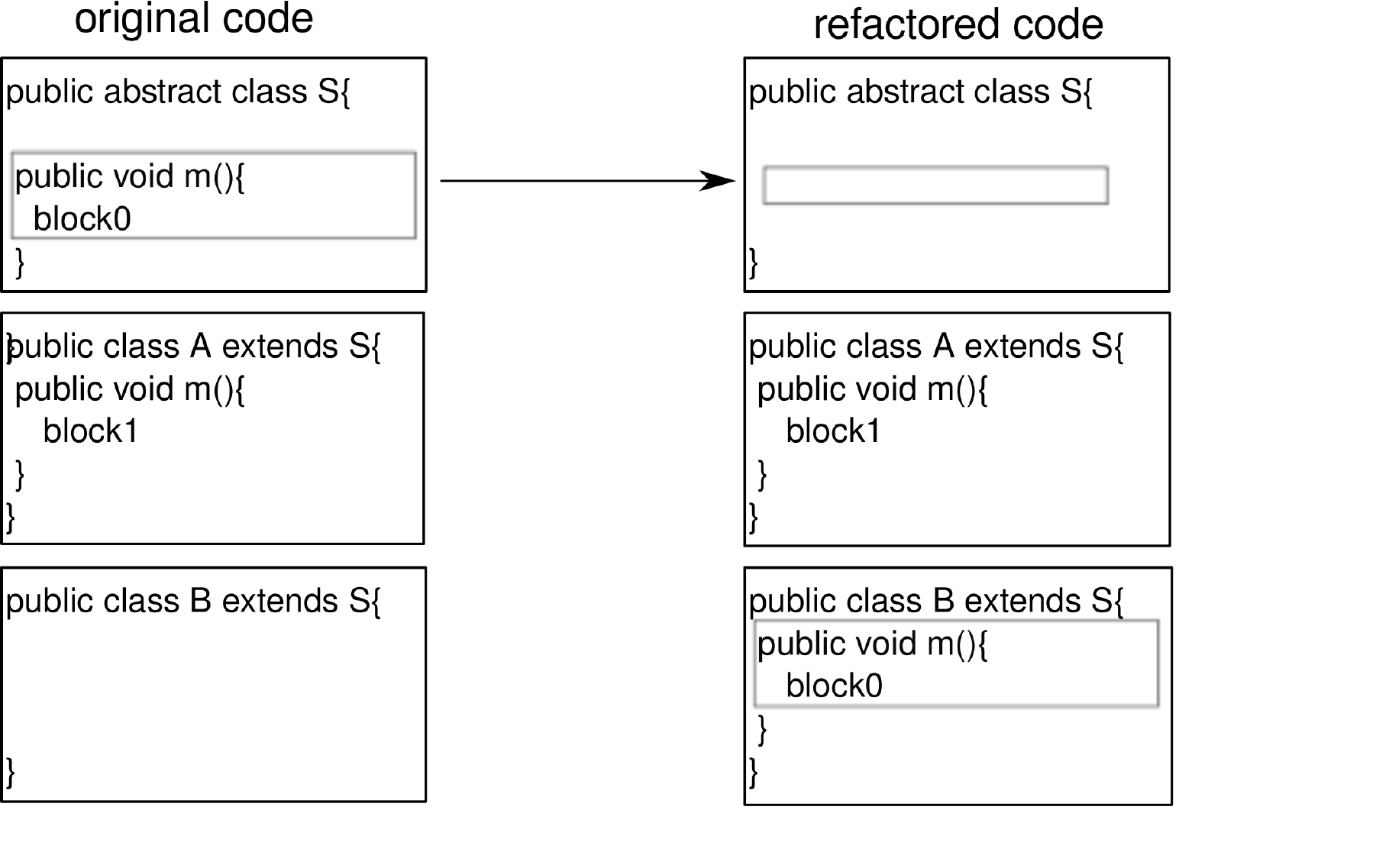}
\end{center}

\tools \emph{Push Down} or \emph{Push member Down} in Eclipse tool and \intellij.

\precondition 
$\\ (\mathtt{ExistsClass}(s)\\ 
\wedge \mathtt{IsAbstractClass}(s)\\ 
\wedge \mathtt{ExistsMethodDefinitionWithParams}(s,m,[t])\\ 
\wedge \neg \mathtt{IsUsedMethod}(s,m,[t])\\ 
\wedge \mathtt{AllSubclasses}(s,[a;b])\\
 \wedge \neg \mathtt{IsPrivate}(s,m)\\ 
\wedge \mathtt{ExistsMethodDefinitionWithParams}(s,m,[t])\\ 
\wedge \mathtt{ExistsMethodDefinitionWithParams}(s,m,[t])\\ 
\wedge \neg \mathtt{IsPrivate}(s,att1)\\ 
\wedge \neg \mathtt{IsPrivate}(s,att2))$ \\

\backdescr 
$\mathtt{ExistsMethodDefinitionWithParams}(s,m,[t])\, \mapsto \,\bot \\[2pt]
 \mathtt{IsUsedMethodIn}(s,m,C)\, \mapsto \,\bot \\[2pt]
 \mathtt{ExistsMethodDefinition}(s,m)\, \mapsto \,\bot \\[2pt]
 \mathtt{ExistsAbstractMethod}(s,m)\, \mapsto \,\bot \\[2pt]
 \mathtt{IsDelegator}(s,m,V1)\, \mapsto \,\bot \\[2pt]
 \mathtt{HasReturnType}(s,m,V1)\, \mapsto \,\bot \\[2pt]
 \mathtt{AllInvokedMethodsOnObjectOInBodyOfMAreDeclaredInC}(s,m,V1,V2)\, \mapsto \,\bot \\[2pt]
 \mathtt{AllInvokedMethodsWithParameterOInBodyOfMAreNotOverloaded}(s,m,V1)\, \mapsto \,\bot \\[2pt]
 \mathtt{BoundVariableInMethodBody}(s,m,V1)\, \mapsto \,\bot \\[2pt]
 \mathtt{ExistsParameterWithName}(s,m,[t],V1)\, \mapsto \,\bot \\[2pt]
 \mathtt{ExistsParameterWithType}(s,m,[t],V1)\, \mapsto \,\bot \\[2pt]
 \mathtt{ExistsMethodInvocation}(s,m,V1,V2)\, \mapsto \,\bot \\[2pt]
 \mathtt{IsPublic}(s,m)\, \mapsto \,\bot \\[2pt]
 \mathtt{IsProtected}(s,m)\, \mapsto \,\bot \\[2pt]
 \mathtt{IsPrivate}(s,m)\, \mapsto \,\bot \\[2pt]
 \mathtt{IsOverloaded}(s,m)\, \mapsto \,\bot \\[2pt]
 \mathtt{IsUsedAttributeInMethodBody}(s,V1,m)\, \mapsto \,\bot \\[2pt]
 \mathtt{IsRecursiveMethod}(s,m)\, \mapsto \,\bot \\[2pt]
 \mathtt{IsIndirectlyRecursive}(s,m)\, \mapsto \,\bot \\[2pt]
 \mathtt{HasReturnType}(s,m,V1)\, \mapsto \,\bot \\[2pt]
 \mathtt{MethodHasParameterType}(s,m,V1)\, \mapsto \,\bot \\[2pt]
 \mathtt{MethodIsUsedWithType}(s,m,[t],[t])\, \mapsto \,\bot \\[2pt]
 \mathtt{IsUsedConstructorAsMethodParameter}(V1,s,m)\, \mapsto \,\bot \\[2pt]
 \mathtt{IsUsedConstructorAsObjectReceiver}(V1,s,m)\, \mapsto \,\bot \\[2pt]
 \mathtt{ExistsMethodDefinitionWithParams}(a,m,[t])\, \mapsto \,\top \\[2pt]
 \mathtt{ExistsMethodDefinitionWithParams}(b,m,[t])\, \mapsto \,\top \\[2pt]
 \mathtt{ExistsMethodDefinition}(a,m)\, \mapsto \,\top \\[2pt]
 \mathtt{ExistsMethodDefinition}(b,m)\, \mapsto \,\top \\[2pt]
 \mathtt{IsOverriding}(a,m)\, \mapsto \,\bot \\[2pt]
 \mathtt{IsOverriding}(b,m)\, \mapsto \,\bot \\[2pt]
 \mathtt{IsOverridden}(a,m)\, \mapsto \,\bot \\[2pt]
 \mathtt{IsOverridden}(b,m)\, \mapsto \,\bot \\[2pt]
 \mathtt{IsInheritedMethodWithParams}(a,m,[t])\, \mapsto \,\bot \\[2pt]
 \mathtt{IsInheritedMethodWithParams}(b,m,[t])\, \mapsto \,\bot \\[2pt]
 \mathtt{IsVisibleMethod}(s,m,[t],a)\, \mapsto \,\bot \\[2pt]
 \mathtt{IsVisibleMethod}(s,m,[t],b)\, \mapsto \,\bot \\[2pt]
 \mathtt{IsVisible}(s,m,a)\, \mapsto \,\bot \\[2pt]
 \mathtt{IsVisible}(s,m,b)\, \mapsto \,\bot \\[2pt]
 \mathtt{IsInheritedMethod}(a,m)\, \mapsto \,\bot \\[2pt]
 \mathtt{IsInheritedMethod}(b,m)\, \mapsto \,\bot \\[2pt]
 \mathtt{IsPrivate}(s,att1)\, \mapsto \,\bot \\[2pt]
 \mathtt{IsPrivate}(s,att2)\, \mapsto \,\bot \\[2pt]
 $

\subsection{PushDownImplementation}%%%%%%%%%%%%%%%%%%%%%%%%%%%%%%%%%%%%%%%%%%%
\paragraph{Overview:} \textsf{PushDownImplementation (classname s, attributes [att1,att2], subclasses [a,b], method m,paramType [t,t'])} : 
same as \emph{PushDownAll} but keep the method abstract in the superclass.

\begin{center}
\includegraphics[scale=0.7]{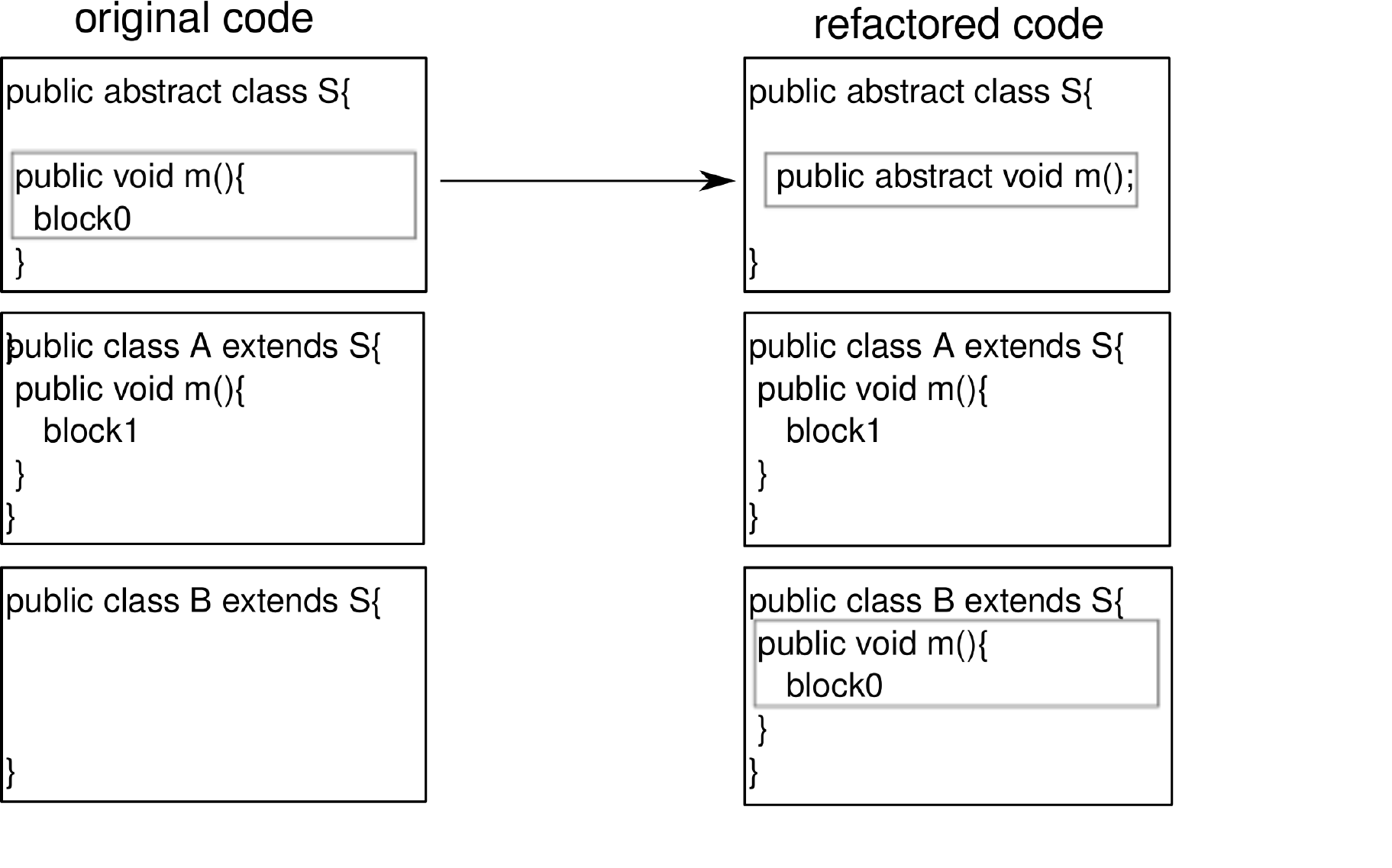}
\end{center}

\precondition 
$\\ (\mathtt{ExistsClass}(s)\\ 
\wedge \mathtt{ExistsMethodDefinition}(s,m)\\ 
\wedge \mathtt{ExistsMethodDefinitionWithParams}(s,m,[t;t'])\\ 
\wedge \neg \mathtt{ExistsAbstractMethod}(s,m)\\ 
\wedge \mathtt{AllSubclasses}(s,[a;b])\\ 
\wedge \neg \mathtt{ExistsMethodDefinitionWithParams}(a,m,[t;t'])\\ 
\wedge \neg \mathtt{ExistsMethodDefinitionWithParams}(b,m,[t;t'])\\ 
\wedge \neg \mathtt{IsPrivate}(s,att1)\\ 
\wedge \neg \mathtt{IsPrivate}(s,att2))$ \\

\backdescr 
 $\mathtt{ExistsAbstractMethod}(s,m)\, \mapsto \,\top \\[2pt]
 \mathtt{AllInvokedMethodsOnObjectOInBodyOfMAreDeclaredInC}(s,m,V1,V2)\, \mapsto \,\top \\[2pt]
 \mathtt{AllInvokedMethodsWithParameterOInBodyOfMAreNotOverloaded}(s,m,V1)\, \mapsto \,\top \\[2pt]
 \mathtt{BoundVariableInMethodBody}(s,m,V1)\, \mapsto \,\bot \\[2pt]
 \mathtt{ExistsMethodInvocation}(s,m,V1,V2)\, \mapsto \,\bot \\[2pt]
 \mathtt{IsInheritedMethodWithParams}(s,m,[t;t'])\, \mapsto \,\bot \\[2pt]
 \mathtt{IsIndirectlyRecursive}(s,m)\, \mapsto \,\bot \\[2pt]
 \mathtt{IsUsedConstructorAsInitializer}(V1,s,m)\, \mapsto \,\bot \\[2pt]
 \mathtt{IsUsedConstructorAsObjectReceiver}(V1,s,m)\, \mapsto \,\bot \\[2pt]
 \mathtt{IsPrivate}(s,m)\, \mapsto \,\bot \\[2pt]
 \mathtt{IsUsedAttributeInMethodBody}(s,V1,m)\, \mapsto \,\bot \\[2pt]
 \mathtt{IsOverridden}(s,m)\, \mapsto \,\bot \\[2pt]
 \mathtt{IsOverriding}(s,m)\, \mapsto \,\bot \\[2pt]
 \mathtt{IsRecursiveMethod}(s,m)\, \mapsto \,\bot \\[2pt]
 \mathtt{MethodHasParameterType}(s,m,V1)\, \mapsto \,\bot \\[2pt]
 \mathtt{MethodIsUsedWithType}(s,m,[t;t'],[t;t'])\, \mapsto \,\bot \\[2pt]
 \mathtt{ExistsMethodDefinitionWithParams}(a,m,[t;t'])\, \mapsto \,\top \\[2pt]
 \mathtt{ExistsMethodDefinitionWithParams}(b,m,[t;t'])\, \mapsto \,\top \\[2pt]
 \mathtt{ExistsMethodDefinition}(a,m)\, \mapsto \,\top \\[2pt]
 \mathtt{ExistsMethodDefinition}(b,m)\, \mapsto \,\top \\[2pt]
 \mathtt{IsPrivate}(s,att1)\, \mapsto \,\bot \\[2pt]
 \mathtt{IsPrivate}(s,att2)\, \mapsto \,\bot \\[2pt]
 $

\subsection{PushDownNotRedefinedMethod}%%%%%%%%%%%%%%%%%%%%%%%%%%%%%%%%%%%%%

\label{def-pushDownNotRedefinedMethod}
 
\textsf{pushDownNotRedefinedMethod(classname c, superclass s ,notredefinedmethods [m1,m2])}

Duplicate the method \textsf{m} of class \texttt{c} into its subclasses.

\begin{center}
\includegraphics[scale=0.7]{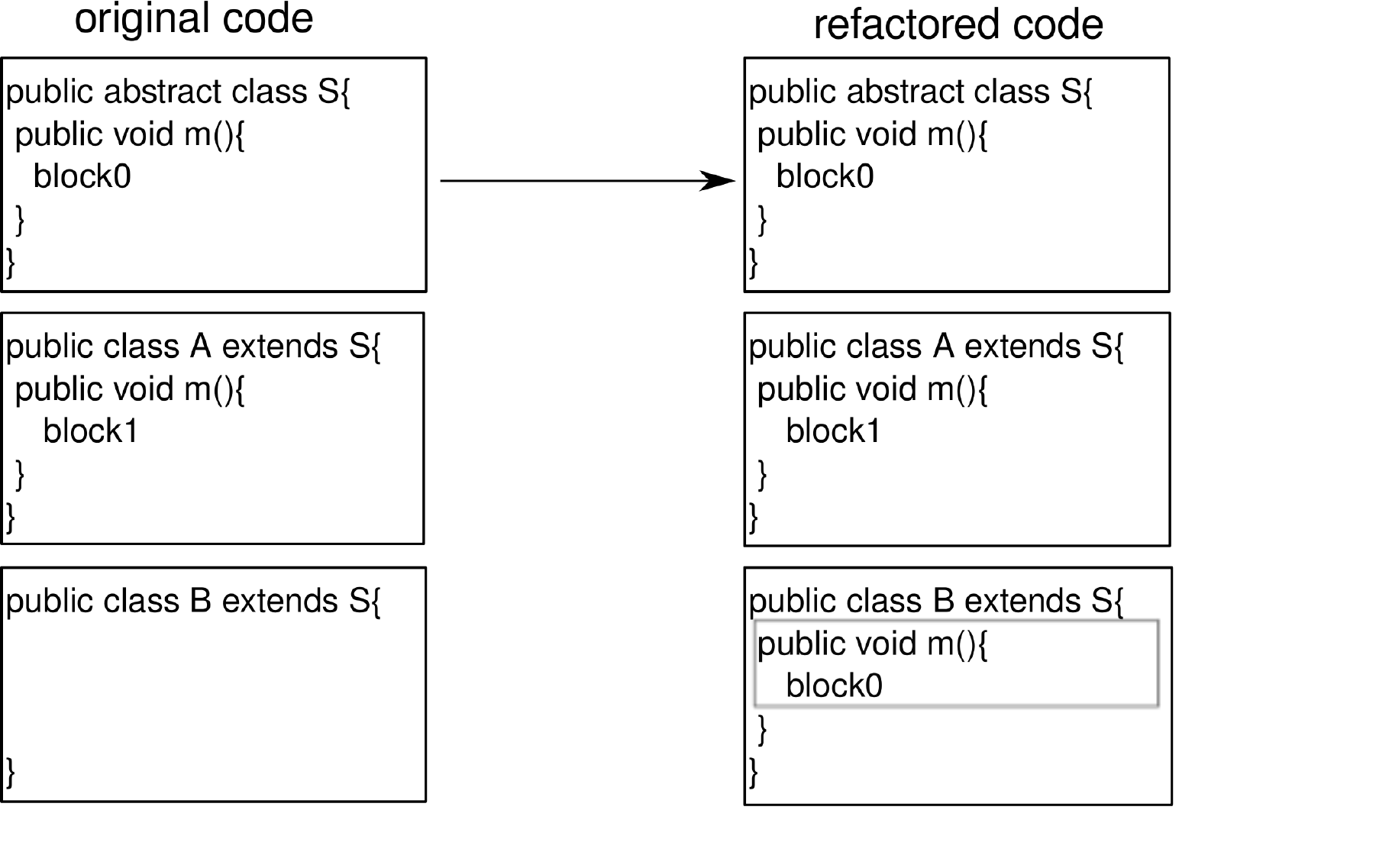}
\end{center}

\tools   \emph{Extract Method}, \emph{Inline}, \emph{Push Down}, \emph{Rename}
 in Eclipse and \intellij.       

\precondition 
$\\ (\mathtt{ExistsType}(c)\\ 
\wedge \mathtt{ExistsClass}(c)\\ 
\wedge \mathtt{IsSubType}(c,s)\\ 
\wedge \neg \mathtt{ExistsMethodDefinition}(c,m1)\\ 
\wedge \neg \mathtt{ExistsMethodDefinition}(c,m2)\\ 
\wedge \mathtt{ExistsMethodDefinition}(s,m1)\\ 
\wedge \mathtt{ExistsMethodDefinition}(s,m2))$ \\ 

\backdescr

$\\ \mathtt{ExistsMethodDefinition}(c,m1)\, \mapsto \,\top \\[2pt]
 \mathtt{ExistsMethodDefinition}(c,m2)\, \mapsto \,\top \\[2pt]
 \mathtt{IsOverriding}(c,m1)\, \mapsto \,\top \\[2pt]
 \mathtt{IsOverriding}(c,m2)\, \mapsto \,\top \\[2pt]
 \mathtt{IsOverridden}(c,m1)\, \mapsto \,\top \\[2pt]
 \mathtt{IsOverridden}(c,m2)\, \mapsto \,\top \\[2pt]
 \mathtt{BoundVariableInMethodBody}(c,m1,V)\, \mapsto \,\mathtt{BoundVariableInMethodBody}(s,m1,V)\\[2pt]
 \mathtt{BoundVariableInMethodBody}(c,m2,V)\, \mapsto \,\mathtt{BoundVariableInMethodBody}(s,m2,V)\\[2pt]
 \mathtt{HasSameBody}(c,m1,s,m1)\, \mapsto \,\top \\[2pt]
 \mathtt{HasSameBody}(c,m2,s,m2)\, \mapsto \,\top \\[2pt]
 \mathtt{AllInvokedMethodsOnObjectOInBodyOfMAreDeclaredInC}(c,m1,this,s)\, \mapsto \,\top \\[2pt]
 \mathtt{AllInvokedMethodsOnObjectOInBodyOfMAreDeclaredInC}(c,m2,this,s)\, \mapsto \,\top \\[2pt]
 $

\subsection{ReplaceMethodDuplication}%%%%%%%%%%%%%%%%%%%%%%%%%%%%%%%%%%%%%%%%%%
\label{def-ReplaceMethodDuplication}

\paragraph{Overview:} \textsf{ ReplaceMethodDuplication (classname s, subclasses [a,b], method m,copy n, paramType [t])} : this operation is used to replace 
any occurrence of method s::m.

\begin{center}
\includegraphics[scale=0.7]{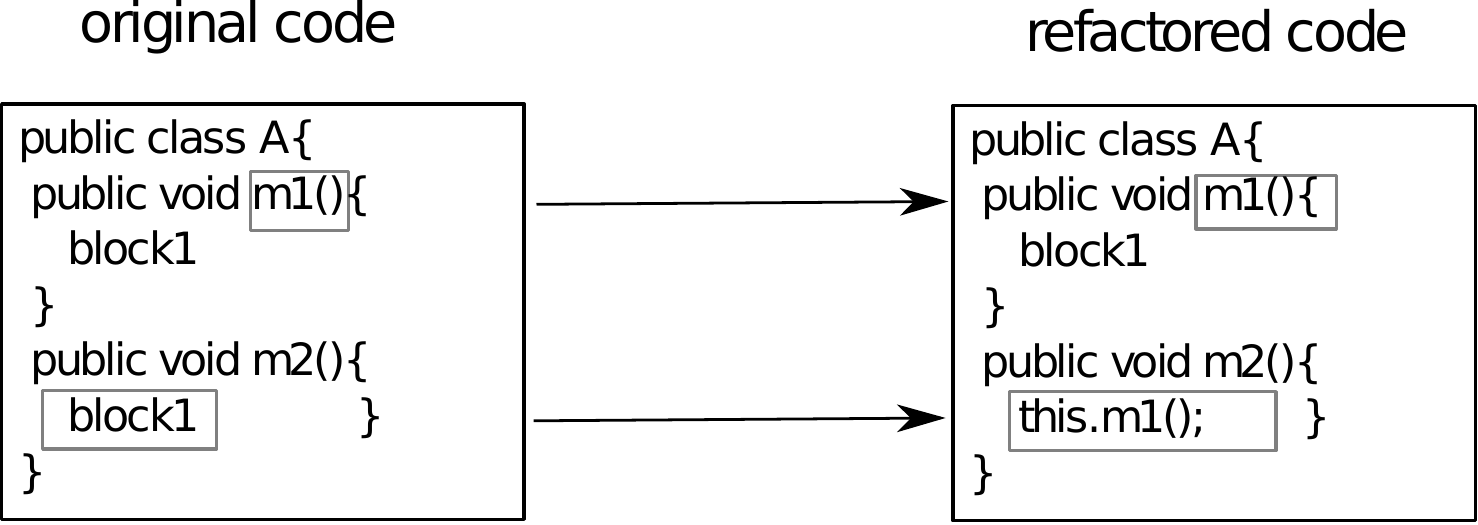}
\end{center}

\tools  \emph{Replace Method Duplication} in \intellij.   

\precondition 
$\\ (\mathtt{ExistsClass}(s)\\ 
\wedge \mathtt{ExistsMethodDefinition}(s,m)\\ 
\wedge \mathtt{ExistsMethodDefinition}(s,n)\\ 
\wedge \mathtt{IsDelegator}(s,n,m)\\ 
\wedge \mathtt{AllSubclasses}(s,[a;b]))$ \\

\backdescr 
 $ \\ \mathtt{IsUsedMethod}(s,n,[t])\, \mapsto \,\bot \\[2pt]
 \mathtt{IsDelegator}(s,n,m)\, \mapsto \,\top \\[2pt]
 \mathtt{ExistsMethodInvocation}(s,n,s,m)\, \mapsto \,\bot \\[2pt]
 \mathtt{IsRecursiveMethod}(s,n)\, \mapsto \,\bot \\[2pt]
 \mathtt{IsRecursiveMethod}(s,n)\, \mapsto \,\bot \\[2pt]
 \mathtt{IsRecursiveMethod}(a,n)\, \mapsto \,\bot \\[2pt]
 \mathtt{IsRecursiveMethod}(b,n)\, \mapsto \,\bot \\[2pt]
 \mathtt{ExistsMethodInvocation}(a,m,a,n)\, \mapsto \,\mathtt{ExistsMethodInvocation}(a,n,a,m)\\[2pt]
 \mathtt{ExistsMethodInvocation}(b,m,b,n)\, \mapsto \,\mathtt{ExistsMethodInvocation}(b,n,b,m)\\[2pt]
 $

\subsection{DeleteClass}%%%%%%%%%%%%%%%%%%%%%%%%%%%%%%%%%%%%%%%%%%%%%%%%%%%%
\label{def-DeleteClassInHierarchy}
\textsf{DeleteClass(class c)}: Delete a class c which is not used.

\paragraph{Overview:} \textsf{ DeleteClass (classname a, classnamesuperclass s ,allclasses [s,a,b], classnamemethods [m,m1],othermethods [m2,n])} : 
this operation is used to delete the class a which is supposed to be not used.

\tools   \emph{Safe Delete} in \intellij, \emph{Delete} in Eclipse. 

\precondition 
 $\\ (\mathtt{ExistsClass}(a)\\ 
\wedge \mathtt{ExistsType}(a)\\ 
\wedge \neg \mathtt{ExistsMethodDefinitionWithParams}(s,m2,[a])\\ 
\wedge \neg \mathtt{ExistsMethodDefinitionWithParams}(s,n,[a])\\ 
\wedge \neg \mathtt{ExistsMethodDefinitionWithParams}(b,m2,[a])\\ 
\wedge \neg \mathtt{ExistsMethodDefinitionWithParams}(b,n,[a])\\ 
\wedge \neg \mathtt{IsUsedMethodIn}(a,m,s)\\ 
\wedge \neg \mathtt{IsUsedMethodIn}(a,m1,s)\\ 
\wedge \neg \mathtt{IsUsedMethodIn}(a,m,b)\\ 
\wedge \neg \mathtt{IsUsedMethodIn}(a,m1,b)\\ 
\wedge \neg \mathtt{IsUsedConstructorAsMethodParameter}(a,s,m2)\\ 
\wedge \neg \mathtt{IsUsedConstructorAsMethodParameter}(a,s,n)\\ 
\wedge \neg \mathtt{IsUsedConstructorAsMethodParameter}(a,b,m2)\\ 
\wedge \neg \mathtt{IsUsedConstructorAsMethodParameter}(a,b,n)\\ 
\wedge \neg \mathtt{IsUsedConstructorAsInitializer}(a,s,m2)\\ 
\wedge \neg \mathtt{IsUsedConstructorAsInitializer}(a,s,n)\\ 
\wedge \neg \mathtt{IsUsedConstructorAsInitializer}(a,b,m2)\\ 
\wedge \neg \mathtt{IsUsedConstructorAsInitializer}(a,b,n)\\ 
\wedge \neg \mathtt{IsUsedConstructorAsObjectReceiver}(a,s,m2)\\ 
\wedge \neg \mathtt{IsUsedConstructorAsObjectReceiver}(a,s,n)\\ 
\wedge \neg \mathtt{IsUsedConstructorAsObjectReceiver}(a,b,m2)\\ 
\wedge \neg \mathtt{IsUsedConstructorAsObjectReceiver}(a,b,n)\\ 
\wedge \neg \mathtt{IsSubType}(s,a)\\ 
\wedge \neg \mathtt{IsSubType}(b,a))$ \\

\backdescr 
 $\\ \mathtt{ExistsType}(a)\, \mapsto \,\bot \\[2pt]
 \mathtt{ExistsClass}(a)\, \mapsto \,\bot \\[2pt]
 \mathtt{IsSubType}(a,s)\, \mapsto \,\bot \\[2pt]
 \mathtt{AllInvokedMethodsOnObjectOInBodyOfMAreDeclaredInC}(a,V1,V2,V3)\, \mapsto \,\bot \\[2pt]
 \mathtt{AllInvokedMethodsWithParameterOInBodyOfMAreNotOverloaded}(a,V1,V2)\, \mapsto \,\bot \\[2pt]
 \mathtt{BoundVariableInMethodBody}(a,V1,V2)\, \mapsto \,\bot \\[2pt]
 \mathtt{ExistsParameterWithName}(a,V1,[V2],V3)\, \mapsto \,\bot \\[2pt]
 \mathtt{ExistsParameterWithType}(a,V1,[V2],V3)\, \mapsto \,\bot \\[2pt]
 \mathtt{ExistsField}(a,V1)\, \mapsto \,\bot \\[2pt]
 \mathtt{ExistsMethodInvocation}(a,V1,V2,V3)\, \mapsto \,\bot \\[2pt]
 \mathtt{ExistsMethodDefinitionWithParams}(a,V1,[V2])\, \mapsto \,\bot \\[2pt]
 \mathtt{ExtendsDirectly}(a,s)\, \mapsto \,\bot \\[2pt]
 \mathtt{ExtendsDirectly}(V1,a)\, \mapsto \,\bot \\[2pt]
 \mathtt{ExistsAbstractMethod}(a,V1)\, \mapsto \,\bot \\[2pt]
 \mathtt{IsInheritedMethodWithParams}(a,V1,[V2])\, \mapsto \,\bot \\[2pt]
 \mathtt{IsIndirectlyRecursive}(a,V1)\, \mapsto \,\bot \\[2pt]
 \mathtt{IsVisibleMethod}(a,V1,[V2],V3)\, \mapsto \,\bot \\[2pt]
 \mathtt{IsInverter}(a,V1,V2,V3)\, \mapsto \,\bot \\[2pt]
 \mathtt{IsDelegator}(a,V1,V2)\, \mapsto \,\bot \\[2pt]
 \mathtt{IsAbstractClass}(a)\, \mapsto \,\bot \\[2pt]
 \mathtt{IsUsedMethod}(a,V1,[V2])\, \mapsto \,\bot \\[2pt]
 \mathtt{IsUsedMethodIn}(a,V1,V2)\, \mapsto \,\bot \\[2pt]
 \mathtt{IsUsedConstructorAsMethodParameter}(V1,a,V2)\, \mapsto \,\bot \\[2pt]
 \mathtt{IsUsedConstructorAsInitializer}(a,V1,V2)\, \mapsto \,\bot \\[2pt]
 \mathtt{IsUsedConstructorAsObjectReceiver}(a,V1,V2)\, \mapsto \,\bot \\[2pt]
 \mathtt{IsUsedConstructorAsInitializer}(V1,a,V2)\, \mapsto \,\bot \\[2pt]
 \mathtt{IsUsedConstructorAsObjectReceiver}(V1,a,V2)\, \mapsto \,\bot \\[2pt]
 \mathtt{IsPrimitiveType}(a)\, \mapsto \,\bot \\[2pt]
 \mathtt{IsPublic}(a,V1)\, \mapsto \,\bot \\[2pt]
 \mathtt{IsProtected}(a,V1)\, \mapsto \,\bot \\[2pt]
 \mathtt{IsPrivate}(a,V1)\, \mapsto \,\bot \\[2pt]
 \mathtt{IsUsedAttributeInMethodBody}(a,V1,V2)\, \mapsto \,\bot \\[2pt]
 \mathtt{IsGenericsSubtype}(a,[V1],s,[V2])\, \mapsto \,\bot \\[2pt]
 \mathtt{IsGenericsSubtype}(V1,[V2],a,[V3])\, \mapsto \,\bot \\[2pt]
 \mathtt{IsGenericsSubtype}(V1,[a],V2,[V3])\, \mapsto \,\bot \\[2pt]
 \mathtt{IsInheritedField}(a,V1)\, \mapsto \,\bot \\[2pt]
 \mathtt{IsOverridden}(a,V1)\, \mapsto \,\bot \\[2pt]
 \mathtt{IsOverloaded}(a,V1)\, \mapsto \,\bot \\[2pt]
 \mathtt{IsOverriding}(a,V1)\, \mapsto \,\bot \\[2pt]
 \mathtt{IsRecursiveMethod}(a,V1)\, \mapsto \,\bot \\[2pt]
 \mathtt{IsRecursiveMethod}(a,V1)\, \mapsto \,\bot \\[2pt]
 \mathtt{HasReturnType}(a,V1,V2)\, \mapsto \,\bot \\[2pt]
 \mathtt{HasParameterType}(a,V1)\, \mapsto \,\bot \\[2pt]
 \mathtt{HasParameterType}(V1,a)\, \mapsto \,\bot \\[2pt]
 \mathtt{MethodHasParameterType}(a,V1,V2)\, \mapsto \,\bot \\[2pt]
 \mathtt{MethodIsUsedWithType}(a,V1,[V2],[V3])\, \mapsto \,\bot \\[2pt]
 \mathtt{MethodIsUsedWithType}(V1,V2,[a],[a])\, \mapsto \,\bot \\[2pt]
 \mathtt{ExistsMethodDefinition}(a,m)\, \mapsto \,\bot \\[2pt]
 \mathtt{ExistsMethodDefinition}(a,m1)\, \mapsto \,\bot \\[2pt]
 \mathtt{IsInheritedMethodWithParams}(V1,m,[V2])\, \mapsto \,\bot \\[2pt]
 \mathtt{IsInheritedMethodWithParams}(V1,m1,[V2])\, \mapsto \,\bot \\[2pt]
 $

\subsection{ExtractGeneralMethod}%%%%%%%%%%%%%%%%%%%%%%%%%%%%%%%%%%%%%%%%%%%%
\label{def-ExtractGeneralMethod}

\begin{center}
\includegraphics[scale=0.7]{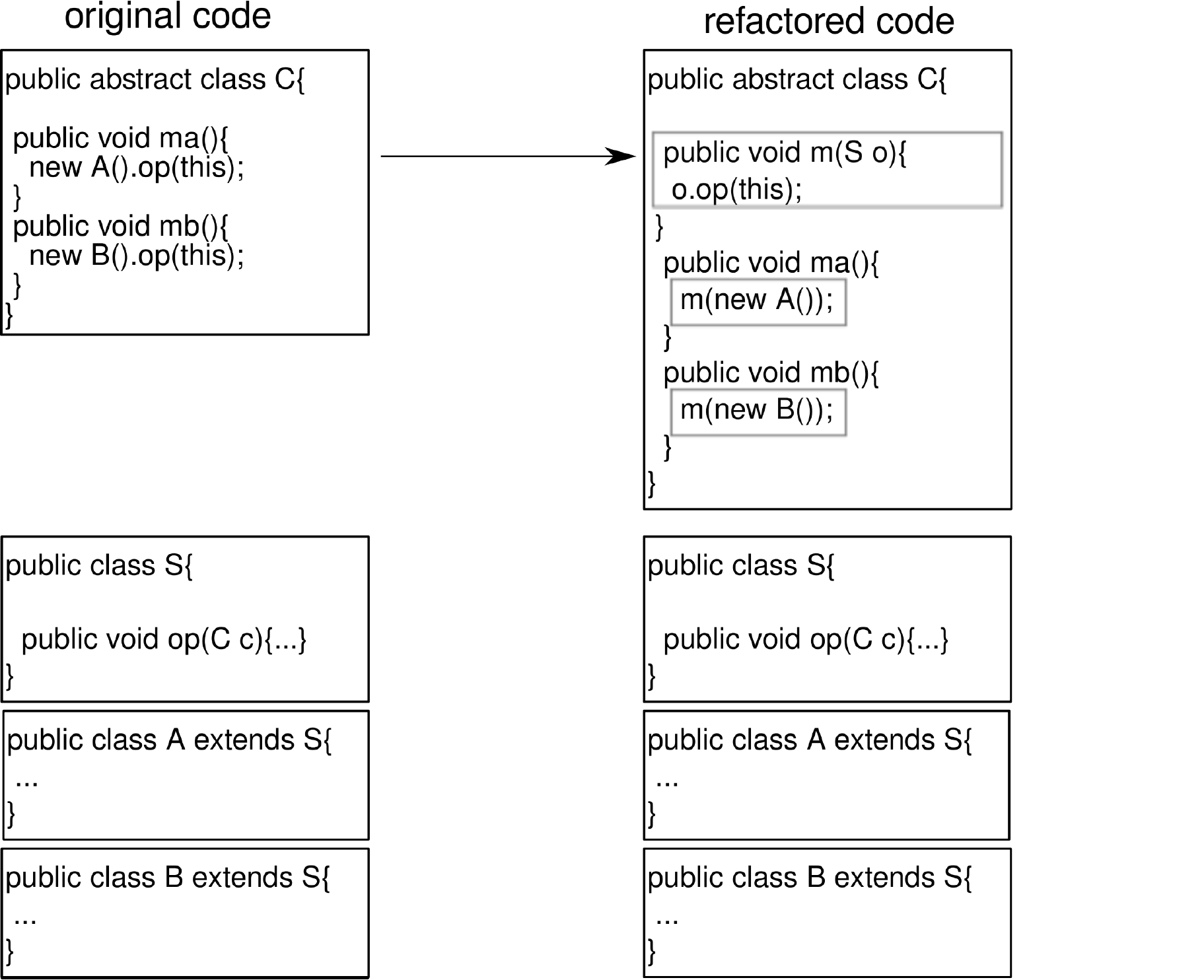}
\end{center}

\subsection{InlineClass}%%%%%%%%%%%%%%%%%%%%%%%%%%%%%%%%%%%%%%%%%%%%

\label{def-InlineClass}
 
\textsf{InlineClass(class c)}:
Inline one or more references to a given class \textsf{c}.

\tools   \emph{Inline} in Eclipse and \intellij.

\subsection{SpecialiseParameter}%%%%%%%%%%%%%%%%%%%%%%%%%%%%%%%%%%%%%%%%

\paragraph{Overview:} \textsf{SpecialiseParameter(classname s, subclasses [a,b], methodname m,paramType t ,paramName p, subtypes [st,q],new~paramType st)}: 
this operation is used to change the type t of the parameter p of the methods s::m, a::m and b::m into one of its subtypes (st).

\precondition 
 $\\ (\mathtt{IsSubType}(t1,t)\\ 
\wedge \neg \mathtt{MethodIsUsedWithType}(s,m,[t],[t])\\ 
\wedge \neg \mathtt{MethodIsUsedWithType}(a,m,[t],[t])\\ 
\wedge \neg \mathtt{MethodIsUsedWithType}(b,m,[t],[t])\\ 
\wedge \mathtt{ExistsClass}(s)\\ 
\wedge \mathtt{ExistsMethodDefinitionWithParams}(s,m,[t])\\ 
\wedge \mathtt{ExistsType}(t)\\ 
\wedge \mathtt{ExistsType}(t1)\\ 
\wedge \neg \mathtt{ExistsMethodDefinitionWithParams}(s,m,[t1])\\ 
\wedge \neg \mathtt{IsInheritedMethodWithParams}(s,m,[t1])\\ 
\wedge \mathtt{AllSubclasses}(s,[a;b])\\ 
\wedge \mathtt{AllInvokedMethodsWithParameterOInBodyOfMAreNotOverloaded}(s,m,p)\\ 
\wedge (\neg \mathtt{MethodIsUsedWithType}(s,m,[t],[t2])\\ 
\tab \vee \mathtt{ExistsMethodDefinitionWithParams}(s,m,[t2])))$ \\ 

\backdescr
$\\ \mathtt{ExistsMethodDefinitionWithParams}(s,m,[t1])\, \mapsto \,\top \\[2pt]
 \mathtt{ExistsMethodDefinitionWithParams}(s,m,[t])\, \mapsto \,\bot \\[2pt]
 \mathtt{ExistsMethodDefinitionWithParams}(a,m,[t])\, \mapsto \,\bot \\[2pt]
 \mathtt{ExistsMethodDefinitionWithParams}(b,m,[t])\, \mapsto \,\bot \\[2pt]
 \mathtt{ExistsMethodDefinitionWithParams}(a,m,[t1])\, \mapsto \,\mathtt{ExistsMethodDefinitionWithParams}(a,m,[t])\\[2pt]
 \mathtt{ExistsMethodDefinitionWithParams}(b,m,[t1])\, \mapsto \,\mathtt{ExistsMethodDefinitionWithParams}(b,m,[t])\\[2pt]
 \mathtt{MethodIsUsedWithType}(s,m,[t],[t])\, \mapsto \,\bot \\[2pt]
 \mathtt{MethodIsUsedWithType}(a,m,[t],[t])\, \mapsto \,\bot \\[2pt]
 \mathtt{MethodIsUsedWithType}(b,m,[t],[t])\, \mapsto \,\bot \\[2pt]
 \mathtt{ExistsParameterWithName}(s,m,[t1],p)\, \mapsto \,\top \\[2pt]
 \mathtt{ExistsParameterWithName}(a,m,[t1],p)\, \mapsto \,\top \\[2pt]
 \mathtt{ExistsParameterWithName}(b,m,[t1],p)\, \mapsto \,\top \\[2pt]
 \mathtt{ExistsParameterWithType}(s,m,[t1],t1)\, \mapsto \,\top \\[2pt]
 \mathtt{ExistsParameterWithType}(a,m,[t1],t1)\, \mapsto \,\top \\[2pt]
 \mathtt{ExistsParameterWithType}(b,m,[t1],t1)\, \mapsto \,\top \\[2pt]
 \mathtt{IsInverter}(s,m,t1,V)\, \mapsto \,\mathtt{IsInverter}(s,m,t,V)\\[2pt]
 \mathtt{IsInverter}(a,m,t1,V)\, \mapsto \,\mathtt{IsInverter}(a,m,t,V)\\[2pt]
 \mathtt{IsInverter}(b,m,t1,V)\, \mapsto \,\mathtt{IsInverter}(b,m,t,V)\\[2pt]
 \mathtt{IsInheritedMethodWithParams}(s,m,[t1])\, \mapsto \,\top \\[2pt]
 \mathtt{IsInheritedMethodWithParams}(a,m,[t1])\, \mapsto \,\top \\[2pt]
 \mathtt{IsInheritedMethodWithParams}(b,m,[t1])\, \mapsto \,\top \\[2pt]
 \mathtt{IsUsedConstructorAsMethodParameter}(t1,s,m)\, \mapsto \,\mathtt{IsUsedConstructorAsMethodParameter}(t,s,m)\\[2pt]
 \mathtt{IsUsedConstructorAsMethodParameter}(t1,a,m)\, \mapsto \,\mathtt{IsUsedConstructorAsMethodParameter}(t,a,m)\\[2pt]
 \mathtt{IsUsedConstructorAsMethodParameter}(t1,b,m)\, \mapsto \,\mathtt{IsUsedConstructorAsMethodParameter}(t,b,m)\\[2pt]
 \mathtt{IsOverridden}(a,m)\, \mapsto \,\mathtt{ExistsMethodDefinition}(a,m)\\[2pt]
 \mathtt{IsOverridden}(b,m)\, \mapsto \,\mathtt{ExistsMethodDefinition}(b,m)\\[2pt]
 \mathtt{IsOverriding}(a,m)\, \mapsto \,\mathtt{ExistsMethodDefinition}(a,m)\\[2pt]
 \mathtt{IsOverriding}(b,m)\, \mapsto \,\mathtt{ExistsMethodDefinition}(b,m)\\[2pt]
 $

\subsection{IntroduceParameterObject}%%%%%%%%%%%%%%%%%%%%%%%%%%%%%%%%%%%%%%%%%%
\label{def-ReplaceMethodDuplication}

\paragraph{Overview:} \textsf{ introduceParmeterObject s [a,b] m [t1,t2] [p1,p2] t n}: introducing the parameter object \texttt{n} of type \texttt{t1} to the method \texttt{s::m(t p)} this creates a class of 
type t1 for instance \texttt{A}, moves the parameter p to A as an instance variable and finally 
changes  \texttt{m(t p)} into \texttt{m(A a)}. Any old access to p in the body of m will be replaced by \texttt{a.p}. 
 
\begin{center}
\includegraphics[scale=0.7]{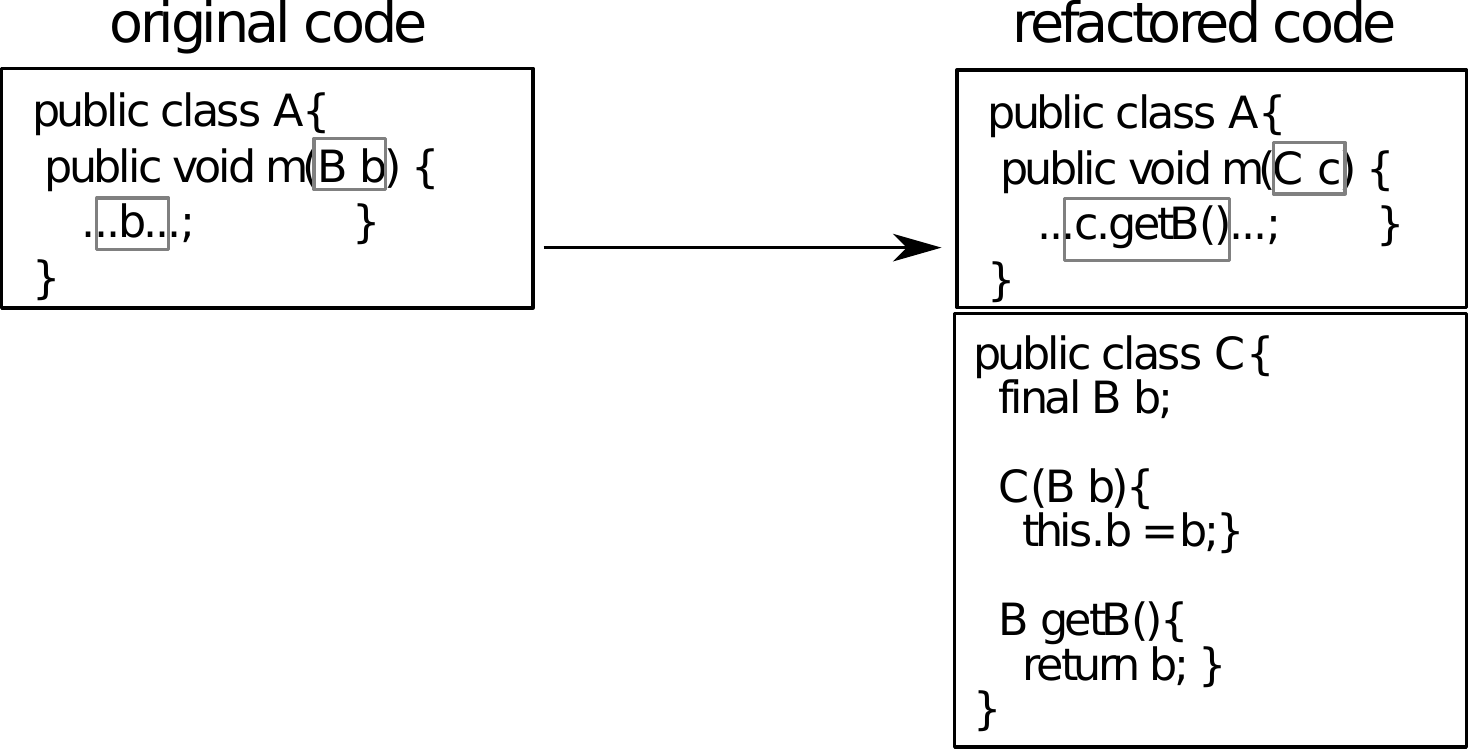}
\end{center}

\tools  \emph{Extract Parameter Object} in \intellij.   

\precondition 
$\\ (\mathtt{ExistsType}(s)\\ 
\wedge \neg \mathtt{ExistsType}(t)\\ 
\wedge \mathtt{ExistsMethodDefinitionWithParams}(s,m,[t1;t2])\\ 
\wedge \mathtt{AllSubclasses}(s,[a;b])\\ 
\wedge \mathtt{BoundVariableInMethodBody}(a,m,v1)\\ 
\wedge \mathtt{BoundVariableInMethodBody}(a,m,v2)\\ 
\wedge \mathtt{BoundVariableInMethodBody}(b,m,v1)\\ 
\wedge \mathtt{BoundVariableInMethodBody}(b,m,v2))$ \\ 

\backdescr 
$\\ \mathtt{ExistsClass}(t)\, \mapsto \,\top \\[2pt]
 \mathtt{ExistsType}(t)\, \mapsto \,\top \\[2pt]
 \mathtt{IsPrimitiveType}(t)\, \mapsto \,\bot \\[2pt]
 \mathtt{IsUsedMethod}(s,m,[t])\, \mapsto \,\top \\[2pt]
 \mathtt{IsUsedMethod}(s,m,[t1;t2])\, \mapsto \,\bot \\[2pt]
 \mathtt{IsUsedMethodIn}(t,C,M)\, \mapsto \,\bot \\[2pt]
 \mathtt{IsUsedMethod}(t,C,[T1])\, \mapsto \,\bot \\[2pt]
 \mathtt{IsUsedMethod}(t,C,[T1;T2])\, \mapsto \,\bot \\[2pt]
 \mathtt{IsUsedMethod}(t,C,[T1;T2;T3])\, \mapsto \,\bot \\[2pt]
 \mathtt{IsUsedMethod}(t,C,[T1;T2;T3;T4])\, \mapsto \,\bot \\[2pt]
 \mathtt{IsUsedConstructorAsMethodParameter}(t,C,m)\, \mapsto \,\bot \\[2pt]
 \mathtt{IsUsedConstructorAsInitializer}(t,C,M)\, \mapsto \,\bot \\[2pt]
 \mathtt{IsUsedConstructorAsMethodParameter}(C,t,M)\, \mapsto \,\bot \\[2pt]
 \mathtt{IsUsedConstructorAsInitializer}(C,t,M)\, \mapsto \,\bot \\[2pt]
 \mathtt{IsUsedConstructorAsObjectReceiver}(C,t,M)\, \mapsto \,\bot \\[2pt]
 \mathtt{IsUsedConstructorAsInitializer}(t,C,M)\, \mapsto \,\bot \\[2pt]
 \mathtt{IsSubType}(C,t)\, \mapsto \,\bot \\[2pt]
 \mathtt{MethodIsUsedWithType}(s,m,[t],[t])\, \mapsto \,\top \\[2pt]
 \mathtt{MethodIsUsedWithType}(s,m,[t1;t2],[t1;t2])\, \mapsto \,\bot \\[2pt]
 \mathtt{IsInheritedMethodWithParams}(s,M,[t])\, \mapsto \,\bot \\[2pt]
 \mathtt{IsInheritedMethodWithParams}(a,m,[t])\, \mapsto \,\mathtt{IsInheritedMethodWithParams}(a,m,[t1;t2])\\[2pt]
 \mathtt{IsInheritedMethodWithParams}(b,m,[t])\, \mapsto \,\mathtt{IsInheritedMethodWithParams}(b,m,[t1;t2])\\[2pt]
 \mathtt{IsInheritedMethod}(t,M)\, \mapsto \,\mathtt{IsVisible}(java.lang.Object,M,t)\\[2pt]
 \mathtt{IsInheritedMethodWithParams}(t,M,[\,])\, \mapsto \,\mathtt{IsVisibleMethod}(java.lang.Object,M,[\,],t)\\[2pt]
 \mathtt{IsInheritedMethodWithParams}(t,M,[T1])\, \mapsto \,\mathtt{IsVisibleMethod}(java.lang.Object,M,[T1],t)\\[2pt]
 \mathtt{IsInheritedMethodWithParams}(t,M,[T1;T2])\, \mapsto \,\mathtt{IsVisibleMethod}(java.lang.Object,M,[T1;T2],t)\\[2pt]
 \mathtt{IsInheritedMethodWithParams}(t,M,[T1;T2;T3])\, \mapsto \,\mathtt{IsVisibleMethod}(java.lang.Object,M,[T1;T2;T3],t)\\[2pt]
 \mathtt{IsInheritedMethodWithParams}(t,M,[T1;T2;T3;T4;T5])\, \mapsto \,\mathtt{IsVisibleMethod}(java.lang.Object,M,[T1;T2;T3;T4;T5],t)\\[2pt]
 \mathtt{ExistsMethodDefinitionWithParams}(t,M,[\,])\, \mapsto \,\bot \\[2pt]
 \mathtt{ExistsMethodDefinitionWithParams}(t,M,[T1])\, \mapsto \,\bot \\[2pt]
 \mathtt{ExistsMethodDefinitionWithParams}(t,M,[T1;T2])\, \mapsto \,\bot \\[2pt]
 \mathtt{ExistsMethodDefinitionWithParams}(t,M,[T1;T2;T3])\, \mapsto \,\bot \\[2pt]
 \mathtt{ExistsMethodDefinitionWithParams}(t,M,[T1;T2;T3;T4])\, \mapsto \,\bot \\[2pt]
 \mathtt{ExistsMethodDefinitionWithParams}(t,M,[T1;T2;T3;T4;T5])\, \mapsto \,\bot \\[2pt]
 \mathtt{IsSubType}(t,X)\, \mapsto \,\bot  (condition) \\[2pt]
 \mathtt{ExtendsDirectly}(t,X)\, \mapsto \,\bot  (condition) \\[2pt]
 \mathtt{IsInheritedMethodWithParams}(C,M,[t;T1])\, \mapsto \,\bot \\[2pt]
 \mathtt{IsInheritedMethodWithParams}(C,M,[t])\, \mapsto \,\bot \\[2pt]
 \mathtt{IsInheritedMethodWithParams}(C,M,[t;T1])\, \mapsto \,\bot \\[2pt]
 \mathtt{IsInheritedMethodWithParams}(C,M,[T1;t])\, \mapsto \,\bot \\[2pt]
 \mathtt{IsInheritedMethodWithParams}(C,M,[T1;T2;t])\, \mapsto \,\bot \\[2pt]
 \mathtt{IsInheritedMethodWithParams}(C,M,[T1;t;T2])\, \mapsto \,\bot \\[2pt]
 \mathtt{IsInheritedMethodWithParams}(C,M,[t;T1;T2])\, \mapsto \,\bot \\[2pt]
 \mathtt{ExtendsDirectly}(t,java.lang.Object)\, \mapsto \,\top \\[2pt]
 \mathtt{IsUsedConstructorAsObjectReceiver}(t,a,m)\, \mapsto \,\mathtt{BoundVariableInMethodBody}(a,m,v1)\\[2pt]
 \mathtt{IsUsedConstructorAsObjectReceiver}(t,a,m)\, \mapsto \,\mathtt{BoundVariableInMethodBody}(a,m,v2)\\[2pt]
 \mathtt{IsUsedConstructorAsObjectReceiver}(t,b,m)\, \mapsto \,\mathtt{BoundVariableInMethodBody}(b,m,v1)\\[2pt]
 \mathtt{IsUsedConstructorAsObjectReceiver}(t,b,m)\, \mapsto \,\mathtt{BoundVariableInMethodBody}(b,m,v2)\\[2pt]
 \mathtt{ExistsParameterWithName}(s,m,[t],n)\, \mapsto \,\top \\[2pt]
 \mathtt{ExistsParameterWithName}(a,m,[t],n)\, \mapsto \,\top \\[2pt]
 \mathtt{ExistsParameterWithName}(b,m,[t],n)\, \mapsto \,\top \\[2pt]
 \mathtt{ExistsParameterWithType}(s,m,[t],t)\, \mapsto \,\top \\[2pt]
 \mathtt{ExistsParameterWithType}(a,m,[t],t)\, \mapsto \,\top \\[2pt]
 \mathtt{ExistsParameterWithType}(b,m,[t],t)\, \mapsto \,\top \\[2pt]
 \mathtt{ExistsMethodDefinitionWithParams}(s,m,[t])\, \mapsto \,\top \\[2pt]
 \mathtt{ExistsMethodDefinitionWithParams}(C,M,[t])\, \mapsto \,\bot  (condition) \\[2pt]
 \mathtt{ExistsMethodDefinitionWithParams}(C,M,[t;T1])\, \mapsto \,\bot \\[2pt]
 \mathtt{ExistsMethodDefinitionWithParams}(C,M,[T1;t])\, \mapsto \,\bot \\[2pt]
 \mathtt{ExistsMethodDefinitionWithParams}(C,M,[T1;T2;t])\, \mapsto \,\bot \\[2pt]
 \mathtt{ExistsMethodDefinitionWithParams}(C,M,[T1;t;T2])\, \mapsto \,\bot \\[2pt]
 \mathtt{ExistsMethodDefinitionWithParams}(C,M,[t;T1;T2])\, \mapsto \,\bot \\[2pt]
 \mathtt{ExistsMethodDefinitionWithParams}(s,m,[t1;t2])\, \mapsto \,\bot \\[2pt]
 \mathtt{ExistsMethodDefinitionWithParams}(a,m,[t1;t2])\, \mapsto \,\bot \\[2pt]
 \mathtt{ExistsMethodDefinitionWithParams}(b,m,[t1;t2])\, \mapsto \,\bot \\[2pt]
 \mathtt{ExistsMethodDefinitionWithParams}(a,m,[t])\, \mapsto \,\top \\[2pt]
 \mathtt{ExistsMethodDefinitionWithParams}(b,m,[t])\, \mapsto \,\top \\[2pt]
 \mathtt{ExistsField}(t,p1)\, \mapsto \,\top \\[2pt]
 \mathtt{ExistsField}(t,p2)\, \mapsto \,\top \\[2pt]
 \mathtt{ExistsMethodDefinition}(t,getp1)\, \mapsto \,\top \\[2pt]
 \mathtt{ExistsMethodDefinition}(t,getp2)\, \mapsto \,\top \\[2pt]
 \mathtt{IsUsedConstructorAsObjectReceiver}(T1,T2,getp1)\, \mapsto \,\bot \\[2pt]
 \mathtt{IsUsedConstructorAsObjectReceiver}(T1,T2,getp2)\, \mapsto \,\bot \\[2pt]
 \mathtt{IsPrivate}(t,p1)\, \mapsto \,\top \\[2pt]
 \mathtt{IsPrivate}(t,p2)\, \mapsto \,\top \\[2pt]
 \mathtt{IsInheritedField}(t,p1)\, \mapsto \,\bot \\[2pt]
 \mathtt{IsInheritedField}(t,p2)\, \mapsto \,\bot \\[2pt]
 \mathtt{ExistsParameterWithName}(s,m,[t1;t2],p1)\, \mapsto \,\bot \\[2pt]
 \mathtt{ExistsParameterWithName}(s,m,[t1;t2],p2)\, \mapsto \,\bot \\[2pt]
 \mathtt{ExistsParameterWithName}(a,m,[t1;t2],p1)\, \mapsto \,\bot \\[2pt]
 \mathtt{ExistsParameterWithName}(a,m,[t1;t2],p2)\, \mapsto \,\bot \\[2pt]
 \mathtt{ExistsParameterWithName}(b,m,[t1;t2],p1)\, \mapsto \,\bot \\[2pt]
 \mathtt{ExistsParameterWithName}(b,m,[t1;t2],p2)\, \mapsto \,\bot \\[2pt]
 \mathtt{ExistsParameterWithType}(s,m,[t1;t2],t1)\, \mapsto \,\bot \\[2pt]
 \mathtt{ExistsParameterWithType}(s,m,[t1;t2],t2)\, \mapsto \,\bot \\[2pt]
 \mathtt{ExistsParameterWithType}(a,m,[t1;t2],t1)\, \mapsto \,\bot \\[2pt]
 \mathtt{ExistsParameterWithType}(a,m,[t1;t2],t2)\, \mapsto \,\bot \\[2pt]
 \mathtt{ExistsParameterWithType}(b,m,[t1;t2],t1)\, \mapsto \,\bot \\[2pt]
 \mathtt{ExistsParameterWithType}(b,m,[t1;t2],t2)\, \mapsto \,\bot \\[2pt]
 \mathtt{BoundVariableInMethodBody}(s,m,n)\, \mapsto \,\mathtt{BoundVariableInMethodBody}(s,m,p1)\\[2pt]
 \mathtt{BoundVariableInMethodBody}(s,m,n)\, \mapsto \,\mathtt{BoundVariableInMethodBody}(s,m,p2)\\[2pt]
 \mathtt{BoundVariableInMethodBody}(s,m,n)\, \mapsto \,\mathtt{BoundVariableInMethodBody}(s,m,p1)\\[2pt]
 \mathtt{BoundVariableInMethodBody}(s,m,n)\, \mapsto \,\mathtt{BoundVariableInMethodBody}(s,m,p2)\\[2pt]
 \mathtt{AllInvokedMethodsOnObjectOInBodyOfMAreDeclaredInC}(s,m,n,T)\, \mapsto \,\top \\[2pt]
 \mathtt{AllInvokedMethodsOnObjectOInBodyOfMAreDeclaredInC}(a,m,n,T)\, \mapsto \,\top \\[2pt]
 \mathtt{AllInvokedMethodsOnObjectOInBodyOfMAreDeclaredInC}(b,m,n,T)\, \mapsto \,\top \\[2pt]
 \mathtt{ExistsParameterWithName}(s,m,[t1;t2;t],n)\, \mapsto \,\top \\[2pt]
 \mathtt{ExistsParameterWithName}(a,m,[t1;t2;t],n)\, \mapsto \,\top \\[2pt]
 \mathtt{ExistsParameterWithName}(b,m,[t1;t2;t],n)\, \mapsto \,\top \\[2pt]
 \mathtt{ExistsParameterWithType}(s,m,[t1;t2;t],t)\, \mapsto \,\top \\[2pt]
 \mathtt{ExistsParameterWithType}(a,m,[t1;t2;t],t)\, \mapsto \,\top \\[2pt]
 \mathtt{ExistsParameterWithType}(b,m,[t1;t2;t],t)\, \mapsto \,\top \\[2pt]
 $

\subsection{DeleteMethod}
\paragraph{Overview:} \textsf{DeleteMethod c  m  [t1,t2]  s}: this operation is used to delete the method c::m which is semantically equivalent to the method m 
which is inherited from the class s which is the super class of c.

\tools
\emph{Delete} in Eclipse and \intellij.

\precondition 
$\\ \mathtt{ExistsClass}(c)\\ 
\wedge \mathtt{ExistsType}(c)\\ 
\wedge \mathtt{ExistsMethodDefinitionWithParams}(c,m,[t1;t2])\\ 
\wedge \mathtt{IsInheritedMethod}(c,m)\\
 \wedge \mathtt{IsDuplicate}(c,m,s,m)\\ 
\wedge \mathtt{AllInvokedMethodsWithParameterOInBodyOfMAreNotOverloaded}(c,m,this)$ \\

\backdescr  

$\mathtt{ExistsParameterWithType}(c,m,[t1;t2],V1)\, \mapsto \,\bot \\[2pt]
 \mathtt{ExistsMethodDefinitionWithParams}(c,m,[t1;t2])\, \mapsto \,\bot \\[2pt]
 \mathtt{ExistsMethodDefinition}(c,m)\, \mapsto \,\bot \\[2pt]
 \mathtt{BoundVariableInMethodBody}(c,m,t1)\, \mapsto \,\bot \\[2pt]
 \mathtt{BoundVariableInMethodBody}(c,m,t2)\, \mapsto \,\bot \\[2pt]
 \mathtt{ExistsParameterWithName}(c,m,[t1;t2],V1)\, \mapsto \,\bot \\[2pt]
 \mathtt{IsIndirectlyRecursive}(c,m)\, \mapsto \,\bot \\[2pt]
 \mathtt{IsVisibleMethod}(c,m,[t1;t2],V1)\, \mapsto \,\bot \\[2pt]
 \mathtt{IsInverter}(c,m,V1,V2)\, \mapsto \,\bot \\[2pt]
 \mathtt{IsDelegator}(c,V1,m)\, \mapsto \,\bot \\[2pt]
 \mathtt{IsOverridden}(c,m)\, \mapsto \,\bot \\[2pt]
 \mathtt{IsOverloaded}(c,m)\, \mapsto \,\bot \\[2pt]
 \mathtt{IsOverriding}(c,m)\, \mapsto \,\bot \\[2pt]
 \mathtt{IsRecursiveMethod}(c,m)\, \mapsto \,\bot \\[2pt]
 \mathtt{HasReturnType}(c,m,V1)\, \mapsto \,\bot \\[2pt]
 \mathtt{MethodHasParameterType}(c,m,V1)\, \mapsto \,\bot \\[2pt]
 \mathtt{MethodIsUsedWithType}(c,m,[t1;t2],[t1;t2])\, \mapsto \,\bot \\[2pt]
 $

%\subsection{pullupAbstractMethod}
 
%\subsection{pushDownAbstractMethod}

\subsection{DuplicateMethodInHierarchyGen} 

\paragraph{Overview:} \textsf{DuplicateMethodInHierarchyGen (class name s, sublclasslist [a;b], methodname m, return types [r1;r2],  invokedmethodsInmethodname [m1;m2],  callermethods [m3;m4], newname n, 
methodnamparameters [t1;t2] }: this operation is used to 
create a duplicate of the method s::m which has a generic type to two methods of name n and having respectively r1 and r2 as return types.

\tools%%%%%%%%%%%%%%
With \intellij:\begin{enumerate}

\item For each implementation of the method \textsf{m} in
  the subclasses of the class \textsf{s}, duplicate \textsf{m} by
  applying \emph{Extract Method} on its body (give the new
  name, specify the desired visibility), apply \emph{change signature} 
to assign the right return type,then inline the 
   invocation of method \textsf{n} that has replaced the method's body.

\item Use \emph{Pull Members Up} to make the new method
  appear in classes where the initial method is declared
  abstract (specify that it must appear as abstract).

\end{enumerate}

\precondition 
$\\ (\mathtt{ExistsClass}(s)\\ 
\wedge \mathtt{ExistsMethodDefinitionWithParams}(s,m,[t1;t2])\\ 
\wedge \mathtt{ExistsMethodDefinition}(s,m)\\ 
\wedge \neg \mathtt{ExistsMethodDefinitionWithParams}(s,n,[t1;t2])\\ 
\wedge \neg \mathtt{ExistsMethodDefinitionWithParams}(a,n,[t1;t2])\\ 
\wedge \neg \mathtt{ExistsMethodDefinitionWithParams}(b,n,[t1;t2])\\ 
\wedge \mathtt{ExistsType}(r1)\\ 
\wedge \mathtt{ExistsType}(r2)\\ 
\wedge \neg \mathtt{IsInheritedMethodWithParams}(s,n,[t1;t2])\\ 
\wedge \mathtt{AllSubclasses}(s,[a;b]))$ \\

\backdescr

$\\ \mathtt{ExistsMethodDefinition}(s,n)\, \mapsto \,\top \\[2pt]
 \mathtt{ExistsMethodDefinitionWithParams}(s,n,[t1;t2])\, \mapsto \,\top \\[2pt]
 \mathtt{AllInvokedMethodsWithParameterOInBodyOfMAreNotOverloaded}(s,n,V)\, \mapsto \,\top  (condition) \\[2pt]
 \mathtt{AllInvokedMethodsOnObjectOInBodyOfMAreDeclaredInC}(s,n,V,V1)\, \mapsto \,\mathtt{AllInvokedMethodsOnObjectOInBodyOfMAreDeclaredInC}(s,m,V,V1)\\[2pt]
 \mathtt{BoundVariableInMethodBody}(s,n,V)\, \mapsto \,\mathtt{BoundVariableInMethodBody}(s,m,V)\\[2pt]
 \mathtt{IsPublic}(s,n)\, \mapsto \,\mathtt{IsPublic}(s,m)\\[2pt]
 \mathtt{ExistsParameterWithName}(s,n,[t1;t2],V)\, \mapsto \,\mathtt{ExistsParameterWithName}(s,m,[t1;t2],V)\\[2pt]
 \mathtt{ExistsParameterWithType}(s,n,[t1;t2],T)\, \mapsto \,\mathtt{ExistsParameterWithType}(s,m,[t1;t2],T)\\[2pt]
 \mathtt{IsIndirectlyRecursive}(s,n)\, \mapsto \,\mathtt{IsIndirectlyRecursive}(s,m)\\[2pt]
 \mathtt{IsRecursiveMethod}(s,n)\, \mapsto \,\mathtt{IsRecursiveMethod}(s,m)\\[2pt]
 \mathtt{IsInverter}(s,n,T,V)\, \mapsto \,\mathtt{IsInverter}(s,m,T,V)\\[2pt]
 \mathtt{IsUsedAttributeInMethodBody}(s,V,n)\, \mapsto \,\mathtt{IsUsedAttributeInMethodBody}(s,V,m)\\[2pt]
 \mathtt{MethodHasParameterType}(s,n,V)\, \mapsto \,\mathtt{MethodHasParameterType}(s,m,V)\\[2pt]
 \mathtt{ExistsMethodDefinitionWithParams}(a,n,[t1;t2])\, \mapsto \,\mathtt{ExistsMethodDefinitionWithParams}(a,m,[t1;t2])\\[2pt]
 \mathtt{ExistsMethodDefinitionWithParams}(b,n,[t1;t2])\, \mapsto \,\mathtt{ExistsMethodDefinitionWithParams}(b,m,[t1;t2])\\[2pt]
 \mathtt{IsDelegator}(s,n,m3)\, \mapsto \,\top \\[2pt]
 \mathtt{IsDelegator}(s,n,m4)\, \mapsto \,\top \\[2pt]
 \mathtt{IsDelegator}(a,n,m3)\, \mapsto \,\top \\[2pt]
 \mathtt{IsDelegator}(a,n,m4)\, \mapsto \,\top \\[2pt]
 \mathtt{IsDelegator}(b,n,m3)\, \mapsto \,\top \\[2pt]
 \mathtt{IsDelegator}(b,n,m4)\, \mapsto \,\top \\[2pt]
 \mathtt{ExistsMethodDefinition}(s,n)\, \mapsto \,\top \\[2pt]
 \mathtt{ExistsMethodDefinition}(a,n)\, \mapsto \,\top \\[2pt]
 \mathtt{ExistsMethodDefinition}(b,n)\, \mapsto \,\top \\[2pt]
 \mathtt{MethodIsUsedWithType}(s,n,[t1;t2],[t1;t2])\, \mapsto \,\bot \\[2pt]
 \mathtt{MethodIsUsedWithType}(a,n,[t1;t2],[t1;t2])\, \mapsto \,\bot \\[2pt]
 \mathtt{MethodIsUsedWithType}(b,n,[t1;t2],[t1;t2])\, \mapsto \,\bot \\[2pt]
 \mathtt{MethodIsUsedWithType}(s,n,[t1;t2],[T])\, \mapsto \,\bot \\[2pt]
 \mathtt{MethodIsUsedWithType}(a,n,[t1;t2],[T])\, \mapsto \,\bot \\[2pt]
 \mathtt{MethodIsUsedWithType}(b,n,[t1;t2],[T])\, \mapsto \,\bot \\[2pt]
 \mathtt{ExistsMethodInvocation}(s,m1,V,n)\, \mapsto \,\top \\[2pt]
 \mathtt{ExistsMethodInvocation}(s,m2,V,n)\, \mapsto \,\top \\[2pt]
 \mathtt{ExistsMethodInvocation}(a,m1,V,n)\, \mapsto \,\top \\[2pt]
 \mathtt{ExistsMethodInvocation}(a,m2,V,n)\, \mapsto \,\top \\[2pt]
 \mathtt{ExistsMethodInvocation}(b,m1,V,n)\, \mapsto \,\top \\[2pt]
 \mathtt{ExistsMethodInvocation}(b,m2,V,n)\, \mapsto \,\top \\[2pt]
 \mathtt{IsInheritedMethodWithParams}(a,n,[t1;t2])\, \mapsto \,\neg \mathtt{ExistsMethodDefinitionWithParams}(a,m,[t1;t2])\\[2pt]
 \mathtt{IsInheritedMethodWithParams}(b,n,[t1;t2])\, \mapsto \,\neg \mathtt{ExistsMethodDefinitionWithParams}(b,m,[t1;t2])\\[2pt]
 \mathtt{IsInheritedMethod}(a,n)\, \mapsto \,\neg \mathtt{ExistsMethodDefinition}(a,m)\\[2pt]
 \mathtt{IsInheritedMethod}(b,n)\, \mapsto \,\neg \mathtt{ExistsMethodDefinition}(b,m)\\[2pt]
 \mathtt{IsOverriding}(a,n)\, \mapsto \,\neg \mathtt{ExistsMethodDefinition}(a,m)\\[2pt]
 \mathtt{IsOverriding}(b,n)\, \mapsto \,\neg \mathtt{ExistsMethodDefinition}(b,m)\\[2pt]
 \mathtt{IsOverridden}(a,n)\, \mapsto \,\neg \mathtt{ExistsMethodDefinition}(a,m)\\[2pt]
 \mathtt{IsOverridden}(b,n)\, \mapsto \,\neg \mathtt{ExistsMethodDefinition}(b,m)\\[2pt]
 $

\subsection{AddSpecializedMethodInHierarchyGen (composed)}

  \paragraph{Overview:} \textsf{AddSpecializedMethodInHierarchyGen(class s, subclasses [a,b], methodname m, returntypes [r1,r2], callermethods [n,o], inkvokedmethods [p,q], paramtype t, paramname pn , 
subtypesOfparamtype [t1,t2],newtype t')}: this operation is similar to the operation \textsf{AddSpecializedMethodInHierarchy}~\ref{def-AddSpecializedMethodInHierarchy} except 
that it is applied to methods having generic types. This operation is also composed (see the following algorithm).

\paragraph{Algorithm of the operation} The operation \textsf{AddSpecializedMethodInHierarchyGen} is based on three steps : \\ \\
%\begin{figure}
%\begin{center}
\begin{boxedminipage}{\columnwidth}%%%%%%%%%%%%%%%%%%%%%%%%%%%%%%%%%%%%%%%%%%%%%%%

\sf
 AddSpecializedMethodInHierarchyGen(class s, subclasses [a,b], methodname m, returntypes [r1,r2], callermethods [n,o], inkvokedmethods [p,q], paramtype t, paramname pn , 
subtypesOfparamtype [t1,t2],newtype t') =

\begin{enumerate}
\item  DuplicateMethodInHierarchyGen  s [a,b] m [r1,r2] [p,q]  [n,o] temporaryName  [t]
\item  SpecialiseParameter s [a,b] temporaryName t pn [t1,t2] t'; 
\item  RenameDelegatorWithOverloading (s, [a,b], temporaryName,t', pn,t,m)

\end{enumerate}

\end{boxedminipage}

%\subsection{createvariablename}

%\paragraph{Overview:} \textsf{CreateEmptyClass (classname c)}: this operation is used t add a new class c.

%\tools
%\emph{new Class} in Eclipse and \intellij.

\subsection{InlineConstructor}

\paragraph{Overview:} \textsf{InlineConstructor (classname s, methodname m, inlinedConstructor c, fields [f1,f2], getters [g1,g2]}: this operation is used to 
inline the constructor c which is used in the method c::m and whose the corresponding class has the fields [f1,f2] and the getter methods [g1,g2].

\tools
\emph{inline} in \intellij.

\precondition 

$\\ (\mathtt{ExistsType}(s)\\ 
\wedge \mathtt{ExistsType}(c)\\ 
\wedge \mathtt{IsUsedConstructorAsObjectReceiver}(c,s,m)\\ 
\wedge (\neg \mathtt{IsInheritedMethodWithParams}(s,m,[c])\\
\tab \vee \neg \mathtt{ExistsMethodDefinitionWithParams}(s,m,[c])))$ \\

\backdescr  
$\\ \mathtt{IsUsedConstructorAsObjectReceiver}(c,s,m)\, \mapsto \,\bot \\[2pt]
 \mathtt{ExistsMethodDefinitionWithParams}(C,M,[c])\, \mapsto \,\bot \\[2pt]
 \mathtt{IsUsedMethodIn}(c,c,s)\, \mapsto \,\bot \\[2pt]
 \mathtt{IsUsedConstructorAsMethodParameter}(c,s,m)\, \mapsto \,\bot \\[2pt]
 \mathtt{IsUsedConstructorAsInitializer}(c,s,m)\, \mapsto \,\bot \\[2pt]
 \mathtt{existsFieldInMethodScope}(s,m,f1)\, \mapsto \,\top \\[2pt]
 \mathtt{existsFieldInMethodScope}(s,m,f2)\, \mapsto \,\top \\[2pt]
 \mathtt{BoundVariableInMethodBody}(s,m,f1)\, \mapsto \,\top \\[2pt]
 \mathtt{BoundVariableInMethodBody}(s,m,f2)\, \mapsto \,\top \\[2pt]
 \mathtt{existslocalVariable}(s,m,f1var)\, \mapsto \,\top \\[2pt]
 \mathtt{existslocalVariable}(s,m,f2var)\, \mapsto \,\top \\[2pt]
 \mathtt{ExistsMethodDefinition}(s,g1)\, \mapsto \,\top \\[2pt]
 \mathtt{ExistsMethodDefinition}(s,g2)\, \mapsto \,\top \\[2pt]
 \mathtt{IsUsedConstructorAsObjectReceiver}(c,s,g1)\, \mapsto \,\bot \\[2pt]
 \mathtt{IsUsedConstructorAsObjectReceiver}(c,s,g2)\, \mapsto \,\bot \\[2pt]
 \mathtt{IsOverriding}(s,g1)\, \mapsto \,\bot \\[2pt]
 \mathtt{IsOverriding}(s,g2)\, \mapsto \,\bot \\[2pt]
 \mathtt{IsOverridden}(s,g1)\, \mapsto \,\bot \\[2pt]
 \mathtt{IsOverridden}(s,g2)\, \mapsto \,\bot \\[2pt]
 \mathtt{IsRecursiveMethod}(s,g1)\, \mapsto \,\bot \\[2pt]
 \mathtt{IsRecursiveMethod}(s,g2)\, \mapsto \,\bot \\[2pt]
 $

\subsection{InlineLocalField}

\paragraph{Overview:} \textsf{inlineLocalField (classname s, methodname m, fieldname f)}: this operation is used to inline the field f which is used in the scope 
of the method s::m.

\tools
\emph{inline} in Eclipse and \intellij.

\precondition $\\ (\mathtt{ExistsType}(s)\\ 
\wedge \mathtt{ExistsMethodDefinition}(s,m)\\ 
\wedge \mathtt{existsFieldInMethodScope}(s,m,f))$ \\

\backdescr  
$\\ \mathtt{existsFieldInMethodScope}(s,m,f)\, \mapsto \,\bot \\[2pt]
 $

\subsection{InlinelocalVariable}

\paragraph{Overview:} \textsf{InlinelocalVariable (classname s, methodname m, variablename v)}: this operation is used to inline the local variable 
declared in the scope of the method s::m.

\tools
\emph{inline} in Eclipse and \intellij.

\precondition 
$\\ (\mathtt{ExistsType}(s)\\ 
\wedge \mathtt{ExistsMethodDefinition}(s,m)\\ 
\wedge \mathtt{existslocalVariable}(s,m,v))$ \\

\backdescr  
$\mathtt{existslocalVariable}(s,m,v)\, \mapsto \,\bot \\[2pt]
 $

\subsection{InlineParmeterObject (composed)}

\paragraph{Overview:} \textsf{InlineParmeterObject (classname s, methodname m, inlinedConstructor c, inlinedgetters [g1,g2], fields [f1,f2]}: this operation is used to inline 
a parameter object. It has the inverse role of the operation \textsf{IntroduceParameterObject}. This operation is composed (see algorithm below).

\paragraph{Algorithm of the operation}
The follwing algorithm describes the mechanics of this composed operation:\\ \\
\begin{boxedminipage}{\columnwidth}%%%%%%%%%%%%%%%%%%%%%%%%%%%%%%%%%%%%%%%%%%%%%%%

\sf
InlineParmeterObject (s, m, c, [g1,g2], [f1,f2]) =

\begin{enumerate}
\item  InlineConstructor s m c $[$g1,g2$]$ $[$f1,f2$]$
\item Forall f in $[$f1,f2$]$ do InlineLocalField  s m f
\item Forall g in $[$g1,g2$]$ do  InlineAnddelete  (s,g,[], m, [], []) 
\item Forall f in $[$f1,f2$]$ do  InlinelocalVariable s m (generate variable name from f)

\end{enumerate}

\end{boxedminipage}

\section{Precondition for all transformations}
\label{sec-precondition-composition}

In this section, we apply a calculus of minimal precondition
to the sequence of basic refactoring operations that compose
our transformations.
We apply the calculus of Kniesel and
Koch~\cite{composition-of-refactorings2004}, based on
the \emph{backward descriptions} given in the previous
appendix.
We use that calculus to compute a minimum precondition that
ensures that the round-trip transformation succeeds, 
which means we determine a set of programs on which we
can ensure that the preconditions of all the component
refactoring operations will be satisfied when applying
the \emph{Composite}$\rightarrow$\emph{Visitor}$\rightarrow$\emph{Composite}
transformation any number of time (as explained
in~\cite{Cohen-Ajouli:2013}).

We give in the following the preconditions for the basic transformation and also those of transformation of the four variations evoked in the 
technical report.

\subsection{Precondition for basic transformation}
The computed precondition corresponding to the basic transformation is given in Fig.~\ref{fig-fix-point-precondition}.
The chains taken into account for the computation are given
in Figs.~\ref{fig-algo-CV-calculus}
and~\ref{fig-algo-VC-calculus}.
The difference between these chains and the algorithms of
previous sections are: 
\begin{itemize}

\item

We add some parameters to the operations that were not
made explicit before (in previous sections, the \emph{project} was
an implicit parameter).

\item
Composite operations are replaced by their component
operations (\textsf{MergeDuplicateMethods} in step 8 of the \emph{Composite}$\rightarrow$\emph{Visitor} chain).

\item
We split some operations of the tool into several
abstract operations when the behavior of an operation
depends on the state of the program. In that case, each
possible behavior is represented by a different abstract
operation.
For instance, we have two operations for method renaming, one for
overloaded methods and one for not-overloaded methods.

\end{itemize}

\begin{figure}[!htp]
\begin{center}
\begin{boxedminipage}{\textwidth}
\relsize{-1}
\sf

  CreateEmptyClass(PrintVisitor) \textbf{;}\\
  CreateEmptyClass(PrettyprintVisitor) \textbf{;}\\
  CreateIndirectionInSuperClass(Graphic,[Ellipse;Composite;], print, [], void, printAux) \textbf{;}\\
  CreateIndirectionInSuperClass(Graphic, [Ellipse;Composite;], prettyprint, [], void, prettyprintAux) \textbf{;}\\
  InlineMethodInvocations(Composite, printAux, [], Graphic, print, []) \textbf{;}\\
  InlineMethodInvocations(Composite, prettyprintAux, [], Graphic, prettyprint, []) \textbf{;}\\
  AddParameterWithReuse(Graphic, [Ellipse;Composite;], printAux, [], PrintVisitor, v) \textbf{;}\\
  AddParameterWithReuse(Graphic, [Ellipse;Composite;], prettyprintAux, [], PrettyprintVisitor, v) \textbf{;}\\
  MoveMethodWithDelegate (Ellipse, [graphics;], PrintVisitor, printAux, [PrintVisitor;], void, visit) \textbf{;}\\
  MoveMethodWithDelegate (Composite, [graphics;], PrintVisitor, printAux, [PrintVisitor;], void, visit) \textbf{;}\\
  MoveMethodWithDelegate (Ellipse, [graphics;], PrettyprintVisitor, prettyprintAux, [PrettyprintVisitor;], void, visit) \textbf{;}\\
  MoveMethodWithDelegate (Composite, [graphics;], PrettyprintVisitor, prettyprintAux, [PrettyprintVisitor;], void, visit) \textbf{;}\\
  ExtractSuperClass ([PrintVisitor;PrettyprintVisitor;], Visitor, [visit;], void) \textbf{;}\\
  GeneraliseParameter (Graphic, [Ellipse;Composite;], printAux, v, PrintVisitor, Visitor) \textbf{;}\\
  GeneraliseParameter (Graphic, [Ellipse;Composite;], prettyprintAux, v, PrettyprintVisitor, Visitor) \textbf{;}\\
  ReplaceMethodcodeDuplicatesInverter (Ellipse, printAux, [prettyprintAux;], Visitor, void) \textbf{;}\\
  ReplaceMethodcodeDuplicatesInverter (Composite, printAux, [prettyprintAux;], Visitor, void) \textbf{;}\\
  PullupConcreteDelegator (Ellipse, [graphics;], prettyprintAux, Graphic) \textbf{;}\\
  SafeDeleteDelegatorWithOverridden (Composite, prettyprintAux, Graphic) \textbf{;}\\
  InlineAndDelete (Graphic, prettyprintAux)
  RenameInHierarchyNoOverloading (Graphic,[Ellipse;Composite;], printAux,[Visitor;], accept)

%\end{itemize}
\end{boxedminipage}
\end{center}
\caption{Chain of refactoring operations of the transformation  Composite$\rightarrow$Visiteur of basic program}
\label{fig-algo-CV-calculus}
\end{figure}

\begin{figure}[!htp]
\begin{center}
\begin{boxedminipage}{\textwidth}
\relsize{-1}
\sf

 DuplicateMethodInHierarchy (Graphic, [Ellipse;Composite;], accept,[visit;], [print;prettyprint;], acceptPrintVisitoraddspecializedMethodtmp, [Visitor;]) \textbf{;}\\
 SpecialiseParameter (Graphic, [Ellipse;Composite;], acceptPrintVisitoraddspecializedMethodtmp, Visitor, [PrintVisitor;PrettyprintVisitor;], PrintVisitor) \textbf{;}\\
 RenameDelegatorWithOverloading (Graphic, [Ellipse;Composite;], acceptPrintVisitoraddspecializedMethodtmp, PrintVisitor, v, Visitor, accept) \textbf{;}\\
 DuplicateMethodInHierarchy (Graphic, [Ellipse;Composite;], accept, [visit;], [print;prettyprint;], acceptPrettyprintVisitoraddspecializedMethodtmp, [Visitor;]) \textbf{;}\\
 SpecialiseParameter (Graphic, [Ellipse;Composite;], acceptPrettyprintVisitoraddspecializedMethodtmp, Visitor, [PrintVisitor;PrettyprintVisitor;], PrettyprintVisitor) \textbf{;}\\
 RenameDelegatorWithOverloading (Graphic, [Ellipse;Composite;], acceptPrettyprintVisitoraddspecializedMethodtmp, PrettyprintVisitor, v, Visitor, accept) \textbf{;}\\
 DeleteMethodInHierarchy (Graphic, [Ellipse;Composite;], accept, [visit;], Visitor)
 PushDownAll (Visitor, [PrintVisitor;PrettyprintVisitor;], visit, [Ellipse;]) \textbf{;}\\
 PushDownAll (Visitor, [PrintVisitor;PrettyprintVisitor;], visit, [Composite;]) \textbf{;}\\
 InlineMethod (Ellipse, visit, PrintVisitor, accept) \textbf{;}\\
 InlineMethod (Composite, visit, PrintVisitor, accept) \textbf{;}\\
 InlineMethod (Ellipse, visit, PrettyprintVisitor, accept) \textbf{;}\\
 InlineMethod (Composite, visit, PrettyprintVisitor, accept)   \textbf{;}\\
 RenameOverloadedMethodInHierarchy (Graphic, [Ellipse;Composite;], accept, printAux, [PrintVisitor;]) \textbf{;}\\
 RenameOverloadedMethodInHierarchy (Graphic, [Ellipse;Composite;], accept, prettyprintAux, [PrettyprintVisitor;]) \textbf{;}\\
 RemoveParameter (Graphic, [Ellipse;Composite;], printAux, [PrintVisitor;], PrintVisitor, v) \textbf{;}\\
 RemoveParameter (Graphic, [Ellipse;Composite;], prettyprintAux, [PrettyprintVisitor;], PrettyprintVisitor, v) \textbf{;}\\
 ReplaceMethodDuplication (Graphic, [Ellipse;Composite;], print, printAux, []) \textbf{;}\\
 ReplaceMethodDuplication (Graphic, [Ellipse;Composite;], prettyprint, prettyprintAux, []) \textbf{;}\\
 PushDownImplementation (Graphic, [], [Ellipse;Composite;], print,  []) \textbf{;}\\
 PushDownImplementation (Graphic, [], [Ellipse;Composite;], prettyprint,  []) \textbf{;}\\
 PushDownAll (Graphic, [Ellipse;Composite;], printAux, []) \textbf{;}\\
 PushDownAll (Graphic, [Ellipse;Composite;], prettyprintAux, []) \textbf{;}\\
 InlineAndDelete (Ellipse, printAux) \textbf{;}\\
 InlineAndDelete (Composite, printAux) \textbf{;}\\
 InlineAndDelete (Ellipse, prettyprintAux) \textbf{;}\\
 InlineAndDelete (Composite, prettyprintAux) \textbf{;}\\
 DeleteClass (PrintVisitor, Visitor, [Ellipse;Composite;PrintVisitor;PrettyprintVisitor;Visitor;\\Graphic;], [visit;], [accept;eval;prettyprint;]) \textbf{;}\\
 DeleteClass (PrettyprintVisitor, Visitor, [Ellipse;Composite;PrintVisitor;PrettyprintVisitor;Visitor;\\Graphic;], [visit;], [accept;eval;prettyprint;]) \textbf{;}\\
 DeleteClass (Visitor, java.lang.Object,[Ellipse;Composite;PrintVisitor;PrettyprintVisitor;Visitor;\\Graphic;], [visit;], [accept;eval;prettyprint;])

\end{boxedminipage}
\end{center}

\caption{Chain of refactoring operations of the transformation Visiteur$\rightarrow$Composite of basic program}
\label{fig-algo-VC-calculus}
\end{figure}

\begin{figure}[!htp]
\relsize{-1}

$\\ \neg \mathtt{ExistsMethod}(Graphic,accept)\\ 
\wedge \neg \mathtt{ExistsMethod}(Ellipse,accept)\\ 
\wedge \neg \mathtt{ExistsMethod}(Composite,accept)\\ 
\wedge \neg \mathtt{IsInheritedMethod}(Graphic,accept)\\ 
\wedge \mathtt{AllInvokedMethodsWithParameterOInBodyOfMAreNotOverloaded}(Composite,prettyprint,this)\\ 
\wedge \mathtt{AllInvokedMethodsOnObjectOInBodyOfMAreDeclaredInC}(Ellipse,prettyprint,this,Graphic)\\ 
\wedge \mathtt{AllInvokedMethodsWithParameterOInBodyOfMAreNotOverloaded}(Ellipse,prettyprint,this)\\ 
\wedge \neg \mathtt{ExistsType}(Visitor)\\ 
\wedge \mathtt{ExistsClass}(Ellipse)\\ 
\wedge \neg \mathtt{BoundVariableInMethodBody}(Graphic,prettyprint,v)\\ 
\wedge \neg \mathtt{BoundVariableInMethodBody}(Graphic,print,v)\\ 
\wedge \mathtt{IsRecursiveMethod}(Composite,prettyprint)\\ 
\wedge \mathtt{ExistsClass}(Composite)\\ 
\wedge \mathtt{IsRecursiveMethod}(Composite,print)\\ 
\wedge \mathtt{ExistsMethodWithParams}(Graphic,prettyprint,[\,])\\ 
\wedge \mathtt{ExistsAbstractMethod}(Graphic,prettyprint)\\ 
\wedge \neg \mathtt{IsInheritedMethod}(Graphic,prettyprintAux)\\ 
\wedge \neg \mathtt{IsInheritedMethodWithParams}(Graphic,prettyprintAux,[\,])\\ 
\wedge \neg \mathtt{ExistsMethodWithParams}(Graphic,prettyprintAux,[\,])\\ 
\wedge \mathtt{HasReturnType}(Graphic,prettyprint,void)\\ 
\wedge \mathtt{ExistsMethod}(Graphic,prettyprint)\\ 
\wedge \mathtt{ExistsMethod}(Ellipse,prettyprint)\\ 
\wedge \mathtt{ExistsMethod}(Composite,prettyprint)\\ 
\wedge \neg \mathtt{ExistsMethod}(Graphic,prettyprintAux)\\ 
\wedge \neg \mathtt{ExistsMethod}(Ellipse,prettyprintAux)\\ 
\wedge \neg \mathtt{ExistsMethod}(Composite,prettyprintAux)\\ 
\wedge \mathtt{ExistsClass}(Graphic)\\ 
\wedge \mathtt{IsAbstractClass}(Graphic)\\ 
\wedge \mathtt{ExistsMethodWithParams}(Graphic,print,[\,])\\ 
\wedge \mathtt{ExistsAbstractMethod}(Graphic,print)\\ 
\wedge \neg \mathtt{IsInheritedMethod}(Graphic,printAux)\\ 
\wedge \neg \mathtt{IsInheritedMethodWithParams}(Graphic,printAux,[\,])\\ 
\wedge \neg \mathtt{ExistsMethodWithParams}(Graphic,printAux,[\,])\\ 
\wedge \mathtt{AllSubclasses}(Graphic,[Ellipse;Composite])\\ 
\wedge \mathtt{HasReturnType}(Graphic,print,void)\\ 
\wedge \neg \mathtt{IsPrivate}(Graphic,print)\\ 
\wedge \neg \mathtt{IsPrivate}(Ellipse,print)\\ 
\wedge \neg \mathtt{IsPrivate}(Composite,print)\\ 
\wedge \mathtt{ExistsMethod}(Graphic,print)\\ 
\wedge \mathtt{ExistsMethod}(Ellipse,print)\\ 
\wedge \mathtt{ExistsMethod}(Composite,print)\\ 
\wedge \neg \mathtt{ExistsMethod}(Graphic,printAux)\\ 
\wedge \neg \mathtt{ExistsMethod}(Ellipse,printAux)\\ 
\wedge \neg \mathtt{ExistsMethod}(Composite,printAux)\\ 
\wedge \neg \mathtt{ExistsType}(PrettyprintVisitor)\\ 
\wedge \neg \mathtt{ExistsType}(PrintVisitor)$
\caption{Computed precondition for the round-trip transformation}
\label{fig-fix-point-precondition}

\end{figure}

\FloatBarrier

\subsection{Precondition for method with parameter variation}
\label{precond-withparam}

\begin{figure}[!htp]
\relsize{-1}

$\\ \neg \mathtt{IsUsedConstructorAsMethodParameter}(SeColorVisitor,Ellipse,print)\\ 
\wedge \neg \mathtt{IsUsedConstructorAsMethodParameter}(SeColorVisitor,CompositeGraphic,print)\\ 
\wedge \neg \mathtt{IsUsedConstructorAsMethodParameter}(SeColorVisitor,Graphic,print)\\ 
\wedge \neg \mathtt{IsUsedConstructorAsMethodParameter}(SeColorVisitor,Graphic,setColor)\\ 
\wedge \neg \mathtt{IsUsedConstructorAsObjectReceiver}(SeColorVisitor,Ellipse,printAux)\\
\wedge  \neg \mathtt{IsUsedConstructorAsObjectReceiver}(SeColorVisitor,Ellipse,print)\\ 
\wedge \neg \mathtt{IsUsedConstructorAsObjectReceiver}(SeColorVisitor,CompositeGraphic,printAux)\\
\wedge \neg \mathtt{IsUsedConstructorAsObjectReceiver}(SeColorVisitor,CompositeGraphic,print)\\ 
\wedge \mathtt{ExistsType}(CompositeGraphic)\\ \wedge \mathtt{ExistsType}(Ellipse)\\ 
\wedge \neg \mathtt{ExistsMethodInvocation}(Ellipse,setColorAux,Ellipse,print)\\ 
\wedge \neg \mathtt{IsUsedMethod}(Graphic,accept_SeColorVisitor_addspecializedMethod_tmp,[SeColorVisitor])\\ 
\wedge \neg \mathtt{ExistsMethodDefinition}(Graphic,accept)\\ 
\wedge \neg \mathtt{ExistsMethodDefinition}(Ellipse,accept)\\ 
\wedge \neg \mathtt{ExistsMethodDefinition}(CompositeGraphic,accept)\\ 
\wedge \neg \mathtt{IsInheritedMethod}(Graphic,accept)\\ 
\wedge \neg \mathtt{IsUsedMethodIn}(Graphic,setColorAux,Ellipse)\\ 
\wedge \neg \mathtt{IsUsedMethodIn}(Graphic,setColorAux,CompositeGraphic)\\ 
\wedge \mathtt{AllInvokedMethodsWithParameterOInBodyOfMAreNotOverloaded}(CompositeGraphic,setColor,this)\\ 
\wedge \mathtt{AllInvokedMethodsOnObjectOInBodyOfMAreDeclaredInC}(Ellipse,setColor,this,Graphic)\\ 
\wedge \mathtt{AllInvokedMethodsWithParameterOInBodyOfMAreNotOverloaded}(Ellipse,setColor,this)\\ 
\wedge \mathtt{AllInvokedMethodsWithParameterOInBodyOfMAreNotOverloaded}(Graphic,setColor,v)\\ 
\wedge \mathtt{AllInvokedMethodsWithParameterOInBodyOfMAreNotOverloaded}(Ellipse,setColor,v)\\ 
\wedge \mathtt{AllInvokedMethodsWithParameterOInBodyOfMAreNotOverloaded}(CompositeGraphic,setColor,v)\\ 
\wedge \neg \mathtt{ExistsType}(Visitor)\\ \wedge \mathtt{ExistsClass}(Ellipse)\\ 
\wedge \mathtt{ExistsType}(Graphic)\\ 
\wedge \neg \mathtt{ExistsType}(SeColorVisitor)\\ 
\wedge \mathtt{BoundVariableInMethodBody}(Ellipse,setColor,int c)\\ 
\wedge \mathtt{BoundVariableInMethodBody}(CompositeGraphic,setColor,int c)\\ 
\wedge \neg \mathtt{BoundVariableInMethodBody}(Graphic,print,v)\\ 
\wedge \mathtt{IsRecursiveMethod}(CompositeGraphic,setColor)\\ 
\wedge \mathtt{ExistsClass}(CompositeGraphic)\\ 
\wedge \mathtt{IsRecursiveMethod}(CompositeGraphic,print)\\ 
\wedge \mathtt{ExistsMethodDefinitionWithParams}(Graphic,setColor,[int c])\\ 
\wedge \mathtt{ExistsAbstractMethod}(Graphic,setColor)\\ 
\wedge \neg \mathtt{IsInheritedMethod}(Graphic,setColorAux)\\ 
\wedge \neg \mathtt{IsInheritedMethodWithParams}(Graphic,setColorAux,[int c])\\ 
\wedge \neg \mathtt{ExistsMethodDefinitionWithParams}(Graphic,setColorAux,[int c])\\ 
\wedge \mathtt{HasReturnType}(Graphic,setColor,void)\\ 
\wedge \mathtt{ExistsMethodDefinition}(Graphic,setColor)\\ 
\wedge \mathtt{ExistsMethodDefinition}(Ellipse,setColor)\\ 
\wedge \mathtt{ExistsMethodDefinition}(CompositeGraphic,setColor)\\ 
\wedge \neg \mathtt{ExistsMethodDefinition}(Graphic,setColorAux)\\ 
\wedge \neg \mathtt{ExistsMethodDefinition}(Ellipse,setColorAux)\\ 
\wedge \neg \mathtt{ExistsMethodDefinition}(CompositeGraphic,setColorAux)\\ 
\wedge \mathtt{ExistsClass}(Graphic)\\ 
\wedge \mathtt{IsAbstractClass}(Graphic)\\ 
\wedge \mathtt{ExistsMethodDefinitionWithParams}(Graphic,print,[\,])\\
 \wedge \mathtt{ExistsAbstractMethod}(Graphic,print)\\ 
\wedge \neg \mathtt{IsInheritedMethod}(Graphic,printAux)\\ 
\wedge \neg \mathtt{IsInheritedMethodWithParams}(Graphic,printAux,[\,])\\ 
\wedge \neg \mathtt{ExistsMethodDefinitionWithParams}(Graphic,printAux,[\,])\\ 
\wedge \mathtt{AllSubclasses}(Graphic,[Ellipse;CompositeGraphic])\\ 
\wedge \mathtt{HasReturnType}(Graphic,print,void)\\ 
\wedge \neg \mathtt{IsPrivate}(Graphic,print)\\ 
\wedge \neg \mathtt{IsPrivate}(Ellipse,print)\\ 
\wedge \neg \mathtt{IsPrivate}(CompositeGraphic,print)\\ 
\wedge \mathtt{ExistsMethodDefinition}(Graphic,print)\\ 
\wedge \mathtt{ExistsMethodDefinition}(Ellipse,print)\\ 
\wedge \mathtt{ExistsMethodDefinition}(CompositeGraphic,print)\\ 
\wedge \neg \mathtt{ExistsMethodDefinition}(Graphic,printAux)\\ 
\wedge \neg \mathtt{ExistsMethodDefinition}(Ellipse,printAux)\\ 
\wedge \neg \mathtt{ExistsMethodDefinition}(CompositeGraphic,printAux)\\
 \wedge \neg \mathtt{ExistsType}(PrintVisitor)$

\caption{Computed precondition for the round-trip transformation of method with parameter variation}
\label{fig-fix-point-precondition-withparam}
\end{figure}

\FloatBarrier

\subsection{Precondition for method with different return types variation}
\label{precond-return-values}

\begin{figure}[!htp]
\relsize{-1}

$\\ \neg \mathtt{ExistsMethodInvocation}(Ellipse,toStringAux,Ellipse,perimeter)\\ 
\wedge \neg \mathtt{ExistsMethodDefinition}(Graphic,accept)\\ 
\wedge \neg \mathtt{ExistsMethodDefinition}(Ellipse,accept)\\ 
\wedge \neg \mathtt{ExistsMethodDefinition}(CompositeGraphic,accept)\\ 
\wedge \neg \mathtt{IsInheritedMethod}(Graphic,accept)\\ 
\wedge \neg \mathtt{IsUsedMethodIn}(Graphic,toStringAux,Ellipse)\\ 
\wedge \neg \mathtt{IsUsedMethodIn}(Graphic,toStringAux,CompositeGraphic)\\ 
\wedge \mathtt{AllInvokedMethodsWithParameterOInBodyOfMAreNotOverloaded}(CompositeGraphic,toString,this)\\ 
\wedge \mathtt{AllInvokedMethodsOnObjectOInBodyOfMAreDeclaredInC}(Ellipse,toString,this,Graphic)\\ 
\wedge \mathtt{AllInvokedMethodsWithParameterOInBodyOfMAreNotOverloaded}(Ellipse,toString,this)\\ 
\wedge \neg \mathtt{IsPrimitiveType}(String)\\ 
\wedge \neg \mathtt{ExistsAbstractMethod}(Visitor,visit)\\ 
\wedge \neg \mathtt{IsPrimitiveType}(Integer)\\ 
\wedge \neg \mathtt{ExistsType}(Visitor)\\ 
\wedge \mathtt{ExistsClass}(Ellipse)\\ 
\wedge \neg \mathtt{BoundVariableInMethodBody}(Graphic,toString,v)\\ 
\wedge \neg \mathtt{BoundVariableInMethodBody}(Graphic,perimeter,v)\\ 
\wedge \mathtt{IsRecursiveMethod}(CompositeGraphic,toString)\\ 
\wedge \mathtt{ExistsClass}(CompositeGraphic)\\ 
\wedge \mathtt{IsRecursiveMethod}(CompositeGraphic,perimeter)\\ 
\wedge \mathtt{ExistsMethodDefinitionWithParams}(Graphic,toString,[\,])\\ 
\wedge \mathtt{ExistsAbstractMethod}(Graphic,toString)\\ 
\wedge \neg \mathtt{IsInheritedMethod}(Graphic,toStringAux)\\ 
\wedge \neg \mathtt{IsInheritedMethodWithParams}(Graphic,toStringAux,[\,])\\ 
\wedge \neg \mathtt{ExistsMethodDefinitionWithParams}(Graphic,toStringAux,[\,])\\ 
\wedge \mathtt{HasReturnType}(Graphic,toString,String)\\ 
\wedge \mathtt{ExistsMethodDefinition}(Graphic,toString)\\ 
\wedge \mathtt{ExistsMethodDefinition}(Ellipse,toString)\\ 
\wedge \mathtt{ExistsMethodDefinition}(CompositeGraphic,toString)\\ 
\wedge \neg \mathtt{ExistsMethodDefinition}(Graphic,toStringAux)\\ 
\wedge \neg \mathtt{ExistsMethodDefinition}(Ellipse,toStringAux)\\ 
\wedge \neg \mathtt{ExistsMethodDefinition}(CompositeGraphic,toStringAux)\\ 
\wedge \mathtt{ExistsClass}(Graphic)\\ 
\wedge \mathtt{IsAbstractClass}(Graphic)\\ 
\wedge \mathtt{ExistsMethodDefinitionWithParams}(Graphic,perimeter,[\,])\\ 
\wedge \mathtt{ExistsAbstractMethod}(Graphic,perimeter)\\ 
\wedge \neg \mathtt{IsInheritedMethod}(Graphic,perimeterAux)\\ 
\wedge \neg \mathtt{IsInheritedMethodWithParams}(Graphic,perimeterAux,[\,])\\ 
\wedge \neg \mathtt{ExistsMethodDefinitionWithParams}(Graphic,perimeterAux,[\,])\\ 
\wedge \mathtt{AllSubclasses}(Graphic,[Ellipse;CompositeGraphic])\\ 
\wedge \mathtt{HasReturnType}(Graphic,perimeter,Integer)\\ 
\wedge \neg \mathtt{IsPrivate}(Graphic,perimeter)\\ 
\wedge \neg \mathtt{IsPrivate}(Ellipse,perimeter)\\ 
\wedge \neg \mathtt{IsPrivate}(CompositeGraphic,perimeter)\\ 
\wedge \mathtt{ExistsMethodDefinition}(Graphic,perimeter)\\ 
\wedge \mathtt{ExistsMethodDefinition}(Ellipse,perimeter)\\ 
\wedge \mathtt{ExistsMethodDefinition}(CompositeGraphic,perimeter)\\ 
\wedge \neg \mathtt{ExistsMethodDefinition}(Graphic,perimeterAux)\\ 
\wedge \neg \mathtt{ExistsMethodDefinition}(Ellipse,perimeterAux)\\ 
\wedge \neg \mathtt{ExistsMethodDefinition}(CompositeGraphic,perimeterAux)\\ 
\wedge \neg \mathtt{ExistsType}(ToStringVisitor)\\ 
\wedge \neg \mathtt{ExistsType}(PerimeterVisitor)$

\caption{Computed precondition for the round-trip transformation of method with different return type variation}
\label{fig-fix-point-precondition-return-types}

\end{figure}

\FloatBarrier

\subsection{Precondition for several level hierarchy variation}
\label{precond-multi-level}

\begin{figure}[!htp]
\relsize{-1}

$\\ \mathtt{IsInheritedMethod}(ColoredRect,prettyprint)\\ 
\wedge \neg \mathtt{ExistsMethodInvocation}(Rect,prettyprintTmpVC,Rect,print)\\ 
\wedge \neg \mathtt{ExistsMethodDefinition}(Graphic,accept)\\ 
\wedge \neg \mathtt{ExistsMethodDefinition}(Rect,accept)\\ 
\wedge \neg \mathtt{ExistsMethodDefinition}(ColoredRect,accept)\\ 
\wedge \neg \mathtt{ExistsMethodDefinition}(CompositeGraphic,accept)\\ 
\wedge \neg \mathtt{IsInheritedMethod}(Graphic,accept)\\ 
\wedge \neg \mathtt{IsUsedMethodIn}(Graphic,prettyprintTmpVC,Rect)\\ 
\wedge \neg \mathtt{IsUsedMethodIn}(Graphic,prettyprintTmpVC,ColoredRect)\\ 
\wedge \neg \mathtt{IsUsedMethodIn}(Graphic,prettyprintTmpVC,CompositeGraphic)\\ 
\wedge \mathtt{AllInvokedMethodsWithParameterOInBodyOfMAreNotOverloaded}(CompositeGraphic,prettyprint,this)\\ 
\wedge \mathtt{AllInvokedMethodsOnObjectOInBodyOfMAreDeclaredInC}(Rect,prettyprint,this,Graphic)\\ 
\wedge \neg \mathtt{ExistsType}(Visitor)\\ 
\wedge \mathtt{ExistsClass}(Rect)\\ 
\wedge \neg \mathtt{BoundVariableInMethodBody}(Graphic,prettyprint,v)\\ 
\wedge \neg \mathtt{BoundVariableInMethodBody}(Graphic,print,v)\\ 
\wedge \mathtt{IsRecursiveMethod}(CompositeGraphic,prettyprint)\\ 
\wedge \mathtt{ExistsClass}(CompositeGraphic)\\ 
\wedge \mathtt{IsRecursiveMethod}(CompositeGraphic,print)\\ 
\wedge \mathtt{ExistsMethodDefinitionWithParams}(Graphic,prettyprint,[\,])\\ 
\wedge \mathtt{ExistsAbstractMethod}(Graphic,prettyprint)\\
 \wedge \neg \mathtt{IsInheritedMethod}(Graphic,prettyprintTmpVC)\\ 
\wedge \neg \mathtt{IsInheritedMethodWithParams}(Graphic,prettyprintTmpVC,[\,])\\ 
\wedge \neg \mathtt{ExistsMethodDefinitionWithParams}(Graphic,prettyprintTmpVC,[\,])\\ 
\wedge \mathtt{HasReturnType}(Graphic,prettyprint,void)\\ 
\wedge \mathtt{ExistsMethodDefinition}(Graphic,prettyprint)\\ 
\wedge \mathtt{ExistsMethodDefinition}(CompositeGraphic,prettyprint)\\ 
\wedge \neg \mathtt{ExistsMethodDefinition}(Graphic,prettyprintTmpVC)\\ 
\wedge \neg \mathtt{ExistsMethodDefinition}(Rect,prettyprintTmpVC)\\ 
\wedge \neg \mathtt{ExistsMethodDefinition}(ColoredRect,prettyprintTmpVC)\\
 \wedge \neg \mathtt{ExistsMethodDefinition}(CompositeGraphic,prettyprintTmpVC)\\ 
\wedge \mathtt{ExistsClass}(Graphic)\\ 
\wedge \mathtt{IsAbstractClass}(Graphic)\\ 
\wedge \mathtt{ExistsMethodDefinitionWithParams}(Graphic,print,[\,])\\ 
\wedge \mathtt{ExistsAbstractMethod}(Graphic,print)\\ 
\wedge \neg \mathtt{IsInheritedMethod}(Graphic,printTmpVC)\\ 
\wedge \neg \mathtt{IsInheritedMethodWithParams}(Graphic,printTmpVC,[\,])\\ 
\wedge \neg \mathtt{ExistsMethodDefinitionWithParams}(Graphic,printTmpVC,[\,])\\
 \wedge \mathtt{AllSubclasses}(Graphic,[Rect;ColoredRect;CompositeGraphic])\\
 \wedge \mathtt{HasReturnType}(Graphic,print,void)\\ 
\wedge \neg \mathtt{IsPrivate}(Graphic,print)\\ 
\wedge \neg \mathtt{IsPrivate}(Rect,print)\\ 
\wedge \neg \mathtt{IsPrivate}(ColoredRect,print)\\ 
\wedge \neg \mathtt{IsPrivate}(CompositeGraphic,print)\\ 
\wedge \mathtt{ExistsMethodDefinition}(Graphic,print)\\ 
\wedge \mathtt{ExistsMethodDefinition}(Rect,print)\\
 \wedge \mathtt{ExistsMethodDefinition}(ColoredRect,print)\\
 \wedge \mathtt{ExistsMethodDefinition}(CompositeGraphic,print)\\ 
\wedge \neg \mathtt{ExistsMethodDefinition}(Graphic,printTmpVC)\\ 
\wedge \neg \mathtt{ExistsMethodDefinition}(Rect,printTmpVC)\\
 \wedge \neg \mathtt{ExistsMethodDefinition}(ColoredRect,printTmpVC)\\
 \wedge \neg \mathtt{ExistsMethodDefinition}(CompositeGraphic,printTmpVC)\\
 \wedge \neg \mathtt{ExistsType}(PrettyPrintVisitor)\\ 
\wedge \neg \mathtt{ExistsType}(PrintVisitor)\\ 
\wedge \mathtt{ExistsType}(ColoredRect)\\ 
\wedge \mathtt{ExistsClass}(ColoredRect)\\ 
\wedge \mathtt{IsSubType}(ColoredRect,Rect)\\ 
\wedge \neg \mathtt{ExistsMethodDefinition}(ColoredRect,prettyprint)\\
 \wedge \mathtt{ExistsMethodDefinition}(Rect,prettyprint)\\ 
\wedge \mathtt{AllInvokedMethodsWithParameterOInBodyOfMAreNotOverloaded}(Rect,prettyprint,this)$

\caption{Computed precondition for the round-trip transformation of several level hierarchy  variation}
\label{fig-fix-point-precondition-multilevel}

\end{figure}

\FloatBarrier

\subsection{Precondition for interface variation}
\label{precond-interface}

\begin{figure}[!htp]
\relsize{-1}

$\\ \neg \mathtt{ExistsMethodDefinition}(Ellipse,accept)\\ 
\wedge \neg \mathtt{ExistsMethodDefinition}(CompositeGraphic,accept)\\ 
\wedge \mathtt{AllInvokedMethodsWithParameterOInBodyOfMAreNotOverloaded}(CompositeGraphic,prettyprint,this)\\ 
\wedge \mathtt{AllInvokedMethodsWithParameterOInBodyOfMAreNotOverloaded}(Ellipse,prettyprint,this)\\ 
\wedge \neg \mathtt{ExistsType}(Visitor)\\ \wedge \mathtt{IsRecursiveMethod}(CompositeGraphic,prettyprint)\\ 
\wedge \mathtt{IsRecursiveMethod}(CompositeGraphic,print)\\ \wedge \mathtt{HasReturnType}(Ellipse,prettyprint,void)\\ 
\wedge \mathtt{HasReturnType}(CompositeGraphic,prettyprint,void)\\ \wedge \mathtt{ExistsMethodDefinition}(Ellipse,prettyprint)\\ 
\wedge \mathtt{ExistsMethodDefinition}(CompositeGraphic,prettyprint)\\ 
\wedge \neg \mathtt{ExistsMethodDefinition}(Ellipse,prettyprintTmpVC)\\ 
\wedge \neg \mathtt{ExistsMethodDefinition}(CompositeGraphic,prettyprintTmpVC)\\ 
\wedge \mathtt{HasReturnType}(Ellipse,print,void)\\ 
\wedge \mathtt{HasReturnType}(CompositeGraphic,print,void)\\ 
\wedge \neg \mathtt{IsPrivate}(Ellipse,print)\\ 
\wedge \neg \mathtt{IsPrivate}(CompositeGraphic,print)\\ 
\wedge \mathtt{ExistsMethodDefinition}(Ellipse,print)\\ 
\wedge \mathtt{ExistsMethodDefinition}(CompositeGraphic,print)\\ 
\wedge \neg \mathtt{ExistsMethodDefinition}(Ellipse,printTmpVC)\\ 
\wedge \neg \mathtt{ExistsMethodDefinition}(CompositeGraphic,printTmpVC)\\ 
\wedge \neg \mathtt{ExistsType}(PrettyPrintvisitor)\\ 
\wedge \neg \mathtt{ExistsType}(PrintVisitor)\\ 
\wedge \neg \mathtt{ExistsType}(AbstractGraphic)\\ 
\wedge \mathtt{IsInterface}(Graphic)\\ 
\wedge \mathtt{ExistsType}(Graphic)\\ 
\wedge \mathtt{ExistsClass}(Ellipse)\\ 
\wedge \mathtt{ExistsClass}(CompositeGraphic)\\ 
\wedge \mathtt{ExtendsDirectly}(Ellipse,java.lang.Object)\\ 
\wedge \mathtt{ExtendsDirectly}(CompositeGraphic,java.lang.Object)\\ 
\wedge \mathtt{implementsDirectly}(Ellipse,Graphic)\\ 
\wedge \mathtt{implementsDirectly}(CompositeGraphic,Graphic))$

\caption{Computed precondition for the round-trip transformation of interface  variation}
\label{fig-fix-point-precondition-interface}

\end{figure}

\FloatBarrier

\section{JHotDraw transformation precondition}
\label{sec-precondition-jhotdraw}
\subsection{Chain of operations applied in the round-trip transformation of JHotDraw Composite}
(Sequence of 944 operations)

\relsize{-1}
\sf

\noindent
\operation{PushDownNotReDefinedMethod}(LineConnectionFigure, BezierFigure, [findFigureInside; setAttribute; contains]) \textbf{; }\\
\operation{PushDownNotReDefinedMethod}(SVGPath, AbstractCompositeFigure, [addNotify; removeNotify; findFigureInside; contains]) \textbf{; }\\
\operation{PushDownNotReDefinedMethod}(LabeledLineConnectionFigure, BezierFigure, [basicTransform; setAttribute; findFigureInside; contains]) \textbf{; }\\
\operation{PushDownNotReDefinedMethod}(DependencyFigure, LineConnectionFigure, [addNotify; basicTransform; setAttribute; findFigureInside; contains;]) \textbf{; }\\
\operation{PushDownNotReDefinedMethod}(NodeFigure, TextFigure, [addNotify; basicTransform; setAttribute; findFigureInside; contains]) \textbf{; }\\
\operation{PushDownNotReDefinedMethod}(GraphicalCompositeFigure, AbstractCompositeFigure, [findFigureInside]) \textbf{; }\\
\operation{Createindirectioninsuperclass}(AbstractFigure, [LabeledLineConnectionFigure; AbstractCompositeFigure; GraphicalCompositeFigure; EllipseFigure; DiamondFigure; RectangleFigure; RoundRectangleFigure; TriangleFigure; TextFigure; BezierFigure; TextAreaFigure; \\NodeFigure; SVGImage; SVGPath; DependencyFigure; LineConnectionFigure], \\basicTransform, [AffineTransform tx], Void, basicTransformTmpVC) \textbf{; }\\
\operation{Createindirectioninsuperclass}(AbstractFigure, [LabeledLineConnectionFigure; AbstractCompositeFigure; GraphicalCompositeFigure; EllipseFigure; DiamondFigure; RectangleFigure; RoundRectangleFigure; TriangleFigure; TextFigure; BezierFigure; TextAreaFigure; NodeFigure; SVGImage; SVGPath; DependencyFigure; LineConnectionFigure], contains, [Point2D.Double p], Boolean, containsTmpVC) \textbf{; }\\
\operation{Createindirectioninsuperclass}(AbstractFigure, [LabeledLineConnectionFigure; AbstractCompositeFigure; GraphicalCompositeFigure; EllipseFigure; DiamondFigure; RectangleFigure; RoundRectangleFigure; TriangleFigure; TextFigure; BezierFigure; TextAreaFigure; NodeFigure; SVGImage; SVGPath; DependencyFigure; LineConnectionFigure], setAttribute, [AttributeKey key; Object value], Void, setAttributeTmpVC) \textbf{; }\\
\operation{Createindirectioninsuperclass}(AbstractFigure, [LabeledLineConnectionFigure; AbstractCompositeFigure; GraphicalCompositeFigure; EllipseFigure; DiamondFigure; RectangleFigure; RoundRectangleFigure; TriangleFigure; TextFigure; BezierFigure; TextAreaFigure; NodeFigure; SVGImage; SVGPath; DependencyFigure; LineConnectionFigure], findFigureInside, [Point2D.Double p], Figure, findFigureInsideTmpVC) \textbf{; }\\
\operation{Createindirectioninsuperclass}(AbstractFigure, [LabeledLineConnectionFigure; AbstractCompositeFigure; GraphicalCompositeFigure; EllipseFigure; DiamondFigure; RectangleFigure; RoundRectangleFigure; TriangleFigure; TextFigure; BezierFigure; TextAreaFigure; NodeFigure; SVGImage; SVGPath; DependencyFigure; LineConnectionFigure], addNotify, [Drawing d], Void, addNotifyTmpVC) \textbf{; }\\
\operation{Createindirectioninsuperclass}(AbstractFigure, [LabeledLineConnectionFigure; AbstractCompositeFigure; GraphicalCompositeFigure; EllipseFigure; DiamondFigure; RectangleFigure; RoundRectangleFigure; TriangleFigure; TextFigure; BezierFigure; TextAreaFigure; NodeFigure; SVGImage; SVGPath; DependencyFigure; LineConnectionFigure], removeNotify, [Drawing d], Void, removeNotifyTmpVC) \textbf{; }\\
\operation{introduceParmeterObject}(AbstractFigure, [LabeledLineConnectionFigure; AbstractCompositeFigure; GraphicalCompositeFigure; EllipseFigure; DiamondFigure; RectangleFigure; RoundRectangleFigure; TriangleFigure; TextFigure; BezierFigure; TextAreaFigure; NodeFigure; SVGImage; SVGPath; DependencyFigure; LineConnectionFigure], basicTransformTmpVC, [AffineTransform tx], [AffineTransform tx], [tx], BasicTransformVisitor, v) \textbf{; }\\
\operation{introduceParmeterObject}(AbstractFigure, [LabeledLineConnectionFigure; AbstractCompositeFigure; GraphicalCompositeFigure; EllipseFigure; DiamondFigure; RectangleFigure; RoundRectangleFigure; TriangleFigure; TextFigure; BezierFigure; TextAreaFigure; NodeFigure; SVGImage; SVGPath; DependencyFigure; LineConnectionFigure], containsTmpVC, [Point2D.Double p], [Point2D.Double p], [p], ContainsVisitor, v) \textbf{; }\\
\operation{introduceParmeterObject}(AbstractFigure, [LabeledLineConnectionFigure; AbstractCompositeFigure; GraphicalCompositeFigure; EllipseFigure; DiamondFigure; RectangleFigure; RoundRectangleFigure; TriangleFigure; TextFigure; BezierFigure; TextAreaFigure; NodeFigure; SVGImage; SVGPath; DependencyFigure; LineConnectionFigure], setAttributeTmpVC, [AttributeKey key; Object value], [AttributeKey key; Object value], [key; value], SetAttributeVisitor, v) \textbf{; }\\
\operation{introduceParmeterObject}(AbstractFigure, [LabeledLineConnectionFigure; AbstractCompositeFigure; GraphicalCompositeFigure; EllipseFigure; DiamondFigure; RectangleFigure; RoundRectangleFigure; TriangleFigure; TextFigure; BezierFigure; TextAreaFigure; NodeFigure; SVGImage; SVGPath; DependencyFigure; LineConnectionFigure], findFigureInsideTmpVC, [Point2D.Double p], [Point2D.Double p], [p], FindFigureInsideVisitor, v) \textbf{; }\\
\operation{introduceParmeterObject}(AbstractFigure, [LabeledLineConnectionFigure; AbstractCompositeFigure; GraphicalCompositeFigure; EllipseFigure; DiamondFigure; RectangleFigure; RoundRectangleFigure; TriangleFigure; TextFigure; BezierFigure; TextAreaFigure; NodeFigure; SVGImage; SVGPath; DependencyFigure; LineConnectionFigure], addNotifyTmpVC, [Drawing d], [Drawing d], [d], AddNotifyVisitor, v) \textbf{; }\\
\operation{introduceParmeterObject}(AbstractFigure, [LabeledLineConnectionFigure; AbstractCompositeFigure; GraphicalCompositeFigure; EllipseFigure; DiamondFigure; RectangleFigure; RoundRectangleFigure; TriangleFigure; TextFigure; BezierFigure; TextAreaFigure; NodeFigure; SVGImage; SVGPath; DependencyFigure; LineConnectionFigure], removeNotifyTmpVC, [Drawing d], [Drawing d], [d], RemoveNotifyVisitor, v) \textbf{; }\\
\operation{Movemethodwithdelegate}(LabeledLineConnectionFigure, [], BasicTransformVisitor, basicTransformTmpVC, [BasicTransformVisitor], Void, visit) \textbf{; }\\
\operation{Movemethodwithdelegate}(AbstractCompositeFigure, [], BasicTransformVisitor, basicTransformTmpVC, [BasicTransformVisitor], Void, visit) \textbf{; }\\
\operation{Movemethodwithdelegate}(GraphicalCompositeFigure, [], BasicTransformVisitor, basicTransformTmpVC, [BasicTransformVisitor], Void, visit) \textbf{; }\\
\operation{Movemethodwithdelegate}(EllipseFigure, [], BasicTransformVisitor, basicTransformTmpVC, [BasicTransformVisitor], Void, visit) \textbf{; }\\
\operation{Movemethodwithdelegate}(DiamondFigure, [], BasicTransformVisitor, basicTransformTmpVC, [BasicTransformVisitor], Void, visit) \textbf{; }\\
\operation{Movemethodwithdelegate}(RectangleFigure, [], BasicTransformVisitor, basicTransformTmpVC, [BasicTransformVisitor], Void, visit) \textbf{; }\\
\operation{Movemethodwithdelegate}(RoundRectangleFigure, [], BasicTransformVisitor, basicTransformTmpVC, [BasicTransformVisitor], Void, visit) \textbf{; }\\
\operation{Movemethodwithdelegate}(TriangleFigure, [], BasicTransformVisitor, basicTransformTmpVC, [BasicTransformVisitor], Void, visit) \textbf{; }\\
\operation{Movemethodwithdelegate}(TextFigure, [], BasicTransformVisitor, basicTransformTmpVC, [BasicTransformVisitor], Void, visit) \textbf{; }\\
\operation{Movemethodwithdelegate}(BezierFigure, [], BasicTransformVisitor, basicTransformTmpVC, [BasicTransformVisitor], Void, visit) \textbf{; }\\
\operation{Movemethodwithdelegate}(TextAreaFigure, [], BasicTransformVisitor, basicTransformTmpVC, [BasicTransformVisitor], Void, visit) \textbf{; }\\
\operation{Movemethodwithdelegate}(NodeFigure, [], BasicTransformVisitor, basicTransformTmpVC, [BasicTransformVisitor], Void, visit) \textbf{; }\\
\operation{Movemethodwithdelegate}(SVGImage, [], BasicTransformVisitor, basicTransformTmpVC, [BasicTransformVisitor], Void, visit) \textbf{; }\\
\operation{Movemethodwithdelegate}(SVGPath, [], BasicTransformVisitor, basicTransformTmpVC, [BasicTransformVisitor], Void, visit) \textbf{; }\\
\operation{Movemethodwithdelegate}(DependencyFigure, [], BasicTransformVisitor, basicTransformTmpVC, [BasicTransformVisitor], Void, visit) \textbf{; }\\
\operation{Movemethodwithdelegate}(LineConnectionFigure, [], BasicTransformVisitor, basicTransformTmpVC, [BasicTransformVisitor], Void, visit) \textbf{; }\\
\operation{Movemethodwithdelegate}(LabeledLineConnectionFigure, [], ContainsVisitor, containsTmpVC, [ContainsVisitor;], Boolean, visit) \textbf{; }\\
\operation{Movemethodwithdelegate}(AbstractCompositeFigure, [], ContainsVisitor, containsTmpVC, [ContainsVisitor], Boolean, visit) \textbf{; }\\
\operation{Movemethodwithdelegate}(GraphicalCompositeFigure, [], ContainsVisitor, containsTmpVC, [ContainsVisitor], Boolean, visit) \textbf{; }\\
\operation{Movemethodwithdelegate}(EllipseFigure, [], ContainsVisitor, containsTmpVC, [ContainsVisitor], Boolean, visit) \textbf{; }\\
\operation{Movemethodwithdelegate}(DiamondFigure, [], ContainsVisitor, containsTmpVC, [ContainsVisitor], Boolean, visit) \textbf{; }\\
\operation{Movemethodwithdelegate}(RectangleFigure, [], ContainsVisitor, containsTmpVC, [ContainsVisitor], Boolean, visit) \textbf{; }\\
\operation{Movemethodwithdelegate}(RoundRectangleFigure, [], ContainsVisitor, containsTmpVC, [ContainsVisitor], Boolean, visit) \textbf{; }\\
\operation{Movemethodwithdelegate}(TriangleFigure, [], ContainsVisitor, containsTmpVC, [ContainsVisitor], Boolean, visit) \textbf{; }\\
\operation{Movemethodwithdelegate}(TextFigure, [], ContainsVisitor, containsTmpVC, [ContainsVisitor], Boolean, visit) \textbf{; }\\
\operation{Movemethodwithdelegate}(BezierFigure, [], ContainsVisitor, containsTmpVC, [ContainsVisitor], Boolean, visit) \textbf{; }\\
\operation{Movemethodwithdelegate}(TextAreaFigure, [], ContainsVisitor, containsTmpVC, [ContainsVisitor], Boolean, visit) \textbf{; }\\
\operation{Movemethodwithdelegate}(NodeFigure, [], ContainsVisitor, containsTmpVC, [ContainsVisitor], Boolean, visit) \textbf{; }\\
\operation{Movemethodwithdelegate}(SVGImage, [], ContainsVisitor, containsTmpVC, [ContainsVisitor], Boolean, visit) \textbf{; }\\
\operation{Movemethodwithdelegate}(SVGPath, [], ContainsVisitor, containsTmpVC, [ContainsVisitor], Boolean, visit) \textbf{; }\\
\operation{Movemethodwithdelegate}(DependencyFigure, [], ContainsVisitor, containsTmpVC, [ContainsVisitor], Boolean, visit) \textbf{; }\\
\operation{Movemethodwithdelegate}(LineConnectionFigure, [], ContainsVisitor, containsTmpVC, [ContainsVisitor], Boolean, visit) \textbf{; }\\
\operation{Movemethodwithdelegate}(LabeledLineConnectionFigure, [], SetAttributeVisitor, setAttributeTmpVC, [SetAttributeVisitor;], Void, visit) \textbf{; }\\
\operation{Movemethodwithdelegate}(AbstractCompositeFigure, [], SetAttributeVisitor, setAttributeTmpVC, [SetAttributeVisitor], Void, visit) \textbf{; }\\
\operation{Movemethodwithdelegate}(GraphicalCompositeFigure, [], SetAttributeVisitor, setAttributeTmpVC, [SetAttributeVisitor], Void, visit) \textbf{; }\\
\operation{Movemethodwithdelegate}(EllipseFigure, [], SetAttributeVisitor, setAttributeTmpVC, [SetAttributeVisitor], Void, visit) \textbf{; }\\
\operation{Movemethodwithdelegate}(DiamondFigure, [], SetAttributeVisitor, setAttributeTmpVC, [SetAttributeVisitor], Void, visit) \textbf{; }\\
\operation{Movemethodwithdelegate}(RectangleFigure, [], SetAttributeVisitor, setAttributeTmpVC, [SetAttributeVisitor], Void, visit) \textbf{; }\\
\operation{Movemethodwithdelegate}(RoundRectangleFigure, [], SetAttributeVisitor, setAttributeTmpVC, [SetAttributeVisitor], Void, visit) \textbf{; }\\
\operation{Movemethodwithdelegate}(TriangleFigure, [], SetAttributeVisitor, setAttributeTmpVC, [SetAttributeVisitor], Void, visit) \textbf{; }\\
\operation{Movemethodwithdelegate}(TextFigure, [], SetAttributeVisitor, setAttributeTmpVC, [SetAttributeVisitor], Void, visit) \textbf{; }\\
\operation{Movemethodwithdelegate}(BezierFigure, [], SetAttributeVisitor, setAttributeTmpVC, [SetAttributeVisitor], Void, visit) \textbf{; }\\
\operation{Movemethodwithdelegate}(TextAreaFigure, [], SetAttributeVisitor, setAttributeTmpVC, [SetAttributeVisitor], Void, visit) \textbf{; }\\
\operation{Movemethodwithdelegate}(NodeFigure, [], SetAttributeVisitor, setAttributeTmpVC, [SetAttributeVisitor], Void, visit) \textbf{; }\\
\operation{Movemethodwithdelegate}(SVGImage, [], SetAttributeVisitor, setAttributeTmpVC, [SetAttributeVisitor], Void, visit) \textbf{; }\\
\operation{Movemethodwithdelegate}(SVGPath, [], SetAttributeVisitor, setAttributeTmpVC, [SetAttributeVisitor], Void, visit) \textbf{; }\\
\operation{Movemethodwithdelegate}(DependencyFigure, [], SetAttributeVisitor, setAttributeTmpVC, [SetAttributeVisitor], Void, visit) \textbf{; }\\
\operation{Movemethodwithdelegate}(LineConnectionFigure, [], SetAttributeVisitor, setAttributeTmpVC, [SetAttributeVisitor], Void, visit) \textbf{; }\\
\operation{Movemethodwithdelegate}(LabeledLineConnectionFigure, [], FindFigureInsideVisitor, findFigureInsideTmpVC, [FindFigureInsideVisitor], Figure, visit) \textbf{; }\\
\operation{Movemethodwithdelegate}(AbstractCompositeFigure, [], FindFigureInsideVisitor, findFigureInsideTmpVC, [FindFigureInsideVisitor], Figure, visit) \textbf{; }\\
\operation{Movemethodwithdelegate}(GraphicalCompositeFigure, [], FindFigureInsideVisitor, findFigureInsideTmpVC, [FindFigureInsideVisitor], Figure, visit) \textbf{; }\\
\operation{Movemethodwithdelegate}(EllipseFigure, [], FindFigureInsideVisitor, findFigureInsideTmpVC, [FindFigureInsideVisitor], Figure, visit) \textbf{; }\\
\operation{Movemethodwithdelegate}(DiamondFigure, [], FindFigureInsideVisitor, findFigureInsideTmpVC, [FindFigureInsideVisitor], Figure, visit) \textbf{; }\\
\operation{Movemethodwithdelegate}(RectangleFigure, [], FindFigureInsideVisitor, findFigureInsideTmpVC, [FindFigureInsideVisitor], Figure, visit) \textbf{; }\\
\operation{Movemethodwithdelegate}(RoundRectangleFigure, [], FindFigureInsideVisitor, findFigureInsideTmpVC, [FindFigureInsideVisitor], Figure, visit) \textbf{; }\\
\operation{Movemethodwithdelegate}(TriangleFigure, [], FindFigureInsideVisitor, findFigureInsideTmpVC, [FindFigureInsideVisitor], Figure, visit) \textbf{; }\\
\operation{Movemethodwithdelegate}(TextFigure, [], FindFigureInsideVisitor, findFigureInsideTmpVC, [FindFigureInsideVisitor], Figure, visit) \textbf{; }\\
\operation{Movemethodwithdelegate}(BezierFigure, [], FindFigureInsideVisitor, findFigureInsideTmpVC, [FindFigureInsideVisitor], Figure, visit) \textbf{; }\\
\operation{Movemethodwithdelegate}(TextAreaFigure, [], FindFigureInsideVisitor, findFigureInsideTmpVC, [FindFigureInsideVisitor], Figure, visit) \textbf{; }\\
\operation{Movemethodwithdelegate}(NodeFigure, [], FindFigureInsideVisitor, findFigureInsideTmpVC, [FindFigureInsideVisitor], Figure, visit) \textbf{; }\\
\operation{Movemethodwithdelegate}(SVGImage, [], FindFigureInsideVisitor, findFigureInsideTmpVC, [FindFigureInsideVisitor], Figure, visit) \textbf{; }\\
\operation{Movemethodwithdelegate}(SVGPath, [], FindFigureInsideVisitor, findFigureInsideTmpVC, [FindFigureInsideVisitor], Figure, visit) \textbf{; }\\
\operation{Movemethodwithdelegate}(DependencyFigure, [], FindFigureInsideVisitor, findFigureInsideTmpVC, [FindFigureInsideVisitor], Figure, visit) \textbf{; }\\
\operation{Movemethodwithdelegate}(LineConnectionFigure, [], FindFigureInsideVisitor, findFigureInsideTmpVC, [FindFigureInsideVisitor], Figure, visit) \textbf{; }\\
\operation{Movemethodwithdelegate}(LabeledLineConnectionFigure, [], AddNotifyVisitor, addNotifyTmpVC, [AddNotifyVisitor], Void, visit) \textbf{; }\\
\operation{Movemethodwithdelegate}(AbstractCompositeFigure, [], AddNotifyVisitor, addNotifyTmpVC, [AddNotifyVisitor], Void, visit) \textbf{; }\\
\operation{Movemethodwithdelegate}(GraphicalCompositeFigure, [], AddNotifyVisitor, addNotifyTmpVC, [AddNotifyVisitor], Void, visit) \textbf{; }\\
\operation{Movemethodwithdelegate}(EllipseFigure, [], AddNotifyVisitor, addNotifyTmpVC, [AddNotifyVisitor], Void, visit) \textbf{; }\\
\operation{Movemethodwithdelegate}(DiamondFigure, [], AddNotifyVisitor, addNotifyTmpVC, [AddNotifyVisitor], Void, visit) \textbf{; }\\
\operation{Movemethodwithdelegate}(RectangleFigure, [], AddNotifyVisitor, addNotifyTmpVC, [AddNotifyVisitor], Void, visit) \textbf{; }\\
\operation{Movemethodwithdelegate}(RoundRectangleFigure, [], AddNotifyVisitor, addNotifyTmpVC, [AddNotifyVisitor], Void, visit) \textbf{; }\\
\operation{Movemethodwithdelegate}(TriangleFigure, [], AddNotifyVisitor, addNotifyTmpVC, [AddNotifyVisitor], Void, visit) \textbf{; }\\
\operation{Movemethodwithdelegate}(TextFigure, [], AddNotifyVisitor, addNotifyTmpVC, [AddNotifyVisitor], Void, visit) \textbf{; }\\
\operation{Movemethodwithdelegate}(BezierFigure, [], AddNotifyVisitor, addNotifyTmpVC, [AddNotifyVisitor], Void, visit) \textbf{; }\\
\operation{Movemethodwithdelegate}(TextAreaFigure, [], AddNotifyVisitor, addNotifyTmpVC, [AddNotifyVisitor], Void, visit) \textbf{; }\\
\operation{Movemethodwithdelegate}(NodeFigure, [], AddNotifyVisitor, addNotifyTmpVC, [AddNotifyVisitor], Void, visit) \textbf{; }\\
\operation{Movemethodwithdelegate}(SVGImage, [], AddNotifyVisitor, addNotifyTmpVC, [AddNotifyVisitor], Void, visit) \textbf{; }\\
\operation{Movemethodwithdelegate}(SVGPath, [], AddNotifyVisitor, addNotifyTmpVC, [AddNotifyVisitor], Void, visit) \textbf{; }\\
\operation{Movemethodwithdelegate}(DependencyFigure, [], AddNotifyVisitor, addNotifyTmpVC, [AddNotifyVisitor], Void, visit) \textbf{; }\\
\operation{Movemethodwithdelegate}(LineConnectionFigure, [], AddNotifyVisitor, addNotifyTmpVC, [AddNotifyVisitor], Void, visit) \textbf{; }\\
\operation{Movemethodwithdelegate}(LabeledLineConnectionFigure, [], RemoveNotifyVisitor, removeNotifyTmpVC, [RemoveNotifyVisitor], Void, visit) \textbf{; }\\
\operation{Movemethodwithdelegate}(AbstractCompositeFigure, [], RemoveNotifyVisitor, removeNotifyTmpVC, [RemoveNotifyVisitor], Void, visit) \textbf{; }\\
\operation{Movemethodwithdelegate}(GraphicalCompositeFigure, [], RemoveNotifyVisitor, removeNotifyTmpVC, [RemoveNotifyVisitor], Void, visit) \textbf{; }\\
\operation{Movemethodwithdelegate}(EllipseFigure, [], RemoveNotifyVisitor, removeNotifyTmpVC, [RemoveNotifyVisitor], Void, visit) \textbf{; }\\
\operation{Movemethodwithdelegate}(DiamondFigure, [], RemoveNotifyVisitor, removeNotifyTmpVC, [RemoveNotifyVisitor], Void, visit) \textbf{; }\\
\operation{Movemethodwithdelegate}(RectangleFigure, [], RemoveNotifyVisitor, removeNotifyTmpVC, [RemoveNotifyVisitor], Void, visit) \textbf{; }\\
\operation{Movemethodwithdelegate}(RoundRectangleFigure, [], RemoveNotifyVisitor, removeNotifyTmpVC, [RemoveNotifyVisitor], Void, visit) \textbf{; }\\
\operation{Movemethodwithdelegate}(TriangleFigure, [], RemoveNotifyVisitor, removeNotifyTmpVC, [RemoveNotifyVisitor], Void, visit) \textbf{; }\\
\operation{Movemethodwithdelegate}(TextFigure, [], RemoveNotifyVisitor, removeNotifyTmpVC, [RemoveNotifyVisitor], Void, visit) \textbf{; }\\
\operation{Movemethodwithdelegate}(BezierFigure, [], RemoveNotifyVisitor, removeNotifyTmpVC, [RemoveNotifyVisitor], Void, visit) \textbf{; }\\
\operation{Movemethodwithdelegate}(TextAreaFigure, [], RemoveNotifyVisitor, removeNotifyTmpVC, [RemoveNotifyVisitor], Void, visit) \textbf{; }\\
\operation{Movemethodwithdelegate}(NodeFigure, [], RemoveNotifyVisitor, removeNotifyTmpVC, [RemoveNotifyVisitor], Void, visit) \textbf{; }\\
\operation{Movemethodwithdelegate}(SVGImage, [], RemoveNotifyVisitor, removeNotifyTmpVC, [RemoveNotifyVisitor], Void, visit) \textbf{; }\\
\operation{Movemethodwithdelegate}(SVGPath, [], RemoveNotifyVisitor, removeNotifyTmpVC, [RemoveNotifyVisitor], Void, visit) \textbf{; }\\
\operation{Movemethodwithdelegate}(DependencyFigure, [], RemoveNotifyVisitor, removeNotifyTmpVC, [RemoveNotifyVisitor], Void, visit) \textbf{; }\\
\operation{Movemethodwithdelegate}(LineConnectionFigure, [], RemoveNotifyVisitor, removeNotifyTmpVC, [RemoveNotifyVisitor], Void, visit) \textbf{; }\\
\operation{ExtractSuperClassWithoutPullup}([BasicTransformVisitor; ContainsVisitor; SetAttributeVisitor; FindFigureInsideVisitor; AddNotifyVisitor; RemoveNotifyVisitor], Visitor) \textbf{; }\\
\operation{PullupWithGenerics}(Visitor, BasicTransformVisitor, [], visit, Void) \textbf{; }\\
\operation{PullupWithGenerics}(Visitor, ContainsVisitor, [], visit, Boolean) \textbf{; }\\
\operation{PullupWithGenerics}(Visitor, SetAttributeVisitor, [], visit, Void) \textbf{; }\\
\operation{PullupWithGenerics}(Visitor, FindFigureInsideVisitor, [], visit, Figure) \textbf{; }\\
\operation{PullupWithGenerics}(Visitor, AddNotifyVisitor, [], visit, Void) \textbf{; }\\
\operation{PullupWithGenerics}(Visitor, RemoveNotifyVisitor, [], visit, Void) \textbf{; }\\
\operation{GeneraliseParameter}(AbstractFigure, [LabeledLineConnectionFigure; AbstractCompositeFigure; GraphicalCompositeFigure; EllipseFigure; DiamondFigure; RectangleFigure; RoundRectangleFigure; TriangleFigure; TextFigure; BezierFigure; TextAreaFigure; NodeFigure; SVGImage; SVGPath; DependencyFigure; LineConnectionFigure], basicTransformTmpVC, v, BasicTransformVisitor, Visitor) \textbf{; }\\
\operation{GeneraliseParameter}(AbstractFigure, [LabeledLineConnectionFigure; AbstractCompositeFigure; GraphicalCompositeFigure; EllipseFigure; DiamondFigure; RectangleFigure; RoundRectangleFigure; TriangleFigure; TextFigure; BezierFigure; TextAreaFigure; NodeFigure; SVGImage; SVGPath; DependencyFigure; LineConnectionFigure], containsTmpVC, v, ContainsVisitor, Visitor) \textbf{; }\\
\operation{GeneraliseParameter}(AbstractFigure, [LabeledLineConnectionFigure; AbstractCompositeFigure; GraphicalCompositeFigure; EllipseFigure; DiamondFigure; RectangleFigure; RoundRectangleFigure; TriangleFigure; TextFigure; BezierFigure; TextAreaFigure; NodeFigure; SVGImage; SVGPath; DependencyFigure; LineConnectionFigure], setAttributeTmpVC, v, SetAttributeVisitor, Visitor) \textbf{; }\\
\operation{GeneraliseParameter}(AbstractFigure, [LabeledLineConnectionFigure; AbstractCompositeFigure; GraphicalCompositeFigure; EllipseFigure; DiamondFigure; RectangleFigure; RoundRectangleFigure; TriangleFigure; TextFigure; BezierFigure; TextAreaFigure; NodeFigure; SVGImage; SVGPath; DependencyFigure; LineConnectionFigure], findFigureInsideTmpVC, v, FindFigureInsideVisitor, Visitor) \textbf{; }\\
\operation{GeneraliseParameter}(AbstractFigure, [LabeledLineConnectionFigure; AbstractCompositeFigure; GraphicalCompositeFigure; EllipseFigure; DiamondFigure; RectangleFigure; RoundRectangleFigure; TriangleFigure; TextFigure; BezierFigure; TextAreaFigure; NodeFigure; SVGImage; SVGPath; DependencyFigure; LineConnectionFigure], addNotifyTmpVC, v, AddNotifyVisitor, Visitor) \textbf{; }\\
\operation{GeneraliseParameter}(AbstractFigure, [LabeledLineConnectionFigure; AbstractCompositeFigure; GraphicalCompositeFigure; EllipseFigure; DiamondFigure; RectangleFigure; RoundRectangleFigure; TriangleFigure; TextFigure; BezierFigure; TextAreaFigure; NodeFigure; SVGImage; SVGPath; DependencyFigure; LineConnectionFigure], removeNotifyTmpVC, v, RemoveNotifyVisitor, Visitor) \textbf{; }\\
\operation{ReplaceMethodCodeDuplicatesInverters}(LabeledLineConnectionFigure, basicTransformTmpVC, [containsTmpVC; setAttributeTmpVC; findFigureInsideTmpVC; addNotifyTmpVC; removeNotifyTmpVC], Visitor, [Void; Boolean; Void; Figure; Void; Void]) \textbf{; }\\
\operation{ReplaceMethodCodeDuplicatesInverters}(AbstractCompositeFigure, basicTransformTmpVC, [containsTmpVC; setAttributeTmpVC; findFigureInsideTmpVC; addNotifyTmpVC; removeNotifyTmpVC], Visitor, [Void; Boolean; Void; Figure; Void; Void]) \textbf{; }\\
\operation{ReplaceMethodCodeDuplicatesInverters}(GraphicalCompositeFigure, basicTransformTmpVC, [containsTmpVC; setAttributeTmpVC; findFigureInsideTmpVC; addNotifyTmpVC; removeNotifyTmpVC], Visitor, [Void; Boolean; Void; Figure; Void; Void]) \textbf{; }\\
\operation{ReplaceMethodCodeDuplicatesInverters}(EllipseFigure, basicTransformTmpVC, [containsTmpVC; setAttributeTmpVC; findFigureInsideTmpVC; addNotifyTmpVC; removeNotifyTmpVC], Visitor, [Void; Boolean; Void; Figure; Void; Void]) \textbf{; }\\
\operation{ReplaceMethodCodeDuplicatesInverters}(DiamondFigure, basicTransformTmpVC, [containsTmpVC; setAttributeTmpVC; findFigureInsideTmpVC; addNotifyTmpVC; removeNotifyTmpVC], Visitor, [Void; Boolean; Void; Figure; Void; Void]) \textbf{; }\\
\operation{ReplaceMethodCodeDuplicatesInverters}(RectangleFigure, basicTransformTmpVC, [containsTmpVC; setAttributeTmpVC; findFigureInsideTmpVC; addNotifyTmpVC; removeNotifyTmpVC], Visitor, [Void; Boolean; Void; Figure; Void; Void]) \textbf{; }\\
\operation{ReplaceMethodCodeDuplicatesInverters}(RoundRectangleFigure, basicTransformTmpVC, [containsTmpVC; setAttributeTmpVC; findFigureInsideTmpVC; addNotifyTmpVC; removeNotifyTmpVC], Visitor, [Void; Boolean; Void; Figure; Void; Void]) \textbf{; }\\
\operation{ReplaceMethodCodeDuplicatesInverters}(TriangleFigure, basicTransformTmpVC, [containsTmpVC; setAttributeTmpVC; findFigureInsideTmpVC; addNotifyTmpVC; removeNotifyTmpVC], Visitor, [Void; Boolean; Void; Figure; Void; Void]) \textbf{; }\\
\operation{ReplaceMethodCodeDuplicatesInverters}(TextFigure, basicTransformTmpVC, [containsTmpVC; setAttributeTmpVC; findFigureInsideTmpVC; addNotifyTmpVC; removeNotifyTmpVC], Visitor, [Void; Boolean; Void; Figure; Void; Void]) \textbf{; }\\
\operation{ReplaceMethodCodeDuplicatesInverters}(BezierFigure, basicTransformTmpVC, [containsTmpVC; setAttributeTmpVC; findFigureInsideTmpVC; addNotifyTmpVC; removeNotifyTmpVC], Visitor, [Void; Boolean; Void; Figure; Void; Void]) \textbf{; }\\
\operation{ReplaceMethodCodeDuplicatesInverters}(TextAreaFigure, basicTransformTmpVC, [containsTmpVC; setAttributeTmpVC; findFigureInsideTmpVC; addNotifyTmpVC; removeNotifyTmpVC], Visitor, [Void; Boolean; Void; Figure; Void; Void]) \textbf{; }\\
\operation{ReplaceMethodCodeDuplicatesInverters}(NodeFigure, basicTransformTmpVC, [containsTmpVC; setAttributeTmpVC; findFigureInsideTmpVC; addNotifyTmpVC; removeNotifyTmpVC], Visitor, [Void; Boolean; Void; Figure; Void; Void]) \textbf{; }\\
\operation{ReplaceMethodCodeDuplicatesInverters}(SVGImage, basicTransformTmpVC, [containsTmpVC; setAttributeTmpVC; findFigureInsideTmpVC; addNotifyTmpVC; removeNotifyTmpVC], Visitor, [Void; Boolean; Void; Figure; Void; Void]) \textbf{; }\\
\operation{ReplaceMethodCodeDuplicatesInverters}(SVGPath, basicTransformTmpVC, [containsTmpVC; setAttributeTmpVC; findFigureInsideTmpVC; addNotifyTmpVC; removeNotifyTmpVC], Visitor, [Void; Boolean; Void; Figure; Void; Void]) \textbf{; }\\
\operation{ReplaceMethodCodeDuplicatesInverters}(DependencyFigure, basicTransformTmpVC, [containsTmpVC; setAttributeTmpVC; findFigureInsideTmpVC; addNotifyTmpVC; removeNotifyTmpVC], Visitor, [Void; Boolean; Void; Figure; Void; Void]) \textbf{; }\\
\operation{ReplaceMethodCodeDuplicatesInverters}(LineConnectionFigure, basicTransformTmpVC, [containsTmpVC; setAttributeTmpVC; findFigureInsideTmpVC; addNotifyTmpVC; removeNotifyTmpVC], Visitor, [Void; Boolean; Void; Figure; Void; Void]) \textbf{; }\\
\operation{PullUpImplementation}(LabeledLineConnectionFigure, [], containsTmpVC, AbstractFigure) \textbf{; }\\
\operation{SafeDeleteDelegatorOverriding}(AbstractCompositeFigure, containsTmpVC, AbstractFigure) \textbf{; }\\
\operation{SafeDeleteDelegatorOverriding}(GraphicalCompositeFigure, containsTmpVC, AbstractFigure) \textbf{; }\\
\operation{SafeDeleteDelegatorOverriding}(EllipseFigure, containsTmpVC, AbstractFigure) \textbf{; }\\
\operation{SafeDeleteDelegatorOverriding}(DiamondFigure, containsTmpVC, AbstractFigure) \textbf{; }\\
\operation{SafeDeleteDelegatorOverriding}(RectangleFigure, containsTmpVC, AbstractFigure) \textbf{; }\\
\operation{SafeDeleteDelegatorOverriding}(RoundRectangleFigure, containsTmpVC, AbstractFigure) \textbf{; }\\
\operation{SafeDeleteDelegatorOverriding}(TriangleFigure, containsTmpVC, AbstractFigure) \textbf{; }\\
\operation{SafeDeleteDelegatorOverriding}(TextFigure, containsTmpVC, AbstractFigure) \textbf{; }\\
\operation{SafeDeleteDelegatorOverriding}(BezierFigure, containsTmpVC, AbstractFigure) \textbf{; }\\
\operation{SafeDeleteDelegatorOverriding}(TextAreaFigure, containsTmpVC, AbstractFigure) \textbf{; }\\
\operation{SafeDeleteDelegatorOverriding}(NodeFigure, containsTmpVC, AbstractFigure) \textbf{; }\\
\operation{SafeDeleteDelegatorOverriding}(SVGImage, containsTmpVC, AbstractFigure) \textbf{; }\\
\operation{SafeDeleteDelegatorOverriding}(SVGPath, containsTmpVC, AbstractFigure) \textbf{; }\\
\operation{SafeDeleteDelegatorOverriding}(DependencyFigure, containsTmpVC, AbstractFigure) \textbf{; }\\
\operation{SafeDeleteDelegatorOverriding}(LineConnectionFigure, containsTmpVC, AbstractFigure) \textbf{; }\\
\operation{PullUpImplementation}(LabeledLineConnectionFigure, [], setAttributeTmpVC, AbstractFigure) \textbf{; }\\
\operation{SafeDeleteDelegatorOverriding}(AbstractCompositeFigure, setAttributeTmpVC, AbstractFigure) \textbf{; }\\
\operation{SafeDeleteDelegatorOverriding}(GraphicalCompositeFigure, setAttributeTmpVC, AbstractFigure) \textbf{; }\\
\operation{SafeDeleteDelegatorOverriding}(EllipseFigure, setAttributeTmpVC, AbstractFigure) \textbf{; }\\
\operation{SafeDeleteDelegatorOverriding}(DiamondFigure, setAttributeTmpVC, AbstractFigure) \textbf{; }\\
\operation{SafeDeleteDelegatorOverriding}(RectangleFigure, setAttributeTmpVC, AbstractFigure) \textbf{; }\\
\operation{SafeDeleteDelegatorOverriding}(RoundRectangleFigure, setAttributeTmpVC, AbstractFigure) \textbf{; }\\
\operation{SafeDeleteDelegatorOverriding}(TriangleFigure, setAttributeTmpVC, AbstractFigure) \textbf{; }\\
\operation{SafeDeleteDelegatorOverriding}(TextFigure, setAttributeTmpVC, AbstractFigure) \textbf{; }\\
\operation{SafeDeleteDelegatorOverriding}(BezierFigure, setAttributeTmpVC, AbstractFigure) \textbf{; }\\
\operation{SafeDeleteDelegatorOverriding}(TextAreaFigure, setAttributeTmpVC, AbstractFigure) \textbf{; }\\
\operation{SafeDeleteDelegatorOverriding}(NodeFigure, setAttributeTmpVC, AbstractFigure) \textbf{; }\\
\operation{SafeDeleteDelegatorOverriding}(SVGImage, setAttributeTmpVC, AbstractFigure) \textbf{; }\\
\operation{SafeDeleteDelegatorOverriding}(SVGPath, setAttributeTmpVC, AbstractFigure) \textbf{; }\\
\operation{SafeDeleteDelegatorOverriding}(DependencyFigure, setAttributeTmpVC, AbstractFigure) \textbf{; }\\
\operation{SafeDeleteDelegatorOverriding}(LineConnectionFigure, setAttributeTmpVC, AbstractFigure) \textbf{; }\\
\operation{PullUpImplementation}(LabeledLineConnectionFigure, [], findFigureInsideTmpVC, AbstractFigure) \textbf{; }\\
\operation{SafeDeleteDelegatorOverriding}(AbstractCompositeFigure, findFigureInsideTmpVC, AbstractFigure) \textbf{; }\\
\operation{SafeDeleteDelegatorOverriding}(GraphicalCompositeFigure, findFigureInsideTmpVC, AbstractFigure) \textbf{; }\\
\operation{SafeDeleteDelegatorOverriding}(EllipseFigure, findFigureInsideTmpVC, AbstractFigure) \textbf{; }\\
\operation{SafeDeleteDelegatorOverriding}(DiamondFigure, findFigureInsideTmpVC, AbstractFigure) \textbf{; }\\
\operation{SafeDeleteDelegatorOverriding}(RectangleFigure, findFigureInsideTmpVC, AbstractFigure) \textbf{; }\\
\operation{SafeDeleteDelegatorOverriding}(RoundRectangleFigure, findFigureInsideTmpVC, AbstractFigure) \textbf{; }\\
\operation{SafeDeleteDelegatorOverriding}(TriangleFigure, findFigureInsideTmpVC, AbstractFigure) \textbf{; }\\
\operation{SafeDeleteDelegatorOverriding}(TextFigure, findFigureInsideTmpVC, AbstractFigure) \textbf{; }\\
\operation{SafeDeleteDelegatorOverriding}(BezierFigure, findFigureInsideTmpVC, AbstractFigure) \textbf{; }\\
\operation{SafeDeleteDelegatorOverriding}(TextAreaFigure, findFigureInsideTmpVC, AbstractFigure) \textbf{; }\\
\operation{SafeDeleteDelegatorOverriding}(NodeFigure, findFigureInsideTmpVC, AbstractFigure) \textbf{; }\\
\operation{SafeDeleteDelegatorOverriding}(SVGImage, findFigureInsideTmpVC, AbstractFigure) \textbf{; }\\
\operation{SafeDeleteDelegatorOverriding}(SVGPath, findFigureInsideTmpVC, AbstractFigure) \textbf{; }\\
\operation{SafeDeleteDelegatorOverriding}(DependencyFigure, findFigureInsideTmpVC, AbstractFigure) \textbf{; }\\
\operation{SafeDeleteDelegatorOverriding}(LineConnectionFigure, findFigureInsideTmpVC, AbstractFigure) \textbf{; }\\
\operation{PullUpImplementation}(LabeledLineConnectionFigure, [], addNotifyTmpVC, AbstractFigure) \textbf{; }\\
\operation{SafeDeleteDelegatorOverriding}(AbstractCompositeFigure, addNotifyTmpVC, AbstractFigure) \textbf{; }\\
\operation{SafeDeleteDelegatorOverriding}(GraphicalCompositeFigure, addNotifyTmpVC, AbstractFigure) \textbf{; }\\
\operation{SafeDeleteDelegatorOverriding}(EllipseFigure, addNotifyTmpVC, AbstractFigure) \textbf{; }\\
\operation{SafeDeleteDelegatorOverriding}(DiamondFigure, addNotifyTmpVC, AbstractFigure) \textbf{; }\\
\operation{SafeDeleteDelegatorOverriding}(RectangleFigure, addNotifyTmpVC, AbstractFigure) \textbf{; }\\
\operation{SafeDeleteDelegatorOverriding}(RoundRectangleFigure, addNotifyTmpVC, AbstractFigure) \textbf{; }\\
\operation{SafeDeleteDelegatorOverriding}(TriangleFigure, addNotifyTmpVC, AbstractFigure) \textbf{; }\\
\operation{SafeDeleteDelegatorOverriding}(TextFigure, addNotifyTmpVC, AbstractFigure) \textbf{; }\\
\operation{SafeDeleteDelegatorOverriding}(BezierFigure, addNotifyTmpVC, AbstractFigure) \textbf{; }\\
\operation{SafeDeleteDelegatorOverriding}(TextAreaFigure, addNotifyTmpVC, AbstractFigure) \textbf{; }\\
\operation{SafeDeleteDelegatorOverriding}(NodeFigure, addNotifyTmpVC, AbstractFigure) \textbf{; }\\
\operation{SafeDeleteDelegatorOverriding}(SVGImage, addNotifyTmpVC, AbstractFigure) \textbf{; }\\
\operation{SafeDeleteDelegatorOverriding}(SVGPath, addNotifyTmpVC, AbstractFigure) \textbf{; }\\
\operation{SafeDeleteDelegatorOverriding}(DependencyFigure, addNotifyTmpVC, AbstractFigure) \textbf{; }\\
\operation{SafeDeleteDelegatorOverriding}(LineConnectionFigure, addNotifyTmpVC, AbstractFigure) \textbf{; }\\
\operation{PullUpImplementation}(LabeledLineConnectionFigure, [], removeNotifyTmpVC, AbstractFigure) \textbf{; }\\
\operation{SafeDeleteDelegatorOverriding}(AbstractCompositeFigure, removeNotifyTmpVC, AbstractFigure) \textbf{; }\\
\operation{SafeDeleteDelegatorOverriding}(GraphicalCompositeFigure, removeNotifyTmpVC, AbstractFigure) \textbf{; }\\
\operation{SafeDeleteDelegatorOverriding}(EllipseFigure, removeNotifyTmpVC, AbstractFigure) \textbf{; }\\
\operation{SafeDeleteDelegatorOverriding}(DiamondFigure, removeNotifyTmpVC, AbstractFigure) \textbf{; }\\
\operation{SafeDeleteDelegatorOverriding}(RectangleFigure, removeNotifyTmpVC, AbstractFigure) \textbf{; }\\
\operation{SafeDeleteDelegatorOverriding}(RoundRectangleFigure, removeNotifyTmpVC, AbstractFigure) \textbf{; }\\
\operation{SafeDeleteDelegatorOverriding}(TriangleFigure, removeNotifyTmpVC, AbstractFigure) \textbf{; }\\
\operation{SafeDeleteDelegatorOverriding}(TextFigure, removeNotifyTmpVC, AbstractFigure) \textbf{; }\\
\operation{SafeDeleteDelegatorOverriding}(BezierFigure, removeNotifyTmpVC, AbstractFigure) \textbf{; }\\
\operation{SafeDeleteDelegatorOverriding}(TextAreaFigure, removeNotifyTmpVC, AbstractFigure) \textbf{; }\\
\operation{SafeDeleteDelegatorOverriding}(NodeFigure, removeNotifyTmpVC, AbstractFigure) \textbf{; }\\
\operation{SafeDeleteDelegatorOverriding}(SVGImage, removeNotifyTmpVC, AbstractFigure) \textbf{; }\\
\operation{SafeDeleteDelegatorOverriding}(SVGPath, removeNotifyTmpVC, AbstractFigure) \textbf{; }\\
\operation{SafeDeleteDelegatorOverriding}(DependencyFigure, removeNotifyTmpVC, AbstractFigure) \textbf{; }\\
\operation{SafeDeleteDelegatorOverriding}(LineConnectionFigure, removeNotifyTmpVC, AbstractFigure) \textbf{; }\\
\operation{InlineAndDelete} (AbstractFigure, containsTmpVC, [Visitor], contains, [setAttribute; findFigureInside; addNotify; removeNotify], [LabeledLineConnectionFigure; AbstractCompositeFigure; GraphicalCompositeFigure; EllipseFigure; DiamondFigure; RectangleFigure; RoundRectangleFigure; TriangleFigure; TextFigure; BezierFigure; TextAreaFigure; NodeFigure; SVGImage; SVGPath; DependencyFigure; LineConnectionFigure]) \textbf{; }\\
\operation{InlineAndDelete} (AbstractFigure, setAttributeTmpVC, [Visitor], setAttribute, [contains; findFigureInside; addNotify; removeNotify], [LabeledLineConnectionFigure; AbstractCompositeFigure; GraphicalCompositeFigure; EllipseFigure; DiamondFigure; RectangleFigure; RoundRectangleFigure; TriangleFigure; TextFigure; BezierFigure; TextAreaFigure; NodeFigure; SVGImage; SVGPath; DependencyFigure; LineConnectionFigure]) \textbf{; }\\
\operation{InlineAndDelete} (AbstractFigure, findFigureInsideTmpVC, [Visitor], findFigureInside, [contains; setAttribute; addNotify; removeNotify], [LabeledLineConnectionFigure; AbstractCompositeFigure; GraphicalCompositeFigure; EllipseFigure; DiamondFigure; RectangleFigure; RoundRectangleFigure; TriangleFigure; TextFigure; BezierFigure; TextAreaFigure; NodeFigure; SVGImage; SVGPath; DependencyFigure; LineConnectionFigure]) \textbf{; }\\
\operation{InlineAndDelete} (AbstractFigure, addNotifyTmpVC, [Visitor], addNotify, [contains; setAttribute; findFigureInside; removeNotify], [LabeledLineConnectionFigure; AbstractCompositeFigure; GraphicalCompositeFigure; EllipseFigure; DiamondFigure; RectangleFigure; RoundRectangleFigure; TriangleFigure; TextFigure; BezierFigure; TextAreaFigure; NodeFigure; SVGImage; SVGPath; DependencyFigure; LineConnectionFigure]) \textbf{; }\\
\operation{InlineAndDelete} (AbstractFigure, removeNotifyTmpVC, [Visitor], removeNotify, [contains; setAttribute; findFigureInside; addNotify], [LabeledLineConnectionFigure; AbstractCompositeFigure; GraphicalCompositeFigure; EllipseFigure; DiamondFigure; RectangleFigure; RoundRectangleFigure; TriangleFigure; TextFigure; BezierFigure; TextAreaFigure; NodeFigure; SVGImage; SVGPath; DependencyFigure; LineConnectionFigure]) \textbf{; }\\
\operation{RenameinhierarchyNoOverloading}(AbstractFigure, [LabeledLineConnectionFigure; AbstractCompositeFigure; GraphicalCompositeFigure; EllipseFigure; DiamondFigure; RectangleFigure; RoundRectangleFigure; TriangleFigure; TextFigure; BezierFigure; TextAreaFigure; NodeFigure; SVGImage; SVGPath; DependencyFigure; LineConnectionFigure], basicTransformTmpVC, [Visitor], accept) \textbf{; }\\
\operation{DuplicateMethodInHierarchyGen} (AbstractFigure, [EllipseFigure; DiamondFigure; RectangleFigure; RoundRectangleFigure; TriangleFigure; TextFigure; BezierFigure; TextAreaFigure; NodeFigure; SVGImage; SVGPath; DependencyFigure; LineConnectionFigure; LabeledLineConnectionFigure; AbstractCompositeFigure; GraphicalCompositeFigure], accept, [Void; Boolean; Void; Figure; Void; Void], [visit], [basicTransform; contains; setAttribute; findFigureInside; addNotify; removeNotify], accept\_BasicTransformVisitor\_addspecializedMethod\_tmp, [Visitor]) \textbf{; }\\
\operation{SpecialiseParameter} (AbstractFigure, [EllipseFigure; DiamondFigure; RectangleFigure; RoundRectangleFigure; TriangleFigure; TextFigure; BezierFigure; TextAreaFigure; NodeFigure; SVGImage; SVGPath; DependencyFigure; LineConnectionFigure; LabeledLineConnectionFigure; AbstractCompositeFigure; GraphicalCompositeFigure], accept\_BasicTransformVisitor\_addspecializedMethod\_tmp, Visitor, [BasicTransformVisitor; ContainsVisitor; SetAttributeVisitor; FindFigureInsideVisitor; AddNotifyVisitor; RemoveNotifyVisitor], BasicTransformVisitor) \textbf{; }\\
\operation{RenameDelegatorWithOverloading}(AbstractFigure, [EllipseFigure; DiamondFigure; RectangleFigure; RoundRectangleFigure; TriangleFigure; TextFigure; BezierFigure; TextAreaFigure; NodeFigure; SVGImage; SVGPath; DependencyFigure; LineConnectionFigure; LabeledLineConnectionFigure; AbstractCompositeFigure; GraphicalCompositeFigure], accept\_BasicTransformVisitor\_addspecializedMethod\_tmp, BasicTransformVisitor, v, Visitor, accept) \textbf{; }\\
\operation{DuplicateMethodInHierarchyGen} (AbstractFigure, [EllipseFigure; DiamondFigure; RectangleFigure; RoundRectangleFigure; TriangleFigure; TextFigure; BezierFigure; TextAreaFigure; NodeFigure; SVGImage; SVGPath; DependencyFigure; LineConnectionFigure; LabeledLineConnectionFigure; AbstractCompositeFigure; GraphicalCompositeFigure], accept, [Void; Boolean; Void; Figure; Void; Void], [visit], [basicTransform; contains; setAttribute; findFigureInside; addNotify; removeNotify], accept\_ContainsVisitor\_addspecializedMethod\_tmp, [Visitor]) \textbf{; }\\
\operation{SpecialiseParameter} (AbstractFigure, [EllipseFigure; DiamondFigure; RectangleFigure; RoundRectangleFigure; TriangleFigure; TextFigure; BezierFigure; TextAreaFigure; NodeFigure; SVGImage; SVGPath; DependencyFigure; LineConnectionFigure; LabeledLineConnectionFigure; AbstractCompositeFigure; GraphicalCompositeFigure], accept\_ContainsVisitor\_addspecializedMethod\_tmp, Visitor, [BasicTransformVisitor; ContainsVisitor; SetAttributeVisitor; FindFigureInsideVisitor; AddNotifyVisitor; RemoveNotifyVisitor], ContainsVisitor) \textbf{; }\\
\operation{RenameDelegatorWithOverloading}(AbstractFigure, [EllipseFigure; DiamondFigure; RectangleFigure; RoundRectangleFigure; TriangleFigure; TextFigure; BezierFigure; TextAreaFigure; NodeFigure; SVGImage; SVGPath; DependencyFigure; LineConnectionFigure; LabeledLineConnectionFigure; AbstractCompositeFigure; GraphicalCompositeFigure], accept\_ContainsVisitor\_addspecializedMethod\_tmp, ContainsVisitor, v, Visitor, accept) \textbf{; }\\
\operation{DuplicateMethodInHierarchyGen} (AbstractFigure, [EllipseFigure; DiamondFigure; RectangleFigure; RoundRectangleFigure; TriangleFigure; TextFigure; BezierFigure; TextAreaFigure; NodeFigure; SVGImage; SVGPath; DependencyFigure; LineConnectionFigure; LabeledLineConnectionFigure; AbstractCompositeFigure; GraphicalCompositeFigure], accept, [Void; Boolean; Void; Figure; Void; Void], [visit], [basicTransform; contains; setAttribute; findFigureInside; addNotify; removeNotify], accept\_SetAttributeVisitor\_addspecializedMethod\_tmp, [Visitor]) \textbf{; }\\
\operation{SpecialiseParameter} (AbstractFigure, [EllipseFigure; DiamondFigure; RectangleFigure; RoundRectangleFigure; TriangleFigure; TextFigure; BezierFigure; TextAreaFigure; NodeFigure; SVGImage; SVGPath; DependencyFigure; LineConnectionFigure; LabeledLineConnectionFigure; AbstractCompositeFigure; GraphicalCompositeFigure], accept\_SetAttributeVisitor\_addspecializedMethod\_tmp, Visitor, [BasicTransformVisitor; ContainsVisitor; SetAttributeVisitor; FindFigureInsideVisitor; AddNotifyVisitor; RemoveNotifyVisitor], SetAttributeVisitor) \textbf{; }\\
\operation{RenameDelegatorWithOverloading}(AbstractFigure, [EllipseFigure; DiamondFigure; RectangleFigure; RoundRectangleFigure; TriangleFigure; TextFigure; BezierFigure; TextAreaFigure; NodeFigure; SVGImage; SVGPath; DependencyFigure; LineConnectionFigure; LabeledLineConnectionFigure; AbstractCompositeFigure; GraphicalCompositeFigure], accept\_SetAttributeVisitor\_addspecializedMethod\_tmp, SetAttributeVisitor, v, Visitor, accept) \textbf{; }\\
\operation{DuplicateMethodInHierarchyGen} (AbstractFigure, [EllipseFigure; DiamondFigure; RectangleFigure; RoundRectangleFigure; TriangleFigure; TextFigure; BezierFigure; TextAreaFigure; NodeFigure; SVGImage; SVGPath; DependencyFigure; LineConnectionFigure; LabeledLineConnectionFigure; AbstractCompositeFigure; GraphicalCompositeFigure], accept, [Void; Boolean; Void; Figure; Void; Void], [visit], [basicTransform; contains; setAttribute; findFigureInside; addNotify; removeNotify], accept\_FindFigureInsideVisitor\_addspecializedMethod\_tmp, [Visitor]) \textbf{; }\\
\operation{SpecialiseParameter} (AbstractFigure, [EllipseFigure; DiamondFigure; RectangleFigure; RoundRectangleFigure; TriangleFigure; TextFigure; BezierFigure; TextAreaFigure; NodeFigure; SVGImage; SVGPath; DependencyFigure; LineConnectionFigure; LabeledLineConnectionFigure; AbstractCompositeFigure; GraphicalCompositeFigure], accept\_FindFigureInsideVisitor\_addspecializedMethod\_tmp, Visitor, [BasicTransformVisitor; ContainsVisitor; SetAttributeVisitor; FindFigureInsideVisitor; AddNotifyVisitor; RemoveNotifyVisitor], FindFigureInsideVisitor) \textbf{; }\\
\operation{RenameDelegatorWithOverloading}(AbstractFigure, [EllipseFigure; DiamondFigure; RectangleFigure; RoundRectangleFigure; TriangleFigure; TextFigure; BezierFigure; TextAreaFigure; NodeFigure; SVGImage; SVGPath; DependencyFigure; LineConnectionFigure; LabeledLineConnectionFigure; AbstractCompositeFigure; GraphicalCompositeFigure], accept\_FindFigureInsideVisitor\_addspecializedMethod\_tmp, FindFigureInsideVisitor, v, Visitor, accept) \textbf{; }\\
\operation{DuplicateMethodInHierarchyGen} (AbstractFigure, [EllipseFigure; DiamondFigure; RectangleFigure; RoundRectangleFigure; TriangleFigure; TextFigure; BezierFigure; TextAreaFigure; NodeFigure; SVGImage; SVGPath; DependencyFigure; LineConnectionFigure; LabeledLineConnectionFigure; AbstractCompositeFigure; GraphicalCompositeFigure], accept, [Void; Boolean; Void; Figure; Void; Void], [visit], [basicTransform; contains; setAttribute; findFigureInside; addNotify; removeNotify], accept\_AddNotifyVisitor\_addspecializedMethod\_tmp, [Visitor]) \textbf{; }\\
\operation{SpecialiseParameter} (AbstractFigure, [EllipseFigure; DiamondFigure; RectangleFigure; RoundRectangleFigure; TriangleFigure; TextFigure; BezierFigure; TextAreaFigure; NodeFigure; SVGImage; SVGPath; DependencyFigure; LineConnectionFigure; LabeledLineConnectionFigure; AbstractCompositeFigure; GraphicalCompositeFigure], accept\_AddNotifyVisitor\_addspecializedMethod\_tmp, Visitor, [BasicTransformVisitor; ContainsVisitor; SetAttributeVisitor; FindFigureInsideVisitor; AddNotifyVisitor; RemoveNotifyVisitor], AddNotifyVisitor) \textbf{; }\\
\operation{RenameDelegatorWithOverloading}(AbstractFigure, [EllipseFigure; DiamondFigure; RectangleFigure; RoundRectangleFigure; TriangleFigure; TextFigure; BezierFigure; TextAreaFigure; NodeFigure; SVGImage; SVGPath; DependencyFigure; LineConnectionFigure; LabeledLineConnectionFigure; AbstractCompositeFigure; GraphicalCompositeFigure], accept\_AddNotifyVisitor\_addspecializedMethod\_tmp, AddNotifyVisitor, v, Visitor, accept) \textbf{; }\\
\operation{DuplicateMethodInHierarchyGen} (AbstractFigure, [EllipseFigure; DiamondFigure; RectangleFigure; RoundRectangleFigure; TriangleFigure; TextFigure; BezierFigure; TextAreaFigure; NodeFigure; SVGImage; SVGPath; DependencyFigure; LineConnectionFigure; LabeledLineConnectionFigure; AbstractCompositeFigure; GraphicalCompositeFigure], accept, [Void; Boolean; Void; Figure; Void; Void], [visit], [basicTransform; contains; setAttribute; findFigureInside; addNotify; removeNotify], accept\_RemoveNotifyVisitor\_addspecializedMethod\_tmp, [Visitor]) \textbf{; }\\
\operation{SpecialiseParameter} (AbstractFigure, [EllipseFigure; DiamondFigure; RectangleFigure; RoundRectangleFigure; TriangleFigure; TextFigure; BezierFigure; TextAreaFigure; NodeFigure; SVGImage; SVGPath; DependencyFigure; LineConnectionFigure; LabeledLineConnectionFigure; AbstractCompositeFigure; GraphicalCompositeFigure], accept\_RemoveNotifyVisitor\_addspecializedMethod\_tmp, Visitor, [BasicTransformVisitor; ContainsVisitor; SetAttributeVisitor; FindFigureInsideVisitor; AddNotifyVisitor; RemoveNotifyVisitor], RemoveNotifyVisitor) \textbf{; }\\
\operation{RenameDelegatorWithOverloading}(AbstractFigure, [EllipseFigure; DiamondFigure; RectangleFigure; RoundRectangleFigure; TriangleFigure; TextFigure; BezierFigure; TextAreaFigure; NodeFigure; SVGImage; SVGPath; DependencyFigure; LineConnectionFigure; LabeledLineConnectionFigure; AbstractCompositeFigure; GraphicalCompositeFigure], accept\_RemoveNotifyVisitor\_addspecializedMethod\_tmp, RemoveNotifyVisitor, v, Visitor, accept) \textbf{; }\\
\operation{DeleteMethod}InHierarchy(AbstractFigure, [EllipseFigure; DiamondFigure; RectangleFigure; RoundRectangleFigure; TriangleFigure; TextFigure; BezierFigure; TextAreaFigure; NodeFigure; SVGImage; SVGPath; DependencyFigure; LineConnectionFigure; LabeledLineConnectionFigure; AbstractCompositeFigure; GraphicalCompositeFigure], accept, [visit], Visitor) \textbf{; }\\
\operation{PushDownAll}(Visitor, [BasicTransformVisitor; ContainsVisitor; SetAttributeVisitor; FindFigureInsideVisitor; AddNotifyVisitor; RemoveNotifyVisitor], visit, [EllipseFigure]) \textbf{; }\\
\operation{PushDownAll}(Visitor, [BasicTransformVisitor; ContainsVisitor; SetAttributeVisitor; FindFigureInsideVisitor; AddNotifyVisitor; RemoveNotifyVisitor], visit, [DiamondFigure]) \textbf{; }\\
\operation{PushDownAll}(Visitor, [BasicTransformVisitor; ContainsVisitor; SetAttributeVisitor; FindFigureInsideVisitor; AddNotifyVisitor; RemoveNotifyVisitor], visit, [RectangleFigure]) \textbf{; }\\
\operation{PushDownAll}(Visitor, [BasicTransformVisitor; ContainsVisitor; SetAttributeVisitor; FindFigureInsideVisitor; AddNotifyVisitor; RemoveNotifyVisitor], visit, [RoundRectangleFigure]) \textbf{; }\\
\operation{PushDownAll}(Visitor, [BasicTransformVisitor; ContainsVisitor; SetAttributeVisitor; FindFigureInsideVisitor; AddNotifyVisitor; RemoveNotifyVisitor], visit, [TriangleFigure]) \textbf{; }\\
\operation{PushDownAll}(Visitor, [BasicTransformVisitor; ContainsVisitor; SetAttributeVisitor; FindFigureInsideVisitor; AddNotifyVisitor; RemoveNotifyVisitor], visit, [TextFigure]) \textbf{; }\\
\operation{PushDownAll}(Visitor, [BasicTransformVisitor; ContainsVisitor; SetAttributeVisitor; FindFigureInsideVisitor; AddNotifyVisitor; RemoveNotifyVisitor], visit, [BezierFigure]) \textbf{; }\\
\operation{PushDownAll}(Visitor, [BasicTransformVisitor; ContainsVisitor; SetAttributeVisitor; FindFigureInsideVisitor; AddNotifyVisitor; RemoveNotifyVisitor], visit, [TextAreaFigure]) \textbf{; }\\
\operation{PushDownAll}(Visitor, [BasicTransformVisitor; ContainsVisitor; SetAttributeVisitor; FindFigureInsideVisitor; AddNotifyVisitor; RemoveNotifyVisitor], visit, [NodeFigure]) \textbf{; }\\
\operation{PushDownAll}(Visitor, [BasicTransformVisitor; ContainsVisitor; SetAttributeVisitor; FindFigureInsideVisitor; AddNotifyVisitor; RemoveNotifyVisitor], visit, [SVGImage]) \textbf{; }\\
\operation{PushDownAll}(Visitor, [BasicTransformVisitor; ContainsVisitor; SetAttributeVisitor; FindFigureInsideVisitor; AddNotifyVisitor; RemoveNotifyVisitor], visit, [SVGPath]) \textbf{; }\\
\operation{PushDownAll}(Visitor, [BasicTransformVisitor; ContainsVisitor; SetAttributeVisitor; FindFigureInsideVisitor; AddNotifyVisitor; RemoveNotifyVisitor], visit, [DependencyFigure]) \textbf{; }\\
\operation{PushDownAll}(Visitor, [BasicTransformVisitor; ContainsVisitor; SetAttributeVisitor; FindFigureInsideVisitor; AddNotifyVisitor; RemoveNotifyVisitor], visit, [LineConnectionFigure]) \textbf{; }\\
\operation{PushDownAll}(Visitor, [BasicTransformVisitor; ContainsVisitor; SetAttributeVisitor; FindFigureInsideVisitor; AddNotifyVisitor; RemoveNotifyVisitor], visit, [LabeledLineConnectionFigure]) \textbf{; }\\
\operation{PushDownAll}(Visitor, [BasicTransformVisitor; ContainsVisitor; SetAttributeVisitor; FindFigureInsideVisitor; AddNotifyVisitor; RemoveNotifyVisitor], visit, [AbstractCompositeFigure]) \textbf{; }\\
\operation{PushDownAll}(Visitor, [BasicTransformVisitor; ContainsVisitor; SetAttributeVisitor; FindFigureInsideVisitor; AddNotifyVisitor; RemoveNotifyVisitor], visit, [GraphicalCompositeFigure]) \textbf{; }\\
\operation{InlineMethod}(EllipseFigure, visit, BasicTransformVisitor, accept, [AffineTransform tx]) \textbf{; }\\
\operation{InlineMethod}(DiamondFigure, visit, BasicTransformVisitor, accept, [AffineTransform tx]) \textbf{; }\\
\operation{InlineMethod}(RectangleFigure, visit, BasicTransformVisitor, accept, [AffineTransform tx]) \textbf{; }\\
\operation{InlineMethod}(RoundRectangleFigure, visit, BasicTransformVisitor, accept, [AffineTransform tx]) \textbf{; }\\
\operation{InlineMethod}(TriangleFigure, visit, BasicTransformVisitor, accept, [AffineTransform tx]) \textbf{; }\\
\operation{InlineMethod}(TextFigure, visit, BasicTransformVisitor, accept, [AffineTransform tx]) \textbf{; }\\
\operation{InlineMethod}(BezierFigure, visit, BasicTransformVisitor, accept, [AffineTransform tx]) \textbf{; }\\
\operation{InlineMethod}(TextAreaFigure, visit, BasicTransformVisitor, accept, [AffineTransform tx]) \textbf{; }\\
\operation{InlineMethod}(NodeFigure, visit, BasicTransformVisitor, accept, [AffineTransform tx]) \textbf{; }\\
\operation{InlineMethod}(SVGImage, visit, BasicTransformVisitor, accept, [AffineTransform tx]) \textbf{; }\\
\operation{InlineMethod}(SVGPath, visit, BasicTransformVisitor, accept, [AffineTransform tx]) \textbf{; }\\
\operation{InlineMethod}(DependencyFigure, visit, BasicTransformVisitor, accept, [AffineTransform tx]) \textbf{; }\\
\operation{InlineMethod}(LineConnectionFigure, visit, BasicTransformVisitor, accept, [AffineTransform tx]) \textbf{; }\\
\operation{InlineMethod}(LabeledLineConnectionFigure, visit, BasicTransformVisitor, accept, [AffineTransform tx]) \textbf{; }\\
\operation{InlineMethod}(AbstractCompositeFigure, visit, BasicTransformVisitor, accept, [AffineTransform tx]) \textbf{; }\\
\operation{InlineMethod}(GraphicalCompositeFigure, visit, BasicTransformVisitor, accept, [AffineTransform tx]) \textbf{; }\\
\operation{InlineMethod}(EllipseFigure, visit, ContainsVisitor, accept, [Point2D.Double p]) \textbf{; }\\
\operation{InlineMethod}(DiamondFigure, visit, ContainsVisitor, accept, [Point2D.Double p]) \textbf{; }\\
\operation{InlineMethod}(RectangleFigure, visit, ContainsVisitor, accept, [Point2D.Double p]) \textbf{; }\\
\operation{InlineMethod}(RoundRectangleFigure, visit, ContainsVisitor, accept, [Point2D.Double p]) \textbf{; }\\
\operation{InlineMethod}(TriangleFigure, visit, ContainsVisitor, accept, [Point2D.Double p]) \textbf{; }\\
\operation{InlineMethod}(TextFigure, visit, ContainsVisitor, accept, [Point2D.Double p]) \textbf{; }\\
\operation{InlineMethod}(BezierFigure, visit, ContainsVisitor, accept, [Point2D.Double p]) \textbf{; }\\
\operation{InlineMethod}(TextAreaFigure, visit, ContainsVisitor, accept, [Point2D.Double p]) \textbf{; }\\
\operation{InlineMethod}(NodeFigure, visit, ContainsVisitor, accept, [Point2D.Double p]) \textbf{; }\\
\operation{InlineMethod}(SVGImage, visit, ContainsVisitor, accept, [Point2D.Double p]) \textbf{; }\\
\operation{InlineMethod}(SVGPath, visit, ContainsVisitor, accept, [Point2D.Double p]) \textbf{; }\\
\operation{InlineMethod}(DependencyFigure, visit, ContainsVisitor, accept, [Point2D.Double p]) \textbf{; }\\
\operation{InlineMethod}(LineConnectionFigure, visit, ContainsVisitor, accept, [Point2D.Double p]) \textbf{; }\\
\operation{InlineMethod}(LabeledLineConnectionFigure, visit, ContainsVisitor, accept, [Point2D.Double p]) \textbf{; }\\
\operation{InlineMethod}(AbstractCompositeFigure, visit, ContainsVisitor, accept, [Point2D.Double p]) \textbf{; }\\
\operation{InlineMethod}(GraphicalCompositeFigure, visit, ContainsVisitor, accept, [Point2D.Double p]) \textbf{; }\\
\operation{InlineMethod}(EllipseFigure, visit, SetAttributeVisitor, accept, [AttributeKey key; Object value]) \textbf{; }\\
\operation{InlineMethod}(DiamondFigure, visit, SetAttributeVisitor, accept, [AttributeKey key; Object value]) \textbf{; }\\
\operation{InlineMethod}(RectangleFigure, visit, SetAttributeVisitor, accept, [AttributeKey key; Object value]) \textbf{; }\\
\operation{InlineMethod}(RoundRectangleFigure, visit, SetAttributeVisitor, accept, [AttributeKey key; Object value]) \textbf{; }\\
\operation{InlineMethod}(TriangleFigure, visit, SetAttributeVisitor, accept, [AttributeKey key; Object value]) \textbf{; }\\
\operation{InlineMethod}(TextFigure, visit, SetAttributeVisitor, accept, [AttributeKey key; Object value]) \textbf{; }\\
\operation{InlineMethod}(BezierFigure, visit, SetAttributeVisitor, accept, [AttributeKey key; Object value]) \textbf{; }\\
\operation{InlineMethod}(TextAreaFigure, visit, SetAttributeVisitor, accept, [AttributeKey key; Object value]) \textbf{; }\\
\operation{InlineMethod}(NodeFigure, visit, SetAttributeVisitor, accept, [AttributeKey key; Object value]) \textbf{; }\\
\operation{InlineMethod}(SVGImage, visit, SetAttributeVisitor, accept, [AttributeKey key; Object value]) \textbf{; }\\
\operation{InlineMethod}(SVGPath, visit, SetAttributeVisitor, accept, [AttributeKey key; Object value]) \textbf{; }\\
\operation{InlineMethod}(DependencyFigure, visit, SetAttributeVisitor, accept, [AttributeKey key; Object value]) \textbf{; }\\
\operation{InlineMethod}(LineConnectionFigure, visit, SetAttributeVisitor, accept, [AttributeKey key; Object value]) \textbf{; }\\
\operation{InlineMethod}(LabeledLineConnectionFigure, visit, SetAttributeVisitor, accept, [AttributeKey key; Object value]) \textbf{; }\\
\operation{InlineMethod}(AbstractCompositeFigure, visit, SetAttributeVisitor, accept, [AttributeKey key; Object value]) \textbf{; }\\
\operation{InlineMethod}(GraphicalCompositeFigure, visit, SetAttributeVisitor, accept, [AttributeKey key; Object value]) \textbf{; }\\
\operation{InlineMethod}(EllipseFigure, visit, FindFigureInsideVisitor, accept, [Point2D.Double p]) \textbf{; }\\
\operation{InlineMethod}(DiamondFigure, visit, FindFigureInsideVisitor, accept, [Point2D.Double p]) \textbf{; }\\
\operation{InlineMethod}(RectangleFigure, visit, FindFigureInsideVisitor, accept, [Point2D.Double p]) \textbf{; }\\
\operation{InlineMethod}(RoundRectangleFigure, visit, FindFigureInsideVisitor, accept, [Point2D.Double p]) \textbf{; }\\
\operation{InlineMethod}(TriangleFigure, visit, FindFigureInsideVisitor, accept, [Point2D.Double p]) \textbf{; }\\
\operation{InlineMethod}(TextFigure, visit, FindFigureInsideVisitor, accept, [Point2D.Double p]) \textbf{; }\\
\operation{InlineMethod}(BezierFigure, visit, FindFigureInsideVisitor, accept, [Point2D.Double p]) \textbf{; }\\
\operation{InlineMethod}(TextAreaFigure, visit, FindFigureInsideVisitor, accept, [Point2D.Double p]) \textbf{; }\\
\operation{InlineMethod}(NodeFigure, visit, FindFigureInsideVisitor, accept, [Point2D.Double p]) \textbf{; }\\
\operation{InlineMethod}(SVGImage, visit, FindFigureInsideVisitor, accept, [Point2D.Double p]) \textbf{; }\\
\operation{InlineMethod}(SVGPath, visit, FindFigureInsideVisitor, accept, [Point2D.Double p]) \textbf{; }\\
\operation{InlineMethod}(DependencyFigure, visit, FindFigureInsideVisitor, accept, [Point2D.Double p]) \textbf{; }\\
\operation{InlineMethod}(LineConnectionFigure, visit, FindFigureInsideVisitor, accept, [Point2D.Double p]) \textbf{; }\\
\operation{InlineMethod}(LabeledLineConnectionFigure, visit, FindFigureInsideVisitor, accept, [Point2D.Double p]) \textbf{; }\\
\operation{InlineMethod}(AbstractCompositeFigure, visit, FindFigureInsideVisitor, accept, [Point2D.Double p]) \textbf{; }\\
\operation{InlineMethod}(GraphicalCompositeFigure, visit, FindFigureInsideVisitor, accept, [Point2D.Double p]) \textbf{; }\\
\operation{InlineMethod}(EllipseFigure, visit, AddNotifyVisitor, accept, [Drawing d]) \textbf{; }\\
\operation{InlineMethod}(DiamondFigure, visit, AddNotifyVisitor, accept, [Drawing d]) \textbf{; }\\
\operation{InlineMethod}(RectangleFigure, visit, AddNotifyVisitor, accept, [Drawing d]) \textbf{; }\\
\operation{InlineMethod}(RoundRectangleFigure, visit, AddNotifyVisitor, accept, [Drawing d]) \textbf{; }\\
\operation{InlineMethod}(TriangleFigure, visit, AddNotifyVisitor, accept, [Drawing d]) \textbf{; }\\
\operation{InlineMethod}(TextFigure, visit, AddNotifyVisitor, accept, [Drawing d]) \textbf{; }\\
\operation{InlineMethod}(BezierFigure, visit, AddNotifyVisitor, accept, [Drawing d]) \textbf{; }\\
\operation{InlineMethod}(TextAreaFigure, visit, AddNotifyVisitor, accept, [Drawing d]) \textbf{; }\\
\operation{InlineMethod}(NodeFigure, visit, AddNotifyVisitor, accept, [Drawing d]) \textbf{; }\\
\operation{InlineMethod}(SVGImage, visit, AddNotifyVisitor, accept, [Drawing d]) \textbf{; }\\
\operation{InlineMethod}(SVGPath, visit, AddNotifyVisitor, accept, [Drawing d]) \textbf{; }\\
\operation{InlineMethod}(DependencyFigure, visit, AddNotifyVisitor, accept, [Drawing d]) \textbf{; }\\
\operation{InlineMethod}(LineConnectionFigure, visit, AddNotifyVisitor, accept, [Drawing d]) \textbf{; }\\
\operation{InlineMethod}(LabeledLineConnectionFigure, visit, AddNotifyVisitor, accept, [Drawing d]) \textbf{; }\\
\operation{InlineMethod}(AbstractCompositeFigure, visit, AddNotifyVisitor, accept, [Drawing d]) \textbf{; }\\
\operation{InlineMethod}(GraphicalCompositeFigure, visit, AddNotifyVisitor, accept, [Drawing d]) \textbf{; }\\
\operation{InlineMethod}(EllipseFigure, visit, RemoveNotifyVisitor, accept, [Drawing d]) \textbf{; }\\
\operation{InlineMethod}(DiamondFigure, visit, RemoveNotifyVisitor, accept, [Drawing d]) \textbf{; }\\
\operation{InlineMethod}(RectangleFigure, visit, RemoveNotifyVisitor, accept, [Drawing d]) \textbf{; }\\
\operation{InlineMethod}(RoundRectangleFigure, visit, RemoveNotifyVisitor, accept, [Drawing d]) \textbf{; }\\
\operation{InlineMethod}(TriangleFigure, visit, RemoveNotifyVisitor, accept, [Drawing d]) \textbf{; }\\
\operation{InlineMethod}(TextFigure, visit, RemoveNotifyVisitor, accept, [Drawing d]) \textbf{; }\\
\operation{InlineMethod}(BezierFigure, visit, RemoveNotifyVisitor, accept, [Drawing d]) \textbf{; }\\
\operation{InlineMethod}(TextAreaFigure, visit, RemoveNotifyVisitor, accept, [Drawing d]) \textbf{; }\\
\operation{InlineMethod}(NodeFigure, visit, RemoveNotifyVisitor, accept, [Drawing d]) \textbf{; }\\
\operation{InlineMethod}(SVGImage, visit, RemoveNotifyVisitor, accept, [Drawing d]) \textbf{; }\\
\operation{InlineMethod}(SVGPath, visit, RemoveNotifyVisitor, accept, [Drawing d]) \textbf{; }\\
\operation{InlineMethod}(DependencyFigure, visit, RemoveNotifyVisitor, accept, [Drawing d]) \textbf{; }\\
\operation{InlineMethod}(LineConnectionFigure, visit, RemoveNotifyVisitor, accept, [Drawing d]) \textbf{; }\\
\operation{InlineMethod}(LabeledLineConnectionFigure, visit, RemoveNotifyVisitor, accept, [Drawing d]) \textbf{; }\\
\operation{InlineMethod}(AbstractCompositeFigure, visit, RemoveNotifyVisitor, accept, [Drawing d]) \textbf{; }\\
\operation{InlineMethod}(GraphicalCompositeFigure, visit, RemoveNotifyVisitor, accept, [Drawing d]) \textbf{; }\\
\operation{RenameOverloadedMethodInHierarchy}(AbstractFigure, [EllipseFigure; DiamondFigure; RectangleFigure;  
RoundRectangleFigure; TriangleFigure; TextFigure; BezierFigure; TextAreaFigure; NodeFigure; SVGImage; SVGPath; 
DependencyFigure; LineConnectionFigure; LabeledLineConnectionFigure; AbstractCompositeFigure; GraphicalCompositeFigure], accept, [BasicTransformVisitor], basicTransformTmpVC, [BasicTransformVisitor]) \textbf{; }\\
\operation{RenameOverloadedMethodInHierarchy}(AbstractFigure, [EllipseFigure; DiamondFigure; RectangleFigure; RoundRectangleFigure; TriangleFigure; TextFigure; BezierFigure; TextAreaFigure; NodeFigure; SVGImage; SVGPath; DependencyFigure; LineConnectionFigure; LabeledLineConnectionFigure; AbstractCompositeFigure; GraphicalCompositeFigure], accept, [ContainsVisitor], containsTmpVC, [ContainsVisitor]) \textbf{; }\\
\operation{RenameOverloadedMethodInHierarchy}(AbstractFigure, [EllipseFigure; DiamondFigure; RectangleFigure; RoundRectangleFigure; TriangleFigure; TextFigure; BezierFigure; TextAreaFigure; NodeFigure; SVGImage; SVGPath; DependencyFigure; LineConnectionFigure; LabeledLineConnectionFigure; AbstractCompositeFigure; GraphicalCompositeFigure], accept, [SetAttributeVisitor], setAttributeTmpVC, [SetAttributeVisitor]) \textbf{; }\\
\operation{RenameOverloadedMethodInHierarchy}(AbstractFigure, [EllipseFigure; DiamondFigure; RectangleFigure; RoundRectangleFigure; TriangleFigure; TextFigure; BezierFigure; TextAreaFigure; NodeFigure; SVGImage; SVGPath; DependencyFigure; LineConnectionFigure; LabeledLineConnectionFigure; AbstractCompositeFigure; GraphicalCompositeFigure], accept, [FindFigureInsideVisitor], findFigureInsideTmpVC, [FindFigureInsideVisitor]) \textbf{; }\\
\operation{RenameOverloadedMethodInHierarchy}(AbstractFigure, [EllipseFigure; DiamondFigure; RectangleFigure; RoundRectangleFigure; TriangleFigure; TextFigure; BezierFigure; TextAreaFigure; NodeFigure; SVGImage; SVGPath; DependencyFigure; LineConnectionFigure; LabeledLineConnectionFigure; AbstractCompositeFigure; GraphicalCompositeFigure], accept, [AddNotifyVisitor], addNotifyTmpVC, [AddNotifyVisitor]) \textbf{; }\\
\operation{RenameOverloadedMethodInHierarchy}(AbstractFigure, [EllipseFigure; DiamondFigure; RectangleFigure; RoundRectangleFigure; TriangleFigure; TextFigure; BezierFigure; TextAreaFigure; NodeFigure; SVGImage; SVGPath; DependencyFigure; LineConnectionFigure; LabeledLineConnectionFigure; AbstractCompositeFigure; GraphicalCompositeFigure], accept, [RemoveNotifyVisitor], removeNotifyTmpVC, [RemoveNotifyVisitor]) \textbf{; }\\
\operation{Replace-method-duplication}(AbstractFigure, [EllipseFigure; DiamondFigure; RectangleFigure; RoundRectangleFigure; TriangleFigure; TextFigure; BezierFigure; TextAreaFigure; NodeFigure; SVGImage; SVGPath; DependencyFigure; LineConnectionFigure; LabeledLineConnectionFigure; AbstractCompositeFigure; GraphicalCompositeFigure], basicTransform, basicTransformTmpVC, [AffineTransform tx]) \textbf{; }\\
\operation{Replace-method-duplication}(AbstractFigure, [EllipseFigure; DiamondFigure; RectangleFigure; RoundRectangleFigure; TriangleFigure; TextFigure; BezierFigure; TextAreaFigure; NodeFigure; SVGImage; SVGPath; DependencyFigure; LineConnectionFigure; LabeledLineConnectionFigure; AbstractCompositeFigure; GraphicalCompositeFigure], contains, containsTmpVC, [Point2D.Double p]) \textbf{; }\\
\operation{Replace-method-duplication}(AbstractFigure, [EllipseFigure; DiamondFigure; RectangleFigure; RoundRectangleFigure; TriangleFigure; TextFigure; BezierFigure; TextAreaFigure; NodeFigure; SVGImage; SVGPath; DependencyFigure; LineConnectionFigure; LabeledLineConnectionFigure; AbstractCompositeFigure; GraphicalCompositeFigure], setAttribute, setAttributeTmpVC, [AttributeKey key; Object value]) \textbf{; }\\
\operation{Replace-method-duplication}(AbstractFigure, [EllipseFigure; DiamondFigure; RectangleFigure; RoundRectangleFigure; TriangleFigure; TextFigure; BezierFigure; TextAreaFigure; NodeFigure; SVGImage; SVGPath; DependencyFigure; LineConnectionFigure; LabeledLineConnectionFigure; AbstractCompositeFigure; GraphicalCompositeFigure], findFigureInside, findFigureInsideTmpVC, [Point2D.Double p]) \textbf{; }\\
\operation{Replace-method-duplication}(AbstractFigure, [EllipseFigure; DiamondFigure; RectangleFigure; RoundRectangleFigure; TriangleFigure; TextFigure; BezierFigure; TextAreaFigure; NodeFigure; SVGImage; SVGPath; DependencyFigure; LineConnectionFigure; LabeledLineConnectionFigure; AbstractCompositeFigure; GraphicalCompositeFigure], addNotify, addNotifyTmpVC, [Drawing d]) \textbf{; }\\
\operation{Replace-method-duplication}(AbstractFigure, [EllipseFigure; DiamondFigure; RectangleFigure; RoundRectangleFigure; TriangleFigure; TextFigure; BezierFigure; TextAreaFigure; NodeFigure; SVGImage; SVGPath; DependencyFigure; LineConnectionFigure; LabeledLineConnectionFigure; AbstractCompositeFigure; GraphicalCompositeFigure], removeNotify, removeNotifyTmpVC, [Drawing d]) \textbf{; }\\
\operation{PushDownImplementation}(AbstractFigure, [], [EllipseFigure; DiamondFigure; RectangleFigure; RoundRectangleFigure; TriangleFigure; TextFigure; BezierFigure; TextAreaFigure; NodeFigure; SVGImage; SVGPath; DependencyFigure; LineConnectionFigure; LabeledLineConnectionFigure; AbstractCompositeFigure; GraphicalCompositeFigure], basicTransform,  [AffineTransform tx]) \textbf{; }\\
\operation{PushDownImplementation}(AbstractFigure, [], [EllipseFigure; DiamondFigure; RectangleFigure; RoundRectangleFigure; TriangleFigure; TextFigure; BezierFigure; TextAreaFigure; NodeFigure; SVGImage; SVGPath; DependencyFigure; LineConnectionFigure; LabeledLineConnectionFigure; AbstractCompositeFigure; GraphicalCompositeFigure], contains,  [Point2D.Double p]) \textbf{; }\\
\operation{PushDownImplementation}(AbstractFigure, [], [EllipseFigure; DiamondFigure; RectangleFigure; RoundRectangleFigure; TriangleFigure; TextFigure; BezierFigure; TextAreaFigure; NodeFigure; SVGImage; SVGPath; DependencyFigure; LineConnectionFigure; LabeledLineConnectionFigure; AbstractCompositeFigure; GraphicalCompositeFigure], setAttribute,  [AttributeKey key; Object value]) \textbf{; }\\
\operation{PushDownImplementation}(AbstractFigure, [], [EllipseFigure; DiamondFigure; RectangleFigure; RoundRectangleFigure; TriangleFigure; TextFigure; BezierFigure; TextAreaFigure; NodeFigure; SVGImage; SVGPath; DependencyFigure; LineConnectionFigure; LabeledLineConnectionFigure; AbstractCompositeFigure; GraphicalCompositeFigure], findFigureInside,  [Point2D.Double p]) \textbf{; }\\
\operation{PushDownImplementation}(AbstractFigure, [], [EllipseFigure; DiamondFigure; RectangleFigure; RoundRectangleFigure; TriangleFigure; TextFigure; BezierFigure; TextAreaFigure; NodeFigure; SVGImage; SVGPath; DependencyFigure; LineConnectionFigure; LabeledLineConnectionFigure; AbstractCompositeFigure; GraphicalCompositeFigure], addNotify,  [Drawing d]) \textbf{; }\\
\operation{PushDownImplementation}(AbstractFigure, [], [EllipseFigure; DiamondFigure; RectangleFigure; RoundRectangleFigure; TriangleFigure; TextFigure; BezierFigure; TextAreaFigure; NodeFigure; SVGImage; SVGPath; DependencyFigure; LineConnectionFigure; LabeledLineConnectionFigure; AbstractCompositeFigure; GraphicalCompositeFigure], removeNotify,  [Drawing d]) \textbf{; }\\
\operation{PushDownAll}(AbstractFigure, [EllipseFigure; DiamondFigure; RectangleFigure; RoundRectangleFigure; TriangleFigure; TextFigure; BezierFigure; TextAreaFigure; NodeFigure; SVGImage; SVGPath; DependencyFigure; LineConnectionFigure; LabeledLineConnectionFigure; AbstractCompositeFigure; GraphicalCompositeFigure], basicTransformTmpVC, [BasicTransformVisitor]) \textbf{; }\\
\operation{PushDownAll}(AbstractFigure, [EllipseFigure; DiamondFigure; RectangleFigure; RoundRectangleFigure; TriangleFigure; TextFigure; BezierFigure; TextAreaFigure; NodeFigure; SVGImage; SVGPath; DependencyFigure; LineConnectionFigure; LabeledLineConnectionFigure; AbstractCompositeFigure; GraphicalCompositeFigure], containsTmpVC, [ContainsVisitor]) \textbf{; }\\
\operation{PushDownAll}(AbstractFigure, [EllipseFigure; DiamondFigure; RectangleFigure; RoundRectangleFigure; TriangleFigure; TextFigure; BezierFigure; TextAreaFigure; NodeFigure; SVGImage; SVGPath; DependencyFigure; LineConnectionFigure; LabeledLineConnectionFigure; AbstractCompositeFigure; GraphicalCompositeFigure], setAttributeTmpVC, [SetAttributeVisitor]) \textbf{; }\\
\operation{PushDownAll}(AbstractFigure, [EllipseFigure; DiamondFigure; RectangleFigure; RoundRectangleFigure; TriangleFigure; TextFigure; BezierFigure; TextAreaFigure; NodeFigure; SVGImage; SVGPath; DependencyFigure; LineConnectionFigure; LabeledLineConnectionFigure; AbstractCompositeFigure; GraphicalCompositeFigure], findFigureInsideTmpVC, [FindFigureInsideVisitor]) \textbf{; }\\
\operation{PushDownAll}(AbstractFigure, [EllipseFigure; DiamondFigure; RectangleFigure; RoundRectangleFigure; TriangleFigure; TextFigure; BezierFigure; TextAreaFigure; NodeFigure; SVGImage; SVGPath; DependencyFigure; LineConnectionFigure; LabeledLineConnectionFigure; AbstractCompositeFigure; GraphicalCompositeFigure], addNotifyTmpVC, [AddNotifyVisitor]) \textbf{; }\\
\operation{PushDownAll}(AbstractFigure, [EllipseFigure; DiamondFigure; RectangleFigure; RoundRectangleFigure; TriangleFigure; TextFigure; BezierFigure; TextAreaFigure; NodeFigure; SVGImage; SVGPath; DependencyFigure; LineConnectionFigure; LabeledLineConnectionFigure; AbstractCompositeFigure; GraphicalCompositeFigure], removeNotifyTmpVC, [RemoveNotifyVisitor]) \textbf{; }\\
\operation{InlineAndDelete} (EllipseFigure, basicTransformTmpVC, [BasicTransformVisitor], basicTransform, [contains; setAttribute; findFigureInside; addNotify; removeNotify], [Visitor; AbstractFigure; DiamondFigure; RectangleFigure; RoundRectangleFigure; TriangleFigure; TextFigure; BezierFigure; TextAreaFigure; NodeFigure; SVGImage; SVGPath; DependencyFigure; LineConnectionFigure; LabeledLineConnectionFigure; AbstractCompositeFigure; GraphicalCompositeFigure; BasicTransformVisitor; ContainsVisitor; SetAttributeVisitor; FindFigureInsideVisitor; AddNotifyVisitor; RemoveNotifyVisitor]) \textbf{; }\\
\operation{InlineAndDelete} (DiamondFigure, basicTransformTmpVC, [BasicTransformVisitor], basicTransform, [contains; setAttribute; findFigureInside; addNotify; removeNotify], [Visitor; AbstractFigure; EllipseFigure; RectangleFigure; RoundRectangleFigure; TriangleFigure; TextFigure; BezierFigure; TextAreaFigure; NodeFigure; SVGImage; SVGPath; DependencyFigure; LineConnectionFigure; LabeledLineConnectionFigure; AbstractCompositeFigure; GraphicalCompositeFigure; BasicTransformVisitor; ContainsVisitor; SetAttributeVisitor; FindFigureInsideVisitor; AddNotifyVisitor; RemoveNotifyVisitor]) \textbf{; }\\
\operation{InlineAndDelete} (RectangleFigure, basicTransformTmpVC, [BasicTransformVisitor], basicTransform, [contains; setAttribute; findFigureInside; addNotify; removeNotify], [Visitor; AbstractFigure; EllipseFigure; DiamondFigure; RoundRectangleFigure; TriangleFigure; TextFigure; BezierFigure; TextAreaFigure; NodeFigure; SVGImage; SVGPath; DependencyFigure; LineConnectionFigure; LabeledLineConnectionFigure; AbstractCompositeFigure; GraphicalCompositeFigure; BasicTransformVisitor; ContainsVisitor; SetAttributeVisitor; FindFigureInsideVisitor; AddNotifyVisitor; RemoveNotifyVisitor]) \textbf{; }\\
\operation{InlineAndDelete} (RoundRectangleFigure, basicTransformTmpVC, [BasicTransformVisitor], basicTransform, [contains; setAttribute; findFigureInside; addNotify; removeNotify], [Visitor; AbstractFigure; EllipseFigure; DiamondFigure; RectangleFigure; TriangleFigure; TextFigure; BezierFigure; TextAreaFigure; NodeFigure; SVGImage; SVGPath; DependencyFigure; LineConnectionFigure; LabeledLineConnectionFigure; AbstractCompositeFigure; GraphicalCompositeFigure; BasicTransformVisitor; ContainsVisitor; SetAttributeVisitor; FindFigureInsideVisitor; AddNotifyVisitor; RemoveNotifyVisitor]) \textbf{; }\\
\operation{InlineAndDelete} (TriangleFigure, basicTransformTmpVC, [BasicTransformVisitor], basicTransform, [contains; setAttribute; findFigureInside; addNotify; removeNotify], [Visitor; AbstractFigure; EllipseFigure; DiamondFigure; RectangleFigure; RoundRectangleFigure; TextFigure; BezierFigure; TextAreaFigure; NodeFigure; SVGImage; SVGPath; DependencyFigure; LineConnectionFigure; LabeledLineConnectionFigure; AbstractCompositeFigure; GraphicalCompositeFigure; BasicTransformVisitor; ContainsVisitor; SetAttributeVisitor; FindFigureInsideVisitor; AddNotifyVisitor; RemoveNotifyVisitor]) \textbf{; }\\
\operation{InlineAndDelete} (TextFigure, basicTransformTmpVC, [BasicTransformVisitor], basicTransform, [contains; setAttribute; findFigureInside; addNotify; removeNotify], [Visitor; AbstractFigure; EllipseFigure; DiamondFigure; RectangleFigure; RoundRectangleFigure; TriangleFigure; BezierFigure; TextAreaFigure; NodeFigure; SVGImage; SVGPath; DependencyFigure; LineConnectionFigure; LabeledLineConnectionFigure; AbstractCompositeFigure; GraphicalCompositeFigure; BasicTransformVisitor; ContainsVisitor; SetAttributeVisitor; FindFigureInsideVisitor; AddNotifyVisitor; RemoveNotifyVisitor]) \textbf{; }\\
\operation{InlineAndDelete} (BezierFigure, basicTransformTmpVC, [BasicTransformVisitor], basicTransform, [contains; setAttribute; findFigureInside; addNotify; removeNotify], [Visitor; AbstractFigure; EllipseFigure; DiamondFigure; RectangleFigure; RoundRectangleFigure; TriangleFigure; TextFigure; TextAreaFigure; NodeFigure; SVGImage; SVGPath; DependencyFigure; LineConnectionFigure; LabeledLineConnectionFigure; AbstractCompositeFigure; GraphicalCompositeFigure; BasicTransformVisitor; ContainsVisitor; SetAttributeVisitor; FindFigureInsideVisitor; AddNotifyVisitor; RemoveNotifyVisitor]) \textbf{; }\\
\operation{InlineAndDelete} (TextAreaFigure, basicTransformTmpVC, [BasicTransformVisitor], basicTransform, [contains; setAttribute; findFigureInside; addNotify; removeNotify], [Visitor; AbstractFigure; EllipseFigure; DiamondFigure; RectangleFigure; RoundRectangleFigure; TriangleFigure; TextFigure; BezierFigure; NodeFigure; SVGImage; SVGPath; DependencyFigure; LineConnectionFigure; LabeledLineConnectionFigure; AbstractCompositeFigure; GraphicalCompositeFigure; BasicTransformVisitor; ContainsVisitor; SetAttributeVisitor; FindFigureInsideVisitor; AddNotifyVisitor; RemoveNotifyVisitor]) \textbf{; }\\
\operation{InlineAndDelete} (NodeFigure, basicTransformTmpVC, [BasicTransformVisitor], basicTransform, [contains; setAttribute; findFigureInside; addNotify; removeNotify], [Visitor; AbstractFigure; EllipseFigure; DiamondFigure; RectangleFigure; RoundRectangleFigure; TriangleFigure; TextFigure; BezierFigure; TextAreaFigure; SVGImage; SVGPath; DependencyFigure; LineConnectionFigure; LabeledLineConnectionFigure; AbstractCompositeFigure; GraphicalCompositeFigure; BasicTransformVisitor; ContainsVisitor; SetAttributeVisitor; FindFigureInsideVisitor; AddNotifyVisitor; RemoveNotifyVisitor]) \textbf{; }\\
\operation{InlineAndDelete} (SVGImage, basicTransformTmpVC, [BasicTransformVisitor], basicTransform, [contains; setAttribute; findFigureInside; addNotify; removeNotify], [Visitor; AbstractFigure; EllipseFigure; DiamondFigure; RectangleFigure; RoundRectangleFigure; TriangleFigure; TextFigure; BezierFigure; TextAreaFigure; NodeFigure; SVGPath; DependencyFigure; LineConnectionFigure; LabeledLineConnectionFigure; AbstractCompositeFigure; GraphicalCompositeFigure; BasicTransformVisitor; ContainsVisitor; SetAttributeVisitor; FindFigureInsideVisitor; AddNotifyVisitor; RemoveNotifyVisitor]) \textbf{; }\\
\operation{InlineAndDelete} (SVGPath, basicTransformTmpVC, [BasicTransformVisitor], basicTransform, [contains; setAttribute; findFigureInside; addNotify; removeNotify], [Visitor; AbstractFigure; EllipseFigure; DiamondFigure; RectangleFigure; RoundRectangleFigure; TriangleFigure; TextFigure; BezierFigure; TextAreaFigure; NodeFigure; SVGImage; DependencyFigure; LineConnectionFigure; LabeledLineConnectionFigure; AbstractCompositeFigure; GraphicalCompositeFigure; BasicTransformVisitor; ContainsVisitor; SetAttributeVisitor; FindFigureInsideVisitor; AddNotifyVisitor; RemoveNotifyVisitor]) \textbf{; }\\
\operation{InlineAndDelete} (DependencyFigure, basicTransformTmpVC, [BasicTransformVisitor], basicTransform, [contains; setAttribute; findFigureInside; addNotify; removeNotify], [Visitor; AbstractFigure; EllipseFigure; DiamondFigure; RectangleFigure; RoundRectangleFigure; TriangleFigure; TextFigure; BezierFigure; TextAreaFigure; NodeFigure; SVGImage; SVGPath; LineConnectionFigure; LabeledLineConnectionFigure; AbstractCompositeFigure; GraphicalCompositeFigure; BasicTransformVisitor; ContainsVisitor; SetAttributeVisitor; FindFigureInsideVisitor; AddNotifyVisitor; RemoveNotifyVisitor]) \textbf{; }\\
\operation{InlineAndDelete} (LineConnectionFigure, basicTransformTmpVC, [BasicTransformVisitor], basicTransform, [contains; setAttribute; findFigureInside; addNotify; removeNotify], [Visitor; AbstractFigure; EllipseFigure; DiamondFigure; RectangleFigure; RoundRectangleFigure; TriangleFigure; TextFigure; BezierFigure; TextAreaFigure; NodeFigure; SVGImage; SVGPath; DependencyFigure; LabeledLineConnectionFigure; AbstractCompositeFigure; GraphicalCompositeFigure; BasicTransformVisitor; ContainsVisitor; SetAttributeVisitor; FindFigureInsideVisitor; AddNotifyVisitor; RemoveNotifyVisitor]) \textbf{; }\\
\operation{InlineAndDelete} (LabeledLineConnectionFigure, basicTransformTmpVC, [BasicTransformVisitor], basicTransform, [contains; setAttribute; findFigureInside; addNotify; removeNotify], [Visitor; AbstractFigure; EllipseFigure; DiamondFigure; RectangleFigure; RoundRectangleFigure; TriangleFigure; TextFigure; BezierFigure; TextAreaFigure; NodeFigure; SVGImage; SVGPath; DependencyFigure; LineConnectionFigure; AbstractCompositeFigure; GraphicalCompositeFigure; BasicTransformVisitor; ContainsVisitor; SetAttributeVisitor; FindFigureInsideVisitor; AddNotifyVisitor; RemoveNotifyVisitor]) \textbf{; }\\
\operation{InlineAndDelete} (AbstractCompositeFigure, basicTransformTmpVC, [BasicTransformVisitor], basicTransform, [contains; setAttribute; findFigureInside; addNotify; removeNotify], [Visitor; AbstractFigure; EllipseFigure; DiamondFigure; RectangleFigure; RoundRectangleFigure; TriangleFigure; TextFigure; BezierFigure; TextAreaFigure; NodeFigure; SVGImage; SVGPath; DependencyFigure; LineConnectionFigure; LabeledLineConnectionFigure; GraphicalCompositeFigure; BasicTransformVisitor; ContainsVisitor; SetAttributeVisitor; FindFigureInsideVisitor; AddNotifyVisitor; RemoveNotifyVisitor]) \textbf{; }\\
\operation{InlineAndDelete} (GraphicalCompositeFigure, basicTransformTmpVC, [BasicTransformVisitor], basicTransform, [contains; setAttribute; findFigureInside; addNotify; removeNotify], [Visitor; AbstractFigure; EllipseFigure; DiamondFigure; RectangleFigure; RoundRectangleFigure; TriangleFigure; TextFigure; BezierFigure; TextAreaFigure; NodeFigure; SVGImage; SVGPath; DependencyFigure; LineConnectionFigure; LabeledLineConnectionFigure; AbstractCompositeFigure; BasicTransformVisitor; ContainsVisitor; SetAttributeVisitor; FindFigureInsideVisitor; AddNotifyVisitor; RemoveNotifyVisitor]) \textbf{; }\\
\operation{InlineAndDelete} (EllipseFigure, containsTmpVC, [ContainsVisitor], contains, [basicTransform; setAttribute; findFigureInside; addNotify; removeNotify], [Visitor; AbstractFigure; DiamondFigure; RectangleFigure; RoundRectangleFigure; TriangleFigure; TextFigure; BezierFigure; TextAreaFigure; NodeFigure; SVGImage; SVGPath; DependencyFigure; LineConnectionFigure; LabeledLineConnectionFigure; AbstractCompositeFigure; GraphicalCompositeFigure; BasicTransformVisitor; ContainsVisitor; SetAttributeVisitor; FindFigureInsideVisitor; AddNotifyVisitor; RemoveNotifyVisitor]) \textbf{; }\\
\operation{InlineAndDelete} (DiamondFigure, containsTmpVC, [ContainsVisitor], contains, [basicTransform; setAttribute; findFigureInside; addNotify; removeNotify], [Visitor; AbstractFigure; EllipseFigure; RectangleFigure; RoundRectangleFigure; TriangleFigure; TextFigure; BezierFigure; TextAreaFigure; NodeFigure; SVGImage; SVGPath; DependencyFigure; LineConnectionFigure; LabeledLineConnectionFigure; AbstractCompositeFigure; GraphicalCompositeFigure; BasicTransformVisitor; ContainsVisitor; SetAttributeVisitor; FindFigureInsideVisitor; AddNotifyVisitor; RemoveNotifyVisitor]) \textbf{; }\\
\operation{InlineAndDelete} (RectangleFigure, containsTmpVC, [ContainsVisitor], contains, [basicTransform; setAttribute; findFigureInside; addNotify; removeNotify], [Visitor; AbstractFigure; EllipseFigure; DiamondFigure; RoundRectangleFigure; TriangleFigure; TextFigure; BezierFigure; TextAreaFigure; NodeFigure; SVGImage; SVGPath; DependencyFigure; LineConnectionFigure; LabeledLineConnectionFigure; AbstractCompositeFigure; GraphicalCompositeFigure; BasicTransformVisitor; ContainsVisitor; SetAttributeVisitor; FindFigureInsideVisitor; AddNotifyVisitor; RemoveNotifyVisitor]) \textbf{; }\\
\operation{InlineAndDelete} (RoundRectangleFigure, containsTmpVC, [ContainsVisitor], contains, [basicTransform; setAttribute; findFigureInside; addNotify; removeNotify], [Visitor; AbstractFigure; EllipseFigure; DiamondFigure; RectangleFigure; TriangleFigure; TextFigure; BezierFigure; TextAreaFigure; NodeFigure; SVGImage; SVGPath; DependencyFigure; LineConnectionFigure; LabeledLineConnectionFigure; AbstractCompositeFigure; GraphicalCompositeFigure; BasicTransformVisitor; ContainsVisitor; SetAttributeVisitor; FindFigureInsideVisitor; AddNotifyVisitor; RemoveNotifyVisitor]) \textbf{; }\\
\operation{InlineAndDelete} (TriangleFigure, containsTmpVC, [ContainsVisitor], contains, [basicTransform; setAttribute; findFigureInside; addNotify; removeNotify], [Visitor; AbstractFigure; EllipseFigure; DiamondFigure; RectangleFigure; RoundRectangleFigure; TextFigure; BezierFigure; TextAreaFigure; NodeFigure; SVGImage; SVGPath; DependencyFigure; LineConnectionFigure; LabeledLineConnectionFigure; AbstractCompositeFigure; GraphicalCompositeFigure; BasicTransformVisitor; ContainsVisitor; SetAttributeVisitor; FindFigureInsideVisitor; AddNotifyVisitor; RemoveNotifyVisitor]) \textbf{; }\\
\operation{InlineAndDelete} (TextFigure, containsTmpVC, [ContainsVisitor], contains, [basicTransform; setAttribute; findFigureInside; addNotify; removeNotify], [Visitor; AbstractFigure; EllipseFigure; DiamondFigure; RectangleFigure; RoundRectangleFigure; TriangleFigure; BezierFigure; TextAreaFigure; NodeFigure; SVGImage; SVGPath; DependencyFigure; LineConnectionFigure; LabeledLineConnectionFigure; AbstractCompositeFigure; GraphicalCompositeFigure; BasicTransformVisitor; ContainsVisitor; SetAttributeVisitor; FindFigureInsideVisitor; AddNotifyVisitor; RemoveNotifyVisitor]) \textbf{; }\\
\operation{InlineAndDelete} (BezierFigure, containsTmpVC, [ContainsVisitor], contains, [basicTransform; setAttribute; findFigureInside; addNotify; removeNotify], [Visitor; AbstractFigure; EllipseFigure; DiamondFigure; RectangleFigure; RoundRectangleFigure; TriangleFigure; TextFigure; TextAreaFigure; NodeFigure; SVGImage; SVGPath; DependencyFigure; LineConnectionFigure; LabeledLineConnectionFigure; AbstractCompositeFigure; GraphicalCompositeFigure; BasicTransformVisitor; ContainsVisitor; SetAttributeVisitor; FindFigureInsideVisitor; AddNotifyVisitor; RemoveNotifyVisitor]) \textbf{; }\\
\operation{InlineAndDelete} (TextAreaFigure, containsTmpVC, [ContainsVisitor], contains, [basicTransform; setAttribute; findFigureInside; addNotify; removeNotify], [Visitor; AbstractFigure; EllipseFigure; DiamondFigure; RectangleFigure; RoundRectangleFigure; TriangleFigure; TextFigure; BezierFigure; NodeFigure; SVGImage; SVGPath; DependencyFigure; LineConnectionFigure; LabeledLineConnectionFigure; AbstractCompositeFigure; GraphicalCompositeFigure; BasicTransformVisitor; ContainsVisitor; SetAttributeVisitor; FindFigureInsideVisitor; AddNotifyVisitor; RemoveNotifyVisitor]) \textbf{; }\\
\operation{InlineAndDelete} (NodeFigure, containsTmpVC, [ContainsVisitor], contains, [basicTransform; setAttribute; findFigureInside; addNotify; removeNotify], [Visitor; AbstractFigure; EllipseFigure; DiamondFigure; RectangleFigure; RoundRectangleFigure; TriangleFigure; TextFigure; BezierFigure; TextAreaFigure; SVGImage; SVGPath; DependencyFigure; LineConnectionFigure; LabeledLineConnectionFigure; AbstractCompositeFigure; GraphicalCompositeFigure; BasicTransformVisitor; ContainsVisitor; SetAttributeVisitor; FindFigureInsideVisitor; AddNotifyVisitor; RemoveNotifyVisitor]) \textbf{; }\\
\operation{InlineAndDelete} (SVGImage, containsTmpVC, [ContainsVisitor], contains, [basicTransform; setAttribute; findFigureInside; addNotify; removeNotify], [Visitor; AbstractFigure; EllipseFigure; DiamondFigure; RectangleFigure; RoundRectangleFigure; TriangleFigure; TextFigure; BezierFigure; TextAreaFigure; NodeFigure; SVGPath; DependencyFigure; LineConnectionFigure; LabeledLineConnectionFigure; AbstractCompositeFigure; GraphicalCompositeFigure; BasicTransformVisitor; ContainsVisitor; SetAttributeVisitor; FindFigureInsideVisitor; AddNotifyVisitor; RemoveNotifyVisitor]) \textbf{; }\\
\operation{InlineAndDelete} (SVGPath, containsTmpVC, [ContainsVisitor], contains, [basicTransform; setAttribute; findFigureInside; addNotify; removeNotify], [Visitor; AbstractFigure; EllipseFigure; DiamondFigure; RectangleFigure; RoundRectangleFigure; TriangleFigure; TextFigure; BezierFigure; TextAreaFigure; NodeFigure; SVGImage; DependencyFigure; LineConnectionFigure; LabeledLineConnectionFigure; AbstractCompositeFigure; GraphicalCompositeFigure; BasicTransformVisitor; ContainsVisitor; SetAttributeVisitor; FindFigureInsideVisitor; AddNotifyVisitor; RemoveNotifyVisitor]) \textbf{; }\\
\operation{InlineAndDelete} (DependencyFigure, containsTmpVC, [ContainsVisitor], contains, [basicTransform; setAttribute; findFigureInside; addNotify; removeNotify], [Visitor; AbstractFigure; EllipseFigure; DiamondFigure; RectangleFigure; RoundRectangleFigure; TriangleFigure; TextFigure; BezierFigure; TextAreaFigure; NodeFigure; SVGImage; SVGPath; LineConnectionFigure; LabeledLineConnectionFigure; AbstractCompositeFigure; GraphicalCompositeFigure; BasicTransformVisitor; ContainsVisitor; SetAttributeVisitor; FindFigureInsideVisitor; AddNotifyVisitor; RemoveNotifyVisitor]) \textbf{; }\\
\operation{InlineAndDelete} (LineConnectionFigure, containsTmpVC, [ContainsVisitor], contains, [basicTransform; setAttribute; findFigureInside; addNotify; removeNotify], [Visitor; AbstractFigure; EllipseFigure; DiamondFigure; RectangleFigure; RoundRectangleFigure; TriangleFigure; TextFigure; BezierFigure; TextAreaFigure; NodeFigure; SVGImage; SVGPath; DependencyFigure; LabeledLineConnectionFigure; AbstractCompositeFigure; GraphicalCompositeFigure; BasicTransformVisitor; ContainsVisitor; SetAttributeVisitor; FindFigureInsideVisitor; AddNotifyVisitor; RemoveNotifyVisitor]) \textbf{; }\\
\operation{InlineAndDelete} (LabeledLineConnectionFigure, containsTmpVC, [ContainsVisitor], contains, [basicTransform; setAttribute; findFigureInside; addNotify; removeNotify], [Visitor; AbstractFigure; EllipseFigure; DiamondFigure; RectangleFigure; RoundRectangleFigure; TriangleFigure; TextFigure; BezierFigure; TextAreaFigure; NodeFigure; SVGImage; SVGPath; DependencyFigure; LineConnectionFigure; AbstractCompositeFigure; GraphicalCompositeFigure; BasicTransformVisitor; ContainsVisitor; SetAttributeVisitor; FindFigureInsideVisitor; AddNotifyVisitor; RemoveNotifyVisitor]) \textbf{; }\\
\operation{InlineAndDelete} (AbstractCompositeFigure, containsTmpVC, [ContainsVisitor], contains, [basicTransform; setAttribute; findFigureInside; addNotify; removeNotify], [Visitor; AbstractFigure; EllipseFigure; DiamondFigure; RectangleFigure; RoundRectangleFigure; TriangleFigure; TextFigure; BezierFigure; TextAreaFigure; NodeFigure; SVGImage; SVGPath; DependencyFigure; LineConnectionFigure; LabeledLineConnectionFigure; GraphicalCompositeFigure; BasicTransformVisitor; ContainsVisitor; SetAttributeVisitor; FindFigureInsideVisitor; AddNotifyVisitor; RemoveNotifyVisitor]) \textbf{; }\\
\operation{InlineAndDelete} (GraphicalCompositeFigure, containsTmpVC, [ContainsVisitor], contains, [basicTransform; setAttribute; findFigureInside; addNotify; removeNotify], [Visitor; AbstractFigure; EllipseFigure; DiamondFigure; RectangleFigure; RoundRectangleFigure; TriangleFigure; TextFigure; BezierFigure; TextAreaFigure; NodeFigure; SVGImage; SVGPath; DependencyFigure; LineConnectionFigure; LabeledLineConnectionFigure; AbstractCompositeFigure; BasicTransformVisitor; ContainsVisitor; SetAttributeVisitor; FindFigureInsideVisitor; AddNotifyVisitor; RemoveNotifyVisitor]) \textbf{; }\\
\operation{InlineAndDelete} (EllipseFigure, setAttributeTmpVC, [SetAttributeVisitor], setAttribute, [basicTransform; contains; findFigureInside; addNotify; removeNotify], [Visitor; AbstractFigure; DiamondFigure; RectangleFigure; RoundRectangleFigure; TriangleFigure; TextFigure; BezierFigure; TextAreaFigure; NodeFigure; SVGImage; SVGPath; DependencyFigure; LineConnectionFigure; LabeledLineConnectionFigure; AbstractCompositeFigure; GraphicalCompositeFigure; BasicTransformVisitor; ContainsVisitor; SetAttributeVisitor; FindFigureInsideVisitor; AddNotifyVisitor; RemoveNotifyVisitor]) \textbf{; }\\
\operation{InlineAndDelete} (DiamondFigure, setAttributeTmpVC, [SetAttributeVisitor], setAttribute, [basicTransform; contains; findFigureInside; addNotify; removeNotify], [Visitor; AbstractFigure; EllipseFigure; RectangleFigure; RoundRectangleFigure; TriangleFigure; TextFigure; BezierFigure; TextAreaFigure; NodeFigure; SVGImage; SVGPath; DependencyFigure; LineConnectionFigure; LabeledLineConnectionFigure; AbstractCompositeFigure; GraphicalCompositeFigure; BasicTransformVisitor; ContainsVisitor; SetAttributeVisitor; FindFigureInsideVisitor; AddNotifyVisitor; RemoveNotifyVisitor]) \textbf{; }\\
\operation{InlineAndDelete} (RectangleFigure, setAttributeTmpVC, [SetAttributeVisitor], setAttribute, [basicTransform; contains; findFigureInside; addNotify; removeNotify], [Visitor; AbstractFigure; EllipseFigure; DiamondFigure; RoundRectangleFigure; TriangleFigure; TextFigure; BezierFigure; TextAreaFigure; NodeFigure; SVGImage; SVGPath; DependencyFigure; LineConnectionFigure; LabeledLineConnectionFigure; AbstractCompositeFigure; GraphicalCompositeFigure; BasicTransformVisitor; ContainsVisitor; SetAttributeVisitor; FindFigureInsideVisitor; AddNotifyVisitor; RemoveNotifyVisitor]) \textbf{; }\\
\operation{InlineAndDelete} (RoundRectangleFigure, setAttributeTmpVC, [SetAttributeVisitor], setAttribute, [basicTransform; contains; findFigureInside; addNotify; removeNotify], [Visitor; AbstractFigure; EllipseFigure; DiamondFigure; RectangleFigure; TriangleFigure; TextFigure; BezierFigure; TextAreaFigure; NodeFigure; SVGImage; SVGPath; DependencyFigure; LineConnectionFigure; LabeledLineConnectionFigure; AbstractCompositeFigure; GraphicalCompositeFigure; BasicTransformVisitor; ContainsVisitor; SetAttributeVisitor; FindFigureInsideVisitor; AddNotifyVisitor; RemoveNotifyVisitor]) \textbf{; }\\
\operation{InlineAndDelete} (TriangleFigure, setAttributeTmpVC, [SetAttributeVisitor], setAttribute, [basicTransform; contains; findFigureInside; addNotify; removeNotify], [Visitor; AbstractFigure; EllipseFigure; DiamondFigure; RectangleFigure; RoundRectangleFigure; TextFigure; BezierFigure; TextAreaFigure; NodeFigure; SVGImage; SVGPath; DependencyFigure; LineConnectionFigure; LabeledLineConnectionFigure; AbstractCompositeFigure; GraphicalCompositeFigure; BasicTransformVisitor; ContainsVisitor; SetAttributeVisitor; FindFigureInsideVisitor; AddNotifyVisitor; RemoveNotifyVisitor]) \textbf{; }\\
\operation{InlineAndDelete} (TextFigure, setAttributeTmpVC, [SetAttributeVisitor], setAttribute, [basicTransform; contains; findFigureInside; addNotify; removeNotify], [Visitor; AbstractFigure; EllipseFigure; DiamondFigure; RectangleFigure; RoundRectangleFigure; TriangleFigure; BezierFigure; TextAreaFigure; NodeFigure; SVGImage; SVGPath; DependencyFigure; LineConnectionFigure; LabeledLineConnectionFigure; AbstractCompositeFigure; GraphicalCompositeFigure; BasicTransformVisitor; ContainsVisitor; SetAttributeVisitor; FindFigureInsideVisitor; AddNotifyVisitor; RemoveNotifyVisitor]) \textbf{; }\\
\operation{InlineAndDelete} (BezierFigure, setAttributeTmpVC, [SetAttributeVisitor], setAttribute, [basicTransform; contains; findFigureInside; addNotify; removeNotify], [Visitor; AbstractFigure; EllipseFigure; DiamondFigure; RectangleFigure; RoundRectangleFigure; TriangleFigure; TextFigure; TextAreaFigure; NodeFigure; SVGImage; SVGPath; DependencyFigure; LineConnectionFigure; LabeledLineConnectionFigure; AbstractCompositeFigure; GraphicalCompositeFigure; BasicTransformVisitor; ContainsVisitor; SetAttributeVisitor; FindFigureInsideVisitor; AddNotifyVisitor; RemoveNotifyVisitor]) \textbf{; }\\
\operation{InlineAndDelete} (TextAreaFigure, setAttributeTmpVC, [SetAttributeVisitor], setAttribute, [basicTransform; contains; findFigureInside; addNotify; removeNotify], [Visitor; AbstractFigure; EllipseFigure; DiamondFigure; RectangleFigure; RoundRectangleFigure; TriangleFigure; TextFigure; BezierFigure; NodeFigure; SVGImage; SVGPath; DependencyFigure; LineConnectionFigure; LabeledLineConnectionFigure; AbstractCompositeFigure; GraphicalCompositeFigure; BasicTransformVisitor; ContainsVisitor; SetAttributeVisitor; FindFigureInsideVisitor; AddNotifyVisitor; RemoveNotifyVisitor]) \textbf{; }\\
\operation{InlineAndDelete} (NodeFigure, setAttributeTmpVC, [SetAttributeVisitor], setAttribute, [basicTransform; contains; findFigureInside; addNotify; removeNotify], [Visitor; AbstractFigure; EllipseFigure; DiamondFigure; RectangleFigure; RoundRectangleFigure; TriangleFigure; TextFigure; BezierFigure; TextAreaFigure; SVGImage; SVGPath; DependencyFigure; LineConnectionFigure; LabeledLineConnectionFigure; AbstractCompositeFigure; GraphicalCompositeFigure; BasicTransformVisitor; ContainsVisitor; SetAttributeVisitor; FindFigureInsideVisitor; AddNotifyVisitor; RemoveNotifyVisitor]) \textbf{; }\\
\operation{InlineAndDelete} (SVGImage, setAttributeTmpVC, [SetAttributeVisitor], setAttribute, [basicTransform; contains; findFigureInside; addNotify; removeNotify], [Visitor; AbstractFigure; EllipseFigure; DiamondFigure; RectangleFigure; RoundRectangleFigure; TriangleFigure; TextFigure; BezierFigure; TextAreaFigure; NodeFigure; SVGPath; DependencyFigure; LineConnectionFigure; LabeledLineConnectionFigure; AbstractCompositeFigure; GraphicalCompositeFigure; BasicTransformVisitor; ContainsVisitor; SetAttributeVisitor; FindFigureInsideVisitor; AddNotifyVisitor; RemoveNotifyVisitor]) \textbf{; }\\
\operation{InlineAndDelete} (SVGPath, setAttributeTmpVC, [SetAttributeVisitor], setAttribute, [basicTransform; contains; findFigureInside; addNotify; removeNotify], [Visitor; AbstractFigure; EllipseFigure; DiamondFigure; RectangleFigure; RoundRectangleFigure; TriangleFigure; TextFigure; BezierFigure; TextAreaFigure; NodeFigure; SVGImage; DependencyFigure; LineConnectionFigure; LabeledLineConnectionFigure; AbstractCompositeFigure; GraphicalCompositeFigure; BasicTransformVisitor; ContainsVisitor; SetAttributeVisitor; FindFigureInsideVisitor; AddNotifyVisitor; RemoveNotifyVisitor]) \textbf{; }\\
\operation{InlineAndDelete} (DependencyFigure, setAttributeTmpVC, [SetAttributeVisitor], setAttribute, [basicTransform; contains; findFigureInside; addNotify; removeNotify], [Visitor; AbstractFigure; EllipseFigure; DiamondFigure; RectangleFigure; RoundRectangleFigure; TriangleFigure; TextFigure; BezierFigure; TextAreaFigure; NodeFigure; SVGImage; SVGPath; LineConnectionFigure; LabeledLineConnectionFigure; AbstractCompositeFigure; GraphicalCompositeFigure; BasicTransformVisitor; ContainsVisitor; SetAttributeVisitor; FindFigureInsideVisitor; AddNotifyVisitor; RemoveNotifyVisitor]) \textbf{; }\\
\operation{InlineAndDelete} (LineConnectionFigure, setAttributeTmpVC, [SetAttributeVisitor], setAttribute, [basicTransform; contains; findFigureInside; addNotify; removeNotify], [Visitor; AbstractFigure; EllipseFigure; DiamondFigure; RectangleFigure; RoundRectangleFigure; TriangleFigure; TextFigure; BezierFigure; TextAreaFigure; NodeFigure; SVGImage; SVGPath; DependencyFigure; LabeledLineConnectionFigure; AbstractCompositeFigure; GraphicalCompositeFigure; BasicTransformVisitor; ContainsVisitor; SetAttributeVisitor; FindFigureInsideVisitor; AddNotifyVisitor; RemoveNotifyVisitor]) \textbf{; }\\
\operation{InlineAndDelete} (LabeledLineConnectionFigure, setAttributeTmpVC, [SetAttributeVisitor], setAttribute, [basicTransform; contains; findFigureInside; addNotify; removeNotify], [Visitor; AbstractFigure; EllipseFigure; DiamondFigure; RectangleFigure; RoundRectangleFigure; TriangleFigure; TextFigure; BezierFigure; TextAreaFigure; NodeFigure; SVGImage; SVGPath; DependencyFigure; LineConnectionFigure; AbstractCompositeFigure; GraphicalCompositeFigure; BasicTransformVisitor; ContainsVisitor; SetAttributeVisitor; FindFigureInsideVisitor; AddNotifyVisitor; RemoveNotifyVisitor]) \textbf{; }\\
\operation{InlineAndDelete} (AbstractCompositeFigure, setAttributeTmpVC, [SetAttributeVisitor], setAttribute, [basicTransform; contains; findFigureInside; addNotify; removeNotify], [Visitor; AbstractFigure; EllipseFigure; DiamondFigure; RectangleFigure; RoundRectangleFigure; TriangleFigure; TextFigure; BezierFigure; TextAreaFigure; NodeFigure; SVGImage; SVGPath; DependencyFigure; LineConnectionFigure; LabeledLineConnectionFigure; GraphicalCompositeFigure; BasicTransformVisitor; ContainsVisitor; SetAttributeVisitor; FindFigureInsideVisitor; AddNotifyVisitor; RemoveNotifyVisitor]) \textbf{; }\\
\operation{InlineAndDelete} (GraphicalCompositeFigure, setAttributeTmpVC, [SetAttributeVisitor], setAttribute, [basicTransform; contains; findFigureInside; addNotify; removeNotify], [Visitor; AbstractFigure; EllipseFigure; DiamondFigure; RectangleFigure; RoundRectangleFigure; TriangleFigure; TextFigure; BezierFigure; TextAreaFigure; NodeFigure; SVGImage; SVGPath; DependencyFigure; LineConnectionFigure; LabeledLineConnectionFigure; AbstractCompositeFigure; BasicTransformVisitor; ContainsVisitor; SetAttributeVisitor; FindFigureInsideVisitor; AddNotifyVisitor; RemoveNotifyVisitor]) \textbf{; }\\
\operation{InlineAndDelete} (EllipseFigure, findFigureInsideTmpVC, [FindFigureInsideVisitor], findFigureInside, [basicTransform; contains; setAttribute; addNotify; removeNotify], [Visitor; AbstractFigure; DiamondFigure; RectangleFigure; RoundRectangleFigure; TriangleFigure; TextFigure; BezierFigure; TextAreaFigure; NodeFigure; SVGImage; SVGPath; DependencyFigure; LineConnectionFigure; LabeledLineConnectionFigure; AbstractCompositeFigure; GraphicalCompositeFigure; BasicTransformVisitor; ContainsVisitor; SetAttributeVisitor; FindFigureInsideVisitor; AddNotifyVisitor; RemoveNotifyVisitor]) \textbf{; }\\
\operation{InlineAndDelete} (DiamondFigure, findFigureInsideTmpVC, [FindFigureInsideVisitor], findFigureInside, [basicTransform; contains; setAttribute; addNotify; removeNotify], [Visitor; AbstractFigure; EllipseFigure; RectangleFigure; RoundRectangleFigure; TriangleFigure; TextFigure; BezierFigure; TextAreaFigure; NodeFigure; SVGImage; SVGPath; DependencyFigure; LineConnectionFigure; LabeledLineConnectionFigure; AbstractCompositeFigure; GraphicalCompositeFigure; BasicTransformVisitor; ContainsVisitor; SetAttributeVisitor; FindFigureInsideVisitor; AddNotifyVisitor; RemoveNotifyVisitor]) \textbf{; }\\
\operation{InlineAndDelete} (RectangleFigure, findFigureInsideTmpVC, [FindFigureInsideVisitor], findFigureInside, [basicTransform; contains; setAttribute; addNotify; removeNotify], [Visitor; AbstractFigure; EllipseFigure; DiamondFigure; RoundRectangleFigure; TriangleFigure; TextFigure; BezierFigure; TextAreaFigure; NodeFigure; SVGImage; SVGPath; DependencyFigure; LineConnectionFigure; LabeledLineConnectionFigure; AbstractCompositeFigure; GraphicalCompositeFigure; BasicTransformVisitor; ContainsVisitor; SetAttributeVisitor; FindFigureInsideVisitor; AddNotifyVisitor; RemoveNotifyVisitor]) \textbf{; }\\
\operation{InlineAndDelete} (RoundRectangleFigure, findFigureInsideTmpVC, [FindFigureInsideVisitor], findFigureInside, [basicTransform; contains; setAttribute; addNotify; removeNotify], [Visitor; AbstractFigure; EllipseFigure; DiamondFigure; RectangleFigure; TriangleFigure; TextFigure; BezierFigure; TextAreaFigure; NodeFigure; SVGImage; SVGPath; DependencyFigure; LineConnectionFigure; LabeledLineConnectionFigure; AbstractCompositeFigure; GraphicalCompositeFigure; BasicTransformVisitor; ContainsVisitor; SetAttributeVisitor; FindFigureInsideVisitor; AddNotifyVisitor; RemoveNotifyVisitor]) \textbf{; }\\
\operation{InlineAndDelete} (TriangleFigure, findFigureInsideTmpVC, [FindFigureInsideVisitor], findFigureInside, [basicTransform; contains; setAttribute; addNotify; removeNotify], [Visitor; AbstractFigure; EllipseFigure; DiamondFigure; RectangleFigure; RoundRectangleFigure; TextFigure; BezierFigure; TextAreaFigure; NodeFigure; SVGImage; SVGPath; DependencyFigure; LineConnectionFigure; LabeledLineConnectionFigure; AbstractCompositeFigure; GraphicalCompositeFigure; BasicTransformVisitor; ContainsVisitor; SetAttributeVisitor; FindFigureInsideVisitor; AddNotifyVisitor; RemoveNotifyVisitor]) \textbf{; }\\
\operation{InlineAndDelete} (TextFigure, findFigureInsideTmpVC, [FindFigureInsideVisitor], findFigureInside, [basicTransform; contains; setAttribute; addNotify; removeNotify], [Visitor; AbstractFigure; EllipseFigure; DiamondFigure; RectangleFigure; RoundRectangleFigure; TriangleFigure; BezierFigure; TextAreaFigure; NodeFigure; SVGImage; SVGPath; DependencyFigure; LineConnectionFigure; LabeledLineConnectionFigure; AbstractCompositeFigure; GraphicalCompositeFigure; BasicTransformVisitor; ContainsVisitor; SetAttributeVisitor; FindFigureInsideVisitor; AddNotifyVisitor; RemoveNotifyVisitor]) \textbf{; }\\
\operation{InlineAndDelete} (BezierFigure, findFigureInsideTmpVC, [FindFigureInsideVisitor], findFigureInside, [basicTransform; contains; setAttribute; addNotify; removeNotify], [Visitor; AbstractFigure; EllipseFigure; DiamondFigure; RectangleFigure; RoundRectangleFigure; TriangleFigure; TextFigure; TextAreaFigure; NodeFigure; SVGImage; SVGPath; DependencyFigure; LineConnectionFigure; LabeledLineConnectionFigure; AbstractCompositeFigure; GraphicalCompositeFigure; BasicTransformVisitor; ContainsVisitor; SetAttributeVisitor; FindFigureInsideVisitor; AddNotifyVisitor; RemoveNotifyVisitor]) \textbf{; }\\
\operation{InlineAndDelete} (TextAreaFigure, findFigureInsideTmpVC, [FindFigureInsideVisitor], findFigureInside, [basicTransform; contains; setAttribute; addNotify; removeNotify], [Visitor; AbstractFigure; EllipseFigure; DiamondFigure; RectangleFigure; RoundRectangleFigure; TriangleFigure; TextFigure; BezierFigure; NodeFigure; SVGImage; SVGPath; DependencyFigure; LineConnectionFigure; LabeledLineConnectionFigure; AbstractCompositeFigure; GraphicalCompositeFigure; BasicTransformVisitor; ContainsVisitor; SetAttributeVisitor; FindFigureInsideVisitor; AddNotifyVisitor; RemoveNotifyVisitor]) \textbf{; }\\
\operation{InlineAndDelete} (NodeFigure, findFigureInsideTmpVC, [FindFigureInsideVisitor], findFigureInside, [basicTransform; contains; setAttribute; addNotify; removeNotify], [Visitor; AbstractFigure; EllipseFigure; DiamondFigure; RectangleFigure; RoundRectangleFigure; TriangleFigure; TextFigure; BezierFigure; TextAreaFigure; SVGImage; SVGPath; DependencyFigure; LineConnectionFigure; LabeledLineConnectionFigure; AbstractCompositeFigure; GraphicalCompositeFigure; BasicTransformVisitor; ContainsVisitor; SetAttributeVisitor; FindFigureInsideVisitor; AddNotifyVisitor; RemoveNotifyVisitor]) \textbf{; }\\
\operation{InlineAndDelete} (SVGImage, findFigureInsideTmpVC, [FindFigureInsideVisitor], findFigureInside, [basicTransform; contains; setAttribute; addNotify; removeNotify], [Visitor; AbstractFigure; EllipseFigure; DiamondFigure; RectangleFigure; RoundRectangleFigure; TriangleFigure; TextFigure; BezierFigure; TextAreaFigure; NodeFigure; SVGPath; DependencyFigure; LineConnectionFigure; LabeledLineConnectionFigure; AbstractCompositeFigure; GraphicalCompositeFigure; BasicTransformVisitor; ContainsVisitor; SetAttributeVisitor; FindFigureInsideVisitor; AddNotifyVisitor; RemoveNotifyVisitor]) \textbf{; }\\
\operation{InlineAndDelete} (SVGPath, findFigureInsideTmpVC, [FindFigureInsideVisitor], findFigureInside, [basicTransform; contains; setAttribute; addNotify; removeNotify], [Visitor; AbstractFigure; EllipseFigure; DiamondFigure; RectangleFigure; RoundRectangleFigure; TriangleFigure; TextFigure; BezierFigure; TextAreaFigure; NodeFigure; SVGImage; DependencyFigure; LineConnectionFigure; LabeledLineConnectionFigure; AbstractCompositeFigure; GraphicalCompositeFigure; BasicTransformVisitor; ContainsVisitor; SetAttributeVisitor; FindFigureInsideVisitor; AddNotifyVisitor; RemoveNotifyVisitor]) \textbf{; }\\
\operation{InlineAndDelete} (DependencyFigure, findFigureInsideTmpVC, [FindFigureInsideVisitor], findFigureInside, [basicTransform; contains; setAttribute; addNotify; removeNotify], [Visitor; AbstractFigure; EllipseFigure; DiamondFigure; RectangleFigure; RoundRectangleFigure; TriangleFigure; TextFigure; BezierFigure; TextAreaFigure; NodeFigure; SVGImage; SVGPath; LineConnectionFigure; LabeledLineConnectionFigure; AbstractCompositeFigure; GraphicalCompositeFigure; BasicTransformVisitor; ContainsVisitor; SetAttributeVisitor; FindFigureInsideVisitor; AddNotifyVisitor; RemoveNotifyVisitor]) \textbf{; }\\
\operation{InlineAndDelete} (LineConnectionFigure, findFigureInsideTmpVC, [FindFigureInsideVisitor], findFigureInside, [basicTransform; contains; setAttribute; addNotify; removeNotify], [Visitor; AbstractFigure; EllipseFigure; DiamondFigure; RectangleFigure; RoundRectangleFigure; TriangleFigure; TextFigure; BezierFigure; TextAreaFigure; NodeFigure; SVGImage; SVGPath; DependencyFigure; LabeledLineConnectionFigure; AbstractCompositeFigure; GraphicalCompositeFigure; BasicTransformVisitor; ContainsVisitor; SetAttributeVisitor; FindFigureInsideVisitor; AddNotifyVisitor; RemoveNotifyVisitor]) \textbf{; }\\
\operation{InlineAndDelete} (LabeledLineConnectionFigure, findFigureInsideTmpVC, [FindFigureInsideVisitor], findFigureInside, [basicTransform; contains; setAttribute; addNotify; removeNotify], [Visitor; AbstractFigure; EllipseFigure; DiamondFigure; RectangleFigure; RoundRectangleFigure; TriangleFigure; TextFigure; BezierFigure; TextAreaFigure; NodeFigure; SVGImage; SVGPath; DependencyFigure; LineConnectionFigure; AbstractCompositeFigure; GraphicalCompositeFigure; BasicTransformVisitor; ContainsVisitor; SetAttributeVisitor; FindFigureInsideVisitor; AddNotifyVisitor; RemoveNotifyVisitor]) \textbf{; }\\
\operation{InlineAndDelete} (AbstractCompositeFigure, findFigureInsideTmpVC, [FindFigureInsideVisitor], findFigureInside, [basicTransform; contains; setAttribute; addNotify; removeNotify], [Visitor; AbstractFigure; EllipseFigure; DiamondFigure; RectangleFigure; RoundRectangleFigure; TriangleFigure; TextFigure; BezierFigure; TextAreaFigure; NodeFigure; SVGImage; SVGPath; DependencyFigure; LineConnectionFigure; LabeledLineConnectionFigure; GraphicalCompositeFigure; BasicTransformVisitor; ContainsVisitor; SetAttributeVisitor; FindFigureInsideVisitor; AddNotifyVisitor; RemoveNotifyVisitor]) \textbf{; }\\
\operation{InlineAndDelete} (GraphicalCompositeFigure, findFigureInsideTmpVC, [FindFigureInsideVisitor], findFigureInside, [basicTransform; contains; setAttribute; addNotify; removeNotify], [Visitor; AbstractFigure; EllipseFigure; DiamondFigure; RectangleFigure; RoundRectangleFigure; TriangleFigure; TextFigure; BezierFigure; TextAreaFigure; NodeFigure; SVGImage; SVGPath; DependencyFigure; LineConnectionFigure; LabeledLineConnectionFigure; AbstractCompositeFigure; BasicTransformVisitor; ContainsVisitor; SetAttributeVisitor; FindFigureInsideVisitor; AddNotifyVisitor; RemoveNotifyVisitor]) \textbf{; }\\
\operation{InlineAndDelete} (EllipseFigure, addNotifyTmpVC, [AddNotifyVisitor], addNotify, [basicTransform; contains; setAttribute; findFigureInside; removeNotify], [Visitor; AbstractFigure; DiamondFigure; RectangleFigure; RoundRectangleFigure; TriangleFigure; TextFigure; BezierFigure; TextAreaFigure; NodeFigure; SVGImage; SVGPath; DependencyFigure; LineConnectionFigure; LabeledLineConnectionFigure; AbstractCompositeFigure; GraphicalCompositeFigure; BasicTransformVisitor; ContainsVisitor; SetAttributeVisitor; FindFigureInsideVisitor; AddNotifyVisitor; RemoveNotifyVisitor]) \textbf{; }\\
\operation{InlineAndDelete} (DiamondFigure, addNotifyTmpVC, [AddNotifyVisitor], addNotify, [basicTransform; contains; setAttribute; findFigureInside; removeNotify], [Visitor; AbstractFigure; EllipseFigure; RectangleFigure; RoundRectangleFigure; TriangleFigure; TextFigure; BezierFigure; TextAreaFigure; NodeFigure; SVGImage; SVGPath; DependencyFigure; LineConnectionFigure; LabeledLineConnectionFigure; AbstractCompositeFigure; GraphicalCompositeFigure; BasicTransformVisitor; ContainsVisitor; SetAttributeVisitor; FindFigureInsideVisitor; AddNotifyVisitor; RemoveNotifyVisitor]) \textbf{; }\\
\operation{InlineAndDelete} (RectangleFigure, addNotifyTmpVC, [AddNotifyVisitor], addNotify, [basicTransform; contains; setAttribute; findFigureInside; removeNotify], [Visitor; AbstractFigure; EllipseFigure; DiamondFigure; RoundRectangleFigure; TriangleFigure; TextFigure; BezierFigure; TextAreaFigure; NodeFigure; SVGImage; SVGPath; DependencyFigure; LineConnectionFigure; LabeledLineConnectionFigure; AbstractCompositeFigure; GraphicalCompositeFigure; BasicTransformVisitor; ContainsVisitor; SetAttributeVisitor; FindFigureInsideVisitor; AddNotifyVisitor; RemoveNotifyVisitor]) \textbf{; }\\
\operation{InlineAndDelete} (RoundRectangleFigure, addNotifyTmpVC, [AddNotifyVisitor], addNotify, [basicTransform; contains; setAttribute; findFigureInside; removeNotify], [Visitor; AbstractFigure; EllipseFigure; DiamondFigure; RectangleFigure; TriangleFigure; TextFigure; BezierFigure; TextAreaFigure; NodeFigure; SVGImage; SVGPath; DependencyFigure; LineConnectionFigure; LabeledLineConnectionFigure; AbstractCompositeFigure; GraphicalCompositeFigure; BasicTransformVisitor; ContainsVisitor; SetAttributeVisitor; FindFigureInsideVisitor; AddNotifyVisitor; RemoveNotifyVisitor]) \textbf{; }\\
\operation{InlineAndDelete} (TriangleFigure, addNotifyTmpVC, [AddNotifyVisitor], addNotify, [basicTransform; contains; setAttribute; findFigureInside; removeNotify], [Visitor; AbstractFigure; EllipseFigure; DiamondFigure; RectangleFigure; RoundRectangleFigure; TextFigure; BezierFigure; TextAreaFigure; NodeFigure; SVGImage; SVGPath; DependencyFigure; LineConnectionFigure; LabeledLineConnectionFigure; AbstractCompositeFigure; GraphicalCompositeFigure; BasicTransformVisitor; ContainsVisitor; SetAttributeVisitor; FindFigureInsideVisitor; AddNotifyVisitor; RemoveNotifyVisitor]) \textbf{; }\\
\operation{InlineAndDelete} (TextFigure, addNotifyTmpVC, [AddNotifyVisitor], addNotify, [basicTransform; contains; setAttribute; findFigureInside; removeNotify], [Visitor; AbstractFigure; EllipseFigure; DiamondFigure; RectangleFigure; RoundRectangleFigure; TriangleFigure; BezierFigure; TextAreaFigure; NodeFigure; SVGImage; SVGPath; DependencyFigure; LineConnectionFigure; LabeledLineConnectionFigure; AbstractCompositeFigure; GraphicalCompositeFigure; BasicTransformVisitor; ContainsVisitor; SetAttributeVisitor; FindFigureInsideVisitor; AddNotifyVisitor; RemoveNotifyVisitor]) \textbf{; }\\
\operation{InlineAndDelete} (BezierFigure, addNotifyTmpVC, [AddNotifyVisitor], addNotify, [basicTransform; contains; setAttribute; findFigureInside; removeNotify], [Visitor; AbstractFigure; EllipseFigure; DiamondFigure; RectangleFigure; RoundRectangleFigure; TriangleFigure; TextFigure; TextAreaFigure; NodeFigure; SVGImage; SVGPath; DependencyFigure; LineConnectionFigure; LabeledLineConnectionFigure; AbstractCompositeFigure; GraphicalCompositeFigure; BasicTransformVisitor; ContainsVisitor; SetAttributeVisitor; FindFigureInsideVisitor; AddNotifyVisitor; RemoveNotifyVisitor]) \textbf{; }\\
\operation{InlineAndDelete} (TextAreaFigure, addNotifyTmpVC, [AddNotifyVisitor], addNotify, [basicTransform; contains; setAttribute; findFigureInside; removeNotify], [Visitor; AbstractFigure; EllipseFigure; DiamondFigure; RectangleFigure; RoundRectangleFigure; TriangleFigure; TextFigure; BezierFigure; NodeFigure; SVGImage; SVGPath; DependencyFigure; LineConnectionFigure; LabeledLineConnectionFigure; AbstractCompositeFigure; GraphicalCompositeFigure; BasicTransformVisitor; ContainsVisitor; SetAttributeVisitor; FindFigureInsideVisitor; AddNotifyVisitor; RemoveNotifyVisitor]) \textbf{; }\\
\operation{InlineAndDelete} (NodeFigure, addNotifyTmpVC, [AddNotifyVisitor], addNotify, [basicTransform; contains; setAttribute; findFigureInside; removeNotify], [Visitor; AbstractFigure; EllipseFigure; DiamondFigure; RectangleFigure; RoundRectangleFigure; TriangleFigure; TextFigure; BezierFigure; TextAreaFigure; SVGImage; SVGPath; DependencyFigure; LineConnectionFigure; LabeledLineConnectionFigure; AbstractCompositeFigure; GraphicalCompositeFigure; BasicTransformVisitor; ContainsVisitor; SetAttributeVisitor; FindFigureInsideVisitor; AddNotifyVisitor; RemoveNotifyVisitor]) \textbf{; }\\
\operation{InlineAndDelete} (SVGImage, addNotifyTmpVC, [AddNotifyVisitor], addNotify, [basicTransform; contains; setAttribute; findFigureInside; removeNotify], [Visitor; AbstractFigure; EllipseFigure; DiamondFigure; RectangleFigure; RoundRectangleFigure; TriangleFigure; TextFigure; BezierFigure; TextAreaFigure; NodeFigure; SVGPath; DependencyFigure; LineConnectionFigure; LabeledLineConnectionFigure; AbstractCompositeFigure; GraphicalCompositeFigure; BasicTransformVisitor; ContainsVisitor; SetAttributeVisitor; FindFigureInsideVisitor; AddNotifyVisitor; RemoveNotifyVisitor]) \textbf{; }\\
\operation{InlineAndDelete} (SVGPath, addNotifyTmpVC, [AddNotifyVisitor], addNotify, [basicTransform; contains; setAttribute; findFigureInside; removeNotify], [Visitor; AbstractFigure; EllipseFigure; DiamondFigure; RectangleFigure; RoundRectangleFigure; TriangleFigure; TextFigure; BezierFigure; TextAreaFigure; NodeFigure; SVGImage; DependencyFigure; LineConnectionFigure; LabeledLineConnectionFigure; AbstractCompositeFigure; GraphicalCompositeFigure; BasicTransformVisitor; ContainsVisitor; SetAttributeVisitor; FindFigureInsideVisitor; AddNotifyVisitor; RemoveNotifyVisitor]) \textbf{; }\\
\operation{InlineAndDelete} (DependencyFigure, addNotifyTmpVC, [AddNotifyVisitor], addNotify, [basicTransform; contains; setAttribute; findFigureInside; removeNotify], [Visitor; AbstractFigure; EllipseFigure; DiamondFigure; RectangleFigure; RoundRectangleFigure; TriangleFigure; TextFigure; BezierFigure; TextAreaFigure; NodeFigure; SVGImage; SVGPath; LineConnectionFigure; LabeledLineConnectionFigure; AbstractCompositeFigure; GraphicalCompositeFigure; BasicTransformVisitor; ContainsVisitor; SetAttributeVisitor; FindFigureInsideVisitor; AddNotifyVisitor; RemoveNotifyVisitor]) \textbf{; }\\
\operation{InlineAndDelete} (LineConnectionFigure, addNotifyTmpVC, [AddNotifyVisitor], addNotify, [basicTransform; contains; setAttribute; findFigureInside; removeNotify], [Visitor; AbstractFigure; EllipseFigure; DiamondFigure; RectangleFigure; RoundRectangleFigure; TriangleFigure; TextFigure; BezierFigure; TextAreaFigure; NodeFigure; SVGImage; SVGPath; DependencyFigure; LabeledLineConnectionFigure; AbstractCompositeFigure; GraphicalCompositeFigure; BasicTransformVisitor; ContainsVisitor; SetAttributeVisitor; FindFigureInsideVisitor; AddNotifyVisitor; RemoveNotifyVisitor]) \textbf{; }\\
\operation{InlineAndDelete} (LabeledLineConnectionFigure, addNotifyTmpVC, [AddNotifyVisitor], addNotify, [basicTransform; contains; setAttribute; findFigureInside; removeNotify], [Visitor; AbstractFigure; EllipseFigure; DiamondFigure; RectangleFigure; RoundRectangleFigure; TriangleFigure; TextFigure; BezierFigure; TextAreaFigure; NodeFigure; SVGImage; SVGPath; DependencyFigure; LineConnectionFigure; AbstractCompositeFigure; GraphicalCompositeFigure; BasicTransformVisitor; ContainsVisitor; SetAttributeVisitor; FindFigureInsideVisitor; AddNotifyVisitor; RemoveNotifyVisitor]) \textbf{; }\\
\operation{InlineAndDelete} (AbstractCompositeFigure, addNotifyTmpVC, [AddNotifyVisitor], addNotify, [basicTransform; contains; setAttribute; findFigureInside; removeNotify], [Visitor; AbstractFigure; EllipseFigure; DiamondFigure; RectangleFigure; RoundRectangleFigure; TriangleFigure; TextFigure; BezierFigure; TextAreaFigure; NodeFigure; SVGImage; SVGPath; DependencyFigure; LineConnectionFigure; LabeledLineConnectionFigure; GraphicalCompositeFigure; BasicTransformVisitor; ContainsVisitor; SetAttributeVisitor; FindFigureInsideVisitor; AddNotifyVisitor; RemoveNotifyVisitor]) \textbf{; }\\
\operation{InlineAndDelete} (GraphicalCompositeFigure, addNotifyTmpVC, [AddNotifyVisitor], addNotify, [basicTransform; contains; setAttribute; findFigureInside; removeNotify], [Visitor; AbstractFigure; EllipseFigure; DiamondFigure; RectangleFigure; RoundRectangleFigure; TriangleFigure; TextFigure; BezierFigure; TextAreaFigure; NodeFigure; SVGImage; SVGPath; DependencyFigure; LineConnectionFigure; LabeledLineConnectionFigure; AbstractCompositeFigure; BasicTransformVisitor; ContainsVisitor; SetAttributeVisitor; FindFigureInsideVisitor; AddNotifyVisitor; RemoveNotifyVisitor]) \textbf{; }\\
\operation{InlineAndDelete} (EllipseFigure, removeNotifyTmpVC, [RemoveNotifyVisitor], removeNotify, [basicTransform; contains; setAttribute; findFigureInside; addNotify], [Visitor; AbstractFigure; DiamondFigure; RectangleFigure; RoundRectangleFigure; TriangleFigure; TextFigure; BezierFigure; TextAreaFigure; NodeFigure; SVGImage; SVGPath; DependencyFigure; LineConnectionFigure; LabeledLineConnectionFigure; AbstractCompositeFigure; GraphicalCompositeFigure; BasicTransformVisitor; ContainsVisitor; SetAttributeVisitor; FindFigureInsideVisitor; AddNotifyVisitor; RemoveNotifyVisitor]) \textbf{; }\\
\operation{InlineAndDelete} (DiamondFigure, removeNotifyTmpVC, [RemoveNotifyVisitor], removeNotify, [basicTransform; contains; setAttribute; findFigureInside; addNotify], [Visitor; AbstractFigure; EllipseFigure; RectangleFigure; RoundRectangleFigure; TriangleFigure; TextFigure; BezierFigure; TextAreaFigure; NodeFigure; SVGImage; SVGPath; DependencyFigure; LineConnectionFigure; LabeledLineConnectionFigure; AbstractCompositeFigure; GraphicalCompositeFigure; BasicTransformVisitor; ContainsVisitor; SetAttributeVisitor; FindFigureInsideVisitor; AddNotifyVisitor; RemoveNotifyVisitor]) \textbf{; }\\
\operation{InlineAndDelete} (RectangleFigure, removeNotifyTmpVC, [RemoveNotifyVisitor], removeNotify, [basicTransform; contains; setAttribute; findFigureInside; addNotify], [Visitor; AbstractFigure; EllipseFigure; DiamondFigure; RoundRectangleFigure; TriangleFigure; TextFigure; BezierFigure; TextAreaFigure; NodeFigure; SVGImage; SVGPath; DependencyFigure; LineConnectionFigure; LabeledLineConnectionFigure; AbstractCompositeFigure; GraphicalCompositeFigure; BasicTransformVisitor; ContainsVisitor; SetAttributeVisitor; FindFigureInsideVisitor; AddNotifyVisitor; RemoveNotifyVisitor]) \textbf{; }\\
\operation{InlineAndDelete} (RoundRectangleFigure, removeNotifyTmpVC, [RemoveNotifyVisitor], removeNotify, [basicTransform; contains; setAttribute; findFigureInside; addNotify], [Visitor; AbstractFigure; EllipseFigure; DiamondFigure; RectangleFigure; TriangleFigure; TextFigure; BezierFigure; TextAreaFigure; NodeFigure; SVGImage; SVGPath; DependencyFigure; LineConnectionFigure; LabeledLineConnectionFigure; AbstractCompositeFigure; GraphicalCompositeFigure; BasicTransformVisitor; ContainsVisitor; SetAttributeVisitor; FindFigureInsideVisitor; AddNotifyVisitor; RemoveNotifyVisitor]) \textbf{; }\\
\operation{InlineAndDelete} (TriangleFigure, removeNotifyTmpVC, [RemoveNotifyVisitor], removeNotify, [basicTransform; contains; setAttribute; findFigureInside; addNotify], [Visitor; AbstractFigure; EllipseFigure; DiamondFigure; RectangleFigure; RoundRectangleFigure; TextFigure; BezierFigure; TextAreaFigure; NodeFigure; SVGImage; SVGPath; DependencyFigure; LineConnectionFigure; LabeledLineConnectionFigure; AbstractCompositeFigure; GraphicalCompositeFigure; BasicTransformVisitor; ContainsVisitor; SetAttributeVisitor; FindFigureInsideVisitor; AddNotifyVisitor; RemoveNotifyVisitor]) \textbf{; }\\
\operation{InlineAndDelete} (TextFigure, removeNotifyTmpVC, [RemoveNotifyVisitor], removeNotify, [basicTransform; contains; setAttribute; findFigureInside; addNotify], [Visitor; AbstractFigure; EllipseFigure; DiamondFigure; RectangleFigure; RoundRectangleFigure; TriangleFigure; BezierFigure; TextAreaFigure; NodeFigure; SVGImage; SVGPath; DependencyFigure; LineConnectionFigure; LabeledLineConnectionFigure; AbstractCompositeFigure; GraphicalCompositeFigure; BasicTransformVisitor; ContainsVisitor; SetAttributeVisitor; FindFigureInsideVisitor; AddNotifyVisitor; RemoveNotifyVisitor]) \textbf{; }\\
\operation{InlineAndDelete} (BezierFigure, removeNotifyTmpVC, [RemoveNotifyVisitor], removeNotify, [basicTransform; contains; setAttribute; findFigureInside; addNotify], [Visitor; AbstractFigure; EllipseFigure; DiamondFigure; RectangleFigure; RoundRectangleFigure; TriangleFigure; TextFigure; TextAreaFigure; NodeFigure; SVGImage; SVGPath; DependencyFigure; LineConnectionFigure; LabeledLineConnectionFigure; AbstractCompositeFigure; GraphicalCompositeFigure; BasicTransformVisitor; ContainsVisitor; SetAttributeVisitor; FindFigureInsideVisitor; AddNotifyVisitor; RemoveNotifyVisitor]) \textbf{; }\\
\operation{InlineAndDelete} (TextAreaFigure, removeNotifyTmpVC, [RemoveNotifyVisitor], removeNotify, [basicTransform; contains; setAttribute; findFigureInside; addNotify], [Visitor; AbstractFigure; EllipseFigure; DiamondFigure; RectangleFigure; RoundRectangleFigure; TriangleFigure; TextFigure; BezierFigure; NodeFigure; SVGImage; SVGPath; DependencyFigure; LineConnectionFigure; LabeledLineConnectionFigure; AbstractCompositeFigure; GraphicalCompositeFigure; BasicTransformVisitor; ContainsVisitor; SetAttributeVisitor; FindFigureInsideVisitor; AddNotifyVisitor; RemoveNotifyVisitor]) \textbf{; }\\
\operation{InlineAndDelete} (NodeFigure, removeNotifyTmpVC, [RemoveNotifyVisitor], removeNotify, [basicTransform; contains; setAttribute; findFigureInside; addNotify], [Visitor; AbstractFigure; EllipseFigure; DiamondFigure; RectangleFigure; RoundRectangleFigure; TriangleFigure; TextFigure; BezierFigure; TextAreaFigure; SVGImage; SVGPath; DependencyFigure; LineConnectionFigure; LabeledLineConnectionFigure; AbstractCompositeFigure; GraphicalCompositeFigure; BasicTransformVisitor; ContainsVisitor; SetAttributeVisitor; FindFigureInsideVisitor; AddNotifyVisitor; RemoveNotifyVisitor]) \textbf{; }\\
\operation{InlineAndDelete} (SVGImage, removeNotifyTmpVC, [RemoveNotifyVisitor], removeNotify, [basicTransform; contains; setAttribute; findFigureInside; addNotify], [Visitor; AbstractFigure; EllipseFigure; DiamondFigure; RectangleFigure; RoundRectangleFigure; TriangleFigure; TextFigure; BezierFigure; TextAreaFigure; NodeFigure; SVGPath; DependencyFigure; LineConnectionFigure; LabeledLineConnectionFigure; AbstractCompositeFigure; GraphicalCompositeFigure; BasicTransformVisitor; ContainsVisitor; SetAttributeVisitor; FindFigureInsideVisitor; AddNotifyVisitor; RemoveNotifyVisitor]) \textbf{; }\\
\operation{InlineAndDelete} (SVGPath, removeNotifyTmpVC, [RemoveNotifyVisitor], removeNotify, [basicTransform; contains; setAttribute; findFigureInside; addNotify], [Visitor; AbstractFigure; EllipseFigure; DiamondFigure; RectangleFigure; RoundRectangleFigure; TriangleFigure; TextFigure; BezierFigure; TextAreaFigure; NodeFigure; SVGImage; DependencyFigure; LineConnectionFigure; LabeledLineConnectionFigure; AbstractCompositeFigure; GraphicalCompositeFigure; BasicTransformVisitor; ContainsVisitor; SetAttributeVisitor; FindFigureInsideVisitor; AddNotifyVisitor; RemoveNotifyVisitor]) \textbf{; }\\
\operation{InlineAndDelete} (DependencyFigure, removeNotifyTmpVC, [RemoveNotifyVisitor], removeNotify, [basicTransform; contains; setAttribute; findFigureInside; addNotify], [Visitor; AbstractFigure; EllipseFigure; DiamondFigure; RectangleFigure; RoundRectangleFigure; TriangleFigure; TextFigure; BezierFigure; TextAreaFigure; NodeFigure; SVGImage; SVGPath; LineConnectionFigure; LabeledLineConnectionFigure; AbstractCompositeFigure; GraphicalCompositeFigure; BasicTransformVisitor; ContainsVisitor; SetAttributeVisitor; FindFigureInsideVisitor; AddNotifyVisitor; RemoveNotifyVisitor]) \textbf{; }\\
\operation{InlineAndDelete} (LineConnectionFigure, removeNotifyTmpVC, [RemoveNotifyVisitor], removeNotify, [basicTransform; contains; setAttribute; findFigureInside; addNotify], [Visitor; AbstractFigure; EllipseFigure; DiamondFigure; RectangleFigure; RoundRectangleFigure; TriangleFigure; TextFigure; BezierFigure; TextAreaFigure; NodeFigure; SVGImage; SVGPath; DependencyFigure; LabeledLineConnectionFigure; AbstractCompositeFigure; GraphicalCompositeFigure; BasicTransformVisitor; ContainsVisitor; SetAttributeVisitor; FindFigureInsideVisitor; AddNotifyVisitor; RemoveNotifyVisitor]) \textbf{; }\\
\operation{InlineAndDelete} (LabeledLineConnectionFigure, removeNotifyTmpVC, [RemoveNotifyVisitor], removeNotify, [basicTransform; contains; setAttribute; findFigureInside; addNotify], [Visitor; AbstractFigure; EllipseFigure; DiamondFigure; RectangleFigure; RoundRectangleFigure; TriangleFigure; TextFigure; BezierFigure; TextAreaFigure; NodeFigure; SVGImage; SVGPath; DependencyFigure; LineConnectionFigure; AbstractCompositeFigure; GraphicalCompositeFigure; BasicTransformVisitor; ContainsVisitor; SetAttributeVisitor; FindFigureInsideVisitor; AddNotifyVisitor; RemoveNotifyVisitor]) \textbf{; }\\
\operation{InlineAndDelete} (AbstractCompositeFigure, removeNotifyTmpVC, [RemoveNotifyVisitor], removeNotify, [basicTransform; contains; setAttribute; findFigureInside; addNotify], [Visitor; AbstractFigure; EllipseFigure; DiamondFigure; RectangleFigure; RoundRectangleFigure; TriangleFigure; TextFigure; BezierFigure; TextAreaFigure; NodeFigure; SVGImage; SVGPath; DependencyFigure; LineConnectionFigure; LabeledLineConnectionFigure; GraphicalCompositeFigure; BasicTransformVisitor; ContainsVisitor; SetAttributeVisitor; FindFigureInsideVisitor; AddNotifyVisitor; RemoveNotifyVisitor]) \textbf{; }\\
\operation{InlineAndDelete} (GraphicalCompositeFigure, removeNotifyTmpVC, [RemoveNotifyVisitor], removeNotify, [basicTransform; contains; setAttribute; findFigureInside; addNotify], [Visitor; AbstractFigure; EllipseFigure; DiamondFigure; RectangleFigure; RoundRectangleFigure; TriangleFigure; TextFigure; BezierFigure; TextAreaFigure; NodeFigure; SVGImage; SVGPath; DependencyFigure; LineConnectionFigure; LabeledLineConnectionFigure; AbstractCompositeFigure; BasicTransformVisitor; ContainsVisitor; SetAttributeVisitor; FindFigureInsideVisitor; AddNotifyVisitor; RemoveNotifyVisitor]) \textbf{; }\\
\operation{InlineConstructor}(EllipseFigure, basicTransform, BasicTransformVisitor, [tx], [gettx]) \textbf{; }\\
\operation{InlineLocalField}(EllipseFigure, basicTransform, tx) \textbf{; }\\
\operation{InlineAndDelete} (EllipseFigure, gettx, [], basicTransform, [], []) \textbf{; }\\
\operation{InlinelocalVariable}(EllipseFigure, basicTransform, txvar) \textbf{; }\\
\operation{InlineConstructor}(DiamondFigure, basicTransform, BasicTransformVisitor, [tx], [gettx]) \textbf{; }\\
\operation{InlineLocalField}(DiamondFigure, basicTransform, tx) \textbf{; }\\
\operation{InlineAndDelete} (DiamondFigure, gettx, [], basicTransform, [], []) \textbf{; }\\
\operation{InlinelocalVariable}(DiamondFigure, basicTransform, txvar) \textbf{; }\\
\operation{InlineConstructor}(RectangleFigure, basicTransform, BasicTransformVisitor, [tx], [gettx]) \textbf{; }\\
\operation{InlineLocalField}(RectangleFigure, basicTransform, tx) \textbf{; }\\
\operation{InlineAndDelete} (RectangleFigure, gettx, [], basicTransform, [], []) \textbf{; }\\
\operation{InlinelocalVariable}(RectangleFigure, basicTransform, txvar) \textbf{; }\\
\operation{InlineConstructor}(RoundRectangleFigure, basicTransform, BasicTransformVisitor, [tx], [gettx]) \textbf{; }\\
\operation{InlineLocalField}(RoundRectangleFigure, basicTransform, tx) \textbf{; }\\
\operation{InlineAndDelete} (RoundRectangleFigure, gettx, [], basicTransform, [], []) \textbf{; }\\
\operation{InlinelocalVariable}(RoundRectangleFigure, basicTransform, txvar) \textbf{; }\\
\operation{InlineConstructor}(TriangleFigure, basicTransform, BasicTransformVisitor, [tx], [gettx]) \textbf{; }\\
\operation{InlineLocalField}(TriangleFigure, basicTransform, tx) \textbf{; }\\
\operation{InlineAndDelete} (TriangleFigure, gettx, [], basicTransform, [], []) \textbf{; }\\
\operation{InlinelocalVariable}(TriangleFigure, basicTransform, txvar) \textbf{; }\\
\operation{InlineConstructor}(TextFigure, basicTransform, BasicTransformVisitor, [tx], [gettx]) \textbf{; }\\
\operation{InlineLocalField}(TextFigure, basicTransform, tx) \textbf{; }\\
\operation{InlineAndDelete} (TextFigure, gettx, [], basicTransform, [], []) \textbf{; }\\
\operation{InlinelocalVariable}(TextFigure, basicTransform, txvar) \textbf{; }\\
\operation{InlineConstructor}(BezierFigure, basicTransform, BasicTransformVisitor, [tx], [gettx]) \textbf{; }\\
\operation{InlineLocalField}(BezierFigure, basicTransform, tx) \textbf{; }\\
\operation{InlineAndDelete} (BezierFigure, gettx, [], basicTransform, [], []) \textbf{; }\\
\operation{InlinelocalVariable}(BezierFigure, basicTransform, txvar) \textbf{; }\\
\operation{InlineConstructor}(TextAreaFigure, basicTransform, BasicTransformVisitor, [tx], [gettx]) \textbf{; }\\
\operation{InlineLocalField}(TextAreaFigure, basicTransform, tx) \textbf{; }\\
\operation{InlineAndDelete} (TextAreaFigure, gettx, [], basicTransform, [], []) \textbf{; }\\
\operation{InlinelocalVariable}(TextAreaFigure, basicTransform, txvar) \textbf{; }\\
\operation{InlineConstructor}(NodeFigure, basicTransform, BasicTransformVisitor, [tx], [gettx]) \textbf{; }\\
\operation{InlineLocalField}(NodeFigure, basicTransform, tx) \textbf{; }\\
\operation{InlineAndDelete} (NodeFigure, gettx, [], basicTransform, [], []) \textbf{; }\\
\operation{InlinelocalVariable}(NodeFigure, basicTransform, txvar) \textbf{; }\\
\operation{InlineConstructor}(SVGImage, basicTransform, BasicTransformVisitor, [tx], [gettx]) \textbf{; }\\
\operation{InlineLocalField}(SVGImage, basicTransform, tx) \textbf{; }\\
\operation{InlineAndDelete} (SVGImage, gettx, [], basicTransform, [], []) \textbf{; }\\
\operation{InlinelocalVariable}(SVGImage, basicTransform, txvar) \textbf{; }\\
\operation{InlineConstructor}(SVGPath, basicTransform, BasicTransformVisitor, [tx], [gettx]) \textbf{; }\\
\operation{InlineLocalField}(SVGPath, basicTransform, tx) \textbf{; }\\
\operation{InlineAndDelete} (SVGPath, gettx, [], basicTransform, [], []) \textbf{; }\\
\operation{InlinelocalVariable}(SVGPath, basicTransform, txvar) \textbf{; }\\
\operation{InlineConstructor}(DependencyFigure, basicTransform, BasicTransformVisitor, [tx], [gettx]) \textbf{; }\\
\operation{InlineLocalField}(DependencyFigure, basicTransform, tx) \textbf{; }\\
\operation{InlineAndDelete} (DependencyFigure, gettx, [], basicTransform, [], []) \textbf{; }\\
\operation{InlinelocalVariable}(DependencyFigure, basicTransform, txvar) \textbf{; }\\
\operation{InlineConstructor}(LineConnectionFigure, basicTransform, BasicTransformVisitor, [tx], [gettx]) \textbf{; }\\
\operation{InlineLocalField}(LineConnectionFigure, basicTransform, tx) \textbf{; }\\
\operation{InlineAndDelete} (LineConnectionFigure, gettx, [], basicTransform, [], []) \textbf{; }\\
\operation{InlinelocalVariable}(LineConnectionFigure, basicTransform, txvar) \textbf{; }\\
\operation{InlineConstructor}(LabeledLineConnectionFigure, basicTransform, BasicTransformVisitor, [tx], [gettx]) \textbf{; }\\
\operation{InlineLocalField}(LabeledLineConnectionFigure, basicTransform, tx) \textbf{; }\\
\operation{InlineAndDelete} (LabeledLineConnectionFigure, gettx, [], basicTransform, [], []) \textbf{; }\\
\operation{InlinelocalVariable}(LabeledLineConnectionFigure, basicTransform, txvar) \textbf{; }\\
\operation{InlineConstructor}(AbstractCompositeFigure, basicTransform, BasicTransformVisitor, [tx], [gettx]) \textbf{; }\\
\operation{InlineLocalField}(AbstractCompositeFigure, basicTransform, tx) \textbf{; }\\
\operation{InlineAndDelete} (AbstractCompositeFigure, gettx, [], basicTransform, [], []) \textbf{; }\\
\operation{InlinelocalVariable}(AbstractCompositeFigure, basicTransform, txvar) \textbf{; }\\
\operation{InlineConstructor}(GraphicalCompositeFigure, basicTransform, BasicTransformVisitor, [tx], [gettx]) \textbf{; }\\
\operation{InlineLocalField}(GraphicalCompositeFigure, basicTransform, tx) \textbf{; }\\
\operation{InlineAndDelete} (GraphicalCompositeFigure, gettx, [], basicTransform, [], []) \textbf{; }\\
\operation{InlinelocalVariable}(GraphicalCompositeFigure, basicTransform, txvar) \textbf{; }\\
\operation{InlineConstructor}(EllipseFigure, contains, ContainsVisitor, [p], [getp]) \textbf{; }\\
\operation{InlineLocalField}(EllipseFigure, contains, p) \textbf{; }\\
\operation{InlineAndDelete} (EllipseFigure, getp, [], contains, [], []) \textbf{; }\\
\operation{InlinelocalVariable}(EllipseFigure, contains, pvar) \textbf{; }\\
\operation{InlineConstructor}(DiamondFigure, contains, ContainsVisitor, [p], [getp]) \textbf{; }\\
\operation{InlineLocalField}(DiamondFigure, contains, p) \textbf{; }\\
\operation{InlineAndDelete} (DiamondFigure, getp, [], contains, [], []) \textbf{; }\\
\operation{InlinelocalVariable}(DiamondFigure, contains, pvar) \textbf{; }\\
\operation{InlineConstructor}(RectangleFigure, contains, ContainsVisitor, [p], [getp]) \textbf{; }\\
\operation{InlineLocalField}(RectangleFigure, contains, p) \textbf{; }\\
\operation{InlineAndDelete} (RectangleFigure, getp, [], contains, [], []) \textbf{; }\\
\operation{InlinelocalVariable}(RectangleFigure, contains, pvar) \textbf{; }\\
\operation{InlineConstructor}(RoundRectangleFigure, contains, ContainsVisitor, [p], [getp]) \textbf{; }\\
\operation{InlineLocalField}(RoundRectangleFigure, contains, p) \textbf{; }\\
\operation{InlineAndDelete} (RoundRectangleFigure, getp, [], contains, [], []) \textbf{; }\\
\operation{InlinelocalVariable}(RoundRectangleFigure, contains, pvar) \textbf{; }\\
\operation{InlineConstructor}(TriangleFigure, contains, ContainsVisitor, [p], [getp]) \textbf{; }\\
\operation{InlineLocalField}(TriangleFigure, contains, p) \textbf{; }\\
\operation{InlineAndDelete} (TriangleFigure, getp, [], contains, [], []) \textbf{; }\\
\operation{InlinelocalVariable}(TriangleFigure, contains, pvar) \textbf{; }\\
\operation{InlineConstructor}(TextFigure, contains, ContainsVisitor, [p], [getp]) \textbf{; }\\
\operation{InlineLocalField}(TextFigure, contains, p) \textbf{; }\\
\operation{InlineAndDelete} (TextFigure, getp, [], contains, [], []) \textbf{; }\\
\operation{InlinelocalVariable}(TextFigure, contains, pvar) \textbf{; }\\
\operation{InlineConstructor}(BezierFigure, contains, ContainsVisitor, [p], [getp]) \textbf{; }\\
\operation{InlineLocalField}(BezierFigure, contains, p) \textbf{; }\\
\operation{InlineAndDelete} (BezierFigure, getp, [], contains, [], []) \textbf{; }\\
\operation{InlinelocalVariable}(BezierFigure, contains, pvar) \textbf{; }\\
\operation{InlineConstructor}(TextAreaFigure, contains, ContainsVisitor, [p], [getp]) \textbf{; }\\
\operation{InlineLocalField}(TextAreaFigure, contains, p) \textbf{; }\\
\operation{InlineAndDelete} (TextAreaFigure, getp, [], contains, [], []) \textbf{; }\\
\operation{InlinelocalVariable}(TextAreaFigure, contains, pvar) \textbf{; }\\
\operation{InlineConstructor}(NodeFigure, contains, ContainsVisitor, [p], [getp]) \textbf{; }\\
\operation{InlineLocalField}(NodeFigure, contains, p) \textbf{; }\\
\operation{InlineAndDelete} (NodeFigure, getp, [], contains, [], []) \textbf{; }\\
\operation{InlinelocalVariable}(NodeFigure, contains, pvar) \textbf{; }\\
\operation{InlineConstructor}(SVGImage, contains, ContainsVisitor, [p], [getp]) \textbf{; }\\
\operation{InlineLocalField}(SVGImage, contains, p) \textbf{; }\\
\operation{InlineAndDelete} (SVGImage, getp, [], contains, [], []) \textbf{; }\\
\operation{InlinelocalVariable}(SVGImage, contains, pvar) \textbf{; }\\
\operation{InlineConstructor}(SVGPath, contains, ContainsVisitor, [p], [getp]) \textbf{; }\\
\operation{InlineLocalField}(SVGPath, contains, p) \textbf{; }\\
\operation{InlineAndDelete} (SVGPath, getp, [], contains, [], []) \textbf{; }\\
\operation{InlinelocalVariable}(SVGPath, contains, pvar) \textbf{; }\\
\operation{InlineConstructor}(DependencyFigure, contains, ContainsVisitor, [p], [getp]) \textbf{; }\\
\operation{InlineLocalField}(DependencyFigure, contains, p) \textbf{; }\\
\operation{InlineAndDelete} (DependencyFigure, getp, [], contains, [], []) \textbf{; }\\
\operation{InlinelocalVariable}(DependencyFigure, contains, pvar) \textbf{; }\\
\operation{InlineConstructor}(LineConnectionFigure, contains, ContainsVisitor, [p], [getp]) \textbf{; }\\
\operation{InlineLocalField}(LineConnectionFigure, contains, p) \textbf{; }\\
\operation{InlineAndDelete} (LineConnectionFigure, getp, [], contains, [], []) \textbf{; }\\
\operation{InlinelocalVariable}(LineConnectionFigure, contains, pvar) \textbf{; }\\
\operation{InlineConstructor}(LabeledLineConnectionFigure, contains, ContainsVisitor, [p], [getp]) \textbf{; }\\
\operation{InlineLocalField}(LabeledLineConnectionFigure, contains, p) \textbf{; }\\
\operation{InlineAndDelete} (LabeledLineConnectionFigure, getp, [], contains, [], []) \textbf{; }\\
\operation{InlinelocalVariable}(LabeledLineConnectionFigure, contains, pvar) \textbf{; }\\
\operation{InlineConstructor}(AbstractCompositeFigure, contains, ContainsVisitor, [p], [getp]) \textbf{; }\\
\operation{InlineLocalField}(AbstractCompositeFigure, contains, p) \textbf{; }\\
\operation{InlineAndDelete} (AbstractCompositeFigure, getp, [], contains, [], []) \textbf{; }\\
\operation{InlinelocalVariable}(AbstractCompositeFigure, contains, pvar) \textbf{; }\\
\operation{InlineConstructor}(GraphicalCompositeFigure, contains, ContainsVisitor, [p], [getp]) \textbf{; }\\
\operation{InlineLocalField}(GraphicalCompositeFigure, contains, p) \textbf{; }\\
\operation{InlineAndDelete} (GraphicalCompositeFigure, getp, [], contains, [], []) \textbf{; }\\
\operation{InlinelocalVariable}(GraphicalCompositeFigure, contains, pvar) \textbf{; }\\
\operation{InlineConstructor}(EllipseFigure, setAttribute, SetAttributeVisitor, [key; value], [getkey; getvalue]) \textbf{; }\\
\operation{InlineLocalField}(EllipseFigure, setAttribute, key) \textbf{; }\\
\operation{InlineLocalField}(EllipseFigure, setAttribute, value) \textbf{; }\\
\operation{InlineAndDelete} (EllipseFigure, getkey, [], setAttribute, [], []) \textbf{; }\\
\operation{InlineAndDelete} (EllipseFigure, getvalue, [], setAttribute, [], []) \textbf{; }\\
\operation{InlinelocalVariable}(EllipseFigure, setAttribute, keyvar) \textbf{; }\\
\operation{InlinelocalVariable}(EllipseFigure, setAttribute, valuevar) \textbf{; }\\
\operation{InlineConstructor}(DiamondFigure, setAttribute, SetAttributeVisitor, [key; value], [getkey; getvalue]) \textbf{; }\\
\operation{InlineLocalField}(DiamondFigure, setAttribute, key) \textbf{; }\\
\operation{InlineLocalField}(DiamondFigure, setAttribute, value) \textbf{; }\\
\operation{InlineAndDelete} (DiamondFigure, getkey, [], setAttribute, [], []) \textbf{; }\\
\operation{InlineAndDelete} (DiamondFigure, getvalue, [], setAttribute, [], []) \textbf{; }\\
\operation{InlinelocalVariable}(DiamondFigure, setAttribute, keyvar) \textbf{; }\\
\operation{InlinelocalVariable}(DiamondFigure, setAttribute, valuevar) \textbf{; }\\
\operation{InlineConstructor}(RectangleFigure, setAttribute, SetAttributeVisitor, [key; value], [getkey; getvalue]) \textbf{; }\\
\operation{InlineLocalField}(RectangleFigure, setAttribute, key) \textbf{; }\\
\operation{InlineLocalField}(RectangleFigure, setAttribute, value) \textbf{; }\\
\operation{InlineAndDelete} (RectangleFigure, getkey, [], setAttribute, [], []) \textbf{; }\\
\operation{InlineAndDelete} (RectangleFigure, getvalue, [], setAttribute, [], []) \textbf{; }\\
\operation{InlinelocalVariable}(RectangleFigure, setAttribute, keyvar) \textbf{; }\\
\operation{InlinelocalVariable}(RectangleFigure, setAttribute, valuevar) \textbf{; }\\
\operation{InlineConstructor}(RoundRectangleFigure, setAttribute, SetAttributeVisitor, [key; value], [getkey; getvalue]) \textbf{; }\\
\operation{InlineLocalField}(RoundRectangleFigure, setAttribute, key) \textbf{; }\\
\operation{InlineLocalField}(RoundRectangleFigure, setAttribute, value) \textbf{; }\\
\operation{InlineAndDelete} (RoundRectangleFigure, getkey, [], setAttribute, [], []) \textbf{; }\\
\operation{InlineAndDelete} (RoundRectangleFigure, getvalue, [], setAttribute, [], []) \textbf{; }\\
\operation{InlinelocalVariable}(RoundRectangleFigure, setAttribute, keyvar) \textbf{; }\\
\operation{InlinelocalVariable}(RoundRectangleFigure, setAttribute, valuevar) \textbf{; }\\
\operation{InlineConstructor}(TriangleFigure, setAttribute, SetAttributeVisitor, [key; value], [getkey; getvalue]) \textbf{; }\\
\operation{InlineLocalField}(TriangleFigure, setAttribute, key) \textbf{; }\\
\operation{InlineLocalField}(TriangleFigure, setAttribute, value) \textbf{; }\\
\operation{InlineAndDelete} (TriangleFigure, getkey, [], setAttribute, [], []) \textbf{; }\\
\operation{InlineAndDelete} (TriangleFigure, getvalue, [], setAttribute, [], []) \textbf{; }\\
\operation{InlinelocalVariable}(TriangleFigure, setAttribute, keyvar) \textbf{; }\\
\operation{InlinelocalVariable}(TriangleFigure, setAttribute, valuevar) \textbf{; }\\
\operation{InlineConstructor}(TextFigure, setAttribute, SetAttributeVisitor, [key; value], [getkey; getvalue]) \textbf{; }\\
\operation{InlineLocalField}(TextFigure, setAttribute, key) \textbf{; }\\
\operation{InlineLocalField}(TextFigure, setAttribute, value) \textbf{; }\\
\operation{InlineAndDelete} (TextFigure, getkey, [], setAttribute, [], []) \textbf{; }\\
\operation{InlineAndDelete} (TextFigure, getvalue, [], setAttribute, [], []) \textbf{; }\\
\operation{InlinelocalVariable}(TextFigure, setAttribute, keyvar) \textbf{; }\\
\operation{InlinelocalVariable}(TextFigure, setAttribute, valuevar) \textbf{; }\\
\operation{InlineConstructor}(BezierFigure, setAttribute, SetAttributeVisitor, [key; value], [getkey; getvalue]) \textbf{; }\\
\operation{InlineLocalField}(BezierFigure, setAttribute, key) \textbf{; }\\
\operation{InlineLocalField}(BezierFigure, setAttribute, value) \textbf{; }\\
\operation{InlineAndDelete} (BezierFigure, getkey, [], setAttribute, [], []) \textbf{; }\\
\operation{InlineAndDelete} (BezierFigure, getvalue, [], setAttribute, [], []) \textbf{; }\\
\operation{InlinelocalVariable}(BezierFigure, setAttribute, keyvar) \textbf{; }\\
\operation{InlinelocalVariable}(BezierFigure, setAttribute, valuevar) \textbf{; }\\
\operation{InlineConstructor}(TextAreaFigure, setAttribute, SetAttributeVisitor, [key; value], [getkey; getvalue]) \textbf{; }\\
\operation{InlineLocalField}(TextAreaFigure, setAttribute, key) \textbf{; }\\
\operation{InlineLocalField}(TextAreaFigure, setAttribute, value) \textbf{; }\\
\operation{InlineAndDelete} (TextAreaFigure, getkey, [], setAttribute, [], []) \textbf{; }\\
\operation{InlineAndDelete} (TextAreaFigure, getvalue, [], setAttribute, [], []) \textbf{; }\\
\operation{InlinelocalVariable}(TextAreaFigure, setAttribute, keyvar) \textbf{; }\\
\operation{InlinelocalVariable}(TextAreaFigure, setAttribute, valuevar) \textbf{; }\\
\operation{InlineConstructor}(NodeFigure, setAttribute, SetAttributeVisitor, [key; value], [getkey; getvalue]) \textbf{; }\\
\operation{InlineLocalField}(NodeFigure, setAttribute, key) \textbf{; }\\
\operation{InlineLocalField}(NodeFigure, setAttribute, value) \textbf{; }\\
\operation{InlineAndDelete} (NodeFigure, getkey, [], setAttribute, [], []) \textbf{; }\\
\operation{InlineAndDelete} (NodeFigure, getvalue, [], setAttribute, [], []) \textbf{; }\\
\operation{InlinelocalVariable}(NodeFigure, setAttribute, keyvar) \textbf{; }\\
\operation{InlinelocalVariable}(NodeFigure, setAttribute, valuevar) \textbf{; }\\
\operation{InlineConstructor}(SVGImage, setAttribute, SetAttributeVisitor, [key; value], [getkey; getvalue]) \textbf{; }\\
\operation{InlineLocalField}(SVGImage, setAttribute, key) \textbf{; }\\
\operation{InlineLocalField}(SVGImage, setAttribute, value) \textbf{; }\\
\operation{InlineAndDelete} (SVGImage, getkey, [], setAttribute, [], []) \textbf{; }\\
\operation{InlineAndDelete} (SVGImage, getvalue, [], setAttribute, [], []) \textbf{; }\\
\operation{InlinelocalVariable}(SVGImage, setAttribute, keyvar) \textbf{; }\\
\operation{InlinelocalVariable}(SVGImage, setAttribute, valuevar) \textbf{; }\\
\operation{InlineConstructor}(SVGPath, setAttribute, SetAttributeVisitor, [key; value], [getkey; getvalue]) \textbf{; }\\
\operation{InlineLocalField}(SVGPath, setAttribute, key) \textbf{; }\\
\operation{InlineLocalField}(SVGPath, setAttribute, value) \textbf{; }\\
\operation{InlineAndDelete} (SVGPath, getkey, [], setAttribute, [], []) \textbf{; }\\
\operation{InlineAndDelete} (SVGPath, getvalue, [], setAttribute, [], []) \textbf{; }\\
\operation{InlinelocalVariable}(SVGPath, setAttribute, keyvar) \textbf{; }\\
\operation{InlinelocalVariable}(SVGPath, setAttribute, valuevar) \textbf{; }\\
\operation{InlineConstructor}(DependencyFigure, setAttribute, SetAttributeVisitor, [key; value], [getkey; getvalue]) \textbf{; }\\
\operation{InlineLocalField}(DependencyFigure, setAttribute, key) \textbf{; }\\
\operation{InlineLocalField}(DependencyFigure, setAttribute, value) \textbf{; }\\
\operation{InlineAndDelete} (DependencyFigure, getkey, [], setAttribute, [], []) \textbf{; }\\
\operation{InlineAndDelete} (DependencyFigure, getvalue, [], setAttribute, [], []) \textbf{; }\\
\operation{InlinelocalVariable}(DependencyFigure, setAttribute, keyvar) \textbf{; }\\
\operation{InlinelocalVariable}(DependencyFigure, setAttribute, valuevar) \textbf{; }\\
\operation{InlineConstructor}(LineConnectionFigure, setAttribute, SetAttributeVisitor, [key; value], [getkey; getvalue]) \textbf{; }\\
\operation{InlineLocalField}(LineConnectionFigure, setAttribute, key) \textbf{; }\\
\operation{InlineLocalField}(LineConnectionFigure, setAttribute, value) \textbf{; }\\
\operation{InlineAndDelete} (LineConnectionFigure, getkey, [], setAttribute, [], []) \textbf{; }\\
\operation{InlineAndDelete} (LineConnectionFigure, getvalue, [], setAttribute, [], []) \textbf{; }\\
\operation{InlinelocalVariable}(LineConnectionFigure, setAttribute, keyvar) \textbf{; }\\
\operation{InlinelocalVariable}(LineConnectionFigure, setAttribute, valuevar) \textbf{; }\\
\operation{InlineConstructor}(LabeledLineConnectionFigure, setAttribute, SetAttributeVisitor, [key; value], [getkey; getvalue]) \textbf{; }\\
\operation{InlineLocalField}(LabeledLineConnectionFigure, setAttribute, key) \textbf{; }\\
\operation{InlineLocalField}(LabeledLineConnectionFigure, setAttribute, value) \textbf{; }\\
\operation{InlineAndDelete} (LabeledLineConnectionFigure, getkey, [], setAttribute, [], []) \textbf{; }\\
\operation{InlineAndDelete} (LabeledLineConnectionFigure, getvalue, [], setAttribute, [], []) \textbf{; }\\
\operation{InlinelocalVariable}(LabeledLineConnectionFigure, setAttribute, keyvar) \textbf{; }\\
\operation{InlinelocalVariable}(LabeledLineConnectionFigure, setAttribute, valuevar) \textbf{; }\\
\operation{InlineConstructor}(AbstractCompositeFigure, setAttribute, SetAttributeVisitor, [key; value], [getkey; getvalue]) \textbf{; }\\
\operation{InlineLocalField}(AbstractCompositeFigure, setAttribute, key) \textbf{; }\\
\operation{InlineLocalField}(AbstractCompositeFigure, setAttribute, value) \textbf{; }\\
\operation{InlineAndDelete} (AbstractCompositeFigure, getkey, [], setAttribute, [], []) \textbf{; }\\
\operation{InlineAndDelete} (AbstractCompositeFigure, getvalue, [], setAttribute, [], []) \textbf{; }\\
\operation{InlinelocalVariable}(AbstractCompositeFigure, setAttribute, keyvar) \textbf{; }\\
\operation{InlinelocalVariable}(AbstractCompositeFigure, setAttribute, valuevar) \textbf{; }\\
\operation{InlineConstructor}(GraphicalCompositeFigure, setAttribute, SetAttributeVisitor, [key; value], [getkey; getvalue]) \textbf{; }\\
\operation{InlineLocalField}(GraphicalCompositeFigure, setAttribute, key) \textbf{; }\\
\operation{InlineLocalField}(GraphicalCompositeFigure, setAttribute, value) \textbf{; }\\
\operation{InlineAndDelete} (GraphicalCompositeFigure, getkey, [], setAttribute, [], []) \textbf{; }\\
\operation{InlineAndDelete} (GraphicalCompositeFigure, getvalue, [], setAttribute, [], []) \textbf{; }\\
\operation{InlinelocalVariable}(GraphicalCompositeFigure, setAttribute, keyvar) \textbf{; }\\
\operation{InlinelocalVariable}(GraphicalCompositeFigure, setAttribute, valuevar) \textbf{; }\\
\operation{InlineConstructor}(EllipseFigure, findFigureInside, FindFigureInsideVisitor, [p], [getp]) \textbf{; }\\
\operation{InlineLocalField}(EllipseFigure, findFigureInside, p) \textbf{; }\\
\operation{InlineAndDelete} (EllipseFigure, getp, [], findFigureInside, [], []) \textbf{; }\\
\operation{InlinelocalVariable}(EllipseFigure, findFigureInside, pvar) \textbf{; }\\
\operation{InlineConstructor}(DiamondFigure, findFigureInside, FindFigureInsideVisitor, [p], [getp]) \textbf{; }\\
\operation{InlineLocalField}(DiamondFigure, findFigureInside, p) \textbf{; }\\
\operation{InlineAndDelete} (DiamondFigure, getp, [], findFigureInside, [], []) \textbf{; }\\
\operation{InlinelocalVariable}(DiamondFigure, findFigureInside, pvar) \textbf{; }\\
\operation{InlineConstructor}(RectangleFigure, findFigureInside, FindFigureInsideVisitor, [p], [getp]) \textbf{; }\\
\operation{InlineLocalField}(RectangleFigure, findFigureInside, p) \textbf{; }\\
\operation{InlineAndDelete} (RectangleFigure, getp, [], findFigureInside, [], []) \textbf{; }\\
\operation{InlinelocalVariable}(RectangleFigure, findFigureInside, pvar) \textbf{; }\\
\operation{InlineConstructor}(RoundRectangleFigure, findFigureInside, FindFigureInsideVisitor, [p], [getp]) \textbf{; }\\
\operation{InlineLocalField}(RoundRectangleFigure, findFigureInside, p) \textbf{; }\\
\operation{InlineAndDelete} (RoundRectangleFigure, getp, [], findFigureInside, [], []) \textbf{; }\\
\operation{InlinelocalVariable}(RoundRectangleFigure, findFigureInside, pvar) \textbf{; }\\
\operation{InlineConstructor}(TriangleFigure, findFigureInside, FindFigureInsideVisitor, [p], [getp]) \textbf{; }\\
\operation{InlineLocalField}(TriangleFigure, findFigureInside, p) \textbf{; }\\
\operation{InlineAndDelete} (TriangleFigure, getp, [], findFigureInside, [], []) \textbf{; }\\
\operation{InlinelocalVariable}(TriangleFigure, findFigureInside, pvar) \textbf{; }\\
\operation{InlineConstructor}(TextFigure, findFigureInside, FindFigureInsideVisitor, [p], [getp]) \textbf{; }\\
\operation{InlineLocalField}(TextFigure, findFigureInside, p) \textbf{; }\\
\operation{InlineAndDelete} (TextFigure, getp, [], findFigureInside, [], []) \textbf{; }\\
\operation{InlinelocalVariable}(TextFigure, findFigureInside, pvar) \textbf{; }\\
\operation{InlineConstructor}(BezierFigure, findFigureInside, FindFigureInsideVisitor, [p], [getp]) \textbf{; }\\
\operation{InlineLocalField}(BezierFigure, findFigureInside, p) \textbf{; }\\
\operation{InlineAndDelete} (BezierFigure, getp, [], findFigureInside, [], []) \textbf{; }\\
\operation{InlinelocalVariable}(BezierFigure, findFigureInside, pvar) \textbf{; }\\
\operation{InlineConstructor}(TextAreaFigure, findFigureInside, FindFigureInsideVisitor, [p], [getp]) \textbf{; }\\
\operation{InlineLocalField}(TextAreaFigure, findFigureInside, p) \textbf{; }\\
\operation{InlineAndDelete} (TextAreaFigure, getp, [], findFigureInside, [], []) \textbf{; }\\
\operation{InlinelocalVariable}(TextAreaFigure, findFigureInside, pvar) \textbf{; }\\
\operation{InlineConstructor}(NodeFigure, findFigureInside, FindFigureInsideVisitor, [p], [getp]) \textbf{; }\\
\operation{InlineLocalField}(NodeFigure, findFigureInside, p) \textbf{; }\\
\operation{InlineAndDelete} (NodeFigure, getp, [], findFigureInside, [], []) \textbf{; }\\
\operation{InlinelocalVariable}(NodeFigure, findFigureInside, pvar) \textbf{; }\\
\operation{InlineConstructor}(SVGImage, findFigureInside, FindFigureInsideVisitor, [p], [getp]) \textbf{; }\\
\operation{InlineLocalField}(SVGImage, findFigureInside, p) \textbf{; }\\
\operation{InlineAndDelete} (SVGImage, getp, [], findFigureInside, [], []) \textbf{; }\\
\operation{InlinelocalVariable}(SVGImage, findFigureInside, pvar) \textbf{; }\\
\operation{InlineConstructor}(SVGPath, findFigureInside, FindFigureInsideVisitor, [p], [getp]) \textbf{; }\\
\operation{InlineLocalField}(SVGPath, findFigureInside, p) \textbf{; }\\
\operation{InlineAndDelete} (SVGPath, getp, [], findFigureInside, [], []) \textbf{; }\\
\operation{InlinelocalVariable}(SVGPath, findFigureInside, pvar) \textbf{; }\\
\operation{InlineConstructor}(DependencyFigure, findFigureInside, FindFigureInsideVisitor, [p], [getp]) \textbf{; }\\
\operation{InlineLocalField}(DependencyFigure, findFigureInside, p) \textbf{; }\\
\operation{InlineAndDelete} (DependencyFigure, getp, [], findFigureInside, [], []) \textbf{; }\\
\operation{InlinelocalVariable}(DependencyFigure, findFigureInside, pvar) \textbf{; }\\
\operation{InlineConstructor}(LineConnectionFigure, findFigureInside, FindFigureInsideVisitor, [p], [getp]) \textbf{; }\\
\operation{InlineLocalField}(LineConnectionFigure, findFigureInside, p) \textbf{; }\\
\operation{InlineAndDelete} (LineConnectionFigure, getp, [], findFigureInside, [], []) \textbf{; }\\
\operation{InlinelocalVariable}(LineConnectionFigure, findFigureInside, pvar) \textbf{; }\\
\operation{InlineConstructor}(LabeledLineConnectionFigure, findFigureInside, FindFigureInsideVisitor, [p], [getp]) \textbf{; }\\
\operation{InlineLocalField}(LabeledLineConnectionFigure, findFigureInside, p) \textbf{; }\\
\operation{InlineAndDelete} (LabeledLineConnectionFigure, getp, [], findFigureInside, [], []) \textbf{; }\\
\operation{InlinelocalVariable}(LabeledLineConnectionFigure, findFigureInside, pvar) \textbf{; }\\
\operation{InlineConstructor}(AbstractCompositeFigure, findFigureInside, FindFigureInsideVisitor, [p], [getp]) \textbf{; }\\
\operation{InlineLocalField}(AbstractCompositeFigure, findFigureInside, p) \textbf{; }\\
\operation{InlineAndDelete} (AbstractCompositeFigure, getp, [], findFigureInside, [], []) \textbf{; }\\
\operation{InlinelocalVariable}(AbstractCompositeFigure, findFigureInside, pvar) \textbf{; }\\
\operation{InlineConstructor}(GraphicalCompositeFigure, findFigureInside, FindFigureInsideVisitor, [p], [getp]) \textbf{; }\\
\operation{InlineLocalField}(GraphicalCompositeFigure, findFigureInside, p) \textbf{; }\\
\operation{InlineAndDelete} (GraphicalCompositeFigure, getp, [], findFigureInside, [], []) \textbf{; }\\
\operation{InlinelocalVariable}(GraphicalCompositeFigure, findFigureInside, pvar) \textbf{; }\\
\operation{InlineConstructor}(EllipseFigure, addNotify, AddNotifyVisitor, [d], [getd]) \textbf{; }\\
\operation{InlineLocalField}(EllipseFigure, addNotify, d) \textbf{; }\\
\operation{InlineAndDelete} (EllipseFigure, getd, [], addNotify, [], []) \textbf{; }\\
\operation{InlinelocalVariable}(EllipseFigure, addNotify, dvar) \textbf{; }\\
\operation{InlineConstructor}(DiamondFigure, addNotify, AddNotifyVisitor, [d], [getd]) \textbf{; }\\
\operation{InlineLocalField}(DiamondFigure, addNotify, d) \textbf{; }\\
\operation{InlineAndDelete} (DiamondFigure, getd, [], addNotify, [], []) \textbf{; }\\
\operation{InlinelocalVariable}(DiamondFigure, addNotify, dvar) \textbf{; }\\
\operation{InlineConstructor}(RectangleFigure, addNotify, AddNotifyVisitor, [d], [getd]) \textbf{; }\\
\operation{InlineLocalField}(RectangleFigure, addNotify, d) \textbf{; }\\
\operation{InlineAndDelete} (RectangleFigure, getd, [], addNotify, [], []) \textbf{; }\\
\operation{InlinelocalVariable}(RectangleFigure, addNotify, dvar) \textbf{; }\\
\operation{InlineConstructor}(RoundRectangleFigure, addNotify, AddNotifyVisitor, [d], [getd]) \textbf{; }\\
\operation{InlineLocalField}(RoundRectangleFigure, addNotify, d) \textbf{; }\\
\operation{InlineAndDelete} (RoundRectangleFigure, getd, [], addNotify, [], []) \textbf{; }\\
\operation{InlinelocalVariable}(RoundRectangleFigure, addNotify, dvar) \textbf{; }\\
\operation{InlineConstructor}(TriangleFigure, addNotify, AddNotifyVisitor, [d], [getd]) \textbf{; }\\
\operation{InlineLocalField}(TriangleFigure, addNotify, d) \textbf{; }\\
\operation{InlineAndDelete} (TriangleFigure, getd, [], addNotify, [], []) \textbf{; }\\
\operation{InlinelocalVariable}(TriangleFigure, addNotify, dvar) \textbf{; }\\
\operation{InlineConstructor}(TextFigure, addNotify, AddNotifyVisitor, [d], [getd]) \textbf{; }\\
\operation{InlineLocalField}(TextFigure, addNotify, d) \textbf{; }\\
\operation{InlineAndDelete} (TextFigure, getd, [], addNotify, [], []) \textbf{; }\\
\operation{InlinelocalVariable}(TextFigure, addNotify, dvar) \textbf{; }\\
\operation{InlineConstructor}(BezierFigure, addNotify, AddNotifyVisitor, [d], [getd]) \textbf{; }\\
\operation{InlineLocalField}(BezierFigure, addNotify, d) \textbf{; }\\
\operation{InlineAndDelete} (BezierFigure, getd, [], addNotify, [], []) \textbf{; }\\
\operation{InlinelocalVariable}(BezierFigure, addNotify, dvar) \textbf{; }\\
\operation{InlineConstructor}(TextAreaFigure, addNotify, AddNotifyVisitor, [d], [getd]) \textbf{; }\\
\operation{InlineLocalField}(TextAreaFigure, addNotify, d) \textbf{; }\\
\operation{InlineAndDelete} (TextAreaFigure, getd, [], addNotify, [], []) \textbf{; }\\
\operation{InlinelocalVariable}(TextAreaFigure, addNotify, dvar) \textbf{; }\\
\operation{InlineConstructor}(NodeFigure, addNotify, AddNotifyVisitor, [d], [getd]) \textbf{; }\\
\operation{InlineLocalField}(NodeFigure, addNotify, d) \textbf{; }\\
\operation{InlineAndDelete} (NodeFigure, getd, [], addNotify, [], []) \textbf{; }\\
\operation{InlinelocalVariable}(NodeFigure, addNotify, dvar) \textbf{; }\\
\operation{InlineConstructor}(SVGImage, addNotify, AddNotifyVisitor, [d], [getd]) \textbf{; }\\
\operation{InlineLocalField}(SVGImage, addNotify, d) \textbf{; }\\
\operation{InlineAndDelete} (SVGImage, getd, [], addNotify, [], []) \textbf{; }\\
\operation{InlinelocalVariable}(SVGImage, addNotify, dvar) \textbf{; }\\
\operation{InlineConstructor}(SVGPath, addNotify, AddNotifyVisitor, [d], [getd]) \textbf{; }\\
\operation{InlineLocalField}(SVGPath, addNotify, d) \textbf{; }\\
\operation{InlineAndDelete} (SVGPath, getd, [], addNotify, [], []) \textbf{; }\\
\operation{InlinelocalVariable}(SVGPath, addNotify, dvar) \textbf{; }\\
\operation{InlineConstructor}(DependencyFigure, addNotify, AddNotifyVisitor, [d], [getd]) \textbf{; }\\
\operation{InlineLocalField}(DependencyFigure, addNotify, d) \textbf{; }\\
\operation{InlineAndDelete} (DependencyFigure, getd, [], addNotify, [], []) \textbf{; }\\
\operation{InlinelocalVariable}(DependencyFigure, addNotify, dvar) \textbf{; }\\
\operation{InlineConstructor}(LineConnectionFigure, addNotify, AddNotifyVisitor, [d], [getd]) \textbf{; }\\
\operation{InlineLocalField}(LineConnectionFigure, addNotify, d) \textbf{; }\\
\operation{InlineAndDelete} (LineConnectionFigure, getd, [], addNotify, [], []) \textbf{; }\\
\operation{InlinelocalVariable}(LineConnectionFigure, addNotify, dvar) \textbf{; }\\
\operation{InlineConstructor}(LabeledLineConnectionFigure, addNotify, AddNotifyVisitor, [d], [getd]) \textbf{; }\\
\operation{InlineLocalField}(LabeledLineConnectionFigure, addNotify, d) \textbf{; }\\
\operation{InlineAndDelete} (LabeledLineConnectionFigure, getd, [], addNotify, [], []) \textbf{; }\\
\operation{InlinelocalVariable}(LabeledLineConnectionFigure, addNotify, dvar) \textbf{; }\\
\operation{InlineConstructor}(AbstractCompositeFigure, addNotify, AddNotifyVisitor, [d], [getd]) \textbf{; }\\
\operation{InlineLocalField}(AbstractCompositeFigure, addNotify, d) \textbf{; }\\
\operation{InlineAndDelete} (AbstractCompositeFigure, getd, [], addNotify, [], []) \textbf{; }\\
\operation{InlinelocalVariable}(AbstractCompositeFigure, addNotify, dvar) \textbf{; }\\
\operation{InlineConstructor}(GraphicalCompositeFigure, addNotify, AddNotifyVisitor, [d], [getd]) \textbf{; }\\
\operation{InlineLocalField}(GraphicalCompositeFigure, addNotify, d) \textbf{; }\\
\operation{InlineAndDelete} (GraphicalCompositeFigure, getd, [], addNotify, [], []) \textbf{; }\\
\operation{InlinelocalVariable}(GraphicalCompositeFigure, addNotify, dvar) \textbf{; }\\
\operation{InlineConstructor}(EllipseFigure, removeNotify, RemoveNotifyVisitor, [d], [getd]) \textbf{; }\\
\operation{InlineLocalField}(EllipseFigure, removeNotify, d) \textbf{; }\\
\operation{InlineAndDelete} (EllipseFigure, getd, [], removeNotify, [], []) \textbf{; }\\
\operation{InlinelocalVariable}(EllipseFigure, removeNotify, dvar) \textbf{; }\\
\operation{InlineConstructor}(DiamondFigure, removeNotify, RemoveNotifyVisitor, [d], [getd]) \textbf{; }\\
\operation{InlineLocalField}(DiamondFigure, removeNotify, d) \textbf{; }\\
\operation{InlineAndDelete} (DiamondFigure, getd, [], removeNotify, [], []) \textbf{; }\\
\operation{InlinelocalVariable}(DiamondFigure, removeNotify, dvar) \textbf{; }\\
\operation{InlineConstructor}(RectangleFigure, removeNotify, RemoveNotifyVisitor, [d], [getd]) \textbf{; }\\
\operation{InlineLocalField}(RectangleFigure, removeNotify, d) \textbf{; }\\
\operation{InlineAndDelete} (RectangleFigure, getd, [], removeNotify, [], []) \textbf{; }\\
\operation{InlinelocalVariable}(RectangleFigure, removeNotify, dvar) \textbf{; }\\
\operation{InlineConstructor}(RoundRectangleFigure, removeNotify, RemoveNotifyVisitor, [d], [getd]) \textbf{; }\\
\operation{InlineLocalField}(RoundRectangleFigure, removeNotify, d) \textbf{; }\\
\operation{InlineAndDelete} (RoundRectangleFigure, getd, [], removeNotify, [], []) \textbf{; }\\
\operation{InlinelocalVariable}(RoundRectangleFigure, removeNotify, dvar) \textbf{; }\\
\operation{InlineConstructor}(TriangleFigure, removeNotify, RemoveNotifyVisitor, [d], [getd]) \textbf{; }\\
\operation{InlineLocalField}(TriangleFigure, removeNotify, d) \textbf{; }\\
\operation{InlineAndDelete} (TriangleFigure, getd, [], removeNotify, [], []) \textbf{; }\\
\operation{InlinelocalVariable}(TriangleFigure, removeNotify, dvar) \textbf{; }\\
\operation{InlineConstructor}(TextFigure, removeNotify, RemoveNotifyVisitor, [d], [getd]) \textbf{; }\\
\operation{InlineLocalField}(TextFigure, removeNotify, d) \textbf{; }\\
\operation{InlineAndDelete} (TextFigure, getd, [], removeNotify, [], []) \textbf{; }\\
\operation{InlinelocalVariable}(TextFigure, removeNotify, dvar) \textbf{; }\\
\operation{InlineConstructor}(BezierFigure, removeNotify, RemoveNotifyVisitor, [d], [getd]) \textbf{; }\\
\operation{InlineLocalField}(BezierFigure, removeNotify, d) \textbf{; }\\
\operation{InlineAndDelete} (BezierFigure, getd, [], removeNotify, [], []) \textbf{; }\\
\operation{InlinelocalVariable}(BezierFigure, removeNotify, dvar) \textbf{; }\\
\operation{InlineConstructor}(TextAreaFigure, removeNotify, RemoveNotifyVisitor, [d], [getd]) \textbf{; }\\
\operation{InlineLocalField}(TextAreaFigure, removeNotify, d) \textbf{; }\\
\operation{InlineAndDelete} (TextAreaFigure, getd, [], removeNotify, [], []) \textbf{; }\\
\operation{InlinelocalVariable}(TextAreaFigure, removeNotify, dvar) \textbf{; }\\
\operation{InlineConstructor}(NodeFigure, removeNotify, RemoveNotifyVisitor, [d], [getd]) \textbf{; }\\
\operation{InlineLocalField}(NodeFigure, removeNotify, d) \textbf{; }\\
\operation{InlineAndDelete} (NodeFigure, getd, [], removeNotify, [], []) \textbf{; }\\
\operation{InlinelocalVariable}(NodeFigure, removeNotify, dvar) \textbf{; }\\
\operation{InlineConstructor}(SVGImage, removeNotify, RemoveNotifyVisitor, [d], [getd]) \textbf{; }\\
\operation{InlineLocalField}(SVGImage, removeNotify, d) \textbf{; }\\
\operation{InlineAndDelete} (SVGImage, getd, [], removeNotify, [], []) \textbf{; }\\
\operation{InlinelocalVariable}(SVGImage, removeNotify, dvar) \textbf{; }\\
\operation{InlineConstructor}(SVGPath, removeNotify, RemoveNotifyVisitor, [d], [getd]) \textbf{; }\\
\operation{InlineLocalField}(SVGPath, removeNotify, d) \textbf{; }\\
\operation{InlineAndDelete} (SVGPath, getd, [], removeNotify, [], []) \textbf{; }\\
\operation{InlinelocalVariable}(SVGPath, removeNotify, dvar) \textbf{; }\\
\operation{InlineConstructor}(DependencyFigure, removeNotify, RemoveNotifyVisitor, [d], [getd]) \textbf{; }\\
\operation{InlineLocalField}(DependencyFigure, removeNotify, d) \textbf{; }\\
\operation{InlineAndDelete} (DependencyFigure, getd, [], removeNotify, [], []) \textbf{; }\\
\operation{InlinelocalVariable}(DependencyFigure, removeNotify, dvar) \textbf{; }\\
\operation{InlineConstructor}(LineConnectionFigure, removeNotify, RemoveNotifyVisitor, [d], [getd]) \textbf{; }\\
\operation{InlineLocalField}(LineConnectionFigure, removeNotify, d) \textbf{; }\\
\operation{InlineAndDelete} (LineConnectionFigure, getd, [], removeNotify, [], []) \textbf{; }\\
\operation{InlinelocalVariable}(LineConnectionFigure, removeNotify, dvar) \textbf{; }\\
\operation{InlineConstructor}(LabeledLineConnectionFigure, removeNotify, RemoveNotifyVisitor, [d], [getd]) \textbf{; }\\
\operation{InlineLocalField}(LabeledLineConnectionFigure, removeNotify, d) \textbf{; }\\
\operation{InlineAndDelete} (LabeledLineConnectionFigure, getd, [], removeNotify, [], []) \textbf{; }\\
\operation{InlinelocalVariable}(LabeledLineConnectionFigure, removeNotify, dvar) \textbf{; }\\
\operation{InlineConstructor}(AbstractCompositeFigure, removeNotify, RemoveNotifyVisitor, [d], [getd]) \textbf{; }\\
\operation{InlineLocalField}(AbstractCompositeFigure, removeNotify, d) \textbf{; }\\
\operation{InlineAndDelete} (AbstractCompositeFigure, getd, [], removeNotify, [], []) \textbf{; }\\
\operation{InlinelocalVariable}(AbstractCompositeFigure, removeNotify, dvar) \textbf{; }\\
\operation{InlineConstructor}(GraphicalCompositeFigure, removeNotify, RemoveNotifyVisitor, [d], [getd]) \textbf{; }\\
\operation{InlineLocalField}(GraphicalCompositeFigure, removeNotify, d) \textbf{; }\\
\operation{InlineAndDelete} (GraphicalCompositeFigure, getd, [], removeNotify, [], []) \textbf{; }\\
\operation{InlinelocalVariable}(GraphicalCompositeFigure, removeNotify, dvar) \textbf{; }\\
\operation{DeleteClass}(BasicTransformVisitor, Visitor, [EllipseFigure; DiamondFigure; RectangleFigure; RoundRectangleFigure;  TriangleFigure; TextFigure; BezierFigure;  TextAreaFigure;  NodeFigure;  SVGImage;  
SVGPath;  DependencyFigure;  LineConnectionFigure;  LabeledLineConnectionFigure;  AbstractCompositeFigure;  GraphicalCompositeFigure;  BasicTransformVisitor;  ContainsVisitor;  SetAttributeVisitor;  FindFigureInsideVisitor;  AddNotifyVisitor;  RemoveNotifyVisitor;  Visitor;  AbstractFigure;  ], [visit;  ], [accept;  basicTransform;  contains;  setAttribute;  findFigureInside; addNotify; removeNotify]) \textbf{; }\\
\operation{DeleteClass}(ContainsVisitor, Visitor, [EllipseFigure; DiamondFigure; RectangleFigure; RoundRectangleFigure; TriangleFigure; TextFigure; BezierFigure; TextAreaFigure; NodeFigure; SVGImage; SVGPath; DependencyFigure; LineConnectionFigure; LabeledLineConnectionFigure; AbstractCompositeFigure; GraphicalCompositeFigure; BasicTransformVisitor; ContainsVisitor; SetAttributeVisitor; FindFigureInsideVisitor; AddNotifyVisitor; RemoveNotifyVisitor; Visitor; AbstractFigure], [visit], [accept; basicTransform; contains; setAttribute; findFigureInside; addNotify; removeNotify]) \textbf{; }\\
\operation{DeleteClass}(SetAttributeVisitor, Visitor, [EllipseFigure; DiamondFigure; RectangleFigure; RoundRectangleFigure; TriangleFigure; TextFigure; BezierFigure; TextAreaFigure; NodeFigure; SVGImage; SVGPath; DependencyFigure; LineConnectionFigure; LabeledLineConnectionFigure; AbstractCompositeFigure; GraphicalCompositeFigure; BasicTransformVisitor; ContainsVisitor; SetAttributeVisitor; FindFigureInsideVisitor; AddNotifyVisitor; RemoveNotifyVisitor; Visitor; AbstractFigure], [visit], [accept; basicTransform; contains; setAttribute; findFigureInside; addNotify; removeNotify]) \textbf{; }\\
\operation{DeleteClass}(FindFigureInsideVisitor, Visitor, [EllipseFigure; DiamondFigure; RectangleFigure; RoundRectangleFigure; TriangleFigure; TextFigure; BezierFigure; TextAreaFigure; NodeFigure; SVGImage; SVGPath; DependencyFigure; LineConnectionFigure; LabeledLineConnectionFigure; AbstractCompositeFigure; GraphicalCompositeFigure; BasicTransformVisitor; ContainsVisitor; SetAttributeVisitor; FindFigureInsideVisitor; AddNotifyVisitor; RemoveNotifyVisitor; Visitor; AbstractFigure], [visit], [accept; basicTransform; contains; setAttribute; findFigureInside; addNotify; removeNotify]) \textbf{; }\\
\operation{DeleteClass}(AddNotifyVisitor, Visitor, [EllipseFigure; DiamondFigure; RectangleFigure; RoundRectangleFigure; TriangleFigure; TextFigure; BezierFigure; TextAreaFigure; NodeFigure; SVGImage; SVGPath; DependencyFigure; LineConnectionFigure; LabeledLineConnectionFigure; AbstractCompositeFigure; GraphicalCompositeFigure; BasicTransformVisitor; ContainsVisitor; SetAttributeVisitor; FindFigureInsideVisitor; AddNotifyVisitor; RemoveNotifyVisitor; Visitor; AbstractFigure], [visit], [accept; basicTransform; contains; setAttribute; findFigureInside; addNotify; removeNotify]) \textbf{; }\\
\operation{DeleteClass}(RemoveNotifyVisitor, Visitor, [EllipseFigure; DiamondFigure; RectangleFigure; RoundRectangleFigure; TriangleFigure; TextFigure; BezierFigure; TextAreaFigure; NodeFigure; SVGImage; SVGPath; DependencyFigure; LineConnectionFigure; LabeledLineConnectionFigure; AbstractCompositeFigure; GraphicalCompositeFigure; BasicTransformVisitor; ContainsVisitor; SetAttributeVisitor; FindFigureInsideVisitor; AddNotifyVisitor; RemoveNotifyVisitor; Visitor; AbstractFigure], [visit], [accept; basicTransform; contains; setAttribute; findFigureInside; addNotify; removeNotify]) \textbf{; }\\
\operation{DeleteClass}(Visitor, java.lang.Object, [EllipseFigure; DiamondFigure; RectangleFigure; RoundRectangleFigure; TriangleFigure; TextFigure; BezierFigure; TextAreaFigure; NodeFigure; SVGImage; SVGPath; DependencyFigure; LineConnectionFigure; LabeledLineConnectionFigure; AbstractCompositeFigure; GraphicalCompositeFigure; BasicTransformVisitor; ContainsVisitor; SetAttributeVisitor; FindFigureInsideVisitor; AddNotifyVisitor; RemoveNotifyVisitor; Visitor; AbstractFigure], [visit], [accept; basicTransform; contains; setAttribute; findFigureInside; addNotify; removeNotify]) \textbf{; }\\
\operation{DeleteMethod}(LineConnectionFigure, findFigureInside, [Point2D.Double p], BezierFigure, findFigureInside) \textbf{; }\\
\operation{DeleteMethod}(LineConnectionFigure, setAttribute, [AttributeKey key; Object value], BezierFigure, setAttribute) \textbf{; }\\
\operation{DeleteMethod}(LineConnectionFigure, contains, [Point2D.Double p], BezierFigure, contains) \textbf{; }\\
\operation{DeleteMethod}(SVGPath, addNotify, [Drawing d], AbstractCompositeFigure, addNotify) \textbf{; }\\
\operation{DeleteMethod}(SVGPath, removeNotify, [Drawing d], AbstractCompositeFigure, removeNotify) \textbf{; }\\
\operation{DeleteMethod}(SVGPath, findFigureInside, [Point2D.Double p], AbstractCompositeFigure, findFigureInside) \textbf{; }\\
\operation{DeleteMethod}(SVGPath, contains, [Point2D.Double p], AbstractCompositeFigure, contains) \textbf{; }\\
\operation{DeleteMethod}(LabeledLineConnectionFigure, basicTransform, [AffineTransform tx], BezierFigure, basicTransform) \textbf{; }\\
\operation{DeleteMethod}(LabeledLineConnectionFigure, setAttribute, [AttributeKey key; Object value], BezierFigure, setAttribute) \textbf{; }\\
\operation{DeleteMethod}(LabeledLineConnectionFigure, findFigureInside, [Point2D.Double p], BezierFigure, findFigureInside) \textbf{; }\\
\operation{DeleteMethod}(LabeledLineConnectionFigure, contains, [Point2D.Double p], BezierFigure, contains) \textbf{; }\\
\operation{DeleteMethod}(DependencyFigure, addNotify, [Drawing d], LineConnectionFigure, addNotify) \textbf{; }\\
\operation{DeleteMethod}(DependencyFigure, basicTransform, [AffineTransform tx], LineConnectionFigure, basicTransform) \textbf{; }\\
\operation{DeleteMethod}(DependencyFigure, setAttribute, [AttributeKey key; Object value], LineConnectionFigure, setAttribute) \textbf{; }\\
\operation{DeleteMethod}(DependencyFigure, findFigureInside, [Point2D.Double p], LineConnectionFigure, findFigureInside) \textbf{; }\\
\operation{DeleteMethod}(DependencyFigure, contains, [Point2D.Double p], LineConnectionFigure, contains) \textbf{; }\\
\operation{DeleteMethod}(NodeFigure, addNotify, [Drawing d], TextFigure, addNotify) \textbf{; }\\
\operation{DeleteMethod}(NodeFigure, basicTransform, [AffineTransform tx], TextFigure, basicTransform) \textbf{; }\\
\operation{DeleteMethod}(NodeFigure, setAttribute, [AttributeKey key; Object value], TextFigure, setAttribute) \textbf{; }\\
\operation{DeleteMethod}(NodeFigure, findFigureInside, [Point2D.Double p], TextFigure, findFigureInside) \textbf{; }\\
\operation{DeleteMethod}(NodeFigure, contains, [Point2D.Double p], TextFigure, contains) \textbf{; }\\
\operation{DeleteMethod}(GraphicalCompositeFigure, findFigureInside, [Point2D.Double p], AbstractCompositeFigure, findFigureInside)

\subsection{Computed precondition for a round-trip transformation of JHotDraw}
(Conjunction of 1852 propositions)

\noindent $\mathtt{IsInheritedMethod}(GraphicalCompositeFigure, findFigureInside)\\
 \wedge \mathtt{IsInheritedMethod}(NodeFigure, contains)\\
 \wedge \mathtt{IsInheritedMethod}(NodeFigure, findFigureInside)\\
 \wedge \mathtt{IsInheritedMethod}(NodeFigure, setAttribute)\\
 \wedge \mathtt{IsInheritedMethod}(NodeFigure, basicTransform)\\
 \wedge \mathtt{IsInheritedMethod}(NodeFigure, addNotify)\\
 \wedge \mathtt{IsInheritedMethod}(DependencyFigure, contains)\\
 \wedge \mathtt{IsInheritedMethod}(DependencyFigure, findFigureInside)\\
 \wedge \mathtt{IsInheritedMethod}(DependencyFigure, setAttribute)\\
 \wedge \mathtt{IsInheritedMethod}(DependencyFigure, basicTransform)\\
 \wedge \mathtt{IsInheritedMethod}(DependencyFigure, addNotify)\\
 \wedge \mathtt{IsInheritedMethod}(LabeledLineConnectionFigure, contains)\\
 \wedge \mathtt{IsInheritedMethod}(LabeledLineConnectionFigure, findFigureInside)\\
 \wedge \mathtt{IsInheritedMethod}(LabeledLineConnectionFigure, setAttribute)\\
 \wedge \mathtt{IsInheritedMethod}(LabeledLineConnectionFigure, basicTransform)\\
 \wedge \mathtt{IsInheritedMethod}(SVGPath, contains)\\
 \wedge \mathtt{IsInheritedMethod}(SVGPath, findFigureInside)\\
 \wedge \mathtt{IsInheritedMethod}(SVGPath, removeNotify)\\
 \wedge \mathtt{IsInheritedMethod}(SVGPath, addNotify)\\
 \wedge \mathtt{IsInheritedMethod}(LineConnectionFigure, contains)\\
 \wedge \mathtt{IsInheritedMethod}(LineConnectionFigure, setAttribute)\\
 \wedge \mathtt{IsInheritedMethod}(LineConnectionFigure, findFigureInside)\\
 \wedge \neg \mathtt{IsUsedConstructorAsMethodParameter}(RemoveNotifyVisitor, EllipseFigure, basicTransform)\\
 \wedge \neg \mathtt{IsUsedConstructorAsMethodParameter}(RemoveNotifyVisitor, EllipseFigure, contains)\\
 \wedge \neg \mathtt{IsUsedConstructorAsMethodParameter}(RemoveNotifyVisitor, EllipseFigure, setAttribute)\\
 \wedge \neg \mathtt{IsUsedConstructorAsMethodParameter}(RemoveNotifyVisitor, EllipseFigure, findFigureInside)\\
 \wedge \neg \mathtt{IsUsedConstructorAsMethodParameter}(RemoveNotifyVisitor, EllipseFigure, addNotify)\\
 \wedge \neg \mathtt{IsUsedConstructorAsMethodParameter}(RemoveNotifyVisitor, DiamondFigure, basicTransform)\\
 \wedge \neg \mathtt{IsUsedConstructorAsMethodParameter}(RemoveNotifyVisitor, DiamondFigure, contains)\\
 \wedge \neg \mathtt{IsUsedConstructorAsMethodParameter}(RemoveNotifyVisitor, DiamondFigure, setAttribute)\\
 \wedge \neg \mathtt{IsUsedConstructorAsMethodParameter}(RemoveNotifyVisitor, DiamondFigure, findFigureInside)\\
 \wedge \neg \mathtt{IsUsedConstructorAsMethodParameter}(RemoveNotifyVisitor, DiamondFigure, addNotify)\\
 \wedge \neg \mathtt{IsUsedConstructorAsMethodParameter}(RemoveNotifyVisitor, RectangleFigure, basicTransform)\\
 \wedge \neg \mathtt{IsUsedConstructorAsMethodParameter}(RemoveNotifyVisitor, RectangleFigure, contains)\\
 \wedge \neg \mathtt{IsUsedConstructorAsMethodParameter}(RemoveNotifyVisitor, RectangleFigure, setAttribute)\\
 \wedge \neg \mathtt{IsUsedConstructorAsMethodParameter}(RemoveNotifyVisitor, RectangleFigure, findFigureInside)\\
 \wedge \neg \mathtt{IsUsedConstructorAsMethodParameter}(RemoveNotifyVisitor, RectangleFigure, addNotify)\\
 \wedge \neg \mathtt{IsUsedConstructorAsMethodParameter}(RemoveNotifyVisitor, RoundRectangleFigure, basicTransform)\\
 \wedge \neg \mathtt{IsUsedConstructorAsMethodParameter}(RemoveNotifyVisitor, RoundRectangleFigure, contains)\\
 \wedge \neg \mathtt{IsUsedConstructorAsMethodParameter}(RemoveNotifyVisitor, RoundRectangleFigure, setAttribute)\\
 \wedge \neg \mathtt{IsUsedConstructorAsMethodParameter}(RemoveNotifyVisitor, RoundRectangleFigure, findFigureInside)\\
 \wedge \neg \mathtt{IsUsedConstructorAsMethodParameter}(RemoveNotifyVisitor, RoundRectangleFigure, addNotify)\\
 \wedge \neg \mathtt{IsUsedConstructorAsMethodParameter}(RemoveNotifyVisitor, TriangleFigure, basicTransform)\\
 \wedge \neg \mathtt{IsUsedConstructorAsMethodParameter}(RemoveNotifyVisitor, TriangleFigure, contains)\\
 \wedge \neg \mathtt{IsUsedConstructorAsMethodParameter}(RemoveNotifyVisitor, TriangleFigure, setAttribute)\\
 \wedge \neg \mathtt{IsUsedConstructorAsMethodParameter}(RemoveNotifyVisitor, TriangleFigure, findFigureInside)\\
 \wedge \neg \mathtt{IsUsedConstructorAsMethodParameter}(RemoveNotifyVisitor, TriangleFigure, addNotify)\\
 \wedge \neg \mathtt{IsUsedConstructorAsMethodParameter}(RemoveNotifyVisitor, TextFigure, basicTransform)\\
 \wedge \neg \mathtt{IsUsedConstructorAsMethodParameter}(RemoveNotifyVisitor, TextFigure, contains)\\
 \wedge \neg \mathtt{IsUsedConstructorAsMethodParameter}(RemoveNotifyVisitor, TextFigure, setAttribute)\\
 \wedge \neg \mathtt{IsUsedConstructorAsMethodParameter}(RemoveNotifyVisitor, TextFigure, findFigureInside)\\
 \wedge \neg \mathtt{IsUsedConstructorAsMethodParameter}(RemoveNotifyVisitor, TextFigure, addNotify)\\
 \wedge \neg \mathtt{IsUsedConstructorAsMethodParameter}(RemoveNotifyVisitor, BezierFigure, basicTransform)\\
 \wedge \neg \mathtt{IsUsedConstructorAsMethodParameter}(RemoveNotifyVisitor, BezierFigure, contains)\\
 \wedge \neg \mathtt{IsUsedConstructorAsMethodParameter}(RemoveNotifyVisitor, BezierFigure, setAttribute)\\
 \wedge \neg \mathtt{IsUsedConstructorAsMethodParameter}(RemoveNotifyVisitor, BezierFigure, findFigureInside)\\
 \wedge \neg \mathtt{IsUsedConstructorAsMethodParameter}(RemoveNotifyVisitor, BezierFigure, addNotify)\\
 \wedge \neg \mathtt{IsUsedConstructorAsMethodParameter}(RemoveNotifyVisitor, TextAreaFigure, basicTransform)\\
 \wedge \neg \mathtt{IsUsedConstructorAsMethodParameter}(RemoveNotifyVisitor, TextAreaFigure, contains)\\
 \wedge \neg \mathtt{IsUsedConstructorAsMethodParameter}(RemoveNotifyVisitor, TextAreaFigure, setAttribute)\\
 \wedge \neg \mathtt{IsUsedConstructorAsMethodParameter}(RemoveNotifyVisitor, TextAreaFigure, findFigureInside)\\
 \wedge \neg \mathtt{IsUsedConstructorAsMethodParameter}(RemoveNotifyVisitor, TextAreaFigure, addNotify)\\
 \wedge \neg \mathtt{IsUsedConstructorAsMethodParameter}(RemoveNotifyVisitor, NodeFigure, basicTransform)\\
 \wedge \neg \mathtt{IsUsedConstructorAsMethodParameter}(RemoveNotifyVisitor, NodeFigure, contains)\\
 \wedge \neg \mathtt{IsUsedConstructorAsMethodParameter}(RemoveNotifyVisitor, NodeFigure, setAttribute)\\
 \wedge \neg \mathtt{IsUsedConstructorAsMethodParameter}(RemoveNotifyVisitor, NodeFigure, findFigureInside)\\
 \wedge \neg \mathtt{IsUsedConstructorAsMethodParameter}(RemoveNotifyVisitor, NodeFigure, addNotify)\\
 \wedge \neg \mathtt{IsUsedConstructorAsMethodParameter}(RemoveNotifyVisitor, SVGImage, basicTransform)\\
 \wedge \neg \mathtt{IsUsedConstructorAsMethodParameter}(RemoveNotifyVisitor, SVGImage, contains)\\
 \wedge \neg \mathtt{IsUsedConstructorAsMethodParameter}(RemoveNotifyVisitor, SVGImage, setAttribute)\\
 \wedge \neg \mathtt{IsUsedConstructorAsMethodParameter}(RemoveNotifyVisitor, SVGImage, findFigureInside)\\
 \wedge \neg \mathtt{IsUsedConstructorAsMethodParameter}(RemoveNotifyVisitor, SVGImage, addNotify)\\
 \wedge \neg \mathtt{IsUsedConstructorAsMethodParameter}(RemoveNotifyVisitor, SVGPath, basicTransform)\\
 \wedge \neg \mathtt{IsUsedConstructorAsMethodParameter}(RemoveNotifyVisitor, SVGPath, contains)\\
 \wedge \neg \mathtt{IsUsedConstructorAsMethodParameter}(RemoveNotifyVisitor, SVGPath, setAttribute)\\
 \wedge \neg \mathtt{IsUsedConstructorAsMethodParameter}(RemoveNotifyVisitor, SVGPath, findFigureInside)\\
 \wedge \neg \mathtt{IsUsedConstructorAsMethodParameter}(RemoveNotifyVisitor, SVGPath, addNotify)\\
 \wedge \neg \mathtt{IsUsedConstructorAsMethodParameter}(RemoveNotifyVisitor, DependencyFigure, basicTransform)\\
 \wedge \neg \mathtt{IsUsedConstructorAsMethodParameter}(RemoveNotifyVisitor, DependencyFigure, contains)\\
 \wedge \neg \mathtt{IsUsedConstructorAsMethodParameter}(RemoveNotifyVisitor, DependencyFigure, setAttribute)\\
 \wedge \neg \mathtt{IsUsedConstructorAsMethodParameter}(RemoveNotifyVisitor, DependencyFigure, findFigureInside)\\
 \wedge \neg \mathtt{IsUsedConstructorAsMethodParameter}(RemoveNotifyVisitor, DependencyFigure, addNotify)\\
 \wedge \neg \mathtt{IsUsedConstructorAsMethodParameter}(RemoveNotifyVisitor, LineConnectionFigure, basicTransform)\\
 \wedge \neg \mathtt{IsUsedConstructorAsMethodParameter}(RemoveNotifyVisitor, LineConnectionFigure, contains)\\
 \wedge \neg \mathtt{IsUsedConstructorAsMethodParameter}(RemoveNotifyVisitor, LineConnectionFigure, setAttribute)\\
 \wedge \neg \mathtt{IsUsedConstructorAsMethodParameter}(RemoveNotifyVisitor, LineConnectionFigure, findFigureInside)\\
 \wedge \neg \mathtt{IsUsedConstructorAsMethodParameter}(RemoveNotifyVisitor, LineConnectionFigure, addNotify)\\
 \wedge \neg \mathtt{IsUsedConstructorAsMethodParameter}(RemoveNotifyVisitor, LabeledLineConnectionFigure, basicTransform)\\
 \wedge \neg \mathtt{IsUsedConstructorAsMethodParameter}(RemoveNotifyVisitor, LabeledLineConnectionFigure, contains)\\
 \wedge \neg \mathtt{IsUsedConstructorAsMethodParameter}(RemoveNotifyVisitor, LabeledLineConnectionFigure, setAttribute)\\
 \wedge \neg \mathtt{IsUsedConstructorAsMethodParameter}(RemoveNotifyVisitor, LabeledLineConnectionFigure, findFigureInside)\\
 \wedge \neg \mathtt{IsUsedConstructorAsMethodParameter}(RemoveNotifyVisitor, LabeledLineConnectionFigure, addNotify)\\
 \wedge \neg \mathtt{IsUsedConstructorAsMethodParameter}(RemoveNotifyVisitor, AbstractCompositeFigure, basicTransform)\\
 \wedge \neg \mathtt{IsUsedConstructorAsMethodParameter}(RemoveNotifyVisitor, AbstractCompositeFigure, contains)\\
 \wedge \neg \mathtt{IsUsedConstructorAsMethodParameter}(RemoveNotifyVisitor, AbstractCompositeFigure, setAttribute)\\
 \wedge \neg \mathtt{IsUsedConstructorAsMethodParameter}(RemoveNotifyVisitor, AbstractCompositeFigure, findFigureInside)\\
 \wedge \neg \mathtt{IsUsedConstructorAsMethodParameter}(RemoveNotifyVisitor, AbstractCompositeFigure, addNotify)\\
 \wedge \neg \mathtt{IsUsedConstructorAsMethodParameter}(RemoveNotifyVisitor, GraphicalCompositeFigure, basicTransform)\\
 \wedge \neg \mathtt{IsUsedConstructorAsMethodParameter}(RemoveNotifyVisitor, GraphicalCompositeFigure, contains)\\
 \wedge \neg \mathtt{IsUsedConstructorAsMethodParameter}(RemoveNotifyVisitor, GraphicalCompositeFigure, setAttribute)\\
 \wedge \neg \mathtt{IsUsedConstructorAsMethodParameter}(RemoveNotifyVisitor, GraphicalCompositeFigure, findFigureInside)\\
 \wedge \neg \mathtt{IsUsedConstructorAsMethodParameter}(RemoveNotifyVisitor, GraphicalCompositeFigure, addNotify)\\
 \wedge \neg \mathtt{IsUsedConstructorAsMethodParameter}(RemoveNotifyVisitor, AbstractFigure, basicTransform)\\
 \wedge \neg \mathtt{IsUsedConstructorAsMethodParameter}(RemoveNotifyVisitor, AbstractFigure, contains)\\
 \wedge \neg \mathtt{IsUsedConstructorAsMethodParameter}(RemoveNotifyVisitor, AbstractFigure, setAttribute)\\
 \wedge \neg \mathtt{IsUsedConstructorAsMethodParameter}(RemoveNotifyVisitor, AbstractFigure, findFigureInside)\\
 \wedge \neg \mathtt{IsUsedConstructorAsMethodParameter}(RemoveNotifyVisitor, AbstractFigure, addNotify)\\
 \wedge \neg \mathtt{IsUsedConstructorAsMethodParameter}(RemoveNotifyVisitor, AbstractFigure, removeNotify)\\
 \wedge \neg (\mathtt{IsUsedConstructorAsObjectReceiver}(RemoveNotifyVisitor, EllipseFigure, basicTransformTmpVC)\\
 \tab \vee  \mathtt{IsUsedConstructorAsObjectReceiver}(RemoveNotifyVisitor, EllipseFigure, basicTransform))\\
 \wedge \neg \mathtt{IsUsedConstructorAsObjectReceiver}(RemoveNotifyVisitor, EllipseFigure, contains)\\
 \wedge \neg \mathtt{IsUsedConstructorAsObjectReceiver}(RemoveNotifyVisitor, EllipseFigure, setAttribute)\\
 \wedge \neg \mathtt{IsUsedConstructorAsObjectReceiver}(RemoveNotifyVisitor, EllipseFigure, findFigureInside)\\
 \wedge \neg \mathtt{IsUsedConstructorAsObjectReceiver}(RemoveNotifyVisitor, EllipseFigure, addNotify)\\
 \wedge \neg (\mathtt{IsUsedConstructorAsObjectReceiver}(RemoveNotifyVisitor, DiamondFigure, basicTransformTmpVC)\\
 \tab \vee  \mathtt{IsUsedConstructorAsObjectReceiver}(RemoveNotifyVisitor, DiamondFigure, basicTransform))\\
 \wedge \neg \mathtt{IsUsedConstructorAsObjectReceiver}(RemoveNotifyVisitor, DiamondFigure, contains)\\
 \wedge \neg \mathtt{IsUsedConstructorAsObjectReceiver}(RemoveNotifyVisitor, DiamondFigure, setAttribute)\\
 \wedge \neg \mathtt{IsUsedConstructorAsObjectReceiver}(RemoveNotifyVisitor, DiamondFigure, findFigureInside)\\
 \wedge \neg \mathtt{IsUsedConstructorAsObjectReceiver}(RemoveNotifyVisitor, DiamondFigure, addNotify)\\
 \wedge \neg (\mathtt{IsUsedConstructorAsObjectReceiver}(RemoveNotifyVisitor, RectangleFigure, basicTransformTmpVC)\\
 \tab \vee  \mathtt{IsUsedConstructorAsObjectReceiver}(RemoveNotifyVisitor, RectangleFigure, basicTransform))\\
 \wedge \neg \mathtt{IsUsedConstructorAsObjectReceiver}(RemoveNotifyVisitor, RectangleFigure, contains)\\
 \wedge \neg \mathtt{IsUsedConstructorAsObjectReceiver}(RemoveNotifyVisitor, RectangleFigure, setAttribute)\\
 \wedge \neg \mathtt{IsUsedConstructorAsObjectReceiver}(RemoveNotifyVisitor, RectangleFigure, findFigureInside)\\
 \wedge \neg \mathtt{IsUsedConstructorAsObjectReceiver}(RemoveNotifyVisitor, RectangleFigure, addNotify)\\
 \wedge \neg (\mathtt{IsUsedConstructorAsObjectReceiver}(RemoveNotifyVisitor, RoundRectangleFigure, basicTransformTmpVC)\\
 \tab \vee  \mathtt{IsUsedConstructorAsObjectReceiver}(RemoveNotifyVisitor, RoundRectangleFigure, basicTransform))\\
 \wedge \neg \mathtt{IsUsedConstructorAsObjectReceiver}(RemoveNotifyVisitor, RoundRectangleFigure, contains)\\
 \wedge \neg \mathtt{IsUsedConstructorAsObjectReceiver}(RemoveNotifyVisitor, RoundRectangleFigure, setAttribute)\\
 \wedge \neg \mathtt{IsUsedConstructorAsObjectReceiver}(RemoveNotifyVisitor, RoundRectangleFigure, findFigureInside)\\
 \wedge \neg \mathtt{IsUsedConstructorAsObjectReceiver}(RemoveNotifyVisitor, RoundRectangleFigure, addNotify)\\
 \wedge \neg (\mathtt{IsUsedConstructorAsObjectReceiver}(RemoveNotifyVisitor, TriangleFigure, basicTransformTmpVC)\\
 \tab \vee  \mathtt{IsUsedConstructorAsObjectReceiver}(RemoveNotifyVisitor, TriangleFigure, basicTransform))\\
 \wedge \neg \mathtt{IsUsedConstructorAsObjectReceiver}(RemoveNotifyVisitor, TriangleFigure, contains)\\
 \wedge \neg \mathtt{IsUsedConstructorAsObjectReceiver}(RemoveNotifyVisitor, TriangleFigure, setAttribute)\\
 \wedge \neg \mathtt{IsUsedConstructorAsObjectReceiver}(RemoveNotifyVisitor, TriangleFigure, findFigureInside)\\
 \wedge \neg \mathtt{IsUsedConstructorAsObjectReceiver}(RemoveNotifyVisitor, TriangleFigure, addNotify)\\
 \wedge \neg (\mathtt{IsUsedConstructorAsObjectReceiver}(RemoveNotifyVisitor, TextFigure, basicTransformTmpVC)\\
 \tab \vee  \mathtt{IsUsedConstructorAsObjectReceiver}(RemoveNotifyVisitor, TextFigure, basicTransform))\\
 \wedge \neg \mathtt{IsUsedConstructorAsObjectReceiver}(RemoveNotifyVisitor, TextFigure, contains)\\
 \wedge \neg \mathtt{IsUsedConstructorAsObjectReceiver}(RemoveNotifyVisitor, TextFigure, setAttribute)\\
 \wedge \neg \mathtt{IsUsedConstructorAsObjectReceiver}(RemoveNotifyVisitor, TextFigure, findFigureInside)\\
 \wedge \neg \mathtt{IsUsedConstructorAsObjectReceiver}(RemoveNotifyVisitor, TextFigure, addNotify)\\
 \wedge \neg (\mathtt{IsUsedConstructorAsObjectReceiver}(RemoveNotifyVisitor, BezierFigure, basicTransformTmpVC)\\
 \tab \vee  \mathtt{IsUsedConstructorAsObjectReceiver}(RemoveNotifyVisitor, BezierFigure, basicTransform))\\
 \wedge \neg \mathtt{IsUsedConstructorAsObjectReceiver}(RemoveNotifyVisitor, BezierFigure, contains)\\
 \wedge \neg \mathtt{IsUsedConstructorAsObjectReceiver}(RemoveNotifyVisitor, BezierFigure, setAttribute)\\
 \wedge \neg \mathtt{IsUsedConstructorAsObjectReceiver}(RemoveNotifyVisitor, BezierFigure, findFigureInside)\\
 \wedge \neg \mathtt{IsUsedConstructorAsObjectReceiver}(RemoveNotifyVisitor, BezierFigure, addNotify)\\
 \wedge \neg (\mathtt{IsUsedConstructorAsObjectReceiver}(RemoveNotifyVisitor, TextAreaFigure, basicTransformTmpVC)\\
 \tab \vee  \mathtt{IsUsedConstructorAsObjectReceiver}(RemoveNotifyVisitor, TextAreaFigure, basicTransform))\\
 \wedge \neg \mathtt{IsUsedConstructorAsObjectReceiver}(RemoveNotifyVisitor, TextAreaFigure, contains)\\
 \wedge \neg \mathtt{IsUsedConstructorAsObjectReceiver}(RemoveNotifyVisitor, TextAreaFigure, setAttribute)\\
 \wedge \neg \mathtt{IsUsedConstructorAsObjectReceiver}(RemoveNotifyVisitor, TextAreaFigure, findFigureInside)\\
 \wedge \neg \mathtt{IsUsedConstructorAsObjectReceiver}(RemoveNotifyVisitor, TextAreaFigure, addNotify)\\
 \wedge \neg (\mathtt{IsUsedConstructorAsObjectReceiver}(RemoveNotifyVisitor, NodeFigure, basicTransformTmpVC)\\
 \tab \vee  \mathtt{IsUsedConstructorAsObjectReceiver}(RemoveNotifyVisitor, NodeFigure, basicTransform))\\
 \wedge \neg \mathtt{IsUsedConstructorAsObjectReceiver}(RemoveNotifyVisitor, NodeFigure, contains)\\
 \wedge \neg \mathtt{IsUsedConstructorAsObjectReceiver}(RemoveNotifyVisitor, NodeFigure, setAttribute)\\
 \wedge \neg \mathtt{IsUsedConstructorAsObjectReceiver}(RemoveNotifyVisitor, NodeFigure, findFigureInside)\\
 \wedge \neg \mathtt{IsUsedConstructorAsObjectReceiver}(RemoveNotifyVisitor, NodeFigure, addNotify)\\
 \wedge \neg (\mathtt{IsUsedConstructorAsObjectReceiver}(RemoveNotifyVisitor, SVGImage, basicTransformTmpVC)\\
 \tab \vee  \mathtt{IsUsedConstructorAsObjectReceiver}(RemoveNotifyVisitor, SVGImage, basicTransform))\\
 \wedge \neg \mathtt{IsUsedConstructorAsObjectReceiver}(RemoveNotifyVisitor, SVGImage, contains)\\
 \wedge \neg \mathtt{IsUsedConstructorAsObjectReceiver}(RemoveNotifyVisitor, SVGImage, setAttribute)\\
 \wedge \neg \mathtt{IsUsedConstructorAsObjectReceiver}(RemoveNotifyVisitor, SVGImage, findFigureInside)\\
 \wedge \neg \mathtt{IsUsedConstructorAsObjectReceiver}(RemoveNotifyVisitor, SVGImage, addNotify)\\
 \wedge \neg (\mathtt{IsUsedConstructorAsObjectReceiver}(RemoveNotifyVisitor, SVGPath, basicTransformTmpVC)\\
 \tab \vee  \mathtt{IsUsedConstructorAsObjectReceiver}(RemoveNotifyVisitor, SVGPath, basicTransform))\\
 \wedge \neg \mathtt{IsUsedConstructorAsObjectReceiver}(RemoveNotifyVisitor, SVGPath, contains)\\
 \wedge \neg \mathtt{IsUsedConstructorAsObjectReceiver}(RemoveNotifyVisitor, SVGPath, setAttribute)\\
 \wedge \neg \mathtt{IsUsedConstructorAsObjectReceiver}(RemoveNotifyVisitor, SVGPath, findFigureInside)\\
 \wedge \neg \mathtt{IsUsedConstructorAsObjectReceiver}(RemoveNotifyVisitor, SVGPath, addNotify)\\
 \wedge \neg (\mathtt{IsUsedConstructorAsObjectReceiver}(RemoveNotifyVisitor, DependencyFigure, basicTransformTmpVC)\\
 \tab \vee  \mathtt{IsUsedConstructorAsObjectReceiver}(RemoveNotifyVisitor, DependencyFigure, basicTransform))\\
 \wedge \neg \mathtt{IsUsedConstructorAsObjectReceiver}(RemoveNotifyVisitor, DependencyFigure, contains)\\
 \wedge \neg \mathtt{IsUsedConstructorAsObjectReceiver}(RemoveNotifyVisitor, DependencyFigure, setAttribute)\\
 \wedge \neg \mathtt{IsUsedConstructorAsObjectReceiver}(RemoveNotifyVisitor, DependencyFigure, findFigureInside)\\
 \wedge \neg \mathtt{IsUsedConstructorAsObjectReceiver}(RemoveNotifyVisitor, DependencyFigure, addNotify)\\
 \wedge \neg (\mathtt{IsUsedConstructorAsObjectReceiver}(RemoveNotifyVisitor, LineConnectionFigure, basicTransformTmpVC)\\
 \tab \vee  \mathtt{IsUsedConstructorAsObjectReceiver}(RemoveNotifyVisitor, LineConnectionFigure, basicTransform))\\
 \wedge \neg \mathtt{IsUsedConstructorAsObjectReceiver}(RemoveNotifyVisitor, LineConnectionFigure, contains)\\
 \wedge \neg \mathtt{IsUsedConstructorAsObjectReceiver}(RemoveNotifyVisitor, LineConnectionFigure, setAttribute)\\
 \wedge \neg \mathtt{IsUsedConstructorAsObjectReceiver}(RemoveNotifyVisitor, LineConnectionFigure, findFigureInside)\\
 \wedge \neg \mathtt{IsUsedConstructorAsObjectReceiver}(RemoveNotifyVisitor, LineConnectionFigure, addNotify)\\
 \wedge \neg (\mathtt{IsUsedConstructorAsObjectReceiver}(RemoveNotifyVisitor, LabeledLineConnectionFigure, basicTransformTmpVC)\\
 \tab \vee  \mathtt{IsUsedConstructorAsObjectReceiver}(RemoveNotifyVisitor, LabeledLineConnectionFigure, basicTransform))\\
 \wedge \neg (\mathtt{IsUsedConstructorAsObjectReceiver}(RemoveNotifyVisitor, LabeledLineConnectionFigure, containsTmpVC)\\
 \tab \vee  \mathtt{IsUsedConstructorAsObjectReceiver}(RemoveNotifyVisitor, LabeledLineConnectionFigure, contains))\\
 \wedge \neg (\mathtt{IsUsedConstructorAsObjectReceiver}(RemoveNotifyVisitor, LabeledLineConnectionFigure, setAttributeTmpVC)\\
 \tab \vee  \mathtt{IsUsedConstructorAsObjectReceiver}(RemoveNotifyVisitor, LabeledLineConnectionFigure, setAttribute))\\
 \wedge \neg (\mathtt{IsUsedConstructorAsObjectReceiver}(RemoveNotifyVisitor, LabeledLineConnectionFigure, findFigureInsideTmpVC)\\
 \tab \vee  \mathtt{IsUsedConstructorAsObjectReceiver}(RemoveNotifyVisitor, LabeledLineConnectionFigure, findFigureInside))\\
 \wedge \neg (\mathtt{IsUsedConstructorAsObjectReceiver}(RemoveNotifyVisitor, LabeledLineConnectionFigure, addNotifyTmpVC)\\
 \tab \vee  \mathtt{IsUsedConstructorAsObjectReceiver}(RemoveNotifyVisitor, LabeledLineConnectionFigure, addNotify))\\
 \wedge \neg (\mathtt{IsUsedConstructorAsObjectReceiver}(RemoveNotifyVisitor, AbstractCompositeFigure, basicTransformTmpVC)\\
 \tab \vee  \mathtt{IsUsedConstructorAsObjectReceiver}(RemoveNotifyVisitor, AbstractCompositeFigure, basicTransform))\\
 \wedge \neg \mathtt{IsUsedConstructorAsObjectReceiver}(RemoveNotifyVisitor, AbstractCompositeFigure, contains)\\
 \wedge \neg \mathtt{IsUsedConstructorAsObjectReceiver}(RemoveNotifyVisitor, AbstractCompositeFigure, setAttribute)\\
 \wedge \neg \mathtt{IsUsedConstructorAsObjectReceiver}(RemoveNotifyVisitor, AbstractCompositeFigure, findFigureInside)\\
 \wedge \neg \mathtt{IsUsedConstructorAsObjectReceiver}(RemoveNotifyVisitor, AbstractCompositeFigure, addNotify)\\
 \wedge \neg (\mathtt{IsUsedConstructorAsObjectReceiver}(RemoveNotifyVisitor, GraphicalCompositeFigure, basicTransformTmpVC)\\
 \tab \vee  \mathtt{IsUsedConstructorAsObjectReceiver}(RemoveNotifyVisitor, GraphicalCompositeFigure, basicTransform))\\
 \wedge \neg \mathtt{IsUsedConstructorAsObjectReceiver}(RemoveNotifyVisitor, GraphicalCompositeFigure, contains)\\
 \wedge \neg \mathtt{IsUsedConstructorAsObjectReceiver}(RemoveNotifyVisitor, GraphicalCompositeFigure, setAttribute)\\
 \wedge \neg \mathtt{IsUsedConstructorAsObjectReceiver}(RemoveNotifyVisitor, GraphicalCompositeFigure, findFigureInside)\\
 \wedge \neg \mathtt{IsUsedConstructorAsObjectReceiver}(RemoveNotifyVisitor, GraphicalCompositeFigure, addNotify)\\
 \wedge \neg \mathtt{IsUsedConstructorAsMethodParameter}(AddNotifyVisitor, EllipseFigure, basicTransform)\\
 \wedge \neg \mathtt{IsUsedConstructorAsMethodParameter}(AddNotifyVisitor, EllipseFigure, contains)\\
 \wedge \neg \mathtt{IsUsedConstructorAsMethodParameter}(AddNotifyVisitor, EllipseFigure, setAttribute)\\
 \wedge \neg \mathtt{IsUsedConstructorAsMethodParameter}(AddNotifyVisitor, EllipseFigure, findFigureInside)\\
 \wedge \neg \mathtt{IsUsedConstructorAsMethodParameter}(AddNotifyVisitor, EllipseFigure, removeNotify)\\
 \wedge \neg \mathtt{IsUsedConstructorAsMethodParameter}(AddNotifyVisitor, DiamondFigure, basicTransform)\\
 \wedge \neg \mathtt{IsUsedConstructorAsMethodParameter}(AddNotifyVisitor, DiamondFigure, contains)\\
 \wedge \neg \mathtt{IsUsedConstructorAsMethodParameter}(AddNotifyVisitor, DiamondFigure, setAttribute)\\
 \wedge \neg \mathtt{IsUsedConstructorAsMethodParameter}(AddNotifyVisitor, DiamondFigure, findFigureInside)\\
 \wedge \neg \mathtt{IsUsedConstructorAsMethodParameter}(AddNotifyVisitor, DiamondFigure, removeNotify)\\
 \wedge \neg \mathtt{IsUsedConstructorAsMethodParameter}(AddNotifyVisitor, RectangleFigure, basicTransform)\\
 \wedge \neg \mathtt{IsUsedConstructorAsMethodParameter}(AddNotifyVisitor, RectangleFigure, contains)\\
 \wedge \neg \mathtt{IsUsedConstructorAsMethodParameter}(AddNotifyVisitor, RectangleFigure, setAttribute)\\
 \wedge \neg \mathtt{IsUsedConstructorAsMethodParameter}(AddNotifyVisitor, RectangleFigure, findFigureInside)\\
 \wedge \neg \mathtt{IsUsedConstructorAsMethodParameter}(AddNotifyVisitor, RectangleFigure, removeNotify)\\
 \wedge \neg \mathtt{IsUsedConstructorAsMethodParameter}(AddNotifyVisitor, RoundRectangleFigure, basicTransform)\\
 \wedge \neg \mathtt{IsUsedConstructorAsMethodParameter}(AddNotifyVisitor, RoundRectangleFigure, contains)\\
 \wedge \neg \mathtt{IsUsedConstructorAsMethodParameter}(AddNotifyVisitor, RoundRectangleFigure, setAttribute)\\
 \wedge \neg \mathtt{IsUsedConstructorAsMethodParameter}(AddNotifyVisitor, RoundRectangleFigure, findFigureInside)\\
 \wedge \neg \mathtt{IsUsedConstructorAsMethodParameter}(AddNotifyVisitor, RoundRectangleFigure, removeNotify)\\
 \wedge \neg \mathtt{IsUsedConstructorAsMethodParameter}(AddNotifyVisitor, TriangleFigure, basicTransform)\\
 \wedge \neg \mathtt{IsUsedConstructorAsMethodParameter}(AddNotifyVisitor, TriangleFigure, contains)\\
 \wedge \neg \mathtt{IsUsedConstructorAsMethodParameter}(AddNotifyVisitor, TriangleFigure, setAttribute)\\
 \wedge \neg \mathtt{IsUsedConstructorAsMethodParameter}(AddNotifyVisitor, TriangleFigure, findFigureInside)\\
 \wedge \neg \mathtt{IsUsedConstructorAsMethodParameter}(AddNotifyVisitor, TriangleFigure, removeNotify)\\
 \wedge \neg \mathtt{IsUsedConstructorAsMethodParameter}(AddNotifyVisitor, TextFigure, basicTransform)\\
 \wedge \neg \mathtt{IsUsedConstructorAsMethodParameter}(AddNotifyVisitor, TextFigure, contains)\\
 \wedge \neg \mathtt{IsUsedConstructorAsMethodParameter}(AddNotifyVisitor, TextFigure, setAttribute)\\
 \wedge \neg \mathtt{IsUsedConstructorAsMethodParameter}(AddNotifyVisitor, TextFigure, findFigureInside)\\
 \wedge \neg \mathtt{IsUsedConstructorAsMethodParameter}(AddNotifyVisitor, TextFigure, removeNotify)\\
 \wedge \neg \mathtt{IsUsedConstructorAsMethodParameter}(AddNotifyVisitor, BezierFigure, basicTransform)\\
 \wedge \neg \mathtt{IsUsedConstructorAsMethodParameter}(AddNotifyVisitor, BezierFigure, contains)\\
 \wedge \neg \mathtt{IsUsedConstructorAsMethodParameter}(AddNotifyVisitor, BezierFigure, setAttribute)\\
 \wedge \neg \mathtt{IsUsedConstructorAsMethodParameter}(AddNotifyVisitor, BezierFigure, findFigureInside)\\
 \wedge \neg \mathtt{IsUsedConstructorAsMethodParameter}(AddNotifyVisitor, BezierFigure, removeNotify)\\
 \wedge \neg \mathtt{IsUsedConstructorAsMethodParameter}(AddNotifyVisitor, TextAreaFigure, basicTransform)\\
 \wedge \neg \mathtt{IsUsedConstructorAsMethodParameter}(AddNotifyVisitor, TextAreaFigure, contains)\\
 \wedge \neg \mathtt{IsUsedConstructorAsMethodParameter}(AddNotifyVisitor, TextAreaFigure, setAttribute)\\
 \wedge \neg \mathtt{IsUsedConstructorAsMethodParameter}(AddNotifyVisitor, TextAreaFigure, findFigureInside)\\
 \wedge \neg \mathtt{IsUsedConstructorAsMethodParameter}(AddNotifyVisitor, TextAreaFigure, removeNotify)\\
 \wedge \neg \mathtt{IsUsedConstructorAsMethodParameter}(AddNotifyVisitor, NodeFigure, basicTransform)\\
 \wedge \neg \mathtt{IsUsedConstructorAsMethodParameter}(AddNotifyVisitor, NodeFigure, contains)\\
 \wedge \neg \mathtt{IsUsedConstructorAsMethodParameter}(AddNotifyVisitor, NodeFigure, setAttribute)\\
 \wedge \neg \mathtt{IsUsedConstructorAsMethodParameter}(AddNotifyVisitor, NodeFigure, findFigureInside)\\
 \wedge \neg \mathtt{IsUsedConstructorAsMethodParameter}(AddNotifyVisitor, NodeFigure, removeNotify)\\
 \wedge \neg \mathtt{IsUsedConstructorAsMethodParameter}(AddNotifyVisitor, SVGImage, basicTransform)\\
 \wedge \neg \mathtt{IsUsedConstructorAsMethodParameter}(AddNotifyVisitor, SVGImage, contains)\\
 \wedge \neg \mathtt{IsUsedConstructorAsMethodParameter}(AddNotifyVisitor, SVGImage, setAttribute)\\
 \wedge \neg \mathtt{IsUsedConstructorAsMethodParameter}(AddNotifyVisitor, SVGImage, findFigureInside)\\
 \wedge \neg \mathtt{IsUsedConstructorAsMethodParameter}(AddNotifyVisitor, SVGImage, removeNotify)\\
 \wedge \neg \mathtt{IsUsedConstructorAsMethodParameter}(AddNotifyVisitor, SVGPath, basicTransform)\\
 \wedge \neg \mathtt{IsUsedConstructorAsMethodParameter}(AddNotifyVisitor, SVGPath, contains)\\
 \wedge \neg \mathtt{IsUsedConstructorAsMethodParameter}(AddNotifyVisitor, SVGPath, setAttribute)\\
 \wedge \neg \mathtt{IsUsedConstructorAsMethodParameter}(AddNotifyVisitor, SVGPath, findFigureInside)\\
 \wedge \neg \mathtt{IsUsedConstructorAsMethodParameter}(AddNotifyVisitor, SVGPath, removeNotify)\\
 \wedge \neg \mathtt{IsUsedConstructorAsMethodParameter}(AddNotifyVisitor, DependencyFigure, basicTransform)\\
 \wedge \neg \mathtt{IsUsedConstructorAsMethodParameter}(AddNotifyVisitor, DependencyFigure, contains)\\
 \wedge \neg \mathtt{IsUsedConstructorAsMethodParameter}(AddNotifyVisitor, DependencyFigure, setAttribute)\\
 \wedge \neg \mathtt{IsUsedConstructorAsMethodParameter}(AddNotifyVisitor, DependencyFigure, findFigureInside)\\
 \wedge \neg \mathtt{IsUsedConstructorAsMethodParameter}(AddNotifyVisitor, DependencyFigure, removeNotify)\\
 \wedge \neg \mathtt{IsUsedConstructorAsMethodParameter}(AddNotifyVisitor, LineConnectionFigure, basicTransform)\\
 \wedge \neg \mathtt{IsUsedConstructorAsMethodParameter}(AddNotifyVisitor, LineConnectionFigure, contains)\\
 \wedge \neg \mathtt{IsUsedConstructorAsMethodParameter}(AddNotifyVisitor, LineConnectionFigure, setAttribute)\\
 \wedge \neg \mathtt{IsUsedConstructorAsMethodParameter}(AddNotifyVisitor, LineConnectionFigure, findFigureInside)\\
 \wedge \neg \mathtt{IsUsedConstructorAsMethodParameter}(AddNotifyVisitor, LineConnectionFigure, removeNotify)\\
 \wedge \neg \mathtt{IsUsedConstructorAsMethodParameter}(AddNotifyVisitor, LabeledLineConnectionFigure, basicTransform)\\
 \wedge \neg \mathtt{IsUsedConstructorAsMethodParameter}(AddNotifyVisitor, LabeledLineConnectionFigure, contains)\\
 \wedge \neg \mathtt{IsUsedConstructorAsMethodParameter}(AddNotifyVisitor, LabeledLineConnectionFigure, setAttribute)\\
 \wedge \neg \mathtt{IsUsedConstructorAsMethodParameter}(AddNotifyVisitor, LabeledLineConnectionFigure, findFigureInside)\\
 \wedge \neg \mathtt{IsUsedConstructorAsMethodParameter}(AddNotifyVisitor, LabeledLineConnectionFigure, removeNotify)\\
 \wedge \neg \mathtt{IsUsedConstructorAsMethodParameter}(AddNotifyVisitor, AbstractCompositeFigure, basicTransform)\\
 \wedge \neg \mathtt{IsUsedConstructorAsMethodParameter}(AddNotifyVisitor, AbstractCompositeFigure, contains)\\
 \wedge \neg \mathtt{IsUsedConstructorAsMethodParameter}(AddNotifyVisitor, AbstractCompositeFigure, setAttribute)\\
 \wedge \neg \mathtt{IsUsedConstructorAsMethodParameter}(AddNotifyVisitor, AbstractCompositeFigure, findFigureInside)\\
 \wedge \neg \mathtt{IsUsedConstructorAsMethodParameter}(AddNotifyVisitor, AbstractCompositeFigure, removeNotify)\\
 \wedge \neg \mathtt{IsUsedConstructorAsMethodParameter}(AddNotifyVisitor, GraphicalCompositeFigure, basicTransform)\\
 \wedge \neg \mathtt{IsUsedConstructorAsMethodParameter}(AddNotifyVisitor, GraphicalCompositeFigure, contains)\\
 \wedge \neg \mathtt{IsUsedConstructorAsMethodParameter}(AddNotifyVisitor, GraphicalCompositeFigure, setAttribute)\\
 \wedge \neg \mathtt{IsUsedConstructorAsMethodParameter}(AddNotifyVisitor, GraphicalCompositeFigure, findFigureInside)\\
 \wedge \neg \mathtt{IsUsedConstructorAsMethodParameter}(AddNotifyVisitor, GraphicalCompositeFigure, removeNotify)\\
 \wedge \neg \mathtt{IsUsedConstructorAsMethodParameter}(AddNotifyVisitor, AbstractFigure, basicTransform)\\
 \wedge \neg \mathtt{IsUsedConstructorAsMethodParameter}(AddNotifyVisitor, AbstractFigure, contains)\\
 \wedge \neg \mathtt{IsUsedConstructorAsMethodParameter}(AddNotifyVisitor, AbstractFigure, setAttribute)\\
 \wedge \neg \mathtt{IsUsedConstructorAsMethodParameter}(AddNotifyVisitor, AbstractFigure, findFigureInside)\\
 \wedge \neg \mathtt{IsUsedConstructorAsMethodParameter}(AddNotifyVisitor, AbstractFigure, addNotify)\\
 \wedge \neg \mathtt{IsUsedConstructorAsMethodParameter}(AddNotifyVisitor, AbstractFigure, removeNotify)\\
 \wedge \neg (\mathtt{IsUsedConstructorAsObjectReceiver}(AddNotifyVisitor, EllipseFigure, basicTransformTmpVC)\\
 \tab \vee  \mathtt{IsUsedConstructorAsObjectReceiver}(AddNotifyVisitor, EllipseFigure, basicTransform))\\
 \wedge \neg \mathtt{IsUsedConstructorAsObjectReceiver}(AddNotifyVisitor, EllipseFigure, contains)\\
 \wedge \neg \mathtt{IsUsedConstructorAsObjectReceiver}(AddNotifyVisitor, EllipseFigure, setAttribute)\\
 \wedge \neg \mathtt{IsUsedConstructorAsObjectReceiver}(AddNotifyVisitor, EllipseFigure, findFigureInside)\\
 \wedge \neg \mathtt{IsUsedConstructorAsObjectReceiver}(AddNotifyVisitor, EllipseFigure, removeNotify)\\
 \wedge \neg (\mathtt{IsUsedConstructorAsObjectReceiver}(AddNotifyVisitor, DiamondFigure, basicTransformTmpVC)\\
 \tab \vee  \mathtt{IsUsedConstructorAsObjectReceiver}(AddNotifyVisitor, DiamondFigure, basicTransform))\\
 \wedge \neg \mathtt{IsUsedConstructorAsObjectReceiver}(AddNotifyVisitor, DiamondFigure, contains)\\
 \wedge \neg \mathtt{IsUsedConstructorAsObjectReceiver}(AddNotifyVisitor, DiamondFigure, setAttribute)\\
 \wedge \neg \mathtt{IsUsedConstructorAsObjectReceiver}(AddNotifyVisitor, DiamondFigure, findFigureInside)\\
 \wedge \neg \mathtt{IsUsedConstructorAsObjectReceiver}(AddNotifyVisitor, DiamondFigure, removeNotify)\\
 \wedge \neg (\mathtt{IsUsedConstructorAsObjectReceiver}(AddNotifyVisitor, RectangleFigure, basicTransformTmpVC)\\
 \tab \vee  \mathtt{IsUsedConstructorAsObjectReceiver}(AddNotifyVisitor, RectangleFigure, basicTransform))\\
 \wedge \neg \mathtt{IsUsedConstructorAsObjectReceiver}(AddNotifyVisitor, RectangleFigure, contains)\\
 \wedge \neg \mathtt{IsUsedConstructorAsObjectReceiver}(AddNotifyVisitor, RectangleFigure, setAttribute)\\
 \wedge \neg \mathtt{IsUsedConstructorAsObjectReceiver}(AddNotifyVisitor, RectangleFigure, findFigureInside)\\
 \wedge \neg \mathtt{IsUsedConstructorAsObjectReceiver}(AddNotifyVisitor, RectangleFigure, removeNotify)\\
 \wedge \neg (\mathtt{IsUsedConstructorAsObjectReceiver}(AddNotifyVisitor, RoundRectangleFigure, basicTransformTmpVC)\\
 \tab \vee  \mathtt{IsUsedConstructorAsObjectReceiver}(AddNotifyVisitor, RoundRectangleFigure, basicTransform))\\
 \wedge \neg \mathtt{IsUsedConstructorAsObjectReceiver}(AddNotifyVisitor, RoundRectangleFigure, contains)\\
 \wedge \neg \mathtt{IsUsedConstructorAsObjectReceiver}(AddNotifyVisitor, RoundRectangleFigure, setAttribute)\\
 \wedge \neg \mathtt{IsUsedConstructorAsObjectReceiver}(AddNotifyVisitor, RoundRectangleFigure, findFigureInside)\\
 \wedge \neg \mathtt{IsUsedConstructorAsObjectReceiver}(AddNotifyVisitor, RoundRectangleFigure, removeNotify)\\
 \wedge \neg (\mathtt{IsUsedConstructorAsObjectReceiver}(AddNotifyVisitor, TriangleFigure, basicTransformTmpVC)\\
 \tab \vee  \mathtt{IsUsedConstructorAsObjectReceiver}(AddNotifyVisitor, TriangleFigure, basicTransform))\\
 \wedge \neg \mathtt{IsUsedConstructorAsObjectReceiver}(AddNotifyVisitor, TriangleFigure, contains)\\
 \wedge \neg \mathtt{IsUsedConstructorAsObjectReceiver}(AddNotifyVisitor, TriangleFigure, setAttribute)\\
 \wedge \neg \mathtt{IsUsedConstructorAsObjectReceiver}(AddNotifyVisitor, TriangleFigure, findFigureInside)\\
 \wedge \neg \mathtt{IsUsedConstructorAsObjectReceiver}(AddNotifyVisitor, TriangleFigure, removeNotify)\\
 \wedge \neg (\mathtt{IsUsedConstructorAsObjectReceiver}(AddNotifyVisitor, TextFigure, basicTransformTmpVC)\\
 \tab \vee  \mathtt{IsUsedConstructorAsObjectReceiver}(AddNotifyVisitor, TextFigure, basicTransform))\\
 \wedge \neg \mathtt{IsUsedConstructorAsObjectReceiver}(AddNotifyVisitor, TextFigure, contains)\\
 \wedge \neg \mathtt{IsUsedConstructorAsObjectReceiver}(AddNotifyVisitor, TextFigure, setAttribute)\\
 \wedge \neg \mathtt{IsUsedConstructorAsObjectReceiver}(AddNotifyVisitor, TextFigure, findFigureInside)\\
 \wedge \neg \mathtt{IsUsedConstructorAsObjectReceiver}(AddNotifyVisitor, TextFigure, removeNotify)\\
 \wedge \neg (\mathtt{IsUsedConstructorAsObjectReceiver}(AddNotifyVisitor, BezierFigure, basicTransformTmpVC)\\
 \tab \vee  \mathtt{IsUsedConstructorAsObjectReceiver}(AddNotifyVisitor, BezierFigure, basicTransform))\\
 \wedge \neg \mathtt{IsUsedConstructorAsObjectReceiver}(AddNotifyVisitor, BezierFigure, contains)\\
 \wedge \neg \mathtt{IsUsedConstructorAsObjectReceiver}(AddNotifyVisitor, BezierFigure, setAttribute)\\
 \wedge \neg \mathtt{IsUsedConstructorAsObjectReceiver}(AddNotifyVisitor, BezierFigure, findFigureInside)\\
 \wedge \neg \mathtt{IsUsedConstructorAsObjectReceiver}(AddNotifyVisitor, BezierFigure, removeNotify)\\
 \wedge \neg (\mathtt{IsUsedConstructorAsObjectReceiver}(AddNotifyVisitor, TextAreaFigure, basicTransformTmpVC)\\
 \tab \vee  \mathtt{IsUsedConstructorAsObjectReceiver}(AddNotifyVisitor, TextAreaFigure, basicTransform))\\
 \wedge \neg \mathtt{IsUsedConstructorAsObjectReceiver}(AddNotifyVisitor, TextAreaFigure, contains)\\
 \wedge \neg \mathtt{IsUsedConstructorAsObjectReceiver}(AddNotifyVisitor, TextAreaFigure, setAttribute)\\
 \wedge \neg \mathtt{IsUsedConstructorAsObjectReceiver}(AddNotifyVisitor, TextAreaFigure, findFigureInside)\\
 \wedge \neg \mathtt{IsUsedConstructorAsObjectReceiver}(AddNotifyVisitor, TextAreaFigure, removeNotify)\\
 \wedge \neg (\mathtt{IsUsedConstructorAsObjectReceiver}(AddNotifyVisitor, NodeFigure, basicTransformTmpVC)\\
 \tab \vee  \mathtt{IsUsedConstructorAsObjectReceiver}(AddNotifyVisitor, NodeFigure, basicTransform))\\
 \wedge \neg \mathtt{IsUsedConstructorAsObjectReceiver}(AddNotifyVisitor, NodeFigure, contains)\\
 \wedge \neg \mathtt{IsUsedConstructorAsObjectReceiver}(AddNotifyVisitor, NodeFigure, setAttribute)\\
 \wedge \neg \mathtt{IsUsedConstructorAsObjectReceiver}(AddNotifyVisitor, NodeFigure, findFigureInside)\\
 \wedge \neg \mathtt{IsUsedConstructorAsObjectReceiver}(AddNotifyVisitor, NodeFigure, removeNotify)\\
 \wedge \neg (\mathtt{IsUsedConstructorAsObjectReceiver}(AddNotifyVisitor, SVGImage, basicTransformTmpVC)\\
 \tab \vee  \mathtt{IsUsedConstructorAsObjectReceiver}(AddNotifyVisitor, SVGImage, basicTransform))\\
 \wedge \neg \mathtt{IsUsedConstructorAsObjectReceiver}(AddNotifyVisitor, SVGImage, contains)\\
 \wedge \neg \mathtt{IsUsedConstructorAsObjectReceiver}(AddNotifyVisitor, SVGImage, setAttribute)\\
 \wedge \neg \mathtt{IsUsedConstructorAsObjectReceiver}(AddNotifyVisitor, SVGImage, findFigureInside)\\
 \wedge \neg \mathtt{IsUsedConstructorAsObjectReceiver}(AddNotifyVisitor, SVGImage, removeNotify)\\
 \wedge \neg (\mathtt{IsUsedConstructorAsObjectReceiver}(AddNotifyVisitor, SVGPath, basicTransformTmpVC)\\
 \tab \vee  \mathtt{IsUsedConstructorAsObjectReceiver}(AddNotifyVisitor, SVGPath, basicTransform))\\
 \wedge \neg \mathtt{IsUsedConstructorAsObjectReceiver}(AddNotifyVisitor, SVGPath, contains)\\
 \wedge \neg \mathtt{IsUsedConstructorAsObjectReceiver}(AddNotifyVisitor, SVGPath, setAttribute)\\
 \wedge \neg \mathtt{IsUsedConstructorAsObjectReceiver}(AddNotifyVisitor, SVGPath, findFigureInside)\\
 \wedge \neg \mathtt{IsUsedConstructorAsObjectReceiver}(AddNotifyVisitor, SVGPath, removeNotify)\\
 \wedge \neg (\mathtt{IsUsedConstructorAsObjectReceiver}(AddNotifyVisitor, DependencyFigure, basicTransformTmpVC)\\
 \tab \vee  \mathtt{IsUsedConstructorAsObjectReceiver}(AddNotifyVisitor, DependencyFigure, basicTransform))\\
 \wedge \neg \mathtt{IsUsedConstructorAsObjectReceiver}(AddNotifyVisitor, DependencyFigure, contains)\\
 \wedge \neg \mathtt{IsUsedConstructorAsObjectReceiver}(AddNotifyVisitor, DependencyFigure, setAttribute)\\
 \wedge \neg \mathtt{IsUsedConstructorAsObjectReceiver}(AddNotifyVisitor, DependencyFigure, findFigureInside)\\
 \wedge \neg \mathtt{IsUsedConstructorAsObjectReceiver}(AddNotifyVisitor, DependencyFigure, removeNotify)\\
 \wedge \neg (\mathtt{IsUsedConstructorAsObjectReceiver}(AddNotifyVisitor, LineConnectionFigure, basicTransformTmpVC)\\
 \tab \vee  \mathtt{IsUsedConstructorAsObjectReceiver}(AddNotifyVisitor, LineConnectionFigure, basicTransform))\\
 \wedge \neg \mathtt{IsUsedConstructorAsObjectReceiver}(AddNotifyVisitor, LineConnectionFigure, contains)\\
 \wedge \neg \mathtt{IsUsedConstructorAsObjectReceiver}(AddNotifyVisitor, LineConnectionFigure, setAttribute)\\
 \wedge \neg \mathtt{IsUsedConstructorAsObjectReceiver}(AddNotifyVisitor, LineConnectionFigure, findFigureInside)\\
 \wedge \neg \mathtt{IsUsedConstructorAsObjectReceiver}(AddNotifyVisitor, LineConnectionFigure, removeNotify)\\
 \wedge \neg (\mathtt{IsUsedConstructorAsObjectReceiver}(AddNotifyVisitor, LabeledLineConnectionFigure, basicTransformTmpVC)\\
 \tab \vee  \mathtt{IsUsedConstructorAsObjectReceiver}(AddNotifyVisitor, LabeledLineConnectionFigure, basicTransform))\\
 \wedge \neg (\mathtt{IsUsedConstructorAsObjectReceiver}(AddNotifyVisitor, LabeledLineConnectionFigure, containsTmpVC)\\
 \tab \vee  \mathtt{IsUsedConstructorAsObjectReceiver}(AddNotifyVisitor, LabeledLineConnectionFigure, contains))\\
 \wedge \neg (\mathtt{IsUsedConstructorAsObjectReceiver}(AddNotifyVisitor, LabeledLineConnectionFigure, setAttributeTmpVC)\\
 \tab \vee  \mathtt{IsUsedConstructorAsObjectReceiver}(AddNotifyVisitor, LabeledLineConnectionFigure, setAttribute))\\
 \wedge \neg (\mathtt{IsUsedConstructorAsObjectReceiver}(AddNotifyVisitor, LabeledLineConnectionFigure, findFigureInsideTmpVC)\\
 \tab \vee  \mathtt{IsUsedConstructorAsObjectReceiver}(AddNotifyVisitor, LabeledLineConnectionFigure, findFigureInside))\\
 \wedge \neg (\mathtt{IsUsedConstructorAsObjectReceiver}(AddNotifyVisitor, LabeledLineConnectionFigure, removeNotifyTmpVC)\\
 \tab \vee  \mathtt{IsUsedConstructorAsObjectReceiver}(AddNotifyVisitor, LabeledLineConnectionFigure, removeNotify))\\
 \wedge \neg (\mathtt{IsUsedConstructorAsObjectReceiver}(AddNotifyVisitor, AbstractCompositeFigure, basicTransformTmpVC)\\
 \tab \vee  \mathtt{IsUsedConstructorAsObjectReceiver}(AddNotifyVisitor, AbstractCompositeFigure, basicTransform))\\
 \wedge \neg \mathtt{IsUsedConstructorAsObjectReceiver}(AddNotifyVisitor, AbstractCompositeFigure, contains)\\
 \wedge \neg \mathtt{IsUsedConstructorAsObjectReceiver}(AddNotifyVisitor, AbstractCompositeFigure, setAttribute)\\
 \wedge \neg \mathtt{IsUsedConstructorAsObjectReceiver}(AddNotifyVisitor, AbstractCompositeFigure, findFigureInside)\\
 \wedge \neg \mathtt{IsUsedConstructorAsObjectReceiver}(AddNotifyVisitor, AbstractCompositeFigure, removeNotify)\\
 \wedge \neg (\mathtt{IsUsedConstructorAsObjectReceiver}(AddNotifyVisitor, GraphicalCompositeFigure, basicTransformTmpVC)\\
 \tab \vee  \mathtt{IsUsedConstructorAsObjectReceiver}(AddNotifyVisitor, GraphicalCompositeFigure, basicTransform))\\
 \wedge \neg \mathtt{IsUsedConstructorAsObjectReceiver}(AddNotifyVisitor, GraphicalCompositeFigure, contains)\\
 \wedge \neg \mathtt{IsUsedConstructorAsObjectReceiver}(AddNotifyVisitor, GraphicalCompositeFigure, setAttribute)\\
 \wedge \neg \mathtt{IsUsedConstructorAsObjectReceiver}(AddNotifyVisitor, GraphicalCompositeFigure, findFigureInside)\\
 \wedge \neg \mathtt{IsUsedConstructorAsObjectReceiver}(AddNotifyVisitor, GraphicalCompositeFigure, removeNotify)\\
 \wedge \neg \mathtt{IsUsedConstructorAsMethodParameter}(FindFigureInsideVisitor, EllipseFigure, basicTransform)\\
 \wedge \neg \mathtt{IsUsedConstructorAsMethodParameter}(FindFigureInsideVisitor, EllipseFigure, contains)\\
 \wedge \neg \mathtt{IsUsedConstructorAsMethodParameter}(FindFigureInsideVisitor, EllipseFigure, setAttribute)\\
 \wedge \neg \mathtt{IsUsedConstructorAsMethodParameter}(FindFigureInsideVisitor, EllipseFigure, addNotify)\\
 \wedge \neg \mathtt{IsUsedConstructorAsMethodParameter}(FindFigureInsideVisitor, EllipseFigure, removeNotify)\\
 \wedge \neg \mathtt{IsUsedConstructorAsMethodParameter}(FindFigureInsideVisitor, DiamondFigure, basicTransform)\\
 \wedge \neg \mathtt{IsUsedConstructorAsMethodParameter}(FindFigureInsideVisitor, DiamondFigure, contains)\\
 \wedge \neg \mathtt{IsUsedConstructorAsMethodParameter}(FindFigureInsideVisitor, DiamondFigure, setAttribute)\\
 \wedge \neg \mathtt{IsUsedConstructorAsMethodParameter}(FindFigureInsideVisitor, DiamondFigure, addNotify)\\
 \wedge \neg \mathtt{IsUsedConstructorAsMethodParameter}(FindFigureInsideVisitor, DiamondFigure, removeNotify)\\
 \wedge \neg \mathtt{IsUsedConstructorAsMethodParameter}(FindFigureInsideVisitor, RectangleFigure, basicTransform)\\
 \wedge \neg \mathtt{IsUsedConstructorAsMethodParameter}(FindFigureInsideVisitor, RectangleFigure, contains)\\
 \wedge \neg \mathtt{IsUsedConstructorAsMethodParameter}(FindFigureInsideVisitor, RectangleFigure, setAttribute)\\
 \wedge \neg \mathtt{IsUsedConstructorAsMethodParameter}(FindFigureInsideVisitor, RectangleFigure, addNotify)\\
 \wedge \neg \mathtt{IsUsedConstructorAsMethodParameter}(FindFigureInsideVisitor, RectangleFigure, removeNotify)\\
 \wedge \neg \mathtt{IsUsedConstructorAsMethodParameter}(FindFigureInsideVisitor, RoundRectangleFigure, basicTransform)\\
 \wedge \neg \mathtt{IsUsedConstructorAsMethodParameter}(FindFigureInsideVisitor, RoundRectangleFigure, contains)\\
 \wedge \neg \mathtt{IsUsedConstructorAsMethodParameter}(FindFigureInsideVisitor, RoundRectangleFigure, setAttribute)\\
 \wedge \neg \mathtt{IsUsedConstructorAsMethodParameter}(FindFigureInsideVisitor, RoundRectangleFigure, addNotify)\\
 \wedge \neg \mathtt{IsUsedConstructorAsMethodParameter}(FindFigureInsideVisitor, RoundRectangleFigure, removeNotify)\\
 \wedge \neg \mathtt{IsUsedConstructorAsMethodParameter}(FindFigureInsideVisitor, TriangleFigure, basicTransform)\\
 \wedge \neg \mathtt{IsUsedConstructorAsMethodParameter}(FindFigureInsideVisitor, TriangleFigure, contains)\\
 \wedge \neg \mathtt{IsUsedConstructorAsMethodParameter}(FindFigureInsideVisitor, TriangleFigure, setAttribute)\\
 \wedge \neg \mathtt{IsUsedConstructorAsMethodParameter}(FindFigureInsideVisitor, TriangleFigure, addNotify)\\
 \wedge \neg \mathtt{IsUsedConstructorAsMethodParameter}(FindFigureInsideVisitor, TriangleFigure, removeNotify)\\
 \wedge \neg \mathtt{IsUsedConstructorAsMethodParameter}(FindFigureInsideVisitor, TextFigure, basicTransform)\\
 \wedge \neg \mathtt{IsUsedConstructorAsMethodParameter}(FindFigureInsideVisitor, TextFigure, contains)\\
 \wedge \neg \mathtt{IsUsedConstructorAsMethodParameter}(FindFigureInsideVisitor, TextFigure, setAttribute)\\
 \wedge \neg \mathtt{IsUsedConstructorAsMethodParameter}(FindFigureInsideVisitor, TextFigure, addNotify)\\
 \wedge \neg \mathtt{IsUsedConstructorAsMethodParameter}(FindFigureInsideVisitor, TextFigure, removeNotify)\\
 \wedge \neg \mathtt{IsUsedConstructorAsMethodParameter}(FindFigureInsideVisitor, BezierFigure, basicTransform)\\
 \wedge \neg \mathtt{IsUsedConstructorAsMethodParameter}(FindFigureInsideVisitor, BezierFigure, contains)\\
 \wedge \neg \mathtt{IsUsedConstructorAsMethodParameter}(FindFigureInsideVisitor, BezierFigure, setAttribute)\\
 \wedge \neg \mathtt{IsUsedConstructorAsMethodParameter}(FindFigureInsideVisitor, BezierFigure, addNotify)\\
 \wedge \neg \mathtt{IsUsedConstructorAsMethodParameter}(FindFigureInsideVisitor, BezierFigure, removeNotify)\\
 \wedge \neg \mathtt{IsUsedConstructorAsMethodParameter}(FindFigureInsideVisitor, TextAreaFigure, basicTransform)\\
 \wedge \neg \mathtt{IsUsedConstructorAsMethodParameter}(FindFigureInsideVisitor, TextAreaFigure, contains)\\
 \wedge \neg \mathtt{IsUsedConstructorAsMethodParameter}(FindFigureInsideVisitor, TextAreaFigure, setAttribute)\\
 \wedge \neg \mathtt{IsUsedConstructorAsMethodParameter}(FindFigureInsideVisitor, TextAreaFigure, addNotify)\\
 \wedge \neg \mathtt{IsUsedConstructorAsMethodParameter}(FindFigureInsideVisitor, TextAreaFigure, removeNotify)\\
 \wedge \neg \mathtt{IsUsedConstructorAsMethodParameter}(FindFigureInsideVisitor, NodeFigure, basicTransform)\\
 \wedge \neg \mathtt{IsUsedConstructorAsMethodParameter}(FindFigureInsideVisitor, NodeFigure, contains)\\
 \wedge \neg \mathtt{IsUsedConstructorAsMethodParameter}(FindFigureInsideVisitor, NodeFigure, setAttribute)\\
 \wedge \neg \mathtt{IsUsedConstructorAsMethodParameter}(FindFigureInsideVisitor, NodeFigure, addNotify)\\
 \wedge \neg \mathtt{IsUsedConstructorAsMethodParameter}(FindFigureInsideVisitor, NodeFigure, removeNotify)\\
 \wedge \neg \mathtt{IsUsedConstructorAsMethodParameter}(FindFigureInsideVisitor, SVGImage, basicTransform)\\
 \wedge \neg \mathtt{IsUsedConstructorAsMethodParameter}(FindFigureInsideVisitor, SVGImage, contains)\\
 \wedge \neg \mathtt{IsUsedConstructorAsMethodParameter}(FindFigureInsideVisitor, SVGImage, setAttribute)\\
 \wedge \neg \mathtt{IsUsedConstructorAsMethodParameter}(FindFigureInsideVisitor, SVGImage, addNotify)\\
 \wedge \neg \mathtt{IsUsedConstructorAsMethodParameter}(FindFigureInsideVisitor, SVGImage, removeNotify)\\
 \wedge \neg \mathtt{IsUsedConstructorAsMethodParameter}(FindFigureInsideVisitor, SVGPath, basicTransform)\\
 \wedge \neg \mathtt{IsUsedConstructorAsMethodParameter}(FindFigureInsideVisitor, SVGPath, contains)\\
 \wedge \neg \mathtt{IsUsedConstructorAsMethodParameter}(FindFigureInsideVisitor, SVGPath, setAttribute)\\
 \wedge \neg \mathtt{IsUsedConstructorAsMethodParameter}(FindFigureInsideVisitor, SVGPath, addNotify)\\
 \wedge \neg \mathtt{IsUsedConstructorAsMethodParameter}(FindFigureInsideVisitor, SVGPath, removeNotify)\\
 \wedge \neg \mathtt{IsUsedConstructorAsMethodParameter}(FindFigureInsideVisitor, DependencyFigure, basicTransform)\\
 \wedge \neg \mathtt{IsUsedConstructorAsMethodParameter}(FindFigureInsideVisitor, DependencyFigure, contains)\\
 \wedge \neg \mathtt{IsUsedConstructorAsMethodParameter}(FindFigureInsideVisitor, DependencyFigure, setAttribute)\\
 \wedge \neg \mathtt{IsUsedConstructorAsMethodParameter}(FindFigureInsideVisitor, DependencyFigure, addNotify)\\
 \wedge \neg \mathtt{IsUsedConstructorAsMethodParameter}(FindFigureInsideVisitor, DependencyFigure, removeNotify)\\
 \wedge \neg \mathtt{IsUsedConstructorAsMethodParameter}(FindFigureInsideVisitor, LineConnectionFigure, basicTransform)\\
 \wedge \neg \mathtt{IsUsedConstructorAsMethodParameter}(FindFigureInsideVisitor, LineConnectionFigure, contains)\\
 \wedge \neg \mathtt{IsUsedConstructorAsMethodParameter}(FindFigureInsideVisitor, LineConnectionFigure, setAttribute)\\
 \wedge \neg \mathtt{IsUsedConstructorAsMethodParameter}(FindFigureInsideVisitor, LineConnectionFigure, addNotify)\\
 \wedge \neg \mathtt{IsUsedConstructorAsMethodParameter}(FindFigureInsideVisitor, LineConnectionFigure, removeNotify)\\
 \wedge \neg \mathtt{IsUsedConstructorAsMethodParameter}(FindFigureInsideVisitor, LabeledLineConnectionFigure, basicTransform)\\
 \wedge \neg \mathtt{IsUsedConstructorAsMethodParameter}(FindFigureInsideVisitor, LabeledLineConnectionFigure, contains)\\
 \wedge \neg \mathtt{IsUsedConstructorAsMethodParameter}(FindFigureInsideVisitor, LabeledLineConnectionFigure, setAttribute)\\
 \wedge \neg \mathtt{IsUsedConstructorAsMethodParameter}(FindFigureInsideVisitor, LabeledLineConnectionFigure, addNotify)\\
 \wedge \neg \mathtt{IsUsedConstructorAsMethodParameter}(FindFigureInsideVisitor, LabeledLineConnectionFigure, removeNotify)\\
 \wedge \neg \mathtt{IsUsedConstructorAsMethodParameter}(FindFigureInsideVisitor, AbstractCompositeFigure, basicTransform)\\
 \wedge \neg \mathtt{IsUsedConstructorAsMethodParameter}(FindFigureInsideVisitor, AbstractCompositeFigure, contains)\\
 \wedge \neg \mathtt{IsUsedConstructorAsMethodParameter}(FindFigureInsideVisitor, AbstractCompositeFigure, setAttribute)\\
 \wedge \neg \mathtt{IsUsedConstructorAsMethodParameter}(FindFigureInsideVisitor, AbstractCompositeFigure, addNotify)\\
 \wedge \neg \mathtt{IsUsedConstructorAsMethodParameter}(FindFigureInsideVisitor, AbstractCompositeFigure, removeNotify)\\
 \wedge \neg \mathtt{IsUsedConstructorAsMethodParameter}(FindFigureInsideVisitor, GraphicalCompositeFigure, basicTransform)\\
 \wedge \neg \mathtt{IsUsedConstructorAsMethodParameter}(FindFigureInsideVisitor, GraphicalCompositeFigure, contains)\\
 \wedge \neg \mathtt{IsUsedConstructorAsMethodParameter}(FindFigureInsideVisitor, GraphicalCompositeFigure, setAttribute)\\
 \wedge \neg \mathtt{IsUsedConstructorAsMethodParameter}(FindFigureInsideVisitor, GraphicalCompositeFigure, addNotify)\\
 \wedge \neg \mathtt{IsUsedConstructorAsMethodParameter}(FindFigureInsideVisitor, GraphicalCompositeFigure, removeNotify)\\
 \wedge \neg \mathtt{IsUsedConstructorAsMethodParameter}(FindFigureInsideVisitor, AbstractFigure, basicTransform)\\
 \wedge \neg \mathtt{IsUsedConstructorAsMethodParameter}(FindFigureInsideVisitor, AbstractFigure, contains)\\
 \wedge \neg \mathtt{IsUsedConstructorAsMethodParameter}(FindFigureInsideVisitor, AbstractFigure, setAttribute)\\
 \wedge \neg \mathtt{IsUsedConstructorAsMethodParameter}(FindFigureInsideVisitor, AbstractFigure, findFigureInside)\\
 \wedge \neg \mathtt{IsUsedConstructorAsMethodParameter}(FindFigureInsideVisitor, AbstractFigure, addNotify)\\
 \wedge \neg \mathtt{IsUsedConstructorAsMethodParameter}(FindFigureInsideVisitor, AbstractFigure, removeNotify)\\
 \wedge \neg (\mathtt{IsUsedConstructorAsObjectReceiver}(FindFigureInsideVisitor, EllipseFigure, basicTransformTmpVC)\\
 \tab \vee  \mathtt{IsUsedConstructorAsObjectReceiver}(FindFigureInsideVisitor, EllipseFigure, basicTransform))\\
 \wedge \neg \mathtt{IsUsedConstructorAsObjectReceiver}(FindFigureInsideVisitor, EllipseFigure, contains)\\
 \wedge \neg \mathtt{IsUsedConstructorAsObjectReceiver}(FindFigureInsideVisitor, EllipseFigure, setAttribute)\\
 \wedge \neg \mathtt{IsUsedConstructorAsObjectReceiver}(FindFigureInsideVisitor, EllipseFigure, addNotify)\\
 \wedge \neg \mathtt{IsUsedConstructorAsObjectReceiver}(FindFigureInsideVisitor, EllipseFigure, removeNotify)\\
 \wedge \neg (\mathtt{IsUsedConstructorAsObjectReceiver}(FindFigureInsideVisitor, DiamondFigure, basicTransformTmpVC)\\
 \tab \vee  \mathtt{IsUsedConstructorAsObjectReceiver}(FindFigureInsideVisitor, DiamondFigure, basicTransform))\\
 \wedge \neg \mathtt{IsUsedConstructorAsObjectReceiver}(FindFigureInsideVisitor, DiamondFigure, contains)\\
 \wedge \neg \mathtt{IsUsedConstructorAsObjectReceiver}(FindFigureInsideVisitor, DiamondFigure, setAttribute)\\
 \wedge \neg \mathtt{IsUsedConstructorAsObjectReceiver}(FindFigureInsideVisitor, DiamondFigure, addNotify)\\
 \wedge \neg \mathtt{IsUsedConstructorAsObjectReceiver}(FindFigureInsideVisitor, DiamondFigure, removeNotify)\\
 \wedge \neg (\mathtt{IsUsedConstructorAsObjectReceiver}(FindFigureInsideVisitor, RectangleFigure, basicTransformTmpVC)\\
 \tab \vee  \mathtt{IsUsedConstructorAsObjectReceiver}(FindFigureInsideVisitor, RectangleFigure, basicTransform))\\
 \wedge \neg \mathtt{IsUsedConstructorAsObjectReceiver}(FindFigureInsideVisitor, RectangleFigure, contains)\\
 \wedge \neg \mathtt{IsUsedConstructorAsObjectReceiver}(FindFigureInsideVisitor, RectangleFigure, setAttribute)\\
 \wedge \neg \mathtt{IsUsedConstructorAsObjectReceiver}(FindFigureInsideVisitor, RectangleFigure, addNotify)\\
 \wedge \neg \mathtt{IsUsedConstructorAsObjectReceiver}(FindFigureInsideVisitor, RectangleFigure, removeNotify)\\
 \wedge \neg (\mathtt{IsUsedConstructorAsObjectReceiver}(FindFigureInsideVisitor, RoundRectangleFigure, basicTransformTmpVC)\\
 \tab \vee  \mathtt{IsUsedConstructorAsObjectReceiver}(FindFigureInsideVisitor, RoundRectangleFigure, basicTransform))\\
 \wedge \neg \mathtt{IsUsedConstructorAsObjectReceiver}(FindFigureInsideVisitor, RoundRectangleFigure, contains)\\
 \wedge \neg \mathtt{IsUsedConstructorAsObjectReceiver}(FindFigureInsideVisitor, RoundRectangleFigure, setAttribute)\\
 \wedge \neg \mathtt{IsUsedConstructorAsObjectReceiver}(FindFigureInsideVisitor, RoundRectangleFigure, addNotify)\\
 \wedge \neg \mathtt{IsUsedConstructorAsObjectReceiver}(FindFigureInsideVisitor, RoundRectangleFigure, removeNotify)\\
 \wedge \neg (\mathtt{IsUsedConstructorAsObjectReceiver}(FindFigureInsideVisitor, TriangleFigure, basicTransformTmpVC)\\
 \tab \vee  \mathtt{IsUsedConstructorAsObjectReceiver}(FindFigureInsideVisitor, TriangleFigure, basicTransform))\\
 \wedge \neg \mathtt{IsUsedConstructorAsObjectReceiver}(FindFigureInsideVisitor, TriangleFigure, contains)\\
 \wedge \neg \mathtt{IsUsedConstructorAsObjectReceiver}(FindFigureInsideVisitor, TriangleFigure, setAttribute)\\
 \wedge \neg \mathtt{IsUsedConstructorAsObjectReceiver}(FindFigureInsideVisitor, TriangleFigure, addNotify)\\
 \wedge \neg \mathtt{IsUsedConstructorAsObjectReceiver}(FindFigureInsideVisitor, TriangleFigure, removeNotify)\\
 \wedge \neg (\mathtt{IsUsedConstructorAsObjectReceiver}(FindFigureInsideVisitor, TextFigure, basicTransformTmpVC)\\
 \tab \vee  \mathtt{IsUsedConstructorAsObjectReceiver}(FindFigureInsideVisitor, TextFigure, basicTransform))\\
 \wedge \neg \mathtt{IsUsedConstructorAsObjectReceiver}(FindFigureInsideVisitor, TextFigure, contains)\\
 \wedge \neg \mathtt{IsUsedConstructorAsObjectReceiver}(FindFigureInsideVisitor, TextFigure, setAttribute)\\
 \wedge \neg \mathtt{IsUsedConstructorAsObjectReceiver}(FindFigureInsideVisitor, TextFigure, addNotify)\\
 \wedge \neg \mathtt{IsUsedConstructorAsObjectReceiver}(FindFigureInsideVisitor, TextFigure, removeNotify)\\
 \wedge \neg (\mathtt{IsUsedConstructorAsObjectReceiver}(FindFigureInsideVisitor, BezierFigure, basicTransformTmpVC)\\
 \tab \vee  \mathtt{IsUsedConstructorAsObjectReceiver}(FindFigureInsideVisitor, BezierFigure, basicTransform))\\
 \wedge \neg \mathtt{IsUsedConstructorAsObjectReceiver}(FindFigureInsideVisitor, BezierFigure, contains)\\
 \wedge \neg \mathtt{IsUsedConstructorAsObjectReceiver}(FindFigureInsideVisitor, BezierFigure, setAttribute)\\
 \wedge \neg \mathtt{IsUsedConstructorAsObjectReceiver}(FindFigureInsideVisitor, BezierFigure, addNotify)\\
 \wedge \neg \mathtt{IsUsedConstructorAsObjectReceiver}(FindFigureInsideVisitor, BezierFigure, removeNotify)\\
 \wedge \neg (\mathtt{IsUsedConstructorAsObjectReceiver}(FindFigureInsideVisitor, TextAreaFigure, basicTransformTmpVC)\\
 \tab \vee  \mathtt{IsUsedConstructorAsObjectReceiver}(FindFigureInsideVisitor, TextAreaFigure, basicTransform))\\
 \wedge \neg \mathtt{IsUsedConstructorAsObjectReceiver}(FindFigureInsideVisitor, TextAreaFigure, contains)\\
 \wedge \neg \mathtt{IsUsedConstructorAsObjectReceiver}(FindFigureInsideVisitor, TextAreaFigure, setAttribute)\\
 \wedge \neg \mathtt{IsUsedConstructorAsObjectReceiver}(FindFigureInsideVisitor, TextAreaFigure, addNotify)\\
 \wedge \neg \mathtt{IsUsedConstructorAsObjectReceiver}(FindFigureInsideVisitor, TextAreaFigure, removeNotify)\\
 \wedge \neg (\mathtt{IsUsedConstructorAsObjectReceiver}(FindFigureInsideVisitor, NodeFigure, basicTransformTmpVC)\\
 \tab \vee  \mathtt{IsUsedConstructorAsObjectReceiver}(FindFigureInsideVisitor, NodeFigure, basicTransform))\\
 \wedge \neg \mathtt{IsUsedConstructorAsObjectReceiver}(FindFigureInsideVisitor, NodeFigure, contains)\\
 \wedge \neg \mathtt{IsUsedConstructorAsObjectReceiver}(FindFigureInsideVisitor, NodeFigure, setAttribute)\\
 \wedge \neg \mathtt{IsUsedConstructorAsObjectReceiver}(FindFigureInsideVisitor, NodeFigure, addNotify)\\
 \wedge \neg \mathtt{IsUsedConstructorAsObjectReceiver}(FindFigureInsideVisitor, NodeFigure, removeNotify)\\
 \wedge \neg (\mathtt{IsUsedConstructorAsObjectReceiver}(FindFigureInsideVisitor, SVGImage, basicTransformTmpVC)\\
 \tab \vee  \mathtt{IsUsedConstructorAsObjectReceiver}(FindFigureInsideVisitor, SVGImage, basicTransform))\\
 \wedge \neg \mathtt{IsUsedConstructorAsObjectReceiver}(FindFigureInsideVisitor, SVGImage, contains)\\
 \wedge \neg \mathtt{IsUsedConstructorAsObjectReceiver}(FindFigureInsideVisitor, SVGImage, setAttribute)\\
 \wedge \neg \mathtt{IsUsedConstructorAsObjectReceiver}(FindFigureInsideVisitor, SVGImage, addNotify)\\
 \wedge \neg \mathtt{IsUsedConstructorAsObjectReceiver}(FindFigureInsideVisitor, SVGImage, removeNotify)\\
 \wedge \neg (\mathtt{IsUsedConstructorAsObjectReceiver}(FindFigureInsideVisitor, SVGPath, basicTransformTmpVC)\\
 \tab \vee  \mathtt{IsUsedConstructorAsObjectReceiver}(FindFigureInsideVisitor, SVGPath, basicTransform))\\
 \wedge \neg \mathtt{IsUsedConstructorAsObjectReceiver}(FindFigureInsideVisitor, SVGPath, contains)\\
 \wedge \neg \mathtt{IsUsedConstructorAsObjectReceiver}(FindFigureInsideVisitor, SVGPath, setAttribute)\\
 \wedge \neg \mathtt{IsUsedConstructorAsObjectReceiver}(FindFigureInsideVisitor, SVGPath, addNotify)\\
 \wedge \neg \mathtt{IsUsedConstructorAsObjectReceiver}(FindFigureInsideVisitor, SVGPath, removeNotify)\\
 \wedge \neg (\mathtt{IsUsedConstructorAsObjectReceiver}(FindFigureInsideVisitor, DependencyFigure, basicTransformTmpVC)\\
 \tab \vee  \mathtt{IsUsedConstructorAsObjectReceiver}(FindFigureInsideVisitor, DependencyFigure, basicTransform))\\
 \wedge \neg \mathtt{IsUsedConstructorAsObjectReceiver}(FindFigureInsideVisitor, DependencyFigure, contains)\\
 \wedge \neg \mathtt{IsUsedConstructorAsObjectReceiver}(FindFigureInsideVisitor, DependencyFigure, setAttribute)\\
 \wedge \neg \mathtt{IsUsedConstructorAsObjectReceiver}(FindFigureInsideVisitor, DependencyFigure, addNotify)\\
 \wedge \neg \mathtt{IsUsedConstructorAsObjectReceiver}(FindFigureInsideVisitor, DependencyFigure, removeNotify)\\
 \wedge \neg (\mathtt{IsUsedConstructorAsObjectReceiver}(FindFigureInsideVisitor, LineConnectionFigure, basicTransformTmpVC)\\
 \tab \vee  \mathtt{IsUsedConstructorAsObjectReceiver}(FindFigureInsideVisitor, LineConnectionFigure, basicTransform))\\
 \wedge \neg \mathtt{IsUsedConstructorAsObjectReceiver}(FindFigureInsideVisitor, LineConnectionFigure, contains)\\
 \wedge \neg \mathtt{IsUsedConstructorAsObjectReceiver}(FindFigureInsideVisitor, LineConnectionFigure, setAttribute)\\
 \wedge \neg \mathtt{IsUsedConstructorAsObjectReceiver}(FindFigureInsideVisitor, LineConnectionFigure, addNotify)\\
 \wedge \neg \mathtt{IsUsedConstructorAsObjectReceiver}(FindFigureInsideVisitor, LineConnectionFigure, removeNotify)\\
 \wedge \neg (\mathtt{IsUsedConstructorAsObjectReceiver}(FindFigureInsideVisitor, LabeledLineConnectionFigure, basicTransformTmpVC)\\
 \tab \vee  \mathtt{IsUsedConstructorAsObjectReceiver}(FindFigureInsideVisitor, LabeledLineConnectionFigure, basicTransform))\\
 \wedge \neg (\mathtt{IsUsedConstructorAsObjectReceiver}(FindFigureInsideVisitor, LabeledLineConnectionFigure, containsTmpVC)\\
 \tab \vee  \mathtt{IsUsedConstructorAsObjectReceiver}(FindFigureInsideVisitor, LabeledLineConnectionFigure, contains))\\
 \wedge \neg (\mathtt{IsUsedConstructorAsObjectReceiver}(FindFigureInsideVisitor, LabeledLineConnectionFigure, setAttributeTmpVC)\\
 \tab \vee  \mathtt{IsUsedConstructorAsObjectReceiver}(FindFigureInsideVisitor, LabeledLineConnectionFigure, setAttribute))\\
 \wedge \neg (\mathtt{IsUsedConstructorAsObjectReceiver}(FindFigureInsideVisitor, LabeledLineConnectionFigure, addNotifyTmpVC)\\
 \tab \vee  \mathtt{IsUsedConstructorAsObjectReceiver}(FindFigureInsideVisitor, LabeledLineConnectionFigure, addNotify))\\
 \wedge \neg (\mathtt{IsUsedConstructorAsObjectReceiver}(FindFigureInsideVisitor, LabeledLineConnectionFigure, removeNotifyTmpVC)\\
 \tab \vee  \mathtt{IsUsedConstructorAsObjectReceiver}(FindFigureInsideVisitor, LabeledLineConnectionFigure, removeNotify))\\
 \wedge \neg (\mathtt{IsUsedConstructorAsObjectReceiver}(FindFigureInsideVisitor, AbstractCompositeFigure, basicTransformTmpVC)\\
 \tab \vee  \mathtt{IsUsedConstructorAsObjectReceiver}(FindFigureInsideVisitor, AbstractCompositeFigure, basicTransform))\\
 \wedge \neg \mathtt{IsUsedConstructorAsObjectReceiver}(FindFigureInsideVisitor, AbstractCompositeFigure, contains)\\
 \wedge \neg \mathtt{IsUsedConstructorAsObjectReceiver}(FindFigureInsideVisitor, AbstractCompositeFigure, setAttribute)\\
 \wedge \neg \mathtt{IsUsedConstructorAsObjectReceiver}(FindFigureInsideVisitor, AbstractCompositeFigure, addNotify)\\
 \wedge \neg \mathtt{IsUsedConstructorAsObjectReceiver}(FindFigureInsideVisitor, AbstractCompositeFigure, removeNotify)\\
 \wedge \neg (\mathtt{IsUsedConstructorAsObjectReceiver}(FindFigureInsideVisitor, GraphicalCompositeFigure, basicTransformTmpVC)\\
 \tab \vee  \mathtt{IsUsedConstructorAsObjectReceiver}(FindFigureInsideVisitor, GraphicalCompositeFigure, basicTransform))\\
 \wedge \neg \mathtt{IsUsedConstructorAsObjectReceiver}(FindFigureInsideVisitor, GraphicalCompositeFigure, contains)\\
 \wedge \neg \mathtt{IsUsedConstructorAsObjectReceiver}(FindFigureInsideVisitor, GraphicalCompositeFigure, setAttribute)\\
 \wedge \neg \mathtt{IsUsedConstructorAsObjectReceiver}(FindFigureInsideVisitor, GraphicalCompositeFigure, addNotify)\\
 \wedge \neg \mathtt{IsUsedConstructorAsObjectReceiver}(FindFigureInsideVisitor, GraphicalCompositeFigure, removeNotify)\\
 \wedge \neg \mathtt{IsUsedConstructorAsMethodParameter}(SetAttributeVisitor, EllipseFigure, basicTransform)\\
 \wedge \neg \mathtt{IsUsedConstructorAsMethodParameter}(SetAttributeVisitor, EllipseFigure, contains)\\
 \wedge \neg \mathtt{IsUsedConstructorAsMethodParameter}(SetAttributeVisitor, EllipseFigure, findFigureInside)\\
 \wedge \neg \mathtt{IsUsedConstructorAsMethodParameter}(SetAttributeVisitor, EllipseFigure, addNotify)\\
 \wedge \neg \mathtt{IsUsedConstructorAsMethodParameter}(SetAttributeVisitor, EllipseFigure, removeNotify)\\
 \wedge \neg \mathtt{IsUsedConstructorAsMethodParameter}(SetAttributeVisitor, DiamondFigure, basicTransform)\\
 \wedge \neg \mathtt{IsUsedConstructorAsMethodParameter}(SetAttributeVisitor, DiamondFigure, contains)\\
 \wedge \neg \mathtt{IsUsedConstructorAsMethodParameter}(SetAttributeVisitor, DiamondFigure, findFigureInside)\\
 \wedge \neg \mathtt{IsUsedConstructorAsMethodParameter}(SetAttributeVisitor, DiamondFigure, addNotify)\\
 \wedge \neg \mathtt{IsUsedConstructorAsMethodParameter}(SetAttributeVisitor, DiamondFigure, removeNotify)\\
 \wedge \neg \mathtt{IsUsedConstructorAsMethodParameter}(SetAttributeVisitor, RectangleFigure, basicTransform)\\
 \wedge \neg \mathtt{IsUsedConstructorAsMethodParameter}(SetAttributeVisitor, RectangleFigure, contains)\\
 \wedge \neg \mathtt{IsUsedConstructorAsMethodParameter}(SetAttributeVisitor, RectangleFigure, findFigureInside)\\
 \wedge \neg \mathtt{IsUsedConstructorAsMethodParameter}(SetAttributeVisitor, RectangleFigure, addNotify)\\
 \wedge \neg \mathtt{IsUsedConstructorAsMethodParameter}(SetAttributeVisitor, RectangleFigure, removeNotify)\\
 \wedge \neg \mathtt{IsUsedConstructorAsMethodParameter}(SetAttributeVisitor, RoundRectangleFigure, basicTransform)\\
 \wedge \neg \mathtt{IsUsedConstructorAsMethodParameter}(SetAttributeVisitor, RoundRectangleFigure, contains)\\
 \wedge \neg \mathtt{IsUsedConstructorAsMethodParameter}(SetAttributeVisitor, RoundRectangleFigure, findFigureInside)\\
 \wedge \neg \mathtt{IsUsedConstructorAsMethodParameter}(SetAttributeVisitor, RoundRectangleFigure, addNotify)\\
 \wedge \neg \mathtt{IsUsedConstructorAsMethodParameter}(SetAttributeVisitor, RoundRectangleFigure, removeNotify)\\
 \wedge \neg \mathtt{IsUsedConstructorAsMethodParameter}(SetAttributeVisitor, TriangleFigure, basicTransform)\\
 \wedge \neg \mathtt{IsUsedConstructorAsMethodParameter}(SetAttributeVisitor, TriangleFigure, contains)\\
 \wedge \neg \mathtt{IsUsedConstructorAsMethodParameter}(SetAttributeVisitor, TriangleFigure, findFigureInside)\\
 \wedge \neg \mathtt{IsUsedConstructorAsMethodParameter}(SetAttributeVisitor, TriangleFigure, addNotify)\\
 \wedge \neg \mathtt{IsUsedConstructorAsMethodParameter}(SetAttributeVisitor, TriangleFigure, removeNotify)\\
 \wedge \neg \mathtt{IsUsedConstructorAsMethodParameter}(SetAttributeVisitor, TextFigure, basicTransform)\\
 \wedge \neg \mathtt{IsUsedConstructorAsMethodParameter}(SetAttributeVisitor, TextFigure, contains)\\
 \wedge \neg \mathtt{IsUsedConstructorAsMethodParameter}(SetAttributeVisitor, TextFigure, findFigureInside)\\
 \wedge \neg \mathtt{IsUsedConstructorAsMethodParameter}(SetAttributeVisitor, TextFigure, addNotify)\\
 \wedge \neg \mathtt{IsUsedConstructorAsMethodParameter}(SetAttributeVisitor, TextFigure, removeNotify)\\
 \wedge \neg \mathtt{IsUsedConstructorAsMethodParameter}(SetAttributeVisitor, BezierFigure, basicTransform)\\
 \wedge \neg \mathtt{IsUsedConstructorAsMethodParameter}(SetAttributeVisitor, BezierFigure, contains)\\
 \wedge \neg \mathtt{IsUsedConstructorAsMethodParameter}(SetAttributeVisitor, BezierFigure, findFigureInside)\\
 \wedge \neg \mathtt{IsUsedConstructorAsMethodParameter}(SetAttributeVisitor, BezierFigure, addNotify)\\
 \wedge \neg \mathtt{IsUsedConstructorAsMethodParameter}(SetAttributeVisitor, BezierFigure, removeNotify)\\
 \wedge \neg \mathtt{IsUsedConstructorAsMethodParameter}(SetAttributeVisitor, TextAreaFigure, basicTransform)\\
 \wedge \neg \mathtt{IsUsedConstructorAsMethodParameter}(SetAttributeVisitor, TextAreaFigure, contains)\\
 \wedge \neg \mathtt{IsUsedConstructorAsMethodParameter}(SetAttributeVisitor, TextAreaFigure, findFigureInside)\\
 \wedge \neg \mathtt{IsUsedConstructorAsMethodParameter}(SetAttributeVisitor, TextAreaFigure, addNotify)\\
 \wedge \neg \mathtt{IsUsedConstructorAsMethodParameter}(SetAttributeVisitor, TextAreaFigure, removeNotify)\\
 \wedge \neg \mathtt{IsUsedConstructorAsMethodParameter}(SetAttributeVisitor, NodeFigure, basicTransform)\\
 \wedge \neg \mathtt{IsUsedConstructorAsMethodParameter}(SetAttributeVisitor, NodeFigure, contains)\\
 \wedge \neg \mathtt{IsUsedConstructorAsMethodParameter}(SetAttributeVisitor, NodeFigure, findFigureInside)\\
 \wedge \neg \mathtt{IsUsedConstructorAsMethodParameter}(SetAttributeVisitor, NodeFigure, addNotify)\\
 \wedge \neg \mathtt{IsUsedConstructorAsMethodParameter}(SetAttributeVisitor, NodeFigure, removeNotify)\\
 \wedge \neg \mathtt{IsUsedConstructorAsMethodParameter}(SetAttributeVisitor, SVGImage, basicTransform)\\
 \wedge \neg \mathtt{IsUsedConstructorAsMethodParameter}(SetAttributeVisitor, SVGImage, contains)\\
 \wedge \neg \mathtt{IsUsedConstructorAsMethodParameter}(SetAttributeVisitor, SVGImage, findFigureInside)\\
 \wedge \neg \mathtt{IsUsedConstructorAsMethodParameter}(SetAttributeVisitor, SVGImage, addNotify)\\
 \wedge \neg \mathtt{IsUsedConstructorAsMethodParameter}(SetAttributeVisitor, SVGImage, removeNotify)\\
 \wedge \neg \mathtt{IsUsedConstructorAsMethodParameter}(SetAttributeVisitor, SVGPath, basicTransform)\\
 \wedge \neg \mathtt{IsUsedConstructorAsMethodParameter}(SetAttributeVisitor, SVGPath, contains)\\
 \wedge \neg \mathtt{IsUsedConstructorAsMethodParameter}(SetAttributeVisitor, SVGPath, findFigureInside)\\
 \wedge \neg \mathtt{IsUsedConstructorAsMethodParameter}(SetAttributeVisitor, SVGPath, addNotify)\\
 \wedge \neg \mathtt{IsUsedConstructorAsMethodParameter}(SetAttributeVisitor, SVGPath, removeNotify)\\
 \wedge \neg \mathtt{IsUsedConstructorAsMethodParameter}(SetAttributeVisitor, DependencyFigure, basicTransform)\\
 \wedge \neg \mathtt{IsUsedConstructorAsMethodParameter}(SetAttributeVisitor, DependencyFigure, contains)\\
 \wedge \neg \mathtt{IsUsedConstructorAsMethodParameter}(SetAttributeVisitor, DependencyFigure, findFigureInside)\\
 \wedge \neg \mathtt{IsUsedConstructorAsMethodParameter}(SetAttributeVisitor, DependencyFigure, addNotify)\\
 \wedge \neg \mathtt{IsUsedConstructorAsMethodParameter}(SetAttributeVisitor, DependencyFigure, removeNotify)\\
 \wedge \neg \mathtt{IsUsedConstructorAsMethodParameter}(SetAttributeVisitor, LineConnectionFigure, basicTransform)\\
 \wedge \neg \mathtt{IsUsedConstructorAsMethodParameter}(SetAttributeVisitor, LineConnectionFigure, contains)\\
 \wedge \neg \mathtt{IsUsedConstructorAsMethodParameter}(SetAttributeVisitor, LineConnectionFigure, findFigureInside)\\
 \wedge \neg \mathtt{IsUsedConstructorAsMethodParameter}(SetAttributeVisitor, LineConnectionFigure, addNotify)\\
 \wedge \neg \mathtt{IsUsedConstructorAsMethodParameter}(SetAttributeVisitor, LineConnectionFigure, removeNotify)\\
 \wedge \neg \mathtt{IsUsedConstructorAsMethodParameter}(SetAttributeVisitor, LabeledLineConnectionFigure, basicTransform)\\
 \wedge \neg \mathtt{IsUsedConstructorAsMethodParameter}(SetAttributeVisitor, LabeledLineConnectionFigure, contains)\\
 \wedge \neg \mathtt{IsUsedConstructorAsMethodParameter}(SetAttributeVisitor, LabeledLineConnectionFigure, findFigureInside)\\
 \wedge \neg \mathtt{IsUsedConstructorAsMethodParameter}(SetAttributeVisitor, LabeledLineConnectionFigure, addNotify)\\
 \wedge \neg \mathtt{IsUsedConstructorAsMethodParameter}(SetAttributeVisitor, LabeledLineConnectionFigure, removeNotify)\\
 \wedge \neg \mathtt{IsUsedConstructorAsMethodParameter}(SetAttributeVisitor, AbstractCompositeFigure, basicTransform)\\
 \wedge \neg \mathtt{IsUsedConstructorAsMethodParameter}(SetAttributeVisitor, AbstractCompositeFigure, contains)\\
 \wedge \neg \mathtt{IsUsedConstructorAsMethodParameter}(SetAttributeVisitor, AbstractCompositeFigure, findFigureInside)\\
 \wedge \neg \mathtt{IsUsedConstructorAsMethodParameter}(SetAttributeVisitor, AbstractCompositeFigure, addNotify)\\
 \wedge \neg \mathtt{IsUsedConstructorAsMethodParameter}(SetAttributeVisitor, AbstractCompositeFigure, removeNotify)\\
 \wedge \neg \mathtt{IsUsedConstructorAsMethodParameter}(SetAttributeVisitor, GraphicalCompositeFigure, basicTransform)\\
 \wedge \neg \mathtt{IsUsedConstructorAsMethodParameter}(SetAttributeVisitor, GraphicalCompositeFigure, contains)\\
 \wedge \neg \mathtt{IsUsedConstructorAsMethodParameter}(SetAttributeVisitor, GraphicalCompositeFigure, findFigureInside)\\
 \wedge \neg \mathtt{IsUsedConstructorAsMethodParameter}(SetAttributeVisitor, GraphicalCompositeFigure, addNotify)\\
 \wedge \neg \mathtt{IsUsedConstructorAsMethodParameter}(SetAttributeVisitor, GraphicalCompositeFigure, removeNotify)\\
 \wedge \neg \mathtt{IsUsedConstructorAsMethodParameter}(SetAttributeVisitor, AbstractFigure, basicTransform)\\
 \wedge \neg \mathtt{IsUsedConstructorAsMethodParameter}(SetAttributeVisitor, AbstractFigure, contains)\\
 \wedge \neg \mathtt{IsUsedConstructorAsMethodParameter}(SetAttributeVisitor, AbstractFigure, setAttribute)\\
 \wedge \neg \mathtt{IsUsedConstructorAsMethodParameter}(SetAttributeVisitor, AbstractFigure, findFigureInside)\\
 \wedge \neg \mathtt{IsUsedConstructorAsMethodParameter}(SetAttributeVisitor, AbstractFigure, addNotify)\\
 \wedge \neg \mathtt{IsUsedConstructorAsMethodParameter}(SetAttributeVisitor, AbstractFigure, removeNotify)\\
 \wedge \neg (\mathtt{IsUsedConstructorAsObjectReceiver}(SetAttributeVisitor, EllipseFigure, basicTransformTmpVC)\\
 \tab \vee  \mathtt{IsUsedConstructorAsObjectReceiver}(SetAttributeVisitor, EllipseFigure, basicTransform))\\
 \wedge \neg \mathtt{IsUsedConstructorAsObjectReceiver}(SetAttributeVisitor, EllipseFigure, contains)\\
 \wedge \neg \mathtt{IsUsedConstructorAsObjectReceiver}(SetAttributeVisitor, EllipseFigure, findFigureInside)\\
 \wedge \neg \mathtt{IsUsedConstructorAsObjectReceiver}(SetAttributeVisitor, EllipseFigure, addNotify)\\
 \wedge \neg \mathtt{IsUsedConstructorAsObjectReceiver}(SetAttributeVisitor, EllipseFigure, removeNotify)\\
 \wedge \neg (\mathtt{IsUsedConstructorAsObjectReceiver}(SetAttributeVisitor, DiamondFigure, basicTransformTmpVC)\\
 \tab \vee  \mathtt{IsUsedConstructorAsObjectReceiver}(SetAttributeVisitor, DiamondFigure, basicTransform))\\
 \wedge \neg \mathtt{IsUsedConstructorAsObjectReceiver}(SetAttributeVisitor, DiamondFigure, contains)\\
 \wedge \neg \mathtt{IsUsedConstructorAsObjectReceiver}(SetAttributeVisitor, DiamondFigure, findFigureInside)\\
 \wedge \neg \mathtt{IsUsedConstructorAsObjectReceiver}(SetAttributeVisitor, DiamondFigure, addNotify)\\
 \wedge \neg \mathtt{IsUsedConstructorAsObjectReceiver}(SetAttributeVisitor, DiamondFigure, removeNotify)\\
 \wedge \neg (\mathtt{IsUsedConstructorAsObjectReceiver}(SetAttributeVisitor, RectangleFigure, basicTransformTmpVC)\\
 \tab \vee  \mathtt{IsUsedConstructorAsObjectReceiver}(SetAttributeVisitor, RectangleFigure, basicTransform))\\
 \wedge \neg \mathtt{IsUsedConstructorAsObjectReceiver}(SetAttributeVisitor, RectangleFigure, contains)\\
 \wedge \neg \mathtt{IsUsedConstructorAsObjectReceiver}(SetAttributeVisitor, RectangleFigure, findFigureInside)\\
 \wedge \neg \mathtt{IsUsedConstructorAsObjectReceiver}(SetAttributeVisitor, RectangleFigure, addNotify)\\
 \wedge \neg \mathtt{IsUsedConstructorAsObjectReceiver}(SetAttributeVisitor, RectangleFigure, removeNotify)\\
 \wedge \neg (\mathtt{IsUsedConstructorAsObjectReceiver}(SetAttributeVisitor, RoundRectangleFigure, basicTransformTmpVC)\\
 \tab \vee  \mathtt{IsUsedConstructorAsObjectReceiver}(SetAttributeVisitor, RoundRectangleFigure, basicTransform))\\
 \wedge \neg \mathtt{IsUsedConstructorAsObjectReceiver}(SetAttributeVisitor, RoundRectangleFigure, contains)\\
 \wedge \neg \mathtt{IsUsedConstructorAsObjectReceiver}(SetAttributeVisitor, RoundRectangleFigure, findFigureInside)\\
 \wedge \neg \mathtt{IsUsedConstructorAsObjectReceiver}(SetAttributeVisitor, RoundRectangleFigure, addNotify)\\
 \wedge \neg \mathtt{IsUsedConstructorAsObjectReceiver}(SetAttributeVisitor, RoundRectangleFigure, removeNotify)\\
 \wedge \neg (\mathtt{IsUsedConstructorAsObjectReceiver}(SetAttributeVisitor, TriangleFigure, basicTransformTmpVC)\\
 \tab \vee  \mathtt{IsUsedConstructorAsObjectReceiver}(SetAttributeVisitor, TriangleFigure, basicTransform))\\
 \wedge \neg \mathtt{IsUsedConstructorAsObjectReceiver}(SetAttributeVisitor, TriangleFigure, contains)\\
 \wedge \neg \mathtt{IsUsedConstructorAsObjectReceiver}(SetAttributeVisitor, TriangleFigure, findFigureInside)\\
 \wedge \neg \mathtt{IsUsedConstructorAsObjectReceiver}(SetAttributeVisitor, TriangleFigure, addNotify)\\
 \wedge \neg \mathtt{IsUsedConstructorAsObjectReceiver}(SetAttributeVisitor, TriangleFigure, removeNotify)\\
 \wedge \neg (\mathtt{IsUsedConstructorAsObjectReceiver}(SetAttributeVisitor, TextFigure, basicTransformTmpVC)\\
 \tab \vee  \mathtt{IsUsedConstructorAsObjectReceiver}(SetAttributeVisitor, TextFigure, basicTransform))\\
 \wedge \neg \mathtt{IsUsedConstructorAsObjectReceiver}(SetAttributeVisitor, TextFigure, contains)\\
 \wedge \neg \mathtt{IsUsedConstructorAsObjectReceiver}(SetAttributeVisitor, TextFigure, findFigureInside)\\
 \wedge \neg \mathtt{IsUsedConstructorAsObjectReceiver}(SetAttributeVisitor, TextFigure, addNotify)\\
 \wedge \neg \mathtt{IsUsedConstructorAsObjectReceiver}(SetAttributeVisitor, TextFigure, removeNotify)\\
 \wedge \neg (\mathtt{IsUsedConstructorAsObjectReceiver}(SetAttributeVisitor, BezierFigure, basicTransformTmpVC)\\
 \tab \vee  \mathtt{IsUsedConstructorAsObjectReceiver}(SetAttributeVisitor, BezierFigure, basicTransform))\\
 \wedge \neg \mathtt{IsUsedConstructorAsObjectReceiver}(SetAttributeVisitor, BezierFigure, contains)\\
 \wedge \neg \mathtt{IsUsedConstructorAsObjectReceiver}(SetAttributeVisitor, BezierFigure, findFigureInside)\\
 \wedge \neg \mathtt{IsUsedConstructorAsObjectReceiver}(SetAttributeVisitor, BezierFigure, addNotify)\\
 \wedge \neg \mathtt{IsUsedConstructorAsObjectReceiver}(SetAttributeVisitor, BezierFigure, removeNotify)\\
 \wedge \neg (\mathtt{IsUsedConstructorAsObjectReceiver}(SetAttributeVisitor, TextAreaFigure, basicTransformTmpVC)\\
 \tab \vee  \mathtt{IsUsedConstructorAsObjectReceiver}(SetAttributeVisitor, TextAreaFigure, basicTransform))\\
 \wedge \neg \mathtt{IsUsedConstructorAsObjectReceiver}(SetAttributeVisitor, TextAreaFigure, contains)\\
 \wedge \neg \mathtt{IsUsedConstructorAsObjectReceiver}(SetAttributeVisitor, TextAreaFigure, findFigureInside)\\
 \wedge \neg \mathtt{IsUsedConstructorAsObjectReceiver}(SetAttributeVisitor, TextAreaFigure, addNotify)\\
 \wedge \neg \mathtt{IsUsedConstructorAsObjectReceiver}(SetAttributeVisitor, TextAreaFigure, removeNotify)\\
 \wedge \neg (\mathtt{IsUsedConstructorAsObjectReceiver}(SetAttributeVisitor, NodeFigure, basicTransformTmpVC)\\
 \tab \vee  \mathtt{IsUsedConstructorAsObjectReceiver}(SetAttributeVisitor, NodeFigure, basicTransform))\\
 \wedge \neg \mathtt{IsUsedConstructorAsObjectReceiver}(SetAttributeVisitor, NodeFigure, contains)\\
 \wedge \neg \mathtt{IsUsedConstructorAsObjectReceiver}(SetAttributeVisitor, NodeFigure, findFigureInside)\\
 \wedge \neg \mathtt{IsUsedConstructorAsObjectReceiver}(SetAttributeVisitor, NodeFigure, addNotify)\\
 \wedge \neg \mathtt{IsUsedConstructorAsObjectReceiver}(SetAttributeVisitor, NodeFigure, removeNotify)\\
 \wedge \neg (\mathtt{IsUsedConstructorAsObjectReceiver}(SetAttributeVisitor, SVGImage, basicTransformTmpVC)\\
 \tab \vee  \mathtt{IsUsedConstructorAsObjectReceiver}(SetAttributeVisitor, SVGImage, basicTransform))\\
 \wedge \neg \mathtt{IsUsedConstructorAsObjectReceiver}(SetAttributeVisitor, SVGImage, contains)\\
 \wedge \neg \mathtt{IsUsedConstructorAsObjectReceiver}(SetAttributeVisitor, SVGImage, findFigureInside)\\
 \wedge \neg \mathtt{IsUsedConstructorAsObjectReceiver}(SetAttributeVisitor, SVGImage, addNotify)\\
 \wedge \neg \mathtt{IsUsedConstructorAsObjectReceiver}(SetAttributeVisitor, SVGImage, removeNotify)\\
 \wedge \neg (\mathtt{IsUsedConstructorAsObjectReceiver}(SetAttributeVisitor, SVGPath, basicTransformTmpVC)\\
 \tab \vee  \mathtt{IsUsedConstructorAsObjectReceiver}(SetAttributeVisitor, SVGPath, basicTransform))\\
 \wedge \neg \mathtt{IsUsedConstructorAsObjectReceiver}(SetAttributeVisitor, SVGPath, contains)\\
 \wedge \neg \mathtt{IsUsedConstructorAsObjectReceiver}(SetAttributeVisitor, SVGPath, findFigureInside)\\
 \wedge \neg \mathtt{IsUsedConstructorAsObjectReceiver}(SetAttributeVisitor, SVGPath, addNotify)\\
 \wedge \neg \mathtt{IsUsedConstructorAsObjectReceiver}(SetAttributeVisitor, SVGPath, removeNotify)\\
 \wedge \neg (\mathtt{IsUsedConstructorAsObjectReceiver}(SetAttributeVisitor, DependencyFigure, basicTransformTmpVC)\\
 \tab \vee  \mathtt{IsUsedConstructorAsObjectReceiver}(SetAttributeVisitor, DependencyFigure, basicTransform))\\
 \wedge \neg \mathtt{IsUsedConstructorAsObjectReceiver}(SetAttributeVisitor, DependencyFigure, contains)\\
 \wedge \neg \mathtt{IsUsedConstructorAsObjectReceiver}(SetAttributeVisitor, DependencyFigure, findFigureInside)\\
 \wedge \neg \mathtt{IsUsedConstructorAsObjectReceiver}(SetAttributeVisitor, DependencyFigure, addNotify)\\
 \wedge \neg \mathtt{IsUsedConstructorAsObjectReceiver}(SetAttributeVisitor, DependencyFigure, removeNotify)\\
 \wedge \neg (\mathtt{IsUsedConstructorAsObjectReceiver}(SetAttributeVisitor, LineConnectionFigure, basicTransformTmpVC)\\
 \tab \vee  \mathtt{IsUsedConstructorAsObjectReceiver}(SetAttributeVisitor, LineConnectionFigure, basicTransform))\\
 \wedge \neg \mathtt{IsUsedConstructorAsObjectReceiver}(SetAttributeVisitor, LineConnectionFigure, contains)\\
 \wedge \neg \mathtt{IsUsedConstructorAsObjectReceiver}(SetAttributeVisitor, LineConnectionFigure, findFigureInside)\\
 \wedge \neg \mathtt{IsUsedConstructorAsObjectReceiver}(SetAttributeVisitor, LineConnectionFigure, addNotify)\\
 \wedge \neg \mathtt{IsUsedConstructorAsObjectReceiver}(SetAttributeVisitor, LineConnectionFigure, removeNotify)\\
 \wedge \neg (\mathtt{IsUsedConstructorAsObjectReceiver}(SetAttributeVisitor, LabeledLineConnectionFigure, basicTransformTmpVC)\\
 \tab \vee  \mathtt{IsUsedConstructorAsObjectReceiver}(SetAttributeVisitor, LabeledLineConnectionFigure, basicTransform))\\
 \wedge \neg (\mathtt{IsUsedConstructorAsObjectReceiver}(SetAttributeVisitor, LabeledLineConnectionFigure, containsTmpVC)\\
 \tab \vee  \mathtt{IsUsedConstructorAsObjectReceiver}(SetAttributeVisitor, LabeledLineConnectionFigure, contains))\\
 \wedge \neg (\mathtt{IsUsedConstructorAsObjectReceiver}(SetAttributeVisitor, LabeledLineConnectionFigure, findFigureInsideTmpVC)\\
 \tab \vee  \mathtt{IsUsedConstructorAsObjectReceiver}(SetAttributeVisitor, LabeledLineConnectionFigure, findFigureInside))\\
 \wedge \neg (\mathtt{IsUsedConstructorAsObjectReceiver}(SetAttributeVisitor, LabeledLineConnectionFigure, addNotifyTmpVC)\\
 \tab \vee  \mathtt{IsUsedConstructorAsObjectReceiver}(SetAttributeVisitor, LabeledLineConnectionFigure, addNotify))\\
 \wedge \neg (\mathtt{IsUsedConstructorAsObjectReceiver}(SetAttributeVisitor, LabeledLineConnectionFigure, removeNotifyTmpVC)\\
 \tab \vee  \mathtt{IsUsedConstructorAsObjectReceiver}(SetAttributeVisitor, LabeledLineConnectionFigure, removeNotify))\\
 \wedge \neg (\mathtt{IsUsedConstructorAsObjectReceiver}(SetAttributeVisitor, AbstractCompositeFigure, basicTransformTmpVC)\\
 \tab \vee  \mathtt{IsUsedConstructorAsObjectReceiver}(SetAttributeVisitor, AbstractCompositeFigure, basicTransform))\\
 \wedge \neg \mathtt{IsUsedConstructorAsObjectReceiver}(SetAttributeVisitor, AbstractCompositeFigure, contains)\\
 \wedge \neg \mathtt{IsUsedConstructorAsObjectReceiver}(SetAttributeVisitor, AbstractCompositeFigure, findFigureInside)\\
 \wedge \neg \mathtt{IsUsedConstructorAsObjectReceiver}(SetAttributeVisitor, AbstractCompositeFigure, addNotify)\\
 \wedge \neg \mathtt{IsUsedConstructorAsObjectReceiver}(SetAttributeVisitor, AbstractCompositeFigure, removeNotify)\\
 \wedge \neg (\mathtt{IsUsedConstructorAsObjectReceiver}(SetAttributeVisitor, GraphicalCompositeFigure, basicTransformTmpVC)\\
 \tab \vee  \mathtt{IsUsedConstructorAsObjectReceiver}(SetAttributeVisitor, GraphicalCompositeFigure, basicTransform))\\
 \wedge \neg \mathtt{IsUsedConstructorAsObjectReceiver}(SetAttributeVisitor, GraphicalCompositeFigure, contains)\\
 \wedge \neg \mathtt{IsUsedConstructorAsObjectReceiver}(SetAttributeVisitor, GraphicalCompositeFigure, findFigureInside)\\
 \wedge \neg \mathtt{IsUsedConstructorAsObjectReceiver}(SetAttributeVisitor, GraphicalCompositeFigure, addNotify)\\
 \wedge \neg \mathtt{IsUsedConstructorAsObjectReceiver}(SetAttributeVisitor, GraphicalCompositeFigure, removeNotify)\\
 \wedge \neg \mathtt{IsUsedConstructorAsMethodParameter}(ContainsVisitor, EllipseFigure, basicTransform)\\
 \wedge \neg \mathtt{IsUsedConstructorAsMethodParameter}(ContainsVisitor, EllipseFigure, setAttribute)\\
 \wedge \neg \mathtt{IsUsedConstructorAsMethodParameter}(ContainsVisitor, EllipseFigure, findFigureInside)\\
 \wedge \neg \mathtt{IsUsedConstructorAsMethodParameter}(ContainsVisitor, EllipseFigure, addNotify)\\
 \wedge \neg \mathtt{IsUsedConstructorAsMethodParameter}(ContainsVisitor, EllipseFigure, removeNotify)\\
 \wedge \neg \mathtt{IsUsedConstructorAsMethodParameter}(ContainsVisitor, DiamondFigure, basicTransform)\\
 \wedge \neg \mathtt{IsUsedConstructorAsMethodParameter}(ContainsVisitor, DiamondFigure, setAttribute)\\
 \wedge \neg \mathtt{IsUsedConstructorAsMethodParameter}(ContainsVisitor, DiamondFigure, findFigureInside)\\
 \wedge \neg \mathtt{IsUsedConstructorAsMethodParameter}(ContainsVisitor, DiamondFigure, addNotify)\\
 \wedge \neg \mathtt{IsUsedConstructorAsMethodParameter}(ContainsVisitor, DiamondFigure, removeNotify)\\
 \wedge \neg \mathtt{IsUsedConstructorAsMethodParameter}(ContainsVisitor, RectangleFigure, basicTransform)\\
 \wedge \neg \mathtt{IsUsedConstructorAsMethodParameter}(ContainsVisitor, RectangleFigure, setAttribute)\\
 \wedge \neg \mathtt{IsUsedConstructorAsMethodParameter}(ContainsVisitor, RectangleFigure, findFigureInside)\\
 \wedge \neg \mathtt{IsUsedConstructorAsMethodParameter}(ContainsVisitor, RectangleFigure, addNotify)\\
 \wedge \neg \mathtt{IsUsedConstructorAsMethodParameter}(ContainsVisitor, RectangleFigure, removeNotify)\\
 \wedge \neg \mathtt{IsUsedConstructorAsMethodParameter}(ContainsVisitor, RoundRectangleFigure, basicTransform)\\
 \wedge \neg \mathtt{IsUsedConstructorAsMethodParameter}(ContainsVisitor, RoundRectangleFigure, setAttribute)\\
 \wedge \neg \mathtt{IsUsedConstructorAsMethodParameter}(ContainsVisitor, RoundRectangleFigure, findFigureInside)\\
 \wedge \neg \mathtt{IsUsedConstructorAsMethodParameter}(ContainsVisitor, RoundRectangleFigure, addNotify)\\
 \wedge \neg \mathtt{IsUsedConstructorAsMethodParameter}(ContainsVisitor, RoundRectangleFigure, removeNotify)\\
 \wedge \neg \mathtt{IsUsedConstructorAsMethodParameter}(ContainsVisitor, TriangleFigure, basicTransform)\\
 \wedge \neg \mathtt{IsUsedConstructorAsMethodParameter}(ContainsVisitor, TriangleFigure, setAttribute)\\
 \wedge \neg \mathtt{IsUsedConstructorAsMethodParameter}(ContainsVisitor, TriangleFigure, findFigureInside)\\
 \wedge \neg \mathtt{IsUsedConstructorAsMethodParameter}(ContainsVisitor, TriangleFigure, addNotify)\\
 \wedge \neg \mathtt{IsUsedConstructorAsMethodParameter}(ContainsVisitor, TriangleFigure, removeNotify)\\
 \wedge \neg \mathtt{IsUsedConstructorAsMethodParameter}(ContainsVisitor, TextFigure, basicTransform)\\
 \wedge \neg \mathtt{IsUsedConstructorAsMethodParameter}(ContainsVisitor, TextFigure, setAttribute)\\
 \wedge \neg \mathtt{IsUsedConstructorAsMethodParameter}(ContainsVisitor, TextFigure, findFigureInside)\\
 \wedge \neg \mathtt{IsUsedConstructorAsMethodParameter}(ContainsVisitor, TextFigure, addNotify)\\
 \wedge \neg \mathtt{IsUsedConstructorAsMethodParameter}(ContainsVisitor, TextFigure, removeNotify)\\
 \wedge \neg \mathtt{IsUsedConstructorAsMethodParameter}(ContainsVisitor, BezierFigure, basicTransform)\\
 \wedge \neg \mathtt{IsUsedConstructorAsMethodParameter}(ContainsVisitor, BezierFigure, setAttribute)\\
 \wedge \neg \mathtt{IsUsedConstructorAsMethodParameter}(ContainsVisitor, BezierFigure, findFigureInside)\\
 \wedge \neg \mathtt{IsUsedConstructorAsMethodParameter}(ContainsVisitor, BezierFigure, addNotify)\\
 \wedge \neg \mathtt{IsUsedConstructorAsMethodParameter}(ContainsVisitor, BezierFigure, removeNotify)\\
 \wedge \neg \mathtt{IsUsedConstructorAsMethodParameter}(ContainsVisitor, TextAreaFigure, basicTransform)\\
 \wedge \neg \mathtt{IsUsedConstructorAsMethodParameter}(ContainsVisitor, TextAreaFigure, setAttribute)\\
 \wedge \neg \mathtt{IsUsedConstructorAsMethodParameter}(ContainsVisitor, TextAreaFigure, findFigureInside)\\
 \wedge \neg \mathtt{IsUsedConstructorAsMethodParameter}(ContainsVisitor, TextAreaFigure, addNotify)\\
 \wedge \neg \mathtt{IsUsedConstructorAsMethodParameter}(ContainsVisitor, TextAreaFigure, removeNotify)\\
 \wedge \neg \mathtt{IsUsedConstructorAsMethodParameter}(ContainsVisitor, NodeFigure, basicTransform)\\
 \wedge \neg \mathtt{IsUsedConstructorAsMethodParameter}(ContainsVisitor, NodeFigure, setAttribute)\\
 \wedge \neg \mathtt{IsUsedConstructorAsMethodParameter}(ContainsVisitor, NodeFigure, findFigureInside)\\
 \wedge \neg \mathtt{IsUsedConstructorAsMethodParameter}(ContainsVisitor, NodeFigure, addNotify)\\
 \wedge \neg \mathtt{IsUsedConstructorAsMethodParameter}(ContainsVisitor, NodeFigure, removeNotify)\\
 \wedge \neg \mathtt{IsUsedConstructorAsMethodParameter}(ContainsVisitor, SVGImage, basicTransform)\\
 \wedge \neg \mathtt{IsUsedConstructorAsMethodParameter}(ContainsVisitor, SVGImage, setAttribute)\\
 \wedge \neg \mathtt{IsUsedConstructorAsMethodParameter}(ContainsVisitor, SVGImage, findFigureInside)\\
 \wedge \neg \mathtt{IsUsedConstructorAsMethodParameter}(ContainsVisitor, SVGImage, addNotify)\\
 \wedge \neg \mathtt{IsUsedConstructorAsMethodParameter}(ContainsVisitor, SVGImage, removeNotify)\\
 \wedge \neg \mathtt{IsUsedConstructorAsMethodParameter}(ContainsVisitor, SVGPath, basicTransform)\\
 \wedge \neg \mathtt{IsUsedConstructorAsMethodParameter}(ContainsVisitor, SVGPath, setAttribute)\\
 \wedge \neg \mathtt{IsUsedConstructorAsMethodParameter}(ContainsVisitor, SVGPath, findFigureInside)\\
 \wedge \neg \mathtt{IsUsedConstructorAsMethodParameter}(ContainsVisitor, SVGPath, addNotify)\\
 \wedge \neg \mathtt{IsUsedConstructorAsMethodParameter}(ContainsVisitor, SVGPath, removeNotify)\\
 \wedge \neg \mathtt{IsUsedConstructorAsMethodParameter}(ContainsVisitor, DependencyFigure, basicTransform)\\
 \wedge \neg \mathtt{IsUsedConstructorAsMethodParameter}(ContainsVisitor, DependencyFigure, setAttribute)\\
 \wedge \neg \mathtt{IsUsedConstructorAsMethodParameter}(ContainsVisitor, DependencyFigure, findFigureInside)\\
 \wedge \neg \mathtt{IsUsedConstructorAsMethodParameter}(ContainsVisitor, DependencyFigure, addNotify)\\
 \wedge \neg \mathtt{IsUsedConstructorAsMethodParameter}(ContainsVisitor, DependencyFigure, removeNotify)\\
 \wedge \neg \mathtt{IsUsedConstructorAsMethodParameter}(ContainsVisitor, LineConnectionFigure, basicTransform)\\
 \wedge \neg \mathtt{IsUsedConstructorAsMethodParameter}(ContainsVisitor, LineConnectionFigure, setAttribute)\\
 \wedge \neg \mathtt{IsUsedConstructorAsMethodParameter}(ContainsVisitor, LineConnectionFigure, findFigureInside)\\
 \wedge \neg \mathtt{IsUsedConstructorAsMethodParameter}(ContainsVisitor, LineConnectionFigure, addNotify)\\
 \wedge \neg \mathtt{IsUsedConstructorAsMethodParameter}(ContainsVisitor, LineConnectionFigure, removeNotify)\\
 \wedge \neg \mathtt{IsUsedConstructorAsMethodParameter}(ContainsVisitor, LabeledLineConnectionFigure, basicTransform)\\
 \wedge \neg \mathtt{IsUsedConstructorAsMethodParameter}(ContainsVisitor, LabeledLineConnectionFigure, setAttribute)\\
 \wedge \neg \mathtt{IsUsedConstructorAsMethodParameter}(ContainsVisitor, LabeledLineConnectionFigure, findFigureInside)\\
 \wedge \neg \mathtt{IsUsedConstructorAsMethodParameter}(ContainsVisitor, LabeledLineConnectionFigure, addNotify)\\
 \wedge \neg \mathtt{IsUsedConstructorAsMethodParameter}(ContainsVisitor, LabeledLineConnectionFigure, removeNotify)\\
 \wedge \neg \mathtt{IsUsedConstructorAsMethodParameter}(ContainsVisitor, AbstractCompositeFigure, basicTransform)\\
 \wedge \neg \mathtt{IsUsedConstructorAsMethodParameter}(ContainsVisitor, AbstractCompositeFigure, setAttribute)\\
 \wedge \neg \mathtt{IsUsedConstructorAsMethodParameter}(ContainsVisitor, AbstractCompositeFigure, findFigureInside)\\
 \wedge \neg \mathtt{IsUsedConstructorAsMethodParameter}(ContainsVisitor, AbstractCompositeFigure, addNotify)\\
 \wedge \neg \mathtt{IsUsedConstructorAsMethodParameter}(ContainsVisitor, AbstractCompositeFigure, removeNotify)\\
 \wedge \neg \mathtt{IsUsedConstructorAsMethodParameter}(ContainsVisitor, GraphicalCompositeFigure, basicTransform)\\
 \wedge \neg \mathtt{IsUsedConstructorAsMethodParameter}(ContainsVisitor, GraphicalCompositeFigure, setAttribute)\\
 \wedge \neg \mathtt{IsUsedConstructorAsMethodParameter}(ContainsVisitor, GraphicalCompositeFigure, findFigureInside)\\
 \wedge \neg \mathtt{IsUsedConstructorAsMethodParameter}(ContainsVisitor, GraphicalCompositeFigure, addNotify)\\
 \wedge \neg \mathtt{IsUsedConstructorAsMethodParameter}(ContainsVisitor, GraphicalCompositeFigure, removeNotify)\\
 \wedge \neg \mathtt{IsUsedConstructorAsMethodParameter}(ContainsVisitor, AbstractFigure, basicTransform)\\
 \wedge \neg \mathtt{IsUsedConstructorAsMethodParameter}(ContainsVisitor, AbstractFigure, contains)\\
 \wedge \neg \mathtt{IsUsedConstructorAsMethodParameter}(ContainsVisitor, AbstractFigure, setAttribute)\\
 \wedge \neg \mathtt{IsUsedConstructorAsMethodParameter}(ContainsVisitor, AbstractFigure, findFigureInside)\\
 \wedge \neg \mathtt{IsUsedConstructorAsMethodParameter}(ContainsVisitor, AbstractFigure, addNotify)\\
 \wedge \neg \mathtt{IsUsedConstructorAsMethodParameter}(ContainsVisitor, AbstractFigure, removeNotify)\\
 \wedge \neg (\mathtt{IsUsedConstructorAsObjectReceiver}(ContainsVisitor, EllipseFigure, basicTransformTmpVC)\\
 \tab \vee  \mathtt{IsUsedConstructorAsObjectReceiver}(ContainsVisitor, EllipseFigure, basicTransform))\\
 \wedge \neg \mathtt{IsUsedConstructorAsObjectReceiver}(ContainsVisitor, EllipseFigure, setAttribute)\\
 \wedge \neg \mathtt{IsUsedConstructorAsObjectReceiver}(ContainsVisitor, EllipseFigure, findFigureInside)\\
 \wedge \neg \mathtt{IsUsedConstructorAsObjectReceiver}(ContainsVisitor, EllipseFigure, addNotify)\\
 \wedge \neg \mathtt{IsUsedConstructorAsObjectReceiver}(ContainsVisitor, EllipseFigure, removeNotify)\\
 \wedge \neg (\mathtt{IsUsedConstructorAsObjectReceiver}(ContainsVisitor, DiamondFigure, basicTransformTmpVC)\\
 \tab \vee  \mathtt{IsUsedConstructorAsObjectReceiver}(ContainsVisitor, DiamondFigure, basicTransform))\\
 \wedge \neg \mathtt{IsUsedConstructorAsObjectReceiver}(ContainsVisitor, DiamondFigure, setAttribute)\\
 \wedge \neg \mathtt{IsUsedConstructorAsObjectReceiver}(ContainsVisitor, DiamondFigure, findFigureInside)\\
 \wedge \neg \mathtt{IsUsedConstructorAsObjectReceiver}(ContainsVisitor, DiamondFigure, addNotify)\\
 \wedge \neg \mathtt{IsUsedConstructorAsObjectReceiver}(ContainsVisitor, DiamondFigure, removeNotify)\\
 \wedge \neg (\mathtt{IsUsedConstructorAsObjectReceiver}(ContainsVisitor, RectangleFigure, basicTransformTmpVC)\\
 \tab \vee  \mathtt{IsUsedConstructorAsObjectReceiver}(ContainsVisitor, RectangleFigure, basicTransform))\\
 \wedge \neg \mathtt{IsUsedConstructorAsObjectReceiver}(ContainsVisitor, RectangleFigure, setAttribute)\\
 \wedge \neg \mathtt{IsUsedConstructorAsObjectReceiver}(ContainsVisitor, RectangleFigure, findFigureInside)\\
 \wedge \neg \mathtt{IsUsedConstructorAsObjectReceiver}(ContainsVisitor, RectangleFigure, addNotify)\\
 \wedge \neg \mathtt{IsUsedConstructorAsObjectReceiver}(ContainsVisitor, RectangleFigure, removeNotify)\\
 \wedge \neg (\mathtt{IsUsedConstructorAsObjectReceiver}(ContainsVisitor, RoundRectangleFigure, basicTransformTmpVC)\\
 \tab \vee  \mathtt{IsUsedConstructorAsObjectReceiver}(ContainsVisitor, RoundRectangleFigure, basicTransform))\\
 \wedge \neg \mathtt{IsUsedConstructorAsObjectReceiver}(ContainsVisitor, RoundRectangleFigure, setAttribute)\\
 \wedge \neg \mathtt{IsUsedConstructorAsObjectReceiver}(ContainsVisitor, RoundRectangleFigure, findFigureInside)\\
 \wedge \neg \mathtt{IsUsedConstructorAsObjectReceiver}(ContainsVisitor, RoundRectangleFigure, addNotify)\\
 \wedge \neg \mathtt{IsUsedConstructorAsObjectReceiver}(ContainsVisitor, RoundRectangleFigure, removeNotify)\\
 \wedge \neg (\mathtt{IsUsedConstructorAsObjectReceiver}(ContainsVisitor, TriangleFigure, basicTransformTmpVC)\\
 \tab \vee  \mathtt{IsUsedConstructorAsObjectReceiver}(ContainsVisitor, TriangleFigure, basicTransform))\\
 \wedge \neg \mathtt{IsUsedConstructorAsObjectReceiver}(ContainsVisitor, TriangleFigure, setAttribute)\\
 \wedge \neg \mathtt{IsUsedConstructorAsObjectReceiver}(ContainsVisitor, TriangleFigure, findFigureInside)\\
 \wedge \neg \mathtt{IsUsedConstructorAsObjectReceiver}(ContainsVisitor, TriangleFigure, addNotify)\\
 \wedge \neg \mathtt{IsUsedConstructorAsObjectReceiver}(ContainsVisitor, TriangleFigure, removeNotify)\\
 \wedge \neg (\mathtt{IsUsedConstructorAsObjectReceiver}(ContainsVisitor, TextFigure, basicTransformTmpVC)\\
 \tab \vee  \mathtt{IsUsedConstructorAsObjectReceiver}(ContainsVisitor, TextFigure, basicTransform))\\
 \wedge \neg \mathtt{IsUsedConstructorAsObjectReceiver}(ContainsVisitor, TextFigure, setAttribute)\\
 \wedge \neg \mathtt{IsUsedConstructorAsObjectReceiver}(ContainsVisitor, TextFigure, findFigureInside)\\
 \wedge \neg \mathtt{IsUsedConstructorAsObjectReceiver}(ContainsVisitor, TextFigure, addNotify)\\
 \wedge \neg \mathtt{IsUsedConstructorAsObjectReceiver}(ContainsVisitor, TextFigure, removeNotify)\\
 \wedge \neg (\mathtt{IsUsedConstructorAsObjectReceiver}(ContainsVisitor, BezierFigure, basicTransformTmpVC)\\
 \tab \vee  \mathtt{IsUsedConstructorAsObjectReceiver}(ContainsVisitor, BezierFigure, basicTransform))\\
 \wedge \neg \mathtt{IsUsedConstructorAsObjectReceiver}(ContainsVisitor, BezierFigure, setAttribute)\\
 \wedge \neg \mathtt{IsUsedConstructorAsObjectReceiver}(ContainsVisitor, BezierFigure, findFigureInside)\\
 \wedge \neg \mathtt{IsUsedConstructorAsObjectReceiver}(ContainsVisitor, BezierFigure, addNotify)\\
 \wedge \neg \mathtt{IsUsedConstructorAsObjectReceiver}(ContainsVisitor, BezierFigure, removeNotify)\\
 \wedge \neg (\mathtt{IsUsedConstructorAsObjectReceiver}(ContainsVisitor, TextAreaFigure, basicTransformTmpVC)\\
 \tab \vee  \mathtt{IsUsedConstructorAsObjectReceiver}(ContainsVisitor, TextAreaFigure, basicTransform))\\
 \wedge \neg \mathtt{IsUsedConstructorAsObjectReceiver}(ContainsVisitor, TextAreaFigure, setAttribute)\\
 \wedge \neg \mathtt{IsUsedConstructorAsObjectReceiver}(ContainsVisitor, TextAreaFigure, findFigureInside)\\
 \wedge \neg \mathtt{IsUsedConstructorAsObjectReceiver}(ContainsVisitor, TextAreaFigure, addNotify)\\
 \wedge \neg \mathtt{IsUsedConstructorAsObjectReceiver}(ContainsVisitor, TextAreaFigure, removeNotify)\\
 \wedge \neg (\mathtt{IsUsedConstructorAsObjectReceiver}(ContainsVisitor, NodeFigure, basicTransformTmpVC)\\
 \tab \vee  \mathtt{IsUsedConstructorAsObjectReceiver}(ContainsVisitor, NodeFigure, basicTransform))\\
 \wedge \neg \mathtt{IsUsedConstructorAsObjectReceiver}(ContainsVisitor, NodeFigure, setAttribute)\\
 \wedge \neg \mathtt{IsUsedConstructorAsObjectReceiver}(ContainsVisitor, NodeFigure, findFigureInside)\\
 \wedge \neg \mathtt{IsUsedConstructorAsObjectReceiver}(ContainsVisitor, NodeFigure, addNotify)\\
 \wedge \neg \mathtt{IsUsedConstructorAsObjectReceiver}(ContainsVisitor, NodeFigure, removeNotify)\\
 \wedge \neg (\mathtt{IsUsedConstructorAsObjectReceiver}(ContainsVisitor, SVGImage, basicTransformTmpVC)\\
 \tab \vee  \mathtt{IsUsedConstructorAsObjectReceiver}(ContainsVisitor, SVGImage, basicTransform))\\
 \wedge \neg \mathtt{IsUsedConstructorAsObjectReceiver}(ContainsVisitor, SVGImage, setAttribute)\\
 \wedge \neg \mathtt{IsUsedConstructorAsObjectReceiver}(ContainsVisitor, SVGImage, findFigureInside)\\
 \wedge \neg \mathtt{IsUsedConstructorAsObjectReceiver}(ContainsVisitor, SVGImage, addNotify)\\
 \wedge \neg \mathtt{IsUsedConstructorAsObjectReceiver}(ContainsVisitor, SVGImage, removeNotify)\\
 \wedge \neg (\mathtt{IsUsedConstructorAsObjectReceiver}(ContainsVisitor, SVGPath, basicTransformTmpVC)\\
 \tab \vee  \mathtt{IsUsedConstructorAsObjectReceiver}(ContainsVisitor, SVGPath, basicTransform))\\
 \wedge \neg \mathtt{IsUsedConstructorAsObjectReceiver}(ContainsVisitor, SVGPath, setAttribute)\\
 \wedge \neg \mathtt{IsUsedConstructorAsObjectReceiver}(ContainsVisitor, SVGPath, findFigureInside)\\
 \wedge \neg \mathtt{IsUsedConstructorAsObjectReceiver}(ContainsVisitor, SVGPath, addNotify)\\
 \wedge \neg \mathtt{IsUsedConstructorAsObjectReceiver}(ContainsVisitor, SVGPath, removeNotify)\\
 \wedge \neg (\mathtt{IsUsedConstructorAsObjectReceiver}(ContainsVisitor, DependencyFigure, basicTransformTmpVC)\\
 \tab \vee  \mathtt{IsUsedConstructorAsObjectReceiver}(ContainsVisitor, DependencyFigure, basicTransform))\\
 \wedge \neg \mathtt{IsUsedConstructorAsObjectReceiver}(ContainsVisitor, DependencyFigure, setAttribute)\\
 \wedge \neg \mathtt{IsUsedConstructorAsObjectReceiver}(ContainsVisitor, DependencyFigure, findFigureInside)\\
 \wedge \neg \mathtt{IsUsedConstructorAsObjectReceiver}(ContainsVisitor, DependencyFigure, addNotify)\\
 \wedge \neg \mathtt{IsUsedConstructorAsObjectReceiver}(ContainsVisitor, DependencyFigure, removeNotify)\\
 \wedge \neg (\mathtt{IsUsedConstructorAsObjectReceiver}(ContainsVisitor, LineConnectionFigure, basicTransformTmpVC)\\
 \tab \vee  \mathtt{IsUsedConstructorAsObjectReceiver}(ContainsVisitor, LineConnectionFigure, basicTransform))\\
 \wedge \neg \mathtt{IsUsedConstructorAsObjectReceiver}(ContainsVisitor, LineConnectionFigure, setAttribute)\\
 \wedge \neg \mathtt{IsUsedConstructorAsObjectReceiver}(ContainsVisitor, LineConnectionFigure, findFigureInside)\\
 \wedge \neg \mathtt{IsUsedConstructorAsObjectReceiver}(ContainsVisitor, LineConnectionFigure, addNotify)\\
 \wedge \neg \mathtt{IsUsedConstructorAsObjectReceiver}(ContainsVisitor, LineConnectionFigure, removeNotify)\\
 \wedge \neg (\mathtt{IsUsedConstructorAsObjectReceiver}(ContainsVisitor, LabeledLineConnectionFigure, basicTransformTmpVC)\\
 \tab \vee  \mathtt{IsUsedConstructorAsObjectReceiver}(ContainsVisitor, LabeledLineConnectionFigure, basicTransform))\\
 \wedge \neg (\mathtt{IsUsedConstructorAsObjectReceiver}(ContainsVisitor, LabeledLineConnectionFigure, setAttributeTmpVC)\\
 \tab \vee  \mathtt{IsUsedConstructorAsObjectReceiver}(ContainsVisitor, LabeledLineConnectionFigure, setAttribute))\\
 \wedge \neg (\mathtt{IsUsedConstructorAsObjectReceiver}(ContainsVisitor, LabeledLineConnectionFigure, findFigureInsideTmpVC)\\
 \tab \vee  \mathtt{IsUsedConstructorAsObjectReceiver}(ContainsVisitor, LabeledLineConnectionFigure, findFigureInside))\\
 \wedge \neg (\mathtt{IsUsedConstructorAsObjectReceiver}(ContainsVisitor, LabeledLineConnectionFigure, addNotifyTmpVC)\\
 \tab \vee  \mathtt{IsUsedConstructorAsObjectReceiver}(ContainsVisitor, LabeledLineConnectionFigure, addNotify))\\
 \wedge \neg (\mathtt{IsUsedConstructorAsObjectReceiver}(ContainsVisitor, LabeledLineConnectionFigure, removeNotifyTmpVC)\\
 \tab \vee  \mathtt{IsUsedConstructorAsObjectReceiver}(ContainsVisitor, LabeledLineConnectionFigure, removeNotify))\\
 \wedge \neg (\mathtt{IsUsedConstructorAsObjectReceiver}(ContainsVisitor, AbstractCompositeFigure, basicTransformTmpVC)\\
 \tab \vee  \mathtt{IsUsedConstructorAsObjectReceiver}(ContainsVisitor, AbstractCompositeFigure, basicTransform))\\
 \wedge \neg \mathtt{IsUsedConstructorAsObjectReceiver}(ContainsVisitor, AbstractCompositeFigure, setAttribute)\\
 \wedge \neg \mathtt{IsUsedConstructorAsObjectReceiver}(ContainsVisitor, AbstractCompositeFigure, findFigureInside)\\
 \wedge \neg \mathtt{IsUsedConstructorAsObjectReceiver}(ContainsVisitor, AbstractCompositeFigure, addNotify)\\
 \wedge \neg \mathtt{IsUsedConstructorAsObjectReceiver}(ContainsVisitor, AbstractCompositeFigure, removeNotify)\\
 \wedge \neg (\mathtt{IsUsedConstructorAsObjectReceiver}(ContainsVisitor, GraphicalCompositeFigure, basicTransformTmpVC)\\
 \tab \vee  \mathtt{IsUsedConstructorAsObjectReceiver}(ContainsVisitor, GraphicalCompositeFigure, basicTransform))\\
 \wedge \neg \mathtt{IsUsedConstructorAsObjectReceiver}(ContainsVisitor, GraphicalCompositeFigure, setAttribute)\\
 \wedge \neg \mathtt{IsUsedConstructorAsObjectReceiver}(ContainsVisitor, GraphicalCompositeFigure, findFigureInside)\\
 \wedge \neg \mathtt{IsUsedConstructorAsObjectReceiver}(ContainsVisitor, GraphicalCompositeFigure, addNotify)\\
 \wedge \neg \mathtt{IsUsedConstructorAsObjectReceiver}(ContainsVisitor, GraphicalCompositeFigure, removeNotify)\\
 \wedge \neg \mathtt{IsUsedConstructorAsMethodParameter}(BasicTransformVisitor, EllipseFigure, contains)\\
 \wedge \neg \mathtt{IsUsedConstructorAsMethodParameter}(BasicTransformVisitor, EllipseFigure, setAttribute)\\
 \wedge \neg \mathtt{IsUsedConstructorAsMethodParameter}(BasicTransformVisitor, EllipseFigure, findFigureInside)\\
 \wedge \neg \mathtt{IsUsedConstructorAsMethodParameter}(BasicTransformVisitor, EllipseFigure, addNotify)\\
 \wedge \neg \mathtt{IsUsedConstructorAsMethodParameter}(BasicTransformVisitor, EllipseFigure, removeNotify)\\
 \wedge \neg \mathtt{IsUsedConstructorAsMethodParameter}(BasicTransformVisitor, DiamondFigure, contains)\\
 \wedge \neg \mathtt{IsUsedConstructorAsMethodParameter}(BasicTransformVisitor, DiamondFigure, setAttribute)\\
 \wedge \neg \mathtt{IsUsedConstructorAsMethodParameter}(BasicTransformVisitor, DiamondFigure, findFigureInside)\\
 \wedge \neg \mathtt{IsUsedConstructorAsMethodParameter}(BasicTransformVisitor, DiamondFigure, addNotify)\\
 \wedge \neg \mathtt{IsUsedConstructorAsMethodParameter}(BasicTransformVisitor, DiamondFigure, removeNotify)\\
 \wedge \neg \mathtt{IsUsedConstructorAsMethodParameter}(BasicTransformVisitor, RectangleFigure, contains)\\
 \wedge \neg \mathtt{IsUsedConstructorAsMethodParameter}(BasicTransformVisitor, RectangleFigure, setAttribute)\\
 \wedge \neg \mathtt{IsUsedConstructorAsMethodParameter}(BasicTransformVisitor, RectangleFigure, findFigureInside)\\
 \wedge \neg \mathtt{IsUsedConstructorAsMethodParameter}(BasicTransformVisitor, RectangleFigure, addNotify)\\
 \wedge \neg \mathtt{IsUsedConstructorAsMethodParameter}(BasicTransformVisitor, RectangleFigure, removeNotify)\\
 \wedge \neg \mathtt{IsUsedConstructorAsMethodParameter}(BasicTransformVisitor, RoundRectangleFigure, contains)\\
 \wedge \neg \mathtt{IsUsedConstructorAsMethodParameter}(BasicTransformVisitor, RoundRectangleFigure, setAttribute)\\
 \wedge \neg \mathtt{IsUsedConstructorAsMethodParameter}(BasicTransformVisitor, RoundRectangleFigure, findFigureInside)\\
 \wedge \neg \mathtt{IsUsedConstructorAsMethodParameter}(BasicTransformVisitor, RoundRectangleFigure, addNotify)\\
 \wedge \neg \mathtt{IsUsedConstructorAsMethodParameter}(BasicTransformVisitor, RoundRectangleFigure, removeNotify)\\
 \wedge \neg \mathtt{IsUsedConstructorAsMethodParameter}(BasicTransformVisitor, TriangleFigure, contains)\\
 \wedge \neg \mathtt{IsUsedConstructorAsMethodParameter}(BasicTransformVisitor, TriangleFigure, setAttribute)\\
 \wedge \neg \mathtt{IsUsedConstructorAsMethodParameter}(BasicTransformVisitor, TriangleFigure, findFigureInside)\\
 \wedge \neg \mathtt{IsUsedConstructorAsMethodParameter}(BasicTransformVisitor, TriangleFigure, addNotify)\\
 \wedge \neg \mathtt{IsUsedConstructorAsMethodParameter}(BasicTransformVisitor, TriangleFigure, removeNotify)\\
 \wedge \neg \mathtt{IsUsedConstructorAsMethodParameter}(BasicTransformVisitor, TextFigure, contains)\\
 \wedge \neg \mathtt{IsUsedConstructorAsMethodParameter}(BasicTransformVisitor, TextFigure, setAttribute)\\
 \wedge \neg \mathtt{IsUsedConstructorAsMethodParameter}(BasicTransformVisitor, TextFigure, findFigureInside)\\
 \wedge \neg \mathtt{IsUsedConstructorAsMethodParameter}(BasicTransformVisitor, TextFigure, addNotify)\\
 \wedge \neg \mathtt{IsUsedConstructorAsMethodParameter}(BasicTransformVisitor, TextFigure, removeNotify)\\
 \wedge \neg \mathtt{IsUsedConstructorAsMethodParameter}(BasicTransformVisitor, BezierFigure, contains)\\
 \wedge \neg \mathtt{IsUsedConstructorAsMethodParameter}(BasicTransformVisitor, BezierFigure, setAttribute)\\
 \wedge \neg \mathtt{IsUsedConstructorAsMethodParameter}(BasicTransformVisitor, BezierFigure, findFigureInside)\\
 \wedge \neg \mathtt{IsUsedConstructorAsMethodParameter}(BasicTransformVisitor, BezierFigure, addNotify)\\
 \wedge \neg \mathtt{IsUsedConstructorAsMethodParameter}(BasicTransformVisitor, BezierFigure, removeNotify)\\
 \wedge \neg \mathtt{IsUsedConstructorAsMethodParameter}(BasicTransformVisitor, TextAreaFigure, contains)\\
 \wedge \neg \mathtt{IsUsedConstructorAsMethodParameter}(BasicTransformVisitor, TextAreaFigure, setAttribute)\\
 \wedge \neg \mathtt{IsUsedConstructorAsMethodParameter}(BasicTransformVisitor, TextAreaFigure, findFigureInside)\\
 \wedge \neg \mathtt{IsUsedConstructorAsMethodParameter}(BasicTransformVisitor, TextAreaFigure, addNotify)\\
 \wedge \neg \mathtt{IsUsedConstructorAsMethodParameter}(BasicTransformVisitor, TextAreaFigure, removeNotify)\\
 \wedge \neg \mathtt{IsUsedConstructorAsMethodParameter}(BasicTransformVisitor, NodeFigure, contains)\\
 \wedge \neg \mathtt{IsUsedConstructorAsMethodParameter}(BasicTransformVisitor, NodeFigure, setAttribute)\\
 \wedge \neg \mathtt{IsUsedConstructorAsMethodParameter}(BasicTransformVisitor, NodeFigure, findFigureInside)\\
 \wedge \neg \mathtt{IsUsedConstructorAsMethodParameter}(BasicTransformVisitor, NodeFigure, addNotify)\\
 \wedge \neg \mathtt{IsUsedConstructorAsMethodParameter}(BasicTransformVisitor, NodeFigure, removeNotify)\\
 \wedge \neg \mathtt{IsUsedConstructorAsMethodParameter}(BasicTransformVisitor, SVGImage, contains)\\
 \wedge \neg \mathtt{IsUsedConstructorAsMethodParameter}(BasicTransformVisitor, SVGImage, setAttribute)\\
 \wedge \neg \mathtt{IsUsedConstructorAsMethodParameter}(BasicTransformVisitor, SVGImage, findFigureInside)\\
 \wedge \neg \mathtt{IsUsedConstructorAsMethodParameter}(BasicTransformVisitor, SVGImage, addNotify)\\
 \wedge \neg \mathtt{IsUsedConstructorAsMethodParameter}(BasicTransformVisitor, SVGImage, removeNotify)\\
 \wedge \neg \mathtt{IsUsedConstructorAsMethodParameter}(BasicTransformVisitor, SVGPath, contains)\\
 \wedge \neg \mathtt{IsUsedConstructorAsMethodParameter}(BasicTransformVisitor, SVGPath, setAttribute)\\
 \wedge \neg \mathtt{IsUsedConstructorAsMethodParameter}(BasicTransformVisitor, SVGPath, findFigureInside)\\
 \wedge \neg \mathtt{IsUsedConstructorAsMethodParameter}(BasicTransformVisitor, SVGPath, addNotify)\\
 \wedge \neg \mathtt{IsUsedConstructorAsMethodParameter}(BasicTransformVisitor, SVGPath, removeNotify)\\
 \wedge \neg \mathtt{IsUsedConstructorAsMethodParameter}(BasicTransformVisitor, DependencyFigure, contains)\\
 \wedge \neg \mathtt{IsUsedConstructorAsMethodParameter}(BasicTransformVisitor, DependencyFigure, setAttribute)\\
 \wedge \neg \mathtt{IsUsedConstructorAsMethodParameter}(BasicTransformVisitor, DependencyFigure, findFigureInside)\\
 \wedge \neg \mathtt{IsUsedConstructorAsMethodParameter}(BasicTransformVisitor, DependencyFigure, addNotify)\\
 \wedge \neg \mathtt{IsUsedConstructorAsMethodParameter}(BasicTransformVisitor, DependencyFigure, removeNotify)\\
 \wedge \neg \mathtt{IsUsedConstructorAsMethodParameter}(BasicTransformVisitor, LineConnectionFigure, contains)\\
 \wedge \neg \mathtt{IsUsedConstructorAsMethodParameter}(BasicTransformVisitor, LineConnectionFigure, setAttribute)\\
 \wedge \neg \mathtt{IsUsedConstructorAsMethodParameter}(BasicTransformVisitor, LineConnectionFigure, findFigureInside)\\
 \wedge \neg \mathtt{IsUsedConstructorAsMethodParameter}(BasicTransformVisitor, LineConnectionFigure, addNotify)\\
 \wedge \neg \mathtt{IsUsedConstructorAsMethodParameter}(BasicTransformVisitor, LineConnectionFigure, removeNotify)\\
 \wedge \neg \mathtt{IsUsedConstructorAsMethodParameter}(BasicTransformVisitor, LabeledLineConnectionFigure, contains)\\
 \wedge \neg \mathtt{IsUsedConstructorAsMethodParameter}(BasicTransformVisitor, LabeledLineConnectionFigure, setAttribute)\\
 \wedge \neg \mathtt{IsUsedConstructorAsMethodParameter}(BasicTransformVisitor, LabeledLineConnectionFigure, findFigureInside)\\
 \wedge \neg \mathtt{IsUsedConstructorAsMethodParameter}(BasicTransformVisitor, LabeledLineConnectionFigure, addNotify)\\
 \wedge \neg \mathtt{IsUsedConstructorAsMethodParameter}(BasicTransformVisitor, LabeledLineConnectionFigure, removeNotify)\\
 \wedge \neg \mathtt{IsUsedConstructorAsMethodParameter}(BasicTransformVisitor, AbstractCompositeFigure, contains)\\
 \wedge \neg \mathtt{IsUsedConstructorAsMethodParameter}(BasicTransformVisitor, AbstractCompositeFigure, setAttribute)\\
 \wedge \neg \mathtt{IsUsedConstructorAsMethodParameter}(BasicTransformVisitor, AbstractCompositeFigure, findFigureInside)\\
 \wedge \neg \mathtt{IsUsedConstructorAsMethodParameter}(BasicTransformVisitor, AbstractCompositeFigure, addNotify)\\
 \wedge \neg \mathtt{IsUsedConstructorAsMethodParameter}(BasicTransformVisitor, AbstractCompositeFigure, removeNotify)\\
 \wedge \neg \mathtt{IsUsedConstructorAsMethodParameter}(BasicTransformVisitor, GraphicalCompositeFigure, contains)\\
 \wedge \neg \mathtt{IsUsedConstructorAsMethodParameter}(BasicTransformVisitor, GraphicalCompositeFigure, setAttribute)\\
 \wedge \neg \mathtt{IsUsedConstructorAsMethodParameter}(BasicTransformVisitor, GraphicalCompositeFigure, findFigureInside)\\
 \wedge \neg \mathtt{IsUsedConstructorAsMethodParameter}(BasicTransformVisitor, GraphicalCompositeFigure, addNotify)\\
 \wedge \neg \mathtt{IsUsedConstructorAsMethodParameter}(BasicTransformVisitor, GraphicalCompositeFigure, removeNotify)\\
 \wedge \neg \mathtt{IsUsedConstructorAsMethodParameter}(BasicTransformVisitor, AbstractFigure, basicTransform)\\
 \wedge \neg \mathtt{IsUsedConstructorAsMethodParameter}(BasicTransformVisitor, AbstractFigure, contains)\\
 \wedge \neg \mathtt{IsUsedConstructorAsMethodParameter}(BasicTransformVisitor, AbstractFigure, setAttribute)\\
 \wedge \neg \mathtt{IsUsedConstructorAsMethodParameter}(BasicTransformVisitor, AbstractFigure, findFigureInside)\\
 \wedge \neg \mathtt{IsUsedConstructorAsMethodParameter}(BasicTransformVisitor, AbstractFigure, addNotify)\\
 \wedge \neg \mathtt{IsUsedConstructorAsMethodParameter}(BasicTransformVisitor, AbstractFigure, removeNotify)\\
 \wedge \neg \mathtt{IsUsedConstructorAsObjectReceiver}(BasicTransformVisitor, EllipseFigure, contains)\\
 \wedge \neg \mathtt{IsUsedConstructorAsObjectReceiver}(BasicTransformVisitor, EllipseFigure, setAttribute)\\
 \wedge \neg \mathtt{IsUsedConstructorAsObjectReceiver}(BasicTransformVisitor, EllipseFigure, findFigureInside)\\
 \wedge \neg \mathtt{IsUsedConstructorAsObjectReceiver}(BasicTransformVisitor, EllipseFigure, addNotify)\\
 \wedge \neg \mathtt{IsUsedConstructorAsObjectReceiver}(BasicTransformVisitor, EllipseFigure, removeNotify)\\
 \wedge \neg \mathtt{IsUsedConstructorAsObjectReceiver}(BasicTransformVisitor, DiamondFigure, contains)\\
 \wedge \neg \mathtt{IsUsedConstructorAsObjectReceiver}(BasicTransformVisitor, DiamondFigure, setAttribute)\\
 \wedge \neg \mathtt{IsUsedConstructorAsObjectReceiver}(BasicTransformVisitor, DiamondFigure, findFigureInside)\\
 \wedge \neg \mathtt{IsUsedConstructorAsObjectReceiver}(BasicTransformVisitor, DiamondFigure, addNotify)\\
 \wedge \neg \mathtt{IsUsedConstructorAsObjectReceiver}(BasicTransformVisitor, DiamondFigure, removeNotify)\\
 \wedge \neg \mathtt{IsUsedConstructorAsObjectReceiver}(BasicTransformVisitor, RectangleFigure, contains)\\
 \wedge \neg \mathtt{IsUsedConstructorAsObjectReceiver}(BasicTransformVisitor, RectangleFigure, setAttribute)\\
 \wedge \neg \mathtt{IsUsedConstructorAsObjectReceiver}(BasicTransformVisitor, RectangleFigure, findFigureInside)\\
 \wedge \neg \mathtt{IsUsedConstructorAsObjectReceiver}(BasicTransformVisitor, RectangleFigure, addNotify)\\
 \wedge \neg \mathtt{IsUsedConstructorAsObjectReceiver}(BasicTransformVisitor, RectangleFigure, removeNotify)\\
 \wedge \neg \mathtt{IsUsedConstructorAsObjectReceiver}(BasicTransformVisitor, RoundRectangleFigure, contains)\\
 \wedge \neg \mathtt{IsUsedConstructorAsObjectReceiver}(BasicTransformVisitor, RoundRectangleFigure, setAttribute)\\
 \wedge \neg \mathtt{IsUsedConstructorAsObjectReceiver}(BasicTransformVisitor, RoundRectangleFigure, findFigureInside)\\
 \wedge \neg \mathtt{IsUsedConstructorAsObjectReceiver}(BasicTransformVisitor, RoundRectangleFigure, addNotify)\\
 \wedge \neg \mathtt{IsUsedConstructorAsObjectReceiver}(BasicTransformVisitor, RoundRectangleFigure, removeNotify)\\
 \wedge \neg \mathtt{IsUsedConstructorAsObjectReceiver}(BasicTransformVisitor, TriangleFigure, contains)\\
 \wedge \neg \mathtt{IsUsedConstructorAsObjectReceiver}(BasicTransformVisitor, TriangleFigure, setAttribute)\\
 \wedge \neg \mathtt{IsUsedConstructorAsObjectReceiver}(BasicTransformVisitor, TriangleFigure, findFigureInside)\\
 \wedge \neg \mathtt{IsUsedConstructorAsObjectReceiver}(BasicTransformVisitor, TriangleFigure, addNotify)\\
 \wedge \neg \mathtt{IsUsedConstructorAsObjectReceiver}(BasicTransformVisitor, TriangleFigure, removeNotify)\\
 \wedge \neg \mathtt{IsUsedConstructorAsObjectReceiver}(BasicTransformVisitor, TextFigure, contains)\\
 \wedge \neg \mathtt{IsUsedConstructorAsObjectReceiver}(BasicTransformVisitor, TextFigure, setAttribute)\\
 \wedge \neg \mathtt{IsUsedConstructorAsObjectReceiver}(BasicTransformVisitor, TextFigure, findFigureInside)\\
 \wedge \neg \mathtt{IsUsedConstructorAsObjectReceiver}(BasicTransformVisitor, TextFigure, addNotify)\\
 \wedge \neg \mathtt{IsUsedConstructorAsObjectReceiver}(BasicTransformVisitor, TextFigure, removeNotify)\\
 \wedge \neg \mathtt{IsUsedConstructorAsObjectReceiver}(BasicTransformVisitor, BezierFigure, contains)\\
 \wedge \neg \mathtt{IsUsedConstructorAsObjectReceiver}(BasicTransformVisitor, BezierFigure, setAttribute)\\
 \wedge \neg \mathtt{IsUsedConstructorAsObjectReceiver}(BasicTransformVisitor, BezierFigure, findFigureInside)\\
 \wedge \neg \mathtt{IsUsedConstructorAsObjectReceiver}(BasicTransformVisitor, BezierFigure, addNotify)\\
 \wedge \neg \mathtt{IsUsedConstructorAsObjectReceiver}(BasicTransformVisitor, BezierFigure, removeNotify)\\
 \wedge \neg \mathtt{IsUsedConstructorAsObjectReceiver}(BasicTransformVisitor, TextAreaFigure, contains)\\
 \wedge \neg \mathtt{IsUsedConstructorAsObjectReceiver}(BasicTransformVisitor, TextAreaFigure, setAttribute)\\
 \wedge \neg \mathtt{IsUsedConstructorAsObjectReceiver}(BasicTransformVisitor, TextAreaFigure, findFigureInside)\\
 \wedge \neg \mathtt{IsUsedConstructorAsObjectReceiver}(BasicTransformVisitor, TextAreaFigure, addNotify)\\
 \wedge \neg \mathtt{IsUsedConstructorAsObjectReceiver}(BasicTransformVisitor, TextAreaFigure, removeNotify)\\
 \wedge \neg \mathtt{IsUsedConstructorAsObjectReceiver}(BasicTransformVisitor, NodeFigure, contains)\\
 \wedge \neg \mathtt{IsUsedConstructorAsObjectReceiver}(BasicTransformVisitor, NodeFigure, setAttribute)\\
 \wedge \neg \mathtt{IsUsedConstructorAsObjectReceiver}(BasicTransformVisitor, NodeFigure, findFigureInside)\\
 \wedge \neg \mathtt{IsUsedConstructorAsObjectReceiver}(BasicTransformVisitor, NodeFigure, addNotify)\\
 \wedge \neg \mathtt{IsUsedConstructorAsObjectReceiver}(BasicTransformVisitor, NodeFigure, removeNotify)\\
 \wedge \neg \mathtt{IsUsedConstructorAsObjectReceiver}(BasicTransformVisitor, SVGImage, contains)\\
 \wedge \neg \mathtt{IsUsedConstructorAsObjectReceiver}(BasicTransformVisitor, SVGImage, setAttribute)\\
 \wedge \neg \mathtt{IsUsedConstructorAsObjectReceiver}(BasicTransformVisitor, SVGImage, findFigureInside)\\
 \wedge \neg \mathtt{IsUsedConstructorAsObjectReceiver}(BasicTransformVisitor, SVGImage, addNotify)\\
 \wedge \neg \mathtt{IsUsedConstructorAsObjectReceiver}(BasicTransformVisitor, SVGImage, removeNotify)\\
 \wedge \neg \mathtt{IsUsedConstructorAsObjectReceiver}(BasicTransformVisitor, SVGPath, contains)\\
 \wedge \neg \mathtt{IsUsedConstructorAsObjectReceiver}(BasicTransformVisitor, SVGPath, setAttribute)\\
 \wedge \neg \mathtt{IsUsedConstructorAsObjectReceiver}(BasicTransformVisitor, SVGPath, findFigureInside)\\
 \wedge \neg \mathtt{IsUsedConstructorAsObjectReceiver}(BasicTransformVisitor, SVGPath, addNotify)\\
 \wedge \neg \mathtt{IsUsedConstructorAsObjectReceiver}(BasicTransformVisitor, SVGPath, removeNotify)\\
 \wedge \neg \mathtt{IsUsedConstructorAsObjectReceiver}(BasicTransformVisitor, DependencyFigure, contains)\\
 \wedge \neg \mathtt{IsUsedConstructorAsObjectReceiver}(BasicTransformVisitor, DependencyFigure, setAttribute)\\
 \wedge \neg \mathtt{IsUsedConstructorAsObjectReceiver}(BasicTransformVisitor, DependencyFigure, findFigureInside)\\
 \wedge \neg \mathtt{IsUsedConstructorAsObjectReceiver}(BasicTransformVisitor, DependencyFigure, addNotify)\\
 \wedge \neg \mathtt{IsUsedConstructorAsObjectReceiver}(BasicTransformVisitor, DependencyFigure, removeNotify)\\
 \wedge \neg \mathtt{IsUsedConstructorAsObjectReceiver}(BasicTransformVisitor, LineConnectionFigure, contains)\\
 \wedge \neg \mathtt{IsUsedConstructorAsObjectReceiver}(BasicTransformVisitor, LineConnectionFigure, setAttribute)\\
 \wedge \neg \mathtt{IsUsedConstructorAsObjectReceiver}(BasicTransformVisitor, LineConnectionFigure, findFigureInside)\\
 \wedge \neg \mathtt{IsUsedConstructorAsObjectReceiver}(BasicTransformVisitor, LineConnectionFigure, addNotify)\\
 \wedge \neg \mathtt{IsUsedConstructorAsObjectReceiver}(BasicTransformVisitor, LineConnectionFigure, removeNotify)\\
 \wedge \neg (\mathtt{IsUsedConstructorAsObjectReceiver}(BasicTransformVisitor, LabeledLineConnectionFigure, containsTmpVC)\\
 \tab \vee  \mathtt{IsUsedConstructorAsObjectReceiver}(BasicTransformVisitor, LabeledLineConnectionFigure, contains))\\
 \wedge \neg (\mathtt{IsUsedConstructorAsObjectReceiver}(BasicTransformVisitor, LabeledLineConnectionFigure, setAttributeTmpVC)\\
 \tab \vee  \mathtt{IsUsedConstructorAsObjectReceiver}(BasicTransformVisitor, LabeledLineConnectionFigure, setAttribute))\\
 \wedge \neg (\mathtt{IsUsedConstructorAsObjectReceiver}(BasicTransformVisitor, LabeledLineConnectionFigure, findFigureInsideTmpVC)\\
 \tab \vee  \mathtt{IsUsedConstructorAsObjectReceiver}(BasicTransformVisitor, LabeledLineConnectionFigure, findFigureInside))\\
 \wedge \neg (\mathtt{IsUsedConstructorAsObjectReceiver}(BasicTransformVisitor, LabeledLineConnectionFigure, addNotifyTmpVC)\\
 \tab \vee  \mathtt{IsUsedConstructorAsObjectReceiver}(BasicTransformVisitor, LabeledLineConnectionFigure, addNotify))\\
 \wedge \neg (\mathtt{IsUsedConstructorAsObjectReceiver}(BasicTransformVisitor, LabeledLineConnectionFigure, removeNotifyTmpVC)\\
 \tab \vee  \mathtt{IsUsedConstructorAsObjectReceiver}(BasicTransformVisitor, LabeledLineConnectionFigure, removeNotify))\\
 \wedge \neg \mathtt{IsUsedConstructorAsObjectReceiver}(BasicTransformVisitor, AbstractCompositeFigure, contains)\\
 \wedge \neg \mathtt{IsUsedConstructorAsObjectReceiver}(BasicTransformVisitor, AbstractCompositeFigure, setAttribute)\\
 \wedge \neg \mathtt{IsUsedConstructorAsObjectReceiver}(BasicTransformVisitor, AbstractCompositeFigure, findFigureInside)\\
 \wedge \neg \mathtt{IsUsedConstructorAsObjectReceiver}(BasicTransformVisitor, AbstractCompositeFigure, addNotify)\\
 \wedge \neg \mathtt{IsUsedConstructorAsObjectReceiver}(BasicTransformVisitor, AbstractCompositeFigure, removeNotify)\\
 \wedge \neg \mathtt{IsUsedConstructorAsObjectReceiver}(BasicTransformVisitor, GraphicalCompositeFigure, contains)\\
 \wedge \neg \mathtt{IsUsedConstructorAsObjectReceiver}(BasicTransformVisitor, GraphicalCompositeFigure, setAttribute)\\
 \wedge \neg \mathtt{IsUsedConstructorAsObjectReceiver}(BasicTransformVisitor, GraphicalCompositeFigure, findFigureInside)\\
 \wedge \neg \mathtt{IsUsedConstructorAsObjectReceiver}(BasicTransformVisitor, GraphicalCompositeFigure, addNotify)\\
 \wedge \neg \mathtt{IsUsedConstructorAsObjectReceiver}(BasicTransformVisitor, GraphicalCompositeFigure, removeNotify)\\
 \wedge \mathtt{ExistsType}(AbstractCompositeFigure)\\
 \wedge \mathtt{ExistsType}(SVGImage)\\
 \wedge \mathtt{ExistsType}(TextAreaFigure)\\
 \wedge \mathtt{ExistsType}(BezierFigure)\\
 \wedge \mathtt{ExistsType}(TextFigure)\\
 \wedge \mathtt{ExistsType}(TriangleFigure)\\
 \wedge \mathtt{ExistsType}(RoundRectangleFigure)\\
 \wedge \mathtt{ExistsType}(RectangleFigure)\\
 \wedge \mathtt{ExistsType}(DiamondFigure)\\
 \wedge \mathtt{ExistsType}(EllipseFigure)\\
 \wedge \neg \mathtt{ExistsMethodInvocation}(LabeledLineConnectionFigure, removeNotifyTmpVC, LabeledLineConnectionFigure, basicTransform)\\
 \wedge \neg \mathtt{ExistsMethodInvocation}(LabeledLineConnectionFigure, removeNotifyTmpVC, LabeledLineConnectionFigure, contains)\\
 \wedge \neg \mathtt{ExistsMethodInvocation}(LabeledLineConnectionFigure, removeNotifyTmpVC, LabeledLineConnectionFigure, setAttribute)\\
 \wedge \neg \mathtt{ExistsMethodInvocation}(LabeledLineConnectionFigure, removeNotifyTmpVC, LabeledLineConnectionFigure, findFigureInside)\\
 \wedge \neg \mathtt{ExistsMethodInvocation}(LabeledLineConnectionFigure, removeNotifyTmpVC, LabeledLineConnectionFigure, addNotify)\\
 \wedge \neg \mathtt{ExistsMethodInvocation}(LabeledLineConnectionFigure, addNotifyTmpVC, LabeledLineConnectionFigure, basicTransform)\\
 \wedge \neg \mathtt{ExistsMethodInvocation}(LabeledLineConnectionFigure, addNotifyTmpVC, LabeledLineConnectionFigure, contains)\\
 \wedge \neg \mathtt{ExistsMethodInvocation}(LabeledLineConnectionFigure, addNotifyTmpVC, LabeledLineConnectionFigure, setAttribute)\\
 \wedge \neg \mathtt{ExistsMethodInvocation}(LabeledLineConnectionFigure, addNotifyTmpVC, LabeledLineConnectionFigure, findFigureInside)\\
 \wedge \neg \mathtt{ExistsMethodInvocation}(LabeledLineConnectionFigure, addNotifyTmpVC, LabeledLineConnectionFigure, removeNotify)\\
 \wedge \neg \mathtt{ExistsMethodInvocation}(LabeledLineConnectionFigure, findFigureInsideTmpVC, LabeledLineConnectionFigure, basicTransform)\\
 \wedge \neg \mathtt{ExistsMethodInvocation}(LabeledLineConnectionFigure, findFigureInsideTmpVC, LabeledLineConnectionFigure, contains)\\
 \wedge \neg \mathtt{ExistsMethodInvocation}(LabeledLineConnectionFigure, findFigureInsideTmpVC, LabeledLineConnectionFigure, setAttribute)\\
 \wedge \neg \mathtt{ExistsMethodInvocation}(LabeledLineConnectionFigure, findFigureInsideTmpVC, LabeledLineConnectionFigure, addNotify)\\
 \wedge \neg \mathtt{ExistsMethodInvocation}(LabeledLineConnectionFigure, findFigureInsideTmpVC, LabeledLineConnectionFigure, removeNotify)\\
 \wedge \neg \mathtt{ExistsMethodInvocation}(LabeledLineConnectionFigure, setAttributeTmpVC, LabeledLineConnectionFigure, basicTransform)\\
 \wedge \neg \mathtt{ExistsMethodInvocation}(LabeledLineConnectionFigure, setAttributeTmpVC, LabeledLineConnectionFigure, contains)\\
 \wedge \neg \mathtt{ExistsMethodInvocation}(LabeledLineConnectionFigure, setAttributeTmpVC, LabeledLineConnectionFigure, findFigureInside)\\
 \wedge \neg \mathtt{ExistsMethodInvocation}(LabeledLineConnectionFigure, setAttributeTmpVC, LabeledLineConnectionFigure, addNotify)\\
 \wedge \neg \mathtt{ExistsMethodInvocation}(LabeledLineConnectionFigure, setAttributeTmpVC, LabeledLineConnectionFigure, removeNotify)\\
 \wedge \neg \mathtt{ExistsMethodInvocation}(LabeledLineConnectionFigure, containsTmpVC, LabeledLineConnectionFigure, basicTransform)\\
 \wedge \neg \mathtt{ExistsMethodInvocation}(LabeledLineConnectionFigure, containsTmpVC, LabeledLineConnectionFigure, setAttribute)\\
 \wedge \neg \mathtt{ExistsMethodInvocation}(LabeledLineConnectionFigure, containsTmpVC, LabeledLineConnectionFigure, findFigureInside)\\
 \wedge \neg \mathtt{ExistsMethodInvocation}(LabeledLineConnectionFigure, containsTmpVC, LabeledLineConnectionFigure, addNotify)\\
 \wedge \neg \mathtt{ExistsMethodInvocation}(LabeledLineConnectionFigure, containsTmpVC, LabeledLineConnectionFigure, removeNotify)\\
 \wedge \neg \mathtt{IsUsedMethod}(AbstractFigure, accept\_RemoveNotifyVisitor\_addspecializedMethod\_tmp, [RemoveNotifyVisitor])\\
 \wedge \neg \mathtt{IsUsedMethod}(AbstractFigure, accept\_AddNotifyVisitor\_addspecializedMethod\_tmp, [AddNotifyVisitor])\\
 \wedge \neg \mathtt{IsUsedMethod}(AbstractFigure, accept\_FindFigureInsideVisitor\_addspecializedMethod\_tmp, [FindFigureInsideVisitor])\\
 \wedge \neg \mathtt{IsUsedMethod}(AbstractFigure, accept\_SetAttributeVisitor\_addspecializedMethod\_tmp, [SetAttributeVisitor])\\
 \wedge \neg \mathtt{IsUsedMethod}(AbstractFigure, accept\_ContainsVisitor\_addspecializedMethod\_tmp, [ContainsVisitor])\\
 \wedge \neg \mathtt{IsUsedMethod}(AbstractFigure, accept\_BasicTransformVisitor\_addspecializedMethod\_tmp, [BasicTransformVisitor])\\
 \wedge \mathtt{AllSubclasses}(AbstractFigure, [EllipseFigure; DiamondFigure; RectangleFigure; RoundRectangleFigure; TriangleFigure; TextFigure; BezierFigure; TextAreaFigure; NodeFigure; SVGImage; SVGPath; DependencyFigure; LineConnectionFigure; LabeledLineConnectionFigure; AbstractCompositeFigure; GraphicalCompositeFigure])\\
 \wedge \neg \mathtt{ExistsMethodDefinition}(AbstractFigure, accept)\\
 \wedge \neg \mathtt{ExistsMethodDefinition}(LabeledLineConnectionFigure, accept)\\
 \wedge \neg \mathtt{ExistsMethodDefinition}(AbstractCompositeFigure, accept)\\
 \wedge \neg \mathtt{ExistsMethodDefinition}(GraphicalCompositeFigure, accept)\\
 \wedge \neg \mathtt{ExistsMethodDefinition}(EllipseFigure, accept)\\
 \wedge \neg \mathtt{ExistsMethodDefinition}(DiamondFigure, accept)\\
 \wedge \neg \mathtt{ExistsMethodDefinition}(RectangleFigure, accept)\\
 \wedge \neg \mathtt{ExistsMethodDefinition}(RoundRectangleFigure, accept)\\
 \wedge \neg \mathtt{ExistsMethodDefinition}(TriangleFigure, accept)\\
 \wedge \neg \mathtt{ExistsMethodDefinition}(TextFigure, accept)\\
 \wedge \neg \mathtt{ExistsMethodDefinition}(BezierFigure, accept)\\
 \wedge \neg \mathtt{ExistsMethodDefinition}(TextAreaFigure, accept)\\
 \wedge \neg \mathtt{ExistsMethodDefinition}(NodeFigure, accept)\\
 \wedge \neg \mathtt{ExistsMethodDefinition}(SVGImage, accept)\\
 \wedge \neg \mathtt{ExistsMethodDefinition}(SVGPath, accept)\\
 \wedge \neg \mathtt{ExistsMethodDefinition}(DependencyFigure, accept)\\
 \wedge \neg \mathtt{ExistsMethodDefinition}(LineConnectionFigure, accept)\\
 \wedge \neg \mathtt{IsInheritedMethod}(AbstractFigure, accept)\\
 \wedge \neg \mathtt{ExistsMethodInvocation}(AbstractFigure, removeNotifyTmpVC, AbstractFigure, contains)\\
 \wedge \neg \mathtt{ExistsMethodInvocation}(AbstractFigure, removeNotifyTmpVC, AbstractFigure, setAttribute)\\
 \wedge \neg \mathtt{ExistsMethodInvocation}(AbstractFigure, removeNotifyTmpVC, AbstractFigure, findFigureInside)\\
 \wedge \neg \mathtt{ExistsMethodInvocation}(AbstractFigure, removeNotifyTmpVC, AbstractFigure, addNotify)\\
 \wedge \neg \mathtt{IsUsedMethodIn}(AbstractFigure, removeNotifyTmpVC, LabeledLineConnectionFigure)\\
 \wedge \neg \mathtt{IsUsedMethodIn}(AbstractFigure, removeNotifyTmpVC, AbstractCompositeFigure)\\
 \wedge \neg \mathtt{IsUsedMethodIn}(AbstractFigure, removeNotifyTmpVC, GraphicalCompositeFigure)\\
 \wedge \neg \mathtt{IsUsedMethodIn}(AbstractFigure, removeNotifyTmpVC, EllipseFigure)\\
 \wedge \neg \mathtt{IsUsedMethodIn}(AbstractFigure, removeNotifyTmpVC, DiamondFigure)\\
 \wedge \neg \mathtt{IsUsedMethodIn}(AbstractFigure, removeNotifyTmpVC, RectangleFigure)\\
 \wedge \neg \mathtt{IsUsedMethodIn}(AbstractFigure, removeNotifyTmpVC, RoundRectangleFigure)\\
 \wedge \neg \mathtt{IsUsedMethodIn}(AbstractFigure, removeNotifyTmpVC, TriangleFigure)\\
 \wedge \neg \mathtt{IsUsedMethodIn}(AbstractFigure, removeNotifyTmpVC, TextFigure)\\
 \wedge \neg \mathtt{IsUsedMethodIn}(AbstractFigure, removeNotifyTmpVC, BezierFigure)\\
 \wedge \neg \mathtt{IsUsedMethodIn}(AbstractFigure, removeNotifyTmpVC, TextAreaFigure)\\
 \wedge \neg \mathtt{IsUsedMethodIn}(AbstractFigure, removeNotifyTmpVC, NodeFigure)\\
 \wedge \neg \mathtt{IsUsedMethodIn}(AbstractFigure, removeNotifyTmpVC, SVGImage)\\
 \wedge \neg \mathtt{IsUsedMethodIn}(AbstractFigure, removeNotifyTmpVC, SVGPath)\\
 \wedge \neg \mathtt{IsUsedMethodIn}(AbstractFigure, removeNotifyTmpVC, DependencyFigure)\\
 \wedge \neg \mathtt{IsUsedMethodIn}(AbstractFigure, removeNotifyTmpVC, LineConnectionFigure)\\
 \wedge \neg \mathtt{ExistsMethodInvocation}(AbstractFigure, addNotifyTmpVC, AbstractFigure, contains)\\
 \wedge \neg \mathtt{ExistsMethodInvocation}(AbstractFigure, addNotifyTmpVC, AbstractFigure, setAttribute)\\
 \wedge \neg \mathtt{ExistsMethodInvocation}(AbstractFigure, addNotifyTmpVC, AbstractFigure, findFigureInside)\\
 \wedge \neg \mathtt{ExistsMethodInvocation}(AbstractFigure, addNotifyTmpVC, AbstractFigure, removeNotify)\\
 \wedge \neg \mathtt{IsUsedMethodIn}(AbstractFigure, addNotifyTmpVC, LabeledLineConnectionFigure)\\
 \wedge \neg \mathtt{IsUsedMethodIn}(AbstractFigure, addNotifyTmpVC, AbstractCompositeFigure)\\
 \wedge \neg \mathtt{IsUsedMethodIn}(AbstractFigure, addNotifyTmpVC, GraphicalCompositeFigure)\\
 \wedge \neg \mathtt{IsUsedMethodIn}(AbstractFigure, addNotifyTmpVC, EllipseFigure)\\
 \wedge \neg \mathtt{IsUsedMethodIn}(AbstractFigure, addNotifyTmpVC, DiamondFigure)\\
 \wedge \neg \mathtt{IsUsedMethodIn}(AbstractFigure, addNotifyTmpVC, RectangleFigure)\\
 \wedge \neg \mathtt{IsUsedMethodIn}(AbstractFigure, addNotifyTmpVC, RoundRectangleFigure)\\
 \wedge \neg \mathtt{IsUsedMethodIn}(AbstractFigure, addNotifyTmpVC, TriangleFigure)\\
 \wedge \neg \mathtt{IsUsedMethodIn}(AbstractFigure, addNotifyTmpVC, TextFigure)\\
 \wedge \neg \mathtt{IsUsedMethodIn}(AbstractFigure, addNotifyTmpVC, BezierFigure)\\
 \wedge \neg \mathtt{IsUsedMethodIn}(AbstractFigure, addNotifyTmpVC, TextAreaFigure)\\
 \wedge \neg \mathtt{IsUsedMethodIn}(AbstractFigure, addNotifyTmpVC, NodeFigure)\\
 \wedge \neg \mathtt{IsUsedMethodIn}(AbstractFigure, addNotifyTmpVC, SVGImage)\\
 \wedge \neg \mathtt{IsUsedMethodIn}(AbstractFigure, addNotifyTmpVC, SVGPath)\\
 \wedge \neg \mathtt{IsUsedMethodIn}(AbstractFigure, addNotifyTmpVC, DependencyFigure)\\
 \wedge \neg \mathtt{IsUsedMethodIn}(AbstractFigure, addNotifyTmpVC, LineConnectionFigure)\\
 \wedge \neg \mathtt{ExistsMethodInvocation}(AbstractFigure, findFigureInsideTmpVC, AbstractFigure, contains)\\
 \wedge \neg \mathtt{ExistsMethodInvocation}(AbstractFigure, findFigureInsideTmpVC, AbstractFigure, setAttribute)\\
 \wedge \neg \mathtt{ExistsMethodInvocation}(AbstractFigure, findFigureInsideTmpVC, AbstractFigure, addNotify)\\
 \wedge \neg \mathtt{ExistsMethodInvocation}(AbstractFigure, findFigureInsideTmpVC, AbstractFigure, removeNotify)\\
 \wedge \neg \mathtt{IsUsedMethodIn}(AbstractFigure, findFigureInsideTmpVC, LabeledLineConnectionFigure)\\
 \wedge \neg \mathtt{IsUsedMethodIn}(AbstractFigure, findFigureInsideTmpVC, AbstractCompositeFigure)\\
 \wedge \neg \mathtt{IsUsedMethodIn}(AbstractFigure, findFigureInsideTmpVC, GraphicalCompositeFigure)\\
 \wedge \neg \mathtt{IsUsedMethodIn}(AbstractFigure, findFigureInsideTmpVC, EllipseFigure)\\
 \wedge \neg \mathtt{IsUsedMethodIn}(AbstractFigure, findFigureInsideTmpVC, DiamondFigure)\\
 \wedge \neg \mathtt{IsUsedMethodIn}(AbstractFigure, findFigureInsideTmpVC, RectangleFigure)\\
 \wedge \neg \mathtt{IsUsedMethodIn}(AbstractFigure, findFigureInsideTmpVC, RoundRectangleFigure)\\
 \wedge \neg \mathtt{IsUsedMethodIn}(AbstractFigure, findFigureInsideTmpVC, TriangleFigure)\\
 \wedge \neg \mathtt{IsUsedMethodIn}(AbstractFigure, findFigureInsideTmpVC, TextFigure)\\
 \wedge \neg \mathtt{IsUsedMethodIn}(AbstractFigure, findFigureInsideTmpVC, BezierFigure)\\
 \wedge \neg \mathtt{IsUsedMethodIn}(AbstractFigure, findFigureInsideTmpVC, TextAreaFigure)\\
 \wedge \neg \mathtt{IsUsedMethodIn}(AbstractFigure, findFigureInsideTmpVC, NodeFigure)\\
 \wedge \neg \mathtt{IsUsedMethodIn}(AbstractFigure, findFigureInsideTmpVC, SVGImage)\\
 \wedge \neg \mathtt{IsUsedMethodIn}(AbstractFigure, findFigureInsideTmpVC, SVGPath)\\
 \wedge \neg \mathtt{IsUsedMethodIn}(AbstractFigure, findFigureInsideTmpVC, DependencyFigure)\\
 \wedge \neg \mathtt{IsUsedMethodIn}(AbstractFigure, findFigureInsideTmpVC, LineConnectionFigure)\\
 \wedge \neg \mathtt{ExistsMethodInvocation}(AbstractFigure, setAttributeTmpVC, AbstractFigure, contains)\\
 \wedge \neg \mathtt{ExistsMethodInvocation}(AbstractFigure, setAttributeTmpVC, AbstractFigure, findFigureInside)\\
 \wedge \neg \mathtt{ExistsMethodInvocation}(AbstractFigure, setAttributeTmpVC, AbstractFigure, addNotify)\\
 \wedge \neg \mathtt{ExistsMethodInvocation}(AbstractFigure, setAttributeTmpVC, AbstractFigure, removeNotify)\\
 \wedge \neg \mathtt{IsUsedMethodIn}(AbstractFigure, setAttributeTmpVC, LabeledLineConnectionFigure)\\
 \wedge \neg \mathtt{IsUsedMethodIn}(AbstractFigure, setAttributeTmpVC, AbstractCompositeFigure)\\
 \wedge \neg \mathtt{IsUsedMethodIn}(AbstractFigure, setAttributeTmpVC, GraphicalCompositeFigure)\\
 \wedge \neg \mathtt{IsUsedMethodIn}(AbstractFigure, setAttributeTmpVC, EllipseFigure)\\
 \wedge \neg \mathtt{IsUsedMethodIn}(AbstractFigure, setAttributeTmpVC, DiamondFigure)\\
 \wedge \neg \mathtt{IsUsedMethodIn}(AbstractFigure, setAttributeTmpVC, RectangleFigure)\\
 \wedge \neg \mathtt{IsUsedMethodIn}(AbstractFigure, setAttributeTmpVC, RoundRectangleFigure)\\
 \wedge \neg \mathtt{IsUsedMethodIn}(AbstractFigure, setAttributeTmpVC, TriangleFigure)\\
 \wedge \neg \mathtt{IsUsedMethodIn}(AbstractFigure, setAttributeTmpVC, TextFigure)\\
 \wedge \neg \mathtt{IsUsedMethodIn}(AbstractFigure, setAttributeTmpVC, BezierFigure)\\
 \wedge \neg \mathtt{IsUsedMethodIn}(AbstractFigure, setAttributeTmpVC, TextAreaFigure)\\
 \wedge \neg \mathtt{IsUsedMethodIn}(AbstractFigure, setAttributeTmpVC, NodeFigure)\\
 \wedge \neg \mathtt{IsUsedMethodIn}(AbstractFigure, setAttributeTmpVC, SVGImage)\\
 \wedge \neg \mathtt{IsUsedMethodIn}(AbstractFigure, setAttributeTmpVC, SVGPath)\\
 \wedge \neg \mathtt{IsUsedMethodIn}(AbstractFigure, setAttributeTmpVC, DependencyFigure)\\
 \wedge \neg \mathtt{IsUsedMethodIn}(AbstractFigure, setAttributeTmpVC, LineConnectionFigure)\\
 \wedge \neg \mathtt{ExistsMethodInvocation}(AbstractFigure, containsTmpVC, AbstractFigure, setAttribute)\\
 \wedge \neg \mathtt{ExistsMethodInvocation}(AbstractFigure, containsTmpVC, AbstractFigure, findFigureInside)\\
 \wedge \neg \mathtt{ExistsMethodInvocation}(AbstractFigure, containsTmpVC, AbstractFigure, addNotify)\\
 \wedge \neg \mathtt{ExistsMethodInvocation}(AbstractFigure, containsTmpVC, AbstractFigure, removeNotify)\\
 \wedge \neg \mathtt{IsUsedMethodIn}(AbstractFigure, containsTmpVC, LabeledLineConnectionFigure)\\
 \wedge \neg \mathtt{IsUsedMethodIn}(AbstractFigure, containsTmpVC, AbstractCompositeFigure)\\
 \wedge \neg \mathtt{IsUsedMethodIn}(AbstractFigure, containsTmpVC, GraphicalCompositeFigure)\\
 \wedge \neg \mathtt{IsUsedMethodIn}(AbstractFigure, containsTmpVC, EllipseFigure)\\
 \wedge \neg \mathtt{IsUsedMethodIn}(AbstractFigure, containsTmpVC, DiamondFigure)\\
 \wedge \neg \mathtt{IsUsedMethodIn}(AbstractFigure, containsTmpVC, RectangleFigure)\\
 \wedge \neg \mathtt{IsUsedMethodIn}(AbstractFigure, containsTmpVC, RoundRectangleFigure)\\
 \wedge \neg \mathtt{IsUsedMethodIn}(AbstractFigure, containsTmpVC, TriangleFigure)\\
 \wedge \neg \mathtt{IsUsedMethodIn}(AbstractFigure, containsTmpVC, TextFigure)\\
 \wedge \neg \mathtt{IsUsedMethodIn}(AbstractFigure, containsTmpVC, BezierFigure)\\
 \wedge \neg \mathtt{IsUsedMethodIn}(AbstractFigure, containsTmpVC, TextAreaFigure)\\
 \wedge \neg \mathtt{IsUsedMethodIn}(AbstractFigure, containsTmpVC, NodeFigure)\\
 \wedge \neg \mathtt{IsUsedMethodIn}(AbstractFigure, containsTmpVC, SVGImage)\\
 \wedge \neg \mathtt{IsUsedMethodIn}(AbstractFigure, containsTmpVC, SVGPath)\\
 \wedge \neg \mathtt{IsUsedMethodIn}(AbstractFigure, containsTmpVC, DependencyFigure)\\
 \wedge \neg \mathtt{IsUsedMethodIn}(AbstractFigure, containsTmpVC, LineConnectionFigure)\\
 \wedge \mathtt{AllInvokedMethodsWithParameterOInBodyOfMAreNotOverloaded}(LineConnectionFigure, removeNotify, this)\\
 \wedge \mathtt{AllInvokedMethodsWithParameterOInBodyOfMAreNotOverloaded}(DependencyFigure, removeNotify, this)\\
 \wedge \mathtt{AllInvokedMethodsWithParameterOInBodyOfMAreNotOverloaded}(SVGImage, removeNotify, this)\\
 \wedge \mathtt{AllInvokedMethodsWithParameterOInBodyOfMAreNotOverloaded}(NodeFigure, removeNotify, this)\\
 \wedge \mathtt{AllInvokedMethodsWithParameterOInBodyOfMAreNotOverloaded}(TextAreaFigure, removeNotify, this)\\
 \wedge \mathtt{AllInvokedMethodsWithParameterOInBodyOfMAreNotOverloaded}(BezierFigure, removeNotify, this)\\
 \wedge \mathtt{AllInvokedMethodsWithParameterOInBodyOfMAreNotOverloaded}(TextFigure, removeNotify, this)\\
 \wedge \mathtt{AllInvokedMethodsWithParameterOInBodyOfMAreNotOverloaded}(TriangleFigure, removeNotify, this)\\
 \wedge \mathtt{AllInvokedMethodsWithParameterOInBodyOfMAreNotOverloaded}(RoundRectangleFigure, removeNotify, this)\\
 \wedge \mathtt{AllInvokedMethodsWithParameterOInBodyOfMAreNotOverloaded}(RectangleFigure, removeNotify, this)\\
 \wedge \mathtt{AllInvokedMethodsWithParameterOInBodyOfMAreNotOverloaded}(DiamondFigure, removeNotify, this)\\
 \wedge \mathtt{AllInvokedMethodsWithParameterOInBodyOfMAreNotOverloaded}(EllipseFigure, removeNotify, this)\\
 \wedge \mathtt{AllInvokedMethodsWithParameterOInBodyOfMAreNotOverloaded}(GraphicalCompositeFigure, removeNotify, this)\\
 \wedge \mathtt{AllInvokedMethodsOnObjectOInBodyOfMAreDeclaredInC}(LabeledLineConnectionFigure, removeNotify, this, AbstractFigure)\\
 \wedge \mathtt{AllInvokedMethodsWithParameterOInBodyOfMAreNotOverloaded}(LabeledLineConnectionFigure, removeNotify, this)\\
 \wedge \mathtt{AllInvokedMethodsWithParameterOInBodyOfMAreNotOverloaded}(SVGImage, addNotify, this)\\
 \wedge \mathtt{AllInvokedMethodsWithParameterOInBodyOfMAreNotOverloaded}(TextAreaFigure, addNotify, this)\\
 \wedge \mathtt{AllInvokedMethodsWithParameterOInBodyOfMAreNotOverloaded}(BezierFigure, addNotify, this)\\
 \wedge \mathtt{AllInvokedMethodsWithParameterOInBodyOfMAreNotOverloaded}(TriangleFigure, addNotify, this)\\
 \wedge \mathtt{AllInvokedMethodsWithParameterOInBodyOfMAreNotOverloaded}(RoundRectangleFigure, addNotify, this)\\
 \wedge \mathtt{AllInvokedMethodsWithParameterOInBodyOfMAreNotOverloaded}(RectangleFigure, addNotify, this)\\
 \wedge \mathtt{AllInvokedMethodsWithParameterOInBodyOfMAreNotOverloaded}(DiamondFigure, addNotify, this)\\
 \wedge \mathtt{AllInvokedMethodsWithParameterOInBodyOfMAreNotOverloaded}(EllipseFigure, addNotify, this)\\
 \wedge \mathtt{AllInvokedMethodsWithParameterOInBodyOfMAreNotOverloaded}(GraphicalCompositeFigure, addNotify, this)\\
 \wedge \mathtt{AllInvokedMethodsOnObjectOInBodyOfMAreDeclaredInC}(LabeledLineConnectionFigure, addNotify, this, AbstractFigure)\\
 \wedge \mathtt{AllInvokedMethodsWithParameterOInBodyOfMAreNotOverloaded}(LabeledLineConnectionFigure, addNotify, this)\\
 \wedge \mathtt{AllInvokedMethodsWithParameterOInBodyOfMAreNotOverloaded}(SVGImage, findFigureInside, this)\\
 \wedge \mathtt{AllInvokedMethodsWithParameterOInBodyOfMAreNotOverloaded}(TextAreaFigure, findFigureInside, this)\\
 \wedge \mathtt{AllInvokedMethodsWithParameterOInBodyOfMAreNotOverloaded}(TriangleFigure, findFigureInside, this)\\
 \wedge \mathtt{AllInvokedMethodsWithParameterOInBodyOfMAreNotOverloaded}(RoundRectangleFigure, findFigureInside, this)\\
 \wedge \mathtt{AllInvokedMethodsWithParameterOInBodyOfMAreNotOverloaded}(RectangleFigure, findFigureInside, this)\\
 \wedge \mathtt{AllInvokedMethodsWithParameterOInBodyOfMAreNotOverloaded}(DiamondFigure, findFigureInside, this)\\
 \wedge \mathtt{AllInvokedMethodsWithParameterOInBodyOfMAreNotOverloaded}(EllipseFigure, findFigureInside, this)\\
 \wedge \mathtt{AllInvokedMethodsOnObjectOInBodyOfMAreDeclaredInC}(LabeledLineConnectionFigure, findFigureInside, this, AbstractFigure)\\
 \wedge \mathtt{AllInvokedMethodsWithParameterOInBodyOfMAreNotOverloaded}(SVGPath, setAttribute, this)\\
 \wedge \mathtt{AllInvokedMethodsWithParameterOInBodyOfMAreNotOverloaded}(SVGImage, setAttribute, this)\\
 \wedge \mathtt{AllInvokedMethodsWithParameterOInBodyOfMAreNotOverloaded}(TextAreaFigure, setAttribute, this)\\
 \wedge \mathtt{AllInvokedMethodsWithParameterOInBodyOfMAreNotOverloaded}(TriangleFigure, setAttribute, this)\\
 \wedge \mathtt{AllInvokedMethodsWithParameterOInBodyOfMAreNotOverloaded}(RoundRectangleFigure, setAttribute, this)\\
 \wedge \mathtt{AllInvokedMethodsWithParameterOInBodyOfMAreNotOverloaded}(RectangleFigure, setAttribute, this)\\
 \wedge \mathtt{AllInvokedMethodsWithParameterOInBodyOfMAreNotOverloaded}(DiamondFigure, setAttribute, this)\\
 \wedge \mathtt{AllInvokedMethodsWithParameterOInBodyOfMAreNotOverloaded}(EllipseFigure, setAttribute, this)\\
 \wedge \mathtt{AllInvokedMethodsWithParameterOInBodyOfMAreNotOverloaded}(GraphicalCompositeFigure, setAttribute, this)\\
 \wedge \mathtt{AllInvokedMethodsWithParameterOInBodyOfMAreNotOverloaded}(AbstractCompositeFigure, setAttribute, this)\\
 \wedge \mathtt{AllInvokedMethodsOnObjectOInBodyOfMAreDeclaredInC}(LabeledLineConnectionFigure, setAttribute, this, AbstractFigure)\\
 \wedge \mathtt{AllInvokedMethodsWithParameterOInBodyOfMAreNotOverloaded}(SVGImage, contains, this)\\
 \wedge \mathtt{AllInvokedMethodsWithParameterOInBodyOfMAreNotOverloaded}(TextAreaFigure, contains, this)\\
 \wedge \mathtt{AllInvokedMethodsWithParameterOInBodyOfMAreNotOverloaded}(TriangleFigure, contains, this)\\
 \wedge \mathtt{AllInvokedMethodsWithParameterOInBodyOfMAreNotOverloaded}(RoundRectangleFigure, contains, this)\\
 \wedge \mathtt{AllInvokedMethodsWithParameterOInBodyOfMAreNotOverloaded}(RectangleFigure, contains, this)\\
 \wedge \mathtt{AllInvokedMethodsWithParameterOInBodyOfMAreNotOverloaded}(DiamondFigure, contains, this)\\
 \wedge \mathtt{AllInvokedMethodsWithParameterOInBodyOfMAreNotOverloaded}(EllipseFigure, contains, this)\\
 \wedge \mathtt{AllInvokedMethodsWithParameterOInBodyOfMAreNotOverloaded}(GraphicalCompositeFigure, contains, this)\\
 \wedge \mathtt{AllInvokedMethodsOnObjectOInBodyOfMAreDeclaredInC}(LabeledLineConnectionFigure, contains, this, AbstractFigure)\\
 \wedge \mathtt{AllInvokedMethodsWithParameterOInBodyOfMAreNotOverloaded}(AbstractFigure, removeNotify, v)\\
 \wedge \mathtt{AllInvokedMethodsWithParameterOInBodyOfMAreNotOverloaded}(LabeledLineConnectionFigure, removeNotify, v)\\
 \wedge \mathtt{AllInvokedMethodsWithParameterOInBodyOfMAreNotOverloaded}(AbstractCompositeFigure, removeNotify, v)\\
 \wedge \mathtt{AllInvokedMethodsWithParameterOInBodyOfMAreNotOverloaded}(GraphicalCompositeFigure, removeNotify, v)\\
 \wedge \mathtt{AllInvokedMethodsWithParameterOInBodyOfMAreNotOverloaded}(EllipseFigure, removeNotify, v)\\
 \wedge \mathtt{AllInvokedMethodsWithParameterOInBodyOfMAreNotOverloaded}(DiamondFigure, removeNotify, v)\\
 \wedge \mathtt{AllInvokedMethodsWithParameterOInBodyOfMAreNotOverloaded}(RectangleFigure, removeNotify, v)\\
 \wedge \mathtt{AllInvokedMethodsWithParameterOInBodyOfMAreNotOverloaded}(RoundRectangleFigure, removeNotify, v)\\
 \wedge \mathtt{AllInvokedMethodsWithParameterOInBodyOfMAreNotOverloaded}(TriangleFigure, removeNotify, v)\\
 \wedge \mathtt{AllInvokedMethodsWithParameterOInBodyOfMAreNotOverloaded}(TextFigure, removeNotify, v)\\
 \wedge \mathtt{AllInvokedMethodsWithParameterOInBodyOfMAreNotOverloaded}(BezierFigure, removeNotify, v)\\
 \wedge \mathtt{AllInvokedMethodsWithParameterOInBodyOfMAreNotOverloaded}(TextAreaFigure, removeNotify, v)\\
 \wedge \mathtt{AllInvokedMethodsWithParameterOInBodyOfMAreNotOverloaded}(NodeFigure, removeNotify, v)\\
 \wedge \mathtt{AllInvokedMethodsWithParameterOInBodyOfMAreNotOverloaded}(SVGImage, removeNotify, v)\\
 \wedge \mathtt{AllInvokedMethodsWithParameterOInBodyOfMAreNotOverloaded}(SVGPath, removeNotify, v)\\
 \wedge \mathtt{AllInvokedMethodsWithParameterOInBodyOfMAreNotOverloaded}(DependencyFigure, removeNotify, v)\\
 \wedge \mathtt{AllInvokedMethodsWithParameterOInBodyOfMAreNotOverloaded}(LineConnectionFigure, removeNotify, v)\\
 \wedge \mathtt{AllInvokedMethodsWithParameterOInBodyOfMAreNotOverloaded}(AbstractFigure, addNotify, v)\\
 \wedge \mathtt{AllInvokedMethodsWithParameterOInBodyOfMAreNotOverloaded}(LabeledLineConnectionFigure, addNotify, v)\\
 \wedge \mathtt{AllInvokedMethodsWithParameterOInBodyOfMAreNotOverloaded}(AbstractCompositeFigure, addNotify, v)\\
 \wedge \mathtt{AllInvokedMethodsWithParameterOInBodyOfMAreNotOverloaded}(GraphicalCompositeFigure, addNotify, v)\\
 \wedge \mathtt{AllInvokedMethodsWithParameterOInBodyOfMAreNotOverloaded}(EllipseFigure, addNotify, v)\\
 \wedge \mathtt{AllInvokedMethodsWithParameterOInBodyOfMAreNotOverloaded}(DiamondFigure, addNotify, v)\\
 \wedge \mathtt{AllInvokedMethodsWithParameterOInBodyOfMAreNotOverloaded}(RectangleFigure, addNotify, v)\\
 \wedge \mathtt{AllInvokedMethodsWithParameterOInBodyOfMAreNotOverloaded}(RoundRectangleFigure, addNotify, v)\\
 \wedge \mathtt{AllInvokedMethodsWithParameterOInBodyOfMAreNotOverloaded}(TriangleFigure, addNotify, v)\\
 \wedge \mathtt{AllInvokedMethodsWithParameterOInBodyOfMAreNotOverloaded}(TextFigure, addNotify, v)\\
 \wedge \mathtt{AllInvokedMethodsWithParameterOInBodyOfMAreNotOverloaded}(BezierFigure, addNotify, v)\\
 \wedge \mathtt{AllInvokedMethodsWithParameterOInBodyOfMAreNotOverloaded}(TextAreaFigure, addNotify, v)\\
 \wedge \mathtt{AllInvokedMethodsWithParameterOInBodyOfMAreNotOverloaded}(NodeFigure, addNotify, v)\\
 \wedge \mathtt{AllInvokedMethodsWithParameterOInBodyOfMAreNotOverloaded}(SVGImage, addNotify, v)\\
 \wedge \mathtt{AllInvokedMethodsWithParameterOInBodyOfMAreNotOverloaded}(SVGPath, addNotify, v)\\
 \wedge \mathtt{AllInvokedMethodsWithParameterOInBodyOfMAreNotOverloaded}(DependencyFigure, addNotify, v)\\
 \wedge \mathtt{AllInvokedMethodsWithParameterOInBodyOfMAreNotOverloaded}(LineConnectionFigure, addNotify, v)\\
 \wedge \mathtt{AllInvokedMethodsWithParameterOInBodyOfMAreNotOverloaded}(AbstractFigure, findFigureInside, v)\\
 \wedge \mathtt{AllInvokedMethodsWithParameterOInBodyOfMAreNotOverloaded}(LabeledLineConnectionFigure, findFigureInside, v)\\
 \wedge \mathtt{AllInvokedMethodsWithParameterOInBodyOfMAreNotOverloaded}(AbstractCompositeFigure, findFigureInside, v)\\
 \wedge \mathtt{AllInvokedMethodsWithParameterOInBodyOfMAreNotOverloaded}(GraphicalCompositeFigure, findFigureInside, v)\\
 \wedge \mathtt{AllInvokedMethodsWithParameterOInBodyOfMAreNotOverloaded}(EllipseFigure, findFigureInside, v)\\
 \wedge \mathtt{AllInvokedMethodsWithParameterOInBodyOfMAreNotOverloaded}(DiamondFigure, findFigureInside, v)\\
 \wedge \mathtt{AllInvokedMethodsWithParameterOInBodyOfMAreNotOverloaded}(RectangleFigure, findFigureInside, v)\\
 \wedge \mathtt{AllInvokedMethodsWithParameterOInBodyOfMAreNotOverloaded}(RoundRectangleFigure, findFigureInside, v)\\
 \wedge \mathtt{AllInvokedMethodsWithParameterOInBodyOfMAreNotOverloaded}(TriangleFigure, findFigureInside, v)\\
 \wedge \mathtt{AllInvokedMethodsWithParameterOInBodyOfMAreNotOverloaded}(TextFigure, findFigureInside, v)\\
 \wedge \mathtt{AllInvokedMethodsWithParameterOInBodyOfMAreNotOverloaded}(BezierFigure, findFigureInside, v)\\
 \wedge \mathtt{AllInvokedMethodsWithParameterOInBodyOfMAreNotOverloaded}(TextAreaFigure, findFigureInside, v)\\
 \wedge \mathtt{AllInvokedMethodsWithParameterOInBodyOfMAreNotOverloaded}(NodeFigure, findFigureInside, v)\\
 \wedge \mathtt{AllInvokedMethodsWithParameterOInBodyOfMAreNotOverloaded}(SVGImage, findFigureInside, v)\\
 \wedge \mathtt{AllInvokedMethodsWithParameterOInBodyOfMAreNotOverloaded}(SVGPath, findFigureInside, v)\\
 \wedge \mathtt{AllInvokedMethodsWithParameterOInBodyOfMAreNotOverloaded}(DependencyFigure, findFigureInside, v)\\
 \wedge \mathtt{AllInvokedMethodsWithParameterOInBodyOfMAreNotOverloaded}(LineConnectionFigure, findFigureInside, v)\\
 \wedge \mathtt{AllInvokedMethodsWithParameterOInBodyOfMAreNotOverloaded}(AbstractFigure, setAttribute, v)\\
 \wedge \mathtt{AllInvokedMethodsWithParameterOInBodyOfMAreNotOverloaded}(LabeledLineConnectionFigure, setAttribute, v)\\
 \wedge \mathtt{AllInvokedMethodsWithParameterOInBodyOfMAreNotOverloaded}(AbstractCompositeFigure, setAttribute, v)\\
 \wedge \mathtt{AllInvokedMethodsWithParameterOInBodyOfMAreNotOverloaded}(GraphicalCompositeFigure, setAttribute, v)\\
 \wedge \mathtt{AllInvokedMethodsWithParameterOInBodyOfMAreNotOverloaded}(EllipseFigure, setAttribute, v)\\
 \wedge \mathtt{AllInvokedMethodsWithParameterOInBodyOfMAreNotOverloaded}(DiamondFigure, setAttribute, v)\\
 \wedge \mathtt{AllInvokedMethodsWithParameterOInBodyOfMAreNotOverloaded}(RectangleFigure, setAttribute, v)\\
 \wedge \mathtt{AllInvokedMethodsWithParameterOInBodyOfMAreNotOverloaded}(RoundRectangleFigure, setAttribute, v)\\
 \wedge \mathtt{AllInvokedMethodsWithParameterOInBodyOfMAreNotOverloaded}(TriangleFigure, setAttribute, v)\\
 \wedge \mathtt{AllInvokedMethodsWithParameterOInBodyOfMAreNotOverloaded}(TextFigure, setAttribute, v)\\
 \wedge \mathtt{AllInvokedMethodsWithParameterOInBodyOfMAreNotOverloaded}(BezierFigure, setAttribute, v)\\
 \wedge \mathtt{AllInvokedMethodsWithParameterOInBodyOfMAreNotOverloaded}(TextAreaFigure, setAttribute, v)\\
 \wedge \mathtt{AllInvokedMethodsWithParameterOInBodyOfMAreNotOverloaded}(NodeFigure, setAttribute, v)\\
 \wedge \mathtt{AllInvokedMethodsWithParameterOInBodyOfMAreNotOverloaded}(SVGImage, setAttribute, v)\\
 \wedge \mathtt{AllInvokedMethodsWithParameterOInBodyOfMAreNotOverloaded}(SVGPath, setAttribute, v)\\
 \wedge \mathtt{AllInvokedMethodsWithParameterOInBodyOfMAreNotOverloaded}(DependencyFigure, setAttribute, v)\\
 \wedge \mathtt{AllInvokedMethodsWithParameterOInBodyOfMAreNotOverloaded}(LineConnectionFigure, setAttribute, v)\\
 \wedge \mathtt{AllInvokedMethodsWithParameterOInBodyOfMAreNotOverloaded}(AbstractFigure, contains, v)\\
 \wedge \mathtt{AllInvokedMethodsWithParameterOInBodyOfMAreNotOverloaded}(LabeledLineConnectionFigure, contains, v)\\
 \wedge \mathtt{AllInvokedMethodsWithParameterOInBodyOfMAreNotOverloaded}(AbstractCompositeFigure, contains, v)\\
 \wedge \mathtt{AllInvokedMethodsWithParameterOInBodyOfMAreNotOverloaded}(GraphicalCompositeFigure, contains, v)\\
 \wedge \mathtt{AllInvokedMethodsWithParameterOInBodyOfMAreNotOverloaded}(EllipseFigure, contains, v)\\
 \wedge \mathtt{AllInvokedMethodsWithParameterOInBodyOfMAreNotOverloaded}(DiamondFigure, contains, v)\\
 \wedge \mathtt{AllInvokedMethodsWithParameterOInBodyOfMAreNotOverloaded}(RectangleFigure, contains, v)\\
 \wedge \mathtt{AllInvokedMethodsWithParameterOInBodyOfMAreNotOverloaded}(RoundRectangleFigure, contains, v)\\
 \wedge \mathtt{AllInvokedMethodsWithParameterOInBodyOfMAreNotOverloaded}(TriangleFigure, contains, v)\\
 \wedge \mathtt{AllInvokedMethodsWithParameterOInBodyOfMAreNotOverloaded}(TextFigure, contains, v)\\
 \wedge \mathtt{AllInvokedMethodsWithParameterOInBodyOfMAreNotOverloaded}(BezierFigure, contains, v)\\
 \wedge \mathtt{AllInvokedMethodsWithParameterOInBodyOfMAreNotOverloaded}(TextAreaFigure, contains, v)\\
 \wedge \mathtt{AllInvokedMethodsWithParameterOInBodyOfMAreNotOverloaded}(NodeFigure, contains, v)\\
 \wedge \mathtt{AllInvokedMethodsWithParameterOInBodyOfMAreNotOverloaded}(SVGImage, contains, v)\\
 \wedge \mathtt{AllInvokedMethodsWithParameterOInBodyOfMAreNotOverloaded}(SVGPath, contains, v)\\
 \wedge \mathtt{AllInvokedMethodsWithParameterOInBodyOfMAreNotOverloaded}(DependencyFigure, contains, v)\\
 \wedge \mathtt{AllInvokedMethodsWithParameterOInBodyOfMAreNotOverloaded}(LineConnectionFigure, contains, v)\\
 \wedge \mathtt{AllInvokedMethodsWithParameterOInBodyOfMAreNotOverloaded}(AbstractFigure, basicTransform, v)\\
 \wedge \mathtt{AllInvokedMethodsWithParameterOInBodyOfMAreNotOverloaded}(LabeledLineConnectionFigure, basicTransform, v)\\
 \wedge \mathtt{AllInvokedMethodsWithParameterOInBodyOfMAreNotOverloaded}(AbstractCompositeFigure, basicTransform, v)\\
 \wedge \mathtt{AllInvokedMethodsWithParameterOInBodyOfMAreNotOverloaded}(GraphicalCompositeFigure, basicTransform, v)\\
 \wedge \mathtt{AllInvokedMethodsWithParameterOInBodyOfMAreNotOverloaded}(EllipseFigure, basicTransform, v)\\
 \wedge \mathtt{AllInvokedMethodsWithParameterOInBodyOfMAreNotOverloaded}(DiamondFigure, basicTransform, v)\\
 \wedge \mathtt{AllInvokedMethodsWithParameterOInBodyOfMAreNotOverloaded}(RectangleFigure, basicTransform, v)\\
 \wedge \mathtt{AllInvokedMethodsWithParameterOInBodyOfMAreNotOverloaded}(RoundRectangleFigure, basicTransform, v)\\
 \wedge \mathtt{AllInvokedMethodsWithParameterOInBodyOfMAreNotOverloaded}(TriangleFigure, basicTransform, v)\\
 \wedge \mathtt{AllInvokedMethodsWithParameterOInBodyOfMAreNotOverloaded}(TextFigure, basicTransform, v)\\
 \wedge \mathtt{AllInvokedMethodsWithParameterOInBodyOfMAreNotOverloaded}(BezierFigure, basicTransform, v)\\
 \wedge \mathtt{AllInvokedMethodsWithParameterOInBodyOfMAreNotOverloaded}(TextAreaFigure, basicTransform, v)\\
 \wedge \mathtt{AllInvokedMethodsWithParameterOInBodyOfMAreNotOverloaded}(NodeFigure, basicTransform, v)\\
 \wedge \mathtt{AllInvokedMethodsWithParameterOInBodyOfMAreNotOverloaded}(SVGImage, basicTransform, v)\\
 \wedge \mathtt{AllInvokedMethodsWithParameterOInBodyOfMAreNotOverloaded}(SVGPath, basicTransform, v)\\
 \wedge \mathtt{AllInvokedMethodsWithParameterOInBodyOfMAreNotOverloaded}(DependencyFigure, basicTransform, v)\\
 \wedge \mathtt{AllInvokedMethodsWithParameterOInBodyOfMAreNotOverloaded}(LineConnectionFigure, basicTransform, v)\\
 \wedge \neg \mathtt{IsPrivate}(RemoveNotifyVisitor, visit)\\
 \wedge \neg \mathtt{HasParameterType}(RemoveNotifyVisitor, Void)\\
 \wedge \neg \mathtt{IsPrivate}(AddNotifyVisitor, visit)\\
 \wedge \neg \mathtt{HasParameterType}(AddNotifyVisitor, Void)\\
 \wedge \neg \mathtt{IsPrimitiveType}(Figure)\\
 \wedge \neg \mathtt{IsPrivate}(FindFigureInsideVisitor, visit)\\
 \wedge \neg \mathtt{HasParameterType}(FindFigureInsideVisitor, Figure)\\
 \wedge \neg \mathtt{IsPrivate}(SetAttributeVisitor, visit)\\
 \wedge \neg \mathtt{HasParameterType}(SetAttributeVisitor, Void)\\
 \wedge \neg \mathtt{IsPrimitiveType}(Boolean)\\
 \wedge \neg \mathtt{IsPrivate}(ContainsVisitor, visit)\\
 \wedge \neg \mathtt{HasParameterType}(ContainsVisitor, Boolean)\\
 \wedge \neg \mathtt{ExistsAbstractMethod}(Visitor, visit)\\
 \wedge \neg \mathtt{IsPrimitiveType}(Void)\\
 \wedge \neg \mathtt{IsPrivate}(BasicTransformVisitor, visit)\\
 \wedge \neg \mathtt{HasParameterType}(BasicTransformVisitor, Void)\\
 \wedge \neg \mathtt{ExistsType}(Visitor)\\
 \wedge \mathtt{ExistsClass}(SVGImage)\\
 \wedge \mathtt{ExistsClass}(TextAreaFigure)\\
 \wedge \mathtt{ExistsClass}(BezierFigure)\\
 \wedge \mathtt{ExistsClass}(TextFigure)\\
 \wedge \mathtt{ExistsClass}(TriangleFigure)\\
 \wedge \mathtt{ExistsClass}(RoundRectangleFigure)\\
 \wedge \mathtt{ExistsClass}(RectangleFigure)\\
 \wedge \mathtt{ExistsClass}(DiamondFigure)\\
 \wedge \mathtt{ExistsClass}(EllipseFigure)\\
 \wedge \mathtt{ExistsClass}(AbstractCompositeFigure)\\
 \wedge \neg \mathtt{ExistsType}(RemoveNotifyVisitor)\\
 \wedge \mathtt{BoundVariableInMethodBody}(LabeledLineConnectionFigure, removeNotify, Drawing d)\\
 \wedge \mathtt{BoundVariableInMethodBody}(GraphicalCompositeFigure, removeNotify, Drawing d)\\
 \wedge \mathtt{BoundVariableInMethodBody}(EllipseFigure, removeNotify, Drawing d)\\
 \wedge \mathtt{BoundVariableInMethodBody}(DiamondFigure, removeNotify, Drawing d)\\
 \wedge \mathtt{BoundVariableInMethodBody}(RectangleFigure, removeNotify, Drawing d)\\
 \wedge \mathtt{BoundVariableInMethodBody}(RoundRectangleFigure, removeNotify, Drawing d)\\
 \wedge \mathtt{BoundVariableInMethodBody}(TriangleFigure, removeNotify, Drawing d)\\
 \wedge \mathtt{BoundVariableInMethodBody}(TextFigure, removeNotify, Drawing d)\\
 \wedge \mathtt{BoundVariableInMethodBody}(BezierFigure, removeNotify, Drawing d)\\
 \wedge \mathtt{BoundVariableInMethodBody}(TextAreaFigure, removeNotify, Drawing d)\\
 \wedge \mathtt{BoundVariableInMethodBody}(NodeFigure, removeNotify, Drawing d)\\
 \wedge \mathtt{BoundVariableInMethodBody}(SVGImage, removeNotify, Drawing d)\\
 \wedge \mathtt{BoundVariableInMethodBody}(AbstractCompositeFigure, removeNotify, Drawing d)\\
 \wedge \mathtt{BoundVariableInMethodBody}(DependencyFigure, removeNotify, Drawing d)\\
 \wedge \mathtt{BoundVariableInMethodBody}(LineConnectionFigure, removeNotify, Drawing d)\\
 \wedge \neg \mathtt{ExistsType}(AddNotifyVisitor)\\
 \wedge \mathtt{BoundVariableInMethodBody}(LabeledLineConnectionFigure, addNotify, Drawing d)\\
 \wedge \mathtt{BoundVariableInMethodBody}(GraphicalCompositeFigure, addNotify, Drawing d)\\
 \wedge \mathtt{BoundVariableInMethodBody}(EllipseFigure, addNotify, Drawing d)\\
 \wedge \mathtt{BoundVariableInMethodBody}(DiamondFigure, addNotify, Drawing d)\\
 \wedge \mathtt{BoundVariableInMethodBody}(RectangleFigure, addNotify, Drawing d)\\
 \wedge \mathtt{BoundVariableInMethodBody}(RoundRectangleFigure, addNotify, Drawing d)\\
 \wedge \mathtt{BoundVariableInMethodBody}(TriangleFigure, addNotify, Drawing d)\\
 \wedge \mathtt{BoundVariableInMethodBody}(BezierFigure, addNotify, Drawing d)\\
 \wedge \mathtt{BoundVariableInMethodBody}(TextAreaFigure, addNotify, Drawing d)\\
 \wedge \mathtt{BoundVariableInMethodBody}(TextFigure, addNotify, Drawing d)\\
 \wedge \mathtt{BoundVariableInMethodBody}(SVGImage, addNotify, Drawing d)\\
 \wedge \mathtt{BoundVariableInMethodBody}(AbstractCompositeFigure, addNotify, Drawing d)\\
 \wedge \mathtt{BoundVariableInMethodBody}(LineConnectionFigure, addNotify, Drawing d)\\
 \wedge \neg \mathtt{ExistsType}(FindFigureInsideVisitor)\\
 \wedge \mathtt{BoundVariableInMethodBody}(EllipseFigure, findFigureInside, Point2D.Double p)\\
 \wedge \mathtt{BoundVariableInMethodBody}(DiamondFigure, findFigureInside, Point2D.Double p)\\
 \wedge \mathtt{BoundVariableInMethodBody}(RectangleFigure, findFigureInside, Point2D.Double p)\\
 \wedge \mathtt{BoundVariableInMethodBody}(RoundRectangleFigure, findFigureInside, Point2D.Double p)\\
 \wedge \mathtt{BoundVariableInMethodBody}(TriangleFigure, findFigureInside, Point2D.Double p)\\
 \wedge \mathtt{BoundVariableInMethodBody}(TextAreaFigure, findFigureInside, Point2D.Double p)\\
 \wedge \mathtt{BoundVariableInMethodBody}(TextFigure, findFigureInside, Point2D.Double p)\\
 \wedge \mathtt{BoundVariableInMethodBody}(SVGImage, findFigureInside, Point2D.Double p)\\
 \wedge \mathtt{BoundVariableInMethodBody}(AbstractCompositeFigure, findFigureInside, Point2D.Double p)\\
 \wedge \mathtt{BoundVariableInMethodBody}(BezierFigure, findFigureInside, Point2D.Double p)\\
 \wedge \neg \mathtt{ExistsType}(SetAttributeVisitor)\\
 \wedge \mathtt{BoundVariableInMethodBody}(AbstractCompositeFigure, setAttribute, AttributeKey key)\\
 \wedge \mathtt{BoundVariableInMethodBody}(AbstractCompositeFigure, setAttribute, Object value)\\
 \wedge \mathtt{BoundVariableInMethodBody}(GraphicalCompositeFigure, setAttribute, AttributeKey key)\\
 \wedge \mathtt{BoundVariableInMethodBody}(GraphicalCompositeFigure, setAttribute, Object value)\\
 \wedge \mathtt{BoundVariableInMethodBody}(EllipseFigure, setAttribute, AttributeKey key)\\
 \wedge \mathtt{BoundVariableInMethodBody}(EllipseFigure, setAttribute, Object value)\\
 \wedge \mathtt{BoundVariableInMethodBody}(DiamondFigure, setAttribute, AttributeKey key)\\
 \wedge \mathtt{BoundVariableInMethodBody}(DiamondFigure, setAttribute, Object value)\\
 \wedge \mathtt{BoundVariableInMethodBody}(RectangleFigure, setAttribute, AttributeKey key)\\
 \wedge \mathtt{BoundVariableInMethodBody}(RectangleFigure, setAttribute, Object value)\\
 \wedge \mathtt{BoundVariableInMethodBody}(RoundRectangleFigure, setAttribute, AttributeKey key)\\
 \wedge \mathtt{BoundVariableInMethodBody}(RoundRectangleFigure, setAttribute, Object value)\\
 \wedge \mathtt{BoundVariableInMethodBody}(TriangleFigure, setAttribute, AttributeKey key)\\
 \wedge \mathtt{BoundVariableInMethodBody}(TriangleFigure, setAttribute, Object value)\\
 \wedge \mathtt{BoundVariableInMethodBody}(TextAreaFigure, setAttribute, AttributeKey key)\\
 \wedge \mathtt{BoundVariableInMethodBody}(TextAreaFigure, setAttribute, Object value)\\
 \wedge \mathtt{BoundVariableInMethodBody}(TextFigure, setAttribute, AttributeKey key)\\
 \wedge \mathtt{BoundVariableInMethodBody}(TextFigure, setAttribute, Object value)\\
 \wedge \mathtt{BoundVariableInMethodBody}(SVGImage, setAttribute, AttributeKey key)\\
 \wedge \mathtt{BoundVariableInMethodBody}(SVGImage, setAttribute, Object value)\\
 \wedge \mathtt{BoundVariableInMethodBody}(SVGPath, setAttribute, AttributeKey key)\\
 \wedge \mathtt{BoundVariableInMethodBody}(SVGPath, setAttribute, Object value)\\
 \wedge \mathtt{BoundVariableInMethodBody}(BezierFigure, setAttribute, AttributeKey key)\\
 \wedge \mathtt{BoundVariableInMethodBody}(BezierFigure, setAttribute, Object value)\\
 \wedge \neg \mathtt{ExistsType}(ContainsVisitor)\\
 \wedge \mathtt{BoundVariableInMethodBody}(GraphicalCompositeFigure, contains, Point2D.Double p)\\
 \wedge \mathtt{BoundVariableInMethodBody}(EllipseFigure, contains, Point2D.Double p)\\
 \wedge \mathtt{BoundVariableInMethodBody}(DiamondFigure, contains, Point2D.Double p)\\
 \wedge \mathtt{BoundVariableInMethodBody}(RectangleFigure, contains, Point2D.Double p)\\
 \wedge \mathtt{BoundVariableInMethodBody}(RoundRectangleFigure, contains, Point2D.Double p)\\
 \wedge \mathtt{BoundVariableInMethodBody}(TriangleFigure, contains, Point2D.Double p)\\
 \wedge \mathtt{BoundVariableInMethodBody}(TextAreaFigure, contains, Point2D.Double p)\\
 \wedge \mathtt{BoundVariableInMethodBody}(TextFigure, contains, Point2D.Double p)\\
 \wedge \mathtt{BoundVariableInMethodBody}(SVGImage, contains, Point2D.Double p)\\
 \wedge \mathtt{BoundVariableInMethodBody}(AbstractCompositeFigure, contains, Point2D.Double p)\\
 \wedge \mathtt{BoundVariableInMethodBody}(BezierFigure, contains, Point2D.Double p)\\
 \wedge \mathtt{ExistsType}(AbstractFigure)\\
 \wedge \neg \mathtt{ExistsType}(BasicTransformVisitor)\\
 \wedge \mathtt{BoundVariableInMethodBody}(AbstractCompositeFigure, basicTransform, AffineTransform tx)\\
 \wedge \mathtt{BoundVariableInMethodBody}(GraphicalCompositeFigure, basicTransform, AffineTransform tx)\\
 \wedge \mathtt{BoundVariableInMethodBody}(EllipseFigure, basicTransform, AffineTransform tx)\\
 \wedge \mathtt{BoundVariableInMethodBody}(DiamondFigure, basicTransform, AffineTransform tx)\\
 \wedge \mathtt{BoundVariableInMethodBody}(RectangleFigure, basicTransform, AffineTransform tx)\\
 \wedge \mathtt{BoundVariableInMethodBody}(RoundRectangleFigure, basicTransform, AffineTransform tx)\\
 \wedge \mathtt{BoundVariableInMethodBody}(TriangleFigure, basicTransform, AffineTransform tx)\\
 \wedge \mathtt{BoundVariableInMethodBody}(BezierFigure, basicTransform, AffineTransform tx)\\
 \wedge \mathtt{BoundVariableInMethodBody}(TextAreaFigure, basicTransform, AffineTransform tx)\\
 \wedge \mathtt{BoundVariableInMethodBody}(TextFigure, basicTransform, AffineTransform tx)\\
 \wedge \mathtt{BoundVariableInMethodBody}(SVGImage, basicTransform, AffineTransform tx)\\
 \wedge \mathtt{BoundVariableInMethodBody}(SVGPath, basicTransform, AffineTransform tx)\\
 \wedge \mathtt{BoundVariableInMethodBody}(LineConnectionFigure, basicTransform, AffineTransform tx)\\
 \wedge \mathtt{ExistsMethodDefinitionWithParams}(AbstractFigure, removeNotify, [Drawing d])\\
 \wedge \mathtt{ExistsAbstractMethod}(AbstractFigure, removeNotify)\\
 \wedge \neg \mathtt{IsInheritedMethod}(AbstractFigure, removeNotifyTmpVC)\\
 \wedge \neg \mathtt{IsInheritedMethodWithParams}(AbstractFigure, removeNotifyTmpVC, [Drawing d])\\
 \wedge \neg \mathtt{ExistsMethodDefinitionWithParams}(AbstractFigure, removeNotifyTmpVC, [Drawing d])\\
 \wedge \mathtt{HasReturnType}(AbstractFigure, removeNotify, Void)\\
 \wedge \mathtt{ExistsMethodDefinition}(AbstractFigure, removeNotify)\\
 \wedge \mathtt{ExistsMethodDefinition}(LabeledLineConnectionFigure, removeNotify)\\
 \wedge \mathtt{ExistsMethodDefinition}(GraphicalCompositeFigure, removeNotify)\\
 \wedge \mathtt{ExistsMethodDefinition}(EllipseFigure, removeNotify)\\
 \wedge \mathtt{ExistsMethodDefinition}(DiamondFigure, removeNotify)\\
 \wedge \mathtt{ExistsMethodDefinition}(RectangleFigure, removeNotify)\\
 \wedge \mathtt{ExistsMethodDefinition}(RoundRectangleFigure, removeNotify)\\
 \wedge \mathtt{ExistsMethodDefinition}(TriangleFigure, removeNotify)\\
 \wedge \mathtt{ExistsMethodDefinition}(TextFigure, removeNotify)\\
 \wedge \mathtt{ExistsMethodDefinition}(BezierFigure, removeNotify)\\
 \wedge \mathtt{ExistsMethodDefinition}(TextAreaFigure, removeNotify)\\
 \wedge \mathtt{ExistsMethodDefinition}(NodeFigure, removeNotify)\\
 \wedge \mathtt{ExistsMethodDefinition}(SVGImage, removeNotify)\\
 \wedge \mathtt{ExistsMethodDefinition}(DependencyFigure, removeNotify)\\
 \wedge \mathtt{ExistsMethodDefinition}(LineConnectionFigure, removeNotify)\\
 \wedge \neg \mathtt{ExistsMethodDefinition}(AbstractFigure, removeNotifyTmpVC)\\
 \wedge \neg \mathtt{ExistsMethodDefinition}(LabeledLineConnectionFigure, removeNotifyTmpVC)\\
 \wedge \neg \mathtt{ExistsMethodDefinition}(AbstractCompositeFigure, removeNotifyTmpVC)\\
 \wedge \neg \mathtt{ExistsMethodDefinition}(GraphicalCompositeFigure, removeNotifyTmpVC)\\
 \wedge \neg \mathtt{ExistsMethodDefinition}(EllipseFigure, removeNotifyTmpVC)\\
 \wedge \neg \mathtt{ExistsMethodDefinition}(DiamondFigure, removeNotifyTmpVC)\\
 \wedge \neg \mathtt{ExistsMethodDefinition}(RectangleFigure, removeNotifyTmpVC)\\
 \wedge \neg \mathtt{ExistsMethodDefinition}(RoundRectangleFigure, removeNotifyTmpVC)\\
 \wedge \neg \mathtt{ExistsMethodDefinition}(TriangleFigure, removeNotifyTmpVC)\\
 \wedge \neg \mathtt{ExistsMethodDefinition}(TextFigure, removeNotifyTmpVC)\\
 \wedge \neg \mathtt{ExistsMethodDefinition}(BezierFigure, removeNotifyTmpVC)\\
 \wedge \neg \mathtt{ExistsMethodDefinition}(TextAreaFigure, removeNotifyTmpVC)\\
 \wedge \neg \mathtt{ExistsMethodDefinition}(NodeFigure, removeNotifyTmpVC)\\
 \wedge \neg \mathtt{ExistsMethodDefinition}(SVGImage, removeNotifyTmpVC)\\
 \wedge \neg \mathtt{ExistsMethodDefinition}(SVGPath, removeNotifyTmpVC)\\
 \wedge \neg \mathtt{ExistsMethodDefinition}(DependencyFigure, removeNotifyTmpVC)\\
 \wedge \neg \mathtt{ExistsMethodDefinition}(LineConnectionFigure, removeNotifyTmpVC)\\
 \wedge \mathtt{ExistsMethodDefinitionWithParams}(AbstractFigure, addNotify, [Drawing d])\\
 \wedge \mathtt{ExistsAbstractMethod}(AbstractFigure, addNotify)\\
 \wedge \neg \mathtt{IsInheritedMethod}(AbstractFigure, addNotifyTmpVC)\\
 \wedge \neg \mathtt{IsInheritedMethodWithParams}(AbstractFigure, addNotifyTmpVC, [Drawing d])\\
 \wedge \neg \mathtt{ExistsMethodDefinitionWithParams}(AbstractFigure, addNotifyTmpVC, [Drawing d])\\
 \wedge \mathtt{HasReturnType}(AbstractFigure, addNotify, Void)\\
 \wedge \mathtt{ExistsMethodDefinition}(AbstractFigure, addNotify)\\
 \wedge \mathtt{ExistsMethodDefinition}(LabeledLineConnectionFigure, addNotify)\\
 \wedge \mathtt{ExistsMethodDefinition}(GraphicalCompositeFigure, addNotify)\\
 \wedge \mathtt{ExistsMethodDefinition}(EllipseFigure, addNotify)\\
 \wedge \mathtt{ExistsMethodDefinition}(DiamondFigure, addNotify)\\
 \wedge \mathtt{ExistsMethodDefinition}(RectangleFigure, addNotify)\\
 \wedge \mathtt{ExistsMethodDefinition}(RoundRectangleFigure, addNotify)\\
 \wedge \mathtt{ExistsMethodDefinition}(TriangleFigure, addNotify)\\
 \wedge \mathtt{ExistsMethodDefinition}(BezierFigure, addNotify)\\
 \wedge \mathtt{ExistsMethodDefinition}(TextAreaFigure, addNotify)\\
 \wedge \mathtt{ExistsMethodDefinition}(SVGImage, addNotify)\\
 \wedge \neg \mathtt{ExistsMethodDefinition}(AbstractFigure, addNotifyTmpVC)\\
 \wedge \neg \mathtt{ExistsMethodDefinition}(LabeledLineConnectionFigure, addNotifyTmpVC)\\
 \wedge \neg \mathtt{ExistsMethodDefinition}(AbstractCompositeFigure, addNotifyTmpVC)\\
 \wedge \neg \mathtt{ExistsMethodDefinition}(GraphicalCompositeFigure, addNotifyTmpVC)\\
 \wedge \neg \mathtt{ExistsMethodDefinition}(EllipseFigure, addNotifyTmpVC)\\
 \wedge \neg \mathtt{ExistsMethodDefinition}(DiamondFigure, addNotifyTmpVC)\\
 \wedge \neg \mathtt{ExistsMethodDefinition}(RectangleFigure, addNotifyTmpVC)\\
 \wedge \neg \mathtt{ExistsMethodDefinition}(RoundRectangleFigure, addNotifyTmpVC)\\
 \wedge \neg \mathtt{ExistsMethodDefinition}(TriangleFigure, addNotifyTmpVC)\\
 \wedge \neg \mathtt{ExistsMethodDefinition}(TextFigure, addNotifyTmpVC)\\
 \wedge \neg \mathtt{ExistsMethodDefinition}(BezierFigure, addNotifyTmpVC)\\
 \wedge \neg \mathtt{ExistsMethodDefinition}(TextAreaFigure, addNotifyTmpVC)\\
 \wedge \neg \mathtt{ExistsMethodDefinition}(NodeFigure, addNotifyTmpVC)\\
 \wedge \neg \mathtt{ExistsMethodDefinition}(SVGImage, addNotifyTmpVC)\\
 \wedge \neg \mathtt{ExistsMethodDefinition}(SVGPath, addNotifyTmpVC)\\
 \wedge \neg \mathtt{ExistsMethodDefinition}(DependencyFigure, addNotifyTmpVC)\\
 \wedge \neg \mathtt{ExistsMethodDefinition}(LineConnectionFigure, addNotifyTmpVC)\\
 \wedge \mathtt{ExistsMethodDefinitionWithParams}(AbstractFigure, findFigureInside, [Point2D.Double p])\\
 \wedge \mathtt{ExistsAbstractMethod}(AbstractFigure, findFigureInside)\\
 \wedge \neg \mathtt{IsInheritedMethod}(AbstractFigure, findFigureInsideTmpVC)\\
 \wedge \neg \mathtt{IsInheritedMethodWithParams}(AbstractFigure, findFigureInsideTmpVC, [Point2D.Double p])\\
 \wedge \neg \mathtt{ExistsMethodDefinitionWithParams}(AbstractFigure, findFigureInsideTmpVC, [Point2D.Double p])\\
 \wedge \mathtt{HasReturnType}(AbstractFigure, findFigureInside, Figure)\\
 \wedge \mathtt{ExistsMethodDefinition}(AbstractFigure, findFigureInside)\\
 \wedge \mathtt{ExistsMethodDefinition}(EllipseFigure, findFigureInside)\\
 \wedge \mathtt{ExistsMethodDefinition}(DiamondFigure, findFigureInside)\\
 \wedge \mathtt{ExistsMethodDefinition}(RectangleFigure, findFigureInside)\\
 \wedge \mathtt{ExistsMethodDefinition}(RoundRectangleFigure, findFigureInside)\\
 \wedge \mathtt{ExistsMethodDefinition}(TriangleFigure, findFigureInside)\\
 \wedge \mathtt{ExistsMethodDefinition}(TextAreaFigure, findFigureInside)\\
 \wedge \mathtt{ExistsMethodDefinition}(SVGImage, findFigureInside)\\
 \wedge \neg \mathtt{ExistsMethodDefinition}(AbstractFigure, findFigureInsideTmpVC)\\
 \wedge \neg \mathtt{ExistsMethodDefinition}(LabeledLineConnectionFigure, findFigureInsideTmpVC)\\
 \wedge \neg \mathtt{ExistsMethodDefinition}(AbstractCompositeFigure, findFigureInsideTmpVC)\\
 \wedge \neg \mathtt{ExistsMethodDefinition}(GraphicalCompositeFigure, findFigureInsideTmpVC)\\
 \wedge \neg \mathtt{ExistsMethodDefinition}(EllipseFigure, findFigureInsideTmpVC)\\
 \wedge \neg \mathtt{ExistsMethodDefinition}(DiamondFigure, findFigureInsideTmpVC)\\
 \wedge \neg \mathtt{ExistsMethodDefinition}(RectangleFigure, findFigureInsideTmpVC)\\
 \wedge \neg \mathtt{ExistsMethodDefinition}(RoundRectangleFigure, findFigureInsideTmpVC)\\
 \wedge \neg \mathtt{ExistsMethodDefinition}(TriangleFigure, findFigureInsideTmpVC)\\
 \wedge \neg \mathtt{ExistsMethodDefinition}(TextFigure, findFigureInsideTmpVC)\\
 \wedge \neg \mathtt{ExistsMethodDefinition}(BezierFigure, findFigureInsideTmpVC)\\
 \wedge \neg \mathtt{ExistsMethodDefinition}(TextAreaFigure, findFigureInsideTmpVC)\\
 \wedge \neg \mathtt{ExistsMethodDefinition}(NodeFigure, findFigureInsideTmpVC)\\
 \wedge \neg \mathtt{ExistsMethodDefinition}(SVGImage, findFigureInsideTmpVC)\\
 \wedge \neg \mathtt{ExistsMethodDefinition}(SVGPath, findFigureInsideTmpVC)\\
 \wedge \neg \mathtt{ExistsMethodDefinition}(DependencyFigure, findFigureInsideTmpVC)\\
 \wedge \neg \mathtt{ExistsMethodDefinition}(LineConnectionFigure, findFigureInsideTmpVC)\\
 \wedge \mathtt{ExistsMethodDefinitionWithParams}(AbstractFigure, setAttribute, [AttributeKey key; Object value])\\
 \wedge \mathtt{ExistsAbstractMethod}(AbstractFigure, setAttribute)\\
 \wedge \neg \mathtt{IsInheritedMethod}(AbstractFigure, setAttributeTmpVC)\\
 \wedge \neg \mathtt{IsInheritedMethodWithParams}(AbstractFigure, setAttributeTmpVC, [AttributeKey key; Object value])\\
 \wedge \neg \mathtt{ExistsMethodDefinitionWithParams}(AbstractFigure, setAttributeTmpVC, [AttributeKey key; Object value])\\
 \wedge \mathtt{HasReturnType}(AbstractFigure, setAttribute, Void)\\
 \wedge \mathtt{ExistsMethodDefinition}(AbstractFigure, setAttribute)\\
 \wedge \mathtt{ExistsMethodDefinition}(AbstractCompositeFigure, setAttribute)\\
 \wedge \mathtt{ExistsMethodDefinition}(GraphicalCompositeFigure, setAttribute)\\
 \wedge \mathtt{ExistsMethodDefinition}(EllipseFigure, setAttribute)\\
 \wedge \mathtt{ExistsMethodDefinition}(DiamondFigure, setAttribute)\\
 \wedge \mathtt{ExistsMethodDefinition}(RectangleFigure, setAttribute)\\
 \wedge \mathtt{ExistsMethodDefinition}(RoundRectangleFigure, setAttribute)\\
 \wedge \mathtt{ExistsMethodDefinition}(TriangleFigure, setAttribute)\\
 \wedge \mathtt{ExistsMethodDefinition}(TextAreaFigure, setAttribute)\\
 \wedge \mathtt{ExistsMethodDefinition}(SVGImage, setAttribute)\\
 \wedge \mathtt{ExistsMethodDefinition}(SVGPath, setAttribute)\\
 \wedge \neg \mathtt{ExistsMethodDefinition}(AbstractFigure, setAttributeTmpVC)\\
 \wedge \neg \mathtt{ExistsMethodDefinition}(LabeledLineConnectionFigure, setAttributeTmpVC)\\
 \wedge \neg \mathtt{ExistsMethodDefinition}(AbstractCompositeFigure, setAttributeTmpVC)\\
 \wedge \neg \mathtt{ExistsMethodDefinition}(GraphicalCompositeFigure, setAttributeTmpVC)\\
 \wedge \neg \mathtt{ExistsMethodDefinition}(EllipseFigure, setAttributeTmpVC)\\
 \wedge \neg \mathtt{ExistsMethodDefinition}(DiamondFigure, setAttributeTmpVC)\\
 \wedge \neg \mathtt{ExistsMethodDefinition}(RectangleFigure, setAttributeTmpVC)\\
 \wedge \neg \mathtt{ExistsMethodDefinition}(RoundRectangleFigure, setAttributeTmpVC)\\
 \wedge \neg \mathtt{ExistsMethodDefinition}(TriangleFigure, setAttributeTmpVC)\\
 \wedge \neg \mathtt{ExistsMethodDefinition}(TextFigure, setAttributeTmpVC)\\
 \wedge \neg \mathtt{ExistsMethodDefinition}(BezierFigure, setAttributeTmpVC)\\
 \wedge \neg \mathtt{ExistsMethodDefinition}(TextAreaFigure, setAttributeTmpVC)\\
 \wedge \neg \mathtt{ExistsMethodDefinition}(NodeFigure, setAttributeTmpVC)\\
 \wedge \neg \mathtt{ExistsMethodDefinition}(SVGImage, setAttributeTmpVC)\\
 \wedge \neg \mathtt{ExistsMethodDefinition}(SVGPath, setAttributeTmpVC)\\
 \wedge \neg \mathtt{ExistsMethodDefinition}(DependencyFigure, setAttributeTmpVC)\\
 \wedge \neg \mathtt{ExistsMethodDefinition}(LineConnectionFigure, setAttributeTmpVC)\\
 \wedge \mathtt{ExistsMethodDefinitionWithParams}(AbstractFigure, contains, [Point2D.Double p])\\
 \wedge \mathtt{ExistsAbstractMethod}(AbstractFigure, contains)\\
 \wedge \neg \mathtt{IsInheritedMethod}(AbstractFigure, containsTmpVC)\\
 \wedge \neg \mathtt{IsInheritedMethodWithParams}(AbstractFigure, containsTmpVC, [Point2D.Double p])\\
 \wedge \neg \mathtt{ExistsMethodDefinitionWithParams}(AbstractFigure, containsTmpVC, [Point2D.Double p])\\
 \wedge \mathtt{HasReturnType}(AbstractFigure, contains, Boolean)\\
 \wedge \mathtt{ExistsMethodDefinition}(AbstractFigure, contains)\\
 \wedge \mathtt{ExistsMethodDefinition}(GraphicalCompositeFigure, contains)\\
 \wedge \mathtt{ExistsMethodDefinition}(EllipseFigure, contains)\\
 \wedge \mathtt{ExistsMethodDefinition}(DiamondFigure, contains)\\
 \wedge \mathtt{ExistsMethodDefinition}(RectangleFigure, contains)\\
 \wedge \mathtt{ExistsMethodDefinition}(RoundRectangleFigure, contains)\\
 \wedge \mathtt{ExistsMethodDefinition}(TriangleFigure, contains)\\
 \wedge \mathtt{ExistsMethodDefinition}(TextAreaFigure, contains)\\
 \wedge \mathtt{ExistsMethodDefinition}(SVGImage, contains)\\
 \wedge \neg \mathtt{ExistsMethodDefinition}(AbstractFigure, containsTmpVC)\\
 \wedge \neg \mathtt{ExistsMethodDefinition}(LabeledLineConnectionFigure, containsTmpVC)\\
 \wedge \neg \mathtt{ExistsMethodDefinition}(AbstractCompositeFigure, containsTmpVC)\\
 \wedge \neg \mathtt{ExistsMethodDefinition}(GraphicalCompositeFigure, containsTmpVC)\\
 \wedge \neg \mathtt{ExistsMethodDefinition}(EllipseFigure, containsTmpVC)\\
 \wedge \neg \mathtt{ExistsMethodDefinition}(DiamondFigure, containsTmpVC)\\
 \wedge \neg \mathtt{ExistsMethodDefinition}(RectangleFigure, containsTmpVC)\\
 \wedge \neg \mathtt{ExistsMethodDefinition}(RoundRectangleFigure, containsTmpVC)\\
 \wedge \neg \mathtt{ExistsMethodDefinition}(TriangleFigure, containsTmpVC)\\
 \wedge \neg \mathtt{ExistsMethodDefinition}(TextFigure, containsTmpVC)\\
 \wedge \neg \mathtt{ExistsMethodDefinition}(BezierFigure, containsTmpVC)\\
 \wedge \neg \mathtt{ExistsMethodDefinition}(TextAreaFigure, containsTmpVC)\\
 \wedge \neg \mathtt{ExistsMethodDefinition}(NodeFigure, containsTmpVC)\\
 \wedge \neg \mathtt{ExistsMethodDefinition}(SVGImage, containsTmpVC)\\
 \wedge \neg \mathtt{ExistsMethodDefinition}(SVGPath, containsTmpVC)\\
 \wedge \neg \mathtt{ExistsMethodDefinition}(DependencyFigure, containsTmpVC)\\
 \wedge \neg \mathtt{ExistsMethodDefinition}(LineConnectionFigure, containsTmpVC)\\
 \wedge \mathtt{ExistsClass}(AbstractFigure)\\
 \wedge \mathtt{IsAbstractClass}(AbstractFigure)\\
 \wedge \mathtt{ExistsMethodDefinitionWithParams}(AbstractFigure, basicTransform, [AffineTransform tx])\\
 \wedge \mathtt{ExistsAbstractMethod}(AbstractFigure, basicTransform)\\
 \wedge \neg \mathtt{IsInheritedMethod}(AbstractFigure, basicTransformTmpVC)\\
 \wedge \neg \mathtt{IsInheritedMethodWithParams}(AbstractFigure, basicTransformTmpVC, [AffineTransform tx])\\
 \wedge \neg \mathtt{ExistsMethodDefinitionWithParams}(AbstractFigure, basicTransformTmpVC, [AffineTransform tx])\\
 \wedge \mathtt{AllSubclasses}(AbstractFigure, [LabeledLineConnectionFigure; AbstractCompositeFigure; GraphicalCompositeFigure; EllipseFigure; DiamondFigure; RectangleFigure; RoundRectangleFigure; TriangleFigure; TextFigure; BezierFigure; TextAreaFigure; NodeFigure; SVGImage; SVGPath; DependencyFigure; LineConnectionFigure])\\
 \wedge \mathtt{HasReturnType}(AbstractFigure, basicTransform, Void)\\
 \wedge \neg \mathtt{IsPrivate}(AbstractFigure, basicTransform)\\
 \wedge \neg \mathtt{IsPrivate}(LabeledLineConnectionFigure, basicTransform)\\
 \wedge \neg \mathtt{IsPrivate}(AbstractCompositeFigure, basicTransform)\\
 \wedge \neg \mathtt{IsPrivate}(GraphicalCompositeFigure, basicTransform)\\
 \wedge \neg \mathtt{IsPrivate}(EllipseFigure, basicTransform)\\
 \wedge \neg \mathtt{IsPrivate}(DiamondFigure, basicTransform)\\
 \wedge \neg \mathtt{IsPrivate}(RectangleFigure, basicTransform)\\
 \wedge \neg \mathtt{IsPrivate}(RoundRectangleFigure, basicTransform)\\
 \wedge \neg \mathtt{IsPrivate}(TriangleFigure, basicTransform)\\
 \wedge \neg \mathtt{IsPrivate}(TextFigure, basicTransform)\\
 \wedge \neg \mathtt{IsPrivate}(BezierFigure, basicTransform)\\
 \wedge \neg \mathtt{IsPrivate}(TextAreaFigure, basicTransform)\\
 \wedge \neg \mathtt{IsPrivate}(NodeFigure, basicTransform)\\
 \wedge \neg \mathtt{IsPrivate}(SVGImage, basicTransform)\\
 \wedge \neg \mathtt{IsPrivate}(SVGPath, basicTransform)\\
 \wedge \neg \mathtt{IsPrivate}(DependencyFigure, basicTransform)\\
 \wedge \neg \mathtt{IsPrivate}(LineConnectionFigure, basicTransform)\\
 \wedge \mathtt{ExistsMethodDefinition}(AbstractFigure, basicTransform)\\
 \wedge \mathtt{ExistsMethodDefinition}(AbstractCompositeFigure, basicTransform)\\
 \wedge \mathtt{ExistsMethodDefinition}(GraphicalCompositeFigure, basicTransform)\\
 \wedge \mathtt{ExistsMethodDefinition}(EllipseFigure, basicTransform)\\
 \wedge \mathtt{ExistsMethodDefinition}(DiamondFigure, basicTransform)\\
 \wedge \mathtt{ExistsMethodDefinition}(RectangleFigure, basicTransform)\\
 \wedge \mathtt{ExistsMethodDefinition}(RoundRectangleFigure, basicTransform)\\
 \wedge \mathtt{ExistsMethodDefinition}(TriangleFigure, basicTransform)\\
 \wedge \mathtt{ExistsMethodDefinition}(TextAreaFigure, basicTransform)\\
 \wedge \mathtt{ExistsMethodDefinition}(SVGImage, basicTransform)\\
 \wedge \mathtt{ExistsMethodDefinition}(SVGPath, basicTransform)\\
 \wedge \neg \mathtt{ExistsMethodDefinition}(AbstractFigure, basicTransformTmpVC)\\
 \wedge \neg \mathtt{ExistsMethodDefinition}(LabeledLineConnectionFigure, basicTransformTmpVC)\\
 \wedge \neg \mathtt{ExistsMethodDefinition}(AbstractCompositeFigure, basicTransformTmpVC)\\
 \wedge \neg \mathtt{ExistsMethodDefinition}(GraphicalCompositeFigure, basicTransformTmpVC)\\
 \wedge \neg \mathtt{ExistsMethodDefinition}(EllipseFigure, basicTransformTmpVC)\\
 \wedge \neg \mathtt{ExistsMethodDefinition}(DiamondFigure, basicTransformTmpVC)\\
 \wedge \neg \mathtt{ExistsMethodDefinition}(RectangleFigure, basicTransformTmpVC)\\
 \wedge \neg \mathtt{ExistsMethodDefinition}(RoundRectangleFigure, basicTransformTmpVC)\\
 \wedge \neg \mathtt{ExistsMethodDefinition}(TriangleFigure, basicTransformTmpVC)\\
 \wedge \neg \mathtt{ExistsMethodDefinition}(TextFigure, basicTransformTmpVC)\\
 \wedge \neg \mathtt{ExistsMethodDefinition}(BezierFigure, basicTransformTmpVC)\\
 \wedge \neg \mathtt{ExistsMethodDefinition}(TextAreaFigure, basicTransformTmpVC)\\
 \wedge \neg \mathtt{ExistsMethodDefinition}(NodeFigure, basicTransformTmpVC)\\
 \wedge \neg \mathtt{ExistsMethodDefinition}(SVGImage, basicTransformTmpVC)\\
 \wedge \neg \mathtt{ExistsMethodDefinition}(SVGPath, basicTransformTmpVC)\\
 \wedge \neg \mathtt{ExistsMethodDefinition}(DependencyFigure, basicTransformTmpVC)\\
 \wedge \neg \mathtt{ExistsMethodDefinition}(LineConnectionFigure, basicTransformTmpVC)\\
 \wedge \mathtt{ExistsType}(GraphicalCompositeFigure)\\
 \wedge \mathtt{ExistsClass}(GraphicalCompositeFigure)\\
 \wedge \mathtt{IsSubType}(GraphicalCompositeFigure, AbstractCompositeFigure)\\
 \wedge \neg \mathtt{ExistsMethodDefinition}(GraphicalCompositeFigure, findFigureInside)\\
 \wedge \mathtt{ExistsType}(NodeFigure)\\
 \wedge \mathtt{ExistsClass}(NodeFigure)\\
 \wedge \mathtt{IsSubType}(NodeFigure, TextFigure)\\
 \wedge \neg \mathtt{ExistsMethodDefinition}(NodeFigure, addNotify)\\
 \wedge \neg \mathtt{ExistsMethodDefinition}(NodeFigure, basicTransform)\\
 \wedge \neg \mathtt{ExistsMethodDefinition}(NodeFigure, setAttribute)\\
 \wedge \neg \mathtt{ExistsMethodDefinition}(NodeFigure, findFigureInside)\\
 \wedge \neg \mathtt{ExistsMethodDefinition}(NodeFigure, contains)\\
 \wedge \mathtt{ExistsMethodDefinition}(TextFigure, addNotify)\\
 \wedge \mathtt{ExistsMethodDefinition}(TextFigure, basicTransform)\\
 \wedge \mathtt{ExistsMethodDefinition}(TextFigure, setAttribute)\\
 \wedge \mathtt{ExistsMethodDefinition}(TextFigure, findFigureInside)\\
 \wedge \mathtt{ExistsMethodDefinition}(TextFigure, contains)\\
 \wedge \mathtt{AllInvokedMethodsWithParameterOInBodyOfMAreNotOverloaded}(TextFigure, addNotify, this)\\
 \wedge \mathtt{AllInvokedMethodsWithParameterOInBodyOfMAreNotOverloaded}(TextFigure, basicTransform, this)\\
 \wedge \mathtt{AllInvokedMethodsWithParameterOInBodyOfMAreNotOverloaded}(TextFigure, setAttribute, this)\\
 \wedge \mathtt{AllInvokedMethodsWithParameterOInBodyOfMAreNotOverloaded}(TextFigure, findFigureInside, this)\\
 \wedge \mathtt{AllInvokedMethodsWithParameterOInBodyOfMAreNotOverloaded}(TextFigure, contains, this)\\
 \wedge \mathtt{ExistsType}(DependencyFigure)\\
 \wedge \mathtt{ExistsClass}(DependencyFigure)\\
 \wedge \mathtt{IsSubType}(DependencyFigure, LineConnectionFigure)\\
 \wedge \neg \mathtt{ExistsMethodDefinition}(DependencyFigure, addNotify)\\
 \wedge \neg \mathtt{ExistsMethodDefinition}(DependencyFigure, basicTransform)\\
 \wedge \neg \mathtt{ExistsMethodDefinition}(DependencyFigure, setAttribute)\\
 \wedge \neg \mathtt{ExistsMethodDefinition}(DependencyFigure, findFigureInside)\\
 \wedge \neg \mathtt{ExistsMethodDefinition}(DependencyFigure, contains)\\
 \wedge \mathtt{ExistsMethodDefinition}(LineConnectionFigure, addNotify)\\
 \wedge \mathtt{ExistsMethodDefinition}(LineConnectionFigure, basicTransform)\\
 \wedge \mathtt{AllInvokedMethodsWithParameterOInBodyOfMAreNotOverloaded}(LineConnectionFigure, addNotify, this)\\
 \wedge \mathtt{AllInvokedMethodsWithParameterOInBodyOfMAreNotOverloaded}(LineConnectionFigure, basicTransform, this)\\
 \wedge \mathtt{ExistsType}(LabeledLineConnectionFigure)\\
 \wedge \mathtt{ExistsClass}(LabeledLineConnectionFigure)\\
 \wedge \mathtt{IsSubType}(LabeledLineConnectionFigure, BezierFigure)\\
 \wedge \neg \mathtt{ExistsMethodDefinition}(LabeledLineConnectionFigure, basicTransform)\\
 \wedge \neg \mathtt{ExistsMethodDefinition}(LabeledLineConnectionFigure, setAttribute)\\
 \wedge \neg \mathtt{ExistsMethodDefinition}(LabeledLineConnectionFigure, findFigureInside)\\
 \wedge \neg \mathtt{ExistsMethodDefinition}(LabeledLineConnectionFigure, contains)\\
 \wedge \mathtt{ExistsMethodDefinition}(BezierFigure, basicTransform)\\
 \wedge \mathtt{AllInvokedMethodsWithParameterOInBodyOfMAreNotOverloaded}(BezierFigure, basicTransform, this)\\
 \wedge \mathtt{ExistsType}(SVGPath)\\
 \wedge \mathtt{ExistsClass}(SVGPath)\\
 \wedge \mathtt{IsSubType}(SVGPath, AbstractCompositeFigure)\\
 \wedge \neg \mathtt{ExistsMethodDefinition}(SVGPath, addNotify)\\
 \wedge \neg \mathtt{ExistsMethodDefinition}(SVGPath, removeNotify)\\
 \wedge \neg \mathtt{ExistsMethodDefinition}(SVGPath, findFigureInside)\\
 \wedge \neg \mathtt{ExistsMethodDefinition}(SVGPath, contains)\\
 \wedge \mathtt{ExistsMethodDefinition}(AbstractCompositeFigure, addNotify)\\
 \wedge \mathtt{ExistsMethodDefinition}(AbstractCompositeFigure, removeNotify)\\
 \wedge \mathtt{ExistsMethodDefinition}(AbstractCompositeFigure, findFigureInside)\\
 \wedge \mathtt{ExistsMethodDefinition}(AbstractCompositeFigure, contains)\\
 \wedge \mathtt{AllInvokedMethodsWithParameterOInBodyOfMAreNotOverloaded}(AbstractCompositeFigure, addNotify, this)\\
 \wedge \mathtt{AllInvokedMethodsWithParameterOInBodyOfMAreNotOverloaded}(AbstractCompositeFigure, removeNotify, this)\\
 \wedge \mathtt{AllInvokedMethodsWithParameterOInBodyOfMAreNotOverloaded}(AbstractCompositeFigure, findFigureInside, this)\\
 \wedge \mathtt{AllInvokedMethodsWithParameterOInBodyOfMAreNotOverloaded}(AbstractCompositeFigure, contains, this)\\
 \wedge \mathtt{ExistsType}(LineConnectionFigure)\\
 \wedge \mathtt{ExistsClass}(LineConnectionFigure)\\
 \wedge \mathtt{IsSubType}(LineConnectionFigure, BezierFigure)\\
 \wedge \neg \mathtt{ExistsMethodDefinition}(LineConnectionFigure, findFigureInside)\\
 \wedge \neg \mathtt{ExistsMethodDefinition}(LineConnectionFigure, setAttribute)\\
 \wedge \neg \mathtt{ExistsMethodDefinition}(LineConnectionFigure, contains)\\
 \wedge \mathtt{ExistsMethodDefinition}(BezierFigure, findFigureInside)\\
 \wedge \mathtt{ExistsMethodDefinition}(BezierFigure, setAttribute)\\
 \wedge \mathtt{ExistsMethodDefinition}(BezierFigure, contains)\\
 \wedge \mathtt{AllInvokedMethodsWithParameterOInBodyOfMAreNotOverloaded}(BezierFigure, findFigureInside, this)\\
 \wedge \mathtt{AllInvokedMethodsWithParameterOInBodyOfMAreNotOverloaded}(BezierFigure, setAttribute, this)\\
 \wedge \mathtt{AllInvokedMethodsWithParameterOInBodyOfMAreNotOverloaded}(BezierFigure, contains, this)$

\end{document}